\documentclass[11pt,titlepage]{article}
\usepackage{amssymb}

\usepackage{graphicx}
\usepackage{amsmath}
\usepackage[dvips]{geometry}
\def\tprod{\mathop{\textstyle \prod }}%
\def\tsum{\mathop{\textstyle \sum }}%

\def\binom#1#2{{#1 \choose #2}}%
\def\tbinom#1#2{{\textstyle {#1 \choose #2}}}%
\def\QATOP#1#2{{#1 \atop #2}}%
\def\QOVERD#1#2#3#4{{#3 \overwithdelims#1#2 #4}}%
\def\QATOPD#1#2#3#4{{#3 \atopwithdelims#1#2 #4}}%
\newtheorem{theorem}{Theorem}

\newenvironment{proof}[1][Proof]{\textbf{#1.} }{\ \rule{0.5em}{0.5em}}
\input{tcilatex}
\begin{document}
\ifx\href\undefined\else\hypersetup{linktocpage=true}\fi
\title{Classical and Quantum Causality\\
in Quantum Field Theory\\
{\small or}\\
{\textit{\small The Quantum Universe}}}
\author{by Jonathan Simon Eakins, M.Sci.}
\date{Thesis submitted to the University of Nottingham\\
for the degree of Doctor of Philosophy, April 2004}
\maketitle

\begin{abstract}
\ \ \ \ Based on a number of experimentally verified physical observations,
it is argued that the standard principles of quantum mechanics should be
applied to the Universe as a whole. Thus, a paradigm is proposed in which
the entire Universe is represented by a pure state wavefunction contained in
a factorisable Hilbert space of enormous dimension, and where this
statevector is developed by successive applications of operators that
correspond to unitary rotations and Hermitian tests. Moreover, because by
definition the Universe contains everything, it is argued that these
operators must be chosen self-referentially; the overall dynamics of the
system is envisaged to be analogous to a gigantic, self-governing, quantum
computation. The issue of how the Universe could choose these operators
without requiring or referring to a fictitious external observer is
addressed, and this in turn rephrases and removes the traditional
Measurement Problem inherent in the Copenhagen interpretation of quantum
mechanics.

The processes by which conventional physics might be recovered from this
fundamental, mathematical and global description of reality are particularly
investigated. Specifically, it is demonstrated that by considering the
changing properties, separabilities and factorisations of both the state and
the operators as the Universe proceeds though a sequence of discrete
computations, familiar notions such as classical distinguishability,
particle physics, space, time, special relativity and endo-physical
experiments can all begin to emerge from the proposed picture. A
pregeometric vision of cosmology is therefore discussed, with all of physics
ultimately arising from the relationships occurring between the elements of
the underlying mathematical structure. The possible origins of observable
physics, including physical objects positioned at definite locations in an
arena of apparently continuous space and time, are consequently investigated
for a Universe that incorporates quantum theory as a fundamental feature.

Overall, a framework for quantum cosmology is introduced and explored which
attempts to account for the existence of time, space, matter and,
eventually, everything else in the Universe, from a physically consistent
perspective.

\begin{itemize}
\item[\textbf{Keywords:}]  \textit{Quantum cosmology, Quantum computation,
Pregeometry, Emergence, Factorisation and Entanglement, Qubit field theory,
Quantum Causal sets, Discrete time, Information Exchange, Subregisters,
Endo-physics, Self-Referential Quantum automata}.
\end{itemize}
\end{abstract}

\tableofcontents
\listoffigures

\newpage

\begin{equation*}
\end{equation*}

\bigskip

\bigskip

\begin{center}
{\Large Acknowledgements}

\bigskip

\bigskip
\end{center}

Due to his continual guidance, ideas, discussions and suggestions during the
entire course of my research, I first thank my supervisor, George
Jaroszkiewicz.

I would also especially like to thank my family for their neverending
support and encouragement, not only over the past few years of my Ph.D., but
also throughout the whole of my education and life.

On a further note, I am additionally grateful to Dr. Christopher Hide for
help with LATEX code and document formatting, and to Drs. Nikolaos Demiris,
Cornelius J. Griffin, Hilmar Hauer and Andrew Whitehead for illuminating
discussions on many different aspects of mathematics. Appreciation also to
Professor A. Sudbery and Professor V. Belavkin for constructive criticism
and comments on the content of the thesis.

Lastly, special thanks to all those friends, colleagues, acquaintances and
others who gave me the inspiration and often unexpected moments of insight
that made this what it is; if you don't know who you are, at least I do.

I acknowledge the EPSRC (UK) for a research studentship.

\newpage

\begin{equation*}
\end{equation*}

\bigskip

\bigskip

\begin{equation*}
\end{equation*}

\qquad \qquad \qquad \textit{To-morrow, and to-morrow, and to-morrow,}

\textit{\qquad \qquad \qquad Creeps in this petty pace from day to day,}

\textit{\qquad \qquad \qquad To the last syllable of recorded time;}

\bigskip

\qquad \qquad \qquad \qquad \qquad \qquad \qquad \qquad \qquad - \ \ \textbf{%
W. Shakespeare}

\qquad \qquad \qquad \qquad \qquad \qquad \qquad \qquad \qquad\ \ \ \ Macbeth

\bigskip

\bigskip \newpage

\section{Introduction}

\pagenumbering{arabic} \setcounter{page}{1} \bigskip \renewcommand{%
\theequation}{1.\arabic{equation}} \setcounter{equation}{0} %
\renewcommand{\thetheorem}{1.\arabic{theorem}} \setcounter{theorem}{0}

Time permeates just about every sphere of life. Indeed, human civilisations
are dominated by regularising and synchronising the events in the
surrounding World, and it is easy to argue that without agreeing on a common
standard of time, society would be unable to function in the way that it
does.

In fact, time has played a crucial role in all of Mankind's development.
From agricultural dependencies on the cyclicity of the Sun, to the machine
timings that structured the Industrial Revolution; from the variable `$t$'
present in most of the equations used in science and engineering, to the
parameter vital in rationalising the study of history; from the timetables
essential for efficient national and global travel, to the Time Machines of
literature and imagination; from the rhythms governing the lives of animals
and plants, to the cadences of music and speech; from the measurements
necessary for road safety laws, to the quantity specified with pinpoint
accuracy when coordinating extraterrestrial exploration; time plays a role
in everything.

Moreover, human life often appears obsessed with the passage of time. Modern
society frequently revolves around questions such as ``When is...?'', ``What
time did...?'', ``How long until...?'', and so on, and the ever present
threat of mortality seems to heighten the sense that time is a precious
commodity to be `saved'\ or `made the most of' wherever possible. Time is
something that employers buy, and the cautious bide. Time is something that
`waits for no man', but can appear to `fly', drag' or `stand still'. Time is
even something whose effects medical research attempts to `hold back'.

Even primitive Man recognised the importance of the nature of time for his
continued existence. Basic subsistence and the quest for food relied heavily
upon an understanding of the temporal durations of the seasons, and many
archaeologists now believe that ancient monuments such as Stongehenge and
the Egyptian Pyramids were used partly as astronomical calendars.
Furthermore, the mystical significance often attached to the lunar cycle,
and the magical rituals associated with mid-summer and mid-winter in,
amongst others, druid and pagan cultures, indicate just how important the
continual `re-birth' of time was taken to be in early society. Indeed, over
3000 years ago Greek mythology talked of Chronos, the personification of
time and father to all Olympian gods. In fact, time still plays crucial
roles in the sacred texts of the current major World religions: the Creation
story of the Judaic texts is told to take place over a period of seven days,
whilst the Eastern religions featured in the Hindu and Buddhist scriptures
describe the eternal, cyclic nature of existence. Indeed, the act of
Creation itself concerns the very origins of time; the Bible even starts
with the words ``In the beginning...''.

The actual quantification of time is also something that has proved
essential in the development of society. Its accurate measurement has
therefore keenly concerned inventors and engineers since ancient
civilisation, from the historical use of sundials, waterclocks, hour glasses
and candles, through to the mechanical world of clockwork gears and springs,
and finally ending up in the modern age of digital chronographs and
precision atomic clocks. Indeed, the second is considered one of the most
important units of science, and its definition is given as one of the seven
base quantities provided in the International System of physical standards.

In human development, too, are time and temporality important. Some
psychologists believe that young children, for example, have very little
temporal awareness, and it is not until about the age of two before a key
stage in their cognitive growth occurs and they can appreciate the abstract
concepts involved in `today', `tomorrow' and `yesterday'. Indeed, even the
development of language, a skill that has given humans a unique advantage
over every other organism on Earth and is a central milestone in a child's
progression towards maturity, is intrinsically linked with temporality, and
whole sets of tenses are required in order for people to express events,
ideas and plans that have happened, had happened, were happening, are
happening, and will happen.\bigskip

Overall, the concepts and measurements of time, rate and duration seem vital
in humans' understanding, description and control of the reality in which
they live.

Despite this, however, still nobody really knows what time actually \textit{%
is.}\bigskip

One ultimate aim of this thesis is to investigate the nature and origin of
time in the physical Universe. In particular, a direction is taken that is
based entirely upon empirical evidence, and is hence fully consistent with
the experimentally verified principles of standard quantum mechanics.

To this end, it actually turns out that in attempting the above goal, a
paradigm is proposed and developed that describes the entire Universe
according to quantum principles. As will become apparent, the vision is of a
fully quantum universe free from external observers, represented by a
complex statevector in an enormous, but finite, factorisable Hilbert space,
and developed by the successive applications of quantum operators.

Now, physicists do not of course perceive the Universe to be a single,
complex vector in a Hilbert space. Instead, the Universe generally appears
to contain an enormous number of classically distinct looking objects,
consisting of molecules, atoms and, ultimately, elementary particles.
Moreover, these objects appear to interact with one another in particular
ways, and appear to exist at unique locations in a background of apparently
continuous and classical space and time. The question immediately faced
therefore asks how all of these features could arise from the proposed
quantum state description, and it is by attempting to address this issue
that a possible origin for a variety of physically observed phenomena is
provided.

So, from the initial aim of describing the possible mechanisms responsible
for the existence of time in physics, a framework is proposed from which
every feature of physical reality is hoped to emerge. Overall, the
suggestion is that by adopting a fully quantum view of the Universe, the
origins of time, space, matter and classicity in physics might be accounted
for from a certain fundamental perspective.

Exactly how this could be achieved is the subject of this work.

\bigskip

The structure of this thesis is as follows. In Chapter 2, a number of the
different traditional interpretations of time are introduced. Historical
perspectives on the subject, including the role of time in various
scientific disciplines, are briefly mentioned first, and this is then
followed by a discussion of the nature of time in conventional physics.

In Chapter 3, the framework for the paradigm featured throughout this work
is introduced, justified and discussed. The Universe is argued to be
represented by a quantum state, and the constraints placed on this by
physics are considered; the necessary features required for its dynamics are
then defined. The paradigm was originally proposed in \cite{EJ}, and was
also developed in \cite{EakinsSLOV}.

In Chapter 4, the issue of obtaining classically distinguishable objects
from the perspective of the fully quantum Universe is discussed.
Specifically, the mathematical properties of factorisable Hilbert spaces and
entangled/separable states are investigated; these ideas will form the basis
for much of the work featured in the following chapters. Most of Chapter 4
was developed with G. Jaroszkiewicz, and is strongly similar to the work
presented in \cite{EJ1}.

In Chapter 5, the emergence of spatial degrees of freedom from the quantum
universe paradigm is discussed. In particular, it is shown how quantum
causal sets may be generated as the Universe proceeds through a sequence of
stages, from considerations of both the changes in separability of the
state, and from the changes in factorisability of the operators used to
develop this state. The possibility of obtaining continuous, classical
spacetime from such a treatment is suggested. As with Chapter 4, this work
was developed with G. Jaroszkiewicz, and is congruent in content to \cite
{EJ2}.

Chapter 6 is split into two parts. First, a discussion of the development of
low dimensional bit and qubit systems using CNOT operators is given; the
parallels between computation, quantum computation and the quantum universe
paradigm proposed in this thesis are then drawn. The second part of Chapter
6 addresses the issue of information flow in self-contained quantum systems,
thereby defining the concepts of information change and information
exchange. The definition of endo-physical measurements is then given, with
the goal being to investigate quantum experiments in the circumstance where
the observer is part of the subject under observation. Such a discussion is
mandatory for the proposed fully quantum paradigm, because physicists are by
definition part of the Universe they attempt to measure. Much of Chapter 6
was developed in collaboration with G. Jaroszkiewicz, with the second part
related to \cite{EJ3}. Chapter 6 is also supported by Appendix A, in which
conventional (exo-physical) classical and quantum computation are discussed
for completeness and comparison.

In Chapter 7, the possible emergence of quantum field theoretic concepts
from the proposed paradigm is discussed; the generation of particle physics
in a fully quantum Universe is hence suggested. Specifically, the Dirac
field is investigated, and it is shown how the corresponding ladder,
Hamiltonian, momentum and charge operators used in traditional collider
physics may arise from the conjectured vision. The potential links between
quantum computation, quantum field theory, and the quantum universe paradigm
are thus explored. Much of this research was completed with G.
Jaroszkiewicz, and will feature in a forthcoming article currently being
prepared for publication. Chapter 7 is accompanied by Appendix B, which
derives from first principles the conventional (phase space) Hamiltonian,
momentum and charge operators for the relativistic Dirac equation.

Finally, in Chapter 8 the dynamics of developing quantum universes are
discussed. On the basis that its state is developed by applying a quantum
operator, and that there is no external physicist present to choose such an
object, the various types of method that the Universe might employ to select
this operator are first classified. Then, some of these types are
investigated more thoroughly, with the aim being to discuss how the actual
next operator used in a universe's development could depend on its current
state. The physical results and limitations of the suggested
`self-referential' mechanisms are duly considered in turn, with the
conclusion drawn that only certain types of method are able to provide valid
dynamics for a universe's development. Lastly, a particular type of
mechanism is proposed that appears able to automatically examine the current
state, and then develop it according to what it is and what properties it
might have; the application of this to the issue of endo-physical
experimentation is discussed. Summarising, in Chapter 8 a number of
algorithms are defined for the development of the quantum universe,\ and
these are effectively seen as analogous to self-referential versions of the
rules used in conventional physics to govern quantum computations.\bigskip

A self-contained, developing, fully quantum Universe is thus proposed in the
following, and the possible ways in which time, space, physics and matter
could emerge from this paradigm are demonstrated.

\bigskip \newpage

\section{History and Background}

\renewcommand{\theequation}{2.\arabic{equation}} \setcounter{equation}{0} %
\renewcommand{\thetheorem}{2.\arabic{theorem}} \setcounter{theorem}{0}

\bigskip

It is almost impossible to even attempt to provide a complete account of the
background of the study of time. Indeed, time's `ultimate nature' is a
question that has concerned scientists and philosophers alike for many
thousands of years. Moreover, because time pervades almost every aspect of
human nature,\ it is perhaps not surprising that its study has drawn from a
number of academic disciplines, and that many of these consider its
definition in many different ways.

It is therefore beyond the scope of this thesis to detail every idea from
such a wide range of sources. However, whilst this may be the case, in this
chapter a number of the conventional perspectives on time are briefly
introduced.

\bigskip

\subsection{Time in Mathematics, Philosophy and Biology}

\bigskip

In mathematics, time is generally just the `parameter $t$' that is used
simply as a suitable variable when describing the changes of a developing
system. So, from absolute Newtonian time to proper time in relativity,
temporality is normally only employed merely as a convenient label to
distinguish events. In Newton's \textit{Principia Mathematica, }for example,
time is barely mentioned apart from as the $t$ in the equations, and is
regarded as an almost ethereal notion without further discussion or
justification; to quote: ``\textit{Absolute time flows equably, without
regard to anything else}''.

Time in mathematics is therefore most often used simply as a coordinate
reference, that is, as a linear, real axis stretching from $-\infty $ (the
past) to $+\infty $ (the future) via the origin (now). Although this idea
may be extended to cover imaginary time as well (for example, in the path
integral approach in quantum field theory, where time $t$ is mapped to $%
i\tau $ in order to prevent divergences \cite{Rothe}), the linearity of the
coordinate axis is normally still preserved. In fact, by reversing this line
of thinking, since time is often interpreted as a linear sequence of
moments, which it generally is despite the possibility of some Zeno-type
paradoxes, it is easy to see why it is so frequently linked to the
number-line. Indeed, Kant even thought that time and number were
inseparable.\bigskip

A mathematical construction of continuous time may be built up from some
fairly basic, logical arguments. One particular method is given in \cite
{Whitrow}; if $T$ is a `temporal continuum', and if $p,$ $q,$ and $r$ are
`instants' or `moments' defined to be members of this set, then the
following statements are assumed to hold true:

\begin{enumerate}
\item  Mutual exclusivity: either $p$ and $q$ are simultaneous, or $p$
precedes $q,$ or $q$ precedes $p.$

\item  If $q$ precedes $r$ and $p$ precedes $q,$ then $p$ precedes $r.$

\item  $T$ is a \textit{dense set; }if $p$ precedes $r,$ there exists at
least one event $q$ between $p$ and $r.$

\item  If $T$ contains non-empty subsets $T_{1}$ and $T_{2},$ where $T\equiv
T_{1}\cup T_{2}$ and all elements of $T_{1}$ precede all elements of $T_{2},$
there exists at least one `instant' $t$ for which any instant preceding $t$
is in $T_{1},$ but any instant after $t$ is in $T_{2}.$

\item  Between any two members of $T$ there is at least one instant which is
a member of a denumerable subset of $T;$ the relation between time and the
number line is again drawn.
\end{enumerate}

From a set of axioms such as these, the suggestion is that a mathematical
framework for past, present and future may be derived; a linear temporal
parameter may consequently be generated. The logical framework, however,
does not actually begin to describe what this time might actually \textit{be}%
.

\bigskip

The logical approach to time may be argued to have its roots in philosophy.
Indeed, the definition of time has naturally been a subject pertinent to
many philosophers, too numerous to mention, since at least the World of
Ancient Greece. Plato, for example, thought that time could not actually be
described by mathematics, because he believed that only things that exist
eternally were real, and time, unlike numbers, is transient. His student
Aristotle, on the other hand, thought that time was just ``\textit{a measure
of motion}''.

Historically downstream, William of Alnwick suggested in the Fourteenth
century the idea of a discrete time, referring to an ``\textit{indivisible
of motion}''. One hundred and fifty years later, Leonardo Da Vinci contended
that: ``\textit{an instant has no time. Time is made of movements of the
instant, and instants are the boundaries of time}''. Later still, Immanuel
Kant believed that time can neither have a beginning nor be eternal, and
that Mankind's understanding of what time might be will always, ultimately,
be inadequate.

Some of the great Eighteenth and Nineteenth century mathematicians and
physicists also contributed to the philosophical interpretation of time.
Hamilton, for example, linked time with algebra in an analogous way to how
space is linked with geometry. Conversely, Leibniz and Laplace lived in a
deterministic world, so effectively `removed' the need for time because they
believed that everything could be determined from initial conditions.

Leibniz and Laplace's opinions are not reconciled from every modern
perspective, particularly from the current belief that the universe
incorporates stochastic quantum laws and so is not strictly and classically
determined. Indeed, Penrose, for example, even argues that classical
determinism is broken by quantum mechanical effects in the brain, and this
has profound implications for discussions of time. This `consciousness
debate' is moreover congruent to the belief of Hobbes, who contended that
time is a decay of the Before and After in the mind. In fact in many ways,
this type of interpretation may be summarised by the words of Einstein: ``%
\textit{Objective reality }is\textit{\ and does not happen. Only by
consciousness does the world come to be as an image in space continuously
changing in time}''.

\bigskip

It is by invoking such notions of consciousness that provides a bridge
between philosophical discussions of time, and those present in the
biological sciences.

Time in biology splits broadly into two categories: on the one hand, there
exists the temporal rhythms of nature; on the other, there is the
subjective, conscious experience.

Temporal rhythms are generally governed by the responses to external time
stimuli, and imply that living organisms often appear to possess inbuilt
`biological clocks'. Such bio-clocks are often synchronised to well
regulated outside `cues': these may be variations in light intensity, length
of daylight, average temperature, lunar cycles, tidal effects, etc. So,
examples include diurnal rhythms (e.g. differences in day/night mental
activity), monthly rhythms (e.g. menstrual cycle), annual rhythms (e.g. in
perennial plants), and so on. As a consequence of these periodic patterns,
human beings are often able to get the false impression that time,
ultimately, is cyclic in nature.

`Conscious time' is more complicated, partly perhaps because it is uncertain
exactly what consciousness is, and hence partly also because scientists
cannot easily ascertain its existence in other organisms. Moreover, it is
also experienced in different ways that are subject to context; a human's
perception of duration, simultaneity and time elapse are all highly
dependent on the particular individual, her state of mind, her age, her
memory, and the physical stimulus itself, etc.

There is also, of course, clearly a fine line between scientific evaluation
of psychological and neurophysiological time, and the philosophical question
of mind itself.\smallskip

\bigskip

\subsection{Time in Physics}

\bigskip

Despite the appearance of time in mathematics, philosophy and biology, it is
perhaps the physical sciences that should be most concerned with providing a
definition for what time actually is. After all, time evidently seems to be
a physical phenomenon.

In addition to this, it is also noted that time is present in most of the
equations of physics. Indeed, some might argue that it is the purpose of
physics to either predict the future outcome of an experiment from a given
set of initial conditions, or to reconstruct the past from results that
exist in the present. The question, then, of what past, present and future
actually are should consequently be taken to be of prime importance. In
fact, given that time is surely one of the most fundamental physical
phenomena there is, its ultimate origin and definition would have to be
explained by any law that pertains to be a Theory of Everything in order for
such a suggestion be accepted as truly satisfactory.\bigskip

So, having mentioned these points, it is perhaps surprising to consider just
how little physicists seem to understand about the true nature of time.
Going further, it is perhaps equally surprising to observe just how seldom
this issue is even addressed. In only a few areas of physics, for example,
is time actually viewed as a fundamental quantity, instead of just as a
convenient `yardstick' to measure against or label events. Rarely does
physics really consider what this yardstick might actually be, or what this
label might mean.

One possible reason for this lack of definition might be because it does not
normally appear to matter what time actually is, as long as its effects may
be accounted for. Newton's laws, for example, are entirely symmetric with
respect to a reversal of time: a ball rolling without friction from $A$ to $%
B $ and back to $A$ appears exactly equivalent in `reverse'; so, in this
situation time is reduced to a mere coordinate that provides a convenient
parameter useful in defining dynamics. In short, as long as the continuous
variable $`t$' may be employed in the equations of motion with accurate
results, the issues concerning what it might actually be are often ignored.

However, not every phenomenon of science is time symmetric. Thus, time might
most interestingly be discussed in situations where its direction does seem
to play a distinguishing part. As examples, cosmology, particle theory,
thermodynamics and quantum mechanics each contain such asymmetries, and each
of these appears to introduce important comments regarding the role of time.
So, it is these issues that are discussed in turn now.\bigskip

The cosmological development of the Universe is intrisically linked to a
number of questions regarding time. Indeed for a start, its evolution as a
whole is manifestly time asymmetric.

Specifically, current thinking is that the Universe began as a Big Bang
about 12 billion years ago, and has been expanding\ ever since; indeed, most
relativistic cosmologists believe that the Big Bang actually marked the very
beginning of time. Now, this scenario has two implications for the present
discussion. The first point concernes the Universe's fate and future: either
it will stay expanded forever, or else it will collapse back to a Big
Crunch, depending on its density. If the former is true, there is an
immediate asymmetry associated with a finite past but an infinite future,
and the question is provoked as to how and why the Universe actually began.
Moreover, given that this question might naturally be rephrased as ``what
was it that changed and caused the Universe?'', it is remarked both that the
notion of `change' itself implies a reference to an external time, and that
the concepts of cause and effect rely on a sense of before and after,
whereas none of these are defined at the Big Bang.

Conversely, if the latter situation is true and the Universe will eventually
collapse back on itself, the conclusion may be drawn that either the Big
Crunch is different to the Big Bang, leading to another asymmetry, or that
time is somehow reversed during the contraction phase of the Universe. This
second point is immediately undesirable, because it could imply that entropy
might decrease, stars would `suck in' light, particles should disappear from
event horizons, etc.

The remaining implication of an expanding Universe scenario is that the
frame of reference in which it is expanding may be considered to be
`preferred'. In this case, such a frame's time component could then
naturally be linked with an absolute or universal time, and this appears at
odds with the generally accepted principles of relativity. Furthermore,
although such a hypothetical frame is often taken in the literature to be
the frame in which the Cosmic Microwave Background Radiation (CMBR) is
isotropic, it is still debateable as to whether this really provides a
genuinely preferred frame; it is hence unclear as to what the consequences
of this might mean.\bigskip

As discussed above, problems concerning time exist on the largest scales of
physics. However, difficulties also arise at some of the smallest scales.

As an example, it is noted that the equations describing elementary
particles are expected to be invariant under the combined operations of
Charge conjugation, Parity reversal and Time reversal (CPT). It appears to
be an empirical fact, however, that the decay of the electrically neutral
kaons $K^{0}$ and $\overline{K}^{0}$ via the weak interaction appears to
violate Charge-Parity (CP) conservation, and so this decay is also expected
to violate time reversal if the overall CPT operation is to remain
invariant. There is currently no satisfactory explanation for this effect,
and it is therefore believed by some physicists that its investigation might
shed light on the true nature of time.\smallskip \bigskip

Discussions of time, however, are not just limited to the scales of
cosmology or fundamental particles. In fact, one important area of
`everyday' physics that exhibits time asymmetry occurs in thermodynamics.
Indeed, even from the outset, the equations of thermodynamics do not
obviously appear reversible: heat always flows from a hot body to a cooler
one.

Of course, the above observation may be phrased more precisely by stating
that the entropy of a system always increases with time. In other words, a
system that is initially macroscopically heterogeneous becomes
microscopically heterogeneous (or, equivalently, macroscopically
homogeneous) over time. Moreover, the converse of this is not in general
true, and this has lead some authors to conject that it is such an effect
that defines the arrow of time. Thus, the `direction' of increasing entropy
is consequently taken to define the `direction' of the increase in time.

Others authors \cite{Prig}, however, contend that classical entropy is, in
fact, really reversible (at least in principle), because its microscopic
scale is still governed by deterministic kinetic theory, and hence time
symmetric laws. The argument is then that real irreversibility only comes
from quantum effects, by introducing a random `ingredient' into an
observer's knowledge of the kinematics. In short, because the particles'
positions and velocities are ultimately indefinite in the quantum case, the
argument is that they can no longer develop deterministically. An
irreversibility is therefore introduced into the dynamics, and it is this
that is eventually taken to provide a direction for the `flow' of time.

So, the suggestion here is that it is quantum mechanics that ultimately
provides an explanation for the origin of asymmetric time.

\bigskip

The above comment introduces perhaps the most important conflict in the
history of the study of time.

Without exaggeration, much of fundamental physics in the Twentieth century
was founded on two great pillars, namely, relativity and quantum mechanics.
Each of these tremendous theories says something profound about the nature
of time, and, moreover, each is ultimately incompatible with the other.

Special and General relativity are based on the notion of coordinate time.
In other words, time is assumed to be a dimension, as real and linear as
length, breadth and width. In fact, the temporal parameter is placed on a
completely equal footing to the spatial coordinates, and this has led to a
vision of physics existing in a four dimensional spacetime. So, in the
relativistic approach, spacetime is imagined to be a four dimensional
`fabric' which is then curved and distorted by the presence of matter.
Moreover, the resulting `shape' of this fabric may be described by a metric,
and this is a continuous function of the temporal and spatial coordinates.
Physical objects consequently describe trajectories, or worldlines, through
this background arena of spacetime.

Thus, relativity adopts a `\textit{Block universe}' perspective: time and
space are effectively equivalent, extended dimensions. It appears, moreover,
to just be a `biological accident' that humans appear to perceive a three
dimensional space evolving temporally; according to relativity, each of the
temporal and spatial dimensions is just as special as the others, with
metric signatures providing the only difference.

Of course, this interpretation leaves a number of questions unanswered. Why
does time appear to be special for humans? Why can objects move back and
forth in space with complete freedom, but appear unable to travel backwards
in time at all? Why can humans only `go forwards' in time at an apparently
fixed rate? Indeed, why can an object not remain at one position in time,
just as it appears able to rest at a single spatial location?\bigskip

As an outcome of the association of time with dimension, in the Block
universe approach past, present and future all exist in an equal way. The
only distinction between them, in fact, arises from the point of view of a
particular observer: two observers in different frames of reference
witnessing the same set of events cannot necessarily agree on their order.
Specifically, an event that lies in the future of one observer (or more
technically, in the future lightcone of one observer) could be the present
for a second observer, but could lie in the past (lightcone) of a third
observer. Thus, two events that may appear simultaneous in one frame of
reference may be temporally separated in a second

Moreover, this analysis then implies that the relativistic universe is
effectively deterministic: any object in such a universe has its future
`mapped out', because future events on its worldline might be in the past of
an observer in a different frame of reference. For any given moment in the
Block universe approach, the past exists just as much as the present does,
and a pre-determined future is already `out there' waiting to be
discovered.\bigskip

In quantum mechanics, however, the situation is a little different.

In classical mechanics, on which relativity is based, it is acceptable to
observe an object and expect it to remain unchanged. Thus, two observers can
not only measure the same event, but they can also measure it
simultaneously, confident in the knowledge that neither is affecting or
altering it.

In quantum mechanics, however, the same is not true: the act of measurement
generally destroys the state under observation, and replaces it with a new
state that is an eigenvector of whichever operator is used to represent the
test. Thus, quantum mechanics provides another example of a situation in
which physics behaves time asymmetrically. Specifically, although in the
absence of observation the evolution of a quantum state is deterministic and
time reversible according to the Schr\"{o}dinger equation, the state vector
reduction (or `collapse') occurring at the act of measurement is time
asymmetric: the wavefunction discontinuously, randomly and irreversibly
`jumps' into one of the eigenstates of the Hermitian operator used to test
the state.\bigskip

As a consequence of state collapse, quantum theory introduces a problem into
the earlier discussion on relativity. Namely, since by measuring a quantum
object its state is irrevocably changed, this act automatically specifies a
definite `time of observation'. Certainly, it could not be observed again in
its original state by a second observer after it has been altered. The
temporal order of other events can then be compared to this known moment,
thereby implying a strict causal order.

Herein lies the problem. Consider two spacelike separated observers, Alice, $%
A,$ and Bob, $B,$ and consider an `object' initially prepared in a state $O$
(which could, for the sake of illustration, be imagined to be an entangled
state extended across a large region of space). Consider also the point of
view of Alice, and assume that\ if she were to measure $O,$ she would
randomly change its state to $O_{a},$ which is one of the eigenvectors of
some test.

Now consider Bob's point of view. To start, assume that the event
representing the measurement of the object by Alice is later than the event
representing the measurement of the object by Bob, from the perspective of a
particular extended frame of reference (and as given in the standard
literature on relativity). Then, Bob would be able to measure the state $O,$
a course of action that could cause it to collapse to $O_{b}.$ However, this
would then mean that Alice would be unable to observe the object in its
original form $O,$ and would instead only be able to measure the changed
state $O_{b}.$

Moreover, and by applying again the usual relativistic arguments, it could
be the case that in a different frame of reference, Alice and Bob are such
that the event representing the measurement of the object by $B$ is later
than the event representing the measurement of the object by $A.$ So, the
conclusion would consequently be that Alice is able to measure the object
first, thereby changing its state from $O$ to $O_{a},$ such that Bob cannot
therefore measure the object in its original form $O.$

The point is that each of Alice and Bob could believe that the event
representing the measurement of the object by the other observer is in their
own personal future, from the point of view of different frames of
reference. Each would therefore believe that they could measure $O,$
consequently changing it, such that the other observer could only measure
the changed state. Clearly, this reasoning leads to a paradox\footnote{%
Relativity theory is in fact riddled with potential contradictions, the
`Grandfather Paradox' of closed timelike curves in general relativity being
a famous example. Perhaps these difficulties are themselves sufficient to
suggest that relativity does not provide a completely consistent paradigm
for physics, and highlights the general principle that just because
something is mathematically possible, it does not make it physical reality.
Introducing \textit{ad hoc} caveats such as Hawking's Chronology Protection
Theory do little to avail these conclusions.}.\bigskip

Now, the above problem would obviously not arise in classical physics. In
classical mechanics, the measurement of the object $O$ by either $A$ or $B$
leaves it in the same state $O,$ such that the other observer may then also
measure it in its original form. In quantum physics, however, this is no
longer the case, because only one of Alice or Bob may measure the original
state; the act of observation destroys the state, thereby enforcing a
strict, absolute and global causal order for the observation events. So,
although classical, relativistic arguments might suggest that neither of the
spacelike separated observers can fundamentally be said to measure a
classical object first (because the order of such events may be different in
different frames of reference), such a symmetry is broken when quantum
effects are included. By incorporating quantum theory, a frame independent
causal order must be placed upon the events that represent the measurements
of the object by the spacelike separated observers, and this appears
contrary to the standard principles of special relativity.

In a Universe containing quantum mechanics, there cannot be a situation in
which in some frames $A$ measures $O$ and $B$ measures $O_{a}$ whilst in
others $B$ measures $O$ and $A$ measures $O_{b};$ in reality, only one of $A$
or $B$ actually measures $O,$ and this reality is independent of the choice
of frame. A conclusion, then, is that a relativistic, Block universe
approach to physics does not seem immediately compatible with the accepted
principles of quantum measurement.\bigskip

A second difficulty faced by attempts to reconcile the principles of quantum
theory with those of relativity is that the existence of a `concrete' past,
present and future is disputed in quantum mechanics.

The results of the Kochen-Specker theorem and the violation of the Bell
inequality (discussed in the next chapter), for example, conclusively
demonstrate that prior to any measurement, a quantum object cannot be
described as possessing any fixed set of pre-existing physical
characteristics, such as those featuring in theories of classical Hidden
Variables. In physics, a quantum state does not have any pre-existing
attributes just waiting to be discovered by observers.

The point is that if a quantum object does not exist in any definite form
prior to an observation, its present cannot be given any real, concrete,
definite existence. In particular, if the Universe contains quantum objects,
which it certainly appears to, it is not possible to know everything about
the moment of the present.

It is consequently difficult to imagine how the future could be granted such
a status either, because it is difficult to accept that the future is
somehow `more real' than the present.

Specifically, then, if Charlie is a quantum object in the Universe, and if
the Block universe approach to physics were to be accepted, he would have to
conclude that even though according to the empirically verified laws of
quantum mechanics his present is undefined, unknown and unknowable, his
future is somehow fixed, definite and pre-determined. Clearly, this would be
a strange position. Moreover, to accept such a conclusion would also imply
that the result of a measurement on a quantum object is deterministic, and
not the random, probabilistic outcome it is experimentally known to
be.\bigskip

Of course, it is perhaps not surprising that standard relativity runs into
difficulties when attempting to describe a Universe incorporating quantum
principles. After all, historically, Einstein's theory of special relativity
pre-dates Schr\"{o}dinger's quantum mechanics by about twenty years.

The theory of relativity is based upon, and generally framed in terms of,
the relationships arising between sets of classical observers as they
witness classical events. But, the overriding lesson learnt from quantum
mechanics is that physicists' notions of observation must be radically
redefined: measurements cannot be performed non-invasively, and sets of
observers cannot observe the same object in the same way. It consequently
seems difficult to believe how any theory based upon classical observation
could really be taken to provide a truly fundamental picture of a quantum
reality, and this therefore obviously raises the question as to whether the
results and conclusions of classical relativity should ever be accepted as
completely reliable, at least as far as constraints on the ultimate physics
of the Universe are concerned.

Quantum theory taught physicists that, fundamentally, reality does not quite
behave as they thought it did. So, any theory originating from the
pre-quantum era of science can really only be an incomplete vision of a
better, quantum perspective of physics, and this includes any classical
view\ or comment regarding the nature of time.\bigskip

A Block universe interpretation of time is unable to account for quantum
principles, because such an approach assumes the presence of an `eternal',
pre-existing and fixed past, present and future. The conclusion, then, is
that the standard, Block universe vision of physics that arises from
relativity is fundamentally incompatible with the standard principles of
quantum theory. Consequently, if quantum mechanics cannot support such
notions, it is suggested that a Block universe model is not the correct way
to analyse a Universe that undoubtedly contains quantum objects. This
perhaps explains why attempts at deriving theories of quantum gravity by
directly quantising classical general relativity have predominantly been so
unsuccessful.

In quantum theory only the moment of the present can be granted any real
existence, and even this is limited. Thus, instead of adopting a Block
universe approach to physics, quantum theory suggests that a `\textit{Process%
}' interpretation of time is required; only `Now' may be given any physical
significance.

Moreover, in fact, any attempt to describe time and physics from a point of
view that \textit{is} compatible with the empirically verified priciples of
quantum mechanics must consequently also assume this interpretation. It is
therefore such a perspective that is adopted in this thesis.

\bigskip \newpage

\section{The Quantum Universe}

\renewcommand{\theequation}{3.\arabic{equation}} \setcounter{equation}{0} %
\renewcommand{\thetheorem}{3.\arabic{theorem}} \setcounter{theorem}{0}

\bigskip

One aim of this thesis is to investigate how the observed physics of the
Universe, especially including the concept of a continuous time parameter,
might arise from a certain fundamental perspective of reality.

The obvious starting point is therefore to specify what this fundamental
perspective might be. In fact, the viewpoint discussed in this thesis will
itself be shown to follow quite naturally from considering a set of
observations regarding the actual nature of the Universe.

To this end, the intention of this chapter is to define, describe and, where
necessary, justify the observations, assumptions, consequences and
conjectures present in the following work.

\bigskip

\subsection{Quantum Mechanics}

The first observation is summed up by the following statement:
\begin{equation*}
\text{Quantum Mechanics is a \textit{valid} theory.}
\end{equation*}

In other words, the argument is that the `standard' quantum theory of Bohr,
Heisenberg, Schr\"{o}dinger, Dirac \textit{et al} is the correct theory to
use when describing certain physical, microscopic systems. Specifically, the
implication is that a physical system may indeed be represented by a
complex, linearly superposed statefunction, that this state may undergo
unitary evolution in some sense, and that by measuring the state it is
`collapsed' into an eigenvector of an observable represented by an Hermitian
operator.

The evidence cited to justify such a statement is the overwhelmingly
universal success of quantum theory in science. In chemistry and biology,
for example, quantum equations have allowed scientists to confidently model
the properties and reactions of many types of molecule and atom. In high
energy physics, the development of quantum field theory has allowed
physicists to accurately predict the characteristics of particles that may
not have existed since the era of the Big Bang. Even in the everyday world,
the essential 21$^{st}$ Century technologies behind optical
telecommunication and computer science would not work if it were not for an
understanding of the quantum laws governing the laser and silicon
chips.\bigskip

A mathematical demonstration of the validity of quantum mechanics was
provided by J. S. Bell \cite{Bell}, based on an analysis of the correlations
produced in a system similar to that described in the thought experiment
proposed by Einstein, Podolsky and Rosen \cite{EPR}. A number of derivations
of Bell's result are available in the literature; the approach outlined
below roughly follows the treatment given in \cite{Rae} and \cite{Wigner}.

Consider the decay of a spinless, neutral pion into an electron-positron
pair $\pi ^{0}\longrightarrow e^{-}+e^{+}.$ Electrons and positrons possess
spin components of $\pm
%TCIMACRO{\UNICODE[m]{0xbd}}%
%BeginExpansion
{\frac12}%
%EndExpansion
$ (denoted by `up' and `down', or equivalently `+' and `-', in some frame),
so by conservation of angular momentum, a spin-up electron is partnered by a
spin-down positron, and vice versa. Consider also a frame of reference $%
\mathcal{F}$ parameterised by Cartesian coordinates $(x,y,z).$ With the aid
of a suitably orientated Stern-Gerlach apparatus, it is possible to measure
the component of angular momentum of either electron or positron in any
direction in $\mathcal{F}.$

Bell's argument is the following. \textit{If} the system ultimately obeyed
classical instead of quantum mechanics, the assertion would be that after
the decay of the pion the electron and positron would\ each have definite
and independently measurable angular momenta, pointing in the general
directions $\underline{n}$ and $-\underline{n}$ respectively. It would also
be possible to non-invasively measure the same particle many times to obtain
its component of angular momentum in any direction.\bigskip

Consider measuring the spin of such a `classical electron' in three
different directions $\underline{a},$ $\underline{b}$ and $\underline{c},$
noting that $\underline{a},$ $\underline{b}$ and $\underline{c}$ need not be
perpendicular. If, without loss of generality, it may be assumed that $%
\underline{n}$ is not orthogonal to any of $\underline{a},$ $\underline{b}$
or $\underline{c},$ the component of $\underline{n}$ in each direction will
either be $+ve$ or $-ve.$ So, by measuring the electron's spin first in the $%
a$-direction, then the $b$-direction, then the $c$-direction, the overall
result will be one of eight possibilities: $\{a$ result, $b$ result, $c$
result$\}=\{+,+,+\}$ or $\{+,+,-\}$ or $\{+,-,+\}$ or $\{+,-,-\}$ or $%
\{-,+,+\}$ or $\{-,+,-\}$ or $\{-,-,+\}$ or $\{-,-,-\}.$ If the orientation
of $\underline{n}$ is random and may point in any direction, then depending
on the choice of $\underline{a},$ $\underline{b}$ and $\underline{c},$ each
of these eight results has a certain probability $P\{\pm ,\pm ,\pm \}$ of
occurring, with total probability summing to unity. \bigskip

Consider now a measurement of the electron by a Stern-Gerlach apparatus
orientated along one of the directions $\underline{a},$ $\underline{b}$ or $%
\underline{c},$ followed by a measurement of the positron by a second
Stern-Gerlach orientated along a different one of the directions $\underline{%
a},$ $\underline{b}$ or $\underline{c}.$ If $P_{c}(+\underline{a},+%
\underline{b})$ is defined as the classical probability that the component
of the electron's spin along $\underline{a}$ is found to be $+ve$ \textit{and%
} that the component of the positron's spin along $\underline{b}$ is also
found to be $+ve,$ then by conservation of angular momentum $P_{c}(+%
\underline{a},$ $+\underline{b})$ is equally defined as the probability that
the component of the electron's spin along $\underline{a}$ is $+ve$ but its
component if measured along $\underline{b}$\ would be $-ve.$ Note that this
is clearly a classical result, as expected, because the implication is that
two spin components of the electron have been measured even though it is
only disturbed once.

By considering the eight possibilities given above, $P_{c}(+\underline{a},+%
\underline{b})$ is given by the sum
\begin{equation}
P_{c}(+\underline{a},+\underline{b})=P\{+,-,+\}+P\{+,-,-\},
\end{equation}
which reflects the experimenter's ignorance of the spin component of the
electron (or positron) in the direction of $\underline{c}.$ Similarly the
relations $P_{c}(+\underline{a},+\underline{c})=P\{+,-,-\}+P\{+,+,-\}$ and $%
P_{c}(+\underline{b},+\underline{c})=P\{+,+,-\}+P\{-,+,-\}$ are readily
obtained.

The results may be summed,
\begin{eqnarray}
P_{c}(+\underline{a},+\underline{b})+P_{c}(+\underline{b},+\underline{c})
&=&\{+,-,+\}+\{+,-,-\}+\{+,+,-\}+\{-,+,-\} \\
&=&\{+,-,+\}+P_{c}(+\underline{a},+\underline{c})+\{-,+,-\},  \notag
\end{eqnarray}
such that, since all probabilities are positive,\ it is possible to produce
the classical inequality
\begin{equation}
P_{c}(+\underline{a},+\underline{b})+P_{c}(+\underline{b},+\underline{c}%
)\geq P_{c}(+\underline{a},+\underline{c}).  \label{bellclass}
\end{equation}

So, if the system is governed fully by classical mechanics, i.e. if prior to
any measurement the electron definitely possesses angular momentum in the
direction of $\underline{n},$ then any set of measurements must necessarily
satisfy this relation. That is, if the spin components of the correlated
electrons and positrons in a large number of identically prepared systems
are measured along any set of directions $\underline{a},$ $\underline{b}$
and $\underline{c},$ the classical probabilities evaluated from the
statistics of the results would obey the inequality (\ref{bellclass}%
).\bigskip

However, it may be shown that if the electron-positron pair instead obey the
laws of quantum mechanics, the probabilities of obtaining certain results
may violate this inequality.

In quantum theory, a system does not have any pre-existing or definite
properties prior to an observation. Before a measurement, a particle's
component of angular momentum only has the potential to be either $+ve$ or $%
-ve$ in some direction, and it is the measurement itself that forces the
system to `choose' one of these states to collapse into. In this sense,
therefore, it may be said that prior to an observation each particle is in
both potential spin states simultaneously, and the system is represented by
an entangled state $|\psi \rangle $ described in obvious notation by the
antisymmetric linear superposition
\begin{equation}
|\psi \rangle =\frac{1}{\sqrt{2}}\left( \mid \uparrow \rangle _{e}\otimes
\mid \downarrow \rangle _{p}-\mid \downarrow \rangle _{e}\otimes \mid
\uparrow \rangle _{p}\right)  \label{wav}
\end{equation}
where, for example, $\mid \downarrow \rangle _{p}$ represents the state of a
positron that is spin-down in some direction.

Now, a measurement of the spin of either the electron or positron destroys
the entanglement. If, for example, the electron is measured and found to be
in the state $\mid \uparrow \rangle _{e},$ it can obviously no longer be
described as potentially being in the state $\mid \downarrow \rangle _{e},$
and the wavefunction of the electron-positron system collapses to $|\phi
_{\uparrow }\rangle =\mid \uparrow \rangle _{e}\otimes \mid \downarrow
\rangle _{p}.$ Subsequent measurements of the spin of the positron in this
direction, with its state now prepared as part of this new state $|\phi
_{\uparrow }\rangle ,$ must then produce the result $\mid \downarrow \rangle
_{p}.$

Alternatively, if the first measurement had instead found the electron to be
in the state $\mid \downarrow \rangle _{e},$ it would imply a collapse of
the initial entangled state into the product state $|\phi _{\downarrow
}\rangle =$ $\mid \downarrow \rangle _{e}\otimes \mid \uparrow \rangle _{p},$
and later observations of the positron would find it to be spin-up in this
direction, $\mid \uparrow \rangle _{p}.$\bigskip

Consider now the quantum probability $P(+\underline{a},+\underline{b}),$
defined analogously to the classical probability $P_{c}(+\underline{a},+%
\underline{b}).$ In quantum mechanics, the evaluation of this requires two
measurements to be performed on each of a statistical number of identically
prepared systems: firstly the electron's spin is measured in the direction
of $\underline{a},$ and secondly the positron's spin is then measured in the
direction of $\underline{b}.$ Consequently, this process necessarily
involves a collapse of the initial entangled state $|\psi \rangle $ into a
product state $|\phi \rangle $ when the electron is measured, followed by a
projection of the `positron part' of this new state $|\phi \rangle $ in the
direction of $\underline{b}$ when the positron is measured. The overall
result $P(+\underline{a},+\underline{b})$ is then given by the products of
the probabilities obtained from these two measurements.

To illustrate how this may be achieved, consider a particular choice of the
vectors $\underline{a}$ and $\underline{b}.$ For simplicity, and without
loss of generality, $\underline{a}$ may be chosen to lie in the direction of
the $z$-axis, and $\underline{b}$ may be chosen to be a vector in the $x-z$
plane that subtends an angle $\theta _{ab}$ to $\underline{a}$ (or $z).$
When the electron is measured, and its spin component in the direction of $%
\underline{a}$ is found to be either $+ve$ or $-ve,$ the entangled state%
\footnote{%
Where no confusion is likely to occur, the notation for vectors, such as $%
\psi ,$ and quantum states, such as $|\psi \rangle ,$ will be used
interchangeably throughout this thesis, i.e. $\psi \Leftrightarrow |\psi
\rangle .$} $\psi $ collapses into either the state $|a_{+}\rangle
=|+\rangle _{e}\otimes |-\rangle _{p}$ or the state $|a_{-}\rangle
=|-\rangle _{e}\otimes |+\rangle _{p}.$ Clearly, the probability that the
spin of the electron is found to be $+ve$ in the $\underline{a}$ direction
is
%TCIMACRO{\UNICODE{0xbd}}%
%BeginExpansion
$\frac12$%
%EndExpansion
, because both the states $|a_{+}\rangle $ and $|a_{-}\rangle $ are equally
likely, as is evident from the initial entangled state.

For later convenience, note that $|+\rangle $ may alternatively be written
in the matrix form $\binom{1}{0},$ whilst $|-\rangle $ may be written as $%
\binom{0}{1}.$\bigskip

The operator $\hat{S}_{\theta _{ab}}$ representing the subsequent
measurement of the positron by the Stern-Gerlach apparatus orientated along
the direction $\underline{b},$ i.e. at an angle $\theta _{ab}$ to the $z$%
-axis, is given by
\begin{equation}
\hat{S}_{\theta _{ab}}=\hat{S}_{z}\cos \theta _{ab}+\hat{S}_{x}\sin \theta
_{ab}
\end{equation}
where $\hat{S}_{z}=%
%TCIMACRO{\UNICODE[m]{0xbd}}%
%BeginExpansion
{\frac12}%
%EndExpansion
\hbar \hat{\sigma}_{z},$ $\hat{S}_{x}=%
%TCIMACRO{\UNICODE[m]{0xbd}}%
%BeginExpansion
{\frac12}%
%EndExpansion
\hbar \hat{\sigma}_{x}$ and $\hat{\sigma}_{z}$ and $\hat{\sigma}_{x}$ are
the Pauli spin matrices in the $z$ and $x$ directions with matrix
representations
\begin{equation}
\hat{\sigma}_{z}=\left(
\begin{array}{cc}
1 & 0 \\
0 & -1
\end{array}
\right) \text{ \ \ and \ \ }\hat{\sigma}_{x}=\left(
\begin{array}{cc}
0 & 1 \\
1 & 0
\end{array}
\right) .
\end{equation}

So, $\hat{S}_{\theta _{ab}}$ is given by
\begin{equation}
\hat{S}_{\theta _{ab}}=\frac{\hbar }{2}\left(
\begin{array}{cc}
\cos \theta _{ab} & \sin \theta _{ab} \\
\sin \theta _{ab} & -\cos \theta _{ab}
\end{array}
\right) ,
\end{equation}
which has eigenvalues $+\hbar /2$ and $-\hbar /2$ corresponding to
eigenvectors $|b_{+}\rangle =\binom{\cos (\theta _{ab}/2)}{\sin (\theta
_{ab}/2)}$ and $|b_{-}\rangle =\binom{-\sin (\theta _{ab}/2)}{\cos (\theta
_{ab}/2)}$ respectively. The eigenstate $|b_{+}\rangle $ is parallel to $%
\underline{b},$ i.e. has a $+ve$ component in the direction of $\underline{b}%
,$ whereas $|b_{-}\rangle $ is anti-parallel with a $-ve$ component.\bigskip

The overall process may now be summarised. An initial entangled state\ $\psi
$ is collapsed into either the state $|a_{+}\rangle $ or the state $%
|a_{-}\rangle $ when the spin of the electron is measured in the direction $%
\underline{a}.$ If the electron's spin component is found to be $+ve,$
corresponding to the state $|a_{+}\rangle ,$ then the subsequent measurement
of the positron will leave the electron-positron system in either the state $%
|+\rangle _{e}\otimes |b_{+}\rangle $ or the state $|+\rangle _{e}\otimes
|b_{-}\rangle .$ Alternatively, if the electron's spin component is found to
be $-ve,$ corresponding to the state $|a_{-}\rangle ,$ then after the
measurement of the positron the electron-positron system will be in either
of the states $|-\rangle _{e}\otimes |b_{+}\rangle $ or $|-\rangle
_{e}\otimes |b_{-}\rangle .$

With the above in mind, it is possible to rewrite $P(+\underline{a},+%
\underline{b})$ as the probability of obtaining the state $|a_{+}\rangle $
when the electron is measured, given that before this measurement the system
is in an entangled state of the form $\psi ,$ multiplied by the probability
of obtaining the state $|b_{+}\rangle $ when the positron is subsequently
measured, given that its state before this second measurement is now $%
|-\rangle _{p}.$ This latter probability is
\begin{equation}
|\langle b_{+}|-\rangle _{p}|^{2}=\left| \left( \cos (\theta _{ab}/2)\text{
, }\sin (\theta _{ab}/2)\right) \binom{0}{1}\right| ^{2}=\sin ^{2}\left(
\frac{\theta _{ab}}{2}\right)
\end{equation}
which leads to an overall probability $P(+\underline{a},+\underline{b})=%
%TCIMACRO{\UNICODE[m]{0xbd}}%
%BeginExpansion
{\frac12}%
%EndExpansion
\sin ^{2}(\theta _{ab}/2).$

By a similar argument, it can be shown that $P(+\underline{a},+\underline{c}%
)=%
%TCIMACRO{\UNICODE[m]{0xbd}}%
%BeginExpansion
{\frac12}%
%EndExpansion
\sin ^{2}(\theta _{ac}/2)$ and $P(+\underline{b},+\underline{c})=%
%TCIMACRO{\UNICODE[m]{0xbd}}%
%BeginExpansion
{\frac12}%
%EndExpansion
\sin ^{2}(\theta _{bc}/2),$ where $\theta _{ac}$ is the angle between $%
\underline{a}$ and $\underline{c},$ and $\theta _{bc}$ is the angle between $%
\underline{b}$ and $\underline{c}.$

Now, if quantum theory is really a disguised version of classical mechanics,
the probabilities derived from treating the electron-positron system
according to quantum principles should obey the same constraints (\ref
{bellclass}) as those derived from a classical treatment of the system.
However, whilst the classical inequality (\ref{bellclass}) holds, the
relation
\begin{equation}
\sin ^{2}\left( \frac{\theta _{ab}}{2}\right) +\sin ^{2}\left( \frac{\theta
_{bc}}{2}\right) \geq \sin ^{2}\left( \frac{\theta _{ac}}{2}\right)
\label{bell}
\end{equation}
formed by substituting the above quantum probabilities into (\ref{bellclass}%
) generally does not. For example, if $\underline{a},$ $\underline{b}$ and $%
\underline{c}$ lie in a plane with $\theta _{ab}=\pi /3,$ $\theta _{bc}=\pi
/3$ and $\theta _{ac}=2\pi /3,$ then (\ref{bell}) becomes $%
%TCIMACRO{\UNICODE[m]{0xbc}}%
%BeginExpansion
{\frac14}%
%EndExpansion
+%
%TCIMACRO{\UNICODE[m]{0xbc}}%
%BeginExpansion
{\frac14}%
%EndExpansion
\geq
%TCIMACRO{\UNICODE[m]{0xbe}}%
%BeginExpansion
{\frac34}%
%EndExpansion
,$ which is clearly false. So, for quantum systems
\begin{equation}
P(+\underline{a},+\underline{b})+P(+\underline{b},+\underline{c})\ngeq P(+%
\underline{a},+\underline{c})
\end{equation}

Thus, it may be argued that quantum and classical mechanics are
fundamentally inequivalent in that they predict different results. The
constraints placed by classical mechanics on a system, calculated by
scientists as relationships between sets of probabilities of obtaining
particular sets of results, are not present if the system is instead
governed by quantum theory.\bigskip

Importantly,\ it has also been empirically shown that such violations of the
classical Bell inequalities occur in physics. Experiments with entangled
pairs of photons \cite{Aspect} have yielded results that agree with quantum
mechanics to better than $1\%,$ but violate the Bell predictions of
classical mechanics by $35\%.$

Summarising, the work of \cite{Bell} and \cite{Aspect} has demonstrated that
quantum theory is not equivalent to classical mechanics, but that physics
obeys quantum principles. From such an viewpoint, all theories that suggest
that quantum mechanics is simply a disguised theory of classical probability
are ruled out, as are any theories pertaining to classical Hidden Variables.
Such mechanisms will not be discussed further in this work.

The conclusion of this sub-section is that in order to describe certain
physical, microscopic systems, it is quantum mechanics, and not classical,
that is the correct and valid theory to use.

\bigskip

\subsection{Quantum Cosmology}

\bigskip

The second observation regarding the empirical nature of the Universe is the
following:
\begin{equation*}
\text{There is no `\textit{Heisenberg Cut'} in physics.}
\end{equation*}

There is no rigid dividing line that segregates the quantum experiment being
observed with the scientist doing the observing. There is equally no
dividing line setting a scale beyond which quantum mechanics is no longer
valid. Whilst most physicists readily accept that every microscopic
sub-system in the Universe obeys the rules of quantum mechanics, there has
never been demonstrated a definite macroscopic size or scale where quantum
laws cease to be the correct theory of dynamics in favour of more
fundamental classical laws.

As an example of this, it has even been shown that huge macroscopic objects
such as quasars can give rise to observable quantum effects \cite{Wheeler1}.
If on the line of sight between a distant quasar and the Earth is some sort
of massive body, such as a galaxy, the gravitational lensing of the quasar's
light induced by this body may give rise to interference patterns analogous
to those arising in a Young's double-slit type device. Even if the quasar is
sufficiently distant and dim such that a telescope on Earth only registers
one photon at a time, the interference fringes still arise, implying that
the entire Earth-body-quasar system is behaving like a huge quantum
`Which-path' experiment.\bigskip

So, if the Universe that physicists observe appears to be an enormous
collection of microscopic sub-systems, i.e. is composed of protons,
electrons etc., and if each of these microscopic sub-systems obeys quantum
mechanics, and if there is no Heisenberg Cut directly separating these
sub-systems from each other or the observer, and if the size of a system
does not fundamentally affect whether it runs according to quantum laws, the
conclusion drawn is that the entire Universe is itself a giant quantum
system. The conjecture, therefore, is that the principles of quantum
mechanics may be applied to the Universe as a whole\footnote{%
As an aside, note that there is also no known evidence for what could
analogously be called a ``Heisenberg Time'' in astronomy: many cosmologists
conjecture that just after the Big Bang the entire Universe should be
represented by a quantum state, but no explanation is generally given as to
exactly when the Universe should then \textit{stop} being treated according
to quantum principles.
\par
The assertion proposed here is that it should not.}.\bigskip

If this conjecture is true, it should\ then be possible to write down a
unique quantum wavefunction $\Psi $ for the Universe that describes its
large scale properties and evolution as a whole (c.f. \cite{deWitt}\cite
{Hartle}\cite{Deutsch}\cite{Everett}). This quantum state must be
complicated enough to not only model a vast, intricate and expanding cosmos,
but also to describe a universe that appears to be comprised of an enormous
number of microscopic quantum sub-systems. Further, it must also allow
physical observers, who believe themselves to be isolated classical states
that are inside the Universe they are trying to understand, to experience
and measure an apparently classical reality. Classical physics must
therefore be a emergent phenomenon which is somehow borne from the quantum
theory as an approximation on certain, presumably macroscopic, scales. The
true quantum nature of reality should always be present, but will only
demonstrate itself in complicated experiments designed to investigate very
refined circumstances. Any formulation of the wavefunction of the Universe
must somehow take account of this.

Further, every large scale characteristic of the Universe, and every
physical property of every sub-system it contains, must be accounted for in
any formulation of $\Psi .$ If the wavefunction of the Universe describes
\textit{everything,} then space, time, energy, particle physics, and even
semi-classical human observers must all emerge somehow from considerations
of the properties of this quantum state.\bigskip

It is therefore a job for physicists to attempt to discover what the
Universe's statefunction might be like. Now, whilst this task may appear
overwhelmingly daunting, by extending the principles of standard quantum
mechanics, a number of inferences can be drawn about the nature of a fully
quantum universe.

Firstly like all states in conventional quantum theory, the wavefunction\ $%
\Psi $ must be a vector in a Hilbert space $\mathcal{H}.$

Secondly, given that by definition there is only one Universe, there\ can be
no classical confusion as to which state it is in. Thus $\Psi $ cannot be a
mixed state of a classical ensemble of Universes, because such a concept is
obviously contradictory. Consequently the wavefunction $\Psi $ must always
be a pure state.

Thirdly, the Hilbert space $\mathcal{H}$ containing the statevector
representing the Universe must be of truly enormous dimension. One
justification here is that classical physics has been ascribed to be an
emergent approximation to quantum physics on certain scales, and the
physical classical Universe seems to possess an almost uncountable number of
degrees of freedom.

In fact, as a na\"{i}ve lowest estimate of this dimension, consider the
suggestion of many authors that there exists a certain minimum unit of
spatial separation beyond which it is meaningless to discuss notions of
classical distance. This resolutional limit is often assumed to be of the
order of the Planck length, $l_{P}=\sqrt{\hbar G/c^{3}}\sim 10^{-35}$
metres, and marks the boundary of where space is assumed to no longer behave
classically and continuously. Thus, given an empty universe of age $\tau
_{U}=15\times 10^{9}$ years expanding spherically at the speed of light, $c,$
\ the current number, $n,$ of Planck volumes in the physical universe is
given by
\begin{equation}
n=\frac{\frac{4}{3}\pi \left( c\tau _{U}\right) ^{3}}{\left( l_{P}\right)
^{3}}\approx 10^{184}.
\end{equation}

Now, if with each of these minimum spatial volumes is associated just a
single two-dimensional degree of freedom, then the total number of
accessible classical states for the universe is clearly $2^{10^{184}}.$ So,
even in the simplest quantum model, the state vector representing the
universe must have a dimension of at least $2^{10^{184}}$ if the classical
degrees of freedom are expected to emerge somehow from a more fundamental
quantum description.

Whilst the dimension of the Hilbert space $\mathcal{H}$ must be huge, it is
still assumed in this work to be finite. This assumption is based, in part,
from a desire to free the dynamics from some of the problems inherent in
infinite dimensional\ models of physics. In quantum field theory, for
example, the ultraviolet and infrared divergences occur specifically because
the momentum space is unbounded. In addition, this infinite dimensional
theory presents conceptual difficulties when confronted with the underlying
physics: a scientist performing a calculation in quantum field theory should
perhaps ask exactly what the notion of a particle of, say, infinite momentum
may mean in a physical universe of bounded size and energy. This strongly
echoes the ideas of Feynman \cite{Feynman}, who questioned the validity of
any infinite theory contained in a Universe of finite volume.

From this point of view, it therefore makes sense to remedy the problem at
the outset by limiting the size of the Hilbert space to a finite dimension.
Realistically, this should not prove to be a problem so long as it is still
sufficiently large such that every possible physically observed phenomena
may be accounted for.

\bigskip

The fourth inference that may be drawn from an extension of the standard
principles of quantum mechanics for the state of the Universe concerns its
dynamics.

In the Schr\"{o}dinger picture of conventional quantum theory, a given
statevector $\psi $ may be developed in two different types of way. The
first way is evolution by an unitary operator $\hat{u},$ which may be
thought of as a length preserving `rotation' of the vector in its Hilbert
space, i.e. $\psi \rightarrow \psi ^{\prime }=\hat{u}\psi $ for $\left| \psi
\right| =\left| \psi ^{\prime }\right| =1.$ The second way is by state
reduction, in which the wavefunction is `tested' in some sense by an
Hermitian operator $\hat{o}.$ The initial state then `collapses' or `jumps'
to a new state, which is one of the eigenstates of $\hat{o}.$

In fact, in the conventional, semi-classical treatment of the Universe, a
physical sub-system described by quantum mechanics often develops through a
series of evolutions and state reductions. Consider, for instance, a
possible ``day in the life'' of a single electron. A free electron may be
created and subsequently allowed to evolve according to the Schr\"{o}dinger
equation. The electron may propagate as a wave, until a later time when it
is measured by some sort of apparatus and observer. As an example, if the
apparatus involves a Stern-Gerlach device, the measurement process will lead
to a collapse of the electron's wavefunction into one of the spin
eigenstates associated with the Stern-Gerlach's orientation. Whichever of
these two eigenstates the electron collapses into is then taken to represent
the new state of the electron. The measurement is hence equivalent to a
preparation of an electron in either a spin-up or spin-down state, in a
particular direction.

The electron, now in a definite spin eigenstate, may then be allowed to
evolve for another length of time until a further measurement occurs. As an
example, the scientist controlling the experiment may decide that this
second measurement also involves a Stern-Gerlach apparatus. Of course, if
this second apparatus is orientated in the same direction as the first, the
result will certainly leave the electron in the same eigenstate as before.
In this case, the second measurement is equivalent to a \textit{null} test
on the electron because the state is left unchanged and no new information
has been extracted from the system. Alternatively however, if the second
apparatus is instead orientated at some angle to the first, then when the
electron is measured it will collapse into a different spin eigenstate, with
a probability dependent on the relative angle between the axes of the two
Stern-Gerlach devices.

Summarising, then, in this example a state representing a free electron has
evolved, before collapsing to a state with a definite spin component, which
has then itself been evolved, before collapsing into another state with a
different spin component. Obviously, the electron may then subsequently go
on to be involved in any number of further tests.

Or course, the development of a single electron state may appear to be a
particularly specialised or contrived example. However in the real Universe,
this sort of sequence goes on all the time. As an illustration, it should be
recalled that whenever somebody switches on a light-bulb they are
effectively starting a long chain of quantum processes, the outcome of which
is the preparation of an ensemble of quantum states that propagate until
their eventual measurement by the person's eye. In fact, this measurement
itself usually goes on to cause many different subsequent chains.\bigskip

The above process may consequently be generalised: a quantum system
initially prepared by a physicist in a state $\psi _{n}$ may proceed through
a series of evolutions $\hat{u}_{n}$ and tests $\hat{o}_{n+1},$ for $%
n=0,1,2,...$ $.$ The unitary operators $\hat{u}_{n}$ that evolve the state
are generally governed by the Schr\"{o}dinger equation and may be of the
form $\hat{u}_{n}=e^{-i\hat{H}_{n}t},$ where $\hat{H}_{n}$ is the
Hamiltonian and $t$ is a continuous time parameter as measured by the
observer. The exact forms of the Hermitian operators $\hat{o}_{n+1}$ are
chosen by the physicist depending on what she hopes to investigate, for
instance in the above example by which particular component of spin is of
interest. The system thus develops through a series of distinct steps: a
state $\psi _{n}$ may be evolved into a state $\psi _{n}\rightarrow \psi
_{n}^{\prime }=\hat{u}_{n}\psi _{n},$ which is tested by an operator $\hat{o}%
_{n+1},$ and therefore collapses into the next state $\psi _{n+1}$ which is
one of the eigenstates of $\hat{o}_{n+1}.$ This new state $\psi _{n+1}$ may
then be evolved by the operator $\hat{u}_{n+1}$ to the state $\psi
_{n+1}\rightarrow \psi _{n+1}^{\prime }=\hat{u}_{n+1}\psi _{n+1},$ which is
then tested by an operator $\hat{o}_{n+2},$ thereby collapsing it into the
next state $\psi _{n+2}$ which is one of the eigenstates of $\hat{o}_{n+2}.$
And so on.\bigskip

Any quantum experiment necessarily involves the concepts of state
preparation, evolution, and measurement. However it is only the state
reductions that are physically observed, and so it is only these collapses
that can, in any real sense, be given a physical significance. This is in
agreement with the conclusions of the Kochen-Specker theorem \cite{Kochen}
(see also \cite{Peres} for a review) and the results of Bell, which both
demonstrate that before a quantum state is measured it cannot \ be said to
have any physical attributes, such as a definite position or momentum. The
observed properties of a state do not have pre-existing values waiting to be
discovered, rather it is the actual measurement procedure and the collapse
of the state that allows physicists to discuss them. This stance was summed
up by Wheeler \cite{Wheeler1}: ``\textit{No elementary phenomenon is a
phenomenon until it is a registered (observed) phenomenon}''.\bigskip

The collapse of the wavefunction necessarily involves\ an element of change,
which in turn implies an extraction of information about the state. This,
after all, is the purpose of experimentation. Certainly, for example, the
measurement of a system does not decrease the physicist's knowledge of it,
and it is only by performing a null test on the state of the type described
earlier that the physicist's knowledge remains the same. An important point
to gain from the above example is therefore that the electron's development
may, in some sense, be parameterised in terms of information extraction.

Two conclusions may be drawn from this. Firstly, because state reduction is
manifestly a discrete process, the information is similarly extracted in
discrete manner. It is this fact that justifies the subscript $n$ on the
state $\psi _{n}$ (and hence on the operators $\hat{u}_{n}$ and $\hat{o}%
_{n+1}),$ because it is possible to directly associate the state $\psi _{n}$
with the result of the $n^{th}$ collapse.

Secondly, it is noted that the `direction' of state reduction and
information extraction is equivalent to the observed `direction' of time in
physics. This follows immediately from the logic that the state $\psi
_{n-1}, $ resulting from the $(n-1)^{th}$ test represented by the operator $%
\hat{o}_{n-1},$ must certainly have existed after the state $\psi _{n-2}$
but prior to the state $\psi _{n.}$ Consequently, the sub-script $n$ may
also be seen as a type of discrete temporal label. Thus from the point of
view of the state, time is a marker of the process of state reduction
associated with information extraction. This point will be discussed to a
great extent later.\bigskip

By extending the standard principles of quantum theory to the Universe as a
whole, the dynamics of the quantum universe are assumed to closely follow
the above analogy of the dynamics of the developing electron. One important
difference, however, is that any choice of test and any measurement of the
Universe's state must be made by the Universe itself, and not by some
external physicist. This is a consequence of the fact that if, by
definition, the Universe does indeed contain everything, the conclusion is
that there can be nothing `outside'. Thus, if it is `closed' in this way
there can be no notion of any sort of external observer engaged in the
process of evolving or measuring its state. The Universe must hence be the
perfect example of a self-developing system.\bigskip

As in the case of the electron sub-system, the development of the state $%
\Psi $ of the Universe is a discrete process due to the discontinuous nature
of the collapse mechanism. It is permissible, therefore, to label the state
immediately after the $n^{th}$ collapse as the $n^{th}$ state $\Psi _{n}.$
Further, it is the ability to label the state in this way that will be shown
to be the origin of time in the quantum Universe. For now, however, it is
noted that time is ultimately a discrete phenomenon in a universe running on
quantum principles, providing perhaps a natural starting point for future
theories of quantised gravity.

The quantum dynamics of the Universe is the way its state changes from $\Psi
_{n}\rightarrow \Psi _{n+1}\rightarrow \Psi _{n+2}\rightarrow \Psi
_{n+3}\rightarrow ...$ Moreover, and as with the above electron example, the
mechanism governing this dynamics is, at least in principle, fairly simple.

First, note that for the sake of clarity, it is possible to imagine
describing the system from the hypothetical point of view of an observer
outside of the Universe, watching the state change. Although such a point of
view is fundamentally unphysical, it is adopted for convenience, and
justified by the condition that the external observer does not interact with
the Universe's state in any way. Thus, such a privileged witness is allowed
to observe the Universe in a completely non-invasive way.

At the $n^{th}$ stage of the Universe's development, its state may be
represented by the unique vector $\Psi _{n}.$ This wavefunction may then be
evolved with some sort of unitary operator $\hat{U}_{n},$ i.e. $\Psi
_{n}\rightarrow \Psi _{n}^{\prime }=\hat{U}_{n}\Psi _{n},$ before being
`tested' by an Hermitian operator $\hat{\Sigma}_{n+1}.$ The `testing'
process is irreversible and the state $\Psi _{n}^{\prime }$ collapses into
one of the eigenstates $\Phi _{n+1}^{i}$ of $\hat{\Sigma}_{n+1}.$ In fact in
general, the operator $\hat{\Sigma}_{n+1}$ will possess $D$ orthonormal
eigenvectors, labelled $\Phi _{n+1}^{i}$ for $i=1,...,D,$ where $D$ is the
dimension of the Hilbert space $\mathcal{H}$ of $\Psi _{n},$ $\forall n.$
From this it follows that $\hat{\Sigma}_{n+1},$ and indeed $\hat{U}_{n},$
may both be represented by $D\times D$ matrices for all $n.$

The relationship between $\Phi _{n+1}^{i}$ and $\hat{\Sigma}_{n+1}$ obeys
the usual eigenvector equation, viz.,
\begin{equation}
\hat{\Sigma}_{n+1}\Phi _{n+1}^{i}=\lambda ^{i}\Phi _{n+1}^{i},
\end{equation}
where $\lambda ^{i}$ is the eigenvalue of the $i^{th}$ eigenvector $\Phi
_{n+1}^{i}$ of $\hat{\Sigma}_{n+1}.$

Further, given a state $\Psi _{n}^{\prime },$ the probability $P(\Psi
_{n+1}=\Phi _{n+1}^{j}|\Psi _{n}^{\prime })$ that the next state $\Psi
_{n+1} $ will be a particular eigenvector $\Phi _{n+1}^{j}$ of $\hat{\Sigma}%
_{n+1}$ is determined in the usual way as the square of the modulus of the
probability amplitude, i.e.
\begin{equation}
P(\Psi _{n+1}=\Phi _{n+1}^{j}|\Psi _{n}^{\prime })=\left| \langle \Phi
_{n+1}^{j}|\Psi _{n}^{\prime }\rangle \right| ^{2}.  \label{stateprob}
\end{equation}

The forms of the operators $\hat{U}_{n}$ and $\hat{\Sigma}_{n+1}$ are
discussed later.

The result $\Phi _{n+1}^{j}$ of the test $\hat{\Sigma}_{n+1}$ is now
associated with the preparation of a new state $\Psi _{n+1},$ which is
subsequently evolved by an operator $\hat{U}_{n+1}$ to the state $\Psi
_{n+1}^{\prime }=\hat{U}_{n+1}\Psi _{n+1},$ before being tested by an
operator $\hat{\Sigma}_{n+2}$ and collapsing to one of its $D$ orthonormal
eigenvectors $\Phi _{n+2}^{i},$ $i=1,..,D.$ And so on.

Summarising, the Universe runs as an automatic process of state preparation,
evolution and collapse. To this end, the Universe is envisaged to be a
completely self-contained quantum automaton.

\bigskip

As noted earlier, if the Universe contains everything, there can be no
notion of any sort of external observer engaged in the process of developing
or measuring its state. At first glance, therefore, this may appear at odds
with the traditional quantum mechanical tenets of state preparation and
testing, and this has prompted some authors to criticise the possibility of
a completely quantum universe. In fact, there are three obvious points that
need addressing in any attempt to treat the Universe as a closed quantum
system.

Firstly, if there are no external observers, then, as argued by Fink and
Leschke \cite{Fink}, how can the Universe be measured? In what sense,
therefore, can it be described as a quantum system?

Secondly, and again from \cite{Fink}, if there is only one Universe and it
only `runs' once, what is the meaning of statistically derived probabilities
of the form (\ref{stateprob})? In particular, by definition a description of
the Universe's state must involve a description of everything contained
within it. Moreover, any measurement of the state of the Universe by some
sort of detection apparatus necessarily changes the detector's state. But,
since this apparatus is part of the Universe, such a measurement immediately
implies that the state of the Universe is itself changed during this
procedure. It is consequently impossible to measure the same state of the
Universe twice. So, from the point of view that quantum mechanics deals with
the probability distributions of the results of repeated measurements of
observables (either the same state measured a number of times, or a number
of identical states\ each measured once) the argument of Fink and Leschke is
that the rules of quantum mechanics are not applicable to the universe as a
whole.

It is also noted that in conventional quantum theory, states evolve
according to the time dependent Schr\"{o}dinger equation. Thirdly, then, if
there are no external parameters such as time, how does the Universe evolve
as a quantum state?

These points will be discussed briefly here, though their explanations will
become clearer throughout the course of this work, and particularly in
Chapter 8. In general, the lesson learnt is that care is needed when
directly applying the quantum mechanics of states in the laboratory to the
special case where the state in question is the state of the entire
Universe.\bigskip

The standard principles of quantum mechanics were discovered by physicists
based on laboratory observations of relatively tiny sub-systems of the
universe, for example from the photoelectric effect induced in a small lump
of metal, or the measurement of the spin of a single electron. The typical
approach to an experiment involving quantum principles is to draw a dividing
line between the observer and the observed: the scientist produces an
isolated quantum state, allows it to evolve, and then chooses an Hermitian
operator with which to test it. Whilst this is manifestly a semi-classical
construction, it is normally a fairly accurate analysis because the
scientist \textit{is} sufficiently large such that classical mechanics
provides a good approximation, and it is not always difficult in practice to
produce a quantum state that \textit{is} effectively isolated from the rest
of the universe.

However, any semi-classical treatment can only ever be just an approximation
to a reality that is fully quantum in nature. After all, recall that the
quantum state under investigation can be arbitrarily large. From this point
of view it is in principle possible to segregate the universe into two
parts: the observer sub-system, and the sub-system comprising everything
else. Given that it is possible to treat the `everything else' sub-system as
a quantum state, it seems unreasonable to expect that the Universe is really
a semi-classical product of an enormous quantum sub-system containing
everything apart from a single classical observer.

Be that as it may, such an approach of an observer standing outside of the
experiment being observed could be described as exo-physical. It is from
this context that the usual rules of quantum mechanics were determined,
including in particular those contributing to the conventional `Measurement
Problem'.\bigskip

However, what this approach does not take into account is the fact that the
physicists performing the experiments are themselves an integral part of the
Universe they are trying to analyse. From this perspective, a laboratory
experiment is actually equivalent to one part of the Universe measuring
another part. Consequently, whether the true nature of the Universe is
fundamentally quantum, classical or anything else, it must be an example of
a system that is able to examine \textit{itself}. This is therefore an
endo-physical perspective, in which the observer is part of the system being
observed.

The point is that what a physicist may conventionally believe to be an
exo-physical measurement of a quantum sub-system of the Universe by an
apparently external semi-classical observer, should perhaps really be viewed
as an endo-physical measurement of one part of the Universe appearing to
observe another part of itself. Thus, such a self-referential quantum system
may not necessarily be restricted to quantum dynamics relying on external
observers, because the dependence of the standard laws of quantum mechanics
on external observers was only ever derived from the potentially incomplete
viewpoint of exo-physical analyses of physical phenomena. These laws may
therefore not be directly applicable to the Universe as whole. If the
dynamics of the state should instead be described from an endo-physical
point of view, the Universe must be a quantum system that relies on internal
observations; there is hence neither a need nor a place for an external
observer to measure and collapse the state.

Of course exactly how a quantum universe observed from the inside by
endo-physical observers may give rise to internal Measurement problem type
phenomena, such as emergent semi-classical physicists believing they are
observing an external quantum reality, is a difficult question to be
addressed. In fact, the endo-physical measurement problem is discussed more
fully in Chapter 6, whilst in Chapter 8 some simple toy-models are given
that describe how a simple endo-physical dynamics may be achieved.

For now, however, note that in answer to the criticism of a fully quantum
universe given in \cite{Fink}, an analogy is drawn with the argument of
G\"{o}del \cite{Godel} (see also \cite{Penrose}\cite{Penrose1}) that it is
impossible to determine whether a given set of mathematical rules is
self-consistent using just those rules alone. Whilst this may be the case,
it does not imply that the rules themselves are wrong, merely that it is
problematical to demonstrate their validity from the `inside'.

Overall, if the Universe must be described by quantum principles, yet cannot
support any external observers, the conclusion must be that it is somehow
able to prepare, evolve and test itself. Further, these measurements are
made by different sub-systems inside the Universe, and indicate a relative
change between them. This point will be discussed in due course.\bigskip

Fink and Leschke's second argument is philosophically identical to asking
about the\ meaning of the probability of obtaining a particular random
result from a set of possibilities if an experiment is only ever performed
once.

Consider as an example a classical coin toss experiment, noting that similar
restrictions apply to any other physical situation, from atomic decays to
measuring the spin of an electron. Ignoring the possibility of the coin
landing on its edge, it may be generally accepted that the probability of
getting a `heads' result is equal to the probability of getting `tails',
that is
%TCIMACRO{\UNICODE{0xbd}}%
%BeginExpansion
$\frac12$%
%EndExpansion
. This probability, however, only arises from a mathematical abstraction. To
actually be empirically sure of the probability either requires the same
coin to be tossed an infinite number of times, or an infinite number of
coins to be tossed once. Of course, this in unphysical. In the first
instance, it would take an infinite length of time to get the result.
Additionally, each flip would undergo slightly alternative conditions, from
different initial forces, to miniscule air currents, or even the possibility
of being deflected slightly by a stray photon. It is even debateable as to
what condition the coin would be in after it had been struck a million
times. In the second instance, it could not be ensured that all the coins
were identical or flipped under the same conditions. Equally, an infinite
number of coins would require an infinite space and would possess an
infinite mass, and so, according to general relativity, would curve infinite
space infinitely.

Nevertheless, such an incomplete knowledge does not prevent a probability
measure being placed on any result. Instead, the probability is defined
relative to obtaining a particular result from a certain number of given
conditions. It may be asked, for example, what the probability is of
obtaining a head, given that there are two potential outcomes and that the
system is not biased towards either one of them. In this sense, the
probability is \textit{defined} as
%TCIMACRO{\UNICODE{0xbd}}%
%BeginExpansion
$\frac12$%
%EndExpansion
. The corresponding unphysical situation is consequently imagined
implicitly, by assuming that \textit{if} an infinite number of coins were
tossed under identical conditions, \textit{then}
%TCIMACRO{\UNICODE{0xbd} }%
%BeginExpansion
$\frac12$%
%EndExpansion
of them would come up heads. If, however, a coin is only ever flipped once
and gives a heads result, it is not immediately concluded that the result
was deterministic, and that probability may not be used.

The same is true in quantum cosmology. In the case of the Universe
represented by the state $\Psi _{n},$ the probability is defined relative to
the set of $D$ potential future states $\Psi _{n+1}=\Phi _{n+1}^{i},$ for $%
i=1,...,D,$ that are the $D$ normed eigenvectors of the operator $\hat{\Sigma%
}_{n+1}$. Again, it is possible to devise hypothetical situations in which
if an infinite number of identical universes in the state $\Psi _{n}$ were
measured, then a fraction $|\langle \Phi _{n+1}^{j}|\Psi _{n}\rangle |^{2}$
of them would give the particular result $\Phi _{n+1}^{j},$ but this is just
an attempt to attach empiricism onto a mathematical definition. Probability
in the Universe is synonymous with potentiality; the fact that only one of
these eigenstates is actually realised as the next state $\Psi _{n+1}$ does
not mean that the universe proceeds deterministically, any more than if the
spin component of just one electron is measured and found to be `up' it
implies that the electron must be described using classical hidden
variables. Indeed if the Universe is not a random, quantum system, the
question would remain as to how it could therefore evolve deterministically
given that the component quantum sub-systems of which it is comprised are
clearly stochastic.\bigskip

As will be expanded upon in the following chapters (particularly in Chapters
5 and 8), continuous time is taken to be a phenomenon that emerges in a
fully quantum universe as its state proceeds through a long series of
evolutions and collapses. Physical space, and the momenta and energies etc.
of particular sub-systems, will also be shown to originate from
considerations of the properties of this state $\Psi ,$ which is assumed to
obey the laws of conventional quantum dynamics. However, whilst conventional
quantum states in the laboratory evolve according to the continuous time
Schr\"{o}dinger equation in a way dependent upon their Hamiltonians, if the
Universe is taken to possess no intrinsic concepts such as time or energy,
exactly what role the Schr\"{o}dinger equation plays in its evolution
becomes an important question.

In response to this third criticism of quantum cosmology, it should be
recalled that the $n^{th}$ state $\Psi _{n}$ of the Universe as discussed so
far is simply defined as nothing but a vector in a Hilbert space $\mathcal{H}%
.$ Its development is consequently only meaningful in terms of mathematical
mappings of this vector, for example by norm preserving `rotations' due to
unitary evolutions, or by discontinuous jumps into another vector in $%
\mathcal{H}$ that is one of the eigenstates of an Hermitian operator. Care
must be taken, therefore, not to attach to this state too many of the
notions normally associated with emergent physical concepts, such as direct
questions of how `spatially long' this vector might be or how much `mass' it
has. In fact, exactly how the state could ultimately give rise to physics is
a central theme of this thesis.

Recalling the discussion of exo-physics given above, it should be remembered
that the Schr\"{o}dinger equation is something scientists have discovered
that appears to describe the evolution of physical quantum sub-systems.
However, physical phenomena tend only to be witnessed by observers in the
emergent, semi-classical regime. Consequently, the Schr\"{o}dinger equation
has only been defined as an emergent construct used to describe other
emergent phenomena evolving in emergent time, namely, physical states in the
laboratory.

So, since the continuous time Schr\"{o}dinger equation was discovered in the
emergent limit, it cannot automatically be expected to describe the
fundamental, pre-emergent dynamics of the state $\Psi _{n};$ its emergent
definition does not necessarily imply that it has to be held as a
fundamental law that describes the development of the Universe as a whole.
In fact, the only constraint placed on whatever laws are chosen to evolve $%
\Psi _{n}$ is that they must correctly reproduce every physically observed
phenomenon in the semi-classical limit. The laws of emergent physics may
themselves emerge from more fundamental laws governing the mathematical
transformations of the state.

So, the state $\Psi _{n}$ of the Universe described in this work is
`quantum' in the sense that it is a vector in a Hilbert space, and is
subject to unitary transformation and to wavefunction collapse by Hermitian
operators. This will be elaborated upon throughout the following.

\bigskip

\subsection{The Stages Paradigm}

\bigskip

The stages paradigm was proposed in \cite{EJ} in an attempt to draw together
the observations and conclusions of the previous two sub-sections into a
mathematical framework that describes the properties and dynamical evolution
of a fully quantum universe. The proposal represents a certain minimum
number of parameters required to describe the development of the quantum
Universe, and follows from the fact\ that a specification of the state $\Psi
_{n}$ alone cannot fully define its dynamics.\bigskip

To illustrate the idea, an analogy is drawn as before with the conventional,
semi-classical treatment of the single electron experiment introduced in the
previous sub-section. A full description of such an experiment necessarily
contains a number of features. Firstly there is the quantum state of the
electron itself, represented by a vector $\psi $ in a Hilbert space.
Secondly, with the experiment is associated some sort of `information
content'. This information may, for example, include details of the
Hamiltonian of the free electron, the choice of the experiment to be
performed on the state (e.g. the possible orientations of the Stern-Gerlach
apparatus), or even a memory of where the particular state came from or how
it was prepared. Lastly a set of rules are required in order to describe
exactly how the system develops, for example how the Schr\"{o}dinger
equation may govern the propagation of the electron as a wave, or a
statement of how the inhomogeneous magnetic field of the Stern-Gerlach
apparatus will perturb the Hamiltonian according to the spin of the
electron. The rules are hence equivalent to the laws of physics relevant to
the current situation.

As the experiment develops it progresses through a number of distinct
stages. The initial stage, for example, might be defined as the one
containing the newly created free electron. The next stage, then, might be
defined as the period in which the electron has been measured by the first
Stern-Gerlach apparatus, but has not yet encountered the second. Finally, in
the third stage of the experiment's development the electron has passed
through the second apparatus also. In such a picture it is the measurement
of the state of the electron that marks the end of one stage and the
beginning of the next; each collapse of a state in one stage is equivalent
to the preparation of the state for the next stage. This is another
manifestation of the principle that only when information is actually
extracted from a state can it be given any real, physical significance.

Each stage of the experiment's development is clearly associated with its
own unique state, an information content and a set of rules describing the
system. Moreover, some or all of these will change as the system progresses
from one stage to the next. For instance, the state of the newly created
electron in the first stage is clearly different from the state representing
the electron in the third stage, because in the latter case the electron has
been prepared in a particular spin eigenstate. Similarly the information
regarding the actual choice of the next test is different from the first
stage to the second, because the orientations of the Stern-Gerlach
apparatuses are not the same. Equally, any information regarding the
previous test is different from the first stage to the second, because the
states in the first and second stages are prepared in different ways. Thus,
each stage of the experiment's development may be said to be completely
parameterised by the current state, information and rules of the system, and
changes in these, when the wavefunction collapses, define the development of
the system from one stage to the next.\bigskip

By extending the above argument, the conjecture is that the Universe also
progresses through a series of distinct stages, with the divide between one
stage and the next occurring as the Universe's state collapses due to its
self-measurement. Given that the state in each stage is necessarily
different from the state contained in the previous stage, it is permissible
to extend the label $n$ defining the $n^{th}$ state $\Psi _{n}$ to the stage
itself. It is hence possible to define the $n^{th}$ stage $\Omega _{n}$ by
\begin{equation}
\Omega _{n}\equiv \Omega (\Psi _{n},\text{ }I_{n},\text{ }R_{n})
\end{equation}
that is, each stage is a function of the current state, $\Psi _{n},$
Information content, $I_{n},$ and the Rules $R_{n}.$ These are explained in
turn.\bigskip

As described previously, the wavefunction $\Psi _{n}$ is a pure state
represented by a complex vector in a Hilbert space of enormous, but finite,
dimension $D.$ From the dynamics of this state is expected to emerge
classical physics and all of the features in the physical Universe
associated with this, including for example time, space, and particle
physics. The state $\Psi _{n}$ is assumed to represent the product of the
sub-states of every quantum sub-system contained in the Universe (as will be
expanded upon in Chapter 4). Thus a change of just one of these sub-states,
for example a tiny part representing a physicist measuring a tinier part
representing an electron, implies a change in the overall state of the
Universe. Consequently, the change of just one sub-state constitutes a
progression of the Universe from one stage to the next, even though nearly
all of the other sub-systems might appear unaffected by the jump.

In practice, it is expected that very many sub-states might change as the
real Universe jumps from the state $\Psi _{n}$ to the state $\Psi _{n+1}\neq
\Psi _{n},$ corresponding to very many physical sub-systems appearing to
simultaneously observe other physical sub-systems. In general, one, some, or
all of the sub-systems might change as the stage develops from $\Omega _{n}$
to $\Omega _{n+1}.$\bigskip

$I_{n}$ contains the necessary information required for the state's
development. Specifically, $I_{n}$ could incorporate a set of $A$ unitary
operators $\{\hat{U}_{n}^{a}:a=1,...,A\},$ one of which might be chosen to
`rotate' the state $\Psi _{n},$ and also a set of $B$ potential Hermitian
operators $\{\hat{O}_{n}^{b}:b=1,...,B\}$ that represent the different
possible ways that $\Psi _{n}$\ could be tested; one of the set $\{\hat{O}%
_{n}^{b}\}$ which will hence become $\hat{\Sigma}_{n+1}.$ Equivalently,
then, because with each of the $\hat{O}_{n}^{b}$ is associated a basis set
of $D$ orthonormal eigenvectors, $I_{n}$ also defines the set of possible
next states $\Psi _{n+1}.$ Paraphrasing, if only one of the $B$ operators
may be selected, and because each of these has $D$ eigenstates due to the
dimensionality of $\Psi _{n},$ the Information content $I_{n}$ dictates that
the next state $\Psi _{n+1}$ will be one of a set of $(B\times D)$ members,
though there may be a great deal of degeneracy in this set because some (but
not all) of the eigenstates of $\hat{O}_{n}^{i}$ might be the same as some
of the eigenstates of $\hat{O}_{n}^{j}.$ Of course, until one of the
operators $\hat{O}_{n}^{i}$ is chosen to be $\hat{\Sigma}_{n+1},$ and until
the state reduction actually occurs, it is completely unknowable as to which
of this set the subsequent state $\Psi _{n+1}$ will be.

It is further possible that $I_{n}$ may also include information about the
properties of previous stages. It might, for example, contain a record of
what the previous state $\Psi _{n-1}$ was like; or possibly the two previous
states $\Psi _{n-1}$ and $\Psi _{n-2},$ or even the states $\Psi
_{n-1},...,\Psi _{n-x}$ for some large $x.$ Likewise, some sort of list
might be present in $I_{n}$ that details the chain of operators, $\hat{U}%
_{n-y}$ and $\hat{\Sigma}_{n+1-z}$ for $y,z\geq 1,$ that were used as the
Universe progressed through the chain of stages. In this sense, the current
Information $I_{n}$ may be seen as a form of memory of earlier stages, and
might be used to track correlations from one stage to the next. An analogy
here is with the human memory, in which\ the current `state' of the brain
often includes a recollection of its past `states', or with a computer that
is able to store information about past steps of a calculation for later use.

Information about the past may be used in the dynamics to enforce
constraints on future states of the Universe. It might, for example,
influence which of the set of $B$ operators $\{\hat{O}_{n}^{b}\}$ is
actually selected to be the next test $\hat{\Sigma}_{n+1}.$ Of course, this
type of development is really no different from how an experiment is often
conducted in the laboratory: given that a physicist knows that she has just
tested a sample with $X$ and $Y$ and obtained a certain state $\psi _{XY},$
she may decide that it must next be investigated with $Z,$ thereby selecting
just one test out of a number of possibilities.\bigskip

Using the past to influence how the present could develop into one of a set
of possible futures has the potential to introduce an element of order into
the dynamics of the Universe. As an illustration, imagine a universe that
chooses a particular operator $\hat{\Sigma}_{n+1}=\hat{O}_{n}^{j}$ to test
its state $\Psi _{n}$ for the sole reason that $\hat{O}_{n}^{j}$ has
eigenstates `similar', in some sense, to the previous operator $\hat{\Sigma}%
_{n}$ that prepared $\Psi _{n}.$ This would perhaps ensure that $\Psi _{n+1}$
resembles $\Psi _{n}$ to some extent, and if the process continued it might
lead to a situation in which features of the universe appeared to persist
from one stage to the next. If physical states did indeed develop in this
manner, with the present incorporating information about the past, the
mechanism might be speculated to be a root of why the real Universe appears
to look so similar over certain scales.

Generalising the above, it is possible to re-parameterise the $n^{th}$ stage
of the Universe as (\ref{stage}), where $a=1,...,A,$ \ $b=1,...,B,$ and $%
x,y,z\geq 0.$
\begin{equation}
\Omega _{n}\equiv \Omega \left(
\begin{array}{c}
\Psi _{n},\text{ \ \ }{\Large [}\{\hat{U}_{n}^{a}\},\text{ }\{\hat{O}%
_{n}^{b}\},\text{ }\{\{\Psi _{n-1},...,\Psi _{n-x}\},\{\hat{U}_{n-1},...,%
\hat{U}_{n-y}\}, \\
\{\hat{\Sigma}_{n+1-1},...,\hat{\Sigma}_{n+1-z}\}\}{\Large ]},\text{ \ \ }%
R_{n}
\end{array}
\right)  \label{stage}
\end{equation}
\bigskip

The Rules $R_{n}$ are the laws dictating the dynamics obeyed by the
Universe. The Rules specify that, given a state $\Psi _{n},$ it will be
evolved with an operator $\hat{U}_{n}$ and tested with a particular operator
$\hat{\Sigma}_{n+1}.$ Equivalently the Rules are used to select, to act on $%
\Psi _{n},$ one of the $A$ possible unitary operators\ $\hat{U}_{n}^{a},$ $%
a=1,...,A,$ and one of the $B$ possible Hermitian operators\ $\hat{O}%
_{n}^{b},$ $b=1,...,B,$ out of the set of all possible operators contained
by $I_{n}.$

Exactly how a particular operator is selected, i.e. what mechanism the Rules
use to determine which member of the $A$ or $B$ possibilities is chosen,
remains a difficult question for the future. Indeed whether this choice is
deterministic, or itself the result of some random quantum process, is an
important issue to be addressed. It is even possible that the Rules make
reference to additional factors included in the Information $I_{n},$ such
that perhaps the presence of a particular $\Psi _{n-r},$ $\hat{U}_{n-s}$ or $%
\hat{\Sigma}_{n+1-t}$ in $I_{n}$ might lead to the selection of a particular
$\hat{U}_{n}^{a}$ or $\hat{\Sigma}_{n+1}.$ This possibility will be
addressed many times throughout this work, and especially in Chapter 8, and
may be necessary to account for many of the features present in the
physically observed Universe.

In fact, it is also conceivable that the Rules $R_{n-1}$ used to choose the
operator $\hat{\Sigma}_{n}$ are not the same as the Rules $R_{n}$ used to
choose the operator $\hat{\Sigma}_{n+1}.$ In other words, the Rules
themselves may be subject to dynamical development according to some higher
order set of ``Rules of the Rules'' \cite{Buccheri}, and in this case such
an additional `Meta-Rule' would also need to be incorporated into the
definition (\ref{stage}) of a stage. In Chapter 8 an attempt has been made
to find simple Rules that reproduce certain required features of
dynamics.\bigskip

All of physics is expected to emerge from the fundamental quantum picture of
the Universe described in this Chapter. It is reasonable to assume,
therefore, that the Rules $R_{n}$ at each stage must be very carefully
refined in order to produce a classical looking Universe that appears to run
according to ordered and well defined laws of physics. The Rules must ensure
that every phenomenon that physicists experience on the emergent scale is
accounted for from the fundamental quantum level as the Universe jumps from
one stage to the next. For example, if from a particular state $\Psi _{n}$
appears to emerge a physical Universe of enormous spatial size that appears
to be describable by $3+1$ dimensional general relativity and appears to
contain very many distinct protons, electrons, neutrons etc. that have
clumped together in huge lumps resembling galaxies, stars and planets, and
if further, on one of these planets, plants, animals, and humans have
appeared and evolved, and that some of these humans have constructed giant
buildings and complicated machinery in order to measure the Universe they
believe they exist in as semi-classical observers, then it is reasonable to
hope that from the next state $\Psi _{n+1}$ all of these features will also
emerge, instead of, say, something totally different or even just complete
disorder. Since it seems to be an observational fact that the physical
Universe appears to change very little from one stage to the next (it will
be shown in Chapter 4 that this is itself perhaps unexpected), it may be
conjectured that the Rules, and consequently the dynamics, must be very
finely tuned in order to choose an operator $\hat{\Sigma}_{n+1}$ with an
eigenvector so similar to $\Psi _{n}.$ In a quantum universe approximated by
continuous and emergent classical laws there must be some sort of underlying
`similarity theorem' that ensures that $\Psi _{n+1}$ is not too different
from $\Psi _{n}.$\bigskip

The standard laws of physics discovered in the laboratory are also
presumably emergent from the Rules $R_{n}$ describing the dynamics of the
Universe. As an example, consider a stage of the Universe in which, at one
instant, it appears that from an emergent, classical and large scale point
of view, part of the state $\Psi _{n}$ may be considered to describe two
electron sub-systems. Further assume that, from this emergent, classical and
large scale point of view, scientists have defined a measure of distance and
observed that the two electrons are in close proximity\footnote{%
The emergence of space and the emergence of particles are investigated in
Chapters 5 and 7 respectively.}. Whatever the dynamics may be that actually
govern the Universe on the fundamental level, they might be expected to
ensure that from the next state $\Psi _{n+1}$ would emerge a picture in
which the two electrons appear slightly farther away from each other, again
from the classical and large scale point of view of a scientist inside the
Universe. Continuing, in the state $\Psi _{n+2}$ following this the two
electrons might be even farther apart. Thus, by observing the way in which
the state of the Universe appears to change from one stage to the next,
emergent physicists are able to derive emergent laws to describe emergent
phenomena, such as ``Like charges repel''.

Conversely, by studying these emergent laws of physics governing the
physically observed Universe, it might be able to place certain constraints
of the actual Rules $R_{n}$ governing the development of the underlying
state\ from stage to stage.

\bigskip

With the above considerations in mind, it is possible to speculate now on
the necessary sequence of events that might define the dynamical development
of the Universe. Consider a fully quantum Universe, completely specified at
the $n^{th}$ step of its development by a stage $\Omega _{n}\equiv \Omega
(\Psi _{n},$ $I_{n},$ $R_{n}).$ The Information content, $I_{n},$ contains\
a set of possible unitary operators, $\{\hat{U}_{n}^{a}:$ $a=1,...,A\},$ and
a set of possible Hermitian operators, $\{\hat{O}_{n}^{b}:b=1,...,B\},$ each
of which is associated with a basis set of $D$ orthonormal eigenvectors, $%
\Phi _{n+1}^{b,k}$ for $k=1,...,D,$ as well perhaps as some sort of `record'
of previous stages.

According to the specific Rules $R_{n}$ governing the Universe, and possibly
making reference to the current Information $I_{n},$ one of the operators $%
\hat{U}_{n}^{i}$ is chosen to act on the wavefunction, and the state
undergoes unitary evolution. This evolution is effectively a rotation of the
$D$ dimensional vector $\Psi _{n}$ in its Hilbert space $\mathcal{H},$ viz. $%
\Psi _{n}\rightarrow \Psi _{n}^{\prime }=\hat{U}_{n}^{i}\Psi _{n}.$ Note
however that depending on the dynamics, the chosen operator $\hat{U}_{n}^{i}$
may be equal to the identity operation $\hat{I}$ such that
\begin{equation}
\Psi _{n}\rightarrow \Psi _{n}^{\prime }=\hat{U}_{n}^{i}\Psi _{n}=\hat{I}%
\Psi _{n}=\Psi _{n}.
\end{equation}

Next the Rules select, from the set $\{\hat{O}_{n}^{b}\}$ defined in $I_{n},$
one of the Hermitian operators; say, $\hat{O}_{n}^{j}.$ This is equivalent
to the Universe choosing a test to perform on its state. The state
consequently collapses into one of the eigenvectors of $\hat{\Sigma}_{n+1}=%
\hat{O}_{n}^{j},$ effectively preparing the next state $\Psi _{n+1}.$ The
probability that the new state $\Psi _{n+1}$ will be the particular $l^{th}$
eigenstate $\Phi _{n+1}^{j,l}$ of $\hat{\Sigma}_{n+1}$ is given by\ $P(\Psi
_{n+1}=\Phi _{n+1}^{j,l}|\Psi _{n}^{\prime })=|\langle \Phi
_{n+1}^{j,l}|\Psi _{n}^{\prime }\rangle |^{2}.$\bigskip

Details about the particular choice of operators $\hat{U}_{n}^{i}$ and $\hat{%
\Sigma}_{n+1}=\hat{O}_{n}^{j}$ may then be included in the new Information
content $I_{n+1},$ which may also provide a record of the previous state $%
\Psi _{n}.$ In fact, some of the `old' Information content $I_{n}$ may also
be subsumed into the new $I_{n+1}.$ This inclusion may be whole, $%
I_{n}\subset I_{n+1},$ partial $I_{n}\cap I_{n+1}\not\equiv I_{n},$ or even
not at all $I_{n}\cap I_{n+1}=\emptyset ,$ where in the last instance the
new stage could be said to contain no knowledge whatsoever of its `history'.

In fact, the cases in which\ $I_{n+1}$ does not completely encompass $I_{n}$
necessarily imply an irreversible loss of information. Evidently, the
`direction' of the loss of information as the Universe develops from one
stage to the next is the same as the `direction' of time in the model,
because both are based on the `direction' of the state collapsing, i.e. from
$\Psi _{n}$ to $\Psi _{n+1}.$ It is noted, moreover, that the idea of an
irreversible loss of information is strongly analogous to the notion of an
increasing entropy, and in this case it is recalled that the `direction' of
increasing entropy (which is equivalent to the `direction' of the
irreversible increase in disorder of a system as it evolves) defines the
arrow of time in thermodynamics. From these viewpoints, the direction of
time in the Universe is seen as identical to the direction of increasing
ignorance of exactly how the Universe came to have the state it has.

A second point about partial inclusions of information is the fact that no
observer in the present can ever be sure of just how complete or reliable
their information is regarding what past stages might have been like. Since
past states can only ever be reconstructed based on whatever information
about them has survived into the present, if this information survival is
incomplete then the reconstruction of the past can, at best, only be
partial. This conclusion reinforces the idea that physicists can only ever
be truly certain of the current stage of the universe. As is consistent with
the idea of Process time, only the present can be given any real
existence.\bigskip

The new information content $I_{n+1}$ will contain a new set of possible
operators $\{\hat{U}_{n+1}^{a^{\prime }}\},$ $\{\hat{O}_{n+1}^{b^{\prime
}}\},$ where $a^{\prime }=1,...,A^{\prime }$ and $b^{\prime
}=1,...,B^{\prime }.$ The actual members of these sets may be based somehow
upon the random choice of the new state $\Psi _{n+1},$ or on parts of
previous operators or states, and will go on to provide the dynamics for the
next stage. This next stage is clearly parameterised as $\Omega
_{n+1}=\Omega (\Psi _{n+1},I_{n+1},R_{n+1}),$ where the Rules $R_{n+1}$
governing the Universe may also have changed, $R_{n+1}\not\equiv R_{n},$
according to any Rules of the Rules.

Overall, the Universe has developed in a discrete quantum manner from one
stage $\Omega _{n}$ to the next $\Omega _{n+1}.$ This process is expected to
continue indefinitely in a completely self-contained and automatic way. All
of physics, including the dynamics of microscopic and macroscopic
sub-systems evolving against a backdrop of continuous space and time in an
apparently classical looking Universe, is expected to emerge from the
dynamics of this self-referentially developing series of stages.

Exactly how this might occur will form the basis of the remaining chapters
of this work.

\bigskip \newpage

\section{Classicity from Quantum}

\renewcommand{\theequation}{4.\arabic{equation}} \setcounter{equation}{0} %
\renewcommand{\thetheorem}{4.\arabic{theorem}} \setcounter{theorem}{0}

\bigskip

Given that the physical Universe appears to look like an enormous collection
of semi-classical sub-systems, yet the conclusion of the previous chapter is
that it is fundamentally quantum in nature, an immediate question to be
addressed is: how does apparent classical physics emerge from the fully
quantum reality? If the Universe is described by a complex statevector $\Psi
,$ what properties of this state might give rise to semi-classical looking
degrees of freedom?

In an attempt to understand this issue, it is beneficial to reverse part of
the question and define what is meant by classicity. To this end, it is
observed that classicity is in some sense synonymous with
distinguishability; if a set of objects $\{A,B,C\}$ are described as
classical, it implies that it is possible to make distinctions between them.
These distinctions may, for example, be in terms of an observer's ability to
determine that the objects have different physical properties or that they
are positioned at different spatial locations.

If $A,$ $B$ and $C$ can be distinguished, it follows that they may each be
assumed to possess an isolated existence, and may be discussed independently
of one another. From this viewpoint, classicity is therefore a way of
expressing the observation that \textit{this} object with \textit{these}
qualities is \textit{here, }whereas\textit{\ that} object with \textit{those}
properties is\textit{\ there}. Certainly this is a criterion met by all
macroscopic semi-classical states in physics, where for example a particular
large scale apparatus is always assumed to be separate from the quantum
state it is measuring, and does always have an independent existence and a
well defined position.\bigskip

As a consequence of the above, a state in classical mechanics representing a
set of classical objects can always be separated into the distinct
sub-states of which it is comprised. The same is not true in quantum theory,
because the phenomenon of entanglement represents a breakdown of this
ability to separate a system into independent and distinct physical
sub-systems. When two (or more) quantum states become entangled they can no
longer be given any independent existence, and instead it is only by taking
the entire state as a whole that the system can be given any physical
significance. The EPR experiment \cite{EPR}\ provides a famous example of
this.

It is, however, an important fact that a class of states exist in quantum
mechanics that are not entangled. Separable states (to be defined below)
represent situations in which it \textit{is} permissible to segregate the
quantum state into a set of sub-states. Further, because it is possible to
develop and measure the factor sub-states of these vectors independently of
each other, such sub-states may be distinguished. Thus, separable states in
quantum mechanics allow physicists to discuss their constituent parts,
because the factors of a separable state possess a degree of individuality.
Since this is one of the requirements for classicity, the conjecture is that
it is separable states that provide a necessary starting point for the
emergence of semi-classical degrees of freedom.\bigskip

As a simple illustration, consider two Hilbert spaces $\mathcal{H}_{\phi }$
and\ $\mathcal{H}_{\varphi }.$ Consider also a third Hilbert space $\mathcal{%
H}_{[\phi \varphi ]}$ formed by taking the tensor product of $\mathcal{H}%
_{\phi }$ and $\mathcal{H}_{\varphi },$ i.e. $\mathcal{H}_{[\phi \varphi ]}=%
\mathcal{H}_{\phi }\otimes \mathcal{H}_{\varphi }.$ This third vector space $%
\mathcal{H}_{[\phi \varphi ]}$ may be described as \textit{factorisable},
with the sub-spaces $\mathcal{H}_{\phi }$ and $\mathcal{H}_{\varphi }$ being
its \textit{factors}.

Now define two states $\psi _{a}$ and $\psi _{b},$%
\begin{eqnarray}
|\psi _{a}\rangle _{\phi \varphi } &=&|\phi _{1}\rangle _{\phi }\otimes
|\varphi _{2}\rangle _{\varphi } \\
|\psi _{b}\rangle _{\phi \varphi } &=&|\phi _{1}\rangle _{\phi }\otimes
|\varphi _{2}\rangle _{\varphi }+|\phi _{2}\rangle _{\phi }\otimes |\varphi
_{1}\rangle _{\varphi },  \notag
\end{eqnarray}
where $\psi _{a}$ and $\psi _{b}$ are vectors in the product space $\mathcal{%
H}_{[\phi \varphi ]},$ i.e. $\psi _{a},\psi _{b}\in $ $\mathcal{H}_{[\phi
\varphi ]},$ but $\phi _{i}\in $ $\mathcal{H}_{\phi }$ and $\varphi _{j}\in
\mathcal{H}_{\varphi }$ for $i,j=1,2.$

Clearly, the state $\psi _{a}$ is separable into a product of factors, one
of which, $\phi _{1},$ is in the Hilbert space $\mathcal{H}_{\phi }$ and the
other, $\varphi _{2},$ is in\ $\mathcal{H}_{\varphi }.$ However, if it is
assumed that $\phi _{1}$ is not a linear multiple of $\phi _{2},$ and
similarly that $\varphi _{2}$ is not a linear multiple of $\varphi _{1},$ no
such separation is possible for the entangled state $\psi _{b}.$

Now, if a quantum system is prepared in the separable state $\psi _{a},$ it
is possible to measure one factor of it whilst leaving the other factor
unchanged. The state $\psi _{a}$ may, for instance, be tested by an operator
$\hat{O}$ which has an eigenstate of the form $|\chi \rangle _{\phi }\otimes
|\varphi _{2}\rangle _{\varphi },$ where $|\chi \rangle _{\phi }\in $ $%
\mathcal{H}_{\phi },$ such that the factor $|\varphi _{2}\rangle _{\varphi
}\in \mathcal{H}_{\varphi }$ appears unaffected by this measurement. In
other words, a physicist may `ask a question' about the sub-state $\phi _{1}$
in the factor space $\mathcal{H}_{\phi }$ without necessarily changing every
part of the state $\psi _{a}.$ It is, for example, permissible to determine
whether the\ component of $\psi _{a}$ in the Hilbert space $\mathcal{H}%
_{\phi }$ is indeed $\phi _{1},$ without destroying $\psi _{a}.$ In fact,
because it is separable, it is\ generally possible to determine the\
component of $\psi _{a}$ in the Hilbert space $\mathcal{H}_{\phi },$ without
in any way affecting the component of $\psi _{a}$ in the Hilbert space $%
\mathcal{H}_{\varphi }.$

However, the same is not true for the entangled state $\psi _{b}.$ Any
attempt to measure the component of $\psi _{b}$ in either of the factor
Hilbert spaces $\mathcal{H}_{\phi }$ or $\mathcal{H}_{\phi }$ destroys the
entanglement, and irreversibly collapses the wavefunction of the system into
a different state, i.e. into a product form.

This difference between $\psi _{a}$ and $\psi _{b}$ may be rephrased in
terms of the role of information. For the entangled state $\psi _{b}$ it is
possible to learn something about the component of the vector in $\mathcal{H}%
_{\varphi }$ by performing a measurement on the component of the vector in $%
\mathcal{H}_{\phi }.$ However, if during the measurement the entangled state
$\psi _{b}$ collapses into the product state $\psi _{a},$ no new information
is gained about the factor state $\varphi _{2}$ in $\mathcal{H}_{\varphi }$
by performing a subsequent measurement on $\phi _{1}$ in $\mathcal{H}_{\phi
}.$ In fact this will always remain the case, with the two factors leading
independent existences, unless the dynamics conspire in such a way as to
re-entangle the system.\bigskip

The testing of the EPR state of the electron-positron system of Section 3.1
provides a physical example of these principles. The initial entangled state
$|\psi \rangle =\frac{1}{\sqrt{2}}(\mid \uparrow \rangle _{e}\otimes \mid
\downarrow \rangle _{p}-$ $\mid \downarrow \rangle _{e}\otimes \mid \uparrow
\rangle _{p})$ of equation (\ref{wav})\ is destroyed by a measurement of the
spin of the electron, and, depending on the result of this, the system after
collapse may be represented by one of two possible product states $\mid
\uparrow \rangle _{e}\otimes \mid \downarrow \rangle _{p}$ or $\mid
\downarrow \rangle _{e}\otimes \mid \uparrow \rangle _{p}.$

Each factor of these product states consequently represents either an
isolated electron or an isolated positron, with a known component of spin.
Any subsequent measurement of the spin of either the electron or positron in
this direction leaves this new product state unaltered (as this is simply a
null test), but additionally, any further measurement in any direction
involving just the positron (by using, say, an operator of the form $\hat{S}%
_{\theta _{ab}}=\hat{S}_{z}\cos \theta _{ab}+\hat{S}_{x}\sin \theta _{ab}$
defined previously) will not affect the state of the electron, and vice
versa. Unlike the initial entangled state, the product state represents a
system comprising of an electron and a positron that are isolated and
independent from each other.\bigskip

The conclusion of the above discussion is that a quantum state separable
into a product of factors is in some sense equivalent to a system comprising
of a number of distinct semi-classical sub-systems. Because it is possible
to examine just one of these factors without affecting the rest of the
state, these sub-states appear isolated and distinguished from each other,
and can be discussed as separate from the rest of the system, exactly as
required for a semi-classical description of physics to begin to emerge.

The reciprocal of this should also be true. Every sub-system that appears
isolated and distinct from the others may be associated with one of the
factors of the state representing the entire quantum system.

Further, by\ extending this argument to the case of a fully quantum Universe
described by a wavefunction $\Psi _{n},$ the conjecture is that every
individual, semi-classical sub-system within it is represented by a unique
factor of this state.\bigskip

Exactly how this may be achieved is a difficult question, and it is noted
that the above statement may contain an element of idealisation. Being a
factor of a state only guarantees that the sub-system it represents may be
granted a degree of individuality. The individual factors still represent
sub-systems governed by the laws and constraints of quantum mechanics, as is
obvious, for example, for the single electron and positron factors of the
earlier EPR product states, $\mid \uparrow \rangle _{e}\otimes \mid
\downarrow \rangle _{p}$ and $\mid \downarrow \rangle _{e}\otimes \mid
\uparrow \rangle _{p},$ which must of course be treated quantum
mechanically. This, after all, is the origin of the lack of a Heisenberg Cut
in the Universe.

The method for achieving `real' classicity, in the traditional sense of the
word, in large macroscopic sub-systems of the Universe is part of the
difficult question of emergence to be addressed in the future. It is here
that statistical theories such as decoherence may play a part, as will be
discussed in Section 4.3.

Suffice to say, however, that even semi-classical, macroscopic sub-systems
must need to be associated with factor sub-states of the Universe's
wavefunction. The alternative, that they are actually entangled with their
surroundings, would imply that they cannot be given any sense of
individuality, and this would lead to the absurd and unsupported suggestion
that conventional semi-classical systems are actually entangled with each
other, contrary to empirical evidence. A classically distinct and isolated
sub-system must be represented by a factor of the Universe's state, but if
the state of the Universe can be separated it does not automatically imply
that every factor may be treated according to the laws of classical
mechanics, even as an approximation. Separability is a necessary condition
for classicity to arise, in that it implies distinguishability, but it is
unclear at this stage as to whether it is also sufficient.

\bigskip

It is possible that one or some of the factor sub-states may themselves be
entangled within their own Hilbert sub-spaces. For example, consider a
`toy-universe' initially in the state $\Psi _{0}=$ $|Z^{0}\rangle $ that
contains nothing but a single $Z^{0}$ boson. Also, assume that the dynamics
selects a particular operator, $\hat{\Sigma}_{1},$ to test $\Psi _{0},$
thereby causing the universe to jump to the state $\Psi _{1}=|\bar{\pi}%
^{0}\rangle \otimes |\pi ^{0}\rangle $ representing a pion/anti-pion pair.
Such a dynamics is analogous to a particle physics experiment in which the
high energy boson spontaneously decays into a neutral pion and anti-pion.

If the pion itself then goes on to decay to an entangled electron/positron
pair (i.e. an EPR-like state), the state of the universe, $\Psi _{2},$ after
this decay may be given by $\Psi _{2}=|\bar{\pi}^{0}\rangle \otimes |\psi
\rangle ,$ where $|\bar{\pi}^{0}\rangle $ represents the sub-state of the
anti-pion, and $|\psi \rangle $ the sub-state of the entangled electron and
positron (\ref{wav}). Clearly, the overall state $\Psi _{2}$ is a separable
product of two factors, one of which is entangled.

The current example shows how the separability of the state representing a
simple system changes as it develops. In fact, if subsequently an operator $%
\hat{\Sigma}_{3}$ is selected that is equivalent to a measurement of the
spin component of the electron (in a particular direction), and if the
result is that it is found to be spin `up', the next state, $\Psi _{3},$
will be of the product form $\Psi _{3}=|\bar{\pi}^{0}\rangle \otimes \mid
\uparrow \rangle _{e}\otimes \mid \downarrow \rangle _{p}.$ Note that the
dimensions of the Hilbert spaces of the states $\Psi _{0},\Psi _{1},\Psi
_{2} $ and $\Psi _{3}$ must be the same, and that for example the sub-state $%
|\psi \rangle $ of $\Psi _{2}$ is in the same factor Hilbert space as the
product of the sub-states $\mid \uparrow \rangle _{e}\otimes \mid \downarrow
\rangle _{p}$ in $\Psi _{3}.$ In this universe, it is evident that the
separability of the system changes during the transition from $\Psi _{2}$ to
$\Psi _{3};$ this will be an important feature in the following.\bigskip

It is now possible to reinterpret the idea of a physical experiment from the
point of view of a universe described fully by quantum mechanics. Recall
that the standard semi-classical treatment of physics is to segregate the
system into the subject under investigation, the various bits of apparatus,
the physicist conducting the experiment, and everything else in the Universe
(the `Environment'). Indeed, it does not seem possible to perform an
experiment on a quantum subject if it cannot be isolated from everything
else.

This semi-classical approach can be incorporated into the quantum picture of
the Universe by assuming that each of these semi-classical and distinct
parts may now be represented by separate factors of the state $\Psi $ of the
Universe. This is inevitable from the viewpoint asserted in this thesis: if
they are classically distinct, it follows that they cannot be entangled with
each other. Hence, the state may be written as
\begin{equation}
|\Psi \rangle =|\psi \rangle \otimes |A\rangle \otimes |O\rangle \otimes
|R\rangle
\end{equation}
where $|\psi \rangle $ represents the sub-state of the subject under
investigation, $|A\rangle $ the sub-state of the apparatus, $|O\rangle $ the
sub-state of the observer, and $|R\rangle $ is the sub-state representing
the rest of the Universe.

Of course, $|R\rangle $ will itself be a product of an enormous number of
sub-states, some of which may themselves be entangled. \ However, for the
sake of studying the tiny sub-state of interest, i.e. $|\psi \rangle ,$ the
conventional procedure is then to ignore all of the factors of $\Psi $ that
do not contribute to the running of the experiment and focus attention on
changes in $|\psi \rangle .$ This is really an exo-physical approach, where
the physicist falsely believes himself to be excluded from the Universe
being measured, and is therefore potentially misleading, but it is a natural
procedure borne from the physicists subjective experience of the `outside'
world. The `real' situation of endo-physical measurements will be addressed
in Section 6.2, and also briefly in Section 4.3.3.

As with the pion experiment described above, the separability of the state
representing the Universe may change as it develops through a series of
stages. Moreover, it is these changes in separability that are ultimately
responsible for the generation of certain classical effects in the Universe,
for example the emergence of continuous space. This will be expanded upon in
the following chapters, but it is remarked here that even in the simplest
quantum model hypothesised earlier, in which the Universe is represented by
a state in a Hilbert space of dimension greater than $2^{10^{184}},$ the
number of ways in which this state may be separated into a product of
factors, some of which may or may not be entangled themselves within their
factor sub-spaces, is enormous.\bigskip

Summarising, the separability of a state allows a classical distinction to
be made between its constituent factors. The conjecture, then, is that
classicity in a fully quantum Universe emerges somehow from considerations
of the separability of its state $\Psi .$ In addition, if separability is
required for classicity, and since it is an observational fact that the
Universe appears to be comprised of a vast number of classically distinct
sub-systems, the conclusion must be that the current state of the Universe
is highly separable. It is therefore a task to investigate how this might
have occurred.

\bigskip

\subsection{Factorisation and Entanglement}

\bigskip

When is an arbitrary state $\Psi $ in a Hilbert space $\mathcal{H}$
separable? What rules determine whether a given vector $\Psi $ can be
written as a product of factor sub-states?

Before answering these questions, it must first be noted that the concept of
a separable state necessarily implies the existence of a factorisable
Hilbert space. By definition, the property that a state $\Psi \in \mathcal{H}
$ is separable in the form $\Psi =\phi _{1}\otimes \varphi _{2},$ for
example, where $\phi _{1}\in \mathcal{H}_{1}$ and $\varphi _{2}\in \mathcal{H%
}_{2},$ explicitly requires that $\mathcal{H}$ can be factorised in the form
$\mathcal{H}\equiv \mathcal{H}_{[12]}=\mathcal{H}_{1}\otimes \mathcal{H}%
_{2}. $ It is therefore a natural starting point for any discussion of the
separability and entanglement properties of vectors to define what is meant
by the factorisability of their vector spaces.

A Hilbert space $\mathcal{H}^{(d)}\equiv \mathcal{H}$ of dimension $d$ is
factorisable into $N$ factors if it can be written in the tensor product
form
\begin{equation}
\mathcal{H}^{(d)}=\mathcal{H}_{1}^{(d_{1})}\otimes \mathcal{H}%
_{2}^{(d_{2})}\otimes ...\otimes \mathcal{H}_{a}^{(d_{a})}\otimes ...\otimes
\mathcal{H}_{N}^{(d_{N})}  \label{N-part}
\end{equation}
where $\mathcal{H}_{a}^{(d_{a})}$ for $a=1,...,N$ is called the $a^{th}$
factor Hilbert space and is of dimension $d_{a}.$ Clearly, $%
d=d_{1}d_{2}...d_{N}.$ Such a factorisation represents a particular `\textit{%
split}' of the Hilbert space $\mathcal{H}^{(d)}$ into $N$ given factors.

For convenience and clarity, note that here and elsewhere a Hilbert space $%
\mathcal{H}^{(d)}$ factorisable into $N$ factors in the form of (\ref{N-part}%
) can be written using the square bracket notation
\begin{equation}
\mathcal{H}^{(d)}\equiv \mathcal{H}_{[(1)(2)(3)...(N)]}^{(d)}\equiv \mathcal{%
H}_{[123...N]}^{(d)}\equiv \mathcal{H}_{[1...N]}^{(d)}.
\end{equation}

Note also that, in general, Hilbert spaces may be referred to as
`factorisable', whereas the states they contain may be referred to as
`separable'. A Hilbert space could also be described as `separable', but in
conventional texts on vector spaces this name is taken to imply that a
countable basis can be found for it; any vector in a separable Hilbert space
may be written as a discrete sum of basis vectors.\bigskip

It is possible now to define the factorisability, $\zeta ,$ of $\mathcal{H}%
^{(d)}$ written in the form (\ref{N-part}) as $\zeta =N/d,$ that is, the
ratio between the number of factors and the overall dimension of the vector
space. Consequently, the case in which $d_{i}\in \mathbb{P}$ $\forall i,$
where $\mathbb{P}$ is the set of prime numbers, represents the maximum
factorisability of $\mathcal{H}^{(d)}$ for a given $d;$ such a split will be
called a `\textit{fundamental}' or `\textit{natural}' factorisation, and the
factor Hilbert spaces will be called `\textit{elementary}'. Obviously, for a
Hilbert space of even dimension, the factorisability $\zeta $\ is clearly
maximised if the dimension of each factor space is two, in which instance $%
\zeta $ is given by $\zeta =N/2^{N}.$

Two dimensional Hilbert spaces are of great interest to many authors, partly
because they are the simplest, and partly because of an analogy with
computational physics. An orthonormal basis set for a Hilbert space $%
\mathcal{H}^{(2)}$ may be given by $\{|0\rangle ,|1\rangle \},$ for $\langle
i|j\rangle =\delta _{ij}$ with $i,j=0,1,$ and these two vectors may be
likened to any set of `opposite' states in elementary binary logic: $%
|0\rangle $ may for example represent `off', `no', `left-polarised',
`spin-down', or `false', whereas $|1\rangle $ may represent the reverse,
i.e. `on', `yes', `right-polarised', `spin-up' or `true'. \ It is this
analogy to classical `bit' logic that earns the quantum space $\mathcal{H}%
^{(2)}$ the title of a qubit Hilbert space, and a vector in this space may
be called a qubit state.\ Qubit states will be discussed a number of times
throughout this thesis.\bigskip

It is important to note that the left-right ordering of the factor Hilbert
spaces is not taken to be significant in this work. Specifically, this
implies that the factorisation (\ref{N-part}) is invariant to any
permutation $i\longrightarrow j_{i}$ of its factors $\mathcal{H}%
_{i}^{(d_{i})},$ such that for example $\mathcal{H}^{(4)}=\mathcal{H}%
_{1}^{(2)}\otimes \mathcal{H}_{2}^{(2)}\equiv $ $\mathcal{H}%
_{2}^{(2)}\otimes \mathcal{H}_{1}^{(2)}.$

Similarly, the same is taken to hold true for the states contained within
these Hilbert spaces; for example, if $\phi _{1}\in \mathcal{H}_{1}^{(2)}$
and $\varphi _{2}\in \mathcal{H}_{2}^{(2)},$ the product state $\Psi =\phi
_{1}\otimes \varphi _{2}\in \mathcal{H}_{1}^{(2)}\otimes \mathcal{H}%
_{2}^{(2)}$ is defined as equivalent to the re-ordered state $\Psi ^{\prime
}=\varphi _{2}\otimes \phi _{1}\in \mathcal{H}_{2}^{(2)}\otimes \mathcal{H}%
_{1}^{(2)}.$\bigskip

If the dimension of a Hilbert space is large, but not prime, the number of
different ways in which it can be factorised might also be large.

For example, consider a four dimensional Hilbert space $\mathcal{H}^{(4)};$
the only non-trivial factorisation of $\mathcal{H}^{(4)}$ splits the Hilbert
space into a product of two sub-spaces, i.e. $\mathcal{H}^{(4)}=\mathcal{H}%
_{1}^{(2)}\otimes \mathcal{H}_{2}^{(2)},$ where as above the sub-script is a
convenient label and the super-script denotes dimension. Such a split may be
called a \textit{bi-partite} factorisation.

Alternatively consider an eight dimensional Hilbert space $\mathcal{H}%
^{(8)}; $ this space may potentially be split into a tri-partite
factorisation of three two-dimensional Hilbert spaces, or a bi-partite
factorisation of one two-dimensional Hilbert space and one four-dimensional
Hilbert space.

Obviously for higher dimensional cases such as $\mathcal{H}^{(24)},$ the
number of ways in which the Hilbert space might be factorisable in this
simple manner is even greater, schematically because $24=2\times 12=2\times
2\times 6=2\times 2\times 2\times 3=2\times 4\times 3=3\times 8=4\times 6.$
In fact, as will be shown in Chapter 5, the actual number of ways of
splitting a Hilbert spaces is much more complicated than this elementary
`dimensional' argument suggests. There are, for example, a number of
different ways of factorising a $24$ dimensional Hilbert space into a
product of a two dimensional factor and a twelve dimensional factor.\bigskip

In a similar vein, the vectors contained in these Hilbert spaces will also
possess different degrees of separability. An arbitrary vector $\Psi $\ in $%
\mathcal{H}^{(4)},$ for example, is either separable in the form $\Psi =\phi
_{1}\otimes \varphi _{2},$ where $\phi _{1}\in \mathcal{H}_{1}^{(2)}$ and $%
\varphi _{2}\in \mathcal{H}_{2}^{(2)},$ or not, in which case it is said to
be entangled relative to the factorisation $\mathcal{H}^{(4)}=\mathcal{H}%
_{1}^{(2)}\otimes \mathcal{H}_{2}^{(2)}.$ Similarly, for the case of an
arbitrary vector $\Phi $ in $\mathcal{H}^{(8)},$ the state might either be
separable into three factors, or into two factors, or into one giant
entangled `factor'. In the case in which $\Phi $ can only be separated into
two factors relative to a tri-partite factorisation of $\mathcal{H}^{(8)},$ $%
\mathcal{H}^{(8)}=\mathcal{H}_{1}^{(2)}\otimes \mathcal{H}_{2}^{(2)}\otimes
\mathcal{H}_{3}^{(2)},$ it is clear that the state is the product of two
sub-states, one of which is entangled. Again, arbitrary states in higher
dimensional Hilbert spaces might potentially be separable into products of
many sub-states of differing dimension; this will be discussed more
thoroughly in Chapter 5.\bigskip

Although a Hilbert space $\mathcal{H}^{(d)}\equiv \mathcal{H}$ might
potentially be split into the $N$-partite factorisation of equation (\ref
{N-part}), it is only whether a state is separable relative to a particular
bi-partite split that is of most interest. Indeed, without loss of
generality, only the separations of vectors relative to bi-partite
factorisations need be investigated, and so in reality, only the possible
rules governing this need be sought. This conclusion follows because it is a
feature of Hilbert space mathematics that when a state is separable into a
product of vectors in different factor Hilbert spaces, the factor sub-states
are effectively independent. It can then be implicitly assumed that any
method used to determine whether a given state $\Psi $ is separable into two
sub-states may be applied again to determine whether one of these sub-states
is itself separable into a product of two sub-sub-states, because the only
difference between the two cases is that the vectors investigated are of
different dimensions.

In other words, any method used to separate the $d$ dimensional vector $\Psi
$ into a product $\phi _{1}\otimes \varphi _{2}$ of a $d_{1}$ dimensional
vector, $\phi _{1},$ and a $d_{2}$ dimensional vector, $\varphi _{2},$ where
$d=d_{1}d_{2},$ is effectively the same as that used to separate a $d_{1}$
dimensional vector $\phi _{1}$ into a product $\phi _{1}=\alpha
_{1_{a}}\otimes \beta _{1_{b}}$ of a $d_{1_{a}}$ dimensional vector, $\alpha
_{1_{a}},$ and a $d_{1_{b}}$ dimensional vector, $\beta _{1_{b}},$ where $%
\alpha _{1_{a}}\in \mathcal{H}_{1_{a}}^{(d_{1_{a}})},$ $\beta _{12}\in
\mathcal{H}_{1_{b}}^{(d_{1_{b}})}$ and $\mathcal{H}_{1}^{(d_{1})}=\mathcal{H}%
_{[1_{a}1_{b}]}^{(d_{1})}=$ $\mathcal{H}_{1_{a}}^{(d_{1_{a}})}\otimes
\mathcal{H}_{1_{b}}^{(d_{1_{b}})}.$

So, a given state $\Psi $ may be separated into a product of $N$ factors by
a process of first separating it into two factors, followed by independently
separating each of these factors into two factors, followed then by
independently separating each of these four\ factors into two factors, and
so on until each of the individual factors can no longer be separated.
Assuming it is known whether it is possible to separate a given vector into
a product of two factors, then by repeated iteration the separation of the
overall state into $N$ sub-states can be found.\bigskip

As an example, consider the tri-partite factorisation of the
eight-dimensional Hilbert space $\mathcal{H}_{[1...3]}^{(8)}=\mathcal{H}%
_{1}^{(2)}\otimes \mathcal{H}_{2}^{(2)}\otimes \mathcal{H}_{3}^{(2)},$ and
also a state $\Phi $ in $\mathcal{H}^{(8)}$ that is known to be separable
into three factors, i.e. can be written in the form $\Phi =\phi _{1}\otimes
\varphi _{2}\otimes \psi _{3},$ where $\phi _{1}\in \mathcal{H}_{1}^{(2)},$ $%
\varphi _{2}\in \mathcal{H}_{2}^{(2)}$ and $\psi _{3}\in \mathcal{H}%
_{3}^{(2)}.$ It follows that $\Phi $ must also be separable into two
factors,
\begin{eqnarray}
\Phi &=&\phi _{1}\otimes (\varphi _{2}\otimes \psi _{3}) \\
&=&\phi _{1}\otimes \chi _{23}  \notag
\end{eqnarray}
where $\chi _{23}\equiv (\varphi _{2}\otimes \psi _{3})$ is an element of $%
\mathcal{H}_{[23]}^{(4)}\equiv \mathcal{H}_{2}^{(2)}\otimes \mathcal{H}%
_{3}^{(2)}.$

This argument can be reversed. In order to show that $\Phi $ is separable
into three factors, it is only necessary to first show that $\Phi $ is
separable into two factors, $\phi _{1}$ and $\chi _{23},$ relative to the
bi-partite factorisation of the Hilbert space $\mathcal{H}^{(8)}=\mathcal{H}%
_{1}^{(2)}\otimes \mathcal{H}_{[23]}^{(4)},$ and then to show that $\phi
_{1} $ is not separable whilst $\chi _{23}$ may be written as a product of
two factors, $\varphi _{2}$ and $\psi _{3},$ relative to the bi-partite
factorisation of the Hilbert sub-space $\mathcal{H}^{(4)}=\mathcal{H}%
_{2}^{(2)}\otimes \mathcal{H}_{3}^{(2)}.$\ It is at this point that the
procedure would terminate, because the factors $\varphi _{2}$ and $\psi _{3}$
cannot further be separated; the state $\Phi $ can be separated into a
product of no more that three factors.

Of course, in this illustration $\phi _{1},$ $\varphi _{2}$ and $\psi _{3}$
cannot be separated further because they are contained in two-dimensional
Hilbert spaces, but in principle the above method could be used even if they
were entangled sub-states of arbitrary dimension. As an example, if it could
be found that a different vector $\Phi ^{\prime }$ in $\mathcal{H}^{(8)}$ is
separable as $\Phi ^{\prime }=\phi _{1}^{\prime }\otimes \chi _{23}^{\prime
},$ with $\phi _{1}^{\prime }\in \mathcal{H}_{1}^{(2)}$ and $\chi
_{23}^{\prime }\in \mathcal{H}_{[23]}^{(4)},$ but that $\chi _{23}^{\prime }$
is entangled relative to $\mathcal{H}_{[23]}^{(4)}=\mathcal{H}%
_{2}^{(2)}\otimes \mathcal{H}_{3}^{(2)},$ this result would be sufficient to
prove that $\Phi ^{\prime }$ cannot be separated into three factors relative
to this split.\bigskip

As an aside, note that analogously to the fundamental splitting of the
Hilbert space described earlier, a fundamental separation of a state $\Psi $
in $\mathcal{H}^{(d)}$ may be defined as that which contains the maximum
number of factors relative to a given factorisation of the Hilbert space.
For example, the state $\Phi $ in $\mathcal{H}^{(8)}=\mathcal{H}%
_{1}^{(2)}\otimes \mathcal{H}_{2}^{(2)}\otimes \mathcal{H}_{3}^{(2)}$
written in the form $\Phi =\phi _{1}\otimes \varphi _{2}\otimes \psi _{3},$
where $\phi _{1}\in \mathcal{H}_{1}^{(2)},$ $\varphi _{2}\in \mathcal{H}%
_{2}^{(2)}$ and $\psi _{3}\in \mathcal{H}_{3}^{(2)},$ is clearly
fundamentally separated, whereas the same state written as $\Phi =\phi
_{1}\otimes \chi _{23},$ where\ $\chi _{23}$ is an element of $\mathcal{H}%
_{[23]}^{(4)}\equiv \mathcal{H}_{2}^{(2)}\otimes \mathcal{H}_{3}^{(2)},$ is
not. Conversely, the state $\Phi ^{\prime }$ written in the form $\Phi
^{\prime }=\phi _{1}^{\prime }\otimes \chi _{23}^{\prime }$ defined above is
fundamentally separated relative to this factorisation of $\mathcal{H}%
^{(8)}. $

The conclusion of the above few paragraphs is that it is only necessary to
investigate whether or not a given state $\Psi \in \mathcal{H}^{(d)}$ is
separable into a product $\Psi =\phi _{1}\otimes \varphi _{2}$ of two
sub-states $\phi _{1}\in \mathcal{H}_{1}^{(d_{1})}$ and $\varphi _{2}\in
\mathcal{H}_{2}^{(d_{2})},$ relative to some bi-partite factorisation $%
\mathcal{H}^{(d)}=\mathcal{H}_{1}^{(d_{1})}\otimes \mathcal{H}_{2}^{(d_{2})}$
of the $d$-dimensional Hilbert space. A test is hence sought to determine
whether an arbitrary state can be separated into two factors, relative to
such a bi-partite split.

\bigskip

Consider a Hilbert space $\mathcal{H}\equiv \mathcal{H}^{(d)}$ of dimension $%
d$ that is factorisable into the bi-partite split $\mathcal{H}^{(d)}=%
\mathcal{H}_{1}^{(d_{1})}\otimes \mathcal{H}_{2}^{(d_{2})},$ where $d_{a}$
is dimension of the $a^{th}$ factor Hilbert space, $a=1,2,$ and $%
d=d_{1}d_{2}.$ It is a standard theorem of vector spaces \cite{Peres} that a
Hilbert space of dimension $D$ is spanned by a set of $D$ orthonormal basis
vectors. Thus, a basis set $\mathcal{B}_{a}$ for the Hilbert space $\mathcal{%
H}_{a}^{(d_{a})}$ may be given by
\begin{equation}
\mathcal{B}_{a}\equiv \{|i\rangle _{a}:i=0,1,...,(d_{a}-1),\text{ }a=1,2\},
\label{Ch2Basis}
\end{equation}
where $\langle i|j\rangle =\delta _{ij}.$ Moreover, it follows from (\ref
{Ch2Basis}) that an orthonormal basis set $\mathcal{B}=\mathcal{B}_{12}$ for
the product Hilbert space $\mathcal{H}^{(d)}$ is given by
\begin{equation}
\mathcal{B}\equiv \{|i\rangle _{1}\otimes |j\rangle _{2}:i=0,1,...,(d_{1}-1),%
\text{ }j=0,1,...,(d_{2}-1)\}
\end{equation}

So, any vector $\Psi $ in $\mathcal{H}$ is composed of a complex linear
superposition of the members of this set, viz.
\begin{equation}
|\Psi \rangle
=\sum\nolimits_{i=0}^{d_{1}-1}\sum\nolimits_{j=0}^{d_{2}-1}C_{ij}|i\rangle
_{1}\otimes |j\rangle _{2}  \label{Ch2state}
\end{equation}
where the $C_{ij}\in \mathbb{C}$ form a $d_{1}\times d_{2}$ complex
coefficient matrix.

Depending on the set of values of $C_{ij}$ for $i=0,1,...,(d_{1}-1)$ and $%
j=0,1,...,(d_{2}-1),$ the state $\Psi $ will be either separable or
entangled relative to the factorisation $\mathcal{H}^{(d)}=\mathcal{H}%
_{1}^{(d_{1})}\otimes \mathcal{H}_{2}^{(d_{2})}.$ For example, if $C_{ij}=1$
for $i=j=0,$ but $C_{ij}=0$ otherwise, then $|\Psi \rangle =|0\rangle
_{1}\otimes |0\rangle _{2},$ which is clearly separable.

In fact:

\begin{theorem}
A state $|\Psi \rangle \in $ $\mathcal{H}^{(d)}$ is separable relative to
the factorisable Hilbert space basis $\mathcal{B}$ iff its coefficient
matrix satisfies the `microsingularity' condition
\begin{equation}
C_{ij}C_{kl}=C_{il}C_{kj}  \label{Ch2Micro}
\end{equation}

for all $0\leqslant i,k,\leqslant (d_{1}-1)$ and $0\leqslant j,l,\leqslant
(d_{2}-1).$
\end{theorem}

The proof of (\ref{Ch2Micro}) is given below, noting that a similar result
is provided by Albeverio \textit{et al} \cite{Albeverio} based on the idea
of `concurrency'.

\begin{proof}
$\Rightarrow $ If $C_{ij}C_{kl}=C_{il}C_{kj}$ and $|\Psi \rangle
=\sum\nolimits_{i=0}^{d_{1}-1}\sum\nolimits_{j=0}^{d_{2}-1}C_{ij}|i\rangle
_{1}\otimes |j\rangle _{2}:$

Suppose, without loss of generality, $C_{uv}\neq 0$ for some $u,v.$ Then
multiplying (\ref{Ch2state}) by this gives
\begin{eqnarray}
C_{uv}|\Psi \rangle
&=&\sum\nolimits_{i=0}^{d_{1}-1}\sum\nolimits_{j=0}^{d_{2}-1}C_{uv}C_{ij}|i%
\rangle _{1}\otimes |j\rangle _{2} \\
&=&\sum\nolimits_{i=0}^{d_{1}-1}\sum\nolimits_{j=0}^{d_{2}-1}C_{uj}C_{iv}|i%
\rangle _{1}\otimes |j\rangle _{2}  \notag \\
&=&\sum\nolimits_{i=0}^{d_{1}-1}C_{iv}|i\rangle _{1}\otimes
\sum\nolimits_{j=0}^{d_{2}-1}C_{uj}|j\rangle _{2}  \notag
\end{eqnarray}

and the state is separable.

$\Leftarrow $ If $|\Psi \rangle $ is separable relative to $\mathcal{H}%
^{(d)}=\mathcal{H}_{1}^{(d_{1})}\otimes \mathcal{H}_{2}^{(d_{2})}:$%
\begin{eqnarray}
|\Psi \rangle &=&\left( \sum\nolimits_{i=0}^{d_{1}-1}a_{i}|i\rangle
_{1}\right) \otimes \left( \sum\nolimits_{j=0}^{d_{2}-1}b_{j}|j\rangle
_{2}\right) \text{ \ \ , \ \ }a_{i},b_{j}\in \mathbb{C} \\
&=&\sum\nolimits_{i=0}^{d_{1}-1}\sum\nolimits_{j=0}^{d_{2}-1}a_{i}b_{j}|i%
\rangle _{1}\otimes |j\rangle _{2}  \notag
\end{eqnarray}

So $a_{i}b_{j}=C_{ij}$ from (\ref{Ch2state}). It follows that
\begin{equation}
C_{ij}C_{kl}=a_{i}b_{j}a_{k}b_{l}=a_{i}b_{l}a_{k}b_{j}=C_{il}C_{kj}
\end{equation}
\bigskip
\end{proof}

As an example, the two qubit state $\Theta \in \mathcal{H}_{[12]}^{(4)}$
given by
\begin{equation}
\Theta =\alpha |0\rangle _{1}\otimes |0\rangle _{2}+\beta |0\rangle
_{1}\otimes |1\rangle _{2}+\gamma |1\rangle _{1}\otimes |0\rangle
_{2}+\delta |1\rangle _{1}\otimes |1\rangle _{2}
\end{equation}
with $\alpha ,\beta ,\gamma ,\delta \in \mathbb{C}$ and orthonormal basis
set $\mathcal{B}_{a}\equiv \{|i\rangle _{a}:i=0,1\}$ for $\mathcal{H}%
_{a}^{(2)}$ and $a=1,2,$ can be written as a separable state of the form
\begin{equation}
\Theta =(a|0\rangle _{1}+b|1\rangle _{1})\otimes (c|0\rangle _{2}+d|1\rangle
_{2})
\end{equation}
for $a,b,c,d\in \mathbb{C}$ if, and only if, $\alpha \delta =\beta \gamma .$

Note that the separability of a state $\Psi $\ in $\mathcal{H}=\mathcal{H}%
_{1}^{(d_{1})}\otimes \mathcal{H}_{2}^{(d_{2})}$ is independent of the
choice of basis for the individual factor spaces $\mathcal{H}_{1}^{(d_{1})}$
and $\mathcal{H}_{2}^{(d_{2})}.$ For example, if $\mathcal{H}_{a}^{(d_{a})}$
has a basis $\mathcal{B}_{a}\equiv \{|i\rangle _{a}:i=0,1,...,(d_{a}-1)\}$
for $a=1,2,$ the separability of $\Psi $ is invariant to any relabelling $%
i\longrightarrow j_{i}$ of the individual elements $|i\rangle _{a}.$
Similarly, $\Psi $ will not be affected by any `rotation' of the members of
this basis set by local unitary operators $\hat{u}_{a},$ i.e. $|i\rangle
_{a}\longrightarrow |i^{\prime }\rangle _{a}=\hat{u}_{a}|i\rangle _{a}.$%
\bigskip

In general, for a state $\Psi $ to be separable relative to the bi-partite
factorisation of the Hilbert space $\mathcal{H}^{(d)}=\mathcal{H}%
_{1}^{(d_{1})}\otimes \mathcal{H}_{2}^{(d_{2})},$ the number $N_{c}$ of
microsingularity equalities that need to be satisfied is given by
\begin{equation}
N_{c}=\frac{1}{4}{\large [}d_{1}(d_{1}-1)d_{2}(d_{2}-1){\large ]}
\end{equation}
or $N_{c}\sim d^{2}/4$ for $d=d_{1}d_{2}\gg 1.$ In addition, the set of
separable states is a set of measure zero relative to the set of all
possible states; the set of separable states effectively form a hypersurface
in the hypervolume representing every set of values of $C_{ij}.$

It might be surprising, therefore, that there is any separability in the
Universe at all. From the earlier `minimum guess' that the dimensionality of
the Hilbert space of the state $\Psi _{n}$ of the Universe is greater than $%
2^{10^{184}},$ the number of microsingularity conditions that are required
to ensure that $\Psi _{n}$ is not entangled is at least $2^{(2\times
10^{184}-2)}.$ It might therefore be expected that if a vector is chosen at
random from a Hilbert space of dimension $2^{10^{184}},$ the probability
that it is separable, relative to a given bi-partite factorisation of the
Hilbert space, is zero. From this argument, the probability that the
statevector representing the Universe is separable might also be expected to
be approximately zero. Further, if the Universe jumps through a series of
states $\Psi _{n-1}\longrightarrow $ $\Psi _{n}\longrightarrow \Psi _{n+1}$
as it develops, it might be expected that the Universe should almost always
proceed from one entangled state to the next.

This, however, does not appear to be what is actually observed in the
physical Universe. If separability is a necessary prerequisite for
classicity, and given that the Universe does seem to look like a giant
collection of classical objects, the state of the Universe must be highly
separable. Given that it appears overwhelmingly likely that a quantum state
chosen at random from the set of states of dimension $2^{10^{184}}$ is
entangled, the question must remain as to how there can ever be any
classicity in a Universe running according to quantum principles.

So in response to this, the conclusion must therefore be that the operator $%
\hat{\Sigma}_{n}$ used to prepare the state $\Psi _{n}$ must be very
carefully constrained in order to ensure that its eigenvectors are almost
universally separable. Equally, the Rules $R_{n}$ themselves must be very
finely tuned to arrange that an operator $\hat{\Sigma}_{n+1}$ with highly
separable eigenvectors is selected to form the basis for the next state $%
\Psi _{n+1}.$ Quite plainly, the operators that the Universe chooses to test
itself must force $\Psi _{n}$ to jump from one highly separable state to the
next. This is analogous to the conclusion presented at the end of Section
3.3, in which Rules are discussed that guarantee that the state $\Psi _{n+1}$
appears so similar to $\Psi _{n},$ and is a point that will be returned to
many times throughout this thesis.

\bigskip

An important feature of the present discussion is that states that are
separable relative to a particular factorisation of a Hilbert space may be
entangled relative to a different one. Consider, for example, a Hilbert
space $\mathcal{H}^{(8)}$ that is the product of three qubit sub-spaces,
that is $\mathcal{H}^{(8)}=\mathcal{H}_{[123]}^{(8)}=\mathcal{H}%
_{1}^{(2)}\otimes \mathcal{H}_{2}^{(2)}\otimes \mathcal{H}_{3}^{(2)}.$

Now define a bi-partite split of $\mathcal{H}^{(8)}$ of the form $\mathcal{H}%
_{[A3]}^{(8)}=\mathcal{H}_{A}^{(4)}\otimes \mathcal{H}_{3}^{(2)},$ where $%
\mathcal{H}_{A}^{(4)}=\mathcal{H}_{[12]}^{(4)}=\mathcal{H}_{1}^{(2)}\otimes
\mathcal{H}_{2}^{(2)},$ with a suitable basis set $\mathcal{B}_{A3}$ given
by $\mathcal{B}_{A3}\equiv \{|ij\rangle _{A}\otimes |k\rangle
_{3}:i,j,k=0,1\}$ for $|ij\rangle _{A}\in \mathcal{H}_{A}^{(4)}$ and $%
|k\rangle _{3}\in $ $\mathcal{H}_{3}^{(2)}.$ Consider also a state $\Phi \in
\mathcal{H}^{(8)}$ defined as (\ref{Ch2Thi}) for $a,b,c,d,\alpha ,\beta \in
\mathbb{C}.$%
\begin{equation}
\Phi =(a|00\rangle _{A}+b|01\rangle _{A}+c|10\rangle _{A}+d|11\rangle
_{A})\otimes (\alpha |0\rangle _{3}+\beta |1\rangle _{3})  \label{Ch2Thi}
\end{equation}
where for convenience here and in the following, the product state $%
|i\rangle _{1}\otimes |j\rangle _{2}$ has been abbreviated by omitting the
tensor symbol and writing
\begin{equation}
|i\rangle _{1}\otimes |j\rangle _{2}\equiv |i\rangle _{1}|j\rangle
_{2}\equiv |ij\rangle _{12}=|ij\rangle .  \label{Ch2abb}
\end{equation}

Note that because in this contracted form the sub-script denoting the factor
Hilbert space is dropped, the left-right ordering of the products must be
preserved, i.e. $|ij\rangle \neq |ji\rangle .$

According to equation (\ref{Ch2state}), $\Phi $ is equivalently specified by
the coefficient matrix $C_{ij}$ given by
\begin{equation}
\begin{tabular}{|l|l|l|}
\hline
$\mathbf{\otimes }$ & $\mathbf{|0\rangle }_{3}$ & $\mathbf{|1\rangle }_{3}$
\\ \hline
$\mathbf{|00\rangle }_{A}$ & $a\alpha $ & $a\beta $ \\ \hline
$\mathbf{|01\rangle }_{A}$ & $b\alpha $ & $b\beta $ \\ \hline
$\mathbf{|10\rangle }_{A}$ & $c\alpha $ & $c\beta $ \\ \hline
$\mathbf{|11\rangle }_{A}$ & $d\alpha $ & $d\beta $ \\ \hline
\end{tabular}
\tag*{Table 4.1}
\end{equation}
where the first column and first row represent the basis vectors for the
bi-partite factorisation $\mathcal{H}^{(8)}=\mathcal{H}_{A}^{(4)}\otimes
\mathcal{H}_{3}^{(2)},$ and the remaining values represent the coefficients
of their tensor products (\textit{1}$^{st}$ \textit{column }$\otimes $%
\textit{\ 1}$^{st}$ \textit{row}). Obviously $\Phi $ is separable relative
to this factorisation, and the coefficient matrix clearly obeys the
microsingularity condition.

Consider now a different bi-partite factorisation of the Hilbert space
defined as $\mathcal{H}_{[1B]}^{(8)}=\mathcal{H}_{1}^{(2)}\otimes \mathcal{H}%
_{B}^{(4)},$ where $\mathcal{H}_{B}^{(4)}=\mathcal{H}_{[23]}^{(4)}=\mathcal{H%
}_{2}^{(2)}\otimes \mathcal{H}_{3}^{(2)}.$ Such a factorisation is spanned
by an orthonormal basis $\mathcal{B}_{1B}$ given by $\mathcal{B}_{1B}\equiv
\{|i\rangle _{1}\otimes |jk\rangle _{B}:i,j,k=0,1\},$ where $|i\rangle
_{1}\in \mathcal{H}_{1}^{(2)}$ and $|jk\rangle _{B}\in \mathcal{H}%
_{B}^{(4)}. $

Now, by expanding (\ref{Ch2Thi}), the vector $\Phi $ may equally be written
as
\begin{eqnarray}
\Phi &=&a\alpha |0\rangle _{1}\otimes |0\rangle _{2}\otimes |0\rangle
_{3}+a\beta |0\rangle _{1}\otimes |0\rangle _{2}\otimes |1\rangle _{3} \\
&&+b\alpha |0\rangle _{1}\otimes |1\rangle _{2}\otimes |0\rangle
_{3}+...+d\beta |1\rangle _{1}\otimes |1\rangle _{2}\otimes |1\rangle _{3}
\notag \\
&=&a\alpha |0\rangle _{1}\otimes |00\rangle _{B}+a\beta |0\rangle
_{1}\otimes |01\rangle _{B}+b\alpha |0\rangle _{1}\otimes |10\rangle
_{B}+...+d\beta |1\rangle _{1}\otimes |11\rangle _{B}  \notag
\end{eqnarray}
with the coefficient matrix
\begin{equation}
\begin{tabular}{|l|l|l|l|l|}
\hline
$\mathbf{\otimes }$ & $\mathbf{|00\rangle }_{B}$ & $\mathbf{|01\rangle }_{B}$
& $\mathbf{|10\rangle }_{B}$ & $\mathbf{|11\rangle }_{B}$ \\ \hline
$\mathbf{|0\rangle }_{1}$ & $a\alpha $ & $a\beta $ & $b\alpha $ & $b\beta $
\\ \hline
$\mathbf{|1\rangle }_{1}$ & $c\alpha $ & $c\beta $ & $d\alpha $ & $d\beta $
\\ \hline
\end{tabular}
\tag*{Table 4.2}
\end{equation}
which clearly might not satisfy each of the six microsingularity equalities.
Evidently, although the state $\Phi $ is separable relative to the first
factorisation of $\mathcal{H}^{(8)},$ i.e. $\mathcal{H}^{(8)}=\mathcal{H}%
_{[A3]}^{(8)},$ it is entangled relative to the second factorisation of $%
\mathcal{H}^{(8)},$ i.e. $\mathcal{H}^{(8)}=\mathcal{H}_{[1B]}^{(8)}.$

This result highlights the conclusion that it is simply not enough to say
that a particular state is separable, but that it must be qualified by the
statement that it is separable relative to a given factorisation of the
Hilbert space. More precisely, if a state $\Psi \in \mathcal{H}$ may be
written in the form $\Psi =\phi \otimes \varphi ,$ where $\phi \in \mathcal{H%
}_{1}$ and $\varphi \in \mathcal{H}_{2}$ for $\mathcal{H}=\mathcal{H}%
_{1}\otimes \mathcal{H}_{2},$ then it may be said that $\Psi $ is separable
relative to $(\mathcal{H}_{1},\mathcal{H}_{2}).$ Alternatively, if $\Psi $
is not separable in this way, it is said that\ $\Psi $ is entangled relative
to $(\mathcal{H}_{1},\mathcal{H}_{2}).$\bigskip

The above result has an important consequence. If any state is only
separable relative to a given factorisation of the Hilbert space, then the
assertion that the Universe's state is highly separable, because the
Universe appears classical, is only meaningful given a certain factorisation
of the Hilbert space $\mathcal{H}$ containing the Universe's state.

In order to discuss consistent physics, it might therefore be suggested that
a preferred split for $\mathcal{H}$ exists, and that the Universe's state
may only be described as separable, entangled, or a separable product of
entangled factor sub-states, relative to this preferred factorisation. As a
conjecture, the fundamental factorisation of $\mathcal{H},$ in which each
factor sub-space is of prime dimension, might perhaps be a possible
candidate for such a preferred split, but such a `natural' assumption
requires a great deal of future work.

\bigskip

\subsection{Basis Sets and Operators}

\bigskip

Whilst much of the discussion in this section has involved the properties of
the states, it is important to remember that the operators themselves also
play a necessary part in the dynamics. After all, it is the Hermitian
operators used to test the state that provide, as their eigenvectors, the
basis set of next possible states.

Recall that every operator $\hat{\Sigma}_{n+1}$ acting in a $d$ dimensional
Hilbert space $\mathcal{H}=\mathcal{H}^{(d)}$ and `testing' a state $\Psi
_{n}$ (or\ $\Psi _{n}^{\prime }=\hat{U}\Psi _{n},$ where $\hat{U}$ is
unitary) is associated with a basis set $\frak{B}=\{|\phi _{i}\rangle
:i=1,..,d\}$ of $d$ orthonormal eigenvectors\footnote{%
Note that throughout this work, the calligraphic symbol $\mathcal{B}$ will
be used to denote the particular basis set of a Hilbert space represented by
$\{|0\rangle ,|1\rangle ,|2\rangle ,...\}$ (i.e. the `natural' basis set),
whereas the fraktur symbol $\frak{B}$ will denote basis sets of orthonormal
eigenvectors $\{|\phi _{1}\rangle ,\{|\phi _{2}\rangle ,\{|\phi _{3}\rangle
,...\}$ of operators. This is really just a convenience, since a basis set
of eigenstates is also a basis set for the Hilbert space, and vice versa.}.
Further, as the Universe develops, its state jumps from $\Psi _{n}$ to $\Psi
_{n+1},$ and this effectively involves a process of randomly selecting one
of these $d$ orthonormal eigenvectors to be the next state $\Psi _{n+1},$
with the probability that a particular eigenstate $|\phi _{i}\rangle $ is
chosen given by the usual Born rule $|\langle \phi _{i}|\Psi _{n}\rangle
|^{2}.$

Now, each member $|\phi _{i}\rangle $ of $\frak{B}$ could be either
entangled or separable, relative to some bi-partite factorisation of\ $%
\mathcal{H}=\mathcal{H}_{1}^{(d_{1})}\otimes \mathcal{H}_{2}^{(d_{2})}.$ In
fact, the overall basis set $\frak{B}$ may contain $p$ entangled states, and
consequently $q=d-p$ separable members; such a set could hence be labelled a
type $(p,q)$ basis. A basis set of type $(0,d)$ may therefore be called
completely separable, whereas a type $(d,0)$ may be described as a
completely entangled basis. All other types may be called partially
separable, or equivalently, partially entangled.\bigskip

As an extension to the above, note that it would be necessary to introduce a
third parameter, $r,$ in order to describe basis sets that may contain
states that are separable relative to a tri-partite factorisation of the
Hilbert space, $\mathcal{H}=\mathcal{H}_{1}^{(d_{1})}\otimes \mathcal{H}%
_{2}^{(d_{2})}\otimes \mathcal{H}_{3}^{(d_{3})}$ for $d=d_{1}d_{2}d_{3}.$
These sets would be described as a $(p,r,q)$ type, because they contain $p$
entangled members, $q$ members that are separable into three factors, and $r$
members that can be separated into just two factors, one of which a
entangled relative to two of the factor Hilbert spaces.

Equivalently, a label $(p,r,q)$ clearly indicates that the set incorporates $%
p$ states that are separable into a product of one factors (totally
entangled vectors), $r$ states that are separable into a product of two
factors (the partially entangled vectors), and $q$ states that are separable
into a product of three factors (the fundamentally separated vectors).
Obviously, the extension generalises in\ a natural way, such that basis sets
discussed relative to an $N$-partite factorisation require $N$
parameters.\bigskip

It is interesting to note that not every type of $\left( p,q\right) $ basis
set exists.

Consider a four dimensional Hilbert space factorisable into a product of two
qubit factor sub-spaces, that is $\mathcal{H}^{(4)}=\mathcal{H}%
_{1}^{(2)}\otimes \mathcal{H}_{2}^{(2)}.$ Let $\mathcal{B}_{a}=\{|i\rangle
_{a}:i=0,1\}$ for $a=1,2$ be an orthonormal basis set for the factor Hilbert
space $\mathcal{H}_{a}^{(2)},$ and turn attention to finding orthonormal
basis sets spanning the total Hilbert space $\mathcal{H}^{(4)}.$

Firstly, it is possible to find basis sets of vectors that are completely
separable relative to the given bi-partite factorisation of $\mathcal{H}%
^{(4)}.$ An example of such a type $(0,4)$ basis is $\frak{B}_{(0,4)},$
defined as
\begin{equation}
\frak{B}_{(0,4)}=\{|00\rangle ,\text{ }|01\rangle ,\text{ }|10\rangle ,\text{
}|11\rangle \},
\end{equation}
with $\langle kl|ij\rangle =\delta _{ik}\delta _{jl}$ and $|ij\rangle \equiv
|i\rangle _{1}\otimes |j\rangle _{2}$ for $i,j=0,1.$

Using the same notation, it is also possible to find type $(2,2)$ basis sets
for $\mathcal{H}^{(4)}.$ One example, $\frak{B}_{(2,2)},$ may be defined as
\begin{equation}
\frak{B}_{(2,2)}=\left\{ |00\rangle ,\text{ }|11\rangle ,\text{ }\frac{%
{\small 1}}{\sqrt{{\small 2}}}(|01\rangle +|10\rangle ),\text{ }\frac{%
{\small 1}}{\sqrt{{\small 2}}}(|01\rangle -|10\rangle )\right\} .
\end{equation}

Similarly, it is possible to find an example, $\frak{B}_{(3,1)},$ of a type $%
(3,1)$ basis,
\begin{equation}
\frak{B}_{(3,1)}=\left\{
\begin{array}{c}
|00\rangle ,\text{ }\frac{{\small 1}}{\sqrt{{\small 2}}}|11\rangle +\frac{%
{\small 1}}{{\small 2}}(|01\rangle +|10\rangle ), \\
\frac{{\small 1}}{\sqrt{{\small 2}}}|11\rangle -\frac{{\small 1}}{{\small 2}}%
(|01\rangle +|10\rangle ),\text{ }\frac{{\small 1}}{\sqrt{{\small 2}}}%
(|01\rangle -|10\rangle )
\end{array}
\right\} ,
\end{equation}
and an example, $\frak{B}_{(4,0)},$ of a completely entangled, type $(4,0)$
basis,
\begin{equation}
\frak{B}_{(4,0)}=\left\{
\begin{array}{c}
\frac{{\small 1}}{\sqrt{{\small 2}}}(|00\rangle +|11\rangle ),\text{ }\frac{%
{\small 1}}{\sqrt{{\small 2}}}(|00\rangle -|11\rangle ), \\
\frac{{\small 1}}{\sqrt{{\small 2}}}(|01\rangle +|10\rangle ),\text{ }\frac{%
{\small 1}}{\sqrt{{\small 2}}}(|01\rangle -|10\rangle )
\end{array}
\right\} .
\end{equation}

However, despite the existence of type $(0,4),$ $(2,2),$ $(3,1),$ and $(4,0)$
basis sets, no example of a type $(1,3)$ basis set can be found. This leads
to the following:

\begin{theorem}
No type $(1,3)$ basis set exists for a four dimensional Hilbert space,
relative to the factorisation of $\mathcal{H}^{(4)}$ into a product of two
qubit sub-spaces.
\end{theorem}

\begin{proof}
Let $\eta _{1},$ $\eta _{2},$ and $\eta _{3}$ be three orthonormal vectors
in $\mathcal{H}^{(4)}$ that are separable relative to the factorisation $%
\mathcal{H}^{(4)}=\mathcal{H}_{1}^{(2)}\otimes \mathcal{H}_{2}^{(2)}.$ Each
vector $\eta _{i},$ $i=1,2,3,$ is of the form $\eta _{i}=\phi _{i}\otimes
\varphi _{i},$ where $\phi _{i}\in $ $\mathcal{H}_{1}^{(2)}$ and $\varphi
_{i}\in $ $\mathcal{H}_{2}^{(2)}.$

From the condition that $\left| \eta _{i}\right| =1,$
\begin{equation}
\left( \langle \phi _{i}|\otimes \langle \varphi _{i}|\right) \left( |\phi
_{i}\rangle \otimes |\varphi _{i}\rangle \right) =\langle \phi _{i}|\phi
_{i}\rangle \langle \varphi _{i}|\varphi _{i}\rangle >0
\end{equation}

$\Longrightarrow $ None of the factors $\phi _{i}\ $or $\varphi _{i}$ can be
zero for $i=1,2,3.$

Moreover, mutual orthogonality, $\langle \eta _{i}|\eta _{j}\rangle =0$ for $%
i\neq j,$ gives
\begin{eqnarray}
\langle \phi _{1}|\phi _{2}\rangle \langle \varphi _{1}|\varphi _{2}\rangle
&=&0 \\
\langle \phi _{1}|\phi _{3}\rangle \langle \varphi _{1}|\varphi _{3}\rangle
&=&0  \notag \\
\langle \phi _{2}|\phi _{3}\rangle \langle \varphi _{2}|\varphi _{3}\rangle
&=&0.  \notag
\end{eqnarray}

For brevity, the product $\langle \phi _{i}|\phi _{j}\rangle $ shall be
defined $A_{ij},$ and the product $\langle \varphi _{i}|\varphi _{j}\rangle
\equiv B_{ij},$ for $1\leqslant i<j\leqslant 3.$ For the above equalities $%
A_{ij}B_{ij}=0$ to hold, $A_{ij}$ and/or $B_{ij}$ must be zero.

It is firstly evident that not all three of the $A_{ij}$ can be zero. If
this were the case, i.e. $A_{12}=A_{13}=A_{23}=0,$ then
\begin{equation}
A_{12}=0\text{ \ \ }\Longrightarrow \text{ \ \ }\langle \phi _{1}|\phi
_{2}\rangle =0,  \label{Ch2A12}
\end{equation}
which would imply, since $\phi _{i}\neq 0$ for $i=1,2,3,$ and since $\phi
_{i}\in $ $\mathcal{H}_{1}^{(2)}$ and $\mathcal{H}_{1}^{(2)}$ is two
dimensional, that $\phi _{1}$ and $\phi _{2}$ form an orthogonal basis for $%
\mathcal{H}_{1}^{(2)}.$ In this case, therefore,
\begin{equation}
\phi _{3}=a\phi _{1}+b\phi _{2}  \label{Ch2Thi3}
\end{equation}
where $a,b\in \mathbb{C}$ and
\begin{equation}
\left| a\right| ^{2}+\left| b\right| ^{2}=1.
\end{equation}
Then,
\begin{eqnarray}
A_{13} &=&0\text{ \ \ }\Longrightarrow \text{ \ \ }\langle \phi _{1}|\phi
_{3}\rangle =0\text{ \ \ }\Longrightarrow \text{ \ \ }a=0  \label{Ch2As} \\
A_{23} &=&0\text{ \ \ }\Longrightarrow \text{ \ \ }\langle \phi _{2}|\phi
_{3}\rangle =0\text{ \ \ }\Longrightarrow \text{ \ \ }b=0.  \notag
\end{eqnarray}
But (\ref{Ch2As}) with (\ref{Ch2Thi3}) contradicts $\phi _{i}\neq 0,$
leading to the conclusion that not all $A_{12},A_{13}$ and $A_{23}$ can be
zero. Similarly, not all three $B_{ij}$ may be zero.

One way of satisfying the mutual orthogonality conditions, $\langle \eta
_{i}|\eta _{j}\rangle =0$ for $i\neq j,$ is to assume $%
A_{12}=A_{13}=B_{23}=0 $ and $A_{23}\neq 0,$ though by symmetry any other
combination for $i<j$ and $(k\neq i)\&(l\neq j)$ of two $A_{ij}$ and one $%
B_{kl}$ being zero, or two $B_{ij}$ and one $A_{kl},$ would also work.

As before, (\ref{Ch2A12}) may be used to deduce that $\phi _{3}=a\phi
_{1}+b\phi _{2},$ such that $\phi _{1}$ and $\phi _{2}$ form an orthogonal
basis for $\mathcal{H}_{1}^{(2)}.$ From the condition $A_{13}=0,$ it is
clear that $a=0,$ and so because $A_{23}\neq 0$ the conclusion is that $%
b\neq 0.$

Similarly to (\ref{Ch2A12}), the condition $B_{23}=0$ with $\varphi _{i}\in
\mathcal{H}_{2}^{(2)}$ implies that $\varphi _{2}$ and $\varphi _{3}$ form
an orthogonal basis for $\mathcal{H}_{2}^{(2)}.$ Hence
\begin{equation}
\varphi _{1}=c\varphi _{2}+d\varphi _{3}
\end{equation}
for $c,d\in \mathbb{C}$ and $\left| c\right| ^{2}+\left| d\right| ^{2}=1.$
Collecting these results gives
\begin{eqnarray}
\eta _{1} &=&\phi _{1}\otimes (c\varphi _{2}+d\varphi _{3}) \\
\eta _{2} &=&\phi _{2}\otimes \varphi _{2}  \notag \\
\eta _{3} &=&b\phi _{2}\otimes \varphi _{3}.  \notag
\end{eqnarray}
which are clearly mutually orthogonal, as required.

Consider now a fourth non-zero vector $\eta _{4}\in \mathcal{H}^{(4)}.$
Given that $\phi _{1}$ and $\phi _{2}$ form an orthogonal basis for $%
\mathcal{H}_{1}^{(2)},$ and $\varphi _{2}$ and $\varphi _{3}$ form an
orthogonal basis for $\mathcal{H}_{2}^{(2)},$ this new vector may be written
as
\begin{equation}
\eta _{4}=\alpha \phi _{1}\otimes \varphi _{2}+\beta \phi _{1}\otimes
\varphi _{3}+\gamma \phi _{2}\otimes \varphi _{2}+\delta \phi _{2}\otimes
\varphi _{3},
\end{equation}
with $\alpha ,\beta ,\gamma ,\delta \in \mathbb{C}$ and $\left| \alpha
\right| ^{2}+\left| \beta \right| ^{2}+\left| \gamma \right| ^{2}+\left|
\delta \right| ^{2}=1.$ Now, if $\eta _{1},$ $\eta _{2},$ $\eta _{3},$ and $%
\eta _{4}$ are to form an orthonormal, type $(1,3)$ basis for $\mathcal{H}%
^{(4)},$ then because $\eta _{1},$ $\eta _{2}$ and $\eta _{3}$ are clearly
separable, $\eta _{4}$ must be entangled relative to the bi-partite
factorisation $\mathcal{H}^{(4)}=\mathcal{H}_{1}^{(2)}\otimes \mathcal{H}%
_{2}^{(2)}.$ So, from the microsingularity condition (\ref{Ch2Micro}) given
earlier, the condition
\begin{equation}
\alpha \delta \neq \beta \gamma
\end{equation}
must therefore hold for $\eta _{4}.$ However, from orthogonality
\begin{eqnarray}
\langle \eta _{2}|\eta _{4}\rangle &=&0\text{ \ \ }\Longrightarrow \text{ \
\ }\gamma =0 \\
\langle \eta _{3}|\eta _{4}\rangle &=&0\text{ \ \ }\Longrightarrow \text{ \
\ }b\delta =0,  \notag
\end{eqnarray}
but since $b\neq 0,$ the last equality implies $\delta $ is zero. So,
\begin{equation}
\alpha \delta =\beta \gamma =0,
\end{equation}
which is inconsistent with (\ref{Ch2Micro}). Hence, $\eta _{4}$\ cannot be
entangled.\bigskip
\end{proof}

Thus, if three mutually orthogonal vectors in $\mathcal{H}^{(4)}$ are
separable relative to the factorisation $\mathcal{H}^{(4)}=\mathcal{H}%
_{1}^{(2)}\otimes \mathcal{H}_{2}^{(2)},$ then a fourth orthogonal vector
must also be separable. There can be no type $(1,3)$ basis set for $\mathcal{%
H}^{(4)}.$\bigskip

It is interesting as to whether such a theorem can be extended to bases in
higher dimensional Hilbert spaces. Whilst no\ proof has been demonstrated,
no type $(1,5)$ basis set has been found that spans the six dimensional
Hilbert space $\mathcal{H}^{(6)}$ factorised as $\mathcal{H}^{(6)}=\mathcal{H%
}_{1}^{(3)}\otimes \mathcal{H}_{2}^{(2)}.$ If it is actually the case that
no such basis set does exist, it may lead to a conjecture that in a $%
d=d_{1}d_{2}$ dimensional Hilbert space there is no type $(1,d-1)$ basis set
relative to any bi-partite factorisation\ $\mathcal{H}^{(d)}=\mathcal{H}%
_{1}^{(d_{1})}\otimes \mathcal{H}_{2}^{(d_{2})}.$ This in addition also
provokes the question as to which, if any, types of basis sets $(p,r,q)$ are
forbidden relative to a tri-partite factorisation of a $d$ dimensional
Hilbert space. Consequently, which types are allowed in an $N$-partite split
of $\mathcal{H}^{(d)}?$\bigskip

The theorem described above also holds an interesting implication for
physics. As discussed previously, every Hermitian operator, $\hat{\Sigma},$
in a four dimensional Hilbert space $\mathcal{H}^{(4)}$ is associated with a
spectrum of four orthonormal eigenvectors. Further, these four eigenstates
effectively form one possible orthonormal basis set for $\mathcal{H}^{(4)}.$
So, since each basis set may be labelled as type $(p,q)$ relative to the
fundamental factorisation of the Hilbert space $\mathcal{H}^{(4)}=\mathcal{H}%
_{1}^{(2)}\otimes \mathcal{H}_{2}^{(2)},$ reciprocality implies that the
operators themselves may also adopt this label. It is hence possible to
discuss a type $(p,q)$ Hermitian operator, $\hat{\Sigma}_{(p,q)},$ based on
the separability of its eigenstates relative to this factorisation.

The conclusion of the above work is therefore that there exists no type $%
(1,3)$ Hermitian operator acting on a two qubit system. There is no
observable that may be represented by an operator possessing one entangled
and three separable eigenstates, relative to $\mathcal{H}^{(4)}=\mathcal{H}%
_{1}^{(2)}\otimes \mathcal{H}_{2}^{(2)}.$

What makes this result particularly important regards the earlier problem of
separability in the Universe. Even in a two qubit system, the number of
separable states form a set of measure zero in comparison to the number of
all possible states. So, as was remarked in the previous sub-section, the
fact that separability does seem to be a common feature of physically
observed quantum states is ascribed to be due to a careful choice of the
operators that act upon the system. The point that can be learnt from the
present discussion is that mathematics itself appears to enforce certain
constraints on the way in which a system develops. For example, if a
hypothetical mini-universe is imagined with a state $\Psi _{n}$ existing in
a Hilbert space of four dimensions, it is certain that its next state $\Psi
_{n+1}$ will not be one of the eigenstates of a type $(1,3)$ operator.
Mathematics ensures that such universes can only ever be developed with $%
\hat{\Sigma}_{(0,4)},$ $\hat{\Sigma}_{(2,2)},$ $\hat{\Sigma}_{(3,1)}$ or $%
\hat{\Sigma}_{(4,0)}$ type Hermitian operators.

Whilst two qubit universes are of, course, trivial compared to a state of
dimension greater than $2^{10^{184}},$ the result highlights the assertion
that the mathematics of operators places important restrictions on the
development of the state. It may readily be speculated, then, on what other
constraints might naturally be enforced by the operators, especially as the
dimensionality of the Hilbert space increases. Specifically, similar such
constraints may ensure that the possibility of obtaining a highly separable
state for the Universe is actually much more likely than might be expected.
Apparent classicity may be an unavoidable and inevitable feature in a fully
quantum Universe because of tight limitations fixed on its dynamics by
mathematics.

\bigskip

\subsection{Decoherence}

\bigskip

Exactly how quantum mechanics gives way to the classical reality that
scientists observe and measure has been one of the great problems of physics
since the earliest days of the theory. In essence, the difficulty has been
in explaining why states on the macroscopic `everyday' scale never appear to
exhibit the properties associated with quantum states. For example, large
semi-classical states in the laboratory always seem to have well defined
spatial locations, and are never found entangled with one other or existing
in linear superpositions.

Although a number of schemes have been proposed to account for this
phenomenon, by far the current most popular `explanation' is the theory of
decoherence. Since the purpose of this chapter has been to investigate some
of the necessary conditions required for apparent classicity to begin to
emerge from a fully quantum description of the Universe, no such study would
therefore be complete without a discussion of this conventional theory.

\bigskip

\subsubsection{The Theory of Decoherence}

\bigskip

The main thrust of decoherence theory is that a quantum state is driven to
classicality by continual interactions with its environment (see \cite{Zurek}
\cite{Joos}\cite{Omnes}, amongst others).

As an example, consider a quantum subject in the laboratory that is
represented by the state $\psi $ in a two dimensional Hilbert space $%
\mathcal{H}_{\psi }$ spanned by an arbitrary orthonormal basis $\mathcal{B}%
_{\psi }=\{|\downarrow \rangle ,|\uparrow \rangle \}.$ For illustration, it
may be imagined that $\psi $ represents the state of a single electron,
whereas $\mathcal{B}_{\psi }$ represents the set of possible outcomes of a
measurement of the electron's spin component in a particular direction.

Consider also the laboratory detection apparatus used to measure the
electron. This is also described by a unique quantum state, and may in this
simple example be represented by a vector $\Phi $ in the two-dimensional
Hilbert space $\mathcal{H}_{\Phi }$ spanned by an orthonormal basis $%
\mathcal{B}_{\Phi }=\{|\Phi _{\downarrow }\rangle ,|\Phi _{\uparrow }\rangle
\}.$

Now, in order for the apparatus to behave as a detector of $\psi ,$ its
state $\Phi $ must somehow be correlated with the spin states of the
electron. To this end, the basis $\mathcal{B}_{\Phi }$ may be chosen such
that if the detector is found to be in the state $|\Phi _{\downarrow
}\rangle ,$ it is taken be imply that the electron is in a spin down state,
whereas if it is found to be in the state $|\Phi _{\uparrow }\rangle $ then
the electron is assumed to be spin up. In such a system, the basis vectors $%
|\Phi _{\downarrow }\rangle $ and $|\Phi _{\uparrow }\rangle $ are defined
as `pointer states', and are ultimately hoped to give rise to the classical
results of the measurements, i.e. what the physicist actually sees.\bigskip

Let the detector initially be in the `ground' state $|\Phi _{\downarrow
}\rangle .$ If it is to work correctly, it may be assumed that an encounter
with a spin up electron induces a transition in the detector from the state $%
|\Phi _{\downarrow }\rangle $ to the state $|\Phi _{\uparrow }\rangle ,$
whereas a spin down electron leaves the apparatus' state unaffected. In
other words, if the electron is initially in one of the eigenstates $%
|\downarrow \rangle $ or $|\uparrow \rangle ,$ the overall system evolves
according to either (\ref{Ch2dec1}) or (\ref{Ch2dec2}),
\begin{eqnarray}
| &\downarrow &\rangle \otimes |\Phi _{\downarrow }\rangle \rightarrow
|\downarrow \rangle \otimes |\Phi _{\downarrow }\rangle  \label{Ch2dec1} \\
| &\uparrow &\rangle \otimes |\Phi _{\downarrow }\rangle \rightarrow
|\uparrow \rangle \otimes |\Phi _{\uparrow }\rangle .  \label{Ch2dec2}
\end{eqnarray}

Such a process implicitly assumes that there is some sort of coupling
between the electron and detector. This generates an interaction term in the
Hamiltonian governing the system's dynamics, which leads to a unitary and
deterministic evolution of the overall state into one of the products $%
|\downarrow \rangle \otimes |\Phi _{\downarrow }\rangle $ or $|\uparrow
\rangle \otimes |\Phi _{\uparrow }\rangle ,$ depending on the state of the
electron.

Now, the above mechanism provides the correct basis for the classically
expected results if the electron is initially in one of the spin eigenstates
$|\downarrow \rangle $ or $|\uparrow \rangle .$ A problem arises, however,
if the initial electron state is in a linear superposition of the form $\psi
=\alpha |\downarrow \rangle +\beta |\uparrow \rangle ,$ where $\alpha ,\beta
\in \mathbb{C}$ and $\left| \alpha \right| ^{2}+\left| \beta \right| ^{2}=1.$
From (\ref{Ch2dec1}) and (\ref{Ch2dec2}), the electron-detector system is
then evolved into the state
\begin{equation}
(\alpha |\downarrow \rangle +\beta |\uparrow \rangle )\otimes |\Phi
_{\downarrow }\rangle \rightarrow \alpha |\downarrow \rangle \otimes |\Phi
_{\downarrow }\rangle +\beta |\uparrow \rangle \otimes |\Phi _{\uparrow
}\rangle ,  \label{Ch2decent}
\end{equation}
which is clearly an entangled linear superposition of two orthogonal
electron-detector product states. But, such an entangled state is
undesirable if it is hoped that the simple two-level apparatus may be
extended to represent a classical detector, because classical objects are
never seen in linear superpositions. So, if decoherence is to be an answer
to the question of how classicity emerges from quantum theory, it must
provide a mechanism for removing the entanglement of (\ref{Ch2decent}%
).\bigskip

The method proposed in decoherence theory incorporates an extension of the
above `von Neumann chain' of correlated systems to an inclusion of the
environment as well, which is also assumed to be a quantum state. Consider
two particular states of the environment $|\Xi _{\downarrow }\rangle $ and $%
|\Xi _{\uparrow }\rangle $ contained in an enormous Hilbert space $\mathcal{H%
}_{\Xi }.$ These two vectors are taken to be the result of an interaction
between the pointer states of the detector with its surroundings; that is,
if the detector is in the state $|\Phi _{\downarrow }\rangle $ then the
environment will be in the state $|\Xi _{\downarrow }\rangle ,$ whereas if
the detector is in the state $|\Phi _{\uparrow }\rangle $ then the
environment will be in the state $|\Xi _{\uparrow }\rangle .$

With this condition in place, then if the environment is initially in some
ground state $|\Xi _{0}\rangle ,$ it is assumed that the
detector-environment system is evolved into one of the following two
correlations
\begin{eqnarray}
|\Phi _{\downarrow }\rangle \otimes |\Xi _{0}\rangle &\rightarrow &|\Phi
_{\downarrow }\rangle \otimes |\Xi _{\downarrow }\rangle \\
|\Phi _{\uparrow }\rangle \otimes |\Xi _{0}\rangle &\rightarrow &|\Phi
_{\uparrow }\rangle \otimes |\Xi _{\uparrow }\rangle .  \notag
\end{eqnarray}

Overall, then, an initial electron-detector-environment system $\Psi
_{i}\equiv \psi \otimes \Phi _{\downarrow }\otimes \Xi _{0},$ where $\psi
=(\alpha |\downarrow \rangle +\beta |\uparrow \rangle ),$ will develop into
a final entangled state $\Psi _{f},$ such that
\begin{equation}
|\Psi _{i}\rangle =|\psi \rangle \otimes |\Phi _{\downarrow }\rangle \otimes
|\Xi _{0}\rangle \rightarrow \alpha |\downarrow \rangle \otimes |\Phi
_{\downarrow }\rangle \otimes |\Xi _{\downarrow }\rangle +\beta |\uparrow
\rangle \otimes |\Phi _{\uparrow }\rangle \otimes |\Xi _{\uparrow }\rangle
=|\Psi _{f}\rangle .
\end{equation}
\bigskip

If the experiment is repeated identically a large number of times, or
alternatively if a large number of hypothetical identical universes are
simultaneously developed in the same way, the ensemble of final states could
be described in terms of the density matrix $\rho $ defined as $\rho =|\Psi
_{f}\rangle \langle \Psi _{f}|.$ Clearly, then,
\begin{equation}
\rho =
\begin{array}{c}
\alpha ^{\ast }\alpha |\downarrow \rangle \langle \downarrow |\otimes |\Phi
_{\downarrow }\rangle \langle \Phi _{\downarrow }|\otimes |\Xi _{\downarrow
}\rangle \langle \Xi _{\downarrow }|+\alpha \beta ^{\ast }|\downarrow
\rangle \langle \uparrow |\otimes |\Phi _{\downarrow }\rangle \langle \Phi
_{\uparrow }|\otimes |\Xi _{\downarrow }\rangle \langle \Xi _{\uparrow }| \\
+\alpha ^{\ast }\beta |\uparrow \rangle \langle \downarrow |\otimes |\Phi
_{\uparrow }\rangle \langle \Phi _{\downarrow }|\otimes |\Xi _{\uparrow
}\rangle \langle \Xi _{\downarrow }|+\beta ^{\ast }\beta |\uparrow \rangle
\langle \uparrow |\otimes |\Phi _{\uparrow }\rangle \langle \Phi _{\uparrow
}|\otimes |\Xi _{\uparrow }\rangle \langle \Xi _{\uparrow }|
\end{array}
.
\end{equation}

The central argument of the decoherence theorists is that if the environment
is sufficiently large and possesses a large number of energy eigenstates,
and if it is never carefully prepared or probed, then it may be ignored. In
this case, it is possible to trace over all the states of the environment to
obtain the reduced density matrix, $\rho _{s},$ of the electron-detector
system of interest. Specifically,
\begin{equation}
\rho _{s}=Tr_{\Xi }[\rho ]=\sum_{\gamma }\langle \Xi _{\gamma }|\rho |\Xi
_{\gamma }\rangle  \label{ch2dens}
\end{equation}
where the index $\gamma $ implies a sum over every possible normalised state
of the environment, including of course $|\Xi _{0}\rangle ,$ $|\Xi
_{\downarrow }\rangle $ and $|\Xi _{\uparrow }\rangle .$

The result of (\ref{ch2dens}) may be split into a sum $\rho _{s}=\rho
_{d}+\rho _{od}$\ of `diagonal' elements, $\rho _{d},$ given by
\begin{equation}
\rho _{d}=\alpha ^{\ast }\alpha |\downarrow \rangle \langle \downarrow
|\otimes |\Phi _{\downarrow }\rangle \langle \Phi _{\downarrow }|+\beta
^{\ast }\beta |\uparrow \rangle \langle \uparrow |\otimes |\Phi _{\uparrow
}\rangle \langle \Phi _{\uparrow }|
\end{equation}
and `off-diagonal' elements, $\rho _{od},$ of the form
\begin{equation}
\rho _{od}=\alpha \beta ^{\ast }|\downarrow \rangle \langle \uparrow
|\otimes |\Phi _{\downarrow }\rangle \langle \Phi _{\uparrow }|\otimes
\langle \Xi _{\downarrow }|\Xi _{\uparrow }\rangle +...
\end{equation}

In this case, the diagonal elements are equivalent to the states predicted
by classical mechanics, whereas the off-diagonal elements represent the
quantum coherences. Evidently, the environment has no effect on the diagonal
elements, but does influence the off-diagonal terms.

However, if the environmental states are assumed to be orthonormal, $\langle
\Xi _{i}|\Xi _{j}\rangle =\delta _{ij}$ for all $i$ and $j,$ then the
off-diagonal elements clearly become zero. The resulting reduced density
matrix, $\rho _{s}=\rho _{d},$ takes the form of a classical ensemble of
states, with no quantum entanglement.

Overall, then, the superposed electron state $\psi =(\alpha |\downarrow
\rangle +\beta |\uparrow \rangle )$ has been unitarily driven to one of its
classically observed basis states $|\downarrow \rangle $ or $|\uparrow
\rangle $ by an interaction with its environment, and which of these two
states is now actually observed is simply a matter of classical probability.
That is, when an observation is eventually made there is a probability of $%
\alpha ^{2}$ that the electron \textit{is already} in the state $|\downarrow
\rangle ,$ and a probability of $\beta ^{2}$ that the electron is already in
the state $|\uparrow \rangle .$ Compare this with the pre-decoherence case $%
\psi =(\alpha |\downarrow \rangle +\beta |\uparrow \rangle ),$ in which
there is a probability of $\alpha ^{2}$ that the electron \textit{might
subsequently}\ be found in the state $|\downarrow \rangle $ if it is tested
by some operator $\hat{B}$ with orthonormal basis $\mathcal{B}_{\psi
}=\{|\downarrow \rangle ,|\uparrow \rangle \},$ and a probability of $\beta
^{2}$ that the electron might similarly be found in the state $|\uparrow
\rangle ,$ but is really in neither of these states until the actual
observation is made.

\bigskip

\subsubsection{Problems with Decoherence}

\bigskip

It is difficult to predict exactly how the theory of decoherence may fit
with the paradigm proposed in this thesis. As is evident from the brief
summary given above, decoherence is assumed to be a macroscopic phenomenon
that would only arise from a consideration of the interactions and dynamics
of an overall system of very large dimension. In this sense, decoherence may
be viewed as an emergent theory that might therefore potentially be used to
describe how classical physics arises as an approximation to quantum theory
in the large scale limit of systems with very many degrees of freedom. From
this point of view, the ideas of decoherence may well play an important part
in the discussion of a quantum system represented by a state of dimension
greater that $2^{10^{184}}.$\bigskip

On the other hand, it is still difficult to see how decoherence theory could
be applied directly to the Universe as a whole. The main point of
decoherence is that a (usually microscopic) quantum system is evolved into a
classical looking system by continual interactions with its external
surroundings. No similar argument can be applied, however, to the case in
which the quantum state in question is the Universe itself, because by
definition the Universe is not contained in any sort of `background'. In
essence, there is no external environment with which the state of the
Universe is able to decohere.

This observation leads, perhaps, to one of three conclusions: either $^{i)}$
decoherence is a valid theory to describe states inside the Universe, but
not the overall state of the Universe itself; $^{ii)}$ the individual
sub-systems of the Universe decohere each other, such that the overall state
of the Universe is driven to classicity; or $^{iii)}$ decoherence is not
really a fundamental theory of physics. The first of these conclusions seems
a little paradoxical, and leaves the question as to where the `line' can be
drawn that specifies the validity of decoherence. The third conclusion is
quite negative, though still, of course, possibly true. The remaining
possibility is more interesting, and might presumably lead to a situation of
the type in which quantum sub-state $A$ is acting as the environment for
quantum sub-state $B,$ whereas the quantum sub-state $B$ is acting as the
environment for quantum sub-state $C,$ but perhaps quantum sub-state $C$ is
acting as the environment for quantum sub-state $A.$ Such a picture would
immediately be in keeping with the assumed self-referential nature of the
Universe, but a great deal of further investigation is required in order to
discover how, or indeed if, such a hypothetical mechanism might work.\bigskip

One problem that still exists in decoherence theory is the issue of
probability. Using the electron experiment described earlier as an example,
the mathematics of decoherence still provides no explanation of how one of
the basis states $|\downarrow \rangle $ or $|\uparrow \rangle $ actually
gets selected, and therefore why a particular one of these two is actually
observed in the laboratory.

During decoherence, the interplay between an initial quantum state and its
environment gives rise to a well specified interaction term in the
Hamiltonian. The system then undergoes unitary evolution according to the
Schr\"{o}dinger equation, which forces the state into a classical looking
state. The Schr\"{o}dinger equation, however, is a deterministic formula,
and as such the drive of the state from quantum to classical must also be
deterministic. Whilst this not only gives philosophical problems, such as
the possibility of a Laplacian style `clockwork' Universe, it also raises
the question as to how the state can deterministically evolve to only one
out of a set of possibilities. Indeed, as remarked by Erich Joos, one of the
proponents of decoherence, ``\textit{Decoherence can not explain quantum
probabilities without (a) introducing a novel definition of observer systems
in quantum mechanical terms (this is usually done tacitly in classical
terms), and (b) postulating the required probability measure (according to
the Hilbert space norm)}''.

The probability measure is normally introduced into conventional quantum
mechanics by the state reduction process. According to this postulate, then
at the point of measurement of a quantum system, the wavefunction
discontinuously jumps into one of the eigenstates of the Hermitian operator
representing the observation. Moreover, it is this process that abruptly
selects, irreversibly and probabilistically, the next state of the system
out of a set of possibilities. Decoherence, however, contains no such
mechanism, so a question must remain as to how similar selections can be
made if the system is always constrained to evolve reversibly, unitarily,
and deterministically.\bigskip

To give an illustration, in quantum mechanics the famous paradox of
Schr\"{o}dinger's Cat \cite{Schrodinger} relies on which of a set of
possibilities a quantum state develops into. Adapting the earlier electron
example to Schr\"{o}dinger's thought experiment, it might be the case that
if the electron is in the spin down state, then a gun is fired and the Cat
in the sealed box is killed. If conversely the electron develops into the
spin up state, then a gun is not fired and the Cat is spared.

In the conventional Copenhagen interpretation of quantum mechanics, until an
observation collapses the quantum wavefunction, the state of the system is
in an entangled superposition of products of a spin-down electron and a
fired gun with a spin-up electron and an un-fired gun. Consequently, and
taking the conclusion to absurdity, it might then be argued that the Cat is
simultaneously both dead and alive. So, the question has therefore always
been: at what point along the chain is the observation made? If the state
reduction relies on a human observation, is the conclusion to be accepted
that the Cat is able to keep one paw in both life and death until physicists
decide to look inside the box?

In decoherence theory, the linear superposition is destroyed by the
environment, so the electron's spin state is definitely either up or down,
with the inevitable consequences. As such, the corresponding reduced density
matrix for the electron-gun-Cat system only reflects an external observer's
classical ignorance as to what has already happened. Paraphrasing, the
`decision' has already been made by the Universe as to what has gone on in
the box, but until the physicist investigates, only classical probabilities
of obtaining certain results can be discussed. This is obviously like
tossing a coin: the coin definitely lands either heads or tails, but until
it is uncovered it is not known which of these two possibilities has
occurred.

But, in decoherence theory the question remains: how does the Universe
decide whether or not the electron evolves to a spin-up or a spin-down
state? How does the Universe decide if the Cat lives?\bigskip

The lack of randomness is not the only problem caused by a removal of the
state reduction postulate from quantum mechanics. Assuming the principle of
cause and effect is to be believed, any physical state in the universe is
the result of some earlier process. If further it is accepted that every
system in the Universe is fundamentally quantum in nature, then every
physical quantum state in the Universe must therefore be the result of some
earlier quantum process.

However, if these quantum processes appear to ensure that quantum
interferences are eradicated, as the decoherence paradigm suggests, it is
unclear as to how any coherent quantum state might be produced in the first
place. In other words, if quantum systems are only able to develop through a
process of unitary evolution, and if these evolutions effectively remove
quantum coherences and superpositions, what unitary process in the
decoherence paradigm can give rise to entangled states? Specifically, in the
example described above, how is it ever possible to create an initially
superposed state of the form $\psi =(\alpha |\downarrow \rangle +\beta
|\uparrow \rangle )$ using only processes constrained to destroy such
features?

Presumably the conclusion to be drawn is that either decoherence theory
requires an additional mechanism in order to produce such superpositions and
entanglements, or else it must be asserted that every such quantum state
currently in existence has come from some sort of `partial decohering' of an
earlier state that was even more entangled and superposed. In this latter
case, not only would it be necessary to specify how this partial decohering
might work, but also the question would remain as to why, if the Universe
has been continuously and smoothly evolving for a period of about 15 billion
years, are there any quantum effects left in the current epoch at all?

Of course, if the state reduction postulate is included into the formalism,
this problem is not encountered because the preparation of a superposed or
entangled quantum state is simply seen as the outcome of a quantum test.
Moreover, because these outcomes are associated with the eigenstates of
Hermitian operators, they are not constrained to be the result of a
continuous unitary process. Thus with the introduction of Hermitian
operators and state reduction into the dynamics, it is possible to generate
superposed entangled states, and these can then go on to be developed in
subsequent ways, for example to collapse and consequently cause or prevent
guns from firing.

As discussed previously, such a viewpoint forms the basis of the paradigm
proposed in this thesis, in which the test $\hat{\Sigma}_{n}$ simultaneously
collapses the `old' state of the Universe $\Psi _{n-1}$ to prepare and
produce the `new' state $\Psi _{n}.$ In this proposal, the state of the
Universe develops through a long chain consisting of a state reduction,
followed by evolution, followed by a state reduction, and so on.\bigskip

In addition to these theoretical difficulties, recent experiments reviewed
in \cite{Plenio}\ seem to indicate that discontinuous wavefunction jumps are
an observed feature of physical quantum systems. If these investigations
prove conclusive, it is natural to wonder as to how such an empirical result
might be reconciled by a theory of decoherence based on continuous, unitary
evolution.

\bigskip

\subsubsection{Schrodinger's Cat's Stages}

\bigskip

As a final comment to complete this discussion, it might briefly be
mentioned as to how the paradigm proposed in this work views the
Schr\"{o}dinger's Cat paradox, noting that a fuller and more technical
account is evident from Chapter 6.

In the schematic picture given here, an initial state $\Psi _{n}$ is
imagined that is separable into a huge number of factors. Simplifying this,
however, $\Psi _{n}$ may be written in the form
\begin{equation}
\Psi _{n}=|\psi \rangle \otimes |G_{u}\rangle \otimes |C_{l}\rangle \otimes
|R\rangle
\end{equation}
where $|\psi \rangle \in \mathcal{H}_{\psi }$ represents the superposed
electron state $\psi =(\alpha |\downarrow \rangle +\beta |\uparrow \rangle
), $ with $|G_{u}\rangle \in \mathcal{H}_{G}$ the un-fired gun, $%
|C_{l}\rangle \in \mathcal{H}_{C}$ the living Cat, and $|R\rangle \in
\mathcal{H}_{R}$ the rest of the Universe.

Obviously, $\Psi _{n}$ is a vector in the total Hilbert space $\mathcal{H}%
_{\Psi }=$ $\mathcal{H}_{\psi }\otimes \mathcal{H}_{G}\otimes \mathcal{H}%
_{C}\otimes \mathcal{H}_{R}.$

The next test $\hat{\Sigma}_{n+1}$ acting on $\Psi _{n}$ has a basis set of
orthonormal eigenvectors. If two of these eigenvectors are $\Phi $ and $%
\Theta ,$ defined by
\begin{eqnarray}
\Phi &=&|\uparrow \rangle \otimes |G_{u}\rangle \otimes |C_{l}\rangle
\otimes |R\rangle \\
\Theta &=&|\downarrow \rangle \otimes |G_{u}\rangle \otimes |C_{l}\rangle
\otimes |R\rangle  \notag
\end{eqnarray}
then the next state of the Universe $\Psi _{n+1}$ may be either $\Phi $ or $%
\Theta ,$ with relative probabilities $\left| \langle \Psi _{n+1}=\Phi |\Psi
_{n}\rangle \right| ^{2}=\left| \langle \uparrow |\psi \rangle \right| ^{2}$
and\ $\left| \langle \Psi _{n+1}=\Theta |\Psi _{n}\rangle \right|
^{2}=\left| \langle \downarrow |\psi \rangle \right| ^{2}$
respectively.\bigskip

Now, assume that $\Psi _{n+1}=\Phi .$ Moreover, assume also that under this
circumstance, the Rules governing the Universe conspire such that the
subsequent states $\Psi _{n+1+j}$ will `resemble' $\Psi _{n+1}$ for a large
but finite number $J$ of further evolutions $\hat{U}_{n+1+j}$ and tests $%
\hat{\Sigma}_{n+1+j},$ for $J\gg 0$ and $0\leq j\leq J.$ That is, assume
that these subsequent tests $\hat{\Sigma}_{n+1+j}$ for $1\leq j\leq J$ have
eigenstates that are separable in the form $\Phi _{n+1+j}=|\uparrow ^{\prime
}\rangle \otimes |G_{u}^{\prime }\rangle \otimes |C_{l}^{\prime }\rangle
\otimes |R^{\prime }\rangle ,$ where $|\psi ^{\prime }\rangle \in \mathcal{H}%
_{\psi },$ $|G_{u}^{\prime }\rangle \in \mathcal{H}_{G},|C_{l}^{\prime
}\rangle \in \mathcal{H}_{C}$ and $|R^{\prime }\rangle \in \mathcal{H}_{R}$
represent, for example, living cats and un-fired guns that have changed
slightly in their \textit{own} Hilbert spaces as the Universe has developed.

The point is that during these $J$ developments the electron, the gun, the
Cat and the rest of the Universe have not interacted with each other in any
way. Specifically, the electron has not interacted with the gun, and so the
Cat lives.\bigskip

Alternatively, consider the case where $\Psi _{n+1}=\Theta ,$ and assume
that the Rules now conspire so that subsequent states $\Psi _{n+1+m-1}$
resemble $\Psi _{n+1}$ for $1\leq m\ll J,$ but that at time $(n+1+m)$ a test
$\hat{\Sigma}_{n+1+m}$ is chosen which has eigenstates of the form $\Theta
_{n+1+m}=|\downarrow \rangle \otimes |G_{f}\rangle \otimes |C_{l}^{\prime
}\rangle \otimes |R^{\prime }\rangle ,$ where $|G_{f}\rangle \in \mathcal{H}%
_{G}$ represents the state of the fired gun, $|C_{l}^{\prime }\rangle \in
\mathcal{H}_{C}$ the living Cat that has evolved slightly and independently
since its earlier state $|C_{l}\rangle ,$ and $|R^{\prime }\rangle \in
\mathcal{H}_{R}$ the rest of the Universe which has also undergone many
developments during the $m$ preceding evolutions and tests.

Moreover, if the experiment is sound, it is further assumed that this
eigenstate $\Theta _{n+1+m}$ occurs with very high probability. In this
case, it is further assumed that an even later time $(n+1+m+p),$ the Rules
conspire to choose a test $\hat{\Sigma}_{n+1+m+p},$ for $1\leq p\ll J,$ that
has an eigenstate of the form $\Theta _{n+1+m+p}=|\downarrow \rangle \otimes
|G_{f}^{\prime }\rangle \otimes |C_{d}\rangle \otimes |R^{\prime \prime
}\rangle ,$ where $|G_{f}^{\prime }\rangle \in \mathcal{H}_{G}$ represents
the state of the gun that has changed slightly since it was fired, $%
|C_{d}\rangle \in \mathcal{H}_{C}$ the Cat that has now been shot dead, and $%
|R^{\prime \prime }\rangle \in \mathcal{H}_{R}$ the rest of the Universe
which has also developed further in the $p$ evolutions and tests since it
was represented by the state $|R^{\prime }\rangle .$ As before, assuming the
experiment is consistent and the gun well aimed, it is accepted that the
eigenstate $\Phi _{n+1+m+p}$ will also occur with a very high
probability.\bigskip

Thus, the two possible outcomes for the initial collapse of the electron
sub-state from $|\psi \rangle $ to $|\uparrow \rangle $ or $|\downarrow
\rangle $ lead to two different `histories' for the Universe's development.
In neither, however, is there any ambiguity in the fate of the Cat.\bigskip

Obviously, the example given here is described only (highly) schematically.
In reality cats and guns are complicated macroscopic states that will
undergo a series of `internal' transitions as the Universe develops, and
will interact with their surroundings in a multitude of different physical
ways. Indeed, it is a fundamental philosophical question regarding the
nature of persistence to ask what it means to describe an object that is
undergoing tiny changes from moment to moment as `the same cat'. In fact,
some of the ideas of decoherence theory may contribute an important part to
this particular discussion.

In principle, however, the main point from the above treatment of the
Schr\"{o}dinger's Cat paradox should be evident. The conjecture is that the
Universe automatically and self-referentially selects an operator $\hat{%
\Sigma}_{n+1}$ to test itself, and it is this self-measurement that
collapses the electron sub-state into one of its basis vectors $|\downarrow
\rangle $ or $|\uparrow \rangle ,$ without the need for a conscious observer.

Perhaps it is this combination of self-referential testing with
discontinuous state reduction, and maybe even macroscopic decoherence
effects, that might save the Cat's life and give it a classical identity.

\bigskip \newpage

\section{A Quantum Origin of Space}

\renewcommand{\thefigure}{5.\arabic{figure}} \setcounter{figure}{0} %
\renewcommand{\theequation}{5.\arabic{equation}} \setcounter{equation}{0} %
\renewcommand{\thetheorem}{5.\arabic{theorem}} \setcounter{theorem}{0}

\bigskip

As discussed at the beginning of the previous chapter, one qualification for
the presence of classicity follows from the observation that ``\textit{this}
object with \textit{these} qualities is \textit{here, }whereas\textit{\ that}
object with \textit{those} properties is\textit{\ there''}. Whilst the main
of the last chapter focused on the issue of when it is possible to specify
`this' or `that' object, it did not examine how the properties of the
complex vector representing the Universe might give rise to the spacetime
concepts of `here' and `there'. This question is addressed now.

\bigskip

\subsection{Background}

\bigskip

When attempting to develop theories to account for the presence of space,
time and matter in the Universe, physicists often adopt one of two opposing
viewpoints. These methods may be described as either \textit{bottom-up} or
\textit{top-down}, and reflect the basic difference between reductionist and
holistic physics. This difference is itself perhaps predictable in a
Universe containing remarkably successful principles such as quantum field
theory, which exhibits both local and global\ features.

Many of the bottom-up approaches proceed generally from the assertion that,
at its most basic level, the Universe can be represented by a vast
collection of discrete events existing in some sort of mathematical space.
Time and space are introduced as arising from the relations between these
events, such that (classical) reality as we understand it emerges on a
macroscopic scale due to the complex connections between these fundamental,
microscopic entities. Wheeler was one of the earliest proponents of this
idea \cite{Wheeler}, by envisaging a Universe full of a pre-geometric
``dust'' from which spatial degrees of freedom emerge. These `ultimate'
notions of pre-geometry have been developed more recently by Stuckey \cite
{Stuckey}.

On the other hand, many of quantum cosmology's top-down approaches hold that
the entire Universe should be treated as a single system. Top-down theorists
often seek to write down a unique state description for the Universe, before
evolving it according to a given set of laws or conditions. From this point
of view, the apparent classical reality that physicists perceive is just an
approximation to that part of the Universe under investigation whenever a
fully quantum mechanical description can be neglected.\bigskip

A selection of some of the contemporary bottom-up and top-down approaches
are reviewed below in Sub-sections 5.1.1 and 5.1.2. Throughout the rest of
this chapter it will then be shown how some of the general points of these
two approaches might be reconciled as being different aspects of the same
theory. That is, in the paradigm proposed in this thesis, the discrete
events postulated on the microscopic pre-geometric scale may be associated
somehow with the factor sub-states of the single state representing the
completely quantum Universe. Thus, such a viewpoint may be labelled a type
of `top-down pregeometry'. It will be argued that it is from the dynamics of
these changing sub-states that familial relations may arise, and that these
relations could be seen as the origin of spatial degrees of freedom in the
appropriate limit.

\bigskip

\subsubsection{Bottom-Up Approaches}

\bigskip

One of the bottom-up theories of the Universe is the Causal Set Hypothesis
\cite{Sorkin}-\cite{Brightwell}, which states that (quoting \cite{Rideout}):
\textit{``...spacetime, ultimately, is discrete and ... its underlying
structure is that of a locally finite, partially ordered set (a causal
set)''. }In this model it is postulated that classical, discrete ``events''
are generated at random, though it is made clear that they are not embedded
into any sort of physical background space. Spacetime may then be recovered
as an emergent consequence of the ordering that results from imposing
certain logical relations between the members of these sets of events.

Overall, actual physical space in this paradigm manifestly consists of a
causal set (or ``causet'') of points which yield a metric structure in the
continuum limit \cite{Brightwell1}. Additionally it may be shown that the
dimension of this spacetime can be a scale dependent quantity, making the
model equally compatible with notions of four dimensional general relativity
and higher dimensional Kaluza-Klein theories, including modern string and
m-brane physics. The exact details of classical causal set theory are
elaborated upon in Section 5.2.\bigskip

A related idea is that of Cellular Networks (CN) \cite{Requardt}\cite
{Requardt1}, which argues that, on the microscopic scale, the geometry of
space may be represented by a mesh of primordial cells or `nodes'
interacting with each other via a series of interconnecting `bonds'. These
nodes are assumed to contain physical information by way of internal state
structures. The bonds themselves may be in one of a number of `bond states',
allowing the strength and types of interaction to be controlled. The
evolution of the Cellular Network is similar to that of a cellular automaton
in that the Universe proceeds as a giant machine, but differs in the respect
that the number and type of bonds in the network may change with time. For
example, one change might be that two cells unconnected in one instance may
be joined by a certain type of bond in the next. The vision is of a
self-organising topology that is ever changing and depends on the physics of
the situation being modelled. As before, metric structures are recovered as
a continuum concept.\bigskip

Zizzi \cite{Zizzi}\cite{Zizzi1} continues the machine principle of Cellular
Networks with the analogy that the Universe behaves in a way similar to
computational information theory, in what she defines as a ``Quantum Growing
Network'' (QGN). The state of the Universe is postulated to be a tensor
product of a vast number of elementary, two-dimensional quantum degrees of
freedom (qubits) which are connected and processed by a set of quantum logic
gates. Further, as time goes by, the number of qubits increases, and hence
so does the dimensionality of the Universe's Hilbert space. Overall, Zizzi
argues that the Quantum Growing Network system forms a `proto-spacetime'
which may give rise to physical spacetime in a manner similar to
Requardt's.\bigskip

In the Spin Network (or `Spinnet') approach proposed by Penrose \cite
{Penrose2}, spacetime is generated from the relations between combinations
of fundamental ``units'', where each unit may be likened to an elementary
particle that possesses no characteristics apart from total angular
momentum. The units may interact with one other, and a system of interacting
units may be represented by a graph. Each edge of the graph denotes a unit
coming into or arising from an interaction, whereas the vertices are the
interactions themselves. Penrose restricts his analysis to tri-valent
graphs, which may be thought of as describing two units joining to form a
third or one unit splitting into two. Note, however, that because there is
no `direction' inherent to the graphs, each is assumed to represent all of
the allowed interactions between the three units. The only constraint
imposed is that the vertices conserve angular momentum, such that whichever
particles are chosen to be the ones `entering' the interaction, the sum of
their angular momenta must equal the sum of angular momenta of the remaining
units (see \cite{Major} for a review).

Given a large number of units, a large number of graphs may be obtained.
Further, if one of the edges of one graph has the same value of angular
momentum as the edge of another graph, they may be joined and the two graphs
connected. By continuing this process, it is possible to create a network of
graphs where lines represent angular momentum carrying particles and
vertices represent their interactions. Penrose shows how an emergent
geometry may arise by considering this network of relations.

Markopoulou and Smolin \cite{Markopoulou} investigated the causal evolution
of such spin networks by combining the Causal Set approach of chains of
events with the Spin Network notion of geometry. Given the set of edges and
vertices comprising a spin network, rules are suggested for generating a new
set from their particular relations. In fact, a number of possible new
networks may be produced by exploiting the fact that each graph may
represent a number of possible interactions. If the rules are repeated a
series of times, a chain of networks may be created with a definite causal
structure existing between them. By considering, in the manner of Causal
Sets, the sums over histories of these chains of spinnets, Markopoulou
\textit{et al} were able to generate amplitudes of transmission from an
initial to a final topology. The model leads to the production of a series
of timelike surfaces, analogous to an evolving spacetime.

\bigskip

\subsubsection{Top-Down Approaches}

\bigskip

One search for a top-down model of the Universe has been the search for a
consistent theory of quantum cosmology. On the basis that the large scale
structure of the Universe is described by general relativity, some
cosmologists \cite{deWitt} have attempted to canonically quantise the
solutions of the Einstein field equations. Given canonical variables, the
Lagrangian and action functional can be defined, and quantum fields can be
introduced; overall a quantum state function $\Psi $ of the Universe is
generated. This method of quantum cosmology involves an investigation of the
evolution of the Universe's wavefunction according to the Wheeler-deWitt
equation, but is associated with the notorious ``Problem of Time''.

Hartle and Hawking \cite{Hartle} progressed quantum cosmology by adding
appropriate constraint conditions to the dynamics, such that the Universe
could appear to be `created from nothing' by a manner analogous to a quantum
fluctuation or tunnelling process. Further developments have also been made
\cite{Linde}\cite{Liddle} by adding inflationary terms to the Lagrangian in
order to account for the observed isotropy, homogeneity and flatness of the
Cosmos in the current epoch. These approaches again assert that the Universe
is described by a single quantum state.\bigskip

Given that the Universe is observed currently to be expanding, many
cosmologists extrapolate back to conclude that it must have begun from a
spacetime singularity. This, however, causes problems in relativity theory
because regions of very high curvature require a theory of quantum gravity,
and the search for a consistent model of this has proved elusive. So, a
proposed alternative to the inevitable Big Bang singularity has been the
Ekpyrotic Universe model \cite{Turok}.

The approach begins with the hypothesis that every point in our four
dimensional Universe is mapped to a point on part of a hypersurface called a
``D-Brane'', which may be thought of as a `thin wall' or membrane existing
in part of a higher dimensional reality. This D-brane, containing the
entirety of our Universe, is separated by some sort of `Bulk' volume from a
second D-brane which may itself contain a second, `hidden' universe.

Time had no beginning in the Ekpyrotic Universe model. In an era that
conventional cosmologists may refer to as pre-Big Bang, i.e. at times
greater than $\sim 15$ billion years ago, our 4-dimensional universe within
its D-brane was cold and empty. It is postulated that at some time during
this period, a light (compared to the two D-branes) `bulk-brane' peeled away
from the D-brane containing the hidden universe, and travelled across the
bulk volume towards our D-brane. When they collided, it is proposed that the
bulk-brane's kinetic energy was transferred into heat and excitations of the
various force and matter fields contained within our D-brane. This marked
the start of what appeared to be a hot big bang in our Universe, which
proceeded to expand and evolve in the way understood by standard astronomy.

The Ekpyrotic model hopes to provide a mechanism for generating the observed
isotropy, flatness and homogeneity of the universe, without appealing to any
artificial inflation fields, and without containing an initial singularity.
Additionally, it may include an explanation for why gravity is weaker than
the other three fundamental forces. Brane (and string) theory, however, is
still a long way from being generally or empirically accepted, and is itself
riddled with unanswered or unaddressed questions.

\bigskip

\subsection{Classical Causal Sets}

\bigskip

As mentioned above, a number of authors have introduced the possibility that
continuous spacetime might emerge from a consideration of the relationships
between the members of a causal set. In this paradigm, the Universe is
envisaged to consist ultimately of an enormous number of `\textit{events'},
where each event is assumed to be a separate, discrete, mathematical object
of some sort.

By definition \cite{Sorkin}, a causal set (or ``causet'') $\mathcal{C}$ is a
locally finite, partially ordered set (or ``poset'') of objects $\mathcal{C}%
=\{x,y,...\}.$ Each member of a partially ordered set either shares, or does
not share, a unique binary relationship with every other member of the set.
Denoting this relationship by the symbol $\prec ,$ which may be seen as a
type of comparison, two members $x$ and $y$ of a poset are hence connected
as $x\prec y$ or $y\prec x,$ or else $x$ and $y$ are said to be incomparable.

The relationship $\prec $ consequently introduces an order between the
members of the set, and this is made consistent by ensuring that it is
transitive $(\mathbf{T})$ and asymmetric $(\mathbf{A})$. In addition, it is
conventionally assumed that $\prec $ is also irreflexive $(\mathbf{I})$. So,
for $x,y,z\in \mathcal{C}$ the following constraints are imposed:
\begin{eqnarray}
(\mathbf{T}) &:&\text{ \ \ }x\prec y\text{ and }y\prec z\Longrightarrow
x\prec z \\
(\mathbf{A}) &:&\text{ \ \ }x\prec y\Longrightarrow y\nprec x  \notag \\
(\mathbf{I}) &:&\text{ \ \ }x\nprec x  \notag
\end{eqnarray}

A poset may be described as locally finite if, between any two members $x$
and $y,$ where $x\prec y,$ there are a finite number of events $a,$ $b,$ $%
c,...$ such that $x\prec a\prec b\prec ...\prec y.$ In other words, only a
limited number of events ``mediate'' \cite{Raptis} between the event $x$ and
the event $y.$ A causal set is defined to be such a locally finite,
partially ordered set.\bigskip

One method of generating a causal set is via a process of `sequential
growth' \cite{Rideout}. At each step of the growth process a new element is
created at random, and the causal set is developed by considering the
relations between this new event and those already in existence.
Specifically, the new event $y$ may either be related to each of the other
events $x$ as $x\prec y,$ or else $x$ and $y$ are said to be unrelated. Thus
the ordering of the events in the causal set is as defined by the symbol $%
\prec ,$ and it is by a succession of these orderings, i.e. the growth of
the causet, that is ultimately ascribed to constitute the passage of time.

The relation $x\prec y$ is hence interpreted as the statement: ``$y$ is to
the future of $x".$ As a consequence of this interpretation, the asymmetric
condition may now be seen as a removal of the possibility that the causet
will contain anything resembling closed time-like curves.

The above association highlights the similarity between the relations $\prec
$ in causal set theory, and the idea of lightcones in relativity. In a
causal set $\mathcal{C},$ the set of elements $y_{i},$ related to an event $%
x $ by the relation $x\prec y_{i},$ represent the causal future of $x.$ This
relationship is analogous to the volume $V_{X}$ contained within the future
lightcone of a point $X$ in a theory of continuous spacetime, examples of
which being general and special relativity. Conversely, an event $z\in
\mathcal{C}$ that is incomparable to $x$ may be said to be causally
disconnected from $x,$ and this is similar to the set of points outside of
the lightcone of $X.$ In classical physics, events outside of this region $%
V_{X}$ are not affected by changes inside the lightcone, for example at $X,$
and are hence causally independent. This places an important physical
constraint on the members of $\mathcal{C},$ since continuous spacetime is
eventually hoped to emerge from a causal set description.

Of course, similar associations exist for points in the past lightcone of $%
X, $ and the objects $y_{j}$ in $\mathcal{C}$ related to $x$ by $y_{j}\prec
x.$\bigskip

A causal set may be represented by a Hasse diagram. Further, the set of
causal sets that may be constructed from a growing number of events can be
represented by a Hasse diagram of Hasse diagrams.

In each Hasse diagram, the events are shown as spots and the relations as
solid lines or links between the events; emergent time runs from bottom to
top, and the direction of the growth process from one causal set to the next
is consequently denoted by the arrowed lines. A typical such set of diagrams
is given in Figure 5.1, which contains the set of causets of less that four
elements (and features as part of Fig. 1 in \cite{Rideout}), where each
large circle represents an individual Hasse diagram, and hence a particular
causal set.
%\FRAME{ftbpFU}{219.5pt}{125.125pt}{0pt}{\Qcb{The Hasse diagram
%of Hasse diagrams featuring those Causal Sets containing up to three members.%
%}}{}{Figure 5.1}{\includegraphics[natwidth=4in,width=4in,bb=0in 0in 3in 2in]{i1kob000.ps}}

\begin{figure}[t]
\begin{center}
\includegraphics[width=4in]{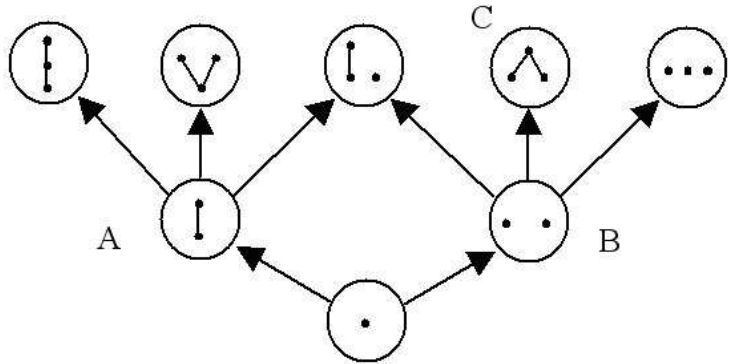}
\caption{The Hasse diagram
of Hasse diagrams featuring those Causal Sets containing up to three members.}\label{Figure 5.1}
\end{center}
\end{figure}

In the example in hand, the initial causal set has just one event, as shown
in the lowest of the large circles. The next event to be born may or may not
share a temporal relation with the first; that is, it may either lie to the
future of the first, or not. Thus, one of two possible causal sets may be
created, as shown by the two Hasse diagrams represented by the large circles
labelled $A$ (temporal relation) and $B$ (no temporal relation). The third
event to be produced may share a temporal relation with all, some, or none
of the previous events, leading to the generation of five possible causal
sets, and hence five possible Hasse diagrams. Figure 5.1 hence represents
the sets of possible causets at three successive times.

The process may obviously be extended indefinitely to create increasingly
longer causets of events, and these may be represented by a growing number
of Hasse diagrams of increasing complexity.\bigskip

With the above in mind, it is possible to introduce familial concepts into
the sequentially growing causal set. Consider as an example a causal set $%
\mathcal{C}_{n}$with $n$ members, $\mathcal{C}_{n}=\{a_{1},a_{2},...,a_{n}%
\}. $ Consider also a second causal set $\mathcal{C}_{n+1}$with $n+1$
members, $\mathcal{C}_{n+1}=\{a_{1},a_{2},...,a_{n},a_{n+1}\},$ `grown' from
$\mathcal{C}_{n}$ by adding the $(n+1)^{th}$ member $a_{n+1}.$ If $a_{n+1}$
is not in the past of any of the elements $a_{i},$ for $i=1,...,n,$ then it
is said to be a ``maximal element''. In this case, the causet $\mathcal{C}%
_{n+1}$ may be called a ``child'' of $\mathcal{C}_{n},$ which in turn may be
named its ``parent''.

In general, a particular causet $\mathcal{C}_{n+1}$ grown from $\mathcal{C}%
_{n}$ could have a number of different topologies, because there are often
very many ways of adding a maximal element $a_{n+1}$ to $\mathcal{C}_{n}$
such that it is to the future of none, one, some or all of the $n$ elements
already in existence. Paraphrasing, there are a number of different Hasse
diagrams that may be drawn by adding just one maximal element to $\mathcal{C}%
_{n},$ depending on `where' it is added. The members of this set of possible
causal sets are hence called ``siblings'' of one another, because they are
all children of a common parent. Such sets of parents and children may be
called ``families''.

This concept can be extended in the obvious way to include, for example,
definitions of grandchildren and great-grandchildren etc. As a simple
illustration of these ideas, the causal sets labelled $A$ and $B$ in Figure
5.1 are clearly siblings, because they are children of the initial parent
causet containing just one member. Similarly, the causal set $C$ is one of
the three children of $B,$ and therefore also one of the grandchildren of
the original (single member) causet.

\bigskip

The crux of causal set theory is that (to quote \cite{Sorkin}): ``...\textit{%
a classical space-time's causal structure comes very close to determining
its entire geometry''. }Thus, in the large scale limit of very many events
the causal sets are hoped to yield the properties of continuous spacetimes.
To this end, metrics, distances and dimension should all be ready features
of the topology.

It is an important feature of causal sets that the events are not taken to
be embedded in any sort of physical background space. The objects themselves
exist in nothing but a mathematical manifold, and it is only by taking
account of the network of relations between a large number of such events
that the actual geometry of the manifold, and hence the spatial relations
familiar to physics, begins to emerge.

The basic methodology employed to generate space from these very large
causal sets is to use the causal order of the set to determine the topology
of the manifold into which it is embedded. This is converse to the standard
procedure employed in continuous geometry in which the properties of the
manifold and metric are used to determine the lightcones of the spacetime,\
and from these the causal order of events may in turn be inferred.\bigskip

Concepts such as timelike geodesics and distances may be introduced into the
analysis of causal sets by considering the length of paths between events
\cite{Brightwell1}.

Consider first a `\textit{chain}', $E,$ of events in a causal set, i.e. an
ordered group of elements $E=\{p,q,...\}$ in $\mathcal{C}$ in which every
two elements of $E$ are somehow related by $\prec .$ By analogy with special
relativity, a chain evidently possesses the causal structure of a spacetime
manifold: each event $r\in E$ is either to the past or future of every other
event $s\in E.$ Moreover, it is possible to define the `\textit{path length}%
' of a given chain between two events $x$ and $y$ in terms of the number of
links in the chain separating these two elements.

Of course depending on the topology of $\mathcal{C}$ there may be a number
of different possible chains `linking' any two events $x$ and $y,$ for $%
x,y\in \mathcal{C}.$

However, from this observation it is possible to define a `\textit{maximal
chain}' $M,$ where $M$ $\subset \mathcal{C},$ as a subset of elements $%
M=\{a_{1},a_{2},...,a_{m}\}$ contained in $\mathcal{C}$ such that $%
a_{i}\prec a_{i+1},$ for $1\leq i\leq m-1,$ and where there is no other
element $b\in \mathcal{C}$ for which $a_{i}\prec b\prec a_{i+1}.$ Clearly, $%
M $ specifies a unique path of events between $a_{1}$ and $a_{m},$ and this
is extremal in $\mathcal{C}.$ Thus, it is immediately possible to define the
path length of a maximal chain $a_{1}\prec $ $a_{2}\prec ...\prec a_{m}$ in
terms of the number of links between $a_{1}$ and $a_{m}.$ In this case, the
path length of $M$ is clearly given by $m-1.$\bigskip

It is from this discussion of path lengths that a notion of timelike
`distance' can arise. Given any two comparable events $x\prec ...\prec y$ in
a causal set, the timelike distance $d(x,y)$ may be defined as the maximum
length of path between them, i.e. the `longest route' allowed by the
topology of the causet to get from $x$\ to $y.$

A number of issues arise from this definition. Firstly, it implies that
(timelike) distance is, at root, manifestly a counting process. As observed
by \cite{Sorkin}, this is in accordance with Riemann's suggestion regarding
the measurement of spatial size.

Secondly, the connection between distance and extremal chains is analogous
to the use of geodesics as extremal path lengths in relativity; recall that
in continuum mechanics a geodesic is defined as the extremal length between
two points, and the distance between them is that length. The proposed
mechanism makes use of the maximal number of objects causally separating two
events, and as such the extremal distances defined in this way are
associated with geodesics in continuous spacetime.

Thirdly, Brightwell \textit{et al} \cite{Brightwell1} remark that the above
definition of distance satisfies a relationship similar\ to the conventional
`Triangle Inequality'. For example, consider three events $x,y,z\in \mathcal{%
C}$ such that $x\prec y\prec z.$ If the distance $d(x,z)$ is given by the
path length of the maximal chain between $x$ and $z,$ then by definition
this distance cannot be shorter than the path length between $x$ and $z$ via
any other possible chain. Specifically, if an alternative route is via $y,$
then this conclusion implies that $d(x,z)\geq d(x,y)+d(y,z),$ with the
equality holding only when $y$ is part of the maximal chain.

Note, however, that such a relationship differs from the standard triangle
inequality of distances, given in obvious notation in the form $%
D(X,Y)+D(Y,Z)\geq D(X,Z).$ Moreover, it is also unclear how the above
theorists would balance this `reversed' result with the conventional case,
an issue made especially pertinent by the fact that the standard version is
generally taken as a pre-requisite for a metric to exist. The physical basis
behind such a reversed inequality relationship needs therefore to be fully
defined by the authors if it is to be used to generate metric-like
structures, and careful physical and mathematical considerations are first
required in order to generate relativistic spacetimes from the underlying
classical causal set ideas.\bigskip

The above definition of timelike distances applies to when quantifying the
separation between comparable events, i.e. between those events $x$ and $y$
in $\mathcal{C}$ for which $x\prec ...\prec y.$ For incomparable events, on
the other hand, no such timelike definition is possible, because
incomparable events instead share the characteristics typically exhibited by
spacelike separated objects in conventional physics.

However, by exploiting this similarity between the incomparable events of
causal set theory and the causally disconnected features of spacelike
separated points in continuous spacetime, it is possible to introduce an
analogous definition of spacelike distances into the causal set description.
Following the lead of \cite{Brightwell1}, it is argued that a method of
measuring spatial distances using light beams and clocks should be employed,
in which the distance between two objects is determined by sending a light
signal from one of these to the other and measuring the time elapsed before
it is returned.

Consider a timelike geodesic $G$ in $\mathcal{C},$ where $G$ is defined as $%
a_{1}\prec $ $a_{2}\prec ...\prec a_{m}.$ From the earlier discussion, $G$
is a maximal chain. Consider also another element $x$ $\in \mathcal{C}$ \
that is not in $G.$ If $a_{k}$ is the highest member of $G$ that is below $%
x, $ then there is no other element $a_{i}$ in $G$ for which $a_{k}\prec $ $%
a_{i}\prec ...\prec x.$ Similarly, if $a_{l}$ is the lowest member of $G$
that is above $x,$ then there is no other element $a_{j}$ in $G$ for which $%
x\prec ...\prec $ $a_{j}\prec a_{l}.$ Then, the spacelike distance $d(x,G)$
between $x$ and $G$ may be defined as $d(x,G)=d(a_{k},a_{l})/2.$

Overall, then, a measure of spacelike separation between members in $%
\mathcal{C}$ is recovered by considering the topology of the temporal
relations over the causal set, analogous to how lightcone structures may be
used in special relativity to determine spatial distances.\bigskip

With the above definitions of timelike and spacelike distances in place, it
is possible to begin a discussion on concepts of velocity \cite{Brightwell1}%
. Specifically, such velocities have meaning in terms of the ratios between
average spatial distances encountered in given lengths of temporal duration.
Since these spatial distances intrinsically involve concepts of geodesics
and basic lightcone structure, it is here that embryonic ideas of special
relativity are expected to emerge from causal set theory.

Also, once a measure of distance has been introduced into the model, it is
possible to discuss concepts of `volume' and `area'. To this end, the
(hyper)volume of the emergent spacetime may be defined in terms of numbers
of events, where a certain quantity of events may specify a certain volume.
As with the definition of a distance in terms of path lengths, volume is
also seen here simply as a counting process. This is perhaps to be expected,
since measurements of distances are in many ways nothing but measurements of
the `size' of a one dimensional volume.

Continuing this logic, the dimension, $d,$ of the causal set may
consequently be obtained in a similar way by considering average lengths of
path, $l,$ in a given volume, $v.$ It may hence be possible to introduce
relational rules of the form $v\sim l^{d},$ in keeping with ideas of
Hausdorff dimension \cite{Hausdorff}. It is from arguments of this type that
the inhomogeneous topology of causal sets may allow different physical
dimensions to emerge at different locations and on different physical scales.

\bigskip\

Whilst the classical causal set hypothesis summarised above is a promising
approach to the origin of space, a number of unresolved questions, problems
and conceptual difficulties arise if it is assumed to provide a complete and
consistent description of the observed Universe. These will be expressed in
turn.\bigskip

The first of these questions regards the physical basis behind the model:
what actually \textit{are} the postulated events that comprise the causal
set? Are they to be taken as some sort of `pregeometric particle', analogous
to the momentum carrying ``units'' peculiar to spin networks? If this is the
case, would it be possible to physically observe them, for example in a
futuristic accelerator-detector experiment? Alternatively, if they are
simply just mathematical objects, by what process is a physical Universe
comprising of fundamental fields and forces expected to emerge?

On a related issue, what exactly \textit{is} the physical mechanism that is
responsible for the events' creation? How do these events, be they physical
objects or mathematical abstracts, suddenly come into existence? Do they
appear from nothing, or are they removed from some sort of \ giant
`reservoir' of pre-existing events before they are added to the causal set
representing the Universe? If this latter supposition is correct then where
is this reservoir, what is its physical basis, and what is it like? If not,
and instead events just constantly appear at random, then what does this
imply for physics in the Universe? Specifically, if the events are hoped to
be the correct `building-blocks' for a Universe that possesses certain
physical properties and characteristics, is it to be accepted that, for
example, the total energy or momentum of the Universe is increasing as more
building blocks are added? If this is not the case, then a paradoxical
situation occurs in which principles such as the conservation of energy,
which appears fundamental for the Universe as a whole, cannot be held as
fundamental for the objects representing the Universe's ultimate description.

Also, is there a physical interpretation for the apparent external time
parameter used to govern \textit{when} events are created?\bigskip

A second problem with the classical causal sets produced is that they are
not quantum. This is obviously not ideal if they are to form the ultimate
description of a physical Universe that does contain quantum theory as a
fundamental ingredient. However, the objection here is not that the model
has simply not yet been extended to the case where, for example, the events
are quantum variables instead of classical objects. Rather, as it stands the
classical causal set description goes against some of the principles present
at the very heart of quantum theory.

As an illustration of why this is so, consider three particular events $%
x,y,z\in \mathcal{C}$ related by $x\prec y\prec z.$ This relational
statement seems to imply that the events $x,$ $y$ and $z$ each possess an
existence independent of each other and of everything else in the causal
set. In other words, in classical causal sets every event is granted just as
much physical significance as every other; in $\mathcal{C}$ the events $%
x,y,z $ all exist to exactly the same extent. Once created an event exists
forever, such that for example $z$ is always in the future of $y$ `for all
time', and $x$ is always in the past of $y.$ Indeed, in order for two events
to be directly compared as $y\prec z,$ it must presumably be accepted that
both $y$ and $z$ exist, at least in some sense. The binary comparisons $%
\prec $ are therefore taken to relate pre-existing relationships between the
events in the causet, and as such are assumed to reflect pre-existing
attributes of the objects.

Such a viewpoint is manifestly a classical `Block Universe' approach. This
is perhaps why classical causal sets are expected to generate continuous
spacetime general relativity so successfully in the emergent limit, because
relativity's overall vision is of a Universe existing in a $3+1$ dimensional
arena in which the temporal parameter is given an equally `eternal'
dimensional footing as the other three.

However, as suggested in Section 2.2, it may be partly because conventional
general relativity relies on Block Universe models of physics that is
preventing its unification with quantum theory.

According to the conclusions of the Kochen-Specker theorem, the results of
the work of Bell discussed in Chapter 3, and the view of Bohr that the
quantum analogues of classical values (such as position and momentum) do not
possess any reality independent of observation, then it is the Process time
approach that is necessary for a consistent description of quantum
mechanics. Assuming, then, that quantum theory is a foundational feature of
the Universe, if the event $x$ exists in the past of the event $y,$ and the
event $z$ exists in the future of the event $y,$ then $x,$ $y$ and $z$
cannot be given equivalent existence. Only one of these, i.e. the `present'
event, can have any physical existence, and even this does not exist in the
sense traditionally assumed by classical physics.

In quantum theory, only the present can be known with any degree of
certainty. It is not possible to discuss \textit{the} future, because no
such concept physically exists, but only \textit{potential} futures in terms
of conditional probabilities. Similarly, the past only has significance in
terms of what observers in the present can recall about where they came from.

It is unclear how this conclusion may be incorporated into a classical
causal set description of physics relying on the equivalence of the
existence of $x,$ $y$ and $z$ across time.\bigskip

A further criticism of causal sets (from the point of view of quantum
theory) comes from an interpretation of what the Hasse diagrams actually
represent. In the growth process from a parent Hasse diagram to one of its
child causets, a new event may be incorporated that is to the future of two
(or more) incomparable events. The problem associated with this is that
without any sort of external agent building the causet, how does this event
`know' that it is to the future of these incomparable events, given that no
information can be exchanged between them? Without a god-like observer, how
are the temporal relations $\prec $ decided? Since time is a phenomenon
expected to emerge from the model, what is the mechanism for deciding how
one event is related to another, so that time can indeed emerge? Indeed,
given that the whole point of causal set theory is that the events are not
embedded in any sort of physical background spacetime, how is any locational
memory contained in the dynamics at all? In the diagram labelled $C$ in
Figure 5.1, for example, it is not clear how the event created last `knows'
which way it is related to the other two: if the first and second events
exist independently of one another, how can they communicate in such a way
to `inform' the third of its temporal position? Is it necessary to postulate
some sort of external source of information, so far ignored in classical
causal set theory, that stores the location of each of the events?

On a related note, since the addition of a maximal element to a parent
causet may give rise to very many possible child sets, how is it that just
one of these new causets actually gets selected to form the basis for the
single reality experienced by the Universe?

\bigskip

Overall, it is argued that the ideas of classical causal sets provide a good
starting point for a discussion of the emergence of space. They do, however,
lack a firm physical basis, and it is unclear how they could be `quantised,'
at least directly.

But, as will be shown in the remainder of this chapter, it is possible to
naturally generate structures resembling causal sets from the fully quantum
description of the Universe proposed in this thesis. Moreover, the `objects'
forming these structures will be shown to have identifiable physical
grounding, and so the quantum causal sets introduced in the following are
not restricted by the problems inherent in their classical counterparts.
Thus, they are ascribed to potentially address the issue of how continuous
space and time may emerge in a fully quantum universe.

Further, it will be shown that many of the Hasse diagrams generated in
classical causal set theory may also be recreated in the proposed quantum
model. However, whilst it may be mathematically possible to produce any
configuration of elements in a classical Hasse diagram, it is argued that
not all types are permissible in physics. So, in the proposed model only
those parts of the Hasse diagrams that are allowed by quantum mechanics, and
are hence physically meaningful, are generated.

\bigskip

\subsection{Splits and Partitions}

\bigskip

In classical causal set theory, continuous spacetime is generated from the
relations between collections of classical objects. Since the intention is
now to investigate how similar relations might arise from a quantum
perspective, an obvious starting point is to examine how the classical
objects of the classical theory might have analogues existing as features of
the quantum paradigm.

As discussed in Chapter 4, classicity is associated with separability of the
quantum state $\Psi _{n}$ representing the Universe. In a state that is
separable relative to a given factorisation of the total Hilbert space, each
factor sub-state may be considered classically isolated from every other
sub-state, in the sense that a measurement of a factor sub-state contained
in one particular factor Hilbert space does not necessarily affect any of
the other factor sub-states contained in other factor Hilbert spaces. This
is unlike the case of entangled states, because their individual components
cannot be measured without destroying the entire state. Therefore, as
concluded previously, factor sub-states may consequently be given a form of
classical identity.

It is asserted, then, that it is the factors of separable states that may be
associated with the classical `events' of conventional causal set theory,
and it is hence from the relations between these factors that physical
spacetime might be generated. The alternative, that continuous space instead
emerges from considerations of entangled states, is contrary to what would
be expected based on empirical observations: entangled states in quantum
theory exhibit characteristics of spatial non-locality. Physical space is in
many ways a classical construct, as expected from the observation that
\textit{this} object is \textit{here}, relative to \textit{that} object
\textit{there}.

Thus, the conjecture that will be discussed in the following is how
structures analogous to those occurring in classical causal sets might arise
naturally by considering the way in which the state of the Universe changes
its separability as it develops through a series of stages, i.e. from $\Psi
_{n}$ to $\Psi _{n+1}$ to $\Psi _{n+2}$ $...$ and so on.

\bigskip

Since classical causal set theory operates in the regime of large numbers of
events, and that in the proposed paradigm these classical events are assumed
to be analogous to the factors of the overall quantum state, it is expected
that quantum causal sets will require highly separable states in order to
yield a picture of continuous spacetime. It is therefore necessary to go
beyond the simple bi- and tri-partite factorisations of the total Hilbert
space discussed previously, and generalise to the case in which large
numbers of factors may be present. So, before a full discussion of quantum
causal sets can properly begin, it is necessary to introduce a precise
notation in order to describe highly separable states in highly factorisable
Hilbert spaces.\bigskip

As before, the situation of interest contains a state $\Psi $ in a Hilbert
space $\mathcal{H}$ of finite dimension. In anticipation of a discussion of
quantum computation in subsequent chapters of this work, it is alternatively
possible to call such a Hilbert space a `quantum register'. Factor Hilbert
sub-spaces of $\mathcal{H}$ may hence be labelled `subregisters'.

If $\mathcal{H}\equiv \mathcal{H}_{[1...N]}$ is defined as a Hilbert space
that may be written as a product of $N$ subregisters, then $\mathcal{H}%
_{[1...N]}$ is clearly given by the tensor product
\begin{equation}
\mathcal{H}_{[1...N]}\equiv \mathcal{H}_{1}\otimes \mathcal{H}_{2}\otimes
\mathcal{H}_{3}\otimes ...\otimes \mathcal{H}_{N}
\end{equation}
where $\mathcal{H}_{i}$ is called the $i^{th}$ factor Hilbert space or
quantum subregister, and $1\leq i\leq N.$ By choice, the notation $\mathcal{H%
}_{[1...N]}$ will generally be used to imply the Hilbert space's fundamental
factorisation, that is, each factor Hilbert space $\mathcal{H}_{i}$ is an
`elementary subregister' of prime dimension, $d_{i}.$ This choice will
henceforth be assumed from now on, unless stated otherwise.

Note that in order for $\mathcal{H}$ to contain the state $\Psi _{n}$
representing the Universe, its dimension $d$ must be huge. The number of
factors $N$ may therefore be in principle very large, with the condition
that $d=\tprod\nolimits_{i=1}^{N}d_{i}.$

As before, it is remarked that the ordering of the factor spaces is not
important in the above use of the tensor product symbol. The mathematics is
invariant to any rearrangement of the individual subregisters, such that for
example $\mathcal{H}_{1}\otimes \mathcal{H}_{2}\equiv \mathcal{H}_{2}\otimes
\mathcal{H}_{1}$ etc. Indeed, if this were not the case the problem would
arise that there are no obvious physical criteria for suggesting why some
factors spaces should either be placed `further away' than others from a
particular subregister, or be given any special position in the tensor
product ordering. In other words, in the factorisable register $\mathcal{H}%
_{[1..3]}\equiv \mathcal{H}_{1}\otimes \mathcal{H}_{2}\otimes \mathcal{H}%
_{3} $ it is meaningless to say that $\mathcal{H}_{1}$ is `nearer' to $%
\mathcal{H}_{2}$ than it is to $\mathcal{H}_{3}$ simply because of the way
the tensor product is written; $\mathcal{H}_{[1..3]}$ may equally well be
expanded as $\mathcal{H}_{[1..3]}\equiv \mathcal{H}_{1}\otimes \mathcal{H}%
_{3}\otimes \mathcal{H}_{2}.$ The factor Hilbert spaces are just vector
spaces, and should therefore not be thought of as embedded in any sort of
physical background space with any pre-existing distance or locational
relationships.

As before, this property is taken to hold for the states in the Hilbert
spaces as well, and is an important feature of their non-locality. After
all, a state such as $\Phi $ in $\mathcal{H}_{[1..3]}\equiv \mathcal{H}%
_{1}\otimes \mathcal{H}_{2}\otimes \mathcal{H}_{3}$ may be separable in the
form $\Phi =\phi _{2}\otimes \varphi _{13},$ where $\phi _{2}\in $ $\mathcal{%
H}_{2}$ with $\varphi _{13}\in \mathcal{H}_{[13]}\equiv \mathcal{H}%
_{1}\otimes \mathcal{H}_{3},$ but $\varphi _{13}$ might be entangled
relative to $\mathcal{H}_{[13]}.$\ In other words, the entanglements can
`stretch across' factor Hilbert spaces.\bigskip

In general, factorisable Hilbert spaces $\mathcal{H}_{[1...N]}$ in quantum
mechanics may contain states that are completely separable, completely
entangled, or a separable product of factors, at least one of which is
entangled relative to the fundamental split of the overall Hilbert space
into its elementary subregisters. It is therefore convenient to define a
notation in order to describe what `type' of fundamental separation an
arbitrary state in $\mathcal{H}_{[1..N]}$ may have.

Consider first a Hilbert space $\mathcal{H}_{[12]}$ factorisable into two
subregisters, $\mathcal{H}_{[12]}=\mathcal{H}_{1}\otimes \mathcal{H}_{2}.$
By axiom, the overall space $\mathcal{H}_{[12]}$ defines the total set of
vectors contained in $\mathcal{H}_{1}\otimes \mathcal{H}_{2}.$

The \textit{separation }$\mathcal{H}_{12},$ then, is defined as the subset
of states contained in $\mathcal{H}_{[12]}$ that are separable relative to $%
\mathcal{H}_{[12]}=\mathcal{H}_{1}\otimes \mathcal{H}_{2}.$ That is,
\begin{equation}
\mathcal{H}_{12}\equiv \{|\phi \rangle _{1}\otimes |\varphi \rangle
_{2}:|\phi \rangle _{1}\in \mathcal{H}_{1}\text{ },\text{\ }|\varphi \rangle
_{2}\in \mathcal{H}_{2}\}.
\end{equation}

Note that as mentioned in Chapter 4, the subset $\mathcal{H}_{12}$ is a set
of measure zero relative to the set $\mathcal{H}_{[12]}.$

For obvious reasons, $\mathcal{H}_{12}$ may be labelled a `rank-2'
separation, and this definition may be extended in a natural way.
Specifically, the rank-$k$ separation $\mathcal{H}_{i_{1}i_{2}...i_{k}}$ is
defined as the subset of vectors contained in the Hilbert space $\mathcal{H}%
_{[i_{1}i_{2}...i_{k}]}=\mathcal{H}_{i_{1}}\otimes \mathcal{H}%
_{i_{2}}\otimes ...\otimes \mathcal{H}_{i_{k}}$ that are separable into $k$
factors, i.e.
\begin{equation}
\mathcal{H}_{i_{1}i_{2}...i_{k}}\equiv \{|\psi _{1}\rangle _{i_{1}}\otimes
|\psi _{2}\rangle _{i_{2}}\otimes ...\otimes |\psi _{k}\rangle
_{i_{k}}:|\psi _{a}\rangle _{i_{a}}\in \mathcal{H}_{i_{a}}\text{ },\text{ }%
1\leq a\leq k\}.
\end{equation}

For convenience, it is also possible to allow the zero vector $\mathbf{0}$
to be a member of $\mathcal{H}_{i_{1}i_{2}...i_{k}},$ because this vector
can always be written in the form
\begin{equation}
\mathbf{0}=\mathbf{0}_{i_{1}}\otimes \mathbf{0}_{i_{2}}\otimes ...\otimes
\mathbf{0}_{i_{k}},
\end{equation}
where $\mathbf{0}_{i_{a}}$ is the zero vector in $\mathcal{H}_{i_{a}}.$
Note, however, that since any vector multiplied by zero is zero, then
although $\mathbf{0}$ could at first glance also appear entangled, for
example
\begin{equation}
\mathbf{0}=(\chi _{i_{1}}\otimes \mathbf{0}_{i_{2}}\otimes ...\otimes
\mathbf{0}_{i_{k}})+(\mathbf{0}_{i_{1}}\otimes \chi _{i_{2}}\otimes
...\otimes \mathbf{0}_{i_{k}})+...
\end{equation}
where $\chi _{i_{a}}\in $ $\mathcal{H}_{i_{a}},$ it could always be
rewritten in the form
\begin{eqnarray}
\mathbf{0} &=&(\mathbf{0}_{i_{1}}\otimes \mathbf{0}_{i_{2}}\otimes
...\otimes \mathbf{0}_{i_{k}})+(\mathbf{0}_{i_{1}}\otimes \mathbf{0}%
_{i_{2}}\otimes ...\otimes \mathbf{0}_{i_{k}})+... \\
&=&C(\mathbf{0}_{i_{1}}\otimes \mathbf{0}_{i_{2}}\otimes ...\otimes \mathbf{0%
}_{i_{k}})=(\mathbf{0}_{i_{1}}\otimes \mathbf{0}_{i_{2}}\otimes ...\otimes
\mathbf{0}_{i_{k}})  \notag
\end{eqnarray}
where $C$ is a constant, so is in fact separable. In other words, the zero
vector never contributes to entanglements in a non-trivial way.

From the above discussion of separations, the convention is adopted from now
on that lower indices on Hilbert spaces denote the subset of $\mathcal{H}$
containing separable states, i.e. the separations, whereas lower indices
within square brackets on Hilbert spaces denote, as before, the overall set
of states, i.e. the tensor product of subregisters.

This leaves free the use of upper indices for a discussion of the \textit{%
entanglements}, which may be defined in terms of the complements of the
separations. For example, in the simplest case in which the Hilbert space $%
\mathcal{H}_{[12]}$ is factorisable into two subregisters, $\mathcal{H}%
_{[12]}=\mathcal{H}_{1}\otimes \mathcal{H}_{2},$ the rank-$2$ entanglement $%
\mathcal{H}^{12}$ is defined as the subset of vectors in $\mathcal{H}_{[12]}$
that are entangled relative to this split. Moreover, since every state in $%
\mathcal{H}_{[12]}$ is either separable into two factors or else completely
entangled, $\mathcal{H}^{12}$ contains all the states that are not
separable, and so may be defined as
\begin{equation}
\mathcal{H}^{12}=\mathcal{H}_{[12]}-\mathcal{H}_{12}
\end{equation}
such that $\mathcal{H}_{[12]}=\mathcal{H}_{12}\cup \mathcal{H}^{12}.$ By
definition, $\mathcal{H}_{12}\cap \mathcal{H}^{12}=\emptyset .$\bigskip

Before the above ideas can be generalised to higher order entanglements it
is necessary to introduce the concept of \textit{separation product}.
Consider a Hilbert space $\mathcal{H}_{[1...N]}$ factorisable in the form $%
\mathcal{H}_{[1...N]}=\mathcal{H}_{a}\otimes \mathcal{H}_{b};$ clearly, if $%
N>2$ then this bi-partite factorisation is not fundamental and the dimension
of at least one of $\mathcal{H}_{a}$ or $\mathcal{H}_{b}$ is not prime. If $%
\mathcal{H}_{a}^{\prime }$ and $\mathcal{H}_{b}^{\prime }$ are arbitrary
subsets of $\mathcal{H}_{a}$ and $\mathcal{H}_{b}$ respectively, then the
separation product $\mathcal{H}_{a}^{\prime }\bullet \mathcal{H}_{b}^{\prime
}$ is defined as the subset of states in $\mathcal{H}_{[1...N]}$ that may be
written as a product of factors, one of which is contained in $\mathcal{H}%
_{a}^{\prime }$ and the other in $\mathcal{H}_{b}^{\prime }.$ Thus, $%
\mathcal{H}_{a}^{\prime }\bullet \mathcal{H}_{b}^{\prime }$ defines the
subset
\begin{equation}
\mathcal{H}_{a}^{\prime }\bullet \mathcal{H}_{b}^{\prime }\equiv \{|\phi
\rangle _{a}^{\prime }\otimes |\varphi \rangle _{b}^{\prime }:|\phi \rangle
_{a}^{\prime }\in \mathcal{H}_{a}^{\prime }\subseteq \mathcal{H}_{a}\text{ },%
\text{\ }|\varphi \rangle _{b}^{\prime }\in \mathcal{H}_{b}^{\prime
}\subseteq \mathcal{H}_{b}\}.
\end{equation}

It is clear that the separation $\mathcal{H}_{12}$ described previously is
just a simple case of this, i.e. $\mathcal{H}_{12}=\mathcal{H}_{1}\bullet
\mathcal{H}_{2}.$

The separation product is associative, commutative and cumulative, i.e.
\begin{eqnarray}
(\mathcal{H}_{i}\bullet \mathcal{H}_{j})\bullet \mathcal{H}_{k} &=&\mathcal{H%
}_{i}\bullet (\mathcal{H}_{j}\bullet \mathcal{H}_{k}) \\
(\mathcal{H}_{i}\bullet \mathcal{H}_{j})\bullet \mathcal{H}_{k} &=&(\mathcal{%
H}_{j}\bullet \mathcal{H}_{i})\bullet \mathcal{H}_{k}  \notag \\
(\mathcal{H}_{i}\bullet \mathcal{H}_{j})\bullet \mathcal{H}_{k} &=&\mathcal{H%
}_{ij}\bullet \mathcal{H}_{k}=\mathcal{H}_{ijk}.  \notag
\end{eqnarray}

The definition of separation product can be used to specify subsets of the
total Hilbert space that contain a product of factors, one or more of which
is entangled. As an example, the separation product $\mathcal{H}^{ij}\bullet
\mathcal{H}_{k}$ labels the subset of $\mathcal{H}_{[ijk]}$ containing
states that are a product of two factors, one of which is entangled relative
to $\mathcal{H}_{i}\otimes \mathcal{H}_{j}.$ Specifically,
\begin{equation}
\mathcal{H}^{ij}\bullet \mathcal{H}_{k}\equiv \{|\phi \rangle \otimes
|\varphi \rangle :|\phi \rangle \in \mathcal{H}^{ij}\text{ },\text{\ }%
|\varphi \rangle \in \mathcal{H}_{k}\}.
\end{equation}

This definition can be extended to higher orders in the obvious way, such
that for example $\mathcal{H}^{ij}\bullet \mathcal{H}_{k}\bullet \mathcal{H}%
_{l}\bullet \mathcal{H}^{mnp}$ identifies the subset of states in $\mathcal{H%
}_{[ijklmnp]}$ separable as
\begin{equation}
\mathcal{H}^{ij}\bullet \mathcal{H}_{k}\bullet \mathcal{H}_{l}\bullet
\mathcal{H}^{mnp}\equiv \{|\phi \rangle \otimes |\varphi \rangle \otimes
|\chi \rangle \otimes |\psi \rangle \}
\end{equation}
where $|\phi \rangle \in \mathcal{H}^{ij}$ $,$\ $|\varphi \rangle \in
\mathcal{H}_{k},$ $|\chi \rangle \in \mathcal{H}_{l}$ and $|\psi \rangle \in
\mathcal{H}^{mnp}.$ Note that the indices of the entanglements are also
commutative, such that for example $\mathcal{H}^{mnp}=\mathcal{H}^{npm},$ as
expected from the property of a tensor product Hilbert space that its
subregisters are not in any definite or particular order.

Of course, the associativity of the entanglements follows directly from the
associativity of the separation product, for example if $\mathcal{H}%
^{ab}\equiv \mathcal{H}_{X}$ etc., then
\begin{equation}
(\mathcal{H}^{ab}\bullet \mathcal{H}^{cd})\bullet \mathcal{H}^{ef}=(\mathcal{%
H}_{X}\bullet \mathcal{H}_{Y})\bullet \mathcal{H}_{Z}=\mathcal{H}_{X}\bullet
(\mathcal{H}_{Y}\bullet \mathcal{H}_{Z})=\mathcal{H}^{ab}\bullet (\mathcal{H}%
^{cd}\bullet \mathcal{H}^{ef}).
\end{equation}

Similar illustrations can be used to demonstrate the commutivity of the
entanglements.

To simplify complicated expressions such as $\mathcal{H}^{ij}\bullet
\mathcal{H}_{k}\bullet \mathcal{H}_{l}\bullet \mathcal{H}^{mnp},$ a single
symbol $\mathcal{H}_{k\bullet l}^{ij\bullet mnp}$ may be employed where the
use of sub-scripts and super-scripts indicates the separations and
entanglements. This symbol can itself be further simplified by making use of
the cumulative property of the separation, i.e. $\mathcal{H}_{a}\bullet
\mathcal{H}_{b}=\mathcal{H}_{a\bullet b}=\mathcal{H}_{ab}.$

Note, however, that no such cumulativity property directly exists for the
entanglements. By way of an illustration of this, consider the observation
that a state $\Phi $ in an entanglement such as $\mathcal{H}^{abcd}$ cannot,
by definition, be separated into a product of entangled states of the form $%
\theta \otimes \eta ,$ where $\theta $ is in the entanglement $\mathcal{H}%
^{ab}$ and $\eta $ is in the entanglement $\mathcal{H}^{cd};$ a vector that
is separable into a product of two entangled factors is not the same as a
vector that is separable into one giant entangled state. In other words, $%
\mathcal{H}^{abcd}\neq \mathcal{H}^{ab}\bullet \mathcal{H}^{cd}$ $(=\mathcal{%
H}^{ab\bullet cd}),$ even though $\mathcal{H}_{abcd}=\mathcal{H}_{ab\bullet
cd}=\mathcal{H}_{ab}\bullet \mathcal{H}_{cd}=\mathcal{H}_{a}\bullet \mathcal{%
H}_{b}\bullet \mathcal{H}_{c}\bullet \mathcal{H}_{d}.$

Overall, separation products such as $\mathcal{H}^{ij}\bullet \mathcal{H}%
_{k}\bullet \mathcal{H}_{l}\bullet \mathcal{H}^{mnp}$ may consequently be
written in a number of alternative ways; for example
\begin{eqnarray}
\mathcal{H}^{ij}\bullet \mathcal{H}_{k}\bullet \mathcal{H}_{l}\bullet
\mathcal{H}^{mnp} &=&\mathcal{H}^{ij}\bullet \mathcal{H}^{mnp}\bullet
\mathcal{H}_{l}\bullet \mathcal{H}_{k}=\mathcal{H}_{l}^{ji}\bullet \mathcal{H%
}_{k}^{mpn} \\
&=&\mathcal{H}_{k\bullet l}^{ij}\bullet \mathcal{H}^{mnp}=\mathcal{H}%
_{k\bullet l}^{ij\bullet mnp}  \notag \\
&=&\mathcal{H}_{kl}^{ij\bullet mnp}\text{ }\mathbf{\neq }\text{ }\mathcal{H}%
_{kl}^{ijmnp}.  \notag
\end{eqnarray}

It is now possible to define rank-$k$ entanglements in terms of their
complements. Starting with the rank-$3$ entanglement in the total Hilbert
space $\mathcal{H}_{[123]}=\mathcal{H}_{1}\otimes \mathcal{H}_{2}\otimes
\mathcal{H}_{3},$ it is immediately noted that $\mathcal{H}^{123}$ is not
simply given by $\mathcal{H}^{123}=\mathcal{H}_{[123]}-\mathcal{H}_{123}.$
Rather the sets of states that are separable into a product of two factors,
one of which is entangled, must also be included. Thus, $\mathcal{H}^{123}$
is given by
\begin{equation}
\mathcal{H}^{123}=\mathcal{H}_{[123]}-\mathcal{H}_{123}\cup \mathcal{H}%
_{1}^{23}\cup \mathcal{H}_{2}^{13}\cup \mathcal{H}_{3}^{12}.
\label{Ch3Rank3}
\end{equation}

Similarly, the rank-$4$ entanglement $\mathcal{H}^{1234}$ in $\mathcal{H}%
_{[1234]}$ is equal to
\begin{eqnarray}
\mathcal{H}^{1234} &=&\mathcal{H}_{[1234]}-\mathcal{H}_{1234}\cup \mathcal{H}%
_{12}^{34}\cup \mathcal{H}_{13}^{24}\cup \mathcal{H}_{14}^{23}\cup \mathcal{H%
}_{23}^{14}\cup \mathcal{H}_{24}^{13}\cup \mathcal{H}_{34}^{12}
\label{Ch3Rank4} \\
&&\cup \mathcal{H}_{1}^{234}\cup \mathcal{H}_{2}^{134}\cup \mathcal{H}%
_{3}^{124}\cup \mathcal{H}_{4}^{123}\cup \mathcal{H}^{12\bullet 34}\cup
\mathcal{H}^{13\bullet 24}\cup \mathcal{H}^{14\bullet 23}.  \notag
\end{eqnarray}

Rank-$k$ entanglements can clearly be defined in similar ways, though their
expressions rapidly become more complicated as $k$ increases.

Equations such as (\ref{Ch3Rank3}) and (\ref{Ch3Rank4}) can be rearranged
such that the overall register is decomposed into a union of disjoint
separations and entanglements, for example $\mathcal{H}_{[123]}=\mathcal{H}%
_{123}\cup \mathcal{H}_{1}^{23}\cup \mathcal{H}_{2}^{13}\cup \mathcal{H}%
_{3}^{12}\cup \mathcal{H}^{123}.$ Making use of the language familiar to set
theory, such a decomposition of a Hilbert space $\mathcal{H}_{[1...N]}$ may
be called its `lattice of partitions' with each subset being called a
`partition'. In general, each partition is a separation product of
separations and entanglements of various ranks, with the condition that the
total number of indices equals the overall number of subregisters. In
addition each subscript index, and each group of superscript indices,
specifies one `block' of the partition, such that for example the partition $%
\mathcal{H}_{14}^{23\bullet 56}$ contains four blocks, denoted by $`1$', $%
`23 $', $`4$' and $`56$'$.$ The union of all partitions of $\mathcal{H}%
_{[1...N]} $ may equivalently be called the `(natural) partitioning' of the
Hilbert space.\bigskip

It is important to realise that none of the partitions in $\mathcal{H}%
_{[1...N]}$ are vector spaces themselves. This conclusion follows from two
reasons. Firstly, the zero vector has been defined to be a member of the
separation $\mathcal{H}_{1...N},$ so only this partition could potentially
be a vector space. Secondly none of the partitions are closed under
arbitrary transformations of the vectors they contain. Given a vector $X$
contained in one partition, it is always possible to add a second vector $x$
to $X$ such that the new vector $Y=X+x$ is contained in a different
partition. Similarly it is possible to find unitary transformations $\hat{U}$
that `rotate' $X$ into the vector $Z=\hat{U}X,$ where $Z$ is also a member
of a different partition from $X.$ Of course, all four vectors $x,$ $X,$ $Y$
and $Z$ are members of $\mathcal{H}_{[1...N]},$ which \textit{is} a vector
space.

\bigskip

The use of upper and lower indices on the symbol $\mathcal{H}$ to denote
subsets of $\mathcal{H}_{[1...N]}$ containing various separations and
entanglements can be extended to the vectors contained within these
partitions. For example, the vector $\Phi _{kl}^{ij\bullet mnp}$ is taken to
be a member of the partition $\mathcal{H}_{kl}^{ij\bullet mnp},$ and implies
that $\Phi $ can be separated into four factors of the form
\begin{equation}
\Phi _{kl}^{ij\bullet mnp}=|\phi \rangle ^{ij}\otimes |\phi ^{\prime
}\rangle _{k}\otimes |\phi ^{\prime \prime }\rangle _{l}\otimes |\phi
^{\prime \prime \prime }\rangle ^{mnp}  \label{Ch3Thi}
\end{equation}
where $|\phi \rangle ^{ij}\in \mathcal{H}^{ij},$ $|\phi ^{\prime }\rangle
_{k}\in \mathcal{H}_{k},$ $|\phi ^{\prime \prime }\rangle _{l}\in \mathcal{H}%
_{l}$ and $|\phi ^{\prime \prime \prime }\rangle ^{mnp}\in \mathcal{H}%
^{mnp}. $ Obviously $\Phi _{kl}^{ij\bullet mnp}$ $\in \mathcal{H}_{[i...p]}$
and $\Phi _{kl}^{ij\bullet mnp}\in $ $\mathcal{H}_{kl}^{ij\bullet mnp}$
because $\mathcal{H}_{kl}^{ij\bullet mnp}\subset \mathcal{H}_{[i...p]}.$

Care is needed when applying this notation, however, because sub-scripts
used in this thesis, and elsewhere, are often context dependent. For example
$\Psi _{12}$ might denote a state in the separation $\mathcal{H}_{12},$ or
an arbitrary state in $\mathcal{H}_{[12]},$ or even the state in the twelfth
stage $\Omega _{12}$ defined as $\Omega _{12}\equiv \Omega (\Psi _{12},$ $%
I_{12},$ $R_{12}).$ The same goes for super-scripts, where the notation $%
\Phi ^{12}$ might perhaps alternatively label a vector in the entanglement $%
\mathcal{H}^{12},$ an arbitrary state $\Phi $ in a twelve dimensional
Hilbert space $\mathcal{H}_{[1...N]}^{12},$ or maybe even one out of $E$
possible eigenvectors $\Phi ^{a}$ of some Hermitian operator $\hat{O}$ for $%
1\leq a\leq E$ where $E\geq 12.$\bigskip

For any given vector $\Psi \in \mathcal{H}_{[1...N]},$ it is possible to
determine which partition it is in by a repeated application of the
microsingularity test (\ref{Ch2Micro}) introduced in Section 4.1. For
example, to show that a state $\Theta \in \mathcal{H}_{[1...3]}$ is
completely entangled, i.e. can be written in the form $\Theta ^{123}$ in the
partition $\mathcal{H}^{123},$ it must be confirmed that $\Theta $ is not in
$\mathcal{H}_{123},$ $\mathcal{H}_{1}^{23},$ $\mathcal{H}_{2}^{13}$ or $%
\mathcal{H}_{3}^{12}.$ This is turn can be proved by demonstrating that $%
\Theta $ is not separable relative to any of the three bi-partite
factorisations of the total Hilbert space, i.e. $\mathcal{H}_{1}\otimes
\mathcal{H}_{[23]},$ $\mathcal{H}_{2}\otimes \mathcal{H}_{[13]}$ and $%
\mathcal{H}_{3}\otimes \mathcal{H}_{[12]},$ because if this is true it also
immediately follows that $\Theta $ is not separable relative to the
tri-partite factorisation $\mathcal{H}_{1}\otimes \mathcal{H}_{2}\otimes
\mathcal{H}_{3}.$

From earlier discussions, any state $\Theta \in \mathcal{H}_{[1...3]}$ can
be expanded in the form
\begin{equation}
\Theta
=\sum\nolimits_{i=0}^{d_{1}-1}\sum\nolimits_{j=0}^{d_{2}-1}\sum%
\nolimits_{k=0}^{d_{3}-1}C_{ijk}|i\rangle _{1}\otimes |j\rangle _{2}\otimes
|k\rangle _{3}  \label{Ch3PsiExp}
\end{equation}
where the Hilbert space $\mathcal{H}_{a}$ of dimension $d_{a}$ is spanned by
the orthonormal basis $\mathcal{B}_{a}=\{|b\rangle _{a}:0\leq b\leq
d_{a}-1\} $ for $a=1,2,3$ and the $C_{ijk}\in \mathbb{C}$ are complex
coefficients. The microsingularity condition can now be used to determine
the separability of $\Theta $ relative to each of the three bi-partite
factorisations of $\mathcal{H}_{[1...3]}.$

For example, to confirm whether $\Theta $ is separable relative to $\mathcal{%
H}_{1}\otimes \mathcal{H}_{[23]},$ equation (\ref{Ch3PsiExp}) should be
rewritten as
\begin{equation}
\Theta
=\sum\nolimits_{i=0}^{d_{1}-1}\sum\nolimits_{x=0}^{(d_{2}d_{3})-1}K_{ix}|i%
\rangle _{1}\otimes |x\rangle _{23}
\end{equation}
where $\mathcal{B}_{23}=\{|x\rangle _{23}:0\leq x\leq (d_{2}d_{3}-1)\}$
forms an orthonormal basis for $\mathcal{H}_{[23]},$ with for example $%
|0\rangle _{23}=|0\rangle _{2}\otimes |0\rangle _{3},$ $|1\rangle
_{23}=|0\rangle _{2}\otimes |1\rangle _{3},...,$ up to $|d_{2}d_{3}-1\rangle
_{23}=|d_{2}-1\rangle _{2}\otimes |d_{3}-1\rangle _{3}.$ The general term $%
|x\rangle _{23}$ in this basis may be given by $|x\rangle _{23}=|j\rangle
_{2}\otimes |k\rangle _{3}$ when $j$ is the integer part of the quotient $%
x/(d_{3})$ and $k$ is the remainder. The coefficients $K_{ix}$ are obtained
from $C_{ijk}$ in the same way, such that for example $K_{i1}$ is equivalent
to $C_{i01}.$

Now, if $K_{ix}K_{yz}=K_{iz}K_{yx}$ for all $0\leq i,y\leq (d_{1}-1)$ and $%
0\leq x,z\leq (d_{2}d_{3}-1),$ then $\Psi $ is separable relative to $%
\mathcal{H}_{1}\otimes \mathcal{H}_{[23]}.$ If this is not the case, $\Theta
$ is entangled relative to $\mathcal{H}_{1}\otimes \mathcal{H}_{[23]},$ and
if the same method shows that $\Theta $ is also entangled relative to $%
\mathcal{H}_{2}\otimes \mathcal{H}_{[13]}$ and $\mathcal{H}_{3}\otimes
\mathcal{H}_{[12]},$ it can be concluded that $\Theta \in \mathcal{H}^{123}.$

Similar procedures can be employed to determine which particular partition
of the lattice of $\mathcal{H}_{[1...N]}$ any given vector $\Psi \in
\mathcal{H}_{[1...N]}$ is in, though the corresponding number of
microsingularity tests that need to be performed increases greatly with $N.$%
\bigskip

The state $\Phi _{kl}^{ij\bullet mnp}\in \mathcal{H}_{kl}^{ij\bullet mnp}$
in (\ref{Ch3Thi}) is an example of a vector that is a separable product of
factors, two of which are entangled relative to the fundamental splitting of
the overall Hilbert space $\mathcal{H}_{[i...p]}$ into its seven
subregisters. In general, however, if an arbitrary state contained within a
Hilbert space $\mathcal{H}_{[1...N]}$ is chosen at random, there are very
many ways in which it might potentially be separated into $F$ factors, where
$1\leq F\leq N,$ because there are in general many different partitions
comprising of $F$ blocks. For example, the state $\Theta _{l}^{ip\bullet
kn\bullet jm}$ in $\mathcal{H}_{[i...p]}$ is also separated into four
factors, but in a completely different manner from $\Phi _{kl}^{ij\bullet
mnp}.$

Of course, if $F=1$ then the state is completely entangled, whereas if $F=N$
it is completely separable, but for all other values of $F$ the state is
separated into a product of factors, at least one of which is entangled.
Further, the number of ways in which an arbitrary state may potentially be
separated into $F$ factors increases rapidly as the number, $N,$ of
subregisters in the fundamental factorisation of the Hilbert space increases.

For example, in a Hilbert space $\mathcal{H}_{[1]}$ of prime dimension,
which is therefore fundamentally split into just one subregister, every
state can obviously only be separated into one factor. States in a Hilbert
space $\mathcal{H}_{[12]}$ that is fundamentally split into two subregisters
$\mathcal{H}_{[12]}=\mathcal{H}_{1}\otimes \mathcal{H}_{2},$ however, are
either entangled relative to this split, or else they are separable into two
factors; it can be said that there are two possible `types' of state
separations in $\mathcal{H}_{[12]}.$

In a Hilbert space $\mathcal{H}_{[123]}$ fundamentally split into three
subregisters $\mathcal{H}_{[123]}=\mathcal{H}_{1}\otimes \mathcal{H}%
_{2}\otimes \mathcal{H}_{3},$ though, a state is either completely entangled
relative to this split, or it is completely separable into three factors, or
else it is separable into one of the forms $\alpha _{1}\otimes \beta _{23},$
$\lambda _{2}\otimes \mu _{13}$ or $\phi _{3}\otimes \varphi _{12},$ where $%
\alpha _{1}\in \mathcal{H}_{1},$ $\lambda _{2}\in \mathcal{H}_{2}$ and $\phi
_{3}\in \mathcal{H}_{3},$ with $\beta _{23},$ $\mu _{13}$ and $\varphi _{12}$
being sub-states that are entangled relative to $\mathcal{H}_{2}\otimes
\mathcal{H}_{3},$ $\mathcal{H}_{1}\otimes \mathcal{H}_{3}$ and $\mathcal{H}%
_{1}\otimes \mathcal{H}_{2}$ respectively. Given an arbitrary state in $%
\mathcal{H}_{[123]},$ there are clearly five different types of way in which
it might be separable relative to $\mathcal{H}_{[123]}$: one of these types
will have one factor, three types will have two factors, and one will have
three factors. Equivalently, every state in $\mathcal{H}_{[123]}$ is in one
of the five partitions that comprise the partitioning of the total Hilbert
space, $\mathcal{H}_{[123]}=\mathcal{H}_{123}\cup \mathcal{H}_{1}^{23}\cup
\mathcal{H}_{2}^{13}\cup \mathcal{H}_{3}^{12}\cup \mathcal{H}^{123}.$

In fact, it can be shown that in Hilbert spaces $\mathcal{H}_{[1...4]}$
fundamentally split into four subregisters, there are $15$ different types
of way in which a given state might possibly be separated, whereas Hilbert
spaces of the form $\mathcal{H}_{[1...5]}$ allow the possibility of $52$
different types of separation. This number grows to $203$ for $\mathcal{H}%
_{[1...6]}.$

Generally, if $h_{N}$ is defined as the number of ways in which an arbitrary
state in $\mathcal{H}_{[1...N]}$ might possibly be separated, then this
number is given by the iterative formula
\begin{equation}
h_{N}=\sum_{i=0}^{N-1}C_{i}^{N-1}h_{(N-1)-i}  \label{Ch3Iter}
\end{equation}
where $C_{b}^{a}$ is the combination function, $C_{b}^{a}=a!/[(a-b)!b!],$
and the initial condition $h_{0}=1$ follows from the assumption that there
is only one way of separating a state contained in zero Hilbert spaces%
\footnote{%
Compare the generally accepted result $0!=1.$ If this argument appears
\textit{ad hoc, }$h_{N}$ may equally be defined as $h_{N}=1+%
\sum_{i=0}^{N-2}C_{i}^{N-1}h_{(N-1)-i}$ without loss of generality.}. The
above relation also specifies the number, $h_{N},$ of partitions comprising
the lattice of $\mathcal{H}_{[1...N]},$ as expected from the fact that every
state in $\mathcal{H}_{[1...N]}$ is in exactly one of the Hilbert space's
partitions, and that it is always possible to find an example of a state in $%
\mathcal{H}_{[1...N]}$ that is a member of a given partition.

Equation (\ref{Ch3Iter}) effectively generates the list of Bell numbers used
in combinatorics to number the set of partitions of a set of size $N,$ and
is equivalently given by Dobinski's formula (see \cite{Wiki-Bell} for an
illustration of these points).\bigskip

An intuitive proof of (\ref{Ch3Iter}) is given from the following. Consider
a factorisable Hilbert space $\mathcal{H}_{[1...N]}.$ Every state in $%
\mathcal{H}_{[1...N]}$ will be associated with its own fundamental
separation, i.e. a way or writing the state into the maximum possible number
of factors relative to $\mathcal{H}_{[1...N]},$ because each state is in
one, and only one, partition of $\mathcal{H}_{[1...N]}.$

Now assume that every state in $\mathcal{H}_{[1...N]}$ is fundamentally
separable into one of $h_{N}$ possible types, where $h_{N}$ is not yet known
and the relation (\ref{Ch3Iter}) is not assumed. For example, it was shown
earlier that every vector in $\mathcal{H}_{[1...3]}$ is fundamentally
separable into one out of five possible types.

Clearly, this number $h_{N}$ of possible types is given by the sum of the
number of ways that vectors in $\mathcal{H}_{[1...N]}$ might be separable
into just one factor, plus the number of ways that vectors might be
separable into just two factors, plus the number of ways that vectors might
be separable into just three factors, plus..., plus the number of ways that
vectors might be separable into just $N$ factors. Thus, $h_{N}$ is also the
total number of possible partitions in the lattice of $\mathcal{H}%
_{[1...N]}, $ or equivalently the total number of types of vector that exist
in $\mathcal{H}_{[1...N]}.$

Of course, there is only one type of way in which vectors in $\mathcal{H}%
_{[1...N]}$ may be fundamentally separated into one factor, and only one
type of way in which vectors in $\mathcal{H}_{[1...N]}$ may be fundamentally
separated into $N$ factors.

Because $\mathcal{H}_{[1...N]}$ is of fixed dimension, every vector it
contains must have a component in every subregister $\mathcal{H}_{i}$ of $%
\mathcal{H}_{[1...N]},$ for $1\leq i\leq N.$ Therefore, every vector in $%
\mathcal{H}_{[1...N]}$ must consequently have a component in the subregister
$\mathcal{H}_{1},$ and this component will be in one of the $F$ factors of
the overall state (whatever $F$ may be). Further, whichever sub-state of the
overall product it is in, the component in $\mathcal{H}_{1}$ will either be
in a factor of the state on its own, or entangled with a component from just
one other subregister, or entangled with the components from two other
subregisters, or..., or entangled with the components from each of the $N-1$
other subregisters (in which case $F=1).$

In other words, a given state $\Phi $ in $\mathcal{H}_{[1...N]}$ might be
fundamentally separable as
\begin{equation}
\Phi =X_{1}\otimes Y_{[2...N]}
\end{equation}
or
\begin{equation}
\Phi =X^{1i}\otimes Y_{[2...(i-1)(i+1)...N]}
\end{equation}
or
\begin{equation}
\Phi =X^{1ij}\otimes Y_{[2...(i-1)(i+1)...(j-1)(j+1)...N]}
\end{equation}
or... etc., for $1<i,j,k,...\leq N$ and $i\neq j\neq k\neq ...$ $.$ Here $%
X_{1}\in $ $\mathcal{H}_{1},$ but $Y_{[2...N]}\in $ $\mathcal{H}_{[2...N]}$
is any vector (completely entangled, completely separable, or a separable
product of entangled factors) in $\mathcal{H}_{[2...N]}.$ Similarly $%
X^{1i}\in $ $\mathcal{H}^{1i}$ and $X^{1ij}\in $ $\mathcal{H}^{1ij},$ but $%
Y_{[2...(i-1)(i+1)...N]}\in $ $\mathcal{H}_{[2...(i-1)(i+1)...N]}$ and $%
Y_{[2...(i-1)(i+1)...(j-1)(j+1)...N]}\in \mathcal{H}%
_{[2...(i-1)(i+1)...(j-1)(j+1)...N]}$ are arbitrary vectors that also may or
may not be separable.

The summation proceeds as follows.

If the component in $\mathcal{H}_{1}$ of a state is in a factor sub-state on
its own, i.e. is not entangled with anything, there are $(N-1)$ components
of the state left `free', corresponding to the remaining $(N-1)$
subregisters $\mathcal{H}_{2},$ $\mathcal{H}_{3},...,$ $\mathcal{H}_{N}.$
This remaining part of the state is a vector in $\mathcal{H}_{[2...N]},$ and
so by assumption this may be separated into one of $h_{N-1}$ different ways.
So, there are $h_{N-1}$ different ways in which states in $\mathcal{H}%
_{[1...N]}$ might be separated in the form $X_{1}\otimes Y_{[2...N]}.$

Now, there are precisely $C_{1}^{N-1}=(N-1)$ ways of selecting just one
component of a vector in $\mathcal{H}_{[2...N]},$ i.e. $(N-1)$ ways of
choosing just one of the components in one of the remaining $(N-1)$ factor
spaces $\mathcal{H}_{i}$ for $2\leq i\leq N.$ There are hence $(N-1)$
different types of factor of the form $X^{1i}$ for states in $\mathcal{H}%
_{[1...N]},$ such that the component in $\mathcal{H}_{1}$ is entangled with
the component in $\mathcal{H}_{i}.$ Further, each of these ways leaves a
remaining vector in $\mathcal{H}_{[2...(i-1)(i+1)...N]},$ with $(N-2)$
`free' components, and this vector may itself be separated into one of $%
h_{N-2}$ different ways. So, overall there are $(N-1)h_{N-2}$ different ways
of separating states in $\mathcal{H}_{[1...N]}$ in the form $X^{1i}\otimes
Y_{[2...(i-1)(i+1)...N]}.$

Continuing, there are $C_{2}^{N-1}$ ways of selecting two components of a
vector in $\mathcal{H}_{[2...N]},$ such that one is in the subregister $%
\mathcal{H}_{i}$ and the other is in the subregister $\mathcal{H}_{j},$ for $%
2\leq i,j\leq N$ and $i\neq j.$ There are hence $C_{2}^{N-1}$ different
types of factor of the form $X^{1ij}$ for states in $\mathcal{H}_{[1...N]},$
such that the component in $\mathcal{H}_{1}$ is entangled with just two of
the other components. This leaves a remaining vector in $\mathcal{H}%
_{[2...(i-1)(i+1)...(j-1)(j+1)...N]},$ which has $(N-3)$ free components,
and this could be separable in one of $h_{N-3}$ different ways. So overall
there are $C_{2}^{N-1}h_{N-3}$ different ways of separating states in $%
\mathcal{H}_{[1...N]}$ in the form $X^{1ij}\otimes
Y_{[2...(i-1)(i+1)...(j-1)(j+1)...N]}.$

This analysis can be continued. In general, there are $C_{x}^{N-1}$ ways of
selecting $x$ components of a vector\ in $\mathcal{H}_{[2...N]},$ such that
the $a_{b}^{th}$ component is in the $a_{b}^{th}$ subregister $\mathcal{H}%
_{a_{b}},$ for $0\leq x\leq (N-1),$ whilst $2\leq a_{b}\leq N$ and $%
b=1,2,...,x,$ with, of course, no two components being in the same
subregister. There are hence $C_{x}^{N-1}$ different types of factor of the
form\ $X^{1a_{1}a_{2}...a_{x}},$ such that the component in $\mathcal{H}_{1}$
is entangled with $x$ of the other components. This leaves a remaining
vector which has $(N-1-x)$ `free' components, and this vector will be
separable in one of $h_{N-1-x}$ different ways. So, there are $%
C_{x}^{N-1}h_{N-1-x}$ different ways of separating states in $\mathcal{H}%
_{[1...N]}$ into a product of factors, one of which is $%
X^{1a_{1}a_{2}...a_{x}}.$\bigskip

Overall, the total number $h_{N}$ of ways in which arbitrary vectors in $%
\mathcal{H}_{[1...N]}$ might potentially be separated into a product of
factors is given by the exhaustive sum of the number of ways in which
vectors in $\mathcal{H}_{[1...N]}$ might be separated such that their
component in $\mathcal{H}_{1}$ is in a factor sub-state on its own, added to
the number of ways in which vectors in $\mathcal{H}_{[1...N]}$ might be
separated such that there is a factor containing the component in $\mathcal{H%
}_{1}$entangled with a component from one other subregister, added to the
number of ways in which vectors in $\mathcal{H}_{[1...N]}$ might be
separated such that there is a factor containing the component in $\mathcal{H%
}_{1}$entangled with the components from two other subregisters, and so on
to the addition of the number of ways in which vectors in $\mathcal{H}%
_{[1...N]}$ might be separated such that there is a factor containing the
component in $\mathcal{H}_{1}$entangled with the components from every other
subregister.

From the above, this gives
\begin{equation}
h_{N}=h_{N-1}+(N-1)h_{N-2}+C_{2}^{N-1}h_{N-3}+...+C_{x}^{N-1}h_{N-1-x}+...+C_{N-1}^{N-1}h_{N-N}
\end{equation}
where the last term is equal to unity because there is only one way of
separating a state into one entangled factor.

Clearly, then, it follows that $h_{N}$ is given by (\ref{Ch3Iter}).\bigskip

As discussed earlier, the first few values for $h_{N}$ are $h_{0}=1,$ $%
h_{1}=1,$ $h_{2}=2,$ $h_{3}=5,$ $h_{4}=15,$ $h_{5}=52,$ $h_{6}=203,$ such
that $h_{N}$ evidently grows quickly for even relatively low values of $N.$
Indeed, note that even for a five qubit system, the number of ways its state
may be separable is greater than the dimension of its Hilbert space: a five
qubit Hilbert space $\mathcal{H}_{[1...5]}^{(32)}$ of $32$ dimensions
contains $h_{5}=52$ partitions.

For the case in which $N$ is of the order $10^{184},$ the value of $h_{N}$
is expected to be truly enormous. So, for a Universe represented by a state
of dimension greater than $2^{10^{184}},$ the number of partitions contained
in the lattice of its Hilbert space $\mathcal{H}_{[1...10^{184}]}$ is
clearly very large. This should consequently provide an incredibly rich
structure, with a tremendous number of different ways in which the state of
the Universe might potentially separate.

As will be shown, this provides a wide scope for the Universe's dynamics.

\bigskip

\subsection{Probability Amplitudes and Quantum Causal Sets}

\bigskip

Now that a notation has been introduced to cope with large dimensional
Hilbert spaces, it is possible to examine how a causal set structure might
arise from a fully quantum description of physics.

From the discussion that the Universe may always be represented by a state $%
\Psi _{n}$ in a Hilbert space $\mathcal{H}$ of enormous dimension, and from
the outcome of the previous section that any vector in a given Hilbert space
is always contained within one, and only one, of the partitions of this
space, the conclusion must be that the Universe's state is always in one of
the partitions of $\mathcal{H}.$ The state $\Psi _{n}$ is separable in a
specific way, and is always a product of between $1$ and $N$ factors, where $%
N$ is the number of subregisters in the fundamental factorisation of $%
\mathcal{H}.$

As the wavefunction of the Universe develops from one state $\Psi _{n}$ to
the next $\Psi _{n+1},$ its pattern of separability might change. That is,
if the state $\Psi _{n}$ may be fundamentally separated into a product of $%
F_{n}$ factors, $1\leq F_{n}\leq N,$ the state $\Psi _{n+1}$ may be
separable into $F_{n+1}$ factors, where $F_{n}$ is not necessarily equal to $%
F_{n+1}.$ In fact, even if $F_{n}=F_{n+1}$ the states $\Psi _{n}$ and $\Psi
_{n+1}$ may have completely different patterns of separability, since there
are in general many different partitions comprising of $F$ blocks. It is
this changing pattern of separability that will be shown to be the origin of
family structures in the quantum universe, and hence the beginning of a
discussion of quantum causal sets.\bigskip

As conjectured at the start of Section 5.3, the individual factors of the
state of the Universe may be analogous to the events of classical causal set
theory. For example, the growth of the events in classical causets satisfies
`internal temporality', in the sense that every new event is born either to
the future of, or unrelated to, every other event; no event is created to
the past of already existing events. The same is true in the present model,
because the next potential state $\Psi _{n+1}$ is an outcome (i.e. one of
the eigenvectors) of a test on the `current' state $\Psi _{n},$ and so any
factor of $\Psi _{n+1}$ cannot in any way be thought of as in the past of
any of the factors of $\Psi _{n}.$

It is important to reiterate, however, that the quantum and classical models
are not completely congruent. For example, as has been discussed previously
the relation $x\prec y\prec z$ between three classical events has no \textit{%
direct} equivalent in the quantum theory. After all, consider three
consecutive states $\Psi _{n-1},\Psi _{n}$ and $\Psi _{n+1}$: whilst $\Psi
_{n+1}$ may indeed be one of the possible outcomes of a test on $\Psi _{n}$
(which is itself one of the outcomes of a test on $\Psi _{n-1}),$ the
successive states $\Psi _{n-1},$ $\Psi _{n}$ and $\Psi _{n+1}$ cannot be
granted equivalent degrees of existence according to the Kochen-Specker
theorem, and so cannot be directly compared. Compared to the current state $%
\Psi _{n}$ only \textit{potential} future states $\Psi _{n+1}$ can be
discussed.\bigskip

One similarity that does still occur between the classical and quantum cases
is the notion of `links', which are defined as being irreducible relations.
In the classical theory described in Section 5.2, for example, two events $x$
and $y$ are linked if $x\prec y$ and there is no other event $z$ such that $%
x\prec z\prec y,$ or if $y\prec x$ and there is no other event $z^{\prime }$
such that $y\prec z^{\prime }\prec x.$ Analogously, in the proposed quantum
scenario the states $\Psi _{n-1}$ and $\Psi _{n}$ could immediately be
described as `linked', because by definition there is no intermediate state
between them.

A further similarity arises from the classical causal set concepts of
families: related notions are also present in the quantum case, based, in
fact, on the factorisability of the probability amplitude. To demonstrate
this, consider the inner product $\langle \Psi _{n+1}=\Phi |\Psi _{n}\rangle
$ between the current state $\Psi _{n}$ and one of the next potential states
$\Psi _{n+1}=\Phi ,$ where $\Phi $ is one of the eigenvectors of some
operator $\hat{\Sigma}_{n+1}.$ The states $\Psi _{n}$ and $\Psi _{n+1}$ are
each contained within particular partitions of the total Hilbert space $%
\mathcal{H}_{[1...N]},$ where as before $\mathcal{H}_{[1...N]}$ is assumed
factorisable into $N$ subregisters. Now, because the factors of one state
can only take inner products with factors of another state if they lie in
the same factor space of some split of the total Hilbert space (where these
factor spaces are not necessarily elementary), then, depending on the
details of the partitions containing $\Psi _{n}$ and $\Psi _{n+1},$ the
probability amplitude may be separable into a number, $r,$ of factors.

Paraphrasing, if $\psi \in $ $\mathcal{H}_{[\psi ]}$ is a factor of $\Psi
_{n}\in \mathcal{H}_{[1...N]},$ where $\mathcal{H}_{[\psi ]}$ is one of the
factors of some split of $\mathcal{H}_{[1...N]}$ and need not be of prime
dimension, and if $\phi \in $ $\mathcal{H}_{[\phi ]}$ is a factor of $\Psi
_{n+1},$ where $\mathcal{H}_{[\phi ]}$ is one of the factors of some split
of $\mathcal{H}_{[1...N]}$ and also need not be of prime dimension, then $%
\langle \phi |\psi \rangle $ will contribute a factor to the overall
probability amplitude $\langle \Psi _{n+1}|\Psi _{n}\rangle $ iff $\mathcal{H%
}_{[\psi ]}=\mathcal{H}_{[\phi ]}.$

This leads to the definition of a `\textit{family}': $\psi $ and $\phi ,$ in
successive states $\Psi _{n}$ and $\Psi _{n+1}$ respectively,\ constitute a
family if $\langle \phi |\psi \rangle $ is a factor of $\langle \Psi
_{n+1}|\Psi _{n}\rangle $ and if $\langle \phi |\psi \rangle $ cannot itself
be factorised further.

The above observation can be generalised, such that $\psi $ might be a
product of $A$ factors, $\psi =\psi _{1}\otimes \psi _{2}\otimes ...\otimes
\psi _{A},$ and $\phi $ might be a product of $B$ factors, $\phi =\phi
_{1}\otimes \phi _{2}\otimes ...\otimes \phi _{B},$ where $A$ is not
necessarily equal to $B.$ In this case, the definition of the family
encompasses the factors of which $\psi $ and $\phi $ are a product.\bigskip

Suppose now that the state of the Universe $\Psi _{n}\in \mathcal{H}%
_{[1...N]}$ is separable into $k$ factors, i.e. $\Psi _{n}=\psi
_{a_{1}}\otimes \psi _{a_{2}}\otimes ...\otimes \psi _{a_{k}},$ where the
individual factors $\psi _{a_{i}},$ for $1\leq i\leq k,$ may, or may not,
themselves be entangled relative to the fundamental factorisation of $%
\mathcal{H}_{[1...N]}.$ Each factor $\psi _{a_{i}}$ is in its own factor
Hilbert space $\mathcal{H}_{[a_{i}]},$ and this may itself be a product of
elementary subregisters with the condition that $\mathcal{H}_{[1...N]}=%
\mathcal{H}_{[a_{1}]}\otimes \mathcal{H}_{[a_{2}]}\otimes ...\otimes
\mathcal{H}_{[a_{k}]}.$ For simplicity, it may also be assumed that the
factor sub-states are normalised within their own factor Hilbert spaces,
i.e. $\langle \psi _{a_{i}}|\psi _{a_{i}}\rangle =1.$

Consider now the next test of the Universe, $\hat{\Sigma}_{n+1}.$ This test
has $d$ orthonormal eigenvectors, where $d$ is the dimension of the Hilbert
space $\mathcal{H}_{[1....N]}.$ Of course, if each elementary subregister of
$\mathcal{H}_{[1...N]}$ is a qubit sub-space then clearly $d=2^{N}.$ If $%
\Phi $ is one of these $d$ eigenvectors, then the conditional probability $%
P(\Psi _{n+1}=\Phi |\Psi _{n},\hat{\Sigma}_{n+1})$ that the next state $\Psi
_{n+1}$ of the Universe is $\Phi ,$ given a test $\hat{\Sigma}_{n+1},$ is
given by the usual Born probability rule $\left| \langle \Phi |\Psi
_{n}\rangle \right| ^{2}.$

Suppose that $\Phi $ is separable into $l$ factors, i.e. $\Phi =\phi
_{b_{1}}\otimes \phi _{b_{2}}\otimes ...\otimes \phi _{b_{l}},$ each of
which is also contained in its own factor Hilbert space $\mathcal{H}%
_{[b_{j}]},$ for $1\leq j\leq l,$ with $\tprod\nolimits_{j=1}^{l}\otimes
\mathcal{H}_{[b_{j}]}=\mathcal{H}_{[1...N]}$ and $\langle \phi _{b_{j}}|\phi
_{b_{j}}\rangle =1.$ Now, depending on the particular partitions of $%
\mathcal{H}_{[1...N]}$ in which $\Psi _{n}$ and $\Psi _{n+1}$ are members,
that is, depending on how the various Hilbert spaces $\mathcal{H}_{[a_{i}]}$
and $\mathcal{H}_{[b_{j}]}$ `overlap' with one another, the probability
amplitude $\langle \Phi |\Psi _{n}\rangle $ may be separable into a product
of factors. In other words,
\begin{equation}
P(\Psi _{n+1}=\Phi |\Psi _{n},\hat{\Sigma}_{n+1})=P_{1}P_{2}...P_{r}
\label{Ch3Prob}
\end{equation}
where the overall probability is factorisable into $r$ factors $P_{s},$ for $%
1\leq s\leq r,$ and each factor can be interpreted as a conditional
transition probability within a particular family.

Assuming that (\ref{Ch3Prob}) represents the `fundamental factorisation' of
the probability $P(\Psi _{n+1}=\Phi |\Psi _{n},\hat{\Sigma}_{n+1}),$ then $r$
represents the maximum number of factors associated with the transition
amplitude, and is constrained by $r\leq \min (k,l).$ Thus, in this case
there are $r$ families involved in the transition of the state from $\Psi
_{n}$ to $\Psi _{n+1}=\Phi .$ Further, because each factor of the transition
amplitude involves a distinct portion of the overall set of quantum
subregisters comprising the total Hilbert space, the complete set of factors
$P_{s}$ specifies a particular $r$-partite split of $\mathcal{H}_{[1...N]}.$

Summarising, then, leads to the following definition. For the quantum
transition from the state $\Psi _{n}$ to a potential state $\Psi _{n+1},$
both of which are vectors in a Hilbert space $\mathcal{H}_{[1...N]}$
factorisable into at least two subregisters, $N\geq 2,$ the number of
families involved is defined as equal to the number of factors in the
fundamental factorisation of the probability amplitude $\langle \Psi
_{n+1}|\Psi _{n}\rangle ,$ as determined from which particular partitions of
$\mathcal{H}_{[1...N]}$ the states $\Psi _{n}$ and $\Psi _{n+1}$ are
in.\bigskip

Analogous to classical causal sets, once a family has been identified it is
possible to define concepts such as \textit{parents, children} and \textit{%
siblings}. Specifically, in a given family transition $\langle \phi
_{b_{j}}|\psi _{a_{i}}\rangle ,$ where $\psi _{a_{i}}$ is a factor of $\Psi
_{n}$ and $\phi _{b_{j}}$ is a factor of $\Psi _{n+1},$ the sub-state $\psi
_{a_{i}}$ may be called the `parent' of $\phi _{b_{j}},$ which is in turn
its `child'. Further, if $\psi _{a_{i}}$ is itself a product of $X$ factors,
$\psi _{a_{i}}=\alpha _{a_{i\_1}}\otimes \alpha _{a_{i\_2}}\otimes
...\otimes \alpha _{a_{1\_X}},$ and if $\phi _{b_{j}}$ is a product of $Y$
factors $\phi _{b_{j}}=\beta _{b_{j\_1}}\otimes \beta _{b_{j\_2}}\otimes
...\otimes \beta _{b_{j\_Y}},$ then each factor $\alpha _{a_{1\_x}}$ for $%
1\leq x\leq X$ is a parent of each factor $\beta _{b_{j\_y}}$ for $1\leq
y\leq Y,$ which are its children or `offspring'. Similarly, every factor $%
\beta _{b_{j\_y}}$ is a sibling of every other factor $\beta _{b_{j\_z}},$
for $1\leq y,z\leq Y$ and $y\neq z,$ because they share a parent. Note,
however, that the fact that the $\beta $'s are siblings does not imply that
the $\alpha $'s must also be siblings. Which, if any, of the factors of $%
\psi _{a_{i}}$ are siblings of each other depends entirely on the
factorisation of the transition amplitude $\langle \Psi _{n}|\Psi
_{n-1}\rangle ,$ as will be discussed shortly.\bigskip

First, however, consider as an illustration of these ideas a Hilbert space $%
\mathcal{H}_{[1...8]}$ factorisable into eight elementary subregisters, $%
\mathcal{H}_{[1...8]}=\mathcal{H}_{1}\otimes ...\otimes \mathcal{H}_{8}.$
Consider also the `current' state $\Psi _{n}\in \mathcal{H}_{[1...8]}$ and
one of the potential next states $\Psi _{n+1},$ defined as $\Psi _{n}=\Theta
_{123}^{456\bullet 78}$ and $\Psi _{n+1}=\Phi _{145}^{23\bullet 678}$
respectively. Clearly, $\Psi _{n}$ is in the partition $\mathcal{H}%
_{123}^{456\bullet 78},$ whereas $\Psi _{n+1}\in \mathcal{H}%
_{145}^{23\bullet 678},$ and the states may be written in the forms
\begin{eqnarray}
|\Psi _{n}\rangle &=&\Theta _{123}^{456\bullet 78}=|\Theta _{1}\rangle
\otimes |\Theta _{2}\rangle \otimes |\Theta _{3}\rangle \otimes |\Theta
^{456}\rangle \otimes |\Theta ^{78}\rangle \\
|\Psi _{n+1}\rangle &=&\Phi _{145}^{23\bullet 678}=|\Phi _{1}\rangle \otimes
|\Phi ^{23}\rangle \otimes |\Phi _{4}\rangle \otimes |\Phi _{5}\rangle
\otimes |\Phi ^{678}\rangle  \notag
\end{eqnarray}
in obvious notation. Note that here, and in the following, the usual Hilbert
space subscripts on the ket factors have been omitted to avoid potentially
confusing clashes in the products of amplitudes; specifically, then, it is
implicitly assumed that
\begin{equation}
|\Theta ^{456}\rangle \equiv |\Theta ^{456}\rangle _{456}\in \mathcal{H}%
_{[456]}
\end{equation}
and $|\Phi _{1}\rangle \in \mathcal{H}_{1},$ etc.

Therefore, the transition amplitude $\langle \Psi _{n+1}|\Psi _{n}\rangle $
takes the form $\langle \Phi _{145}^{23\bullet 678}|\Theta
_{123}^{456\bullet 78}\rangle ,$ which is fundamentally factorised as
\begin{eqnarray}
\langle \Phi _{145}^{23\bullet 678}|\Theta _{123}^{456\bullet 78}\rangle
&=&\langle \Phi _{1}|\Theta _{1}\rangle \langle \Phi ^{23}|\Theta
_{23}\rangle \langle \Phi _{45}^{678}|\Theta ^{456\bullet 78}\rangle \\
&=&\langle \Phi _{1}|\Theta _{1}\rangle \text{ }\langle \Phi ^{23}|(|\Theta
_{2}\rangle \otimes |\Theta _{3}\rangle )  \notag \\
&&\times (\langle \Phi _{4}|\otimes \langle \Phi _{5}|\otimes \langle \Phi
^{678}|)(|\Theta ^{456}\rangle \otimes |\Theta ^{78}\rangle ).  \notag
\end{eqnarray}

So, the probability $P=|\langle \Psi _{n+1}=\Phi _{145}^{23\bullet 678}|\Psi
_{n}=\Theta _{123}^{456\bullet 78}\rangle |^{2}$ can be fundamentally
factorised in the form $P=P_{1}P_{2}P_{3}$ where $P_{1}=\left| \langle \Phi
_{1}|\Theta _{1}\rangle \right| ^{2},$ $P_{2}=\left| \langle \Phi
^{23}|\Theta _{23}\rangle \right| ^{2}$ and $P_{3}=\left| \langle \Phi
_{45}^{678}|\Theta ^{456\bullet 78}\rangle \right| ^{2}.$ Further, it is
evident that $\Psi _{n}=\Theta _{123}^{456\bullet 78}$ has $k=5$ factors, $%
\Psi _{n+1}=\Phi _{145}^{23\bullet 678}$ has $l=5$ factors, and the
probability $P$ has $r=3$ factors, which clearly satisfies the relation $%
r\leq \min (k,l).$

Moreover, in this transition, $\Theta _{1}$ is the (single) parent of $\Phi
_{1},$ which has no siblings. The factor $\Phi ^{23}$ also has no siblings,
and is the child of its parents, namely $\Theta _{2}$ and $\Theta _{3}.$
Lastly, the factors $\Theta ^{456}$ and $\Theta ^{78}$ are the parents of $%
\Phi _{4},$ $\Phi _{5},$ and $\Phi ^{678},$ which are siblings of one
another.\bigskip

Just as the sets of events generated in classical causal sets can be
depicted by Hasse diagrams, so too can the family structures produced by the
quantum transitions also be represented pictorially. The convention adopted
is that every possible factor state present in a transition amplitude is
drawn as a large circle, whilst each factor of the relevant $r$-partite
split of the Hilbert space is denoted by a small circle. These\ two types of
circle are labelled in the obvious way, with, for example, a small circle
labelled as $[isy...b]$ denoting the factor $\mathcal{H}_{[isy...b]}$ of the
total Hilbert space.

The `time' parameter, $n,$ is assigned to run upwards in the diagrams, such
that the large circles representing the ket vectors of the transition
amplitude are below the large circles that represent the bra vectors. In
addition, the large circles are linked to the small circles in a way that
depends on how their factor states are contained in the factor Hilbert
spaces of the $r$-partite split. Specifically, with links drawn as arrows
the convention becomes that those arrows pointing towards the bottom of a
small circle run from a set of parent factor states, whilst those arrows
coming from the top of this small circle point to their corresponding set of
children.

With these conventions adopted, the transition from the state $\Psi
_{n}=\Theta _{123}^{456\bullet 78}$ to the state $\Psi _{n+1}=\Phi
_{145}^{23\bullet 678}$ can be depicted by the diagram shown in Figure
5.2.\bigskip\

%\FRAME{ftbpFU}{305.9375pt}{174.6875pt}{0pt}{\Qcb{The family
%structures present in the transition amplitude from $\Psi _{n}=\Theta
%_{123}^{456\bullet 78}$ to $\Psi _{n+1}=\Phi _{145}^{23\bullet 678}.$}}{}{%
%Figure 5.2}{\special{language "Scientific Word";type
%"GRAPHIC";maintain-aspect-ratio TRUE;display "USEDEF";valid_file "T";width
%305.9375pt;height 174.6875pt;depth 0pt;original-width
%299.6875pt;original-height 170.25pt;cropleft "0";croptop "1";cropright
%"1";cropbottom "0";tempfilename 'I1KOB001.wmf';tempfile-properties "XPR";}}

\begin{figure}[th]
\begin{center}
\includegraphics[width=4in]{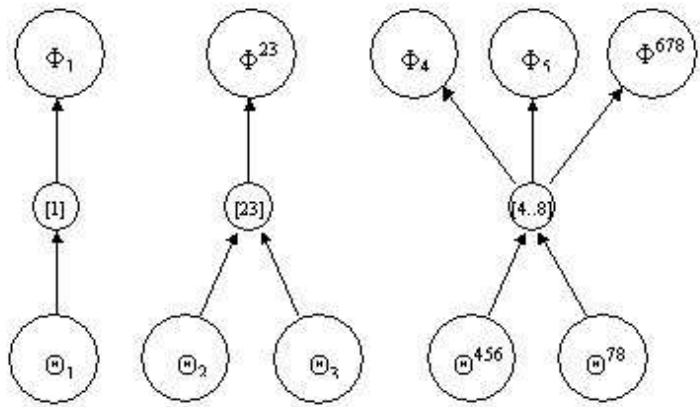}
\caption{The family
structures present in the transition amplitude from $\Psi _{n}=\Theta
_{123}^{456\bullet 78}$ to $\Psi _{n+1}=\Phi _{145}^{23\bullet 678}.$}\label{Figure 5.2}
\end{center}
\end{figure}

In general, familial relations will be generated by every transition
amplitude, and so as the state of the Universe develops through many
transitions, $\Psi _{n}\rightarrow \Psi _{n+1}\rightarrow \Psi
_{n+2}\rightarrow ...$ an extended network of families will begin to emerge.
In addition, definitions of grandparents, grandchildren, cousins etc. will
become apparent, as will identifications of great-grandparents,
great-great-grandparents, and so on. For example, if $A$ is factor of $\Psi
_{n},$ $B$ is factor of $\Psi _{n+1}$ and $C$ is factor of $\Psi _{n+2},$
and if $A$ is a parent of $B$ and $B$ is a parent of $C,$ then $A$ is
necessarily a grandparent of $C.$ Further, as the state develops, individual
families may merge with other families, or may even remain isolated from all
others over a large number of jumps.

Exactly what happens will depend on the specific dynamics that govern the
system, as will be discussed later.\bigskip

The existence of familial relations extending through a number of
transitions gives rise to causal set relationships, with the associated
concepts of lightcones and volume measures. To demonstrate this observation,
consider as an illustrative example a model universe represented by a state
in a Hilbert space fundamentally factorised into six quantum subregisters, $%
\mathcal{H}_{[1...6]}.$ Consider also a possible sequence in the universe's
development, in which five successive states $\Psi _{0},$ $\Psi _{1},...,$ $%
\Psi _{4}$ have the following form
\begin{eqnarray}
\Psi _{0} &=&\varphi ^{123456}\text{ \ \ };\text{ \ \ }\Psi _{1}=\psi
_{1}^{23\bullet 456}\text{ \ \ };\text{ \ \ }\Psi _{2}=\theta
_{16}^{24\bullet 35} \\
\Psi _{3} &=&\eta _{4}^{12\bullet 356}\text{ \ \ \ };\text{ \ \ }\Psi
_{4}=\chi ^{12\bullet 34\bullet 56}.  \notag
\end{eqnarray}

Note the inevitable notational clash here: in this example, subscripts on
the capital Greek letters (e.g. $\Psi _{n})$ will denote temporal ordering,
whereas subscipts on lower case Greek letters will denote separable factors.

With the probability amplitudes given in the usual way, for example
\begin{equation}
\langle \Psi _{n+2}=\theta _{16}^{24\bullet 35}|\Psi _{n+1}=\psi
_{1}^{23\bullet 456}\rangle =\langle \theta _{1}|\psi _{1}\rangle \langle
\theta _{6}^{24\bullet 35}|\psi ^{23\bullet 456}\rangle
\end{equation}
etc., the above sequence of states can be represented by the diagram given
in Figure 5.3.

%\FRAME{ftbpFU}{255.875pt}{365pt}{0pt}{\Qcb{One possible
%network of families produced as a toy-universe develops over five jumps. }}{%
%}{Figure 5.3}{\special{language "Scientific Word";type
%"GRAPHIC";maintain-aspect-ratio TRUE;display "USEDEF";valid_file "T";width
%255.875pt;height 365pt;depth 0pt;original-width 252.5pt;original-height
%361.0625pt;cropleft "0";croptop "1";cropright "1";cropbottom
%"0";tempfilename 'I1KOB002.wmf';tempfile-properties "XPR";}}

\begin{figure}[th!]
\begin{center}
\includegraphics[height=5.6in]{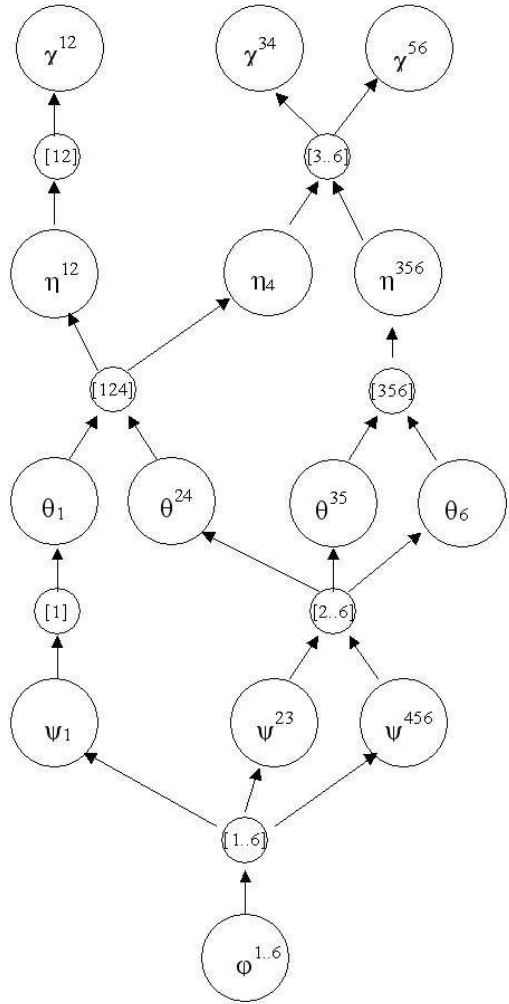}
\caption{One possible
network of families produced as a toy-universe develops over five jumps.}\label{Figure 5.3}
\end{center}
\end{figure}

The universe of this example begins in an initial entangled state $\Psi
_{0}=\varphi ^{123456}.$ Since it is argued throughout this thesis that
separability is a necessary prerequisite for classicity, then at time $n=0$
the universe cannot be given any classical attributes. In fact from
entangled states of the form $\Psi _{0}=\varphi ^{123456},$ no notions of
internal observers, apparatus, or systems under investigation will be able
to emerge. Further, since it has also been argued that the appearance of
space relies on the existence of classicity, then an initial entangled state
cannot contain any sort of spatial relationships.

Separability, and hence the possibility of classicity, occurs in the next
state $\Psi _{1}=\psi _{1}^{23\bullet 456},$ which may be written as a
product of three factors. Of course, it is still not possible to define
measures of distance at this stage, that is to say that the factor $|\psi
_{1}\rangle $ is so many units away from the factor $|\psi ^{456}\rangle ,$
because these factors are nothing but `pregeometric vectors' in a Hilbert
space, whereas physical space is a phenomenon that is only expected to
emerge by considerations of the relationships between large numbers of such
factors over very many transitions. Likewise, there is no immediately
obvious definition of volume on this pregeometric level. However, as with
classical causal sets, embryonic notions of volume may be estimated by a
process of counting; for the quantum causal sets proposed here, measurements
of volume are expected to relate somehow to the number of factors present in
the current state of the universe. As a first approach, it is assumed that
more separable states will generate greater emergent volumes than less
separable states, but it is still unclear at this stage of research exactly
how such a programme should proceed.\bigskip

During the transition from $\Psi _{1}=\psi _{1}^{23\bullet 456}$ to $\Psi
_{2}=\theta _{16}^{24\bullet 35},$ the factor containing the component of
the state in the Hilbert space $\mathcal{H}_{1}$ does not change relative to
the partition structure of the total Hilbert space. In other words, although
the state jumps from one partition of $\mathcal{H}_{[1...6]}$\ to another
during the transition, both $\Psi _{1}$ and $\Psi _{2}$ have a factor in the
same block, i.e.\ $\mathcal{H}_{1}.$ Consequently the component of the state
in the factor Hilbert space $\mathcal{H}_{1}$ changes with no `interaction'
with any other component in $\mathcal{H}_{[2...6]},$ and this may be
physically interpreted as the universe appearing to split into two distinct
sub-universes, neither of which influences the other. Although highly
speculative, this may be the sort of mechanism required to describe the
behaviour of black holes in a fully quantum universe of very many
subregisters, in which an entire region of emergent spacetime appears cut
off from everything else.

A further point can be made if $\theta _{1}$ happens to be the same as $\psi
_{1}.$ In this case, that factor would appear to have been `frozen' in time,
whilst the rest of the universe evolved around it. Such a freezing is a
result of a \textit{local null test,} defined\ in general as an operator $%
\hat{O}_{n+1}$ with eigenvectors of the form $\Phi =\alpha _{a}\otimes
\gamma _{b}$ testing the separable state $\Theta _{n}=\alpha _{a}\otimes
\beta _{b},$ where $\Phi ,\Theta _{n}\in \mathcal{H}_{[ab]},$ $\alpha
_{a}\in \mathcal{H}_{[a]},$ $\beta _{b},\gamma _{b}\in \mathcal{H}_{[b]}$
and $\mathcal{H}_{[a]}$ and $\mathcal{H}_{[b]}$ need not be fundamentally
factorised. Local null tests are often observed in physics, for example when
a spin-%
%TCIMACRO{\UNICODE{0xbd} }%
%BeginExpansion
$\frac12$%
%EndExpansion
particle prepared via the spin-up channel of a Stern-Gerlach apparatus is
passed through an identically orientated Stern-Gerlach device; as discussed
in Chapters 3 and 6, in this type of situation no new information is
acquired about the state by repeating such a test.

Note that \textit{global null tests} could also be a feature of the
dynamics, defined in general as an operator $\hat{O}_{n+1}^{\prime }$ with
an eigenvector $\Theta $ testing the state $\Psi _{n}=\Theta .$ Such a
global null test leaves the entire state unchanged, and is therefore not
physically `noticeable'.\bigskip

The appearance of local null tests, i.e. the persistence of some factors of
the state over a sequence of jumps, has a number of consequences for the
quantum causal sets. Firstly it introduces a concept of endo-time into the
dynamics, that is, the property that over a series of transitions, different
factors of the state will `experience' different durations, where time is
defined in terms of change. In other words, whilst one factor could change $%
a $ times as the state $\Psi _{n}$ develops to $\Psi _{n+m},$ a different
factor may change $b$ times, where $m\geq a,b.$ Consequently this endo-time
is non-integrable, because the number of physically significant jumps that
one particular factor experiences as the Universe develops from the state $%
\Psi _{n}$ to the state $\Psi _{n+m}$ depends on the chain of intermediate
states $\Psi _{n+1},$ $\Psi _{n+2},$ $...,$ $\Psi _{n+m-1}.$ This will be
expanded upon in Chapter 8, but for now note that endo-time is a `route
dependent' concept, analogous to the use of proper time in relativity.
Further, since isolated, classical-looking observers will ultimately be
associated with different groups of factors of the Universe's state, the
possibility that these factors may experience different durations of time
might account for one of the origins of different inertial frames of
reference in emergent relativity, in which different observers witness
different passages of time.

Unlike the physically unobserved (and therefore fictitious) exo-time
parameter, $n,$ endo-time is not necessarily absolute. There is no reason to
assume that any one factor has any more claim to be experiencing the `real'
time than any other. For example if, again, $\theta _{1}=\psi _{1}$ in
Figure 5.3 then $\psi _{1}$ could in principle be regarded as simultaneous
with $\theta ^{24},$ $\theta ^{35},$ and $\theta _{6},$ or instead
simultaneous with $\psi ^{23}$ and $\psi ^{456}.$ Thus, the occurrence of
local null tests also provides a basis for an eventual discussion of
different planes of simultaneity in emergent relativity.\bigskip

Once a notion of endo-time has been introduced it is possible to discuss
timelike distances in a manner analogous to classical causal sets. Generally
speaking, the timelike distance between a factor and one of its `ancestors'
is related to the number of intermediate factors in its family structure.
For example, the factors $\psi ^{456}$ and $\theta _{6}$ may be described as
separated by one `time step' or one `tick' of the `Universe's quantum clock'
( or `\textit{q-tick}' \cite{Jaroszkiewicz}), whereas the factors $\varphi
^{123456}$ and $\theta _{6}$ are separated by two. Likewise, the factors $%
\psi _{1}$ and $\eta ^{12}$ are also separated by two time steps, because
there is one intermediate factor $\theta _{1}.$ However, if again it was the
case that $\theta _{1}=\psi _{1}$ due to a local null test, then in this
instance the factors $\psi _{1}$ and $\eta ^{12}$ would instead be described
as separated by only one q-tick as there are now no physically
distinguishable intermediate factors.

As before, this highlights the fact that endo-time is a concept that depends
on a particular endo-observer's route: if $\theta _{1}=\psi _{1}$ then the
timelike distance between $\varphi ^{123456}$ and\ $\chi ^{34}$ is three
from the point of view of an observer `following' the $\theta _{1}$ path,
but is\ four from the point of view of an observer associated with the
alternative paths via $\theta ^{24},$ $\theta ^{35}$ or $\theta _{6}.$%
\bigskip

A na\"{i}ve concept of lightcone structure can also be gathered from the
above example represented by Figure 5.3. Consider as an illustration the
factor $\theta _{1}$ of the state $\Psi _{2}$ and the factor $\eta ^{356}$
of the state $\Psi _{3}.$ If these factors were simply associated with
classical events, i.e. taken to just be the `objects' of the classical
causal set theory described in Section 5.2, they would be described as
incomparable. That is, there would be no relation of the form $\prec $
linking the events $\theta _{1}$ and $\eta ^{356}$ as $\theta _{1}\prec \eta
^{356}.$ The conclusion is that $\eta ^{356}$ is out of the lightcone of $%
\theta _{1},$ and is hence not in its causal future, and so any change in $%
\theta _{1}$ could not be expected to influence the event $\eta ^{356}.$

This type of lightcone structure is also potentially present in the quantum
causal sets introduced here. It might be possible, for instance, to discuss
whether counterfactual changes in the factors of one state affect particular
factors of later states, simply by a consideration of how the state of
Universe changes from being in one partition to the next as it develops. For
example, in the current toy-universe model it appears that a counterfactual
change in the factor $|\theta _{1}\rangle $ of $\Psi _{2}$ will not affect
the factor $|\eta ^{356}\rangle $ in $\Psi _{3}$ because they are in
completely different blocks of the partition. In other words, because $%
|\theta _{1}\rangle $ is not a parent of $|\eta ^{356}\rangle ,$ a change in
$|\theta _{1}\rangle $ may be expected to leave $|\eta ^{356}\rangle $
invariant. So, by considering how counterfactual changes in one factor of
the universe's state might influence factors in subsequent states, an
embryonic concept analogous to lightcone structure is introduced at the
pregeometric level. Moreover, once such a notion is established, it is
possible to discuss features such as geodesics and spacelike distance, and
ultimately therefore also emergent spacetime, exactly as in the case of
classical causal sets.\bigskip

Note, however, that this line of thinking may be missing an essential point.
In a fully quantum universe with no external observers, the development of
the state $\Psi _{n}$ is achieved by Hermitian operators of the form $\hat{%
\Sigma}_{n+1}$ chosen self-referentially by the Universe itself, as
discussed further in Chapter 8. In other words, there is expected to be some
sort of feedback mechanism in which the current stage somehow affects which
test $\hat{\Sigma}_{n+1}$ is used next. This immediately leads to a serious
problem for the counterfactual argument given above, because any change in
just one factor of a state $\Psi _{n}$ might lead to a completely different
next test $\hat{\Sigma}_{n+1}^{\prime },$ and this may have a completely
different set of eigenvectors.

In the case of the example at hand, the state $\Psi _{2}$ with the factor $%
\theta _{1}$ is assumed to give rise (somehow) to the selection of an
operator $\hat{\Sigma}_{3}$ which has an eigenvector $\Psi _{3}$ that has a
factor $\eta ^{356}.$ If this selection of $\hat{\Sigma}_{3}$ does not
depend on $\Psi _{2},$ then the above simple description of lightcone
structure in terms of counterfactual changes is appropriate. However if
instead the choice of $\hat{\Sigma}_{3}$\ does indeed depend on $\Psi _{2},$
then changing $\Psi _{2}$ may affect $\hat{\Sigma}_{3}.$ So in this case, if
the state $\Psi _{2}$ instead had a factor $\theta _{1}^{\prime },$ the next
test chosen by the universe might be $\hat{\Sigma}_{3}^{\prime },$ and this
alternative operator might have a completely different set of eigenvectors,
with perhaps none of them possessing $\eta ^{356}$ as a factor. In fact,
even a small change from $\theta _{1}$ to $\theta _{1}^{\prime }$\ in $\Psi
_{2}$ might lead to a next state $\Psi _{3}$ that is completely entangled.

Clearly then, in a self-referential Universe developed according to a choice
of operator based upon the current state, an additional mechanism must be
involved in order to ensure that the emergent lightcone structure and
Einstein locality observed in physics is generated. This mechanism will be
shown to involve the operators themselves.

\bigskip

\subsection{Factorisation and Entanglement of Operators}

\bigskip

The previous section showed where a discussion of causal set structure might
begin to emerge from a changing quantum state description of the Universe.
What has not been addressed, however, is how such patterns could arise in
the first place, that is, how and why the separability of the state can
change from one jump to the next.

Since any state is an eigenvector of an Hermitian quantum operator, it is
these tests that must ultimately be responsible for the way in which the
Universe might develop over a series of collapses. Specifically in fact, the
set of eigenvectors belonging to the operator that is chosen to test the
Universe will determine how separable the next potential state will be. For
example, if the rules governing the dynamics dictate that, for all $n,$ an
operator $\hat{\Sigma}_{n+1}$ is chosen that has only entangled
eigenvectors, then the state $\Psi _{n+1}$ of the Universe will always be
entangled and there is no chance that the type of causal set structure
described in the previous section will ever arise. For this reason, the
types of possible Hermitian operator used to develop the Universe must
consequently also be examined.

Up until now, only the separation and entanglement properties of the states
have been investigated. In this section, however, it will be demonstrated
that the operators themselves may also be separable or entangled. Further,
these properties will be shown to also generate structures analogous to
those of causal sets, and this will have far reaching consequences for the
states.\bigskip

The set of Hermitian operators $\mathbb{H}(\mathcal{H}^{(D)})$ of order $D$
is a $D^{2}$-dimensional, real vector space \cite{Joshi}. In general then,
every Hermitian operator $\hat{O}\in \mathbb{H}(\mathcal{H}^{(D)})$ acting
on a state in a $D$ dimensional Hilbert space $\mathcal{H}^{(D)}$ can be
constructed from linear combinations of the $D^{2}$ independent elements
that span this real vector space \cite{Peres}. These $D^{2}$ fundamental
`building blocks' will be called a \textit{skeleton set,} $\mathcal{S},$ of
operators, and are the operators' equivalent in $\mathbb{H}(\mathcal{H}%
^{(D)})$ of the set $\mathcal{B}$ of basis vectors used to construct
arbitrary states in $\mathcal{H}^{(D)}.$

Specifically, if $\mathcal{S}\equiv \{\hat{\sigma}^{\lambda }:\lambda
=0,1,..,(D^{2}-1)\}$ is defined as the skeleton set of operators for a $D$
dimensional Hilbert space $\mathcal{H}^{(D)},$ then any Hermitian operator $%
\hat{O}$ acting upon states in $\mathcal{H}^{(D)}$ may be written in the
form
\begin{equation}
\hat{O}=a_{0}\hat{\sigma}^{0}+a_{1}\hat{\sigma}^{1}+...+a_{D^{2}-1}\hat{%
\sigma}^{D^{2}-1}=\sum\limits_{\lambda =0}^{D^{2}-1}a_{\lambda }\hat{\sigma}%
^{\lambda }
\end{equation}
where $a_{\lambda }$ is a real parameter.

Consider as an illustration a single qubit quantum register, i.e. a
two-dimensional Hilbert space $\mathcal{H}_{A}^{(2)}\equiv \mathcal{H}_{A}$
labelled by the subscript $A,$ with the superscript `$(2)$' that indicates
dimension being now implicitly assumed and hence dropped. Any Hermitian
operator $\hat{o}\in \mathbb{H}(\mathcal{H}_{A})$ acting on a single qubit
state in $\mathcal{H}_{A}$ is composed of a linear sum of the $2^{2}=4$
members of the skeleton set $\mathcal{S}_{A}$ of $\mathbb{H}(\mathcal{H}%
_{A}),$ defined as $\mathcal{S}_{A}\equiv \{\hat{\sigma}_{A}^{\mu }:\mu
=0,1,2,3\},$ where the $\{\hat{\sigma}_{A}^{\mu }\}$ may be associated with
the three Pauli spin operators and the identity operator $\hat{I}=\hat{\sigma%
}_{A}^{0},$ as may be readily verified.

So, the skeleton operators in $\mathcal{S}_{A}$ are taken to obey the
algebraic relations
\begin{eqnarray}
\hat{\sigma}_{A}^{i}\hat{\sigma}_{A}^{j} &=&\delta _{ij}\hat{\sigma}%
_{A}^{0}+i\epsilon _{ijk}\hat{\sigma}_{A}^{k}  \label{Ch3pauli1} \\
\hat{\sigma}_{A}^{i}\hat{\sigma}_{A}^{0} &=&\hat{\sigma}_{A}^{0}\hat{\sigma}%
_{A}^{i}=\hat{\sigma}_{A}^{i}  \notag
\end{eqnarray}
where $i,j,k=1,2,3,$ but $i=\sqrt{-1}$ when it is not used as an index. Here
and below, the Einstein summation convention is assumed over lower case
Latin and Greek indices, and the Levi-Civita tensor is defined in the usual
way:
\begin{equation}
\epsilon _{ijk}=\left\{
\begin{array}{c}
0\text{ \ \ for }i=j,k\text{ or }j=k\text{ } \\
+1\text{ \ \ for }ijk,\text{ }kij,\text{ or }jki\text{ \ \ \ \ \ } \\
-1\text{ \ \ for }ikj,\text{ }jik,\text{ or }kji\text{ \ \ \ \ \ }
\end{array}
\right\} .
\end{equation}

A more compact way of writing the relations (\ref{Ch3pauli1}) is
\begin{equation}
\hat{\sigma}_{A}^{\mu }\hat{\sigma}_{A}^{\upsilon }=C_{\omega }^{\mu
\upsilon }\hat{\sigma}_{A}^{\omega }  \label{Ch3Pauli2}
\end{equation}
where $\mu ,\upsilon ,\omega =0,1,2,3$ and the coefficients $C_{\omega
}^{\mu \upsilon }$ are given by
\begin{eqnarray}
C_{\omega }^{0\upsilon } &=&C_{\omega }^{\upsilon 0}=\delta _{\upsilon
\omega } \\
C_{0}^{ij} &=&\delta _{ij}\text{ \ \ , \ \ }C_{k}^{ij}=i\epsilon _{ijk}.
\notag
\end{eqnarray}

It is possible to obtain a matrix form for the operators $\hat{\sigma}%
_{A}^{\mu }.$ Consider an orthonormal basis set $\mathcal{B}_{A}$ of vectors
spanning $\mathcal{H}_{A},$ defined as $\mathcal{B}_{A}=\{|0\rangle
_{A},|1\rangle _{A}\}.$ A representation of the operators $\hat{\sigma}%
_{A}^{\mu }$ in this basis may be given by
\begin{equation}
\hat{\sigma}_{A}^{\mu }=\sum\limits_{m,n=0}^{1}[\sigma _{A}^{\mu
}]_{(m+1)(n+1)}|m\rangle \langle n|,
\end{equation}
where $m,n=0,1$ and\ $[\sigma _{A}^{\mu }]_{(m+1)(n+1)}$ is the value in the
$(m+1)^{th}$ row of the $(n+1)^{th}$ column of a $2\times 2$ matrix $[\sigma
_{A}^{\mu }].$ As might be expected, one possible such set of matrices may
be defined in the standard way of Pauli:
\begin{eqnarray}
\lbrack \sigma _{A}^{0}] &=&\left(
\begin{array}{cc}
1 & 0 \\
0 & 1
\end{array}
\right) \text{ \ \ , \ \ }[\sigma _{A}^{1}]=\left(
\begin{array}{cc}
0 & 1 \\
1 & 0
\end{array}
\right)  \label{Ch3PauliRep} \\
\lbrack \sigma _{A}^{2}] &=&\left(
\begin{array}{cc}
0 & -i \\
i & 0
\end{array}
\right) \text{ \ \ , \ \ }[\sigma _{A}^{3}]=\left(
\begin{array}{cc}
1 & 0 \\
0 & -1
\end{array}
\right) ,  \notag
\end{eqnarray}
which clearly satisfy (\ref{Ch3Pauli2}).

In addition, it can further be shown that the skeleton set of operators $%
\mathcal{S}_{A}=\{\hat{\sigma}_{A}^{\mu }\}$ may be associated with the
identity $(\mu =0)$ and the generators $(\mu =i=1,2,3)$ of the group $SU(2).$
Consequently, any special unitary operator $\hat{U}$ acting on $\mathcal{H}%
_{A}$ may be written in the form
\begin{equation}
\hat{U}=\exp \left[ i\tsum\limits_{\mu =0}^{3}u_{\mu }\sigma _{A}^{\mu }%
\right]
\end{equation}
where the $u_{\mu }\in \mathbb{R}$ are real parameters.\bigskip

The above arguments can be extended to Hilbert spaces factorisable into more
than one qubit subregister. Consider a Hilbert space $\mathcal{H}_{[1...N]}$
formed from the tensor product of $N$ qubit factor spaces, $\mathcal{H}%
_{[1...N]}=\mathcal{H}_{1}\otimes \mathcal{H}_{2}\otimes ...\otimes \mathcal{%
H}_{N}.$ An orthonormal basis $\mathcal{B}_{a}$ for the $a^{th}$ factor
space $\mathcal{H}_{a}$ $(=\mathcal{H}_{a}^{(2)}$ as before) may be defined
as $\mathcal{B}_{a}=\{|0\rangle _{a},|1\rangle _{a}\},$ where $_{a}\langle
n|m\rangle _{a}=\delta _{mn}$ for $m,n=0,1.$

The skeleton set $\mathcal{S}_{N}$ for the total Hilbert space $\mathcal{H}%
_{[1...N]}$ can be generated by taking the tensor products of the skeleton
operators of the individual qubit spaces, i.e.
\begin{equation}
\mathcal{S}_{N}\equiv \{\hat{\sigma}_{1}^{\mu _{1}}\otimes \hat{\sigma}%
_{2}^{\mu _{2}}\otimes ...\otimes \hat{\sigma}_{N}^{\mu _{N}}:\mu
_{a}=0,1,2,3\text{ for }a=1,2,...,N\},  \label{Ch3Skel}
\end{equation}
which is clearly a set containing $4^{N}=(2^{N})^{2}=D^{2}$ members.

Note that for convenience the skeleton set $\mathcal{S}_{N}$\ may also be
rewritten as $\mathcal{S}_{N}\equiv \{\hat{\sigma}_{1..N}^{\eta }:\eta
=0,1,...,(4^{N}-1)\},$ with the first member $\hat{\sigma}_{1..N}^{0}=\hat{%
\sigma}_{1}^{0}\otimes \hat{\sigma}_{2}^{0}\otimes ...\otimes \hat{\sigma}%
_{N}^{0}$ etc.

The set $\mathcal{S}_{N}$ forms a basis for the real vector space $\mathbb{H}%
(\mathcal{H}_{[1...N]})$ of Hermitian operators in $\mathcal{H}_{[1...N]}.$
Any operator $\hat{A}\in \mathbb{H}(\mathcal{H}_{[1...N]})$ can be written
as a linear sum\footnote{%
For explicitness, the Einstein summation convention has been replaced in
this expression by the `sum' signs.} of the members of $\mathcal{S}_{N}$
\begin{eqnarray}
\hat{A} &=&\sum\limits_{\mu _{1}=0}^{3}\sum\limits_{\mu
_{2}=0}^{3}...\sum\limits_{\mu _{N}=0}^{3}A_{\mu _{1}\mu _{2}...\mu _{N}}%
\hat{\sigma}_{1}^{\mu _{1}}\otimes \hat{\sigma}_{2}^{\mu _{2}}\otimes
...\otimes \hat{\sigma}_{N}^{\mu _{N}} \\
&=&A_{00...0}(\hat{\sigma}_{1}^{0}\otimes \hat{\sigma}_{2}^{0}\otimes
...\otimes \hat{\sigma}_{N}^{0})+A_{10...0}(\hat{\sigma}_{1}^{1}\otimes \hat{%
\sigma}_{2}^{0}\otimes ...\otimes \hat{\sigma}_{N}^{0})+...  \notag \\
&&...+A_{33...3}(\hat{\sigma}_{1}^{3}\otimes \hat{\sigma}_{2}^{3}\otimes
...\otimes \hat{\sigma}_{N}^{3})  \notag
\end{eqnarray}
where the coefficients $A_{\mu _{1}\mu _{2}...\mu _{N}}$ are all real, as
required for Hermicity.

Consider now a second operator $\hat{B}\in \mathbb{H}(\mathcal{H}%
_{[1...N]}), $ such that it is possible to in turn define a third operator $%
\hat{X}$ as the multiplicative product $\hat{X}\equiv \hat{A}\hat{B}.$
Assuming again that the Einstein convention is adopted on repeated Greek
indices, $\hat{A}$ and $\hat{B}$ may be written as $\hat{A}=A_{\mu _{1}\mu
_{2}...\mu _{N}}\hat{\sigma}_{1}^{\mu _{1}}\otimes \hat{\sigma}_{2}^{\mu
_{2}}\otimes ...\otimes \hat{\sigma}_{N}^{\mu _{N}}$ and $\hat{B}%
=B_{\upsilon _{1}\upsilon _{2}...\upsilon _{N}}\hat{\sigma}_{1}^{\upsilon
_{1}}\otimes \hat{\sigma}_{2}^{\upsilon _{2}}\otimes ...\otimes \hat{\sigma}%
_{N}^{\upsilon _{N}}$ for $0\leq \mu _{r}\leq 3$ and $0\leq \upsilon
_{s}\leq 3$ with $r,s=1,2,...,N.$

So, the product $\hat{X}$ is given by
\begin{eqnarray}
\hat{X} &=&(A_{\mu _{1}\mu _{2}...\mu _{N}}\hat{\sigma}_{1}^{\mu
_{1}}\otimes \hat{\sigma}_{2}^{\mu _{2}}\otimes ...\otimes \hat{\sigma}%
_{N}^{\mu _{N}})(B_{\upsilon _{1}\upsilon _{2}...\upsilon _{N}}\hat{\sigma}%
_{1}^{\upsilon _{1}}\otimes \hat{\sigma}_{2}^{\upsilon _{2}}\otimes
...\otimes \hat{\sigma}_{N}^{\upsilon _{N}}) \\
&=&A_{\mu _{1}\mu _{2}...\mu _{N}}B_{\upsilon _{1}\upsilon _{2}...\upsilon
_{N}}(\hat{\sigma}_{1}^{\mu _{1}}\otimes \hat{\sigma}_{2}^{\mu _{2}}\otimes
...\otimes \hat{\sigma}_{N}^{\mu _{N}})(\hat{\sigma}_{1}^{\upsilon
_{1}}\otimes \hat{\sigma}_{2}^{\upsilon _{2}}\otimes ...\otimes \hat{\sigma}%
_{N}^{\upsilon _{N}})  \notag \\
&=&A_{\mu _{1}\mu _{2}...\mu _{N}}B_{\upsilon _{1}\upsilon _{2}...\upsilon
_{N}}C_{\omega _{1}}^{\mu _{1}\upsilon _{1}}C_{\omega _{2}}^{\mu
_{2}\upsilon _{2}}...C_{\omega _{N}}^{\mu _{N}\upsilon _{N}}\hat{\sigma}%
_{1}^{\omega _{1}}\otimes \hat{\sigma}_{2}^{\omega _{2}}\otimes ...\otimes
\hat{\sigma}_{N}^{\omega _{N}}  \notag
\end{eqnarray}
from (\ref{Ch3Pauli2}). The coefficients $A_{\mu _{1}\mu _{2}...\mu
_{N}}B_{\upsilon _{1}\upsilon _{2}...\upsilon _{N}}C_{\omega _{1}}^{\mu
_{1}\upsilon _{1}}C_{\omega _{2}}^{\mu _{2}\upsilon _{2}}...C_{\omega
_{N}}^{\mu _{N}\upsilon _{N}}$ are just products of real parameters, so this
last line may be rewritten in the form
\begin{equation}
\hat{X}=X_{\omega _{1}\omega _{2}...\omega _{N}}\hat{\sigma}_{1}^{\omega
_{1}}\otimes \hat{\sigma}_{2}^{\omega _{2}}\otimes ...\otimes \hat{\sigma}%
_{N}^{\omega _{N}}
\end{equation}
where $X_{\omega _{1}\omega _{2}...\omega _{N}}\in \mathbb{R}$ for $\omega
_{t}=0,1,2,3$ and $t=1,2,...,N.$ The product operator $\hat{X}$ is a linear
sum of the members of $\mathcal{S}_{N}$ with real coefficients, and so is
clearly a member of $\mathbb{H}(\mathcal{H}_{[1...N]}).$ The set $\mathbb{H}(%
\mathcal{H}_{[1...N]})$ is hence confirmed closed under the multiplication
rule, as expected for a vector space, and is an algebra over the real number
field.

\bigskip

Consider again the Hermitian operator $\hat{A}\in \mathbb{H}(\mathcal{H}%
_{[1...N]})$ defined as $\hat{A}=A_{\mu _{1}\mu _{2}...\mu _{N}}\hat{\sigma}%
_{1}^{\mu _{1}}\otimes \hat{\sigma}_{2}^{\mu _{2}}\otimes ...\otimes \hat{%
\sigma}_{N}^{\mu _{N}}.$ Depending on the actual values of the coefficients $%
A_{\mu _{1}\mu _{2}...\mu _{N}},$ this operator may, or may not, factorise
relative to the skeleton set associated with some particular split of the
total Hilbert space $\mathcal{H}_{[1...N]}.$ For instance, if\ $\mathcal{H}%
_{[1...N]}$ can be factorised into the bi-partite split $\mathcal{H}%
_{[1...N]}=\mathcal{H}_{V}\otimes \mathcal{H}_{W},$ where $\mathcal{H}_{V}$
and $\mathcal{H}_{W}$ need not be of prime dimension, it may be the case
that $\hat{A}$ can be written in the form $\hat{A}=\hat{V}\otimes \hat{W},$
where $\hat{V}$ is an Hermitian operator acting in the factor sub-space $%
\mathcal{H}_{V}$ and $\hat{W}$ is an Hermitian operator acting in the factor
sub-space $\mathcal{H}_{W}.$

As an example, in the factorisable two qubit Hilbert space $\mathcal{H}%
_{[12]}=\mathcal{H}_{1}\otimes \mathcal{H}_{2},$ the space of Hermitian
operators $\mathbb{H}(\mathcal{H}_{[12]})$ is spanned by the skeleton set $%
\mathcal{S}_{12}$ defined as $\mathcal{S}_{12}=\{\hat{\sigma}_{1}^{\mu
_{1}}\otimes \hat{\sigma}_{2}^{\mu _{2}}:\mu _{1},\mu _{2}=0,1,2,3\}.$
Clearly then, an Hermitian operator $\hat{E}\in \mathbb{H}(\mathcal{H}%
_{[12]})$ of the form
\begin{equation}
\hat{E}=\frac{1}{2}(3\hat{\sigma}_{1}^{1}\otimes \hat{\sigma}_{2}^{1}+\hat{%
\sigma}_{1}^{2}\otimes \hat{\sigma}_{2}^{2})
\end{equation}
is entangled, whereas an Hermitian operator $\hat{F}\in \mathbb{H}(\mathcal{H%
}_{[12]})$ of the form
\begin{equation}
\hat{F}=3\hat{\sigma}_{1}^{0}\otimes \hat{\sigma}_{2}^{1}-\hat{\sigma}%
_{1}^{3}\otimes \hat{\sigma}_{2}^{1}=(3\hat{\sigma}_{1}^{0}-\hat{\sigma}%
_{1}^{3})\otimes \hat{\sigma}_{2}^{1}  \label{Ch3facOp}
\end{equation}
is factorisable relative to $\mathcal{S}_{12}$ in $\mathcal{H}_{[12]}=%
\mathcal{H}_{1}\otimes \mathcal{H}_{2}.$\bigskip

Whether or not an arbitrary Hermitian operator $\hat{O}$ in $\mathbb{H}(%
\mathcal{H}_{[1...N]})$ is factorisable in the form $\hat{O}=\hat{Y}\otimes
\hat{Z},$ relative to the skeleton set of a particular bi-partite split of
the total Hilbert space\ $\mathcal{H}_{[1...N]}=\mathcal{H}_{Y}\otimes
\mathcal{H}_{Z},$ may be determined in a manner that is similar to the
microsingularity test given in Chapter 4 used to discover whether an
arbitrary state in $\mathcal{H}_{[1...N]}$ is separable relative to $%
\mathcal{H}_{[1...N]}=\mathcal{H}_{Y}\otimes \mathcal{H}_{Z}.$

\begin{theorem}
An arbitrary Hermitian operator $\hat{O}\in \mathbb{H}(\mathcal{H}_{[1...N]})
$ is factorisable in the form $\hat{O}=\hat{Y}\otimes \hat{Z},$ relative to
the skeleton set $\mathcal{S}_{YZ}=\{\hat{\sigma}_{Y}^{\alpha }\otimes \hat{%
\sigma}_{Z}^{\beta }\}$ of a particular bi-partite split of the total
Hilbert space\ $\mathcal{H}_{[1...N]}=\mathcal{H}_{Y}\otimes \mathcal{H}_{Z}$
such that $\hat{Y}\in \mathbb{H}(\mathcal{H}_{Y})$ and $\hat{Z}\in \mathbb{H}%
(\mathcal{H}_{Z})$ if, for $\hat{O}=C_{\alpha \beta }\hat{\sigma}%
_{Y}^{\alpha }\otimes \hat{\sigma}_{Z}^{\beta },$
\begin{equation}
C_{\alpha \beta }C_{\gamma \delta }=C_{\alpha \delta }C_{\gamma \beta }
\label{Ch3MicroOp}
\end{equation}
for all values of the indices $\alpha ,\gamma =0,...,(D_{Y}-1)$ and $\beta
,\delta =0,...,(D_{Z}-1),$ where $C_{\alpha \beta }$ is a real coefficient, $%
\hat{\sigma}_{Y}^{\alpha }$ is a skeleton operator for $\mathbb{H}(\mathcal{H%
}_{Y}),$ $\hat{\sigma}_{Z}^{\beta }$ is a skeleton operator for $\mathbb{H}(%
\mathcal{H}_{Z}),$ and $D_{Y}$ and $D_{Z}$ are the dimensions of $\mathbb{H}(%
\mathcal{H}_{Y})$ and $\mathbb{H}(\mathcal{H}_{Z})$ respectively.
\end{theorem}

This may be shown as follows:\bigskip

\begin{proof}
$\Rightarrow $ Suppose that the coefficients of the operator $\hat{O}%
=C_{\alpha \beta }\hat{\sigma}_{Y}^{\alpha }\otimes \hat{\sigma}_{Z}^{\beta }
$ satisfy the microsingularity condition (\ref{Ch3MicroOp}), and without
loss of generality assume that $\hat{O}$ is not the zero operator. This
implies that at least one coefficient $C_{\alpha \beta }$ must be non-zero
for $\alpha =0,...,(D_{Y}-1)$ and $\beta =0,...,(D_{Z}-1).$ Further, since
any $C_{\gamma \delta }$ is just a real number, the product $C_{\gamma
\delta }\hat{O}$ is just a scalar multiple of $\hat{O},$ so
\begin{eqnarray}
\hat{O} &=&C_{\alpha \beta }\hat{\sigma}_{Y}^{\alpha }\otimes \hat{\sigma}%
_{Z}^{\beta } \\
C_{\gamma \delta }\hat{O} &=&C_{\gamma \delta }C_{\alpha \beta }\hat{\sigma}%
_{Y}^{\alpha }\otimes \hat{\sigma}_{Z}^{\beta }  \notag \\
&=&C_{\alpha \delta }C_{\gamma \beta }\hat{\sigma}_{Y}^{\alpha }\otimes \hat{%
\sigma}_{Z}^{\beta }  \notag \\
&=&(C_{\alpha \delta }\hat{\sigma}_{Y}^{\alpha })\otimes (C_{\gamma \beta }%
\hat{\sigma}_{Z}^{\beta }).  \notag
\end{eqnarray}

Clearly, then, $\hat{O}$ is factorisable with respect to $\mathcal{S}_{YZ}.$%
\bigskip

$\Longleftarrow $ If $\hat{O}\in \mathbb{H}(\mathcal{H}_{[1...N]})$
factorises relative to $\mathcal{S}_{YZ}$ then
\begin{equation}
\hat{O}=(y_{\alpha }\hat{\sigma}_{Y}^{\alpha })\otimes (z_{\beta }\hat{\sigma%
}_{Z}^{\beta })
\end{equation}
where $y_{\alpha }$ and $z_{\beta }$ are real parameters. So
\begin{equation}
\hat{O}=y_{\mu }z_{\upsilon }\hat{\sigma}_{Y}^{\mu }\otimes \hat{\sigma}%
_{Z}^{\upsilon }.
\end{equation}

Taking $C_{\alpha \beta }=y_{\alpha }z_{\beta },$ and similarly $C_{\gamma
\delta }=y_{\gamma }z_{\delta },$ the microsingularity condition (\ref
{Ch3MicroOp}) is clearly satisfied because
\begin{equation}
C_{\alpha \beta }C_{\gamma \delta }=y_{\alpha }z_{\beta }y_{\gamma
}z_{\delta }=y_{\alpha }z_{\delta }y_{\gamma }z_{\beta }=C_{\alpha \delta
}C_{\gamma \beta }.
\end{equation}
\bigskip
\end{proof}

Just as the states can be described as existing in certain entanglements or
separations of the Hilbert space, the operators testing them can also be
placed into similar such sets. Thus, the `partition structure' of the
operators may also be discussed. Again, the use of upper and lower indices
may be adopted in the obvious way, and the symbol ``$\bullet $'' may be used
to denote separable products. For example, in this notation the operator $%
\hat{O}_{a}^{bc\bullet def}$ acting on states in $\mathcal{H}_{[a...f]}$ is
assumed fundamentally factorisable into three sub-operators of the form
\begin{equation}
\hat{O}_{a}^{bc\bullet def}=\hat{A}_{a}\otimes \hat{A}^{bc}\otimes \hat{A}%
^{def}
\end{equation}
where $\hat{A}_{a}$ acts on states in $\mathcal{H}_{a},$ $\hat{A}^{bc}$ is
an entangled sub-operator acting in $\mathcal{H}_{[bc]},$ and $\hat{A}^{def}$
is an entangled sub-operator relative to the skeleton set of $\mathcal{H}%
_{[def]}.$ Moreover, $\hat{O}_{a}^{bc\bullet def}$ is a member of the set $%
\mathbb{H}(\mathcal{H}_{[a...f]})_{a}^{bc\bullet def}$ of Hermitian
operators in $\mathcal{H}_{[a...f]}$ that are fundamentally factorisable
into three factors relative to the skeleton set of the split $\mathcal{H}%
_{[a...f]}=\mathcal{H}_{[a]}\otimes \mathcal{H}_{[bc]}\otimes \mathcal{H}%
_{[d...f]}.$

\bigskip

Whilst Hermitian operators may be factorisable or entangled, it should be
noted that not every type of entangled or factorisable operator is
necessarily Hermitian. This is an important point, since it is only
Hermitian operators that are responsible for physically realisable
observables, and only the eigenvectors of Hermitian operators that make up
the physically realisable states.

Moreover, it should further be noted that not every type of Hermitian
operator can validly be used to test the quantum state of the Universe. This
follows because with every Hermitian operator is associated a set of
eigenvalues, each of which implies\ a corresponding eigenvector. However, if
two (or more) of these eigenvalues are the same, their eigenvectors are not
uniquely determined. This is a standard result of linear algebra \cite
{Kolman}.

In the paradigm proposed in this thesis, the quantum state of the Universe $%
\Psi _{n}$ in its $D$ dimensional Hilbert space is developed by collapsing
into one of the eigenvectors of an Hermitian operator $\hat{\Sigma}_{n+1}.$
In this mechanism, the operator $\hat{\Sigma}_{n+1}$ is assumed to uniquely
provide a complete, orthonormal set of $D$ eigenvectors, $\Phi _{n+1}^{i}$
for $i=1,2,...,D,$ which effectively produces a preferred basis for the next
set of potential states $\Psi _{n+1}.$ It is therefore necessary that this
set of eigenvectors has members that are not only distinguishable, but are
also well defined and specific.

This conclusion is partly because if two eigenvectors have the same
eigenvalue they cannot be distinguished by any sort of measuring apparatus,
since it is generally the eigenvalues that are actually recorded (c.f.
energy eigenvalues in conventional laboratory physics). So, since the jump
from the state $\Psi _{n}$ to the next state $\Psi _{n+1}$ has been ascribed
to be parameterised in terms of information acquisition, any such
uncertainty as to what state this information implies would cause problems
for an interpretation of how the Universe is developing.

Additionally, if the eigenvectors of $\hat{\Sigma}_{n+1}$ are not uniquely
specified, problems arise involving the identification of the members of the
set of potential next states. This, in turn, could lead to an ambiguity
regarding what (pure) state the Universe is actually in.\bigskip

As a very simple illustration of the importance of these ideas, consider a
universe consisting of just a single qutrit, i.e. one represented by a state
in a three dimensional Hilbert space $\mathcal{H}^{(3)}$ spanned by the
orthonormal basis $\mathcal{B}^{(3)}=\{|i\rangle :i=0,1,2\}.$ Consider also
an operator of the form $\hat{P}=|0\rangle \langle 0|,$ denoted in this
representation by the matrix
\begin{equation}
P=\left(
\begin{array}{ccc}
1 & 0 & 0 \\
0 & 0 & 0 \\
0 & 0 & 0
\end{array}
\right)  \label{Ch3Proj}
\end{equation}

It can readily be shown that the states $\Theta _{1}=|0\rangle ,$ $\Theta
_{2}=\frac{1}{\sqrt{2}}(|1\rangle +|2\rangle ),$ and $\Theta _{3}=\frac{1}{%
\sqrt{2}}(|1\rangle -|2\rangle )$ are three orthonormal eigenvectors of $%
\hat{P},$ with eigenvalues $\mu _{1}=1,$ $\mu _{2}=0$ and $\mu _{3}=0$
respectively, because they satisfy $\hat{P}\Theta _{i}=\mu _{i}\Theta _{i}$
and $\langle \Theta _{j}|\Theta _{i}\rangle =\delta _{ij}$ for $i,j=1,2,3.$
These states therefore form an orthonormal basis $\frak{B}^{(3)}$ for $\hat{P%
}.$ But it can also be shown that there is no uniqueness in this
construction of $\frak{B}^{(3)},$ because any other set of states $|0\rangle
,$ $(\alpha |1\rangle +\beta |2\rangle ),$ and $(\beta ^{\ast }|1\rangle
-\alpha ^{\ast }|2\rangle )$ for any other values of $\alpha ,\beta \in
\mathbb{C}$ also comprise an orthogonal basis set $\frak{B}^{(3)\prime }$ of
eigenstates. There is hence an inherent `ambiguity' in the eigenvectors of $%
\hat{P},$ making it an example of the type of Hermitian operator that cannot
be used in the development of the state of the universe. In order to develop
the universe from the state $\Psi _{n}$ to $\Psi _{n+1},$ a unique basis set
$\frak{B}_{n+1}$ must be specified.\bigskip

A suggestion at this point might be to introduce additional ansatz into the
dynamics in order to overcome the above types of problem. In conventional
quantum mechanics, for example, operators with degenerate eigenvalues (and
hence continuous spectra of eigenstates) may be dealt with by an inclusion
of L\"{u}ders' generalised projection postulate \cite{Luders}\cite{Omnes},
and additional procedures may be employed to select a unique preferred basis
set of eigenstates from the infinite set of possibilities possessed by the
degenerate operator. For instance, it may be suggested that upon testing,
the system selects a basis containing the eigenvector that is `nearest' to
the initial state. Paraphrasing such a possibility: if a quantum object
represented by the state $\psi $ is tested by the degenerate operator $\hat{o%
},$ then according to this `selection mechanism' it may be taken to collapse
to a member of a basis set of eigenstates of $\hat{o}$ that contains the
particular eigenvector $\varphi $ for which the value of $|\langle \varphi
|\psi \rangle |^{2}$ is greatest.

Now, it is not clear at this stage exactly how such suggestions could
affect, or be incorporated into, the dynamics proposed in this thesis for
the developing quantum universe. For a start, L\"{u}ders' postulate was
originally phrased in terms of density matrices, whereas such an approach is
not the direct focus of this thesis on the grounds that ensemble
descriptions of the Universe are considered unphysical, as discussed in
Chapter 3. Moreover, L\"{u}ders' idea of generalised projections was also
developed from an exo-physical perspective for quantum systems in the
laboratory, and it is not always obvious whether any such concepts may be
directly applicable to the case where the state describes the entire
universe.

As for the above `selection mechanism', it is not immediately clear how the
other members of the required orthogonal basis set of vectors may be
self-referentially selected during the procedure; recall that in a $D>2$
dimensional Hilbert space, there are an infinite number of $(D-1)$ mutually
orthogonal vectors that are also orthogonal to any given state $\varphi .$
Additionally, it could be expected that there is a high propensity for such
a mechanism to result in null tests on the universe, and these are not
ascribed to play a role in the dynamics of the proposed paradigm.
Elaborating on this last point, the `nearest' eigenstate could be the same
as the initial state, and this would potentially lead to trivial dynamics.
As an example of this possibility, if a single qutrit toy-universe is
initially in a state $\psi =(\alpha |1\rangle +\beta |2\rangle ),$ the
suggested mechanism could imply that the degenerate operator $\hat{P}$
described in (\ref{Ch3Proj}) could leave the universe in the same state $%
\varphi =(\alpha |1\rangle +\beta |2\rangle ).$\bigskip

Two more technical points concerning incorporating the above suggestions
into the quantum universe dynamics are also appropriate. Firstly, even if
additional mechanisms are postulated in conventional quantum mechanics that
select a unique basis set from a continuous spectra of possibilities, they
do\ not necessarily have a place in the work described here. After all, in
the paradigm proposed in this thesis it is the orthogonal basis sets that
have prime importance in the quantum developments, and not the equivalence
class of operators that they imply. In other words, and by reversing the
overall argument, since by definition the dynamics governing the transition
from state $\Psi _{n}$ to $\Psi _{n+1}$ requires the specification of a
unique basis set $\frak{B}_{n+1}$ $(\ni \Psi _{n+1}),$ which \textit{then in
turn} implies the specification of an equivalence class of Hermitian
operators, the question of what happens if the eigenvalues of the operator $%
\hat{\Sigma}_{n+1}$ are degenerate does not automatically arise. By
definition, the operator $\hat{\Sigma}_{n+1}$ is not being used to generate
a unique basis set $\frak{B}_{n+1};$ the unique orthonormal basis set $\frak{%
B}_{n+1}$ is instead used to imply the operator $\hat{\Sigma}_{n+1}.$

Secondly, if the dynamics were to dictate that the current state $\Psi _{n}$
is used to select a particular basis set $\frak{B}_{n+1}$ from the
degenerate operator $\hat{\Sigma}_{n+1}$ (which is identical to arguing:
``if the dynamics were to dictate that the current state $\Psi _{n}$ is used
to select a particular basis set $\frak{B}_{n+1}$ from the infinite set of
possible basis sets of eigenstates of the degenerate operator $\hat{\Sigma}%
_{n+1}"),$ then the question would remain: ``why does the dynamics bother to
define the infinite set in the first place if only one member $\frak{B}%
_{n+1} $ is deterministically picked?''. What is the point in defining a
degenerate operator, and hence an infinite number of possible basis sets, if
additional constraints are then required to select just one of these bases?
Why not instead just define a particular unique basis set $\frak{B}_{n+1},$
and then consider the equivalence class of operators that this
implies?\bigskip

Overall, whilst none of the above issues explicitly forbid the use of
generalised measurements within the framework aimed at in this work, it is
equally evident that their inclusion into the quantum universe dynamics
would require careful attention. Thus, the question of how or whether the
tests of the universe may be allowed to possess degenerate eigenvalues is
left as an area of investigation for the future, and the overall constraint
of only accepting non-degenerate operators $\hat{\Sigma}_{n+1}$ is enforced
for simplicity by definition to avoid possible such considerations.\bigskip

Summarising, in the paradigm proposed here, the orthonormal basis set $\frak{%
B}_{n+1}$ must be uniquely specified if it is to be used in the development
of the state $\Psi _{n}$ of the universe. Thus, the $D$ eigenvectors $\Phi
_{n+1}^{i},$ $i=1,...,D,$ of an operator $\hat{\Sigma}_{n+1}$ that form $%
\frak{B}_{n+1}$ must also be uniquely specified. To ensure this, the
constraint is therefore assumed that only operators $\hat{\Sigma}_{n+1}$
with $D$ non-degenerate and real eigenvalues may be used to test the
universe.

This conclusion leads to a definition of three different types of Hermitian
operator, useful in the following:

\begin{enumerate}
\item[i)]  A \textit{Degenerate} operator is an Hermitian operator with at
least two identical eigenvalues;

\item[ii)]  A \textit{Weak} operator is an Hermitian operator which is
either degenerate, or at least one of its eigenvalues is zero;

\item[iii)]  A \textit{Strong} operator is an Hermitian operator which is
not weak, i.e. all of its eigenvalues are different and none are zero.
\end{enumerate}

It immediately follows that projection operators are weak, as is evident
from, for example, (\ref{Ch3Proj}).

The necessary distinction between Strong and Weak operators will become
apparent when tensor products of operators are considered; it will be shown
later that products of weak operators are in general insufficient to
determine a preferred basis for the developing state, whereas products of
strong operators may be used. Products of strong operators may thus be
associated with the physical tests of the state of the Universe.\bigskip

Note that for an operator $\hat{\Sigma}_{n+1}$ satisfying the eigenvector
equation $\left| (\hat{\Sigma}_{n+1}-\lambda _{i}\hat{I})\right| =0,$ with
eigenvalue $\lambda _{i}\in \mathbb{R},$ $i=1,...,D$ and the identity $\hat{I%
},$ the actual, absolute values of $\lambda _{i}$\ are not important. What
\textit{is} important is the basis set of eigenvectors they represent, and,
specifically for the case of strong operators, that these eigenvectors are
all different. Indeed, given an arbitrary operator $\hat{O}$ with
eigenvalues $\{a,b,c,...\},$ it is possible to find a second arbitrary
operator $\hat{O}^{\prime }$ with eigenvalues $\{k^{\prime }a,k^{\prime
}b,k^{\prime }c,...\},$ where $k^{\prime }\in \mathbb{R}$ is a real,
non-zero constant, that has the same spectrum of eigenvectors as $\hat{O}.$
Moreover, the alternative operator $\hat{O}^{\prime \prime }$ defined as $%
\hat{O}^{\prime \prime }=\hat{O}+k^{\prime \prime }\hat{I}$ also has the
same eigenvectors as $\hat{O},$ even though its eigenvalues $\{(a+k^{\prime
\prime }),(b+k^{\prime \prime }),(c+k^{\prime \prime }),...\}$ are `shifted'
from those of $\hat{O}$ by a constant amount $k^{\prime \prime }.$

Conversely, note that two different strong operators may have the same set
of eigenvalues but different eigenvectors, the Pauli matrices being a good
example.

Of course, these principles are familiar to many physics experiments, where,
for example, energy and momentum eigenvalues often only have relative
significance. Moreover, the three operators\ $\hat{O},$ $\hat{O}^{\prime }$
and $\hat{O}^{\prime \prime }$ would be physically indistinct, in the sense
that if the Universe collapsed to a particular state $\Psi _{n+1}$ that was
one of the members of this set of eigenvectors, an observer would be unable
to determine whether it was $\hat{O},$ $\hat{O}^{\prime }$ or $\hat{O}%
^{\prime \prime }$ that was used to test $\Psi _{n}.$

Summarising, although their actual values are unimportant, it is necessary
that the members of the set of eigenvalues are all different if distinctions
are to be made between the members of the corresponding set of
eigenstates.\bigskip

In addition to the ideas of `weak' and `strong' operators, a further
definition useful in the following is the `\textit{pairwise-product}'.
Consider a set $X\equiv \{x_{1},x_{2},...,x_{M}\}$ with $M$ members, and the
set $Y\equiv \{y_{1},y_{2},...,y_{N}\}$ with $N$ members. The
pairwise-product $XY$ of the sets $X$ and $Y$ is defined as the set of all
the products $XY\equiv \{x_{i}y_{j}:i=1,...,M,$ $j=1,...,N\}.$ Clearly, $XY$
is a set containing $M\times N$ members.

As an extension, the \textit{pairwise tensor product} of two sets can
similarly be defined in an obvious way.\bigskip

Return now to the issue of the separability of operators. Consider a Hilbert
space $\mathcal{H}_{[12]}$ factorisable into two subspaces, $\mathcal{H}%
_{[12]}=\mathcal{H}_{1}\otimes \mathcal{H}_{2},$ where $\mathcal{H}_{1}$ and
$\mathcal{H}_{2}$ are of dimensions $d_{1}$ and $d_{2}$ respectively, which
need not be prime. Consider also the Hermitian operators $\hat{A}_{1}\in
\mathbb{H}(\mathcal{H}_{1})$ and $\hat{B}_{2}\in \mathbb{H}(\mathcal{H}%
_{2}), $ such that the product operator $\hat{O}_{12}=\hat{A}_{1}\otimes
\hat{B}_{2} $ is a factorisable member of $\mathbb{H}(\mathcal{H}_{[12]}),$
i.e. $\hat{O}_{12}\in \mathbb{H}(\mathcal{H}_{[12]})_{12}.$ In addition, let
the set of eigenvalues $V_{A}$ of $\hat{A}_{1}$ be $V_{A}=%
\{a_{1},a_{2},...,a_{d_{1}}\}$ and the set of eigenvalues $V_{B}$ of $\hat{B}%
_{2}$ be $V_{B}=\{b_{1},b_{2},...,b_{d_{2}}\}.$

The set of eigenvalues $V_{AB}$ of the product operator $\hat{O}_{12}$ is
given by the pairwise-product $V_{AB}=V_{A}V_{B}$ of the sets of eigenvalues
of the operators $\hat{A}_{1}$ and $\hat{B}_{2}.$ So, $V_{AB}$ is
\begin{eqnarray}
V_{AB} &=&\{a_{1},a_{2},...,a_{d_{1}}\}\{b_{1},b_{2},...,b_{d_{2}}\} \\
&=&%
\{a_{1}b_{1},a_{1}b_{2},...,a_{1}b_{d_{2}},a_{2}b_{1},a_{2}b_{2},...,a_{d_{1}}b_{d_{2}}\}
\notag \\
&=&\{a_{i}b_{j}:i=1,...,d_{1}\text{ , }j=1,...,d_{2}\}.  \notag
\end{eqnarray}
Then, the following conclusions hold:

\begin{theorem}
If either of $\hat{A}_{1}$ or $\hat{B}_{2}$ is weak, then the product\ $\hat{%
O}_{12}$ has a degenerate set of eigenvalues, and is hence also weak.
\end{theorem}

\begin{proof}
Without loss of generality, let $\hat{A}_{1}$ be weak. Then, at least one
member $a_{x}$ of $V_{A}$ is zero, or else two members $a_{x}$ and $a_{y}$
are equal.

If $a_{x}$ is zero, then the $d_{2}$ members of $V_{AB}$ of the form $%
a_{x}b_{j}$ for $j=1,...,d_{2}$ are also zero. Hence, $\hat{O}_{12}$ has $%
d_{2}$ degenerate eigenvalues, and is consequently a weak operator.

Alternatively, if $a_{x}=a_{y}$ then $a_{x}b_{j}=a_{y}b_{j}$ for all $%
j=1,...,d_{2},$ which means that $V_{AB}$ contains $d_{2}$ sets of
degenerate `pairs'. Hence, $\hat{O}_{12}$ is weak.\bigskip
\end{proof}

The above theorem is logically equivalent to the statement: \textit{only if }%
$\hat{A}_{1}$\textit{\ and }$\hat{B}_{2}$\textit{\ are both strong might the
operator }$\hat{O}_{12}=\hat{A}_{1}\otimes \hat{B}_{2}$\textit{\ be strong.}
Alternatively, \textit{if} $\hat{O}_{12}=\hat{A}_{1}\otimes \hat{B}_{2}$%
\textit{\ is strong, then }$\hat{A}_{1}$\textit{\ and }$\hat{B}_{2}$\textit{%
\ must both be strong.}

However:

\begin{theorem}
If $\hat{A}_{1}$ and $\hat{B}_{2}$ are both strong, then the product\ $\hat{O%
}_{12}$ may be either weak or strong, depending on the actual eigenvalues
of\ $\hat{A}_{1}$ and $\hat{B}_{2}.$
\end{theorem}

\begin{proof}
The proof is obvious. Whether or not the members of
\begin{equation}
V_{AB}=%
\{a_{1}b_{1},a_{1}b_{2},...,a_{1}b_{d_{2}},a_{2}b_{1},a_{2}b_{2},...,a_{d_{1}}b_{d_{2}}\}
\end{equation}
are degenerate (noting that none can be zero if\ $\hat{A}_{1}$ and $\hat{B}%
_{2}$ are strong) clearly depends on the specific values of each of $%
a_{1},a_{2},...,a_{d_{1}}$ and $b_{1},b_{2},...,b_{d_{2}}.$\bigskip
\end{proof}

For example, let $d_{1}=d_{2}=2$ in order to consider a Hilbert space $%
\mathcal{H}_{[12]}$ factorisable into two qubit subregisters, $\mathcal{H}%
_{[12]}=\mathcal{H}_{1}\otimes \mathcal{H}_{2},$ and the skeleton set of
operators $\mathcal{S}_{12}=\{\hat{\sigma}_{1}^{\mu _{1}}\otimes \hat{\sigma}%
_{2}^{\mu _{2}}:\mu _{1},\mu _{2}=0,1,2,3\}$ where $\hat{\sigma}_{1}^{\mu
_{1}}$ and $\hat{\sigma}_{2}^{\mu _{2}}$ are analogous to the Pauli
operators. Then:

\begin{enumerate}
\item  Consider also an operator $\hat{o}$ defined as $\hat{o}\equiv \hat{%
\sigma}_{1}^{1}\otimes \hat{\sigma}_{2}^{2}.$ The skeleton operator $\hat{%
\sigma}_{1}^{1}$ is a strong Hermitian operator, with eigenvalues $+1$ and $%
-1;$ similarly, the skeleton operator $\hat{\sigma}_{2}^{2}$ is also a
strong Hermitian operator, and also has eigenvalues $+1$ and $-1.$ Thus, the
four eigenvalues of the product operator $\hat{o}$ are the products: $%
(1)\times (1)=1,$ $(1)\times (-1)=-1,$ $(-1)\times (1)=-1$ and $(-1)\times
(-1)=1.$ So, $\hat{o}$ clearly has degenerate eigenvalues, and is hence a
weak operator that is the product of strong operators.

\item  Consider instead the Hermitian operator $\hat{F}$ defined as $\hat{F}%
\equiv (3\hat{\sigma}_{1}^{0}-\hat{\sigma}_{1}^{3})\otimes \hat{\sigma}%
_{2}^{1}$ in (\ref{Ch3facOp}). The strong Hermitian operator $(3\hat{\sigma}%
_{1}^{0}-\hat{\sigma}_{1}^{3})$ has eigenvalues of $2$ and $4,$ whereas the
strong Hermitian operator $\hat{\sigma}_{2}^{1}$ has eigenvalues of $+1$ and
$-1,$ so the four eigenvalues of the product operator $\hat{F}$ are: $%
(2)\times (1)=2,$ $(2)\times (-1)=-2,$ $(4)\times (1)=4$ and $(4)\times
(-1)=-4.$ Clearly, $\hat{F}$ has non-degenerate and non-zero eigenvalues,
and is hence a strong operator that is the product of strong operators.
\end{enumerate}

Note that the result of `1.' generalises to higher dimensional cases: every
element of the skeleton set (\ref{Ch3Skel}) associated with an $n$-qubit
register is weak for $n>1.$\ The members of $n$-qubit skeleton sets are
Hermitian, but have degenerate eigenvalues.\bigskip

For clarity, the conclusions of the above two theorems have been summed up
below. If $W$ and $S$ denote weak and strong operators respectively, the
following truth table is generated where the first row denotes the `status'
of $\hat{A}_{1},$ the first column denotes the status of $\hat{B}_{2}$
operator, and the remaining values denote the status of the resulting
product operator $\hat{O}_{12}=\hat{A}_{1}\otimes \hat{B}_{2}$:

\begin{equation}
\begin{tabular}{|l|l|l|}
\hline
$-$ & $\mathbf{\hat{A}}_{1}$ \textbf{is} $\mathbf{W}$ & $\mathbf{\hat{A}}%
_{1} $ \textbf{is} $\mathbf{S}$ \\ \hline
$\mathbf{\hat{B}}_{2}$ \textbf{is} $\mathbf{W}$ & $W$ & $W$ \\ \hline
$\mathbf{\hat{B}}_{2}$\textbf{\ is }$\mathbf{S}$ & $W$ & $S$ or $W$ ? \\
\hline
\end{tabular}
\tag*{Table 5.1}
\end{equation}

The results of the previous two theorems extend to operators that are the
products of more than two factors. The generalisation of the first theorem
implies that if an operator $\hat{O}_{1...M}$ is a product of $M$ factor
operators $\hat{O}_{1...M}=\hat{o}_{1}\otimes \hat{o}_{2}\otimes ...\hat{o}%
_{M},$ then every factor $\hat{o}_{i}$ must be strong if $\hat{O}_{1...M}$
is strong. This follows because operators of the type $\hat{O}_{1...M}$ can
always arbitrarily be rewritten as a product of just two factors: the factor
representing a particular $\hat{o}_{i}$ and the factor containing every
other operator $\hat{o}_{j}$ for $1\leq j\leq M$ and $j\neq i.$ So, if any
of the $\hat{o}_{i}$ are weak, the pair-wise product of the eigenvalues of
these two factors contains either degeneracy or zeroes, and hence $\hat{O}%
_{1...M}$ must also be weak.

Likewise, the extension of the second of the above theorems follows
naturally, since the spectrum of eigenvalues of an operator will always
depend on the set of the products of the eigenvalues of its factors.\bigskip

Attention is now turned to the eigenstates of the operators themselves:

\begin{theorem}
All the eigenstates of a strong, factorisable operator are separable.
\end{theorem}

\begin{proof}
Without loss of generality, consider a strong operator $\hat{O}_{12}$
factorisable into two factor operators, $\hat{O}_{12}=\hat{A}_{1}\otimes
\hat{B}_{2}.$ From the earlier theorem, the factors $\hat{A}_{1}$ and $\hat{B%
}_{2}$ must also both be strong operators.

As before, let the eigenvalues of $\hat{A}_{1}$ be $%
\{a_{1},a_{2},...,a_{d_{1}}\}.$ Each eigenvalue $a_{i}$ corresponds to a
particular normalised eigenvector $|a_{i}\rangle _{1},$ such that the
overall set of eigenvectors for\ $i=1,...,d_{1}$ forms an orthogonal basis
set $\frak{B}_{1}$ of states spanning the $d_{1}$ dimensional factor Hilbert
space $\mathcal{H}_{1},$ i.e.\ $\frak{B}_{1}=\{|a_{i}\rangle
_{1}:i=1,...,d_{1}\}.$

Similarly, if the eigenvalues of $\hat{B}_{2}$ are $%
\{b_{1},b_{2},...,b_{d_{2}}\},$ then each eigenvalue $b_{j}$ corresponds to
a particular normalised eigenvector $|b_{j}\rangle _{2},$ and this set of
eigenvectors for\ $j=1,...,d_{2}$ forms an orthogonal basis set $\frak{B}%
_{2} $ of states spanning the $d_{2}$ dimensional factor Hilbert space $%
\mathcal{H}_{2},$ i.e.\ $\frak{B}_{2}=\{|b_{j}\rangle _{2}:j=1,...,d_{2}\}.$

Consider now the pairwise (tensor) product of $\frak{B}_{1}$ and $\frak{B}%
_{2}$ defined as the set $\{|a_{i}\rangle _{1}\otimes |b_{j}\rangle
_{2}:i=1,...,d_{1}$ $j=1,...,d_{2}\}.$ Clearly, this set has $d_{1}d_{2}$
members.

Now consider one of the members $\psi $ of this set, $\psi =|a_{x}\rangle
_{1}\otimes |b_{y}\rangle _{2}.$ Evidently, $\psi $ is separable, and is a
member of the partition $\psi \in \mathcal{H}_{12}=(\mathcal{H}_{1}\bullet
\mathcal{H}_{2})\subset \mathcal{H}_{[12]}.$ Moreover, $\psi $ is an
eigenstate of $\hat{O}_{12}$ because:
\begin{eqnarray}
\hat{O}_{12}\psi &=&\hat{O}_{12}(|a_{x}\rangle _{1}\otimes |b_{y}\rangle
_{2}) \\
&=&[\hat{A}_{1}\otimes \hat{B}_{2}](|a_{x}\rangle _{1}\otimes |b_{y}\rangle
_{2})  \notag \\
&=&[\hat{A}_{1}|a_{x}\rangle _{1}]\otimes \lbrack \hat{B}_{2}|b_{y}\rangle
_{2}]  \notag \\
&=&a_{x}|a_{x}\rangle _{1}\otimes b_{y}|b_{y}\rangle
_{2}=o_{xy}|a_{x}\rangle _{1}\otimes |b_{y}\rangle _{2}  \notag
\end{eqnarray}
where $o_{xy}=a_{x}b_{y}\in \mathbb{R}^{+}.$

Similarly, every other member of the set $\{|a_{i}\rangle _{1}\otimes
|b_{j}\rangle _{2}:i=1,...,d_{1}$ $j=1,...,d_{2}\}$ is an eigenstate of $%
\hat{O}_{12},$ and is also a member of the separation $\mathcal{H}_{12}.$

However, because $\hat{O}_{12}$ is a strong operator acting on states in a $%
d_{1}d_{2}$ dimensional Hilbert space, it has precisely $d_{1}d_{2}$
independent eigenstates. \ Since there are $d_{1}d_{2}$ independent
eigenstates of $\hat{O}_{12}$ in the set $\{|a_{i}\rangle _{1}\otimes
|b_{j}\rangle _{2}:i=1,...,d_{1}$ $j=1,...,d_{2}\},$ this set must be an
exhaustive, orthonormal basis $\frak{B}_{12}$ for $\hat{O}_{12}.$

Hence, every eigenstate of the strong, factorisable operator $\hat{O}_{12}$
is separable.

The proof extends to strong, separable operators of higher degrees of
factorisation in the obvious way.\bigskip
\end{proof}

In the context of the proposed paradigm that only strong (Hermitian)
operators are used in the development of the Universe's state, the above
theorem can be rephrased as: separable tests only have separable outcomes.

An important consequence of this is that entangled states cannot be the
outcome of separable operators. Paraphrasing: \textit{entangled states can
only be the outcome of entangled operators.}The converse, however, is not
true: entangled operators can have entangled eigenstates, but they can also
have separable eigenstates.

So, overall the `Golden Rule' is that:

\begin{center}
\textit{No operator can have more factors than any of its eigenstates,}

\textit{\ but an eigenstate can have more factors than its operator.\bigskip
}
\end{center}

As illustrations of these ideas, consider again a two qubit toy-universe,
represented by a state in a four dimensional Hilbert space $\mathcal{H}%
_{[12]}^{(4)}=\mathcal{H}_{1}^{(2)}\otimes \mathcal{H}_{2}^{(2)}$ spanned by
the vector basis $\mathcal{B}_{12}=\{|i\rangle _{1}\otimes |j\rangle
_{2}\equiv |ij\rangle :i,j=0,1\},$ with the usual skeleton set $\mathcal{S}%
_{12}=\{\hat{\sigma}_{1}^{\mu _{1}}\otimes \hat{\sigma}_{2}^{\mu _{2}}:\mu
_{1},\mu _{2}=0,1,2,3\}$ for the operators in $\mathbb{H}(\mathcal{H}%
_{[12]}^{(4)}).$ The operator
\begin{equation}
\hat{F}=(3\hat{\sigma}_{1}^{0}-\hat{\sigma}_{1}^{3})\otimes \hat{\sigma}%
_{2}^{1}
\end{equation}
is factorisable, and has four separable eigenstates: $\frac{1}{\sqrt{2}}%
(|01\rangle +|00\rangle ),$ $\frac{1}{\sqrt{2}}(|01\rangle -|00\rangle ),$ $%
\frac{1}{\sqrt{2}}(|11\rangle +|10\rangle ),$ and $\frac{1}{\sqrt{2}}%
(|11\rangle -|10\rangle ).$ Conversely, the operator
\begin{equation}
\hat{E}=\frac{1}{2}(3\hat{\sigma}_{1}^{1}\otimes \hat{\sigma}_{2}^{1}+\hat{%
\sigma}_{1}^{2}\otimes \hat{\sigma}_{2}^{2})
\end{equation}
is entangled and has four entangled eigenstates: $\frac{1}{\sqrt{2}}\left(
|11\rangle +|00\rangle \right) ,$ $\frac{1}{\sqrt{2}}(|11\rangle -|00\rangle
),$ $\frac{1}{\sqrt{2}}(|10\rangle +|01\rangle )$ and $\frac{1}{\sqrt{2}}%
(|10\rangle -|01\rangle ).$ However, the operator
\begin{equation}
\hat{M}=\hat{\sigma}_{1}^{1}\otimes \hat{\sigma}_{2}^{1}+\hat{\sigma}%
_{1}^{2}\otimes \hat{\sigma}_{2}^{2}+(\hat{\sigma}_{1}^{0}\otimes \hat{\sigma%
}_{2}^{3})/2+(\hat{\sigma}_{1}^{3}\otimes \hat{\sigma}_{2}^{0})/2
\end{equation}
is also entangled but has a mixture of separable $|00\rangle ,$ $|11\rangle $
and entangled $\frac{1}{\sqrt{2}}\left( |01\rangle +|10\rangle \right) ,$ $%
\frac{1}{\sqrt{2}}(|01\rangle -|10\rangle )$ eigenstates.

It is interesting to note here that although a set of eigenstates may look
relatively `simple', the operator they come from may still be a non-trivial
combination of skeleton operators. This observation is perhaps a reminder of
how much more complicated the set of skeleton operators is compared to the
set of states; recall that a Hilbert space $\mathcal{H}^{(D)}$ of dimension $%
D$ is spanned by a basis set of $D$ independent vectors, whereas the
corresponding space of Hermitian operators $\mathbb{H}(\mathcal{H}^{(D)})$
is parameterised by a skeleton set containing $D^{2}$ members. In the
present case, although operators in two qubit Hilbert spaces $\mathcal{H}%
_{[12]}^{(4)}$ only have four eigenstates, they will nevertheless comprise
of linear sums of up to sixteen basis operators. It is perhaps not
surprising, then, that the structure of the operators is considerably
`richer' than that of the corresponding sets of eigenstates.\bigskip

Just as the factors of the states can be represented pictorially in ways
analogous to the Hasse diagrams of classical causal set theory, the
operators of which they are outcomes can too. In the chosen convention,
emergent time is taken to run upwards again, and every factor of an operator
will be denoted by a square; so, an operator associated with $k$ factors may
be represented by $k$ squares in a row. Arrows pointing into a square come
from the group of factor states that are tested by the factor operator it
represents, whereas arrows leaving a square point to the set of outcome
factor states of this factor operator.

As an example, a graphical representation of the theorem that `separable
tests only have separable outcomes' is given in Figure 5.4. Figure 5.4a
shows a separable operator $\hat{O}_{12}$ producing an entangled outcome $%
\psi ^{12},$ which is a forbidden process. In Figure 5.4b, however, an
entangled operator $\hat{O}^{12}$ is producing a separable outcome $\psi
_{12},$ whilst in Figure 5.4c an entangled operator $\hat{O}^{12}$ is
producing an entangled outcome $\psi ^{12},$ and these processes are allowed.

%\FRAME{ftbpFU}{350.75pt}{130.3125pt}{0pt}{\Qcb{Pictorial representation of
%the relationship between operators and factor states. The process described
%in Figure a) is forbidden, whereas those of Figures b) and c) are allowed.}}{%
%}{Figure 5.4}{\special{language "Scientific Word";type
%"GRAPHIC";maintain-aspect-ratio TRUE;display "USEDEF";valid_file "T";width
%350.75pt;height 130.3125pt;depth 0pt;original-width
%346.875pt;original-height 127.625pt;cropleft "0";croptop "1";cropright
%"1";cropbottom "0";tempfilename 'I1KOB003.wmf';tempfile-properties "XPR";}}

\begin{figure}[th]
\begin{center}
\includegraphics[height=2in]{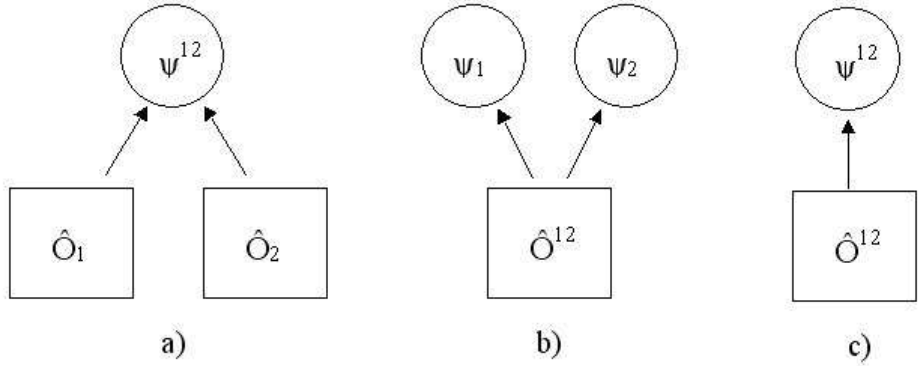}
\caption{Pictorial representation of
the relationship between operators and factor states. The process described
in Figure a) is forbidden, whereas those of Figures b) and c) are allowed.}\label{Figure 5.4}
\end{center}
\end{figure}

The ideas and theorems of this section place important mathematical
constraints on the operators used in the development of the Universe. In a
fully quantum Universe represented by a unique state $\Psi _{n},$ which is
an eigenstate of an operator $\hat{\Sigma}_{n}$ in an enormous and
fundamentally factorised Hilbert space $\mathcal{H}_{[1...N]},$ if $\Psi
_{n} $ is a separable product of factor sub-states, some of which may
themselves be entangled relative to $\mathcal{H}_{[1...N]},$ then the
conclusion must be that the individual factors of the operator $\hat{\Sigma}%
_{n}$ associated with the entangled factors of $\Psi _{n}$ cannot themselves
be factorised any further within the factor Hilbert spaces containing these
entangled factor states. This result will lead to important consequences for
the generation of quantum causal sets, as discussed in the next section.

\bigskip

\subsection{Einstein Locality and Quantum Causal Sets}

\bigskip

Section 5.4 indicated where relationships analogous to those of classical
causal set theory may arise from a consideration of the changes in
separability of the quantum state of the Universe. So, given that Section
5.5 showed that the operators responsible for developing the Universe's
state may also exhibit properties of entanglement and separability, it might
therefore be expected that these operators will also generate causal set
structures. This implies the existence of two different types of causal set
in a fully quantum picture of the Universe, and these may in turn lead to
different results in the large scale limit when considering aspects of
emergent physical spacetime. This discussion will be the focus of the
present section.\bigskip

Before elaborating on this point, however, it should be stressed that any
parallels between tests and states should not necessarily be assumed too
automatically. After all, vectors and operators are mathematically very
different. For example, the states are members of $D$-dimensional, complex
Hilbert spaces $\mathcal{H}^{(D)},$ whereas the corresponding tests that act
upon them exist in $D^{2}$-dimensional, real vector spaces $\mathbb{H}(%
\mathcal{H}^{(D)}).$ Similarly, a bra state $|\Psi \rangle $ may be
represented by a column vector with $D$ elements, whilst the operators may
be represented by self-adjoint $D\times D$ square matrices.

A further difference is evident from an examination of the product structure
of the vector spaces. For two states $\Psi ,\Phi \in \mathcal{H}^{(D)},$ it
is possible to define an inner product of the form $\langle \Phi |\Psi
\rangle \in \mathbb{C},$ which is interpreted in the proposed paradigm as
the probability amplitude for the Universe to develop from the state $\Psi $
to the state $\Phi .$ Conversely, no such inner product is defined for two
operators $\hat{A},\hat{B}\in \mathbb{H}(\mathcal{H}^{(D)}),$ and there is
hence no analogous physical interpretation. However, it is possible to
define a third operator $\hat{C}\in \mathbb{H}(\mathcal{H}^{(D)})$ as the
product $\hat{C}\equiv \hat{A}\hat{B},$ even though this type of
transformation has no equivalent in the space of states. Indeed for vectors,
the product $\Psi \Phi $ is meaningless.\bigskip

There are also more obvious differences between vectors and operators
regarding what they physically represent in quantum theory. The vectors
represent the states of actual quantum systems, and so every phenomenon that
is associated with wavefunctions in the laboratory has also to be applicable
for the vectors. Thus, the vectors may be expected to exist in complex
linear superpositions, and may appear to exhibit non-local and non-classical
correlations that are at odds with emergent views of relativity and general
covariance.

On the other hand, the operators are assigned to represent the observables
of quantum theory, and these tend to have classical analogues that obey
Einstein locality and causality: tests separated by spacelike distances do
not affect one another. In fact, the canonical quantisation procedure
successfully employed in conventional quantum mechanics is a process by
which classical variables are directly replaced with their quantum operator
counterparts. It should not, then, perhaps be too surprising that the
resultant quantum operators\ therefore appear to obey classical laws of
dynamics. An example here is that operators associated with emergent
observables separated by spacelike distances tend to commute.

This point is very much the stance of Peres: quantum mechanics as such does
not strictly have to satisfy covariance in every respect, but its physical
observables do \cite{Peres1}.

A physical illustration of this type of argument is evident in quantum field
theory. Local observables such a energy and momentum density operators
satisfy microscopic causality, because their commutators vanish at spacelike
intervals, but the local quantum fields out of which they are constructed
need not commute at such separations \cite{Bjorken}. In other words,
Einstein locality must always hold for the physical observables, but it need
not for the quantum states themselves. Of course, this may in turn be
because the states are never directly `experienced' \textit{per se}, whereas
it was only ever by experiencing physical observables that the (emergent)
theory of relativity was discovered.

The differences highlighted above between states and operators should
manifest themselves in the type of causal sets they produce. Specifically,
whatever type of structure arises from the state's causal set may be
expected to exhibit characteristics of non-locality, whereas whatever type
of structure arises from the operator's causal set might conversely be
expected to obey Einstein locality. Indeed, if this were not the case it
would be necessary to explain how these features of empirical physics
otherwise emerge in the observed Universe if they are not present on the
underlying pregeometric level.

\bigskip

As discussed a number of times so far in this work, the dynamics proposed
for the Universe is that its state $\Psi _{n}\in \mathcal{H}_{[1..N]}^{(D)}$
is developed by collapsing into one of the $D$ orthonormal eigenvectors $%
\Phi ^{i},$ $i=1,...,D,$ of an Hermitian operator $\hat{\Sigma}_{n+1}\in
\mathbb{H}(\mathcal{H}_{[1..N]}^{(D)}).$ Further, the conditional
probability $P(\Psi _{n+1}=\Phi ^{i}|\Psi _{n},\hat{\Sigma}_{n+1})$ that the
next state $\Psi _{n+1}$ will be the $i^{th}$ eigenvector of $\hat{\Sigma}%
_{n+1},$ given that the Universe is initially in a state $\Psi _{n}$ and is
tested with an operator $\hat{\Sigma}_{n+1},$ is determined by the usual
rule of Born:
\begin{equation}
P(\Psi _{n+1}=\Phi ^{i}|\Psi _{n},\hat{\Sigma}_{n+1})=\left| \langle \Phi
^{i}|\Psi _{n}\rangle \right| ^{2}.  \label{Ch3Prob1}
\end{equation}

The above probability of the Universe collapsing from a state $\Psi _{n}$ to
one of the eigenstates $\Phi ^{i}$ of an operator\ $\hat{\Sigma}_{n+1}$ may
be associated with the concept of entropy. Recall that the Shannon entropy, $%
S,$ attached to a particular probability distribution $%
\{p_{1},p_{2},...,p_{M}\}$ is given by
\begin{equation}
S\equiv -\sum\limits_{r=1}^{M}p_{r}\ln p_{r},
\end{equation}
and is a reflection of a physicist's ignorance of the result prior to a test
that has $M$ outcomes of weighted probability \cite{Peres}.

So, the Shannon entropy associated with the Universe jumping into one out of
a set of $D$ possible outcomes\ $\Phi ^{i}$ of a given test $\hat{\Sigma}%
_{n+1},$ each with probability $P^{i}=P(\Psi _{n+1}=\Phi ^{i}|\Psi _{n},\hat{%
\Sigma}_{n+1}),$ is given by
\begin{eqnarray}
S &=&-\sum\limits_{i=1}^{D}P^{i}\ln P^{i}  \label{Ch3Entropy} \\
&=&-\sum\limits_{i=1}^{D}\left| \langle \Phi ^{i}|\Psi _{n}\rangle \right|
^{2}\ln \left| \langle \Phi ^{i}|\Psi _{n}\rangle \right| ^{2}.  \notag
\end{eqnarray}

Note that this is a classical entropy result, as expected because state
reduction processes do not permit quantum interference terms.

Equation (\ref{Ch3Prob1}) provides the correct probability for obtaining a
particular next state $\Psi _{n+1}=\Phi ^{i}$ as the result of a particular
test $\hat{\Sigma}_{n+1},$ and equation (\ref{Ch3Entropy}) looks at the
corresponding entropy associated with the set of potential outcomes of this
measurement. What these relations do not do, however, is provide an answer
as to why the test $\hat{\Sigma}_{n+1}$ was ever used in the first place;
they say nothing about the Universe's actual selection of this particular
operator. This is perhaps unsettling, because without specifying which
operator $\hat{\Sigma}_{n+1}$ is chosen to test the Universe, the
probability amplitude $\langle \Psi _{n+1}=\Phi ^{i}|\Psi _{n}\rangle $ is
meaningless. Without specifying $\hat{\Sigma}_{n+1}$ it is quite possible
that a different operator $\hat{\Sigma}_{n+1}^{\prime }$ could be used, and
this alternative test may not even have $\Phi ^{i}$ as an eigenstate. In
this case it would then be pointless to ask about the relative probability
of the next state $\Psi _{n+1}$ being $\Phi ^{i}.$

Although the issue is discussed to some extent in Chapter 8, at present
there is no known mechanism for understanding how or why the Universe
selects a particular operator $\hat{\Sigma}_{n+1}$ to test itself, a point
that is summed up by the statement: `\textit{Only some of the rules [of the
Universe] are currently understood; we can calculate answers to quantum
questions, but we do not know why those questions have been asked in the
first place'} \cite{EJ}. It is asserted, then, that any measure of the
entropy associated with the Universe developing through a series of states
should take this additional ignorance into account.

To this end, recall the conjecture of Chapter 3 that the Universe may be
completely parameterised by a unique stage $\Omega _{n}$ defined as $\Omega
_{n}\equiv \Omega (\Psi _{n},$ $I_{n},$ $R_{n}).$ Moreover, recall that the
current `information content' $I_{n}$ was taken to contain the set of
possible operators\ $\{\hat{O}_{n}^{b}:b=1,...,B\}$ that might each provide
a basis for the next potential state $\Psi _{n+1}$ of the Universe in the
next stage $\Omega _{n+1}.$ Then, $\hat{\Sigma}_{n+1}$ will be one of $B$
possibilities, which may be labelled $\hat{\Sigma}_{n+1}^{b}(\equiv \hat{O}%
_{n}^{b}).$ If it may be assumed that there exists a certain probability $%
P^{b}=P(\hat{O}_{n}^{b}|\Omega _{n})$ that a particular operator $\hat{O}%
_{n}^{b}$ is chosen by the Universe at time $n+1$ to be $\hat{\Sigma}_{n+1},$
then $\tsum\nolimits_{b=1}^{B}P^{b}=1,$ noting that possibly $B=1$ if the
operators are selected deterministically.

Thus, if the Universe is initially in the stage $\Omega _{n},$ the
probability $P^{(b,i)}$ that it will be tested by an operator $\hat{\Sigma}%
_{n+1}=\hat{\Sigma}_{n+1}^{b}=\hat{O}_{n}^{b}$ and will then subsequently
jump from the state $\Psi _{n}$ to a particular state $\Psi _{n+1}=\Phi
^{b,i},$ which is referred to as the $i^{th}$ eigenvector of the operator $%
\hat{\Sigma}_{n+1}^{b},$ is given by
\begin{equation}
P^{(b,i)}=P^{b}P^{i}=P(\hat{\Sigma}_{n+1}^{b}|\Omega _{n})\left| \langle
\Phi ^{b,i}|\Psi _{n}\rangle \right| ^{2}
\end{equation}
where $\tsum\nolimits_{i=1}^{D}\left| \langle \Phi ^{b,i}|\Psi _{n}\rangle
\right| ^{2}=1$ as expected.

Further, the Shannon entropy that may be associated with this jump is given
by
\begin{eqnarray}
S^{(1)} &=&-\sum\limits_{b=1}^{B}\sum\limits_{i=1}^{D}P^{(b,i)}\ln P^{(b,i)}
\label{Ch3EntropyS1} \\
&=&-\sum\limits_{b=1}^{B}\sum\limits_{i=1}^{D}P(\hat{\Sigma}%
_{n+1}^{b}|\Omega _{n})\left| \langle \Phi _{n+1}^{b,i}|\Psi _{n}\rangle
\right| ^{2}  \notag \\
&&\times \left\{ \ln P(\hat{\Sigma}_{n+1}^{b}|\Omega _{n})+\ln \left|
\langle \Phi _{n+1}^{b,i}|\Psi _{n}\rangle \right| ^{2}\right\}  \notag \\
&=&-\sum\limits_{b=1}^{B}P(\hat{\Sigma}_{n+1}^{b}|\Omega _{n})\ln P(\hat{%
\Sigma}_{n+1}^{b}|\Omega _{n})  \notag \\
&&-\sum\limits_{b=1}^{B}P(\hat{\Sigma}_{n+1}^{b}|\Omega
_{n})\sum\limits_{i=1}^{D}\left| \langle \Phi ^{b,i}|\Psi _{n}\rangle
\right| ^{2}\ln \left| \langle \Phi ^{b,i}|\Psi _{n}\rangle \right| ^{2}
\notag \\
&=&S^{t}+\sum\limits_{b=1}^{B}P(\hat{\Sigma}_{n+1}^{b}|\Omega _{n})S^{b}
\notag
\end{eqnarray}
where $S^{t}\equiv -\sum\limits_{b=1}^{B}P(\hat{\Sigma}_{n+1}^{b}|\Omega
_{n})\ln P(\hat{\Sigma}_{n+1}^{b}|\Omega _{n})$ is the entropy associated
with the selection of the test, $S\equiv -\sum\limits_{i=1}^{D}\left|
\langle \Phi ^{b,i}|\Psi _{n}\rangle \right| ^{2}\ln \left| \langle \Phi
^{b,i}|\Psi _{n}\rangle \right| ^{2}$ is the entropy associated with the
collapse from the state $\Psi _{n}$ to one of the set of possible
eigenvectors of this test, and the superscript $(1)$ is used to denote that $%
S^{(1)}$ is defined over one jump. Thus, the entropy (\ref{Ch3EntropyS1})
reflects the ignorance associated with how the Universe might develop from
the current stage $\Omega _{n}$ to a potential stage $\Omega _{n+1}.$

It is possible to extend these ideas to gain an appreciation of the entropy
associated with the Universe prior to it developing over a series of jumps.
Define $P_{(b_{n},i_{n})}^{b_{n+1}}=P(\hat{\Sigma}_{n+1}^{b_{n+1}}|\Omega
_{n}^{b_{n},i_{n}})$ as the probability that an operator $\hat{\Sigma}%
_{n+1}^{b_{n+1}}$ will be chosen from a set $\{\hat{\Sigma}%
_{n+1}^{b_{n+1}}:b_{n+1}=1,...,B_{n+1}\}$ of $B_{n+1}$ possibilities, given
that the Universe is currently in the stage $\Omega
_{n}^{b_{n},i_{n}}=\{\Psi
_{n}^{b_{n},i_{n}},I_{n}^{b_{n},i_{n}},R_{n}^{b_{n},i_{n}}\}$ where the
superscript $(b_{b},i_{n})$ implies, for example, that the state\ $\Psi
_{n}^{b_{n},i_{n}}$ is one of the $D$ outcomes $\Psi _{n}^{b_{n},i_{n}}=\Phi
^{b_{n},i_{n}},$ for $i=1,...,D,$ of one of $B_{n}$ possible tests $\hat{%
\Sigma}_{n}^{b_{n}}$ contained in the previous stage $\Omega
_{n-1}^{b_{n-1}}.$

Similarly, the variable
\begin{eqnarray}
P_{(b_{n},i_{n})}^{i_{n+1}} &\equiv &P(\Psi _{n+1}^{b_{n+1},i_{n+1}}=\Phi
^{b_{n+1},i_{n+1}}|\Psi _{n}^{b_{n},i_{n}},\hat{\Sigma}_{n+1}^{b_{n+1}}) \\
&=&\left| \langle \Phi ^{b_{n+1},i_{n+1}}|\Psi _{n}\rangle \right| ^{2}
\notag
\end{eqnarray}
is defined as the probability that the outcome of this chosen test $\hat{%
\Sigma}_{n+1}^{b_{n+1}}$ is $\Psi _{n+1}^{b_{n+1},i_{n+1}}=\Phi
^{b_{n+1},i_{n+1}}.$

Overall then,
\begin{equation}
P_{(b_{n},i_{n})}^{(b_{n+1},i_{n+1})}=P(\hat{\Sigma}_{n+1}^{b_{n+1}}|\Omega
_{n}^{b_{n},i_{n}})P(\Psi _{n+1}^{b_{n+1},i_{n+1}}=\Phi
^{b_{n+1},i_{n+1}}|\Psi _{n}^{b_{n},i_{n}},\hat{\Sigma}_{n+1}^{b_{n+1}})
\end{equation}
is defined as the probability that, given an initial stage $\Omega
_{n}^{b_{n},i_{n}},$ the next test will be $\hat{\Sigma}_{n+1}^{b_{n+1}}$
and the resulting next state will be the eigenvector $\Psi
_{n+1}^{b_{n+1},i_{n+1}}=\Phi ^{b_{n+1},i_{n+1}}$ of $\hat{\Sigma}%
_{n+1}^{b_{n+1}}.$

Since all the probabilities are classical due to the nature of the state
reduction process, it is possible to define chains of jumps in terms of
products of probabilities. Thus,
\begin{equation}
P_{(b_{n},i_{n}),(b_{n+1},i_{n+1}),...,(b_{n+m-1},i_{n+n-1})}^{(b_{n+1},i_{n+1})(b_{n+2},i_{n+2}),...,(b_{n+m},i_{n+m})}=P_{(b_{n},i_{n})}^{(b_{n+1},i_{n+1})}P_{(b_{n+1},i_{n+1})}^{(b_{n+2},i_{n+2})}...P_{(b_{n+m-1},i_{n+m-1})}^{(b_{n+m},i_{n+m})}
\end{equation}
is defined as the probability that the Universe will jump from the state $%
\Psi _{n}^{b_{n},i_{n}}=\Phi ^{b_{n},i_{n}}$ to the state $\Psi
_{n+1}^{b_{n+1},i_{n+1}}=\Phi ^{b_{n+1},i_{n+1}}$ via the test $\hat{\Sigma}%
_{n+1}^{b_{n+1}},$ and that this new state will jump to the state $\Psi
_{n+2}^{b_{n+2},i_{n+2}}=\Phi ^{b_{n+2},i_{n+2}}$ via the test $\hat{\Sigma}%
_{n+2}^{b_{n+2}},$ and so on until the state $\Psi
_{n+m-1}^{b_{n+m-1},i_{n+m-1}}=\Phi ^{b_{n+m-1},i_{n+m-1}}$ finally jumps to
the state $\Psi _{n+m}^{b_{n+m},i_{n+m}}=\Phi ^{b_{n+m},i_{n+m}}$ via the
test $\hat{\Sigma}_{n+m}^{b_{n+m}}.$

Using this notation, the entropy $S^{(1)}$ given in (\ref{Ch3EntropyS1}) may
be rewritten as $S^{(1)}=-\tsum\nolimits_{b_{n+1}=1}^{B_{n+1}}\tsum%
\nolimits_{i_{n+1}=1}^{D}P_{(b_{n},i_{n})}^{(b_{n+1},i_{n+1})}\ln
P_{(b_{n},i_{n})}^{(b_{n+1},i_{n+1})}.$ Similarly, the entropy $S^{(2)}$
over two jumps may be given by
\begin{eqnarray}
S^{(2)}
&=&-\tsum\nolimits_{b_{n+1}=1}^{B_{n+1}}\tsum\nolimits_{i_{n+1}=1}^{D}\tsum%
\nolimits_{b_{n+2}=1}^{B_{n+2}}\tsum\nolimits_{i_{n+2}=1}^{D} \\
&&\left[
P_{(b_{n},i_{n})}^{(b_{n+1},i_{n+1})}P_{(b_{n+1},i_{n+1})}^{(b_{n+2},i_{n+2})}\left\{ \ln (P_{(b_{n},i_{n})}^{(b_{n+1},i_{n+1})}P_{(b_{n+1},i_{n+1})}^{(b_{n+2},i_{n+2})})\right\} %
\right]  \notag
\end{eqnarray}
such that overall, the $m$ jump entropy $S^{(m)}$ may hence be defined as
\begin{eqnarray}
S^{(m)}
&=&-\tsum\nolimits_{b_{n+1}=1}^{B_{n+1}}\tsum\nolimits_{i_{n+1}=1}^{D}\tsum%
\nolimits_{b_{n+2}=1}^{B_{n+2}}\tsum\nolimits_{i_{n+2}=1}^{D}...\tsum%
\nolimits_{b_{n+m}=1}^{B_{n+m}}\tsum\nolimits_{i_{n+m}=1}^{D} \\
&&\left[
\begin{array}{c}
P_{(b_{n},i_{n})}^{(b_{n+1},i_{n+1})}P_{(b_{n+1},i_{n+1})}^{(b_{n+2},i_{n+2})}...P_{(b_{n+m-1},i_{n+m-1})}^{(b_{n+m},i_{n+m})}\times
\\
\left\{ \ln P_{(b_{n},i_{n})}^{(b_{n+1},i_{n+1})}+\ln
P_{(b_{n+1},i_{n+1})}^{(b_{n+2},i_{n+2})}+...+\ln
P_{(b_{n+m-1},i_{n+m-1})}^{(b_{n+m},i_{n+m})}\right\}
\end{array}
\right]  \notag \\
&=&-\tsum\nolimits_{b_{n+1}=1}^{B_{n+1}}\tsum\nolimits_{i_{n+1}=1}^{D}...%
\tsum\nolimits_{b_{n+m}=1}^{B_{n+m}}\tsum\nolimits_{i_{n+m}=1}^{D}  \notag \\
&&\left[
\begin{array}{c}
P(\hat{\Sigma}_{n+1}^{b_{n+1}}|\Omega _{n}^{b_{n},i_{n}})\left| \langle \Phi
^{b_{n+1},i_{n+1}}|\Psi ^{b_{n},i_{n}}\rangle \right| ^{2}\times ... \\
...\times P(\hat{\Sigma}_{n+m}^{b_{n+m}}|\Omega
_{n+m-1}^{b_{n+m-1},i_{n+m-1}})\left| \langle \Phi ^{b_{n+m},i_{n+m}}|\Psi
^{b_{n+m-1},i_{n+m-1}}\rangle \right| ^{2} \\
\times \left\{
\begin{array}{c}
\ln P(\hat{\Sigma}_{n+1}^{b_{n+1}}|\Omega _{n}^{b_{n},i_{n}}) \\
+\ln \left| \langle \Phi ^{b_{n+1},i_{n+1}}|\Psi ^{b_{n},i_{n}}\rangle
\right| ^{2}+... \\
...+\ln P(\hat{\Sigma}_{n+m}^{b_{n+m}}|\Omega _{n+m-1}^{b_{n+m-1},i_{n+m-1}})
\\
+\ln \left| \langle \Phi ^{b_{n+m},i_{n+m}}|\Psi
^{b_{n+m-1},i_{n+m-1}}\rangle \right| ^{2}
\end{array}
\right\}
\end{array}
\right] .  \notag
\end{eqnarray}

\bigskip

Since there is at present no way of knowing how large $B_{n}$ actually is,
or how its value changes with $n,$ the number of potential next states may
be literally gigantic. Moreover, the scope of possible `futures' for the
Universe will clearly increase rapidly over even a relatively small number
of jumps, especially when it is considered how large the dimension $D$ of
the Hilbert space is likely to be, and hence how large the set of orthogonal
eigenstates is for each operator.

It is therefore obvious that the number of possible causal sets that may be
produced over a chain of jumps is also gigantic. This point is analogous to
the Hasse diagram of Hasse diagrams presented in \cite{Rideout} that are
generated by examining the collection of possible classical causal sets that
can be grown by adding one new event to an existing set. In fact, over the $%
m $ jumps from $\Psi _{n}^{b_{n},i_{n}}$ to $\Psi _{n+m}^{b_{n+m},i_{n+m}}$
in the above case, there will in principle be a whole `tree' of $M$
different possible causets produced, where
\begin{equation}
M\leq D^{m}\times (B_{n+1}\times B_{n+2}\times B_{n+3}\times ...\times
B_{n+m}).
\end{equation}

Note that the inequality reflects the fact that there may be some degeneracy
in this set of $M$ members, because two operators\ $\hat{\Sigma}%
_{n+x}^{b_{n+x}}$ and $\hat{\Sigma}_{n+x}^{c_{n+x}}$ for $%
b_{n+x},c_{n+x}=1,...,B_{n+x}$ and $1\leq x\leq m$ may have $y$ eigenvectors
in common, $0\leq y<D.$

\bigskip

Although the probabilities $P_{(b_{n+x},i_{n+x})}^{i_{n+x+1}}$ for
particular state transitions from $\Psi _{n+x}$ to $\Phi
^{b_{n+x+1},i_{n+x+1}}$ given a specific operator $\hat{\Sigma}%
_{n+x+1}^{b_{n+x+1}}$ are evaluated by the Born rule $%
P_{(b_{n+x},i_{n+x})}^{i_{n+x+1}}=\left| \langle \Phi
^{b_{n+x+1},i_{n+x+1}}|\Psi _{n+x}\rangle \right| ^{2},$ as mentioned
earlier there is no similar rule known for specifying the probabilities $%
P_{(b_{n+x},i_{n+x})}^{b_{n+x+1}}=P(\hat{\Sigma}_{n+x+1}^{b_{n+x+1}}|\Omega
_{n+x}^{b_{n+x},i_{n+x}})$ of choosing this particular operator $\hat{\Sigma}%
_{n+x+1}^{b_{n+x+1}}$ from a set of $B_{n+x+1}$ possibilities.

Of course this selection could actually be deterministic, so there is in
fact no choice, and this would give rise to a semi-clockwork Universe in
which quantum state reduction provides the only randomness. In such a
Universe it would always be possible to predict, with certainty, in advance
which test the Universe will choose to test itself with $x$ stages into the
future, assuming that this deterministic rule is known.

Alternatively perhaps, in a Universe free of external observers the choice
of next test may depend somehow on the current state that the Universe is
in. As will be discussed in Chapter 8, the way in which such
self-referential Universes might be developed after $x$ jumps may not be
knowable until it has developed through the $x-1$ preceding stages. This is
possibly how (at least some of) the dynamics of the physical Universe works,
because human scientists, themselves just groups of factors of the state of
the Universe, do appear to be able to exert some sort of influence on how
this state they exist as part of actually gets tested, because they do seem
able to prod and probe those factors that represent their surroundings.

However, even if the physical Universe does develop according to a type of
self-referential mechanism, exactly how its next operator might be
controlled by the current state is still unknown.\bigskip

It is here that an appeal is made to empirical physics. Since it appears to
be the case that the Universe is highly classical, and hence highly
separable, whichever method is used by the Universe to select its next test
seems to be constrained to choose an operator that possesses a highly
separable set of eigenvectors. Since current thinking also indicates that
the Universe has, on average, changed very slowly over the last $10$ or so
billion years, however the Universe actually selects its next operator must
ensure that the test chosen has an outcome that it almost identical to the
present state. In addition, given that it seems an experimentally verified
fact that physical observables in the Universe are constrained by Einstein
locality, it can also be assumed that whatever mechanism the Universe uses
to select the next operator to test itself with, the physically observed
outcomes of this operator must also obey the principles of relativity.

Rephrasing this last point, since the operators are expected to correspond
to physical observables in the emergent limit, their results must eventually
correspond to the outcomes of their classical counterparts. Similarly, and
reversing this line of thinking, if physicists are able to quantise
particular classical variables to get the quantum operator equivalents, the
resulting quantum operators may still ultimately be expected to obey some of
the classical laws. For example, if classical variables are always forced to
obey Einstein locality, and if these variables can be directly quantised to
produce operators that yield accurate physical results, it may be fair to
assume that, in general, quantum operators in the Universe are also forced
to obey Einstein locality. So, their observed outcomes will not permit
features such as superluminal communication. In other words, if observed
physics is limited by Einstein locality, the operators representing these
observables may be too.

Thus, however the Universe selects its quantum operators, the choice made
will ultimately be expected to give the results familiar to classical
experiments. Moreover, since Einstein locality is an important fact of
classical physics, this feature must therefore somehow be reflected by the
operators. So, one way to guarantee this condition would be to argue that
only those operators that are constrained by relativistic relationships are
allowed to be chosen. In other words, any operator selected by the Universe
must have a set of factor operators that do not violate classical causal
laws.

If the above conjectures are correct, they might then suggest that the
causal set structures generated by the changing operators create a pattern
of Einstein locality, in terms of their arrangements of factorisation and
entanglement. Further, since the conclusion of the previous section was that
separable operators can only have separable outcomes, this pattern of
operators would in turn produce an arrangement of factor states that also
frequently share relationships obeying Einstein locality. And, since it is
the states that actually constitute physical reality, the observed
relativity in the physical Universe may hence be seen to be a consequence of
a causal set formed from operators constrained to obey Einstein locality.
Only under certain specially contrived circumstances, such as those
occurring in EPR experiments, would the true quantum non-locality of the
states become apparent.

\bigskip

It is quite possible that the Universe could choose a series of operators to
test itself with that produces a causal set structure that changes very
little from one stage to the next. Indeed, all sorts of patterns of
separations and entanglements could be present in the set, with many
different types of local or global relationships appearing to emerge over a
chain of jumps, and even the possibility of particular groups or families of
factors existing semi-permanently. Such a series of patterns is analogous to
those produced in automata such as Conway's ``Game of Life'' \cite{Conway},
and could ultimately be responsible for all the observed features of the
physical Universe, including, for example, apparent persistence, space,
dimension, particle physics, and even semi-classical looking endo-observers
who are made up of groups of factors seeming to persist in a nearly
unchanged way over very many jumps. Of course, underlying all of this
structure would still be the counting procedures used to estimate the number
of jumps (giving rise to an emergent local temporal parameter) and
estimations of familial relationships (which give rise to emergent spatial
relationships).

As an example, consider a classical cellular automaton in which the values
in the individual cells depend somehow on `nextdoor neighbour' interactions.
Such a dynamics may give rise to zones of causal influence, in which cells
outside of this zone are unable to influence cells inside it, and vice
versa. It is possible that the operators testing the Universe could also
adopt a dynamics that depends on the interaction with `neighbours',
analogous to such classical cellular automata, where neighbouring factors
are defined in terms of the familial relationships formed by the outcome
states. For example, one way of defining two factors of the operator as
neighbours might be if their respective outcome factor eigenstates share a
`parent' factor state.\ Omitting the exact details, the overall point is
that if the patterns of separations produced by the operator's causal set
are somehow forced to look, to some extent, like a type of cellular
automaton whose cells' values change according to nextdoor neighbour
interactions, the effect might be a case in which the resulting states will
possess patterns of separability that incorporate these causal zones.
Further, such zones may strongly resemble the lightcone structures of
relativity, and might produce a set of observed outcomes that are fully
consistent with Einstein causality.\bigskip

The overall conclusion of this section is that there are two different types
of causal set present in a fully quantum Universe. The first is generated by
the changing separability of the operators used to test the Universe. Whilst
it is not known how this set is produced, it does seem to give rise to
observables that respect Einstein locality and everything this entails, such
as a maximum speed for the propagation of physical signals.

The second causal set is generated by the changing separability of the
state. This set incorporates all of the features associated with quantum
states in conventional physics, and can, for example, support the non-local
correlations and apparently superluminal transmission of information
familiar in EPR type experiments. In fact, since such correlations do not
respect Einstein locality, it might be taken as a further indication that
there is an underlying quantum and pregeometric structure lurking beneath
the classical and continuous Lorentzian spacetime manifold. However, the
observation that most of empirical science appears to follow classical
physics does seem to indicate that it is only under rare and special
circumstances that the true `quantum' nature of the states becomes blatantly
apparent. Indeed, scientists generally have to try very hard in order to
prepare a factor of the Universe's state that is entangled, for instance,
and even harder to keep it that way. The repeated efforts of computer
scientists to build a working quantum computer is a good example of this.

The conclusion, then, is that since the states are ultimately the outcome of
the operators, and since in a self-referential Universe the choice of
operator may depend somehow on the current state, there must be a very
careful interplay between the two different causal sets in order to produce
the type of Universe that physicists actually observe.

\bigskip

\subsection{Physical Examples}

\bigskip

The objective of this chapter has been to investigate the types of mechanism
inherent in a quantum Universe that may ultimately be responsible for the
existence of spatial relationships. Whilst there is still a very long way to
go before the details are understood of exactly how the deep and intricate
theory of General Relativity could emerge from the fully quantum picture, it
is still possible to schematically describe how the current line of thinking
might fit into a number of physical situations. To this end, in this final
section a number of physically motivated examples are discussed in terms of
the connections between the states representing them and the operators used
in their development.

\bigskip

\subsubsection{The Quantum Big Bang}

\bigskip

The physical Universe is very large. However, given that it also currently
appears to be expanding, the conventional conclusion is that it was once
very much smaller than it is today. In fact, by observing the acceleration
of its increase in volume\footnote{%
Depending upon different particular models and metrics used to describe the
large scale structure of the Cosmos.}, cosmologists have extrapolated
backwards in time and concluded that the Universe must once have had no size
at all \cite{dInverno}. Further, by measuring the light emitted from far off
clusters of stars in order to determine when they were formed \cite{Krauss},
astronomers have managed to establish that the Universe had no size at a
time between about $10$ and $20$ billion years ago. This is the traditional
Big Bang scenario, and is often taken to imply the very beginning of time,
physics, existence and reality.

In the paradigm proposed in this work, physical space is a marker of
separability of the Universe's state. Thus, the observations of the above
Standard Model of cosmology may be interpreted here as a wavefunction that
is today highly separable, but was considerably less so in the past.
Moreover, if there was a time at which the Universe could be attributed with
no physical size, there could then have been no spatial relationships
existing, and hence by the presented arguments its state must consequently
have been completely entangled.\bigskip

Classical general relativistic cosmology asserts that time must have began
at the Big Bang, because without space there is no spacetime and hence no
time. However this is not a necessary conclusion of the present work, and in
the Stages paradigm it is conjectured that time had no beginning. After all,
physical time is seen as an emergent phenomenon appearing as a complex
vector jumps from one state to the next in its enormous Hilbert space (as
elaborated upon further in Chapter 8). It is hence quite possible that the
state and the Hilbert space can be conjectured as existing eternally,
assuming such a phrase can be used to describe something existing `outside'
of physical time, removing from the dynamics the uncomfortable view of
conventional physics that the Universe suddenly appeared out of nothing and
`no-when'.\bigskip

During an era that might be referred to as pre-Big Bang (i.e. beyond the
time cosmologists have extrapolated a zero size Universe), the Universe's
state would have been completely entangled, from the point of view of the
proposed paradigm. In fact its state may have remained entangled for a large
number of jumps, during which period no classical structures, including
space, could have emerged. From an alternative perspective, whichever
operators were used to develop the Universe through this chain of entangled
states must themselves have been completely entangled, because separable
operators cannot have entangled outcomes.

Consider, however, a case in which the Universe (somehow) eventually chooses
an operator to test itself that has separable outcomes, and further that the
Universe ends up jumping into one of these separable states. This may at
first glance appear unlikely, given the discussion of Chapter 4 that
separable states form a set of measure zero relative to the set of all
states, but is not impossible in a Universe that may have already remained
entangled for a `near-infinite' number of jumps. Besides this, since it is
an empirical fact that the Universe is large, classical looking and
separable, it can be concluded that at some point it must have stopped being
entangled.

After this collapse to a separable state, the Rules governing the way the
operators are chosen may have selected another operator that is also
separable, and the Universe would then have jumped to another separable
state. In fact, this new state could have even more factors than the
previous one. If this procedure is repeated a number of times, a situation
might arise in which the state of the Universe is monotonically becoming
more separable as it develops, and this could eventually give rise to the
observed expansion of continuous space. Overall, the selection of a series
of separable operators drives the Universe to develop through a series of
separable states, and hence to the possible emergence of spatial
relationships.

The initial jump from a fully entangled state to a separable one could now
be called the `Quantum Big Bang', and this may ultimately be what physicists
are really extrapolating back to when then examine the Universe's past and
conclude that it once had no spatial size. However, unlike the Standard
Model scenario, the presented description of the Universe's development has
the desirable feature that there is no initial singularity at the Quantum
Big Bang, and so is not associated with any of the accompanying problems of
quantum gravity. In fact, this line of thinking once again reinforces the
idea that simply quantising space and gravity is the wrong direction to
proceed. Rather, the proposal is that space should perhaps be seen as
something that is meaningless without quantum relations.\bigskip

In addition to the removal of the singularity problem, the suggested
dynamics for the development of the Universe may also provide an origin for
the observed homogeneity and isotropy of the Universe. Just after the
Quantum Big Bang, the individual factors of the Universe's state could still
be highly entangled within their respective factor Hilbert spaces, which may
themselves be of enormous dimension. Since entangled states exhibit the
properties of non-local correlations, when the entangled factor states
eventually develop into separable products of factor states (that are
themselves contained in the factor Hilbert spaces of the larger factor
Hilbert space containing the entangled factor state), these new factors may
end up having similar `properties', even though they may now appear to be
large, emergent spatial distances apart. In other words, the non-local
correlations of entangled factors just after the Quantum Big Bang may
potentially help to solve the Horizon problem of cosmology.\bigskip

Note that it is, in fact, entirely possible that before the chain of
entangled states present in the pre-Big Bang era, there could have been
whole cycles of expansion (i.e. increasing separability) and contraction
(i.e. decreasing separability) back to a `Quantum Big Crunch' of renewed
total entanglement. Indeed, there could also have been any number of `false
starts' in which the Universe chose a series of separable operators, before
suddenly choosing an entangled operator and jumping back to a completely
entangled state. However, if no information regarding these has survived
into the present era, perhaps because no information can be encoded as
relations between factors when a state is fully entangled, there is no way
of knowing about them. A return to full entanglement represents a return to
no familial relationships within the state, and since these are what might
ultimately constitute physical phenomena, all that scientists can ever look
at is what has happened since the last Quantum Big Bang.\bigskip

Consider as an example of these ideas a universe represented by a state $%
\Psi _{n}$ contained within a Hilbert space $\mathcal{H}_{[1...N]}$ of
dimension $D=2^{2^{M}}$ consisting of a vast number $N=2^{M}$ of qubit
subregisters, where $M\in \mathbb{Z}^{+}.$ With $\mathcal{H}_{[1...N]}$ is
associated, in the usual way, a basis set $\mathcal{B}_{[1...N]}$ of
orthonormal vectors and a skeleton set $\mathcal{S}_{N}$ of operators.
Further, for all $n$ the state $\Psi _{n}$ is one of the $D$ eigenstates of
some Hermitian operator $\hat{\Sigma}_{n},$ where $\hat{\Sigma}_{n}\in
\mathbb{H}(\mathcal{H}_{[1..N]}).$

By defining $n=1$ as the Quantum Big Bang, then, if the above discussion is
true, states $\Psi _{n}$ for $n<1$ are fully entangled relative to $\mathcal{%
B}_{[1...N]},$ and are hence the results of operators that are fully
entangled. Conversely, states $\Psi _{n}$ for $n\geq 1$ are separable into a
number of factors.

Now, suppose that the Rules of the universe dictate that for $0\leq n\leq
M-1 $ the operator $\hat{\Sigma}_{n+1}$ has twice as many factors as $\hat{%
\Sigma}_{n}$ according to the scheme:
\begin{eqnarray}
\hat{\Sigma}_{0} &=&\hat{A}^{1...2^{M}} \\
\hat{\Sigma}_{1} &=&\hat{A}^{1...2^{M-1}}\otimes \hat{A}%
^{(2^{M-1}+1)...2^{M}}  \notag \\
\hat{\Sigma}_{2} &=&\hat{A}^{1...2^{M-2}}\otimes \hat{A}^{(2^{M-2}+1)...2%
\times 2^{M-2}}  \notag \\
&&\otimes \hat{A}^{(2\times 2^{M-2}+1)...3\times 2^{M-2}}\otimes \hat{A}%
^{(3\times 2^{M-2}+1)...2^{M}}  \notag \\
&&\vdots  \notag \\
\hat{\Sigma}_{M} &=&\hat{A}_{1}\otimes \hat{A}_{2}\otimes \hat{A}_{3}\otimes
...\otimes \hat{A}_{M}  \notag
\end{eqnarray}
where, for example, $\hat{A}^{1...2^{M-1}}$ is an Hermitian operator in $%
\mathbb{H}(\mathcal{H}_{[1...N]})^{1...2^{M-1}}$ such that $\hat{\Sigma}%
_{1}\in \mathbb{H}(\mathcal{H}_{[1...N]})^{(1...2^{M-1})\bullet
((2^{M-1}+1)...2^{M})}\subset \mathbb{H}(\mathcal{H}_{[1..N]}).$ Further,
any eigenstate of $\hat{A}^{1...2^{M-1}}$ is in $\mathcal{H}_{[1..2^{M-1}]},$
and duly contributes at least one factor to the next state $\Psi _{1}.$

Since separable operators only have separable outcomes, it is certain that $%
\Psi _{1}$ will have at least two factors, whereas $\Psi _{2}$ cannot have
less than four factors, and $\Psi _{3}$ must have at least eight factors,
and so on up to $\Psi _{M}$ which is separable into $M$ factors. So,
whatever the operators $\hat{\Sigma}_{n+1}$ actually are, the resulting
state $\Psi _{n+1}$ of this universe may be expected to have more factors
than the previous state $\Psi _{n},$ for $0\leq n\leq M-1;$ certainly, if $%
M\gg 1$ it may be the case that `on average' the number of factors of the
state $\Psi _{n}$ could possibly increase roughly monotonically with $0\leq
n\leq M.$

Moreover, since separability has already been shown to be a necessary
prerequisite for spatial relationships, this type of development with
deterministically chosen operators may provide a basic starting point for a
discussion on the expansion of space.\bigskip

As a simple illustration of this last point, consider the case where $M=2,$
such that $N=4,$ $D=16,$ and the Hilbert space is denoted by $\mathcal{H}%
_{[1...4]}.$ The operators for $n=0,1,2$ are then of the form: $\hat{\Sigma}%
_{0}=\hat{A}^{1...4},$ $\hat{\Sigma}_{1}=\hat{A}^{12}\otimes \hat{A}^{34},$
and $\hat{\Sigma}_{2}=\hat{A}_{1}\otimes \hat{A}_{2}\otimes \hat{A}%
_{3}\otimes \hat{A}_{4}.$

A corresponding set of states in the development of this universe could
therefore be
\begin{eqnarray}
\Psi _{0} &=&\varphi ^{1234}\text{ \ \ , \ \ }\Psi _{1}=\theta ^{12\bullet
34}\equiv \theta ^{12}\otimes \theta ^{34} \\
\Psi _{2} &=&\psi _{1234}\equiv \psi _{1}\otimes \psi _{2}\otimes \psi
_{3}\otimes \psi _{4}  \notag
\end{eqnarray}
where $\varphi ^{1234}\in $ $\mathcal{H}^{1234},$ $\theta ^{12\bullet
34}(\equiv \theta ^{12}\otimes \theta ^{34})\in $ $\mathcal{H}^{12\bullet
34} $ \ and $\psi _{1234}\in $ $\mathcal{H}_{1234}.$ In this case, the
changing separability of the state would consequently lead to the type of
causal set structure illustrated in Figure 5.5.

%\FRAME{ftbpFU}{282.625pt}{217pt}{0pt}{\Qcb{Causal set structure for the
%state of an expanding universe of four qubits.}}{}{Figure 5.5}{\special%
%{language "Scientific Word";type "GRAPHIC";maintain-aspect-ratio
%TRUE;display "USEDEF";valid_file "T";width 282.625pt;height 217pt;depth
%0pt;original-width 279.125pt;original-height 213.8125pt;cropleft "0";croptop
%"1";cropright "1";cropbottom "0";tempfilename
%'I1KOB004.wmf';tempfile-properties "XPR";}}

\begin{figure}[th]
\begin{center}
\includegraphics[height=4in]{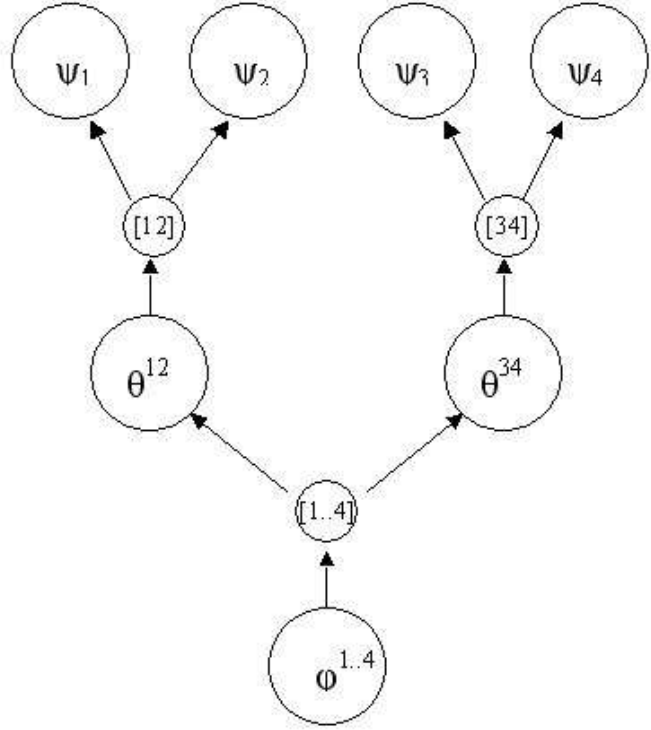}
\caption{Causal set structure for the
state of an expanding universe of four qubits.}\label{Figure 5.5}
\end{center}
\end{figure}

As discussed previously in Section 5.4, such a universe with a deterministic
choice of operator readily permits a discussion of embryonic lightcone
structure, and so ultimately also concepts of distance and metrics. In this
sense, the states\ $\psi _{3}$ and $\psi _{4}$ are `outside' of the causal
future of $\theta ^{12}$ because a counterfactual change in $\theta ^{12}$
will not influence either $\psi _{3}$ or $\psi _{4}.$\bigskip

Note that of the above scheme is not, of course, the only mechanism that
could be used to model an expanding universe. There could instead be a type
of `feedback' mechanism, in which the choice of next operator is influenced
by how separable or entangled the current state of the universe is.
Alternatively, there could be a mechanism in which, for a finite series of
jumps, an operator $\hat{\Sigma}_{n+1}$ is selected that has exponentially
many more factors than the previous test $\hat{\Sigma}_{n}.$ This latter
type of process could cause the state $\Psi _{n+1}$ to have exponentially
more factors than the state $\Psi _{n},$ and this could lead to a period of
rapid expansion analogous to the era of inflation postulated \cite{Linde}\
in the Standard Model of cosmology.

\bigskip

\subsubsection{EPR Paradoxes}

\bigskip

As discussed in Chapter 3 the non-local consequences of quantum entanglement
appear to cause problems for the theory of relativity, because the latter
places physics in a background `arena' of classical and continuous
spacetime. For example, recall the EPR experiment featured earlier involving
an entangled electron and positron. If the electron is measured first and
found to be in a spin-up state then the positron will consequently be found
to be in a spin-down state, and vice versa. Further, the standard priciples
of quantum mechanics (as verified by, amongst others, the Bell inequality)
argue that before the first measurement both the electron and positron may
be thought of as existing in both spin states simultaneously. Relativity's
problem with this can then be summed up by the question: if the electron
detector is $x$ metres away from the positron detector, and if the
positron's spin is measured $t$ seconds after the electron's spin is
measured, then how can any physical signal `inform' the second particle
that, say, the electron has been found in a spin up state such that the
positron must consequently be found to be spin down, if $x/t>c$ where $c$ is
the velocity of light? In other words, the measurement of a particle at one
location appears to be influenced by a measurement of a particle at a
different location, even though these two events are not in causal contact.

In fact, by setting up the system so that $x\gg 1$ and $t\ll 1$ it has been
experimentally shown \cite{Scarani} that if the correlations were arranged
by a signal travelling physically from one particle to the other, this
signal would require a velocity of at least $10^{4}c,$ and this conclusion
appears to be contradict special relativity which asserts that nothing can
travel faster than the speed of light.\bigskip

However, in the paradigm proposed in this work the EPR paradox is not a
problem at all. From the presented viewpoint there is no background space
over which correlations have to cross, and the measurement of the electron
and positron are only spacelike separated from an emergent point of view.
From the point of view of the proposed fully quantum approach, the entangled
electron-postitron state, the two detectors, the physicists and everything
else are just associated with factors of the state representing the
Universe, and so it is not correct to say that when the positron is measured
it is fundamentally $x$ metres away from where the electron was measured. On
the pregeometric quantum level the electron and positron are nothing but
factors of a vector in a Hilbert space, and physical spatial relationships
are meaningless here.\bigskip

As a schematic illustration of how an EPR type experiment might proceed in a
fully quantum Universe, consider the following chain of stages in the
state's development. Note first, however, that as with the Schr\"{o}dinger's
cat discussion of Section 4.3.3 the example\ below is really just a highly
simplified overview; in reality detectors (and the physicists observing
them) are incredibly complicated sets of factors, constantly undergoing many
different types of internal developments and interactions with their
surroundings.

Let a particular split of the Hilbert space $\mathcal{H}$ of the Universe be
of the form
\begin{equation}
\mathcal{H}=\mathcal{H}_{e}\otimes \mathcal{H}_{p}\otimes \mathcal{H}%
_{E}\otimes \mathcal{H}_{P}\otimes \mathcal{H}_{U}
\end{equation}
where $\mathcal{H}_{e}$ represents the factor Hilbert space of an electron, $%
\mathcal{H}_{p}$ represents the factor Hilbert space of a positron, $%
\mathcal{H}_{E}$ represents the factor Hilbert space of an electron
detector, $\mathcal{H}_{P}$ represents the factor Hilbert space of a
positron detector, and $\mathcal{H}_{U}$ represents the factor Hilbert space
containing everything else in the Universe. Note that none of these five
factor Hilbert spaces need be of prime dimension.

Consider now an operator $\hat{\Sigma}_{n}$ factorisable in the form:
\begin{equation}
\hat{\Sigma}_{n}=\hat{A}^{ep}\otimes \hat{A}_{E}\otimes \hat{A}_{P}\otimes
\hat{A}_{U}
\end{equation}
where, for example, $\hat{A}^{ep}\in \mathbb{H}(\mathcal{H})^{ep},$ with the
entanglement $\mathcal{H}^{ep}\subset \mathcal{H}_{[ep]},$ and $\hat{\Sigma}%
_{n}\in \mathbb{H}(\mathcal{H})_{EPU}^{ep}.$ Obviously, this separable
operator $\hat{\Sigma}_{n}$ will have separable eigenstates. So, assume that
the resulting next state of the Universe turns out to be of the form:
\begin{equation}
\Psi _{n}=|\psi \rangle ^{ep}\otimes |D\rangle _{E}\otimes |D\rangle
_{P}\otimes |R\rangle _{U}
\end{equation}
where $|\psi \rangle ^{ep}\in \mathcal{H}^{ep},$ $|D\rangle _{E}\in \mathcal{%
H}_{E},$ $|D\rangle _{P}\in \mathcal{H}_{P}$ and $|R\rangle \in \mathcal{H}%
_{U},$ and hence $\Psi _{n}\in \mathcal{H}_{EPU}^{ep}\subset \mathcal{H},$
etc.

In a Universe represented by a state $\Psi _{n},$ the factor $|\psi \rangle
^{ep}$ may be interpreted as the initial entangled electron-positron
sub-state, with $|D\rangle _{E}$ the initial state of the electron detector
and $|D\rangle _{P}$ the initial state of the positron detector. Of course,
some of these factor states may also be separable relative to a more
fundamental split of their respective factor Hilbert spaces, and some of the
factors of the operators may also be factorised further. Indeed, the factor $%
|R\rangle _{U}$ representing the combined sub-states of everything else in
the Universe is presumably separable into very many factors in order to
account for all of these other parts, but for clarity this issue is ignored
here.

Overall, the operator $\hat{\Sigma}_{n}$ and the subsequent collapse into
the state $\Psi _{n}$ are equivalent to the preparation of a Universe
containing an entangled electron-positron pair.

Assume now that the Rules governing the Universe conspire in such as way as
to choose an operator $\hat{\Sigma}_{n+1}$ to test $\Psi _{n}$ with, defined
as
\begin{equation}
\hat{\Sigma}_{n+1}=\hat{A}_{p}\otimes \hat{A}^{Ee}\otimes \hat{A}_{P}\otimes
\hat{A}_{U},
\end{equation}
and further that this test collapses the Universe into the state $\Psi
_{n+1} $ defined as
\begin{equation}
\Psi _{n+1}=|\downarrow \rangle _{p}\otimes |u\rangle ^{Ee}\otimes |D\rangle
_{P}\otimes |R^{\prime }\rangle _{U}.
\end{equation}

Now, in $\Psi _{n+1}$ the factor $|u\rangle ^{Ee}$ is interpreted as an
entangled sub-state between a spin-up electron and an electron detector.
Similarly, $|\downarrow \rangle _{p}$ may be interpreted as a factor of the
Universe representing a spin-down positron. Note however that the positron
detector is still in its initial condition $|D\rangle _{P}$: the factor
operator $\hat{A}_{P}$ of $\hat{\Sigma}_{n+1}$ is effectively behaving as a
local null test in $\mathcal{H}_{P}$ because it was also a factor of $\hat{%
\Sigma}_{n}.$ The factor $|R^{\prime }\rangle _{U}\in \mathcal{H}_{U}$ is
interpreted as the part of the Universe that has nothing to do with the
electron-positron-detector system developing in its own way, and is again
ignored.

Suppose further that the Rules now conspire to choose an operator $\hat{%
\Sigma}_{n+2}$ of the form
\begin{equation}
\hat{\Sigma}_{n+2}=\hat{A}^{Ee}\otimes \hat{A}^{Pp}\otimes \hat{A}_{U}
\end{equation}
and that the Universe subsequently collapses to the state $\Psi _{n+2}$
defined as
\begin{equation}
\Psi _{n+2}=|u\rangle ^{Ee}\otimes |d\rangle ^{Pp}\otimes |R^{\prime \prime
}\rangle .
\end{equation}

In this case, $|d\rangle ^{Pp}$ might be interpreted as a correlated
sub-state between a spin-down positron and a positron detector.

The sequence of states $\Psi _{n},$ $\Psi _{n+1}$ and $\Psi _{n+2}$ offers a
schematic picture of how a fully quantum Universe might view an EPR type
experiment involving the preparation of an initial entangled
electron-positron pair, through to the measurement of the electron, and then
followed by the measurement of the positron, noting that the issue of the
actual relationship between entanglement, changes of partition and
endophysical measurements will be addressed properly in the next chapter.

In the emergent limit, $|D\rangle _{E}$ is taken to represent that part of
the Universe associated with an electron detector. Moreover, in this limit
the factor $\hat{A}^{Ee}$ of the operator $\hat{\Sigma}_{n+1}$ is associated
with the `interaction' between the electron detector and the component of
the entangled electron-positron pair in the electron's Hilbert space $%
\mathcal{H}_{e}.$ The factor $\hat{A}^{Ee}$ is hence the pregeometric
equivalent of a detector physically testing the spin of the electron, and is
therefore analogous to one of the `usual' Hermitian operators familiar to
conventional physics experiments in which an isolated semi-classical
apparatus measures an isolated system described by quantum mechanics. The
difference between the current work and that of familiar physics is that
these single, isolated experiments of conventional physics are taken in the\
larger context of the whole Universe being developed at once, instead of
just a tiny part of it. As has been discussed previously, this difference
arises from the acknowledgement that because the Universe is \textit{%
everything,} any change in one part of it, no matter how small, necessarily
implies a change in the state of the whole.

The sub-state $|u\rangle ^{Ee}$ may be seen as the outcome of this test $%
\hat{A}^{Ee},$ and would ultimately correspond in the emergent limit to the
physical result of the interaction between an entangled electron and a
detector. Thus, the factor $|u\rangle ^{Ee}$ is taken to be the result of
this measurement, and in this case represents the situation in which the
detector finds the electron to be spin up.

By the argument of Section 3.1, any measurement of the entangled electron
automatically collapses the state of the positron, in this instance into a
spin down factor $|\downarrow \rangle _{p}.$ Consequently, then, the overall
development of the state from $...\otimes |\downarrow \rangle _{p}\otimes
|D\rangle _{P}\otimes ...$ to $...\otimes |d\rangle ^{Pp}\otimes ...$ could
be interpreted in the emergent limit as a semi-classical detector measuring
the positron's spin with a test $\hat{A}^{Pp}$ to give the result $|d\rangle
^{Pp}.$ Thus, the detector duly finds the positron to be spin down.\bigskip

Of course, many other tests $\hat{\Sigma}_{n+2}$ could have been selected by
the Universe to develop $\Psi _{n+1},$ just as semi-classical scientists
appear able to choose many alternative ways of measuring a quantum
sub-system. For example, a particular factor $\hat{B}^{Pp}$ of an
alternative operator $\hat{\Sigma}_{n+2}^{\prime }$ could represent the spin
of the positron being measured along a completely different axis, or it
could even imply a test being performed that may have nothing to do with
spin at all. However, an important constraint is that if the Universe is in
the state $\Psi _{n+1},$ and if it tests itself with an operator $\hat{\Sigma%
}_{n+2}$ containing a factor $\hat{A}^{Pp}$ that, in the emergent limit,
measures the component of spin of the positron in the same emergent
direction as the component of spin of the electron was measured in, only
those states $\Psi _{n+2}$ containing a factor representing a spin-down
positron result will have a non-zero probability of occurring.\bigskip

Now consider the familial relationships present in the causal sets produced
from the network of earlier states $\Psi _{n-m},$ for $m=1,2,...,$ and
relating to what is going on in the rest of the Universe. The result might
be that in the emergent limit one factor $|D\rangle _{E}$ (corresponding to
the factor operator $\hat{A}_{D})$ of the state $\Psi _{n}$ representing the
electron detector seems to be located at one point in emergent space, whilst
another factor $|D\rangle _{P}$ that represents the positron detector (and
corresponding to the factor operator $\hat{A}_{P})$ seems to be located at
another point in emergent space. Moreover, the subsequent factors $|u\rangle
^{Ee}$ and $|d\rangle ^{Pp}$ may also appear to have definite locations in
the emergent limit.

The point is that in this emergent limit, it might therefore appear that the
results of the measurements of the electron and positron are correlated
across emergent spacelike distances, apparently defying relativity. However,
this conflict is resolved by noting that it is only a problem on the
emergent scale: on the `true' quantum level such locational descriptions are
meaningless, and so theories of emergent physics such as Lorentz covariance
cannot be applied there. In this quantum picture the entire experiment is
seen as nothing but a change in the separability of the vector representing
the state of the Universe as it jumps from being in one partition to
another. There are hence no contradictions to superluminality conditions
because velocity is not defined on this pregeometric level. From this point
of view there is no paradox in EPR.

\bigskip

\subsubsection{Superluminal Correlation}

\bigskip

The following simple example illustrates how even a small difference between
two consecutive operators can lead to large consequences for the resulting
two consecutive states.

Consider a Hilbert space $\mathcal{H}_{[1...2N]}$ factorisable into $2N$
qubit subregisters. Consider further the $n^{th}$ operator $\hat{\Sigma}%
_{n}, $ which happens to be factorisable into two entangled sub-operators, $%
\hat{\Sigma}_{n}=\hat{A}^{1...N}\otimes \hat{B}^{(N+1)...2N},$ where $\hat{A}%
^{1...N}\in \mathbb{H}(\mathcal{H}_{[1...N]})^{1...N}$ and $\hat{B}%
^{(N+1)...2N}\in \mathbb{H}(\mathcal{H}_{[(N+1)...2N]})^{(N+1)...2N}.$

Suppose also that the particular eigenstate of $\hat{\Sigma}_{n}$ that
becomes the next state $\Psi _{n}$ is of the form $\Psi _{n}=\psi
^{1...N}\otimes \psi ^{(N+1)...2N},$ such that clearly $\Psi _{n}\in
\mathcal{H}^{(1...N)\bullet ((N+1)...2N)}$ with $\psi ^{1...N}\in \mathcal{H}%
^{1...N}$ and $\psi ^{(N+1)...2N}\in \mathcal{H}^{(N+1)...2N}.$ Evidently,
each factor is entangled relative to its factor subspace, that is, each is
entangled relative to half of the overall quantum register.

Now consider the next operator $\hat{\Sigma}_{n+1},$ and suppose that the
rules governing the universe dictate that this is also a product of
entangled operators, but of the form $\hat{\Sigma}_{n+1}=\hat{C}%
^{1...(N+1)}\otimes \hat{D}^{(N+2)...2N}.$ Roughly speaking, in this type of
development it may be envisaged that the $(N+1)^{th}$ qubit has `gone over'
from one factor of the operator $\hat{\Sigma}_{n}$ to the other in the
selection of $\hat{\Sigma}_{n+1};$ the factor $\hat{A}$ has `gained' a qubit
from the factor $\hat{B}$ as they `became' $\hat{C}$ and $\hat{D}$
respectively.\ So overall the way the operators $\hat{\Sigma}_{n}$ and $\hat{%
\Sigma}_{n+1}$ factorise only differs by one qubit, and if $N\gg 1$ it may
therefore be said that $\hat{\Sigma}_{n}$ and $\hat{\Sigma}_{n+1}$ appear
highly similar from this factorisation point of view.\bigskip

However, given operators $\hat{\Sigma}_{n}$ and $\hat{\Sigma}_{n+1}$ of this
form, then by the discussion of Section 5.4 for any eigenstate $\Theta $ of $%
\hat{\Sigma}_{n+1}$ the probability amplitude $\langle \Psi _{n+1}=\Theta
|\Psi _{n}\rangle $ may not factorise. Thus, the conclusion is that by
making what appears to be a very small change from the perspective of the
operators, the family structure of the state's causal set could be
destroyed. Moreover, for a Universe with a very large number of quantum
subregisters, although this one qubit change in the operators may appear
almost insignificant, it could end up having far reaching consequences
across the entire state. Indeed, since family structure will ultimately
account for the presence of spatial relationships, even small, local changes
in the operator structure could give rise to an emergent situation that
appeared to support superluminal correlations. This again highlights one of
the important differences between states and operators: even by making a
small change in the operator structure that might appear consistent with
Einstein locality and emergent theories of relativity, enormous changes in
the factors of the state could result which might eventually lead to an
apparent violation of these principles.

\bigskip

\subsubsection{Persistence}

\bigskip

As is readily apparent from observing the nature of the Universe, some
physical objects appear to persist over time. A single atom, for example, is
often assumed to be identical from one instant to the next if it is not
interacting with anything, and even macroscopic states such as humans tend
to believe that they continue to be the `same' person for a number of years.

Because time in the proposed paradigm is a concept that is expected to
emerge as the state of the Universe develops through a series of stages, the
existence of persistence is therefore equivalent to the observation that
some features of the state appear to `survive' relatively unchanged from one
jump to the next. Moreover, because it is generally classical objects that
are observed to possess this property of longevity, the concept of
persistence may be seen as evidence that particular factors of the
Universe's can sometimes remain approximately unaltered as it develops.

Now, the appearance of classical features in the Universe has previously
been shown to be a result of the separability of its state. The observation
that there is any persistence at all may therefore seem surprising. After
all, when arguments of microsingularity are taken into account, as well as
the fact that separable states are contained in sets of measure zero, it
appears apparently `inevitable' that the Universe \textit{should} jump from
one completely entangled state to another.

However as has been discussed a number of times in this work, the assertion
that the state jumps from one highly separable vector $\Psi _{n}$ to the
next $\Psi _{n+1}$ is ascribed to be due to the Rules that govern the
Universe's dynamics very carefully selecting the operators $\hat{\Sigma}_{n}$
and $\hat{\Sigma}_{n+1}.$ Further to this, since persistence is clearly a
ready feature of empirical science, it is possible to argue that the Rules
may also be confined to only choose those operators that have outcomes that
are similar, in some sense, to the current state.\bigskip

One way of achieving this result is to consider the earlier conclusion that
the separability of a state may be dictated by the factorisability of the
operator of which it is an eigenvector.

Consider a Universe represented by a state in a Hilbert space $\mathcal{H}%
_{[1...N]}.$ Further, assume that the Rules conspire in such a way that the $%
n^{th}$ Hermitian operator $\hat{\Sigma}_{n}\in \mathbb{H}(\mathcal{H}%
_{[1...N]})$ used to develop the Universe is of the form
\begin{equation}
\hat{\Sigma}_{n}=\hat{A}_{a_{1}}\otimes \hat{A}_{a_{2}}\otimes ...\otimes
\hat{A}_{a_{k}}
\end{equation}
where $\hat{A}_{a_{i}}\in \mathbb{H}(\mathcal{H}_{[a_{i}]})$ and $\mathcal{H}%
_{[a_{i}]}$ need not be of prime dimension. By the conclusion of Section
5.5, whatever eigenvector of $\hat{\Sigma}_{n}$ becomes the next state $\Psi
_{n}$ will therefore have at least $k$ factors. So, $\Psi _{n}$ will be of
the form
\begin{equation}
\Psi _{n}=\psi _{a_{1}}\otimes \psi _{a_{2}}\otimes ...\otimes \psi _{a_{k}}
\end{equation}
where $\psi _{a_{i}}\in \mathcal{H}_{[a_{i}]},$ noting that $\psi _{a_{i}}$
may itself be a product of factors, some of which may be entangled relative
to the fundamental factorisation of $\mathcal{H}_{[a_{i}]}.$

Consider now the next test of the Universe $\hat{\Sigma}_{n+1},$ and assume
that the Rules dictate that it is also factorisable into $k$ sub-operators.
Further, assume that the Rules also specify that each of the $k$
sub-operators of $\hat{\Sigma}_{n+1}$ acts in the same factor Hilbert space
as one of the $k$ individual factors of $\hat{\Sigma}_{n}.$ In other words, $%
\hat{\Sigma}_{n+1}$ is constrained to possess the same sort of `partition
structure' as $\hat{\Sigma}_{n},$ and may hence be of the form:
\begin{equation}
\hat{\Sigma}_{n+1}=\hat{B}_{a_{1}}\otimes \hat{B}_{a_{2}}\otimes ...\otimes
\hat{B}_{a_{k}}
\end{equation}
where $\hat{B}_{a_{i}}\in \mathbb{H}(\mathcal{H}_{[a_{i}]}).$ Now, as before
any eigenvector of $\hat{\Sigma}_{n+1}$ must also have no less than $k$
factors, so whatever the next state $\Psi _{n+1}$ of the Universe actually
is, it clearly has to be of the form
\begin{equation}
\Psi _{n+1}=\phi _{a_{1}}\otimes \phi _{a_{2}}\otimes ...\otimes \phi
_{a_{k}}
\end{equation}
where $\phi _{a_{i}}\in \mathcal{H}_{[a_{i}]},$ noting that $\phi _{a_{i}}$
may also be a product of (possibly entangled) factors.

The point is that in this type of development, the state $\Psi _{n+1}$ has a
very similar structure to the previous state $\Psi _{n}$ in terms of which
partitions of $\mathcal{H}_{[1...N]}$ they are members of. Consequently, the
factor $\psi _{a_{i}}$ of $\Psi _{n}$ may be thought of as developing into
the factor $\phi _{a_{i}}$ of $\Psi _{n+1}$ without `interacting' with any
of the other factors. Thus, this could be an embryonic form of
`semi-persistence' of the sub-state in the factor Hilbert space $\mathcal{H}%
_{[a_{i}]}.$

Moreover, because each sub-state $\psi _{a_{i}}$ could actually be a group
of factors itself, this mechanism allows the possibility for macroscopic
sets of sub-states to survive relatively unchanged from one jump to the
next. Although of course clearly only a schematic model here, the
persistence and apparently isolated nature of semi-classical objects such as
apparatus, laboratories and physicists, each of which is associated with
large groups of factors, may ultimately be a consequence of the
relationships between factorisable operators and separable states.\bigskip

A more definite form of persistence would be evident if the Rules instead
selected the alternative operator $\hat{\Sigma}_{n+1}^{\prime },$ defined as
\begin{equation}
\hat{\Sigma}_{n+1}^{\prime }=\hat{B}_{a_{1}}\otimes \hat{B}_{a_{2}}\otimes
...\otimes \hat{B}_{a_{j-1}}\otimes \hat{A}_{a_{j}}\otimes \hat{B}%
_{a_{j+1}}\otimes ...\otimes \hat{B}_{a_{k}}.
\end{equation}

Any next state $\Psi _{n+1}^{\prime }$ resulting from an eigenvector of $%
\hat{\Sigma}_{n+1}^{\prime }$ is of the form
\begin{equation}
\Psi _{n+1}^{\prime }=\phi _{a_{1}}\otimes \phi _{a_{2}}\otimes ...\otimes
\phi _{a_{j-1}}\otimes \psi _{a_{j}}\otimes \phi _{a_{j+1}}\otimes
...\otimes \phi _{a_{k}}
\end{equation}
which clearly has a factor $\psi _{a_{j}}\in \mathcal{H}_{[a_{j}]}.$ So, the
factor in $\mathcal{H}_{[a_{j}]}$ of both $\Psi _{n}$ and $\Psi
_{n+1}^{\prime }$ is $\psi _{a_{j}},$ such that the sub-operator $\hat{A}%
_{a_{j}}$ of $\hat{\Sigma}_{n+1}^{\prime }$ is acting as a local null test.
Thus, the factor $\psi _{a_{j}}$ has clearly remained unchanged in the
development of the Universe from $\Psi _{n}$ to $\Psi _{n+1}^{\prime }.$
From the point of view of the other factors $\phi _{a_{i}}$ of $\Psi
_{n+1}^{\prime },$ the factor $\psi _{a_{j}}$ can therefore be said to have
persisted during this jump.

Of course, by carefully choosing the subsequent operators $\hat{\Sigma}%
_{n+2},$ $\hat{\Sigma}_{n+3},$ $\hat{\Sigma}_{n+4},...$ the rules could
readily ensure that particular factors persist over many more developments
of the state.

\bigskip

\subsubsection{Position and Dimension}

\bigskip

In this final example it will be shown how positional relationships might be
encoded in terms of factorisation and entanglement. Additionally, it will
also be shown that such relations may also afford a natural inclusion of the
properties of dimension. A simple illustration of these ideas will first be
given, followed then by a generalisation to more complicated
examples.\bigskip

Consider a cubic lattice in three spatial dimensions. In fact for simplicity
consider the smallest such lattice, that is, a single cube formed from only
eight points, where one point is on each of the eight corners of the cube.
Clearly, each `edge' of the cube implies the minimum separation between two
adjacent sites, and may be associated with a length of $1$ unit.

Now, in order to discuss positional relationships within the cube it is
necessary to define a suitable set of axes. This can be achieved by
arbitrarily selecting any one of the corner sites to be an `origin', and
then using the direction of the three corners that are adjacent to this
origin to specify a set of three orthogonal, Cartesian axes. One such choice
is illustrated in Figure 5.6, where it is now possible to label the
positions of the sites according to this set of axes; for example the origin
site is denoted $(0,0,0),$ whereas the site furthest (i.e. $\sqrt{3}$ units)
from the origin is $(1,1,1).$

%\FRAME{ftbpFU}{239.75pt}{164.0625pt}{0pt}{\Qcb{%
%The cubic lattice formed from eight points, with the corners labelled
%according to a set of orthogonal axes through the origin $(0,0,0).$}}{}{%
%Figure 5.6}{\special{language "Scientific Word";type
%"GRAPHIC";maintain-aspect-ratio TRUE;display "USEDEF";valid_file "T";width
%239.75pt;height 164.0625pt;depth 0pt;original-width
%234.4375pt;original-height 159.8125pt;cropleft "0";croptop "1";cropright
%"1";cropbottom "0";tempfilename 'I1KOB105.wmf';tempfile-properties "XPR";}}

\begin{figure}[th]
\begin{center}
\includegraphics[height=4in]{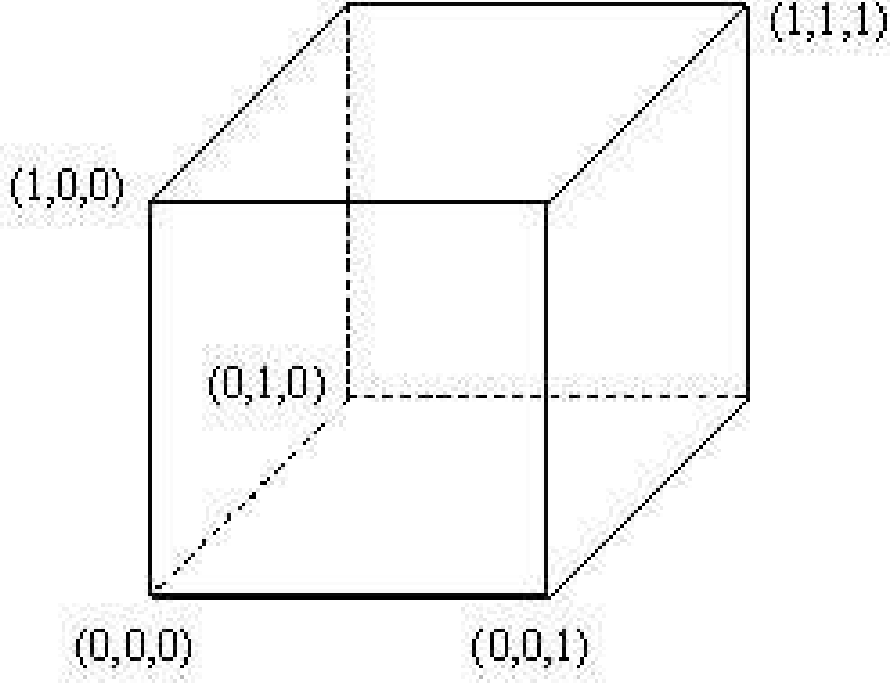}
\caption{The cubic lattice formed from eight points, with the corners labelled
according to a set of orthogonal axes through the origin $(0,0,0).$}\label{Figure 5.6}
\end{center}
\end{figure}

It is possible to label each of the corners with a unique integer $a,$ where
$a=0,1,...,7.$ One way of achieving would be to take the coordinates $%
(x,y,z) $ of the corner as the coefficients in a series expansion of the
powers of two, such that $a$ may be given by the rule
\begin{equation}
a=x(2^{0})+y(2^{1})+z(2^{2}).
\end{equation}

For example, the corner positioned at $(0,0,0)$ corresponds to $a=0,$
whereas the corner $(0,1,0)$ may be denoted by $a=2,$ whilst $(1,1,1)$
corresponds to $a=7$ etc. Indeed, by relabelling the $a^{th}$ site as $b,$
where $b=a+1$ such that $1\leq b\leq 8,$ the eight sites can be numbered
cardinally according to their positions on the lattice.\bigskip

Consider now a model universe represented by a state in a Hilbert space $%
\mathcal{H}_{[1...N]},$ and an Hermitian operator $\hat{o}_{n}\in \mathbb{H}(%
\mathcal{H}_{[1...N]})$ used to develop this state. Suppose that $\hat{o}%
_{n} $ is chosen according to the rule that it can be fundamentally
factorised into eight sub-operators. Suppose further that the rules dictate
that each factor of $\hat{o}_{n}$ is itself entangled relative to the
skeleton sets of $c$ subregisters, where $1\leq c\leq 8,$ and that no two
factors of $\hat{o}_{n}$ may be entangled relative to the same number of
subregisters. Thus, one of the factor operators acts on states in just one
quantum subregister, whereas another factor of $\hat{o}_{n}$ is an entangled
operator acting on states contained in the space of two subregisters, and so
on, up to the eighth factor that is an entangled operator acting upon states
contained in the space of eight subregisters. Evidently, the total number of
subregisters $N$ in $\mathcal{H}_{[1...N]}$ is given by $N=1+2+...+8=36,$
and\ one possible such operator $\hat{o}_{n}$ may be defined as
\begin{equation}
\hat{o}_{n}=\hat{A}_{1}\otimes \hat{A}^{23}\otimes \hat{A}^{456}\otimes \hat{%
A}^{7...(10)}\otimes \hat{A}^{(11)...(15)}\otimes \hat{A}^{(16)...(21)}%
\otimes \hat{A}^{(22)...(28)}\otimes \hat{A}^{(29)...(36)}
\end{equation}
where $\hat{A}^{i...j}\in \mathbb{H}(\mathcal{H}_{[i...j]})^{i...j}\subset
\mathbb{H}(\mathcal{H}_{[i...j]})$ and $\mathcal{H}_{[i...j]}\subset
\mathcal{H}_{[1...N]}.$

It is hence possible to assign a unique number $c$ with each of the factors
of $\hat{o}_{n}$ in terms of the number of subregisters over which it is
entangled. Moreover, since the discussion of the previous paragraph showed
that it is also possible to associate positions on a lattice with numbers,
the individual factors of $\hat{o}_{n}$ can conversely each be associated
with a sort of `position'. In other words, the factor $\hat{A}_{1}$ of $\hat{%
o}_{n}$ may be labelled by the number $c=1$ and so may, in some sense, be
associated with the position $(0,0,0).$ Similarly the factor $\hat{A}^{456}$
corresponds to the number $c=3$ and so may be denoted as $(0,1,0),$ whereas $%
\hat{A}^{(29)...(36)}$ may be denoted by the number $c=8$ and may hence be
associated with $(1,1,1).$\bigskip

Further, any eigenvector of $\hat{o}_{n}$ must be separable into at least
eight factors, and so these outcomes would therefore also follow the pattern
of spatial positioning affiliated with the operator. Hence whichever
eigenvector the universe collapses into, the factors of this next state $%
\psi _{n}$ must share some of the `locational information' of the factors of
the operator $\hat{o}_{n}.$ In other words, whatever factor $\alpha \in
\mathcal{H}_{[i...j]}$ of the next state $\psi _{n}=...\otimes \alpha
\otimes ...$ is the result of a factor $\hat{A}^{i...j}$ of $\hat{o}_{n},$
the location of $\hat{A}^{i...j}$ on the lattice can also be used to
describe the corresponding position of the factor $\alpha $ of the state.

Summarising, although the underlying structure is just a single cube, the
rules selecting the operator $\hat{o}_{n}$ imply that its factors, and those
of its eigenvectors, may be discussed in terms of the position of the
lattice's sites. Number and position are interchangeable, and according to
the rules, so are factor and number.

\bigskip

In order to generalise these ideas, consider a (very large) prime number $p$
and a positive integer $d.$ By analogy with the base $p=2$ expansion of $a$
given above, the p-adic expansion \cite{Gouvea} of any non-negative integer $%
P<p^{d}$ to base $p$ is given by
\begin{equation}
P=i_{0}(p^{0})+i_{1}(p^{1})+i_{2}(p^{2})+...+i_{d-1}(p^{d-1})
\end{equation}
where $i_{j}\in \mathbb{Z}^{\ast }$ is a non-negative integer that is less
that $p,$ and the subscript $j$ identifies which power of $p$ a particular $%
i $ is a coefficient of. Thus, any integer $0\leq P\leq (p^{d}-1)$ can be
uniquely specified by a series of coefficients $i_{j},$ for $%
j=0,1,...,(d-1), $ and a prime number $p.$

Consider now a $d$-dimensional lattice of points, where this lattice may be
imagined to be a $d$-dimensional `cube' with edges of `length' $(p-1)$
units. As before, by picking one corner of the lattice as the origin $%
(0,0,...,0),$ the position of each site can be associated with a specific
number between $0$ and $(p^{d}-1).$ For example, the origin $(0,0,...,0)$
may be associated with the number zero, whereas the number $(p^{d}-1)$ is
related to the position $(p-1,p-1,...,p-1),$ and so on. So by continuing the
analogy of the single cube example of earlier, every integer $P$ in the
appropriate range may be thought of as mapped to a unique site on a lattice
with coordinates $(i_{0},i_{1},...,i_{d-1}).$ As before, these numbers may
be granted cardinality by relabelling them as $P^{\prime }=1,2,...,p^{d}$
defined as $P^{\prime }=P+1.$\bigskip

Now consider a universe represented by a state in $\mathcal{H}_{[1...N]},$
and a particular Hermitian operator $\hat{O}_{n}\in \mathbb{H}(\mathcal{H}%
_{[1...N]})$ factorisable as before into a product of $p^{d}$ sub-operators.
Assuming again the rules are such that each factor sub-operator of $\hat{O}%
_{n}$ is entangled relative to the skeleton sets associated with between one
and $p^{d}$ subregisters, and that no two factors of $\hat{O}_{n}$ are
entangled relative to the same number of subregisters, one possible form of $%
\hat{O}_{n}$ is given by
\begin{equation}
\hat{O}_{n}=\hat{A}_{1}\otimes \hat{A}^{23}\otimes \hat{A}^{456}\otimes
...\otimes \hat{A}^{M...N}.
\end{equation}

Clearly, the total number $N$ of subregisters required for such a
prescription is given by the arithmetic progression
\begin{eqnarray}
N &=&1+2+3+...+p^{d} \\
&=&p^{d}(p^{d}+1)/2  \notag
\end{eqnarray}
and because the `last' factor $\hat{A}^{M...N}$ of $\hat{O}_{n}$ is
entangled relative to $p^{d}$ subregisters, $M$ is given by $%
M=p^{d}(p^{d}-1)/2.$ So, $\hat{O}_{n}\in \mathbb{H}(\mathcal{H}%
_{[1...p^{d}(p^{d}+1)/2]})_{1}^{23\bullet 456\cdot ...\bullet
(M...N)}\subset \mathbb{H}(\mathcal{H}_{[1...N]}),$ with $\hat{A}^{M...N}\in
\mathbb{H}(\mathcal{H})^{(p^{d}(p^{d}-1)/2)...(p^{d}(p^{d}+1)/2)}$ etc.

As before, by assigning each factor of $\hat{O}_{n}$ a unique number
according to how many subregisters it is entangled relative to, and by
associating each of these numbers with a coordinate, the factors of $\hat{O}%
_{n}$ may be associated with `positions' in a lattice. Thus, the factor $%
\hat{A}_{1}$ may be assigned the number $1$ and so may be associated with
the coordinate $(0,0,...,0),$ whereas the factor $\hat{A}^{23}$ may
similarly be associated with $(1,0,...,0),$ whilst $\hat{A}^{456}$ may be
associated\footnote{%
Assuming $p\geq 3.$ Otherwise, say if $p=2,$ $\hat{A}^{456}$ will be
associated with $(0,1,0,...,0)$ etc.} with $(2,0,...,0),$ and so on, until $%
\hat{A}^{M...N}$ is may be associated with $(p-1,p-1,...,p-1).$

Likewise, the eigenstates of $\hat{O}_{n}$ will also share the separability
properties of this operator, and so their factors may similarly also be
considered in terms of these simple positional relationships.\bigskip

It is easy to see how dynamics may be incorporated into this type of model.
As long as the rules governing what type of operators are allowed remain the
same, the lattice structure associated with the operators' factors will be
preserved. For instance, the dynamics may permit permutations of the form $%
r\longrightarrow s_{r}$ in the structure of the operators, where $r,s$ label
different subregisters, and this is equivalent to exchanging the
subregisters over which a factor of the operator is entangled relative to.
For example, if the permutation was such that the next operator $\hat{O}%
_{n+1}$ is of the form
\begin{equation}
\hat{O}_{n+1}=\hat{B}_{3}\otimes \hat{B}^{(1)(12)}\otimes \hat{B}%
^{(7)(8)(14)}\otimes \hat{B}^{(2)(4)(5)(27)}\otimes ...
\end{equation}
the result might be a set of eigenvectors whose factors are completely
different from those of the previous state, but are all still constrained by
the same type of lattice structure.\bigskip

It is important to reiterate that the lattice formed from the operator's
factors does not exist in any sort of background space. The positional
relationships, and hence the corresponding measure of dimension, are simply
a consequence of the way the operator $\hat{O}_{n}$ factorises in terms of
the skeleton set of basis operators spanning the subregisters of the total
Hilbert space, and this is itself just a result of whatever rules dictate
the dynamics. As always throughout this work, physical space is not seen as
absolute but as a marker of distinction between objects in an underlying
mathematical structure.

So as a final remark it should be noted that the present discussion of
position should not be taken too rigidly. As has been a central conclusion
of this chapter, the relationships between the pregeometric quantum register
and the eventual emergence of physical space is a subtle one that requires a
great deal of future work before it is completely understood. Indeed, even
in the above example it is observed that a given number $P$ may potentially
be expanded in many different ways, because many different bases $p$ could
be chosen, and this would lead to a set of alternative lattices of different
dimensions.

Of course, whether or not this last comment has any physical meaning is an
important question to be faced, and might perhaps imply that either: the
current example is too `na\"{i}ve' to describe proper physics; or that it is
missing an important constraint that forces every lattice in the `real'
Universe to be three dimensional; or even that it might possibly allow the
occurrence of multi-dimensional ideas such as Kaluza-Klein and string
theory. What is clear, however, is that in a fully quantum Universe with no
external observers, the Universe must somehow organise itself in such a way
so that internal, semi-classical observers are able to experience a reality
with near-continuous spatial relationships. In such a Universe, the
underlying quantum structure must somehow be responsible for sophisticated
theories such as relativity and four dimensional spacetime to emerge.

\bigskip \newpage

\section{Quantum Registers as Quantum Computers}

\renewcommand{\thefigure}{6.\arabic{figure}} \setcounter{figure}{0} %
\renewcommand{\theequation}{6.\arabic{equation}} \setcounter{equation}{0} %
\renewcommand{\thetheorem}{6.\arabic{theorem}} \setcounter{theorem}{0}

\bigskip

In the previous chapter, it was discussed how spatial degrees of freedom
might begin to emerge if the causal set structure of operators obeys
relationships analogous to those of a classical cellular automaton.
Moreover, starting from the premise that there are no external observers
dictating its development, a central theme of this thesis is that the
Universe is acting as a giant, self-regulating quantum automaton. From these
viewpoints, the development of its state is envisaged to be like an enormous
quantum calculation, such that overall the Universe may be thought of as
behaving like an enormous quantum computer. This conjecture is discussed
now.\bigskip

The present chapter is split into two parts. In Section 6.1 it is shown how
simple quantum computational methods may be applied to a system consisting
of a finite number of Hilbert space subregisters. Since in the proposed
paradigm a state is considered that exists in a large, but finite, number of
such quantum subregisters, it is argued that these principles are equally
applicable to the case where the system is the Universe itself.

Because computation is often seen as synonymous with data manipulation, in
Section 6.2 the role of information change and exchange is defined for
quantum systems. It is then discussed how endo-physical scientists might
obtain `answers' for the Universe's quantum calculation, and how these
answers might be interpreted.

\bigskip

\subsection{Computing with CNOT}

\bigskip

In this section it will be demonstrated how operators can be used to perform
computations in simple quantum systems of qubits. Specifically, in fact, the
example of the Controlled-NOT (or CNOT) operator will be examined. It must
be noted, however, that such computations are not just simply mathematical
exercises; instead, they will be compared with the actual, physical results
of classical computations, namely by a formulation of the Bell inequality.
Some of the implications of this comparison will then be discussed.

Overall, the work described here will serve as a preparation for the
following chapter in which it will be shown how, by treating the state with
quantum computational methods, particle field theoretic concepts may arise
in a fully quantum Universe.

Before quantum computational principles can be applied to a system of
quantum subregisters, though, it is useful to review some of the ideas of
classical computation. Specifically, it will be beneficial to illustrate how
the classical analogue of the quantum CNOT operator, namely the CNOT logic
gate, may be employed in classical computation. This issue is hence
addressed first, noting that a more thorough background description of both
classical and quantum computation is provided in Appendix A.

\bigskip

\subsubsection{Classical Computation}

\bigskip

Broadly speaking, conventional classical computation involves processing the
`values' contained at a sequence of `sites', where each site will definitely
take one, and only one, value from a finite set of possibilities. In fact in
general, classical computation can actually be achieved by a particular
manipulation of a finite set of \textit{bits}, each of which is in one of
two possible states. The workings of modern, digital microelectronic
computers are an example of this. Moreover, because of their binary
property, conventional logic may be applied to these bits, and their states
may consequently be labelled `true' or `false', or perhaps `on' or `off', or
even `0' or `1'. Equally, the processing of these bits may be accomplished
by the use of binary logic gates.

Classical computations generally involve three parts: there is the
specification of the Input, usually given in the form of a string of bits of
which each has a particular value; there is the computation itself, which
involves the processing of these bits according to a particular set of gates
in a certain order; and there is the Output, which is the result of the
computation, and is also usually given in terms of a string of bit values.

Any sequence of 0's and 1's, and consequently any string of bits, denotes a
unique binary number. So from this perspective, a classical computation
involving the transformation of an input series of bit values into an output
sequence may be interpreted as a calculation being performed on an initial
number to generate an `answer'. This answer is also a number, and may itself
go on to be processed in subsequent computations.

Note how this could easily be seen as analogous to the quantum Universe, in
which an initial state $\Psi _{n}$ is developed into the next state $\Psi
_{n+1}$ by some particular combination of unitary and Hermitian
operators.\bigskip

Just as the particular choice of quantum operators dictates the way the
Universe is developed, it is the transformations that determine how a
certain sequence of bits is processed in a computation. It is consequently
the particular choice of logic gates that define which particular classical
computation is performed on the input. As with operators in quantum
mechanics, a number of different types of operation are also possible here.

As an illustration, consider a classical system consisting of just two bits $%
X$ and $Y.$ Each bit may take one of two values, such that $X$ may have the
value $x$ and $Y$ may have the value $y,$ where $x,y=0,1.$ Thus the state $S$
of the system may be denoted by the pair $S=(x,y),$ and this will clearly be
one of four distinct possibilities.

A \textit{local} operation may be defined as any operation that acts on the
bits individually and independently, with no reference to the values of any
of the other bits. Examples in the two bit system are the operations $U$ and
$V,$ defined as
\begin{eqnarray}
U(x,y) &=&(x\oplus x,y)=(0,y) \\
V(x,y) &=&(x,y\oplus 1)  \notag
\end{eqnarray}
where the symbol $\oplus $ denotes addition modulo two, i.e.
\begin{equation}
0\oplus 0=0\text{ \ \ , \ \ }0\oplus 1=1\text{ \ \ , \ \ }1\oplus 0=1\text{
\ \ , \ \ }1\oplus 1=0.
\end{equation}

Conversely, a \textit{global} operation is one that acts on the whole state $%
S$ of the system. In these operations, the way one of the two bits is
processed depends on the value of the other bit.

In fact, note that for all systems with more than one bit it is generally
possible to consider \textit{non-local} operations, that is, those which act
on two or more bits. In such operations, the way a particular bit is
developed may be affected by the value of at least one other bit. Of course,
clearly in a two bit system every non-local operation is also global, but
for systems of more than two bits the global operations are just special,
extreme cases of non-local operations.\bigskip

An example of such a global operation acting on the two bit system $X$ and $%
Y $ is the CNOT logic gate, $C,$ defined as
\begin{equation}
C(x,y)=(x,x\oplus y).  \label{Ch4ClassCNOT}
\end{equation}

For obvious reasons from (\ref{Ch4ClassCNOT}), in the above use of CNOT it
is possible to describe $X$ as the `donor' bit and $Y$ as the `acceptor' bit.

The above gate may be thought of as a type of `question and answer'
operation. Processing a state $S=(x,y)$ with $C$ may be viewed as equivalent
to asking a question of the value $x$ of the bit $X,$ and registering the
answer with a response in the value $y$ of the bit $Y.$

The CNOT computation is reversible. That is, in this case there exists an
inverse operation $C^{-1}$ defined as
\begin{equation}
C^{-1}(x,x\oplus y)=(x,y)
\end{equation}
such that $C^{-1}C(x,y)=(x,y);$ in fact, clearly $C=C^{-1}.$ Analogous to
the gate $C,$ the inverse operation $C^{-1}$ may then be interpreted as the
statement: ``given a particular result, what was the question of which it is
an answer?''. In this case, the answer is $(x,x\oplus y)$ and the question
is $(x,y).$

Note that the inverse gate $C^{-1}$ is different from the `conjugate'
operation $\widetilde{C}$ defined as
\begin{equation}
\widetilde{C}(x,y)=(x\oplus y,y)
\end{equation}
which reverses the role of acceptor and donor.

From this last gate $\widetilde{C}$ it is possible to build a `transpose
gate' $C^{T}$ that swaps the values of the bits $X$ and $Y,$ i.e. $%
C^{T}(x,y)=(y,x).$ With the above descriptions of $C$ and $\widetilde{C}$
this can be achieved by writing $C^{T}$ as $C^{T}=\widetilde{C}C\widetilde{C}%
.$\bigskip

The two bit CNOT gate is a global operation when acting on a system
containing just two bits, but merely a non-local operation for an $N$ bit
system if $N>2.$ Specifically, given a string of $N$ bits $%
Z_{1},Z_{2},...,Z_{N}$ in the state $S=(z_{1},z_{2},...,z_{N}),$ where $%
z_{a}=0,1$ for $a=1,2,...N,$ the CNOT gate $C_{(i,j)}$ may be defined as
\begin{equation}
C_{(i,j)}(z_{1},z_{2},...,z_{i},...,z_{j-1},z_{j},z_{j+1},...,z_{N})=(z_{1},z_{2},...,z_{i},...,z_{j-1},z_{i}\oplus z_{j},z_{j+1},...,z_{N}).
\end{equation}

This last definition will be useful later.

\bigskip

\subsubsection{Quantum Computation}

\bigskip

A classical computation involving operations performed on a classical state
has immediate parallels with the way operators in quantum mechanics act on
quantum states. This latter process may therefore naturally be called a
quantum computation. Furthermore, a classical computation involving
operations performed on a series of bits is itself analogous to the way
operators in quantum mechanics can act on products of qubits. Indeed, just
as a classical bit is defined as some sort of `entity' that can take one of
two possible values, a qubit is defined relative to a basis comprising of
two different (orthogonal) states. However, whilst classical bits are
restricted to always have one value or the other, the states of quantum bits
can exist as complex linear superpositions of their basis vectors.\bigskip

Consider a two dimensional (qubit) Hilbert space, $\mathcal{H}_{a}^{(2)},$
where the super-script may again be assumed and hence dropped, and the
sub-script denotes that this space belongs to the $a^{th}$ qubit, in
preparation for the later discussion of many qubit systems. Assume also that
$\mathcal{H}_{a}$ is spanned by the orthonormal basis set $\mathcal{B}_{a}$
defined as $\mathcal{B}_{a}\equiv \{|0\rangle _{a},|1\rangle _{a}\},$ where $%
_{a}\langle i|j\rangle _{a}=\delta _{ij}$ for $i,j=0,1,$ and note that these
elements may be represented by column vectors of the form $|0\rangle
_{a}\equiv \binom{1}{0}_{a}$ and $|1\rangle _{a}\equiv \binom{0}{1}_{a}.$

Define now the projection operators $\hat{P}_{a}^{0}$ and $\hat{P}_{a}^{1}$
acting on the $a^{th}$ space as
\begin{equation}
\hat{P}_{a}^{0}\equiv |0\rangle _{aa}\langle 0|\text{ \ \ , \ \ }\hat{P}%
_{a}^{1}\equiv |1\rangle _{aa}\langle 1|  \label{Ch4Proj}
\end{equation}
and the `transition' operators $\hat{Q}_{a}$ and $\hat{Q}_{a}^{\dagger }$ as
\begin{equation}
\hat{Q}_{a}\equiv |0\rangle _{aa}\langle 1|\text{ \ \ , \ \ }\hat{Q}%
_{a}^{\dagger }\equiv |1\rangle _{aa}\langle 0|.  \label{Ch4Tran}
\end{equation}

The application of these four operators to a qubit may be interpreted in
particular ways. For example, the projection operator $\hat{P}_{a}^{0}$ may
be thought of as equivalent to the question: ``is the $a^{th}$ qubit in the
state $|0\rangle ?";$ a similar question is appropriate for $\hat{P}%
_{a}^{1}. $ The transition operators $\hat{Q}$ and $\hat{Q}^{\dagger }$ are
analogous to the ladder operators of field theory: $\hat{Q}_{a}$ may be
thought of as an operator which, when applied to a qubit in Hilbert space $%
\mathcal{H}_{a}, $ `annihilates' the state $|1\rangle $ and `replaces' it
with a state $|0\rangle .$ Conversely, $\hat{Q}_{a}^{\dagger }$ may be
considered as an operator that `destroys' $|0\rangle $ and `creates' $%
|1\rangle .$ These connections are explored in Chapter 7.\bigskip

Every operator acting on states in the qubit Hilbert space $\mathcal{H}_{a}$
can be built from complex sums of the four operators $\hat{P}_{a}^{0},$ $%
\hat{P}_{a}^{1},$ $\hat{Q}_{a}$ and $\hat{Q}_{a}^{\dagger }.$ So, if $\hat{A}%
_{a}$ is an arbitrary operator acting in $\mathcal{H}_{a}$ it may be written
\begin{equation}
\hat{A}_{a}=A_{a}^{1}\hat{P}_{a}^{0}+A_{a}^{2}\hat{P}_{a}^{1}+A_{a}^{3}\hat{Q%
}_{a}+A_{a}^{4}\hat{Q}_{a}^{\dagger }
\end{equation}
where $A_{a}^{1},A_{a}^{2},A_{a}^{3},A_{a}^{4}\in \mathbb{C},$ or
alternatively $\hat{A}_{a}=(A_{a}^{1},A_{a}^{2},A_{a}^{3},A_{a}^{4})$ for
brevity. Similarly, the Hermitian conjugate operator $\hat{A}_{a}^{\ast }$
may be given by $\hat{A}_{a}^{\ast }=(A_{a}^{1}{}^{\ast },A_{a}^{2}{}^{\ast
},A_{a}^{4}{}^{\ast },A_{a}^{3}{}^{\ast }).$

As an example, the identity and Pauli operators, $\hat{\sigma}_{a}^{\mu }$
for $\mu =0,1,2,3,$ can be defined as
\begin{eqnarray}
\hat{\sigma}_{a}^{0} &=&\hat{P}_{a}^{0}+\hat{P}_{a}^{1}\text{ \ \ , \ \ }%
\hat{\sigma}_{a}^{1}=\hat{Q}_{a}+\hat{Q}_{a}^{\dagger }  \label{Ch4 PAULI} \\
\hat{\sigma}_{a}^{2} &=&-i(\hat{Q}_{a}-\hat{Q}_{a}^{\dagger })\text{ \ \ , \
\ }\hat{\sigma}_{a}^{3}=\hat{P}_{a}^{0}-\hat{P}_{a}^{1}  \notag
\end{eqnarray}
or equally
\begin{eqnarray}
\hat{\sigma}_{a}^{0} &=&(1,1,0,0)\text{ \ \ , \ \ }\hat{\sigma}%
_{a}^{1}=(0,0,1,1) \\
\hat{\sigma}_{a}^{2} &=&(0,0,-i,i)\text{ \ \ , \ \ }\hat{\sigma}%
_{a}^{3}=(1,-1,0,0)  \notag
\end{eqnarray}
and these clearly satisfy the `standard' algebra (\ref{Ch3pauli1}), and the
representation (\ref{Ch3PauliRep}), as given in Chapter 5.\bigskip

It is possible to define products of operators in the above notation. As an
illustration, consider two operators $\hat{A}_{a}$ and $\hat{B}_{a}$ defined
as $\hat{A}_{a}=(A_{a}^{1},A_{a}^{2},A_{a}^{3},A_{a}^{4})$ and $\hat{B}%
_{a}=(B_{a}^{1},B_{a}^{2},B_{a}^{3},B_{a}^{4}).$ The product $\hat{A}_{a}%
\hat{B}_{a}$ is then given by
\begin{equation}
\hat{A}_{a}\hat{B}%
_{a}=([A_{a}^{1}B_{a}^{1}+A_{a}^{3}B_{a}^{4}],[A_{a}^{2}B_{a}^{2}+A_{a}^{4}B_{a}^{3}],[A_{a}^{1}B_{a}^{3}+A_{a}^{3}B_{a}^{2}],[A_{a}^{2}B_{a}^{4}+A_{a}^{4}B_{a}^{1}]).
\end{equation}

The product algebra of the operators in (\ref{Ch4Proj}) and (\ref{Ch4Tran})
is summarised in Table 6.1. Specifically, the product $XY$ is read as the
member $X$ of the first column multiplied by the member $Y$ of the first
row; for example, $\hat{P}^{0}\hat{Q}=\hat{Q},$ whereas $\hat{Q}\hat{P}%
^{0}=0.$%
\begin{equation}
\begin{tabular}{|l|l|l|l|l|}
\hline
$\mathbf{-}$ & $\mathbf{\hat{P}}^{0}$ & $\mathbf{\hat{P}}^{1}$ & $\mathbf{%
\hat{Q}}$ & $\mathbf{\hat{Q}}^{\mathbf{\dagger }}$ \\ \hline
$\mathbf{\hat{P}}^{0}$ & $\hat{P}^{0}$ & $0$ & $\hat{Q}$ & $0$ \\ \hline
$\mathbf{\hat{P}}^{1}$ & $0$ & $\hat{P}^{1}$ & $0$ & $\hat{Q}^{+}$ \\ \hline
$\mathbf{\hat{Q}}$ & $0$ & $\hat{Q}$ & $0$ & $\hat{P}^{0}$ \\ \hline
$\mathbf{\hat{Q}}^{+}$ & $\hat{Q}^{+}$ & $0$ & $\hat{P}^{1}$ & $0$ \\ \hline
\end{tabular}
\tag*{Table 6.1}
\end{equation}

The results of the products of these operators also comment on the role of
information in quantum processes. The idempotency of the projection
products, $\hat{P}^{0}\hat{P}^{0}=\hat{P}^{0}$ and $\hat{P}^{1}\hat{P}^{1}=%
\hat{P}^{1},$ for example, may be interpreted as the observation that once a
`question' has been asked of a quantum system, no new information can be
extracted by asking the same question again. This reflects the deeper
principle of quantum mechanics that once a state has collapsed into one of
the eigenvectors of a particular operator, testing the system a second time
with the same operator reproduces the same result.

On the other hand, note that the transition operators $\hat{Q}_{a}$ and $%
\hat{Q}_{a}^{\dagger }$ obey the fermionic algebra in the sense that their
squares vanish, $(\hat{Q}_{a})^{2}=(\hat{Q}_{a}^{\dagger })^{2}=0.$ As with
their association with quantum field operators, this property will be useful
in Chapter 7. For now, however, note that any operator $\hat{A}$ that
satisfies the rule $(\hat{A})^{p+1}=0$ may be called a `parafermionic
operator of order $p$', following the language of \cite{Green} and \cite
{Greenberg}, where $p\in \mathbb{Z}^{\ast }$ is the lowest integer required
for this rule to be true. Thus, $\hat{Q}_{a}$ and $\hat{Q}_{a}^{\dagger }$
may be labelled parafermions of order $1.$\bigskip

It is important to note that the construction of Table 6.1 does not rely on
any use of group theory. The sixteen entries in the table come directly from
the logic induced by taking the inner products of the basis vectors from
which the four individual operators $\hat{P}_{a}^{0},$ $\hat{P}_{a}^{1},$ $%
\hat{Q}_{a}$ and $\hat{Q}_{a}^{\dagger }$ are defined. Conversely, in fact,
the rotational symmetry of the $SU(2)$ group can be shown to be preserved as
a natural feature of the underlying qubit perspective.

To demonstrate this, define the `\textit{Transformation}' operators $\hat{T}%
_{a}^{ij}$ acting on the space $\mathcal{H}_{a}$ as $\hat{T}%
_{a}^{ij}=|i\rangle _{aa}\langle j|,$ where $i,j=0,1.$ Clearly, $\hat{T}%
_{a}^{ij}$ is one of four possible operators corresponding to the two
projection and two transition operators defined above in (\ref{Ch4Proj}) and
(\ref{Ch4Tran}). That is,
\begin{eqnarray}
\hat{T}_{a}^{00} &=&\hat{P}_{a}^{0}\text{ \ \ , \ \ }\hat{T}_{a}^{01}=\hat{Q}%
_{a}  \label{CH4Trans} \\
\hat{T}_{a}^{10} &=&\hat{Q}_{a}^{\dagger }\text{ \ \ , \ \ }\hat{T}_{a}^{11}=%
\hat{P}_{a}^{1}.  \notag
\end{eqnarray}

So, Table 6.1 can be summarised for $i,j,k,l=0,1$ by the relation
\begin{equation}
\hat{T}_{a}^{ij}\hat{T}_{a}^{kl}=\delta _{jk}\hat{T}_{a}^{il}.  \label{Ch4ts}
\end{equation}

Consider now a unitary operator $\hat{U}(\mathbf{\theta })$ defined as
\begin{equation}
\hat{U}(\mathbf{\theta })\equiv \exp \left( i\sum_{j=1}^{3}\theta _{j}\hat{%
\sigma}_{a}^{j}\right)
\end{equation}
where $\mathbf{\theta }=(\theta _{1},\theta _{2},\theta _{3}),$ $\theta
_{j}\in \mathbb{R}$ and the $\hat{\sigma}_{a}^{j}$ are generators of the
group $SU(2).$ The set of operators $\hat{U}(\mathbf{\theta })$ form an $%
SU(2)$ rotation group acting on states in $\mathcal{H}_{a}.$

Define now a new transformation operator $(\hat{T}_{a}^{ij})^{\prime }$ as
\begin{equation}
(\hat{T}_{a}^{ij})^{\prime }\equiv \hat{U}^{\ast }(\mathbf{\theta })\hat{T}%
_{a}^{ij}\hat{U}(\mathbf{\theta })
\end{equation}
with $\hat{U}^{\ast }(\mathbf{\theta })\hat{U}(\mathbf{\theta })=\hat{I}_{a}$
where $\hat{I}_{a}$ is the identity in $\mathcal{H}_{a}.$ This new operator
satisfies the product rule $(\hat{T}_{a}^{ij})^{\prime }(\hat{T}%
_{a}^{kl})^{\prime }=\delta _{jk}(\hat{T}_{a}^{il})^{\prime },$ so it may be
concluded that (\ref{Ch4ts}) is invariant to rotations of the operators $%
\hat{T}_{a}^{ij}$ under the group $SU(2).$

The above conclusion might be surprising, given that spatial relationships
were not used in any way in the construction of (\ref{Ch4ts}) or Table 6.1,
and indicates why the language of spin may be used in discussions of qubits
(e.g. spin up, spin down etc.). However, it should be reiterated that the
qubits should not be thought of as fundamentally possessing any sort of
physical spin, and the rotations are really nothing but mathematical
transformations; physical spin is expected to eventually appear in the
emergent limit once space, directions, and frames of reference have been
established.

It is, though, still encouraging perhaps to note that spin relations
automatically seem to emerge as a natural feature of the algebraic rules
inherent in the underlying qubit structure.

\bigskip

Just as it is possible to perform computations on classical bit systems
using a reversible two-bit CNOT gate, it is possible to define a quantum
CNOT operator also. In actual fact this possibility turns out to be of
fundamental importance, because it can be shown that any qubit quantum
computation can be performed using just local unitary operators and the CNOT
gate alone \cite{Barenco}.

As with the classical case, the quantum CNOT asks a `question' of one qubit
and registers the response with a second.

Consider a four dimensional Hilbert space $\mathcal{H}_{[ab]}^{(4)},$
factorisable into two qubit subregisters in the form $\mathcal{H}%
_{[ab]}^{(4)}=\mathcal{H}_{a}^{(2)}\otimes \mathcal{H}_{b}^{(2)},$ where
from now on the dimensional super-scripts shall again be omitted. A
suitable, separable orthonormal basis $\mathcal{B}_{ab}$ for $\mathcal{H}%
_{[ab]}$ may be defined in the usual way as $\mathcal{B}_{ab}\equiv
\{|00\rangle _{ab},|01\rangle _{ab},|10\rangle _{ab},|11\rangle _{ab}\}.$

The quantum CNOT operator $\hat{C}_{(a,b)}$ acting on the qubits $a$ and $b$
may now be defined as
\begin{eqnarray}
\hat{C}_{(a,b)} &=&\hat{P}_{a}^{0}\otimes \hat{\sigma}_{b}^{0}+\hat{P}%
_{a}^{1}\otimes \hat{\sigma}_{b}^{1}  \label{Ch4CNOT} \\
&=&\hat{P}_{a}^{0}\otimes (\hat{P}_{b}^{0}+\hat{P}_{b}^{1})+\hat{P}%
_{a}^{1}\otimes (\hat{Q}_{a}+\hat{Q}_{a}^{\dagger })  \notag
\end{eqnarray}
with $\hat{\sigma}_{b}^{0}$ the identity, and $\hat{\sigma}_{b}^{1}$ the
usual `first' Pauli operator, acting on qubit $b.$ Note that in the
representation $\mathcal{B}_{ab},$ the CNOT operator may equally be written
as the matrix
\begin{equation}
C_{(a,b)}=\binom{\QATOP{1}{0}\QATOP{0}{1}\QATOP{0}{0}\QATOP{0}{0}}{\QATOP{0}{%
0}\QATOP{0}{0}\QATOP{0}{1}\QATOP{1}{0}}.
\end{equation}

CNOT is a unitary operator, and might hence be used to evolve the state of
the two qubit system. To illustrate how, assume that the initial state $\Psi
_{i}$ of the system is in one of the four, separable basis states of $%
\mathcal{B}_{ab}.$ The final state $\Psi _{f}\equiv \hat{C}_{(a,b)}\Psi _{i}$
is then given by Table 6.2
\begin{equation}
\begin{tabular}{|l|l|l|l|l|}
\hline
$\mathbf{\Psi }_{i}$ & $|00\rangle $ & $|01\rangle $ & $|10\rangle $ & $%
|11\rangle $ \\ \hline
$\mathbf{\Psi }_{f}\mathbf{\equiv \hat{C}}_{(a,b)}\mathbf{\Psi }_{i}$ & $%
|00\rangle $ & $|01\rangle $ & $|11\rangle $ & $|10\rangle $ \\ \hline
\end{tabular}
\tag*{Table 6.2}
\end{equation}

From Table 6.2 the result of an application of $\hat{C}_{(a,b)}$ to the
system is apparent: if qubit $a$ is in the state $|0\rangle $ then the state
of qubit $b$ remains the same, whereas if qubit $a$ is in the state $%
|1\rangle $ then the application of $\hat{C}_{(a,b)}$ `flips' the state of
qubit $b$ either from $|0\rangle \rightarrow |1\rangle $ or $|1\rangle
\rightarrow |0\rangle .$\bigskip

Care is needed, however, before physically interpreting exactly how $\hat{C}%
_{(a,b)}$ is working. From one perspective it appears that an application of
$\hat{C}_{(a,b)}$ to $\mathbf{\Psi }_{i}$ does not affect qubit $a.$ The
operator $\hat{C}_{(a,b)}$ appears to be non-invasively determining the
state of qubit $a,$ and then consequently registering the result with qubit $%
b.$ This, however, is a very classical viewpoint, and is therefore not the
best way to proceed. From a quantum perspective it must be noted that any
determination of the state of qubit $a$ necessarily involves an extraction
of information, and from quantum principles it should therefore be expected
that such an extraction will irreversibly alter the state of the system.

It is this conflict between attempted non-invasive, classical information
extraction and the truly invasive, quantum CNOT that forms the basis of the
following example of how quantum computational methods could be applied to
quantum register systems. In the next sub-section it is shown how a
variation of the Bell inequalities can arise from a classical computation
acting on a classical bit system. Then by contrast, in the following
sub-section similar Bell inequalities will be derived for a quantum
computation acting on a product of qubits. However, it is shown further that
once the outcome of a quantum computation is obtained, the state collapses
to a classical-looking system, and as such any potential Bell violations are
irretrievably lost. So, not only should the following sub-sections provide a
physically applicable demonstration of how quantum computational principles
can be used for systems consisting of quantum subregisters, but it will also
illustrate how in a Universe ultimately envisaged to behave as an enormous
quantum computer, the conclusion must be reinforced that the extraction of
information during measurement inevitably comes with a cost.

\bigskip

\subsubsection{Classical Calculations and the Bell Inequality}

\bigskip

Consider a single classical bit that, at any time, can either take the value
$0$ or the value $1.$ Physically, such a bit could in principle represent
any classical bi-level system, for example the off-on states of a digital
switch, or a random number generator programmed to provide one of two
possible outputs. Let the value of the bit at time $T$ be labelled $x_{T},$
and for simplicity further assume that the temporal parameter changes
discretely such that $T=0,1,2,...,$ as is apparent in the case of the
`clock' of a modern microprocessor (c.f. the parameter $``n"$ used to denote
different states $\Psi _{n}$ and stages $\Omega _{n}$ etc.). Moreover,
because it is a classical system that is being discussed, it is reasonable
to conclude that at any particular time $T$ the bit is definitely in a
unique state $x_{T}=0$ or $x_{T}=1$ whether or not it is actually observed;
in classical mechanics every dynamical variable may be said to possess
`independent existence', irrespective of attempts to measure it.\bigskip

Consider now the dynamics, of which there are of course many different
types, that govern the bit's development from its value $x_{T}$ at time $T$
to the value $x_{T+1}$ at time $T+1.$ Generally speaking, any such
development implies some sort of function $F$ that, when applied to the bit
at time $T,$ provides the value $x_{T+1}$ at time $T+1.$ It is consequently
possible to represent this mapping by the operation $x_{T+1}=F(x_{T}).$

One of the simplest set of dynamics could incorporate a deterministic rule
of the form $x_{T+1}=x_{T},$ such that the value of the bit remains constant
for all time. In this situation, after $N$ timesteps the `history' of the
bit's value will either be $(N+1)$ lots of $0$'s or else $(N+1)$ lots of $1$%
's, depending on its value at the initial time. Alternatively, another
simple deterministic rule might be that $x_{T+1}=(x_{T}\oplus 1),$ where $%
\oplus $ denoted addition modulo 2, in which case the history of the bit's
value will be of the form $...\rightarrow 0\rightarrow 1\rightarrow
0\rightarrow 1\rightarrow ...$

Consider instead, however, a dynamics based on randomness, such that at each
time step there is a unique probability that the bit will have the value $0$
and a unique probability that the bit will take the value $1,$ where these
two probabilities sum to unity as expected. The probability $P(x_{T+1}=0)$
that the value of the bit at time $T+1$ is $0$ could be irrespective of the
value $x_{T}$ of the bit at time $T,$ as could the probability $P(x_{T+1}=1)$
that the value of the bit at time $T+1$ is $1.$ On the other hand, it is
conversely possible to consider a case where the dynamics might be based
upon conditional probabilities, incorporating for example the probability $%
P(x_{T+1}=0|x_{T})$ that the value of the bit at time $T+1$ is $0,$ or
similarly the conditional probability $P(x_{T+1}=1|x_{T})$ that the value of
the bit at time $T+1$ is $1,$ \textit{given} that its value at time $T$ was $%
x_{T}.$

Overall, then, it is possible to postulate a `Random Operation' $R$ that
acts on a bit of value $x_{T},$ at time $T,$ to give a value $x_{T+1},$ at
time $T+1,$ with a particular conditional probability; that is $%
R(x_{T})=x_{T+1},$ where $x_{T+1}$ occurs with probability $%
P(x_{T+1}|x_{T}). $\bigskip

Assume now that at time $T=0$ the bit has the initial value $x_{0}=0,$ and
consider such a dynamics based upon conditional probabilities. By assuming
conservation of probability, if the conditional probability $%
P(x_{1}=0|x_{0}=0)$ is denoted
\begin{equation}
P(x_{1}=0|x_{0}=0)=a
\end{equation}
then the probability $P(x_{1}=1|x_{0}=0)$ is given by
\begin{equation}
P(x_{1}=1|x_{0}=0)=\bar{a}\equiv 1-a.
\end{equation}

It is similarly possible to consider the probabilities of obtaining
particular `chains' of results. If the conditional probabilities $%
P(x_{2}=0|x_{1}=0)$ and $P(x_{2}=0|x_{1}=1)$ are given by
\begin{eqnarray}
P(x_{2} &=&0|x_{1}=0)=b \\
P(x_{2} &=&0|x_{1}=1)=c  \notag
\end{eqnarray}
\qquad\ then it immediately follows that
\begin{eqnarray}
P(x_{2} &=&1|x_{1}=0)=\bar{b} \\
P(x_{2} &=&1|x_{1}=1)=\bar{c}  \notag
\end{eqnarray}
so
\begin{eqnarray}
P(x_{2} &=&0\text{ }\&\text{ }%
x_{1}=0|x_{0}=0)=P(x_{1}=0|x_{0}=0)P(x_{2}=0|x_{1}=0)=ab  \label{Ch4Prob} \\
P(x_{2} &=&1\text{ }\&\text{ }%
x_{1}=0|x_{0}=0)=P(x_{1}=0|x_{0}=0)P(x_{2}=1|x_{1}=0)=a\bar{b}  \notag \\
P(x_{2} &=&0\text{ }\&\text{ }%
x_{1}=1|x_{0}=0)=P(x_{1}=1|x_{0}=0)P(x_{2}=0|x_{1}=1)=\bar{a}c  \notag \\
P(x_{2} &=&1\text{ }\&\text{ }%
x_{1}=1|x_{0}=0)=P(x_{1}=1|x_{0}=0)P(x_{2}=1|x_{1}=1)=\bar{a}\bar{c}  \notag
\end{eqnarray}
with as before, and in the following, a `bar' over any variable $\alpha $
implies $\bar{\alpha}=1-\alpha ,$ and noting that
\begin{equation}
\left(
\begin{array}{c}
P(x_{2}=0\text{ }\&\text{ }x_{1}=0|x_{0}=0)+P(x_{2}=1\text{ }\&\text{ }%
x_{1}=0|x_{0}=0) \\
+P(x_{2}=0\text{ }\&\text{ }x_{1}=1|x_{0}=0)+P(x_{2}=1\text{ }\&\text{ }%
x_{1}=1|x_{0}=0)
\end{array}
\right) =1
\end{equation}
as expected. In words, a composite probability of the form $P(x_{2}=1$ $\&$ $%
x_{1}=0|x_{0}=0)$ clearly represents the combined likelihood of the bit
having the values $x_{0}=0,$ $x_{1}=0$ and $x_{2}=1$ at the times $T=0,1$
and $2$ respectively.

As an additional convenience, it is possible to denote consecutive results
as a string of $0$'s and $1$'s, with time running from left to right. As an
illustration, the sequence of results $x_{0}=0,$ $x_{1}=0$ and $x_{2}=1$ is
labelled in this notation by the string $001,$ and from (\ref{Ch4Prob})
occurs with probability $P(001)=a\bar{b}.$ \ Note that, perhaps rather
confusingly, $P(001)\equiv $ $P(x_{2}=1$ $\&$ $x_{1}=0|x_{0}=0);$ as will be
obvious from the quantum case discussed later, the reason for this order
reversal comes from a desire to consider probabilities $P(x_{T+1}|x_{T})$ as
analogous to quantum probability amplitudes of the form $\left| \langle
x_{T+1}|x_{T}\rangle \right| ,$ and sequences $x_{0}x_{1}x_{2}...$ as
analogous to qubit states of the form $|x_{0}\rangle \otimes |x_{1}\rangle
\otimes |x_{2}\rangle \otimes ...$\bigskip

It is important to specify exactly what is meant by `probability' in this
example. To this end, the probabilities are taken to imply here that if a
very large number $N$ of identical bits were all in the same initial state $%
x_{0}=0,$ and if they were all subject to these same dynamics, then at time $%
T=1$ a number $Na$ would be expected to have the value $x_{1}=0,$ and so $%
N(1-a)$ would consequently have the value $x_{1}=1.$ Alternatively, the
probabilities may equally be viewed as the frequencies of obtaining
particular `histories' if the experiment was performed very many times $N.$
So after two time steps, $Nab$ of the experiments would be expected to have
resulted in the history $000,$ whereas $Na\bar{b}$ of the experiments would
be expected to have resulted in the history $001,$ whilst $N\bar{a}c$ of the
experiments would be expected to have resulted in the history $010,$ and $N%
\bar{a}\bar{c}$ of the experiments would be expected to have resulted in the
history $011.$\bigskip

The above process may be continued indefinitely. Clearly, with each time
step the number of different possible sequences of results doubles, such
that at time $T$ the bit will have experienced one of $2^{T}$ possible
histories, each with a specific probability of occurring.

Consider, however, just the set of possible histories for a bit developing
from time $T=0$ to time $T=4.$ By defining the additional probabilities
\begin{eqnarray}
P(0000) &=&abd\text{ \ \ , \ \ }P(0010)=a\bar{b}e\text{ \ \ , \ \ }P(0100)=%
\bar{a}cf\text{ \ \ , \ \ }P(0110)=\bar{a}\bar{c}g \\
P(00000) &=&abdh\text{ \ \ , \ \ }P(00010)=ab\bar{d}i\text{ \ \ , \ \ }%
P(00100)=a\bar{b}ej\text{ \ \ , \ \ }P(00110)=a\bar{b}\bar{e}k  \notag \\
P(01000) &=&\bar{a}cfl\text{ \ \ , \ \ }P(01010)=\bar{a}c\bar{f}m\text{ \ \
, \ \ }P(01100)=\bar{a}\bar{c}gn\text{ \ \ , \ \ }P(01110)=\bar{a}\bar{c}%
\bar{g}o  \notag
\end{eqnarray}
the individual histories can be illustrated as the `branches' in Figure 6.1,
where time runs downwards. At each of the $2^{T}$ individual `forks'
occurring at time $T,$ the value at the fork represents the probability that
the bit at time $T$ will have the history $0...x_{T-2}x_{T-1}x_{T},$ whereas
the value at the bottom of the left hand branch of this fork denotes the
probability that the bit at time $T+1$ will have the history $%
0...x_{T-2}x_{T-1}x_{T}0,$ whilst the value at the bottom of the right hand
branch of this fork denotes the probability that the bit at time $T+1$ will
have the history $0...x_{T-2}x_{T-1}x_{T}1.$

So, for example, if $N$ bits are developed under identical conditions the
diagram shows that $Na\bar{b}\bar{e}$ of them would be expected to have
undergone the history $0011$ (i.e. [left][right] [right]) at time $T=3,$
whereas $N\bar{a}c\bar{f}\bar{m}$ would be expected to have undergone the
history $01011$ (i.e. [right][left][right][right]) at time $T=4,$ etc.

%\FRAME{%
%fhFU}{485.375pt}{223.5pt}{0pt}{\Qcb{Classical probability `tree' for a
%developing bit.}}{}{Figure 6.1}{\special{language "Scientific Word";type
%"GRAPHIC";maintain-aspect-ratio TRUE;display "USEDEF";valid_file "T";width
%485.375pt;height 223.5pt;depth 0pt;original-width 480.75pt;original-height
%220.3125pt;cropleft "0";croptop "1";cropright "1";cropbottom
%"0";tempfilename 'I1KOB106.wmf';tempfile-properties "XPR";}}

\begin{figure}[th]
\begin{center}
\includegraphics[width=6in]{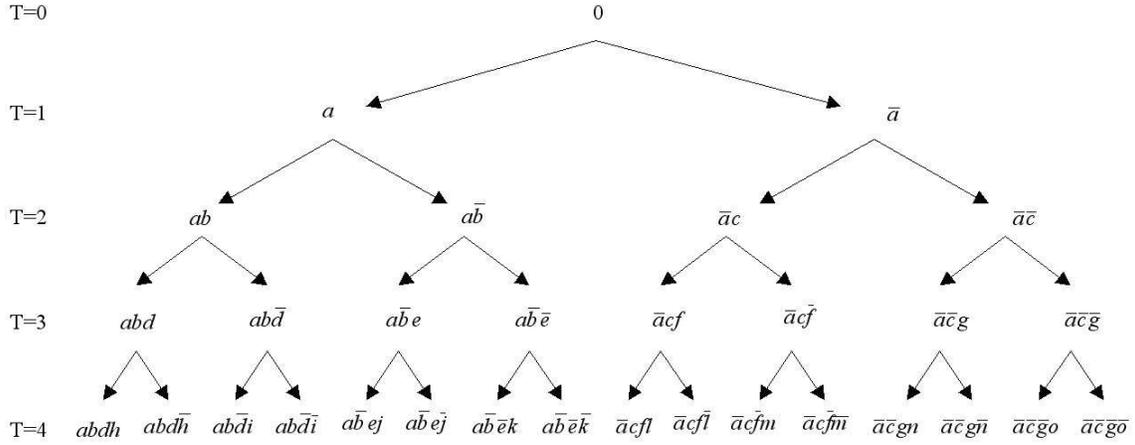}
\caption{Classical probability `tree' for a
developing bit.}\label{Figure 6.1}
\end{center}
\end{figure}

In this model, any information regarding the value of the bit at time $T-1$
is irretrievably lost at time $T.$ Without a form of `memory' recording the
bit's development there is no way of reconstructing any of the histories
featured, and hence there is no way of concluding, for example, at time $T>2$
that the bit had the value $x_{1}$ at time $T=1,$ and had the value $x_{2}$
at time $T=2,$ etc. All that can be known at time $T$ is that the bit has
either the value $x_{T}=0$ or $1$ \textit{now.}

To overcome this it is therefore desirable to incorporate a type of
`information store' into the model. Consider, as a possible method of
achieving such an effect, an $M$ bit system labelled\ at time $T$ by the
string $L_{T}=[x_{0}x_{1}x_{2}...x_{M-1}]_{T},$ where $x_{i}=0,1$ denotes
the value of the $i^{th}$ bit, for $i=0,...,(M-1),$ labelled by left-right
position $i+1.$ Moreover, assume that at initial time $T=0$ all the bits
have the value $0,$ such that $L_{0}=[0_{0}0_{1}0_{2}...0_{M-1}]_{0}.$

Note the change in the use of the sub-script on the variable $x_{i}$ here;
the symbol $x_{i}$ is now used to denote the value $x$ of the $i^{th}$ bit,
and not a value at time $i,$ as it was previously. Instead, the sub-script $%
T $ on $L_{T}$ and outside the square brackets $[...]_{T}$ takes the place
of the temporal parameter. Thus, the expression $%
L_{T}=[x_{0}x_{1}x_{2}...x_{M-1}]_{T}$ denotes a string $%
x_{0}x_{1}x_{2}...x_{M-1}$ of $M$ bits of respective values $x_{0},$ $x_{1},$
$x_{2},$ $...,$ $x_{M-1}$ (for $x_{i}=0,1)$ at time $T,$ labelled by $L_{T}.$
So for example if $L_{3}=[1_{0}0_{1}1_{2}...0_{M-1}]_{3},$ it implies that
at time $T=3$ the $0^{th}$ bit currently has the value $1,$ whilst the $%
1^{st}$ bit has the value $0,$ whereas the $2^{nd}$ bit has the value $1,$
and so on.

In actual fact, this distinction will not matter much in the following,
since the $i^{th}$ bit will be used below to encode information at time $i;$
the underlying change in sub-script nomenclature is, however, nevertheless
apparent.

Let the zeroth bit, $i=0,$ be associated with the single developing bit
discussed so far in this section. So, in this case the previous dynamics is
equivalent to the operation
\begin{equation}
L_{T+1}=R(L_{T})=R([x_{0}x_{1}x_{2}...x_{M-1}]_{T})=R([x_{0}0_{1}0_{2}...0_{M-1}]_{T})=[x_{0}0_{1}0_{2}...0_{M-1}]_{T+1}
\end{equation}
such that for example the string $L_{1}=[0_{0}0_{1}0_{2}...0_{M-1}]_{1}$
occurs with probability $a,$ whereas the string $%
L_{1}=[1_{0}0_{1}0_{2}...0_{M-1}]_{1}$ occurs with probability $\bar{a}.$
Further, the probability that the particular string $%
L_{2}=[1_{0}0_{1}0_{2}...0_{M-1}]_{2}$ occurs at time $2,$ and that the
previous string was $L_{1}=[0_{0}0_{1}0_{2}...0_{M-1}]_{1},$ is clearly $a%
\bar{b}.$ And so on.

To incorporate a `memory' into the system, the rule is now introduced that
at time $T$ the $T^{th}$ bit is assigned the current value of the zeroth
bit, where $T\leq (M-1).$ This result is accomplished by an operation $%
C_{(0,T)},$ where
\begin{equation}
C_{(0,T)}(L_{T})=C_{(0,T)}([x_{0}x_{1}x_{2}...x_{T-1}x_{T}x_{T+1}...x_{M-1}]_{T})=[x_{0}x_{1}x_{2}...x_{T-1}x_{0}x_{T+1}...x_{M-1}]_{T}
\end{equation}
so that the value of the $T^{th}$ bit is therefore providing a record of the
value of the zeroth bit at time $T.$

The rule is obviously based on the relations \{if $x_{0}=0$ at time $T$ then
$x_{T}\rightarrow 0\}$ and \{if $x_{0}=1$ at time $T$ then $x_{T}\rightarrow
1\},$ and given that every bit initially has the value $0$ it is clear to
see that $C_{(0,T)}$ is just the CNOT operator of Section 6.1.1.

Overall, then, the development of this system from time $T$ to $T+1$ follows
the procedure $L_{T+1}=C_{(0,T+1)}(R(L_{T})).$

In order to `remember' the four time steps illustrated in Figure 6.1, a
model involving\ at least five bits is required. At initial time $T=0$ the
system is described by the string $L_{0}=[0_{0}0_{1}0_{2}0_{3}0_{4}]_{0},$
and as it develops the bits labelled $1,2,3$ and $4$ record the history of
bit $0$ at those respective times. As an example, given a large number $N$
of such systems it is expected that at time $T=4$ a fraction $N\bar{a}cf\bar{%
l}$ will have the string $L_{4}=[1_{0}1_{1}0_{2}0_{3}1_{4}]_{4},$ because
the probability $P(01001)$ that the value of bit $0$ develops from $0$ to $1$
to $0$ to $0$ to $1$ over the four times steps is given by $\bar{a}cf\bar{l}%
. $ Similarly, the probability $P(01101)$ that the zeroth bit had the values
$0,$ $1,1,0$ \ and $1$ at times $T=0,1,2,3$ and $4$ respectively is $\bar{a}%
\bar{c}g\bar{n},$ and so if $N$ five bit systems were developed, $N\bar{a}%
\bar{c}g\bar{n}$ of them would be expected to have the configuration $%
L_{4}=[1_{0}1_{1}1_{2}0_{3}1_{4}]_{4}$ at time $T=4.$\bigskip

With the probabilities of obtaining specific histories of results defined in
Figure 6.1, it is possible to examine the probability of obtaining a
particular type of result. For example, a physicist might be interested in
the probability $p_{\alpha }$ that at time $T=3$ the bit, irrespective of
its history, has the value $0,$ and this is readily given by the sum $%
p_{\alpha }=abd+a\bar{b}e+\bar{a}cf+\bar{a}\bar{c}g.$ Alternatively, it
might be of concern to discuss the probability $p_{\beta }$ that at time $%
T=4 $ the bit has the value $1,$ given that at time $T=3$ it also had the
value $1,$ and this is clearly $p_{\beta }=ab\bar{d}\bar{\imath}+a\bar{b}%
\bar{e}\bar{k}+\bar{a}c\bar{f}\bar{m}+\bar{a}\bar{c}\bar{g}\bar{o}.$

One particular probability of interest is the correlation $K_{xy},$ defined
as the probability that the bit had the same value at times $T=x$ and $T=y$
minus the probability that the bit had different values at times $T=x$ and $%
T=y.$ Thus for an experiment performed $N$ times, $K_{xy}$ is equally
defined as the number of histories for which the bit had different,
anti-correlated values at times $x$ and $y,$ subtracted from the number of
histories for which the bit had the same, correlated value at times $x$ and $%
y,$ all divided by $N.$ As an example, the correlation $K_{12}$ is given by
\begin{eqnarray}
K_{12} &=&(ab+\bar{a}\bar{c})-(a\bar{b}+\bar{a}c) \\
&=&ab+(1-a)(1-c)-a(1-b)-(1-a)c  \notag \\
&=&1-2a-2c+2ab+2ac.  \notag
\end{eqnarray}

It is further possible to define `multiple correlations'. One such
possibility, $K,$ is given by
\begin{equation}
K=K_{12}+K_{23}+K_{34}-K_{14}
\end{equation}
which reduces to
\begin{eqnarray}
K &=&2-4(c+g)+4(ac-ae+cg+ag-ak+gn) \\
&&+4(abe-acg-abi+abk+aek+cfl-agn-cgn)  \notag \\
&&+4(abdi-abek-acfl+acgn)  \notag
\end{eqnarray}
given the above probabilities. Moreover, because each individual probability
$a,b,...,o$ is $0\leq a,b,...,o\leq 1$ it can readily be shown that $K$
satisfies the inequality
\begin{equation}
-2\leq K\leq 2.  \label{Ch4ClassBell}
\end{equation}

The importance of (\ref{Ch4ClassBell}) is that it is the temporal equivalent
of the spacelike classical Bell inequality, as given in Section 3.1, for a
bit classically developing over four time steps. It represents the relations
between unequal-time correlations induced in a single bit system, where at
each time the bit has a definite and independently existing value ($0$ or $%
1) $ irrespective of whether or not it is actually measured. Moreover,
because every classical bit must follow one of these 16 possible histories
between times $T=1$ and $4,$ the correlations of any bit that obeys
classical equations of dynamics must necessarily satisfy this inequality.

Furthermore, the unequal-time correlations of a single qubit would also have
to obey this temporal Bell inequality \textit{if} it were to be argued that
its `quantum' dynamics ultimately depended upon any sort of classical hidden
variables, or if the value of the qubit at any time could somehow be known
without disturbing the system. This issue will be discussed now.

\bigskip

\subsubsection{Quantum Calculations and the Bell Inequality.}

\bigskip

In this final part of Section 6.1, the ideas of the previous three
subsections will be drawn together to show how the quantum CNOT operator can
be used to perform a particular quantum computation on a set of qubits.
Moreover, because the CNOT operator can be applied to the qubits a number of
times in succession, the possibility of comparing correlations between
results at unequal times might be expected to arise, just as in the case of
the classical calculation described above. Thus, a `quantised' version of
the model introduced in 6.1.3 is presented, and this leads to a generation
of relations analogous to the classical temporal Bell inequality, which in
turn allows the implications and validity of the suggested quantum method to
be explored.

Note that an alternative analysis of this was introduced in \cite{Kim}. It
is felt, however, that approach lacked clarity, and so in this final
sub-section an improvement and reinterpretation of the issues advocated by
Kim \textit{et al} is sought. Important conclusions may then be drawn about
the nature of quantum computation, and these will consequently provide
insight into some of the limitations inherent in any attempt to treat the
Universe as a giant quantum computer.\bigskip

In the following work, subscripts are generally used to label qubits, with
the exception that the subscript on $\Psi _{T}$ is used to denote a
(discrete) temporal parameter, $T=0,1,2,...$ As a further comment on this,
note that $T$ is taken below as a time parameter that is external to the
qubit system being discussed; thus, $T$ is the `usual' time of conventional
exo-physics, as experienced by an observer who is not part of the system
under investigation. This last point will be an important factor in the
later interpretation of the model.

Consider a five qubit system represented at time $T$ by a state $\Psi _{T}$
in a Hilbert space $\mathcal{H}_{[0...4]}^{(32)}$ spanned by an orthonormal
basis $\mathcal{B}_{0...4}\equiv \{|i\rangle _{0}\otimes |j\rangle
_{1}\otimes |k\rangle _{2}\otimes |l\rangle _{3}\otimes |m\rangle
_{4}:i,j,k,l,m=0,1\}.$ Assume that at initial time $T=0$ every qubit is in
the `down' state $|0\rangle ,$ such that the overall state may be written
\begin{equation}
|\Psi _{0}\rangle =|0\rangle _{0}\otimes |0\rangle _{1}\otimes |0\rangle
_{2}\otimes |0\rangle _{3}\otimes |0\rangle _{4}=|00000\rangle
\end{equation}
where the qubit subscripts may be omitted in favour of left-right
positioning for brevity.

Consider also an unitary operator $\hat{S}_{0}$ acting locally on qubit $0,$
defined as
\begin{equation}
\hat{S}_{0}\equiv \exp \left( -\frac{i}{2}\alpha \hat{\sigma}_{0}^{1}\right)
\label{Ch4S0}
\end{equation}
where $\alpha \in \mathbb{R}$ is a small real parameter and $\hat{\sigma}%
_{0}^{a}$ is the $a^{th}$ Pauli operator $\hat{\sigma}^{a}$ acting in
subregister $\mathcal{H}_{0}$ for $a=1,2,3.$

Note that a unitary operator $\hat{S}_{0}$ acting locally on qubit $0$ is
equivalent to the operator $\hat{S}_{0}^{G}$ acting globally on the entire
state $\Psi ,$ where $\hat{S}_{0}^{G}$ is defined by
\begin{equation}
\hat{S}_{0}^{G}\equiv \hat{S}_{0}\otimes \hat{I}_{1}\otimes \hat{I}%
_{2}\otimes \hat{I}_{3}\otimes \hat{I}_{4}
\end{equation}
with $\hat{I}_{t}$ the identity operator in Hilbert space $\mathcal{H}_{t}$
for $t=0,1,2,3,4.$ This equivalence will be discussed in Chapter 7, but for
now note that the two forms of the operator are used interchangeably.

The exponential (\ref{Ch4S0}) can be expanded, to obtain
\begin{eqnarray}
\hat{S}_{0} &=&\sum\nolimits_{n=0}^{\infty }\left( -\frac{i}{2}\alpha
\right) ^{n}\frac{(\hat{\sigma}_{0}^{1})^{n}}{n!} \\
&=&\sum\nolimits_{n=0}^{\infty }\left( -\frac{i}{2}\alpha \right) ^{2n}\frac{%
(\hat{\sigma}_{0}^{1})^{2n}}{(2n)!}+\sum\nolimits_{n=0}^{\infty }\left( -%
\frac{i}{2}\alpha \right) ^{2n+1}\frac{(\hat{\sigma}_{0}^{1})^{2n+1}}{(2n+1)!%
}  \notag \\
&=&(\hat{\sigma}_{0}^{0})\sum\nolimits_{n=0}^{\infty }\frac{1}{(2n)!}\left( -%
\frac{i}{2}\alpha \right) ^{2n}+(\hat{\sigma}_{0}^{1})\sum\nolimits_{n=0}^{%
\infty }\frac{1}{(2n+1)!}\left( -\frac{i}{2}\alpha \right) ^{2n+1}  \notag \\
&=&\cos \left( \frac{\alpha }{2}\right) \hat{\sigma}_{0}^{0}-i\sin \left(
\frac{\alpha }{2}\right) \hat{\sigma}_{0}^{1}.  \notag
\end{eqnarray}

Consider also the generalised definition of the two-qubit, quantum CNOT
operator $\hat{C}_{(r,s)}$ that acts on subregisters $r$ and $s$%
\begin{equation}
\hat{C}_{(r,s)}\equiv \hat{P}_{r}^{0}\otimes \hat{\sigma}_{s}^{0}+\hat{P}%
_{r}^{1}\otimes \hat{\sigma}_{s}^{1}
\end{equation}
with $\hat{P}_{r}^{z}=|z\rangle _{rr}\langle z|$ for $z=0,1.$ Again, this
operator can also be extended to one that acts globally by taking a suitable
product with $\hat{I}_{t}$ for $t=0,1,...,4$ where $t\neq r,s.$\bigskip

In the example presented in this sub-section, and for reasons to become
apparent, the procedure performed by the quantum computation involves an
application of the operator $\hat{S}_{0}$ to the state $|\Psi _{T}\rangle ,$
followed by the operator $\hat{C}_{(r,s)}.$ Moreover, in the chosen dynamics
attention is restricted to the case where $r=0$ and $s=T+1,$ such that the
CNOT operator may be written $\hat{C}_{(0,T+1)}.$ Thus given a state $|\Psi
_{T}\rangle ,$ the state $|\Psi _{T+1}\rangle $ at time $T+1$ is
\begin{equation}
|\Psi _{T+1}\rangle =\hat{C}_{(0,T+1)}\hat{S}_{0}|\Psi _{T}\rangle .
\end{equation}

The state $|\Psi _{T+1}\rangle $ may itself then be evolved with $\hat{S}%
_{0} $ and $\hat{C}_{(0,T+2)}$ to generate the later state $|\Psi
_{T+2}\rangle .$ So overall, given the initial condition of $\Psi _{0}$ at $%
T=0,$ the state $\Psi _{T}$ at time $T$ will be
\begin{equation}
|\Psi _{T}\rangle =(\hat{C}_{(0,T)}\hat{S}_{0})(\hat{C}_{(0,T-1)}\hat{S}%
_{0})...(\hat{C}_{(0,2)}\hat{S}_{0})(\hat{C}_{(0,1)}\hat{S}_{0})|\Psi
_{0}\rangle =\hat{U}_{T}|\Psi _{0}\rangle  \label{Ch4Unit}
\end{equation}
where $\hat{U}_{T}\equiv (\hat{C}_{(0,T)}\hat{S}_{0})(\hat{C}_{(0,T-1)}\hat{S%
}_{0})...(\hat{C}_{(0,2)}\hat{S}_{0})(\hat{C}_{(0,1)}\hat{S}_{0})$ is also a
unitary operator.\bigskip

Given the above general algorithm, the specific development of the state of
the five qubit system under investigation proceeds as follows. If the
initial state $\Psi _{0}$ is $|00000\rangle ,$ then
\begin{eqnarray}
\hat{S}_{0}|00000\rangle &=&\left( \cos \left( \frac{\alpha }{2}\right) \hat{%
\sigma}_{0}^{0}-i\sin \left( \frac{\alpha }{2}\right) \hat{\sigma}%
_{0}^{1}\right) \otimes |00000\rangle _{01234} \\
&=&\left( \cos \QOVERD( ) {\alpha }{2}|0\rangle _{0}-i\sin \QOVERD( )
{\alpha }{2}|1\rangle _{0}\right) \otimes |0000\rangle _{1234}  \notag
\end{eqnarray}
and this result is subsequently `registered' with qubit $1$ by the CNOT
operator acting in the prescribed way, such that the later state $|\Psi
_{1}\rangle =\hat{C}_{(0,1)}\hat{S}_{0}|00000\rangle $ is given by
\begin{eqnarray}
|\Psi _{1}\rangle &=&\left[ \hat{P}_{0}^{0}\otimes \hat{\sigma}_{1}^{0}+\hat{%
P}_{0}^{1}\otimes \hat{\sigma}_{1}^{1}\right] \left( \cos \QOVERD( ) {\alpha
}{2}|0\rangle _{0}\otimes |0\rangle _{1}-i\sin \QOVERD( ) {\alpha
}{2}|1\rangle _{0}\otimes |0\rangle _{1}\right) \otimes |000\rangle _{234}
\notag \\
&=&\left( \cos \QOVERD( ) {\alpha }{2}|0\rangle _{0}\otimes |0\rangle
_{1}-i\sin \QOVERD( ) {\alpha }{2}|1\rangle _{0}\otimes |1\rangle
_{1}\right) \otimes |000\rangle _{234}.
\end{eqnarray}

Continuing,
\begin{eqnarray}
\hat{S}_{0}|\Psi _{1}\rangle &=&\left[ \cos \left( \frac{\alpha }{2}\right)
\hat{\sigma}_{0}^{0}-i\sin \left( \frac{\alpha }{2}\right) \hat{\sigma}%
_{0}^{1}\right] \left( \cos \QOVERD( ) {\alpha }{2}|00\rangle _{01}-i\sin
\QOVERD( ) {\alpha }{2}|11\rangle _{01}\right) \otimes |000\rangle _{234}
\notag \\
&=&\left(
\begin{array}{c}
\cos ^{2}\QOVERD( ) {\alpha }{2}|00\rangle _{01}-i\cos \left( \frac{\alpha }{%
2}\right) \sin \QOVERD( ) {\alpha }{2}|11\rangle _{01} \\
-i\sin \left( \frac{\alpha }{2}\right) \cos \QOVERD( ) {\alpha
}{2}|10\rangle _{01}-\sin ^{2}\QOVERD( ) {\alpha }{2}|01\rangle _{01}
\end{array}
\right) \otimes |000\rangle _{234}
\end{eqnarray}
and so the subsequent state $|\Psi _{2}\rangle =\hat{C}_{(0,2)}\hat{S}%
_{0}|\Psi _{1}\rangle $ is given by
\begin{eqnarray}
|\Psi _{2}\rangle &=&\left[ \hat{P}_{0}^{0}\otimes \hat{\sigma}_{2}^{0}+\hat{%
P}_{0}^{1}\otimes \hat{\sigma}_{2}^{1}\right] \left(
\begin{array}{c}
\cos ^{2}\QOVERD( ) {\alpha }{2}|000\rangle _{012}-i\cos \left( \frac{\alpha
}{2}\right) \sin \QOVERD( ) {\alpha }{2}|110\rangle _{012} \\
-i\sin \left( \frac{\alpha }{2}\right) \cos \QOVERD( ) {\alpha
}{2}|100\rangle _{012}-\sin ^{2}\QOVERD( ) {\alpha }{2}|010\rangle _{012}
\end{array}
\right) \otimes |00\rangle _{34}  \notag \\
&=&\left(
\begin{array}{c}
\cos ^{2}\QOVERD( ) {\alpha }{2}|000\rangle _{012}-i\cos \left( \frac{\alpha
}{2}\right) \sin \QOVERD( ) {\alpha }{2}|111\rangle _{012} \\
-i\sin \left( \frac{\alpha }{2}\right) \cos \QOVERD( ) {\alpha
}{2}|101\rangle _{012}-\sin ^{2}\QOVERD( ) {\alpha }{2}|010\rangle _{012}
\end{array}
\right) \otimes |00\rangle _{34}.
\end{eqnarray}

The states $|\Psi _{3}\rangle $ and $|\Psi _{4}\rangle $ at times $T=3$ and $%
4$ can generated in a similar way, with the results
\begin{equation}
|\Psi _{3}\rangle =\left(
\begin{array}{c}
\cos ^{3}\QOVERD( ) {\alpha }{2}|0000\rangle _{0123}-i\cos ^{2}\left( \frac{%
\alpha }{2}\right) \sin \QOVERD( ) {\alpha }{2}|1111\rangle _{0123} \\
-i\sin \left( \frac{\alpha }{2}\right) \cos ^{2}\QOVERD( ) {\alpha
}{2}|1011\rangle _{0123}-\cos \QOVERD( ) {\alpha }{2}\sin ^{2}\QOVERD( )
{\alpha }{2}|0100\rangle _{0123} \\
-i\sin \left( \frac{\alpha }{2}\right) \cos ^{2}\QOVERD( ) {\alpha
}{2}|1001\rangle _{0123}-\cos \left( \frac{\alpha }{2}\right) \sin
^{2}\QOVERD( ) {\alpha }{2}|0110\rangle _{0123} \\
-\sin ^{2}\left( \frac{\alpha }{2}\right) \cos \QOVERD( ) {\alpha
}{2}|0010\rangle _{0123}+i\sin ^{3}\QOVERD( ) {\alpha }{2}|1101\rangle
_{0123}
\end{array}
\right) \otimes |0\rangle _{4}
\end{equation}
and
\begin{equation}
|\Psi _{4}\rangle =\left(
\begin{array}{c}
\cos ^{4}\QOVERD( ) {\alpha }{2}|00000\rangle _{01234}-i\cos ^{3}\left(
\frac{\alpha }{2}\right) \sin \QOVERD( ) {\alpha }{2}|11111\rangle _{01234}
\\
-i\sin \left( \frac{\alpha }{2}\right) \cos ^{3}\QOVERD( ) {\alpha
}{2}|10111\rangle _{01234}-\cos ^{2}\QOVERD( ) {\alpha }{2}\sin ^{2}\QOVERD(
) {\alpha }{2}|01000\rangle _{01234} \\
-i\sin \left( \frac{\alpha }{2}\right) \cos ^{3}\QOVERD( ) {\alpha
}{2}|10011\rangle _{01234}-\cos ^{2}\left( \frac{\alpha }{2}\right) \sin
^{2}\QOVERD( ) {\alpha }{2}|01100\rangle _{01234} \\
-\sin ^{2}\left( \frac{\alpha }{2}\right) \cos ^{2}\QOVERD( ) {\alpha
}{2}|00100\rangle _{01234}+i\cos \QOVERD( ) {\alpha }{2}\sin ^{3}\QOVERD( )
{\alpha }{2}|11011\rangle _{01234} \\
-i\sin \left( \frac{\alpha }{2}\right) \cos ^{3}\QOVERD( ) {\alpha
}{2}|10001\rangle _{01234}-\cos ^{2}\left( \frac{\alpha }{2}\right) \sin
^{2}\QOVERD( ) {\alpha }{2}|01110\rangle _{01234} \\
-\sin ^{2}\left( \frac{\alpha }{2}\right) \cos ^{2}\QOVERD( ) {\alpha
}{2}|00110\rangle _{01234}+i\cos \QOVERD( ) {\alpha }{2}\sin ^{3}\QOVERD( )
{\alpha }{2}|11001\rangle _{01234} \\
-\sin ^{2}\left( \frac{\alpha }{2}\right) \cos ^{2}\QOVERD( ) {\alpha
}{2}|00010\rangle _{01234}+i\cos \left( \frac{\alpha }{2}\right) \sin
^{3}\QOVERD( ) {\alpha }{2}|11101\rangle _{01234} \\
+i\sin ^{3}\left( \frac{\alpha }{2}\right) \cos \QOVERD( ) {\alpha
}{2}|10101\rangle _{01234}+\sin ^{4}\QOVERD( ) {\alpha }{2}|01010\rangle
_{01234}
\end{array}
\right) .  \label{Ch4Psi4}
\end{equation}

For the purposes of illustration, it is possible to associate a classical
interpretation to the action of the sequence $(\hat{C}_{(0,1)}\hat{S}_{0}),$
$(\hat{C}_{(0,2)}\hat{S}_{0}),...$ and the consequent development of the
system. Firstly, the operator $\hat{S}_{0}$ may be thought of as one that
locally `rotates' the state of qubit $0$ independently of the other four
qubits. Then, during the development of the system from time $T$ to $T+1$ it
is assumed that the CNOT operator somehow `examines' the state of qubit $0$
before `extracting' this information and registering it with qubit $T+1.$
Moreover, in the current example this registration appears analogous to a
`copying' procedure, because each qubit is initially in the state $|0\rangle
$ and CNOT operates according to the rules $\hat{C}_{(a,b)}(|0\rangle
_{a}\otimes |0\rangle _{b})\rightarrow |0\rangle _{a}\otimes |0\rangle _{b}$
and $\hat{C}_{(a,b)}(|1\rangle _{a}\otimes |0\rangle _{b})\rightarrow
|1\rangle _{a}\otimes |1\rangle _{b}.$

Thus, the action of the operators may be thought of as producing a `wave of
information' that sweeps through the system, moving in time along the chain
of qubits. During the evolution from $\Psi _{T}$ to $\Psi _{T+1}$ only the $%
0^{th}$ and $(T+1)^{th}$ qubits appear affected, and once this classical
looking `wave' has `passed' the $(T+1)^{th}$ qubit its sub-state is never
again altered by the dynamics. So, the qubit $T+1$ appears to serve as a
permanent record of the sub-state of qubit $0$ after the application of $%
\hat{S}$ to $\Psi _{T}.$\bigskip

Because the spins of the qubits $1,2,3$ and $4$ appear to contain
information about the state of qubit $0$ at times $T=1,$ $T=2,$ $T=3$ and $%
T=4$ respectively, it might be natural to expect that these four qubits
could be interrogated\ in order to learn about the `history' of the
development of qubit $0.$ In fact, because once information is encoded into
these qubits it is assumed permanent, the individual spins of these four
correlated qubits $1,2,3$ and $4$ in the final state $\Psi _{4}$ might be
expected to contain a record of the whole history of qubit $0$'s
development. In other words, it may be hoped that by asking a specific
question about the state of the $T^{th}$ qubit of $\Psi _{4},$ insight might
be gained into the state of qubit $0$ at time $T,$ where $1\leq T\leq 4.$

Furthermore, in fact, and following the lead of \cite{Kim}, one possible
such insight might involve the quantum correlation $Q_{xy}$ defined as
\begin{equation}
Q_{xy}=\langle \Psi _{4}|\hat{\sigma}_{x}^{3}\hat{\sigma}_{y}^{3}|\Psi
_{4}\rangle  \label{Ch4Qxy}
\end{equation}
which is the quantum analogue of the classical correlation $K_{xy}$ defined
in Section 6.1.3, and where $x,y=1,2,3,4$ indicate the four `memory' qubits.
Again $\hat{\sigma}_{x}^{3}$ is a Pauli operator acting locally on qubit
space $x,$ and may therefore as before also be associated with an equivalent
global operator by extending it in the obvious way.

The validity of this `insight' is investigated now.\bigskip

In the context of (\ref{Ch4Qxy}), the operator $\hat{\sigma}_{x}^{3}$ may be
interpreted as an object that `asks a question' of the spin of qubit $x$ in $%
\Psi _{4}$ whilst doing nothing to any of the other qubits: if the $x^{th}$
qubit is in the spin-down state $|0\rangle _{x}=\binom{1}{0}_{x}$ the
operation $\hat{\sigma}_{x}^{3}|0\rangle _{x}=\binom{1\text{ \ \ }0}{0\text{
\ }-1}_{x}\binom{1}{0}_{x}$ gives a result $|0\rangle _{x},$ whereas if the $%
x^{th}$ qubit is in the spin-up state $|1\rangle _{x}=\binom{0}{1}_{x}$ then
$\hat{\sigma}_{x}^{3}|1\rangle _{x}$ gives a result $-|1\rangle _{x}.$

So, the correlation $Q_{xy}$ sums the amplitudes of the terms in $|\Psi
_{4}\rangle $ for which the $x^{th}$ and $y^{th}$ qubits have the same spin
state, and subtracts from this the amplitudes of terms in $|\Psi _{4}\rangle
$ for which the $x^{th}$ and $y^{th}$ qubits have opposite spin states.
Moreover, due to the sequential procedure followed in the generation of $%
\Psi _{4},$ the suggestion is then that $Q_{xy}$ may be thought of as the
correlation between the state of qubit $0$ at the times $x$ and $y.$ It is
this suggestion that is now examined.\bigskip

First, though, the correlations $Q_{xy}$ must be evaluated. In the present
representation $Q_{12}=\langle \Psi _{4}|\hat{\sigma}_{1}^{3}\hat{\sigma}%
_{2}^{3}|\Psi _{4}\rangle $ is given by
\begin{eqnarray}
Q_{12} &=&\cos ^{4}\QOVERD( ) {\alpha }{2}\cos ^{4}\QOVERD( ) {\alpha }{2}
\left[ _{01234}\langle 00000|\hat{\sigma}_{1}^{3}\hat{\sigma}%
_{2}^{3}|00000\rangle _{01234}\right] \\
&&+\cos ^{4}\QOVERD( ) {\alpha }{2}(-i\cos ^{3}\left( \frac{\alpha }{2}%
\right) \sin \QOVERD( ) {\alpha }{2})\left[ _{01234}\langle 00000|\hat{\sigma%
}_{1}^{3}\hat{\sigma}_{2}^{3}|11111\rangle _{01234}\right] +...  \notag \\
&&+(+i\cos ^{3}\left( \frac{\alpha }{2}\right) \sin \QOVERD( ) {\alpha
}{2})(-i\cos ^{3}\left( \frac{\alpha }{2}\right) \sin \QOVERD( ) {\alpha
}{2})\left[ _{01234}\langle 11111|\hat{\sigma}_{1}^{3}\hat{\sigma}%
_{2}^{3}|11111\rangle _{01234}\right] +  \notag \\
&&...+\sin ^{4}\QOVERD( ) {\alpha }{2}\sin ^{4}\QOVERD( ) {\alpha }{2}\left[
_{01234}\langle 01010|\hat{\sigma}_{1}^{3}\hat{\sigma}_{2}^{3}|01010\rangle
_{01234}\right]  \notag
\end{eqnarray}
which, by using the orthonormality relation
\begin{equation}
_{01234}\langle i^{\prime }j^{\prime }k^{\prime }l^{\prime }m^{\prime
}|ijklm\rangle _{01234}=\delta _{ii^{\prime }}\delta _{jj^{\prime }}\delta
_{kk^{\prime }}\delta _{ll^{\prime }}\delta _{mm^{\prime }}
\end{equation}
becomes
\begin{eqnarray}
Q_{12} &=&\cos ^{8}\QOVERD( ) {\alpha }{2}+\cos ^{6}\left( \frac{\alpha }{2}%
\right) \sin ^{2}\QOVERD( ) {\alpha }{2}-\sin ^{2}\left( \frac{\alpha }{2}%
\right) \cos ^{6}\QOVERD( ) {\alpha }{2}-\cos ^{4}\QOVERD( ) {\alpha
}{2}\sin ^{4}\QOVERD( ) {\alpha }{2}  \notag \\
&&+\sin ^{2}\left( \frac{\alpha }{2}\right) \cos ^{6}\QOVERD( ) {\alpha
}{2}+\cos ^{4}\left( \frac{\alpha }{2}\right) \sin ^{4}\QOVERD( ) {\alpha
}{2}-\sin ^{4}\left( \frac{\alpha }{2}\right) \cos ^{4}\QOVERD( ) {\alpha
}{2}  \notag \\
&&-\cos ^{2}\QOVERD( ) {\alpha }{2}\sin ^{6}\QOVERD( ) {\alpha }{2}+\sin
^{2}\left( \frac{\alpha }{2}\right) \cos ^{6}\QOVERD( ) {\alpha }{2}+\sin
^{4}\left( \frac{\alpha }{2}\right) \cos ^{4}\QOVERD( ) {\alpha }{2} \\
&&-\sin ^{4}\left( \frac{\alpha }{2}\right) \cos ^{4}\QOVERD( ) {\alpha
}{2}-\cos ^{2}\QOVERD( ) {\alpha }{2}\sin ^{6}\QOVERD( ) {\alpha }{2}+\sin
^{4}\left( \frac{\alpha }{2}\right) \cos ^{4}\QOVERD( ) {\alpha }{2}  \notag
\\
&&+\cos ^{2}\left( \frac{\alpha }{2}\right) \sin ^{6}\QOVERD( ) {\alpha
}{2}-\sin ^{6}\left( \frac{\alpha }{2}\right) \cos ^{2}\QOVERD( ) {\alpha
}{2}-\sin ^{8}\QOVERD( ) {\alpha }{2}  \notag \\
&=&\cos ^{8}\QOVERD( ) {\alpha }{2}+2\sin ^{2}\left( \frac{\alpha }{2}%
\right) \cos ^{6}\QOVERD( ) {\alpha }{2}-2\cos ^{2}\QOVERD( ) {\alpha
}{2}\sin ^{6}\QOVERD( ) {\alpha }{2}-\sin ^{8}\QOVERD( ) {\alpha }{2}  \notag
\\
&=&\cos (\alpha ).  \notag
\end{eqnarray}

Similarly, it can be shown that
\begin{eqnarray}
Q_{23} &=&\cos (\alpha ) \\
Q_{34} &=&\cos (\alpha )  \notag
\end{eqnarray}
but
\begin{equation}
Q_{14}=\cos ^{3}(\alpha ).
\end{equation}

As with the classical result $K$ of Section 6.1.3, it is possible to define
a multiple correlation $Q$ in the manner
\begin{equation}
Q=Q_{12}+Q_{23}+Q_{34}-Q_{14}.
\end{equation}
such that in the present case
\begin{equation}
Q=3\cos (\alpha )-\cos ^{3}(\alpha ).
\end{equation}
\

However, by differentiating it can readily be shown that
\begin{equation}
-2\leq Q\leq 2
\end{equation}
for all $\alpha ,$ exactly as in the case of the classical Bell inequality.
Thus, it appears that the qubit system is obeying classical rules of
dynamics, which initially suggests that something has gone wrong in the
analysis: if the qubits are to obey quantum dynamics, they might be expected
to violate the Bell inequality for at least some values of $\alpha .$

\bigskip

The problem with the above process is that at first glance the operator $%
\hat{C}_{(0,T)}$ seems to be behaving like an information extraction
process. Every time a CNOT operator is used it seems to imply a modification
of the state such that parts of it appear to be `storing' information
regarding the current state of qubit $0.$ An attempt is then made to access
this store at some later time.

This can lead to interpretational difficulties in quantum mechanics. In
classical mechanics it is perfectly reasonable to discuss a system of
individual bits, each of which possesses a definite value at all times.
Moreover, the bits possess these values whether or not they are actually
observed. So, in a classical dynamics it is possible to consider the type of
model described in Section 6.1.3 in which at any time $T$ the `zeroth' bit
has a certain and specific value, and this value is unambiguously and
non-invasively copied by the bit $T$ without affecting anything else.

In quantum theory, however, the same is not true. Firstly, there can be no
analogous copying procedure in quantum mechanics. The No-Cloning theorem
\cite{No-Clone} demonstrates that there is no general unitary operator $\hat{%
u}$ that maps an arbitrary initial product state of the form $\psi
=|A\rangle _{i}\otimes |B\rangle _{j}$ into a final product state $\psi
^{\prime }=|A\rangle _{i}\otimes |A\rangle _{j},$ where $|X\rangle _{a}\in
\mathcal{H}_{a}$ for $a=i,j,$ such that $\psi ^{\prime }=\hat{u}\psi .$ Even
though this is not a direct limitation in the current case because the CNOT
operator does not in general preserve separability, and is therefore not
actually trying to evolve states in this forbidden way, it is evident in
this respect that the classical CNOT operation and its quantum operator
counterpart are not completely equivalent in their action.

Secondly, when a system becomes entangled (a phenomenon unique to quantum
mechanics) it is no longer valid to discuss the components of the
entanglement in different factor Hilbert spaces as having any sort of
individual existence, independent of one another. Just as in the EPR
situation of Chapter 3 and the discussion of separability in Chapter 4, the
introduction of entanglement automatically and directly implies a breakdown
of the ability to state that ``this object with these properties is here''.
From this point of view it is therefore incorrect to say that qubit $0,$
which is initially a factor sub-state of the completely separable state $%
\Psi _{0},$ has any properties on its own, i.e. is either independently up
or down, at the times $T=1,2,3,4$ when it is entangled with the other qubits.

Thirdly, and perhaps more fundamentally, it is also not correct to assume
that the system has any physical properties at all independent of
observation, and so it is inaccurate to argue that the individual qubits are
in any definite state before the measurement. In other words, if the system
is not measured at time $T$ it is not only impossible to say which state it
is in at this time, but also to say that this state actually exists in any
sort of physical sense. It cannot therefore be said that the system is
undergoing any particular `history' or classical `trajectory'. In fact, the
fourth state $\Psi _{4}$ may be seen as an entangled linear superposition of
every potential classical `history', and it is only when a measurement is
finally performed that the system is forced to collapse into a particular
configuration of qubit sub-states. Moreover if the test is of a certain
sort, for example $\hat{\sigma}_{y}^{3},$ it may be natural at this point to
falsely conclude that the resulting product of sub-states indicates a
particular single-valued history for the system (and hence qubit $0),$
because this is what would be inducted in the world of classical physics
familiar to scientists.\bigskip

Strictly, in fact, it is actually misleading to even use the word `history',
and this point leads to an important comment on the role of time in quantum
dynamics. Recall that, normally, quantum probabilities are used to discuss
potential futures. In the present case, however, it might appear that an
attempt is being made to discuss a potential past, and this is contrary to
the usual assertion that the past is a definite and unique, well defined
classical construct.

But, the resolution of this contradiction is to note that the potential
pasts discussed in the superposition of `histories' in $\Psi _{4}$ instead
really form the basis for a set of potential futures. Further, in quantum
mechanics these potential futures are themselves only defined relative to
the eigenstates of whichever Hermitian operator is actually used to test the
state. More accurately, then, the discussion regarding the system's
`history' should perhaps be replaced by the question: ``if an operator $\hat{%
\sigma}_{y}^{3}$ is chosen that has eigenvectors $|00000\rangle _{01234},$ $%
|10000\rangle _{01234},$ $|01000\rangle _{01234},...,$ $|11111\rangle
_{01234}$ what is the probability of projecting $\Psi _{4}$ into one of
these possible future states?''.

Indeed, this point would be clarified further if, instead of $\hat{\sigma}%
_{y}^{3},$ the state $\Psi _{4}$ was tested by an operator that only had
entangled eigenstates; in this instance, none of the separable product
states $|ijklm\rangle _{01234}$ could be an outcome, and so no confusion
would occur by associating the result of this test with an apparent
classical history for the system.\bigskip

The issue can be addressed further. Because of the earlier discussions that
only observed states can be considered physical, it is not strictly
meaningful to consider the state having any physical reality whatsoever
between the preparation of the state $\Psi _{0}$ at time $0$ and the
measurement of $\Psi _{4}$ after time $4.$ The quantum system is not
proceeding through a definite sequence of states $\Psi _{0}\rightarrow \Psi
_{1}\rightarrow \Psi _{2}\rightarrow \Psi _{3}\rightarrow \Psi _{4},$ as
would be expected in a semi-classical model evolving along a specific
trajectory; instead, only initially prepared states $\Psi _{0}$ and measured
outcomes $\langle \Psi _{4}|\hat{\sigma}_{x}^{3}\hat{\sigma}_{y}^{3}|\Psi
_{4}\rangle $ are physically relevant.

In other words, from the point of view of the state there are not four
individual steps existing between $\Psi _{0}$ and $\Psi _{4},$ and this
conclusion may be highlighted by rewriting $\Psi _{4}=\hat{U}_{4}\Psi _{0}$
as in equation (\ref{Ch4Unit}) and noting that, because $\hat{U}_{4}\equiv
\hat{C}_{(0,4)}\hat{S}_{0}\hat{C}_{(0,3)}\hat{S}_{0}\hat{C}_{(0,2)}\hat{S}%
_{0}\hat{C}_{(0,1)}\hat{S}_{0}$ is just a unitary operator, the state $\Psi
_{4}$ is really only `one evolution' away from the state $\Psi _{0}.$

In fact, an immediate analogy may be drawn here to the quantum universe
model proposed throughout this thesis: two states $\Psi _{n}$ and $\Psi
_{n+1}$ of the Universe are deemed successive if there is only one collapse
`separating' them. The states $\Psi _{0}$ and $\Psi _{4}$ of the present
model, however, are not separated by any collapses; $\Psi _{4}$ is simply an
evolved version of $\Psi _{0},$ and as such it is not true to say that the
system has undergone four distinct developments.\bigskip

It might perhaps be more honest, then, to relabel $\Psi _{4}$ as $\Psi
_{0}^{\prime },$ that is, an evolved version of $\Psi _{0}.$ The `temporal
parameter' $4$ on $\Psi _{4}$ should really only be seen as a marker of the
external processes occurring during the unitary Schr\"{o}dinger evolution
from $\Psi _{0}\rightarrow \Psi _{0}^{\prime },$ for $\Psi _{0}^{\prime
}\equiv \Psi _{4}=\hat{U}_{4}\Psi _{0},$ such that the state in question can
only be said to have a `history' between $T=0$ and $T=4$ relative to the
development of the rest of the Universe during this time. This `external
history' occurs because an external scientist, who is assumed isolated from
the state, has physically applied the operator $\hat{S}_{0},$ followed by
the operator $\hat{C}_{(0,1)},$ then $\hat{S}_{0}$ again, then $\hat{C}%
_{(0,2)},$ and so on until she applies $\hat{C}_{(0,4)}.$ Indeed, recall
that $T$ was originally defined just as an external time parameter, and is
hence only valid relative to the observer who can remember `doing something'
in the interval during which the state was evolving, and thereby noticing
that, relative to other external processes, four time steps (or `(q-) ticks'
of the observer's clock \cite{Jaroszkiewicz}) appeared to elapse between the
preparation of the state $\Psi _{0}$ and the eventual measurement of $%
\langle \Psi _{4}|\hat{\sigma}_{x}^{3}\hat{\sigma}_{y}^{3}|\Psi _{4}\rangle
. $ This point reinforces the discussions in Chapters 5 and 8 that physical
time in the quantum Universe is not absolute but contextual, and should only
be discussed relative to change and the `path' taken by endo-observers.

Summarising, from the `internal' point of view of the isolated quantum
system, the state $\Psi _{0}$ develops to $\Psi _{4}\equiv \Psi _{0}^{\prime
}$ in one `rotation', whereas from an external viewpoint of an observer
developing and interacting with her surroundings in her own personal time $T$
the process appears to occur in four distinct steps. Moreover, it is by
falsely granting the external time parameter $T$ an unphysical, internal
significance that may be seen to form an origin of the current
difficulties.\bigskip

The problem is additionally complicated in the present situation by
misinterpreting the result of $\langle \Psi _{4}|\hat{\sigma}_{x}^{3}\hat{%
\sigma}_{y}^{3}|\Psi _{4}\rangle $ as seeming to indicate not only that the
system followed a particular classical path, but also as to what this path
was. Specifically, the inherent error is to assume that even if the state
could physically be discussed between external times $0$ and $4,$ the result
of $\langle \Psi _{4}|\hat{\sigma}_{x}^{3}\hat{\sigma}_{y}^{3}|\Psi
_{4}\rangle $ would actually provide insight into what it was doing. In this
case, the mistake lies in assuming that both the operators $\hat{\sigma}%
_{x}^{3}$ and $\hat{\sigma}_{y}^{3}$ are measuring the same state $\Psi
_{4}. $ This, however, is not true, because $\Psi _{4}$ cannot be measured
non-invasively: the measurement of $\Psi _{4}$ by the operator $\hat{\sigma}%
_{y}^{3}$ collapses the entangled state into one of the $16$ superposed
states\footnote{%
Given $\Psi _{4}$ and an operator with 32 orthonormal eigenstates of the
form $|ijklm\rangle _{01234}$ for $i,j,k,l,m=0,1,$ there are 16 non-zero
amplitudes $\langle ijklm|\Psi _{4}\rangle ,$ and hence 16 possible next
states.} given in equation (\ref{Ch4Psi4}), but by doing so destroys all of
the quantum interferences exhibited by $\Psi _{4}.$ Thus the system is
projected into a classical-looking product of qubit sub-states.

Any subsequent measurement with an operator $\hat{\sigma}_{x}^{3}$ then
produces with certainty either an up result or a down result, because qubit $%
x$ is no longer in an entangled superposition of the two. So, contrary to
what might be hoped, this further investigation of the system by the
operator $\hat{\sigma}_{x}^{3}$ is not asking a question about what the
state of qubit $x$ was \textit{before} the measurement of $\Psi _{4}$ by $%
\hat{\sigma}_{y}^{3},$ but is asking about what the state of qubit $x$ is%
\textit{\ afterwards}. Similarly, and from above, any such further
investigation by $\hat{\sigma}_{x}^{3}$ is not asking a question about what
the state of qubit $x$ (or equivalently qubit $0)$ \textit{was} at an
earlier (external) time $T=x,$ but is asking about what the state of qubit $%
x $ \textit{is} \textit{now}.

As an illustration of this, if $y=4$ and $\hat{\sigma}_{4}^{3}$ finds qubit $%
4$ to be up, and thereby collapses the system into, say, the state $\Phi $
given by $\Phi \equiv $ $|11001\rangle _{01234},$ the subsequent measurement
of $\Phi $ by $\hat{\sigma}_{1}^{3}$ seems at first glance to indicate that
qubit $1$ was definitely in an up state at external time $T=1.$ This,
however, is not the correct analysis: at external time $T=1$ qubit $1$ was
really in an entangled linear superposition of both up and down states,
assuming of course that it is possible to give any existence at all to the
state at a time when it was not measured. In short, the mistake is then to
conclude that $|11001\rangle _{01234}$ represents the history of the system
over an external time span $1\leq T\leq 4,$ and not just the outcome
eigenstate of a particular test.

Overall, therefore, the correlation $Q_{xy}=\langle \Psi _{4}|\hat{\sigma}%
_{x}^{3}\hat{\sigma}_{y}^{3}|\Psi _{4}\rangle $ should not be viewed as
asking about how many possible `histories' of the state between external
times $T=0$ and $4$ shared certain characteristics, but is asking about how
many from a set of sixteen classical looking eigenstates of the form $%
|ijklm\rangle _{01234}$ share them.\bigskip

It is hardly surprising, then, that by ascribing to a state such as $\Phi
\equiv $ $|11001\rangle _{01234}$ the semi-classical status of representing
what actually happened, i.e. the perspective of a single-valued classical
reality for the state's history, the correlations $Q$ do not violate the
classical Bell inequality. In this sense the above method is just a form of
`dressed-up' classical probability, with statistical correlations compared
between states that have undergone well defined histories. In effect, the
classical probabilities $a,$ $ab,$ $a\bar{b}\bar{e},...$ etc. of Section
6.1.3 have been replaced by sines and cosines, such that for example the
probability $P(01100)=\bar{a}\bar{c}gn$ that the bits have the final
configuration $01100$ in the classical case is directly equivalent to the
quantum case where there is a probability $\cos ^{4}\left( \frac{\alpha }{2}%
\right) \sin ^{4}\QOVERD( ) {\alpha }{2}$ that the state collapses to $%
|01100\rangle _{01234}.$

As an aside, note that it is of course always possible to examine an
ensemble of identically prepared and evolved states $\Psi _{4}$ and consider
a density matrix of possible eigenstates, but it must be stressed that this
is still just a classical, probabilistic result due to the nature of the
collapse process, and as such would hence still be expected to obey Bell
relations in the corresponding classical way.

\bigskip

Summarising, the problem of the initial analysis of this system was
two-fold. Firstly, it is incorrect to assume that $\hat{\sigma}_{x}^{3}$ and
$\hat{\sigma}_{y}^{3}$ are both measuring the state $\Psi _{4},$ because
neither acts non-invasively. Secondly, it is wrong to apply an external time
parameter $T$ internally to the state. So, any outcome resulting from these
should not be thought of as containing information about the `historical
development' of the state between $1\leq T\leq 4.$

It is beneficial to rephrase this synopsis in the context of the quantum
Universe. In the fully quantum reality proposed in this thesis, the above
type of `experiment' would ultimately have to be viewed on the emergent
level as one group of factors of the Universe's state (representing a
scientist) appearing to prepare, evolve and test another group of factors
(the five qubit system), even though this perspective was not specified
\textit{per se}. Moreover, and as discussed in Chapter 5 and elsewhere, each
of these groups of factors would be capable of experiencing their own
passages of endo-time, relative to their own internal transitions and
changes, as the Universe jumps from one state to the next.

Of course, in the current case the variable $T$ was defined as the endo-time
of the observer. The problem then arose because this time parameter $T$ was
taken to be absolute and universal, even though it is only relative,
endo-times that can be given any actual physical significance. After all,
recall that the endo-time of an observer is the exo-time of the observed,
and vice versa. $T$ cannot therefore also be taken to be the endo-time for
the qubit system, and it is by incorrectly doing so that results in a
misinterpretation of (\ref{Ch4Qxy}).\bigskip

Overall,\ since the development of the state of the Universe is ultimately
taken to be responsible for the development of the sub-states of everything
in it (including the observer and the qubits), it must be assumed to change
in a very special way if the observer appears to experience four distinct
time steps between the preparation of the qubits and their measurement,
whilst the qubit system itself appears to experience none. So, apart from
the described comments and constraints on the role of time in this type of
quantum computation, a general conclusion of the above discussion should
therefore be that care is clearly needed when attempting to analyse the sort
of endo-physical experiment presented here. It is this issue that is the
focus of the next section.

\bigskip

\subsection{Information Flow in the Quantum Universe}

\bigskip

In many ways, computation may be described as the manipulation of\textit{\
information}. After all, classical computer science generically involves the
encoding of some sort of physical input into a series of `symbols', the
meaning of which is only valid relative to the information regarding what
they actually represent. Furthermore, during an actual computation these
symbols are processed to generate a final sequence, and the information
contained in this can itself then be decoded to describe the properties of a
physical output. Of course, it does not matter what form these symbols take:
$0$'s and $1$'s, low-high voltages, squares and circles, offs or ons; what
is important is the information content they represent. In fact, it is worth
noting here that even a hard drive, the largest part of a modern personal
computer, is specifically designed for the storage of data that is not
related in any obvious way to the physical input it represents that was
entered into the machine.

In other words, it may in some ways be imagined that in computational
procedures the `properties' of the physical input are directly translated
into information, for example as a specific series of $0$'s and $1$'s, and
then it is the information itself that is actually processed.\bigskip

For a Universe running as an enormous quantum computation, the same ideas
might be expected to be true. In fact, the central theme of Chapters 4 and 5
was to demonstrate how physical concepts such as classical identity and
spatial location might be \textit{encoded} as certain features of the
quantum state representing the Universe, in this case in terms of
separability and the ensuing relationships between the various factors.
Indeed, the possibility of encoding space through informational methods
should perhaps not be surprising; after all recall that space and positional
relations may themselves be envisaged as a type of information storage
process: a spatial separation between two objects is ultimately equivalent
to the information that they are semi-classical, individual and not in
contact.

So, in the sense proposed in the previous chapters, the separability of the
state may be said to comprise a part of the information it intrinsically
carries. Moreover, due to this relationship a change in separability from
one jump to the next consequently implies a change in information. Thus, by
defining the separability of the state as part of its information content,
it is evident that some of this information can be used (i.e. `decoded') to
deduce particular physical properties, such as a quantum origin of
space.\bigskip

It might ultimately be expected, then, that any observed changes in physical
systems involve changes in the information carried by the state. In fact, by
reversing this argument due to the assumed `primacy' of the quantum state,
it might equally be expected that changes in the information contained in
the state could result in physical changes occurring on the emergent level.
Going further, since in the paradigm proposed in this thesis the passage of
time and the jump from the state $\Psi _{n}$ to $\Psi _{n+1}$ is
parameterised in terms of the acquisition of information, a change in the
information content of the state might be considered essential for any
suggested dynamics of the Universe.

As a highly schematic example of this, it could be imagined that a
particular change in the state's information content from one jump to the
next could, somehow, eventually result in one collection of factors of the
Universe's state (representing a human observer) being led to believe that
another group of factors of the Universe's state (representing particle $A)$
has reacted with a third group of factors (representing particle $B).$
Although the details are left deliberately vague here, the point is that by
a change in the information carried by the state of the Universe, perhaps
involving a change in its separability, it might on the emergent level
appear to the semi-classical observer that an experiment has been performed
between particles $A$ and $B,$ or even perhaps that particles $A$ and $B$
collided.\bigskip

The purpose of this section is to examine what it actually means to say that
the information content of a state has changed. The issues will be examined
as to what the necessary conditions for this are, what it implies, and how
this relates to physics. The nature of physical experimentation and
endo-physical measurement in the quantum Universe will then be explored.

\bigskip

\subsubsection{Types of Transformation}

\bigskip

Consider a procedure $\mathbf{\Pi }$ that relates two states $\Theta $ and $%
\Phi ,$ both of which are contained in the Hilbert space $\mathcal{H}.$ As
will be explained below in a variety of contexts, the relationship here is
defined such that the state $\Phi $ is the result of the procedure $\mathbf{%
\Pi }$ being performed on the state $\Theta ;$ this procedure could perhaps
involve a state reduction process, or even be some sort of mapping of the
form $\mathbf{\Pi }:\Theta \rightarrow \Theta ^{\prime }=\Phi .$ In fact,
since information changing procedures are taken to provide the basis for
dynamical development of the quantum Universe,\ in cases incorporating
information change it might be possible to view $\Theta $ as $\Psi _{n}$ and
$\Phi $ as $\Psi _{n+1},$ as will be discussed later.

The issue of current interest is now to determine when a given procedure $%
\mathbf{\Pi }$ may be said to result in a change in\ the information carried
by the state. In short, for what types of process $\mathbf{\Pi }$ is the
outcome $\Phi $ noticeably and physically different from $\Theta ?$\bigskip

Before explaining what is meant to say that information has changed in a
quantum system, it is perhaps easier to first demonstrate what it means to
say that it has not. Consider, for example, a null test on the state $\Theta
,$ defined previously as an operator $\hat{N}$ with an eigenvector $\Phi $
where $\Phi \equiv \Theta ,$ such that $\hat{N}\Theta =\lambda \Theta $ with
$\lambda $ an eigenvalue. In this case, the procedure $\hat{N}$ gives an
outcome $\Phi $ that is exactly the same as the initial state $\Theta ,$
such that this type of test leads to no overall change: the resulting state $%
\Phi $ is indistinguishable from the initial state $\Theta ,$ and the action
of the test is as if nothing has happened. Clearly, then, the development of
a quantum system with these sorts of null tests is effectively trivial,
because only differences can be physically observed. Consequently, and as
discussed in Chapter 3, a quantum universe in the state $\Psi _{n}$ may be
developed any number of times by operators $\hat{\Sigma}$ that possess the
eigenvector $\Psi _{n},$ but it is only when an operator $\hat{\Sigma}%
^{\prime }$ is encountered which does not have this eigenvector that the
universe jumps to a different state. Under such circumstances, it is now
possible to label whichever eigenvector the universe happens to jump to as
the new state, $\Psi _{n+1}.$

Of course, it is also possible to discuss \textit{local} null tests. If $%
\Theta $ is separable in the form $\Theta \equiv \alpha \otimes \beta ,$
where $\alpha \in \mathcal{H}_{\alpha },$ $\beta \in \mathcal{H}_{\beta }$
and $\Theta \in \mathcal{H}_{[\alpha \beta ]}\equiv \mathcal{H},$ then an
operator $\hat{N}_{\beta }$ may be said to be local null test on $\beta $ if
it has eigenvectors of the form $\Phi =\gamma \otimes \beta ,$ where $\gamma
\in \mathcal{H}_{\alpha }.$ In this case, a development from $\Theta $ to $%
\Phi $ leaves the factor $\beta $ invariant even though the sub-state in
Hilbert space $\mathcal{H}_{\alpha }$ has changed.

As discussed in Chapter 5, local null tests play an important role in the
dynamics of the quantum Universe, for example as the origin of a
`route-dependent endo-time' experienced by different factors of the
developing state. Further, in this instance $\Theta \neq \Phi $ so it might
be expected that the information carried by the states may have changed in
some respect. This assertion is to be investigated.\bigskip

In global null tests the information contained in the outcome state $\Phi
(=\Theta )$ is obviously always the same as that carried by the initial
state $\Theta ,$ whereas in local null tests the information carried by the
`before' and `after' states may be different. This type of comparison leads
naturally to a discussion of `passive' and `active' transformations, both of
which will be defined below. As will become evident, a global null test
provides a trivial example of a passive transformation, whereas local null
tests may conversely cause active transformations.

Passive and active transformations are defined \cite{Wiki-Act-Pass} by the
statement: ``\textit{An active transformation is one which actually changes
the physical state of a system and makes sense even in the absence of a
coordinate system, whereas a passive transformation is merely a change in
the coordinate system of no physical significance}.'' Furthermore, on the
grounds that every transformation either changes the physical state of a
system, or else it does not, the definition of passive transformations is
revised in the following to include every transformation that is not active.
Thus, passive transformations are taken to be effectively synonymous with
`non-active' transformations, and hence every transformation is assumed to
be either passive or active.

So, after a passive transformation from one state of a system to another, no
observable changes have occurred. Any apparent differences in `before' and
`after' states are merely superficial, and may only arise due to a change in
the way the system is being described. Alternatively, since generally in
physics, and especially in the endo-physical approach advocated in this
thesis, only relative differences between the parts of a system are
measurable, an example of a passive transformation is therefore one in which
every part of the system is altered in exactly the same way.

Cases of such transformations include:

\begin{itemize}
\item  The relabelling of the vacuum ground state in quantum field theory by
the\ addition of a constant term to the energy eigenvalues. This effectively
forms the basis of the renormalisation program, and is `valid' (in some
senses) because only relative differences are measurable in the laboratory;

\item  The addition of a constant term $a$ to two numbers $x$ and $y$ under
the subtraction operation. That is, if $x\rightarrow x^{\prime }=x+a$ and $%
y\rightarrow y^{\prime }=y+a,$ then $x-y=x^{\prime }-y^{\prime
}=(x+a)-(y+a); $

\item  The rotation of an entire space through some angle, such that no
changes occur in the relative positional relationships between any of the
objects inside this space. Such a transformation is unobservable from the
perspective of an observer contained entirely within the space;

\item  A change in the coordinate system of, say, a 3-dimensional space from
Cartesian to cylindrical axes. Indeed, much of general relativity is based
upon this type of invariance;

\item  A change in basis for a Hilbert space. For example, if $\mathcal{B}%
_{a}=\{|i\rangle _{a}:i=0,1,...,d_{a}\}$ is an orthonormal basis for Hilbert
space $\mathcal{H}_{a},$ and if $\mathcal{B}_{b}=\{|j\rangle
_{b}:j=0,1,...,d_{b}\}$ is an orthonormal basis for Hilbert space $\mathcal{H%
}_{b},$ then a state $\psi \in \mathcal{H}_{[ab]}$ given by $\psi
=\tsum\nolimits_{x=0}^{d_{a}}\tsum\nolimits_{y=0}^{d_{b}}C_{xy}|x\rangle
_{a}\otimes $ $|y\rangle _{b}$ with complex coefficient matrix $C_{xy}$ is
invariant to any relabelling $\mathcal{B}_{a}\rightarrow \mathcal{B}%
_{a}^{\prime }=\{|i^{\prime }\rangle _{a}:i^{\prime }=0,1,...,d_{a}\}$ and $%
\mathcal{B}_{b}\rightarrow \mathcal{B}_{b}^{\prime }=\{|j^{\prime }\rangle
_{a}:j^{\prime }=0,1,...,d_{a}\}$ of the individual subregister bases, where
$|i^{\prime }\rangle _{a}\neq |i\rangle _{a}$ and $|j^{\prime }\rangle
_{b}\neq |j\rangle _{b};$
\end{itemize}

%TCIMACRO{
%\TeXButton{noindent}{\noindent%
%}}%
%BeginExpansion
\noindent%
%
%EndExpansion
and so on. The point is that although a mathematical change may appear to
have occurred during a passive transformation, there are no intrinsic
physical consequences. In short, many passive transformations may be
realised or removed simply by relabelling the `axes'.\bigskip

Conversely, an active transformation is one for which differences do become
apparent when `before' and `after' states are compared. In other words, if $%
\Phi $ cannot simply be rewritten as a relabelled version of $\Theta ,$ then
it must be an active transformation that relates $\Theta $ to $\Phi .$

In fact, in active transformations different parts of the system may
actually change relative to each other. In this sense, then, it is active
transformations that are seen to occur in real physics experiments, because
in these situations the physicist notices that the state under investigation
has changed relative to herself (which she often believes has not changed).
A good example here is the measurement of an entangled EPR state and its
subsequent collapse into a product of factors: physically these `before' and
`after' states are completely different, and this fact is observable.

As discussed earlier, the separability of a state may be described as being
part of the `information' it contains. So, a change in separability of a
state must therefore result in a change in this information. As a
consequence of this, another example of an active transformation between $%
\Theta $ and $\Phi $ is one for which these two states lie in different
partitions of the total Hilbert space $\mathcal{H}.$ In this case, $\Theta $
and $\Phi $ must both be separable in different ways, and so by the above
description must duly represent different information contents. As an
illustration, it is evident that the EPR experiment mentioned above
satisfies this condition.\bigskip

Active and passive transformations may readily be seen in the context of the
quantum Universe. For instance, an unitary operator $\hat{U}$ acting
globally on every element of the Hilbert space $\mathcal{H}$ leaves all
inner products between these elements invariant. That is, if $\Psi
_{n}^{\prime }=\hat{U}\Psi _{n}$ and $\Psi _{n+1}^{\prime }=\hat{U}\Psi
_{n+1},$ then $\langle \Psi _{n+1}|\Psi _{n}\rangle =\langle \Psi
_{n+1}^{\prime }|\Psi _{n}^{\prime }\rangle ,$ and such a transformation
could not physically be detected. On the other hand, because (by assumption)
$\Psi _{n+1}\neq \Psi _{n}$ then the jump from state $\Psi _{n}$ to the
state $\Psi _{n+1}$ must be regarded as a physically realisable, active
transformation. Active and passive transformations consequently play
different roles in dynamics.

However, given that in an endo-physical description of reality both the
`experiment' and the `experimentee' are seen as different parts of the same
quantum Universe, what is really of interest in such a picture are the
relative changes occurring between the sub-states of this system. Further,
it is evident that such real physical results, where one system is observed
to change relative to another, must not be `explainable away' simply by a
passive relabelling of axes, or by a global transformation of every part of
the Universe in the same manner. These changes are physically observed, and
so because a passive transformation leads to no physical change in a system,
the conclusion is then that passive transformations cannot be responsible
for changes in the information contained in the state of the quantum
Universe as it develops. On the contrary, because active transformations do
result in a change of the state of a system, it is these that are
conjectured to ultimately form the basis for information change and
exchange, and the nature of endo-physical observation.

\bigskip

It is possible to further investigate what is meant by a passive change of
information in quantum mechanics by using the concept of local
transformations. In fact, because by definition a transformation that is not
passive must be active, a converse exploration is equally demonstrated.

As described above, a simple relabelling of the basis implies a passive
transformation. In such relabellings there are no non-trivial differences in
the state of the system before and after the relabelling, and hence no
observable changes. The information contained by the state is therefore
invariant to such a change. Furthermore, because of the asserted link
between information content and separability, in passive transformations the
partition containing the state is not expected to change: separable states
remain separable, and entangled factors remain entangled, etc.

Reversing these lines of argument provides a definition for passive
transformation. Consider a Hilbert space $\mathcal{H}_{[1...N]}$
factorisable into $N$ subregisters of prime dimension, and further assume
that the states $\Theta $ and $\Phi $ are contained in this space. Consider
also a basis set $\mathcal{B}_{A}$ of vectors spanning $\mathcal{H}%
_{[1...N]},$ and assume that each member of $\mathcal{B}_{A}$ is completely
separable relative to the $N$ subregisters of $\mathcal{H}_{[1...N]}.$ By
defining
\begin{equation}
\mathcal{B}_{A,m}=\{|i_{m}\rangle _{m}:i_{m}=0,1,...,(d_{m}-1)\}
\end{equation}
as an orthonormal basis for factor Hilbert space $\mathcal{H}_{m}$ of
dimension $d_{m},$ where $m=1,2,...,N,$ the set $\mathcal{B}_{A}$ may be
defined as
\begin{equation}
\mathcal{B}_{A}\equiv \{|i_{1}\rangle _{1}\otimes |i_{2}\rangle _{2}\otimes
...\otimes |i_{N}\rangle _{N}:i_{m}=0,1,...,(d_{m}-1);m=1,2,...,N\}.
\end{equation}

Of course, this construction is not unique. It is equally possible to define
a different orthonormal basis set $\mathcal{B}_{B,m}$ for the factor Hilbert
space $\mathcal{H}_{m}$ as $\mathcal{B}_{B,m}=\{|i_{m}^{\prime }\rangle
_{m}:i_{m}^{\prime }=0,1,...,(d_{m}-1)\},$ and similarly define another
completely separable basis set $\mathcal{B}_{B}$ of vectors spanning $%
\mathcal{H}_{[1...N]}$ as
\begin{equation}
\mathcal{B}_{B}\equiv \{|i_{1}^{\prime }\rangle _{1}\otimes |i_{2}^{\prime
}\rangle _{2}\otimes ...\otimes |i_{N}^{\prime }\rangle _{N}:i_{m}^{\prime
}=0,1,...,(d_{m}-1);m=1,2,...,N\}.
\end{equation}

Consider now a state $\Theta ,$ defined relative to $\mathcal{B}_{A}$ as
\begin{equation}
\Theta
=\sum_{i_{1}=0}^{(d_{1}-1)}\sum_{i_{2}=0}^{(d_{2}-1)}...%
\sum_{i_{N}=0}^{(d_{N}-1)}C_{i_{1}i_{2}...i_{N}}|i_{1}\rangle _{1}\otimes
|i_{2}\rangle _{2}\otimes ...\otimes |i_{N}\rangle _{N},
\end{equation}
and a state $\Phi $ defined relative to $\mathcal{B}_{B}$ as
\begin{equation}
\Phi =\sum_{i_{1}^{\prime }=0}^{(d_{1}-1)}\sum_{i_{2}^{\prime
}=0}^{(d_{2}-1)}...\sum_{i_{N}^{\prime }=0}^{(d_{N}-1)}C_{i_{1}^{\prime
}i_{2}^{\prime }...i_{N}^{\prime }}|i_{1}^{\prime }\rangle _{1}\otimes
|i_{2}^{\prime }\rangle _{2}\otimes ...\otimes |i_{N}^{\prime }\rangle _{N}
\end{equation}
where the $C_{i_{1}i_{2}...i_{N}}$ and $C_{i_{1}^{\prime }i_{2}^{\prime
}...i_{N}^{\prime }}$ give rise to complex coefficient matrices.

It is desirable to examine whether $\Theta $ and $\Phi $ are intrinsically
different vectors, or whether $\Phi $ is instead just a `relabelled' version
of $\Theta $ in an alternative basis. To this end, if it is possible to
rotate (in a strictly mathematical sense) the $N$ individual components $%
|i_{m}\rangle _{m}\in \mathcal{H}_{m}$ of $\Theta $ into the components $%
|i_{m}^{\prime }\rangle _{m}$ to give a new state $\Theta ^{\prime },$ then
it is evident that $\Theta $ and $\Phi $ are equivalent if $\Theta ^{\prime
}=\Phi .$

Similarly, if the basis set $\mathcal{B}_{A,m}$ is transformed into the
basis set\ $\mathcal{B}_{B,m}$ for each $m,$ such that $\mathcal{B}_{A}$ is
therefore transformed into $\mathcal{B}_{B},$ and if $\Theta $ is rewritten
in this new basis and called $\Theta ^{\prime },$ then the states $\Theta $
and $\Phi $ are again mathematically equivalent if $\Theta ^{\prime }=\Phi .$
In both cases $\Phi $ is just a different version of $\Theta ,$ but
expressed in an alternative basis. In such a circumstance it may then be
said that $\Theta $ can be passively transformed into $\Phi .$\bigskip

This observation can be stated more precisely by noting that the individual
rotations can be achieved by the use of unitary operators acting locally. If
the unitary operator $\hat{U}_{L}$ is defined as a product of local unitary
operators, $\hat{U}_{L}=\hat{u}_{1}\otimes \hat{u}_{2}\otimes ...\otimes
\hat{u}_{N}$ where $\hat{u}_{m}$ is a unitary operator acting locally in
factor Hilbert space $\mathcal{H}_{m},$ then the state $\Phi $ is just a
relabelled version of $\Theta $ if
\begin{equation}
\Theta ^{\prime }\equiv \hat{U}_{L}\Theta =(\hat{u}_{1}\otimes \hat{u}%
_{2}\otimes ...\otimes \hat{u}_{N})\Theta =\Phi .
\end{equation}

Paraphrasing, $\Theta $ and $\Phi $ are equivalent if there exists a locally
acting, unitary operator $\hat{U}_{L}$ relating them in this way.

As an aside, note that by definition any unitary transformation $\hat{U}$
acting locally in the individual factors of $\mathcal{H}_{[1...N]}$ can be
written in the factorisable form $\hat{U}=\hat{u}_{1}\otimes \hat{u}%
_{2}\otimes ...\otimes \hat{u}_{N},$ where $\hat{u}_{m}$ is an unitary
operator in $\mathcal{H}_{m}$ for $m=1,2,...,N.$ Equally, if it acts locally
the operator $\hat{U}$ must necessarily be factorisable. Moreover, because
it is acting locally on every subregister it must be completely or
fundamentally factorised, i.e. factorisable into $N$ factors.\bigskip

The above results may be compared to global transformations, that is, those
in which the \textit{overall} basis sets $\mathcal{B}_{A}$ and $\mathcal{B}%
_{B}$ are rotated instead of their \textit{individual} subregister basis
sets $\mathcal{B}_{A,m}$ and $\mathcal{B}_{B,m},$ or equivalently those for
which the overall state $\Theta $ is transformed `at once' instead of its
components in $\mathcal{H}_{m}$ being transformed individually. It is always
mathematically possible to find a unitary operator $\hat{U}_{G}$ that
transforms a given state $\Theta $ globally into any other state $\Phi ,$
such that $\Phi =\hat{U}_{G}\Theta .$ However, such a transformation may not
go unnoticed: for example, $\Theta $ could be completely separable whilst $%
\Phi $ could be completely entangled, and these such states are
fundamentally and physically different. Physically, separations and
entanglements are completely different entities.

The same is not true for states that can be transformed locally into each
other. If a state $\Theta $ can be locally transformed into the state $\Phi
, $ then not only are these two states mathematically equivalent but they
are also physically indistinct. In this case, the physical information
contained in the state has remained unaffected by the rotation because such
states can be interchanged simply by a suitable and trivial relabelling of
the basis. Paraphrasing, because a local unitary transformation just leads
to a rotation or relabelling of the basis, it is unobservable in physics,
and it may hence be described as passive. Conversely, recall that it is the
active transformations, i.e. the processes that cause relative changes, that
are of interest to endo-physics.\bigskip

As an example of these ideas, consider a four dimensional Hilbert space $%
\mathcal{H}_{[12]}$ factorisable as $\mathcal{H}_{[12]}=\mathcal{H}%
_{1}\otimes \mathcal{H}_{2},$ and spanned by the orthonormal basis $\mathcal{%
B}_{12}=\{|i\rangle _{1}\otimes |j\rangle _{2}:i,j=0,1\}.$ Consider also the
separable states $\mu =|0\rangle _{1}\otimes |1\rangle _{2}$ and $\chi
=(a|0\rangle _{1}+b|1\rangle _{1})\otimes (c|0\rangle _{2}+d|1\rangle _{2})$
for $a,b,c,d\in \mathbb{C},$ as well as the entangled state $\omega
=|0\rangle _{1}\otimes |0\rangle _{2}+|1\rangle _{1}\otimes |1\rangle _{2}.$
It is always possible to find unitary transformations $\hat{U}_{G}$ and $%
\hat{U}_{G}^{\prime }$ that act globally on one of the states $\mu ,$ $\chi $
or $\omega ,$ and not locally on its individual components, that transform
it into one of the others. That is, it is always possible to find global,
unitary operators of the form $\hat{U}_{G}$ and $\hat{U}_{G}^{\prime }$ that
satisfy $\mu =\hat{U}_{G}\chi =\hat{U}_{G}^{\prime }\omega .$

It is also possible to find a (factorisable) local unitary transformation $%
\hat{U}_{L}^{\prime \prime }=\hat{u}_{1}\otimes \hat{u}_{2}$ that relates $%
\mu $ to $\chi ,$ where $\hat{u}_{1}$ and $\hat{u}_{2}$ are unitary
operators acting locally on the components of $\mu $ and $\chi $ in the
subregister Hilbert spaces $\mathcal{H}_{1}$ and $\mathcal{H}_{2}$
respectively.\ That is, there exists an operator $\hat{U}_{L}^{\prime \prime
}$ such that
\begin{eqnarray}
\hat{U}_{L}^{\prime \prime }\mu &=&[\hat{u}_{1}\otimes \hat{u}%
_{2}](|0\rangle _{1}\otimes |1\rangle _{2})=[\hat{u}_{1}|0\rangle
_{1}]\otimes \lbrack \hat{u}_{2}|1\rangle _{2}] \\
&=&(a|0\rangle _{1}+b|1\rangle _{1})\otimes (c|0\rangle _{2}+d|1\rangle
_{2})=\chi .  \notag
\end{eqnarray}

However it is not possible to find a factorisable, unitary operator of the
form $\hat{U}_{L}^{\prime \prime \prime }=\hat{v}_{1}\otimes \hat{v}_{2}$
that transforms $\omega $ to $\mu ,$ where $\hat{v}_{1}$ and $\hat{v}_{2}$
are local unitary operators in\ $\mathcal{H}_{1}$ and $\mathcal{H}_{2}.$
That is, there is no unitary operator $\hat{U}_{L}^{\prime \prime \prime }$
acting locally such that $\mu =\hat{U}_{L}^{\prime \prime \prime }\omega =(%
\hat{v}_{1}\otimes \hat{v}_{2})\omega .$ Equally, the separable state $\chi $
cannot be transformed into the entangled state $\omega $ by unitary
operators acting locally in the individual subregisters.\bigskip

It seems that the outcome of a unitary operator acting locally on a state is
in the same partition as the original. This observation leads to the
following theorem.\bigskip

\begin{theorem}
Separations and entanglements are preserved by unitary transformations
acting locally in the individual subregisters.
\end{theorem}

\begin{proof}
The theorem is proved first for the separations, and this is then used for
the entanglements.

\begin{itemize}
\item  \textbf{Separations:}
\end{itemize}

Consider a Hilbert space $\mathcal{H}_{[1...N]}$ factorisable in $N$
subregisters $\mathcal{H}_{m}$ of prime dimension. Consider also a unitary
operator $\hat{U}_{1...N}$ acting locally in the individual subregisters $%
\mathcal{H}_{m}.$ Then, by definition, $\hat{U}_{1...N}$ must be completely
factorisable, such that $\hat{U}_{1...N}\equiv \hat{u}_{1}\otimes \hat{u}%
_{2}\otimes ...\otimes \hat{u}_{N}.$

Consider further a completely separable state $\varphi _{1...N}\in \mathcal{H%
}_{1...N}\subset \mathcal{H}_{[1...N]}$ written as $\varphi _{1...N}\equiv
\varphi _{1}\otimes \varphi _{2}\otimes ...\otimes \varphi _{N}$ where $%
\varphi _{m}\in \mathcal{H}_{m}.$ The evolution of $\varphi _{1...N}$ by $%
\hat{U}_{1...N}$ is given by
\begin{eqnarray}
\hat{U}_{1...N}\varphi _{1...N} &=&\left( \hat{u}_{1}\otimes \hat{u}%
_{2}\otimes ...\otimes \hat{u}_{N}\right) \left( \varphi _{1}\otimes \varphi
_{2}\otimes ...\otimes \varphi _{N}\right) \\
&=&\left( \hat{u}_{1}\varphi _{1}\right) \otimes \left( \hat{u}_{2}\varphi
_{2}\right) \otimes ...\otimes \left( \hat{u}_{N}\varphi _{N}\right)  \notag
\\
&=&\varphi _{1}^{\prime }\otimes \varphi _{2}^{\prime }\otimes ...\otimes
\varphi _{N}^{\prime }  \notag
\end{eqnarray}
where $\varphi _{m}^{\prime }\equiv \hat{u}_{m}\varphi _{m}\in \mathcal{H}%
_{m}.$ Clearly the outcome $\varphi _{1}^{\prime }\otimes \varphi
_{2}^{\prime }\otimes ...\otimes \varphi _{N}^{\prime }$ is also a member of
the separation $\mathcal{H}_{1...N},$ as required for the theorem.

\begin{itemize}
\item  \textbf{Entanglements:}
\end{itemize}

The proof here is by contradiction. Consider now a completely entangled
state $\psi ^{1...N}$ $\in \mathcal{H}^{1...N}\subset \mathcal{H}_{[1...N]},$
and consider a hypothetical unitary operator $\hat{U}_{1...N}$ that, when
acting locally upon $\psi ^{1...N},$ evolves it into a state $\psi ^{\prime
} $ that is not in the partition $\mathcal{H}^{1...N}.$ Because $\psi
^{\prime }\notin \mathcal{H}^{1...N}$ it follows that $\psi ^{\prime }$ must
be separable into at least two factors, and so must be a member of the
complement set $\psi ^{\prime }$ $\in (\mathcal{H}_{[1...N]}-\mathcal{H}%
^{1...N}).$

In this case, it would be possible to write
\begin{equation}
\hat{U}_{1...N}\psi ^{1...N}=\psi ^{\prime }=\alpha \otimes \beta
\end{equation}
with $\alpha \in \mathcal{H}_{\alpha }\equiv \mathcal{H}_{[1...M]}$ and $%
\beta \in \mathcal{H}_{\beta }\equiv \mathcal{H}_{[(M+1)...N]},$ where for
simplification a suitable relabelling of the subregisters has been performed
to adopt an ascending order, and it is noted that $\alpha $ and $\beta $ may
themselves be separable further relative to their factor Hilbert spaces.

Now, because $\hat{U}_{1...N}$ is unitary it has a unique inverse denoted by
$(\hat{U}_{1...N})^{-1},$ such that $(\hat{U}_{1...N})^{-1}(\hat{U}_{1...N})=%
\hat{I}_{[1...N]}$ where $\hat{I}_{[1...N]}$ is the identity operator in $%
\mathcal{H}_{[1...N]}.$ To discover the form of $(\hat{U}_{1...N})^{-1},$
observe that if it may be written as a completely factorisable product
\begin{equation}
(\hat{U}_{1...N})^{-1}=\hat{u}_{1}^{-1}\otimes \hat{u}_{2}^{-1}\otimes
...\otimes \hat{u}_{N}^{-1}
\end{equation}
then
\begin{eqnarray}
(\hat{U}_{1...N})^{-1}(\hat{U}_{1...N}) &=&\left( \hat{u}_{1}^{-1}\otimes
\hat{u}_{2}^{-1}\otimes ...\otimes \hat{u}_{N}^{-1}\right) \left( \hat{u}%
_{1}\otimes \hat{u}_{2}\otimes ...\otimes \hat{u}_{N}\right) \\
&=&\left( \hat{u}_{1}^{-1}\hat{u}_{1}\right) \otimes \left( \hat{u}_{2}^{-1}%
\hat{u}_{2}\right) \otimes ...\otimes \left( \hat{u}_{N}^{-1}\hat{u}%
_{N}\right)  \notag \\
&=&\left( \hat{I}_{1}\right) \otimes \left( \hat{I}_{2}\right) \otimes
...\otimes \left( \hat{I}_{N}\right) =\hat{I}_{1...N}  \notag
\end{eqnarray}
where $\hat{I}_{m}$ is the identity operator for $\mathcal{H}_{m}.$
Moreover, because by definition there can only be one unitary operator $(%
\hat{U}_{1...N})^{-1}$ satisfying $(\hat{U}_{1...N})^{-1}(\hat{U}_{1...N})=%
\hat{I}_{[1...N]}$ (i.e. $\hat{U}_{1...N}$ can only have one inverse), the
inverse of $\hat{U}_{1...N}$ must be this completely factorisable operator.

Because $(\hat{U}_{1...N})^{-1}$ is completely factorisable, it must also
factorise in the form
\begin{equation}
(\hat{U}_{1...N})^{-1}=(\hat{U}_{1...M})^{-1}\otimes (\hat{U}%
_{(M+1)...N})^{-1}=(\hat{U}_{\alpha })^{-1}\otimes (\hat{U}_{\beta })^{-1},
\end{equation}
with $\hat{U}_{\alpha }$ and $\hat{U}_{\beta }$ acting locally in Hilbert
spaces $\mathcal{H}_{\alpha }$ and $\mathcal{H}_{\beta }$ respectively. So,
applying $(\hat{U}_{1...N})^{-1}$ to $\psi ^{\prime }$ gives
\begin{eqnarray}
(\hat{U}_{1...N})^{-1}\psi ^{\prime } &=&\left( (\hat{U}_{\alpha
})^{-1}\otimes (\hat{U}_{\beta })^{-1}\right) \left( \alpha \otimes \beta
\right) \\
&=&\left( (\hat{U}_{\alpha })^{-1}\alpha \right) \otimes \left( (\hat{U}%
_{\beta })^{-1}\beta \right) .  \notag
\end{eqnarray}

However, $(\hat{U}_{\alpha })^{-1}$ and $(\hat{U}_{\beta })^{-1}$ are
themselves just unitary operators, so $(\hat{U}_{\alpha })^{-1}\alpha
=\alpha ^{\prime }$ and $(\hat{U}_{\beta })^{-1}\beta =\beta ^{\prime },$
where $\alpha ^{\prime }\in \mathcal{H}_{\alpha }\equiv \mathcal{H}%
_{[1...M]} $ and $\beta ^{\prime }\in \mathcal{H}_{\beta }\equiv \mathcal{H}%
_{[(M+1)...N]}.$ So
\begin{equation}
(\hat{U}_{1...N})^{-1}\psi ^{\prime }=\alpha ^{\prime }\otimes \beta
^{\prime }
\end{equation}
but note that also
\begin{equation}
(\hat{U}_{1...N})^{-1}\psi ^{\prime }=(\hat{U}_{1...N})^{-1}[\hat{U}%
_{1...N}\psi ^{1...N}]=\psi ^{1...N}
\end{equation}
such that equating leads to
\begin{equation}
\psi ^{1...N}=\alpha ^{\prime }\otimes \beta ^{\prime }
\end{equation}
which is clearly a contradiction. Thus, the equation $\hat{U}_{1...N}\psi
^{1...N}=\psi ^{\prime }=\alpha \otimes \beta $ cannot be true, so the
conclusion must be that local unitary operators $\hat{U}_{1...N}$ preserve
entanglements: $\hat{U}_{1...N}\psi ^{1...N}=\psi ^{\prime \prime },$ where $%
\psi ^{\prime \prime }\in \mathcal{H}^{1...N}.$%
\begin{equation*}
\end{equation*}

Note that the proof is readily extended to states that are a separable
product of entangled sub-states, because the above argument is equally true
for each of the individual factors.\bigskip
\end{proof}

Thus, unitary operators acting locally on the components of a state in its
fundamental subregisters do not affect its partition. Conversely, partition
changing processes may not be accomplished by local unitary operators, and
so cannot be `removed' by a simple relabelling of the basis.

It is consequently now possible to specify what is meant by information
change, as discussed in the next sub-section.

\bigskip

\subsubsection{Information Change and Exchange}

\bigskip

Attention is now turned to transformations between the states of the quantum
Universe. For what potential transitions from $\Psi _{n}$ to a next possible
state is there an inherent and intrinsic change of information? Further, in
the context of the `measurement problem', for what transitions from $\Psi
_{n}$ to $\Psi _{n+1}$ is there an actual exchange of information between
different parts of the state?\bigskip

From the work of the previous sub-section, it is now possible to argue that
the transition from $\Theta $ and $\Phi $ implies a change of information if
these two states are fundamentally different. The question then becomes:
when might a difference be described as fundamental, to which an answer may
rely on whether or not it can be `transformed away' by a simple relabelling
of the basis. Specifically, because a basis relabelling may be accomplished
by the use of local unitary operators, it is evident that $\Theta $ and $%
\Phi $ are indistinct if $\Theta $ may be locally transformed into $\Phi .$

Going further, it has been demonstrated that unitary operators acting
locally on a state do not affect its partition. Thus, if $\Theta $ and $\Phi
$ are in different partitions it is clear that $\Theta $ cannot be
transformed into $\Phi $ by the use of local unitary operators. Moreover,
this implies that $\Theta $ may not then be rewritten as $\Phi $ simply by a
passive relabelling of the basis. Consequently, such a partition changing
transition must imply a change in the information carried by the state, and
this point is reinforced by recalling that because a state's separability is
related to its information content, any change in partition necessarily
results in a change of information.

Summarising, partition changing processes necessarily imply an information
change, and hence represent active transformations. Conversely, it might
automatically be expected that all partition preserving transformations are
accordingly passive.\bigskip

Care is needed, however, when applying this rule to the situation where the
state is not just an arbitrary vector in an abstract mathematical space, but
is instead representing an actual physical system. The following examples
demonstrate this point.

For instance, does the change from the state $\Theta $ defined as $\Theta
\equiv |0\rangle _{1}\otimes |1\rangle _{2}-$ $|1\rangle _{1}\otimes
|0\rangle _{2}$ to the state $\Phi \equiv |0\rangle _{1}\otimes |1\rangle
_{2}+|1\rangle _{1}\otimes |0\rangle _{2}$ constitute a change in
information? It can be achieved by the local unitary operator $\hat{\sigma}%
_{1}^{3}\otimes \hat{I}_{2},$ such that $\Phi =[\hat{\sigma}_{1}^{3}\otimes
\hat{I}_{2}]\Theta $ where $\hat{\sigma}_{1}^{3}$ is a Pauli operator in $%
\mathcal{H}_{1}$ and $\hat{I}_{2}$ is the identity in $\mathcal{H}_{2},$ and
so may be thought of simply as a rotation of the state of qubit $1,$ or
alternatively as a relabelling of the\ $|1\rangle _{1}$ basis of $\mathcal{H}%
_{2}$ to $-|1\rangle _{1}.$ Clearly, $\Theta $ and $\Phi $ are in the same
partition here, but in this circumstance are the `before' and `after'
entangled states fundamentally the same?

Indeed, could such an apparently passive transformation ever be physically
allowed in nature? Say, for example, that $\mathcal{H}_{1}$ represents the
Hilbert space of an electron and $\mathcal{H}_{2}$ the Hilbert space of a
positron, with $|0\rangle $ representing a spin down state and $|1\rangle $
a spin up (in some direction). On the subregister level the two states $%
\Theta $ and $\Phi $ are related by a local unitary change of basis, but if
the wavefunction is instead ascribed to represent entangled up and down
electrons and positrons the result of the above sort of change would be an
anti-symmetric state becoming a symmetric one, and these are physically
different. Is this then meant to imply that transformations that are defined
as mathematically passive could potentially lead to observable physical
consequences?\bigskip

As a second example of this type of problem, consider a state $\Psi _{n}\in
\mathcal{H}_{[12R]}$ of the universe prepared such that
\begin{equation}
\Psi _{n}=[|0\rangle _{1}\otimes |0\rangle _{2}-|1\rangle _{1}\otimes
|1\rangle _{2}]\otimes |R\rangle _{R}
\end{equation}
where $\mathcal{H}_{1}$ and $\mathcal{H}_{2}$ are qubit subregisters spanned
by orthonormal bases $\{|0\rangle _{1},|1\rangle _{1}\}$ and $\{|0\rangle
_{2},|1\rangle _{2}\},$ and $R$ represents the rest of the universe in
Hilbert sub-space $\mathcal{H}_{R},$ presumably itself of enormous dimension
and a product of very many factors. Obviously, qubits $1$ and $2$ are in a
correlated state here, so an arbitrary measurement of qubit $1$ with the
result $|i\rangle _{1},$ for $i=0,1,$ projects qubit $2$ into the same state
$|i\rangle _{2},$ and vice versa.

Assume now that the next state $\Psi _{n+1}$ of the universe turns out to be
\begin{equation}
\Psi _{n+1}=|0\rangle _{1}\otimes |r\rangle _{2R}
\end{equation}
where $|r\rangle _{2R}$ is some element of\ $\mathcal{H}_{[2R]}.$ On the
emergent level, and as discussed later in Section 6.2.3, the jump from $\Psi
_{n}$ to $\Psi _{n+1}$ might correspond to an apparent endo-physical
measurement of qubit $2$ by an `apparatus' contained in $|R\rangle _{R},$
with the result that it must have been found in a down state, $|0\rangle
_{2},$ because qubit $1$ has been left in the state $|0\rangle _{1}.$ So, in
this case it would be expected that if qubit $1$ is measured next it will
also be found to be spin down, $|0\rangle _{1}.$

However, if local transformations are always unobservable, it is possible to
find a local unitary operator\ $\hat{u}_{1}$ that results in an effective
relabelling of the basis of $\mathcal{H}_{1}$ as $|0\rangle _{1}\rightarrow
|1\rangle _{1}$ and $|1\rangle _{1}\rightarrow |0\rangle _{1},$ i.e. $\hat{u}%
_{1}\{|0\rangle _{1},|1\rangle _{1}\}=\{|1\rangle _{1},|0\rangle _{1}\}.$ In
other words, applying this rotation to $\Psi _{n+1}$ gives $\Psi
_{n+1}^{\prime },$ where
\begin{equation}
\Psi _{n+1}\rightarrow \Psi _{n+1}^{\prime }=[\hat{u}_{1}\otimes \hat{I}%
_{2R}]\Psi _{n+1}=[\hat{u}_{1}\otimes \hat{I}_{2R}](|0\rangle _{1}\otimes
|r\rangle _{2R})=|1\rangle _{1}\otimes |r\rangle _{2R}.
\end{equation}
\ \ and $\hat{I}_{2R}$ is the identity in $\mathcal{H}_{[2R]}.$ Now when it
is observed by emergent apparatus in $\mathcal{H}_{[2R]},$ qubit $1$ will be
found in the up state $|1\rangle _{1}.$ This appears to violate what would
be expected from an initially correlated state, such that an apparent
passive transformation has again led to an observed physical discrepancy.

In fact, this type of example can be given even greater significance.
Consider a state $\Psi _{n}=|\nu _{L}\rangle _{\nu }\otimes |R^{\prime
}\rangle _{R^{\prime }},$ where $|\nu _{L}\rangle _{\nu }\in \mathcal{H}%
_{\nu }$ represents a (left-handed) neutrino, $|R^{\prime }\rangle
_{R^{\prime }}\in \mathcal{H}_{R^{\prime }}$ represents the rest of the
Universe, and $\Psi _{n}\in \mathcal{H}_{[\nu R^{\prime }]}.$ In this case
it might appear possible to passively transform the sub-state of $\Psi _{n}$
in $\mathcal{H}_{\nu }$ with a local unitary operator $\hat{u}_{\nu }$ such
that
\begin{equation}
\Psi _{n}\rightarrow \Psi _{n}^{\prime }=[\hat{u}_{\nu }\otimes \hat{I}%
_{R^{\prime }}]\Psi _{n}=[\hat{u}_{\nu }\otimes \hat{I}_{R^{\prime }}]|\nu
_{L}\rangle _{\nu }\otimes |R^{\prime }\rangle _{R^{\prime }}=|\nu
_{R}\rangle _{\nu }\otimes |R^{\prime }\rangle _{R^{\prime }}
\end{equation}
where $|\nu _{R}\rangle _{\nu }$ represents the state of a right-handed
neutrino, with $\hat{I}_{R^{\prime }}$ the identity in $\mathcal{H}%
_{R^{\prime }}.$ However, right-handed neutrinos are thought not to exist in
nature\footnote{%
Or at least, no right handed neutrinos have ever been observed.}, so the
local, passive, non-partition changing and unobservable unitary
transformation of the type $|\nu _{L}\rangle \rightarrow |\nu _{R}\rangle $
has ultimately lead to unphysical consequences.\bigskip

So, how is it that apparently passive, local relabellings of basis can be
reconciled with such drastic resultant changes in the physical properties of
the quantum system?

Firstly, one suggestion might be\ perhaps that the Universe forbids certain
unitary processes. It may not be valid, for example, to rotate just some
basis sets and not others; perhaps the bases of qubits $1$ or $2$ in the
second example may not be relabelled without also relabelling the basis of $%
R $ in a similar way. In fact, as discussed in previous chapters, the
suggestion here would be that this is potentially another case of the
Universe being highly selective when deciding which operators it chooses to
develop itself with. This conclusion, however, is not a sufficient argument.
Ignoring the fact that changing every basis set in the same way is really
just equivalent to a global transformation, it must also be noted that the
local unitary transformations discussed above are seen as mathematical
relabellings and not necessarily as direct physical evolutions of the
system. In other words, although it might be possible to impose the
constraint that, say, $|\nu _{L}\rangle $ cannot be \textit{evolved} into $%
|\nu _{R}\rangle $ (e.g. by any sort of Schr\"{o}dinger dynamics) it may not
equally be imposed that $|\nu _{L}\rangle $ cannot be \textit{relabelled} as
$|\nu _{R}\rangle .$

A second potential argument could then be to conject that physics on the
subregister level might obey different or additional constraints from that
in the emergent limit of real particles, for example electrons or neutrinos.
This solution fails, however, because all physical characteristics are
expected to emerge from the proposed pregeometric description, and not the
other way around, and so any considerations or invariances on the
pregeometric level might also be expected to exist at its large scale limit.
At the very least, if this argument was correct a good reason would be
required to explain why such pregeometric equivalencies disappear on
emergent scales.

Alternatively, it might perhaps be argued that although local unitary
operations lead to mathematical invariances, the physical characteristics
exhibited by a particular (sub-) state, for example its asymmetry or
left-handedness, are actually bestowed upon it by external influences. A
neutrino may be left handed, for example, only relative to a frame of
reference defined on the emergent scale by $R,$ and as such this type of
description may be meaningless on the pregeometric, sub-register level.
Whilst this `lack of properties' explanation may be very much in the spirit
of quantum theory, it still misses the essential point that relative to
these external frames\ an observable change in the state does actually seem
to be caused by an apparent, passive (local) transformation.\bigskip

So, a better conclusion from the observation that the above types of passive
relabellings are not seen in nature might then be to suggest that the
Universe selects its own preferred basis, such that all ups and downs and
rights and lefts are defined relative to this. In this case, local rotations
would represent a physical change in the information carried by the state,
because the new, transformed state could be compared to this absolute basis.
Thus, such rotations would be observable, and are hence inequivalent to
simple passive relabellings.

Further, in fact, a more endo-physical suggestion might be that the
preferred basis need not actually be defined by the entire universe, but by
an internal endo-observer. In other words, a preferred basis for a sub-state
under investigation may emerge by considering the basis of the sub-state(s)
representing the observer.

Consider, for instance, the second example of above. The initial measurement
of qubit $2$ by the apparatus may be seen as defining a preferred up-down
`axis', because in order to find qubit $2$ to be either up or down it is
necessary to specify what these `directions' are relative to. Moreover, in
order to initially know that qubits $1$ and $2$ have been prepared in the
correlated way of above, their bases must also be correlated. Thus, the
`fixing' of the basis for qubit $2$ also necessarily fixes qubit $1$'s
basis, such that any subsequent investigation of qubit $1$ can only be valid
relative to the preferred `direction' defined by the measurement of qubit $%
2. $ So, when qubit $2$ is found to be `down', an up-down axis is
automatically defined, and it is this axis that must then be used for qubit $%
1$ if a consistent description of the system is to be used.

Of course, any other basis could have been chosen before the measurement of
qubit $2.$ However, once a `preferred axis' has been chosen it must remain
fixed, and must also be used for qubit $1$ if previously correlated
sub-states are to be compared.

A similar argument applies for the case of the neutrinos. In order to
contest that a neutrino state has been prepared as left-handed, a preferred
left-right basis must first be agreed upon.

Summarising, then, by finding qubit $2$ to be down, i.e. $|0\rangle _{2},$
relative to the basis $\{|0\rangle _{2},|1\rangle _{2}\},$ the corresponding
basis $\{|0\rangle _{1},|1\rangle _{1}\}$ immediately becomes `preferred'
for qubit $1.$ Any subsequent rotation of qubit $1$ (say from $|0\rangle
_{1} $ to $|1\rangle _{1})$ now changes its state relative to this chosen
basis, and as such would imply an active transformation. It is not
surprising, then, that this active transformation leads to a difference in
the information attributed to the state. In fact, taking this point to
absurdity for the sake of clarity, once qubit $2$ has been found to be
`down' in some basis, it could obviously not itself then be passively
relabelled as `up' relative to the same axis.

Note that none of this is saying that $|0\rangle _{1}$ is the only possible
outcome of a \textit{measurement} on qubit $1.$ What it does mean, however,
is that prior to such a measurement the initial state of qubit $1$ must be $%
|0\rangle _{1},$ and not an arbitrarily relabelled version.\bigskip

Overall, then, it is the state collapse processes, i.e. the measurements,
that cause information to be `extracted' from the system and hence a
preferred basis to be defined. Before a measurement it is possible to
arbitrarily relabel $0$'s and $1$'s (i.e. $\Psi _{n}\rightarrow \Psi
_{n}^{\prime }),$ but after information has been extracted and the state has
irreversibly changed it is then too late to consider further changes of the
basis. Paraphrasing, after a measurement the `direction' of the basis
becomes fixed, so any additional rotation would be a transformation relative
to this chosen direction. Moreover, because comparisons are now possible,
this sort of rotation is no longer un-observable, and so any such
`relabelling' of the basis sets becomes an active process.

Of course, the differences caused by rotating a sub-state relative to the
preferred basis of the observer could themselves be passively removed by
transforming this basis in the same way as the sub-state. However, this then
becomes effectively equivalent to a global transformation, and is therefore
not relevant to the present discussion.

Thus, it is the state collapse mechanism that prohibits passive
transformations from leading to observable physical consequences, and
conversely prevents actual changes in individual factors from being
passively `transformed away'. Moreover, this conclusion again highlights the
fact that it is the state reduction postulate of quantum mechanics that
introduces non-trivial dynamics into a system, and thereby parameterises
physical changes in terms of information acquisition.

\bigskip

From the above discussions, it is evident that real, physical information
change is a concept that is meaningful relative to the comparison of states
against the same basis. This is perhaps not too surprising: after all, the
ability to physically compare objects is a fundamental prerequisite to any
discussion involving change. Moreover, in a fully quantum Universe this
basis is defined by an internal, endo-physical `observer' engaged in a
process of apparent, emergent measurements. In fact, these two points
represent the very essence of the measurement problem: in quantum theory a
state under investigation changes when it interacts with an observer. It is
important, then, to specify what it really means to talk of an endo-physical
measurement.

Of course, from an exo-physical perspective it is always valid to discuss
the measurement of an isolated state by an external observer. From the
endo-physical perspective of a fully quantum Universe, however, what is
generally of issue is how the entire state of the Universe changes in such a
way that it appears (on emergent scales) like one of its sub-states has
measured a second. In other words, what is more of interest in an
endo-physical discussion of measurement is not whether information has
\textit{changed} during a transition from $\Psi _{n}$ to $\Psi _{n+1},$ but
whether information has been \textit{exchanged. }Paraphrasing, of more
concern than information change, where separate factors may be taken to
change individually and independently of one other, is information exchange,
in which the \textit{relationship} between different parts of the state is
altered.\bigskip

It is possible to formally define what is meant by a change in information,
and what is meant by an exchange of information. As is evident from earlier,
a change in information occurs during a transition between $\Theta $ and $%
\Phi $ if these two states are fundamentally different. Moreover, this
difference must not just simply be mathematical, but must also take account
of certain physical constraints, such as prior measurements giving rise to
preferred bases.

However, within this set of information changing transitions are the
information exchanging processes, defined below as those procedures in which
parts of the state appear to interact with each other. Specifically, if $%
\Theta $ and $\Phi $ are both contained in the Hilbert space $\mathcal{H}%
_{[1...N]},$ then during a transition from $\Theta $ to $\Phi $ the
component of the state in subregister $\mathcal{H}_{i}$ may be defined as
having exchanged information with the component of the state in $\mathcal{H}%
_{j}$ if the `relationship' between these two components has changed. As an
example, if these two components were perhaps separate in $\Theta $ but are
entangled with one another in $\Phi ,$ it is evident that the relationship
between them has been altered. Note, then, that an exchange of information
necessarily implies a change in information, but a change in information may
not necessarily have to imply an exchange.

What is ideally sought, therefore, is a test of whether two particular
sub-states appear to have exchanged information with each other during a
particular transition. In other words, does the relationship between the
components of the state in factor Hilbert spaces $\mathcal{H}_{i}$ and $%
\mathcal{H}_{j}$ change during the jump from $\Psi _{n}$ to $\Psi _{n+1}?$
Thus, do the components appear to `interact' with each other in any way?
Indeed, could it ultimately be possible to consider the component in $%
\mathcal{H}_{i}$ as `measuring' the component in $\mathcal{H}_{j}$ during
the transition, at least in an emergent sense?\bigskip

To begin to answer these questions, recall that the dynamics of the quantum
Universe relies on the principle that, given an Hermitian operator $\hat{%
\Sigma}_{n+1},$ whichever eigenvector is selected automatically becomes the
initial state for the subsequent transition involving $\hat{\Sigma}_{n+2}.$
That is, the development of the Universe is viewed as a giant and autonomous
process of quantum testing and retesting. Moreover, because it is assumed
that the actual outcome $\Psi _{n+1}$ of a test is necessarily different
from the previous state $\Psi _{n},$ the operator $\hat{\Sigma}_{n+1}$ used
in the Universe's development must induce a change in information, and so
must represent an active transformation. Further, because of the lack of an
external agent deciding upon a preferred basis for the quantum Universe, any
change of information within the system cannot arise simply from a
relabelling of the `axes'. This is another reason why the transition from $%
\Psi _{n}$ to $\Psi _{n+1}$ cannot therefore be a consequence of any sort of
passive transformation.

The previous sub-section demonstrated that unitary operators acting locally
in the individual subregisters have unobservable consequences, and that
these transformations preserve the separation and entanglement properties of
the state. The opposite is also considered: any change in partition of the
state of the Universe must imply an active transformation because it cannot
be removed by a local relabelling of the basis. Consequently, a change of
partition results in a change in the information carried by the state. Of
course, this last point might be expected immediately: given that the
information content of a state has already been related to the ways its
separates, any change in partition must automatically imply a change in
information.\bigskip

This can be presented formally. Consider two successive states $\Psi _{n}$
and $\Psi _{n+1}$ in the Hilbert space $\mathcal{H}_{[1...N]}.$ The state $%
\Psi _{n}$ will have $F_{n}$ factors and lies in the partition $\mathcal{P}%
_{n},$ which has $F_{n}$ blocks, where $1\leq F_{n}\leq N.$ Similarly the
state $\Psi _{n+1}$ will have $F_{n+1}$ factors and lies in the partition $%
\mathcal{P}_{n+1}$ with $F_{n+1}$ blocks, noting that $F_{n+1}$ is not
necessarily equal to $F_{n},$ and that even if $F_{n}=F_{n+1}$ the partition
$\mathcal{P}_{n+1}$ is not necessarily the same as $\mathcal{P}_{n}$ (e.g. $%
\mathcal{H}_{1}^{234}\not\equiv \mathcal{H}^{12\bullet 34},$ but both have
two blocks).

From the conclusions of the earlier discussions, it is now possible to
conject that:

\begin{itemize}
\item  Information has been exchanged during the transition from $\Psi _{n}$
to $\Psi _{n+1}$ if $\mathcal{P}_{n}\neq \mathcal{P}_{n+1}.$
\end{itemize}

The converse is also true:\ an exchange of information implies a change in
partition. The above conjecture follows from the very definition of a
partition; if the state $\Psi _{n}$ is in a different partition from the
state $\Psi _{n+1},$ it means that at least two components of the state have
changed their block during the transition. Moreover, it must then be the
case that at least one of the components of $\Psi _{n},$ in a particular
subregister $\mathcal{H}_{i},$ must have changed its relationship (i.e. its
entanglement) with at least one other component, in a different subregister $%
\mathcal{H}_{j},$ when the state became $\Psi _{n+1}.$\bigskip

To rephrase this conjecture, consider the probability amplitude $P=\langle
\Psi _{n+1}|\Psi _{n}\rangle .$ As shown in Section 5.4 this probability
amplitude will have $F_{P}$ factors, where $F_{P}\leq \min (F_{n},F_{n+1}).$
For example, if $\Psi _{n}$ has two factors whilst $\Psi _{n+1}$ is fully
entangled, then $F_{n}=2$ and $F_{n+1}=1,$ such that $F_{P}=1.$

Now, by the above argument, information exchange has occurred in the
transition from $\Psi _{n}$ to $\Psi _{n+1}$ if
\begin{equation}
F_{P}<F_{n}\text{ \ \ or \ \ }F_{P}<F_{n+1}.
\end{equation}

In other words, if the number of factors of the probability amplitude is
less than the number of factors of either the initial or final states, then $%
\Psi _{n}$ and $\Psi _{n+1}$ are in different partitions, and the transition
is an information exchanging process.

Of course, this condition is immediately satisfied if $F_{n}\neq F_{n+1}.$%
\bigskip

To go further, consider the general form of $P,$ fundamentally factorised as
\begin{equation}
P=\langle \Psi _{n+1}^{(1)}|\Psi _{n}^{(1)}\rangle \langle \Psi
_{n+1}^{(2)}|\Psi _{n}^{(2)}\rangle ...\langle \Psi _{n+1}^{(F_{P})}|\Psi
_{n}^{(F_{P})}\rangle
\end{equation}
where $\Psi _{n}^{(p)}$ and $\Psi _{n+1}^{(p)}$ may themselves be products
of $k_{n}^{(p)}$ and $k_{n+1}^{(p)}$ factors respectively, and $%
p=1,2,...,F_{P}$ for $k_{n}^{(p)},k_{n+1}^{(p)}\in \mathbb{Z}^{+}$ and $%
\sum_{p=1}^{F_{p}}k_{n}^{(p)}=F_{n},$ but $k_{n}^{(p)}$ is not necessarily
equal to $k_{n+1}^{(p)}.$

Moreover each factor of $\Psi _{n}^{(p)}$ is in some block of $\mathcal{P}%
_{n},$ whereas each factor of $\Psi _{n+1}^{(p)}$ is in some block of $%
\mathcal{P}_{n+1},$ with the proviso that for $P$ to factorise in the above
way the two sub-states $\Psi _{n}^{(p)}$ and $\Psi _{n+1}^{(p)}$ must be
contained in exactly the same set of subregisters, such that $\Psi
_{n}^{(p)},\Psi _{n+1}^{(p)}\in \mathcal{H}_{[p]}$ for $\tprod%
\nolimits_{p=1}^{F_{P}}(\otimes \mathcal{H}_{[p]})=\mathcal{H}_{[1...N]}.$

From this, it is now possible to assert that during the jump from $\Psi _{n}$
to $\Psi _{n+1}$:

\begin{itemize}
\item  The component of the state in\ factor Hilbert space $\mathcal{H}_{i}$
exchanges information with the component in\ factor Hilbert space $\mathcal{H%
}_{j},$ for $i\neq j,$ if the components of $\Psi _{n}$ in $\mathcal{H}_{i}$
and $\mathcal{H}_{j}$ are in the same block $B$ but the components of $\Psi
_{n+1}$ in $\mathcal{H}_{i}$ and $\mathcal{H}_{j}$ are not in $B,$ \textit{or%
} if the components of $\Psi _{n+1}$ in $\mathcal{H}_{i}$ and $\mathcal{H}%
_{j}$ are in the same block $B^{\prime }$ but the components of $\Psi _{n}$
in $\mathcal{H}_{i}$ and $\mathcal{H}_{j}$ are not in $B^{\prime }.$
\end{itemize}

Clearly, this statement is equivalent to the condition that:

\begin{itemize}
\item  The component of the state in\ factor Hilbert space $\mathcal{H}_{i}$
exchanges information with the component in\ factor Hilbert space $\mathcal{H%
}_{j},$ for $i\neq j,$ if the components of $\Psi _{n}$ in $\mathcal{H}_{i}$
and $\mathcal{H}_{j}$ are in the same block $B$ of $\mathcal{P}_{n}$ but $%
\mathcal{P}_{n+1}$ does not possess this block, \textit{or} if the
components of $\Psi _{n+1}$ in $\mathcal{H}_{i}$ and $\mathcal{H}_{j}$ are
in the same block $B^{\prime }$ of $\mathcal{P}_{n+1}$ but $\mathcal{P}_{n}$
does not possess this block.\bigskip
\end{itemize}

As an illustration, consider a three qubit universe in the Hilbert space $%
\mathcal{H}_{[123]}$ spanned by the orthonormal basis
\begin{equation}
\mathcal{B}_{123}=\{|a\rangle _{1}\otimes |b\rangle _{2}\otimes |c\rangle
_{3}=|abc\rangle _{123}:a,b,c=0,1\}.
\end{equation}

If two consecutive states $\psi _{n}$ and $\psi _{n+1}$ turn out to be
\begin{eqnarray}
\psi _{n} &=&|000\rangle _{123}=|0\rangle _{1}\otimes |0\rangle _{2}\otimes
|0\rangle _{3} \\
\psi _{n+1} &=&|000\rangle _{123}+|011\rangle _{123}=|0\rangle _{1}\otimes
(|00\rangle _{23}+|11\rangle _{23})  \notag
\end{eqnarray}
it is evident that $\psi _{n}\in \mathcal{H}_{123}$ whilst $\psi _{n+1}\in
\mathcal{H}_{1}^{23}.$ Clearly, the component of $\psi _{n}$ in\ factor
Hilbert space $\mathcal{H}_{1}$ is in the same block as the component of $%
\psi _{n+1}$ in this subregister, and so may not be said to exchange
information with any of the other components during the transition from $%
\psi _{n}$ to $\psi _{n+1}.$ On the other hand, the components of $\psi _{n}$
in\ factor Hilbert spaces $\mathcal{H}_{2}$ and\ $\mathcal{H}_{3}$ are in
different blocks from the components of $\psi _{n+1}$ in $\mathcal{H}_{2}$
and\ $\mathcal{H}_{3},$ and clearly the relationship between these
components changes during the jump. In this case, it may be said that
information is exchanged between qubits $2$ and $3$ during the transition
from $\psi _{n}$ to $\psi _{n+1}.$

\bigskip

\subsubsection{The Ideal Physics Experiment}

\bigskip

Discussions of information exchange during the development of a system lead
naturally onto questions of the measurement problem in quantum mechanics and
the nature of endo-physical experimentation. After all, the acquisition of
information is the very purpose of measurement.

The concept of measurement is generally well understood in exo-physics, with
a famous exception being the problem of state reduction in quantum mechanics
and the corresponding conflicts of interpretation regarding what this
actually means. In exo-physics, discussions often involve large,
semi-classical observers surrounded by an even larger environment, who are
observing the isolated and microscopic quantum state under investigation.
Moreover, during this process the observers and environment are often
assumed to be unaffected, or at least any changes in them are taken to be
insignificant.

These perspectives, however, are incompatible with the notion of
endo-physics, and are therefore also ultimately incompatible with the nature
of a physical Universe incorporating quantum theory as the fundamental
ingredient. As has been remarked previously, in a fully quantum treatment of
the Universe the system under investigation, the observer and every element
of the environment should really just be viewed as factors or groups of
factors of the single state representing all of physical reality. An
endo-physical experiment consequently involves one part of the Universe's
state appearing to measure another part of it. Moreover, this also implies
that the state itself must be developing in a highly organised way if
emergent, internal scientists are to gain the illusion that they are
independent, classical and isolated observers who can investigate and
develop their surroundings with apparent free will.\bigskip

The question of endo-physical measurement is an enormous subject,
undoubtedly worthy of a research programme in its own right. It is therefore
not attempted here to provide a complete and self-contained study of how
this process actually works. After all, recall that there is currently no
mechanism known for determining how and why the Universe selects which
operators it uses to test itself with, so a theory of how emergent, internal
physicists gain the impression that they can decide how the Universe around
them develops must be even further away.

What can be described, however, are the essential points that this unknown
theory should incorporate, or alternatively what conditions such a mechanism
might be expected to satisfy. In other words, in the following the necessary
features and constraints for endo-physical experimentation are discussed.

\bigskip

Consider the sort of experiment performed by physicists everyday. Such a
situation necessarily incorporates at least two parts: there must be a
`subject' to be measured, and there must be some sort of `device' to do the
measuring\footnote{%
Of course, for the sake of the present argument it does not matter what
either the `device' or the `subject' actually are. For example, in different
contexts the generic word `device' could be taken to imply a large piece of
equipment, or a single `pointer-state', or even the eye of a human observer;
it is the fact that a physical experiment necessarily contains both a
subject and a device that is important.}. Furthermore, before any
measurement takes place these two parts must have some sort of independent
existence, and so must be classically isolated from each other.

Now, in the context of the paradigm proposed in this thesis, the above
conditions are achieved by recalling that in the quantum Universe every
classically isolated physical system is associated with a factor of the
Universe's state. Thus, in order for any physical experiment to occur it is
required that the Universe's state must be separable into at least two
factors, one of which is ultimately taken to represent the `subject' and the
other is taken to represent the `device'.

Consider the Universe at `time' $n,$ represented by a state $\Psi _{n}$
contained in a Hilbert space $\mathcal{H}_{[1...N]}$ factorisable into $N$
subregisters $\mathcal{H}_{m},$ for $m=1,2,...N.$ Since all of physical
reality is expected to emerge from this fundamental state description, it
may be assumed that some `portion' of $\Psi _{n}$ accounts for the subject,
some of it accounts for the device, and the rest accounts for everything
else in the Universe.

Thus, without loss of generality assume that the factor of $\Psi _{n}$
representing the subject is contained in subregisters $1$ to $x.$ That is,
assume that it is the components of $\Psi _{n}$ in $\mathcal{H}_{[1...x]}$
that are (somehow) responsible for the physical appearance of the subject on
emergent scales. Similarly, assume that the device emerges from some sort of
consideration of those components of $\Psi _{n}$ contained in the
subregisters $(x+1)$ to $y,$ where again the exact mechanism of this
origination lies in the realm of an unknown theory of emergence. This leaves
subregisters $(y+1)$ to $N$ to account for the emergence of everything else
in the Universe, noting that in the above the subregisters have been
arbitrarily labelled in ascending order for clarity. Of course, generally in
physics experiments the subjects are much smaller than the devices that
measure them, which are in turn dwarfed by the scale of everything else in
the Universe. It may therefore naturally be expected that $x\ll y\ll N,$ but
this is not essential.

The above division of $\mathcal{H}_{[1...N]}$ into subject, device and
everything else denotes a particular split of the Hilbert space. That is,
\begin{equation}
\mathcal{H}_{[1...N]}=\mathcal{H}_{[1...x]}\otimes \mathcal{H}%
_{[(x+1)...y]}\otimes \mathcal{H}_{[(y+1)...N]}.  \label{Ch4Split}
\end{equation}

The present discussion is concerned with the relationships occurring in a
single, isolated endo-measurement between a `subject' and a `device'. It is
therefore possible to further simplify the above situation by considering a
`toy-universe' containing nothing but these two features, that is, one for
which $y=N.$ Such a universe defines the alternative split
\begin{equation}
\mathcal{H}_{[1...N]}=\mathcal{H}_{[1...y]}=\mathcal{H}_{[1...x]}\otimes
\mathcal{H}_{[(x+1)...y]}=\mathcal{H}_{[A]}\otimes \mathcal{H}_{[B]}
\label{Ch4Split1}
\end{equation}
where the sub-scripts $A$ and $B$ are adopted for brevity to denote
respective subject and device factor Hilbert spaces.

Now, in order to consider a classically distinct subject and device, argued
as above to be essential pre-conditions to any discussion of experiments,
the state $\psi _{n}$ of the toy-universe must be separable relative to the
split (\ref{Ch4Split1}). Thus, if $|\phi _{n}\rangle $ represents the
sub-state of the subject at time $n,$ and $|\Lambda _{n}\rangle $ the
sub-state of the device at time $n,$ then clearly $|\phi _{n}\rangle \in
\mathcal{H}_{[1...x]}\equiv \mathcal{H}_{[A]}$ and $|\Lambda _{n}\rangle \in
\mathcal{H}_{[(x+1)...y]}\equiv \mathcal{H}_{[B]},$ such that
\begin{equation}
\psi _{n}=|\phi _{n}\rangle _{\lbrack 1...x]}\otimes |\Lambda _{n}\rangle
_{\lbrack (x+1)...y]}.
\end{equation}

Clearly, $\psi _{n}$ is in the separation given by
\begin{equation}
\psi _{n}\in \mathcal{H}_{[1...x]\bullet \lbrack (x+1)...y]}=\mathcal{H}%
_{AB}.
\end{equation}

It is important that no mention has been made so far as to the actual nature
of either the `subject' or `device'. Indeed, $|\phi _{n}\rangle _{\lbrack
1...x]}$ could itself be separable, entangled, or a separable product of
entangled factors relative to its factor Hilbert space $\mathcal{H}%
_{[1...x]},$ and $|\Lambda _{n}\rangle _{\lbrack (x+1)...y]}$\ could
similarly be in any of the partitions of $\mathcal{H}_{[(x+1)...y]},$
recalling that each of the factors $\mathcal{H}_{[1...x]}$ and $\mathcal{H}%
_{[(x+1)...y]}$ are themselves vector spaces. In fact, in the `real world'
case where $y\neq N$ it is almost taken for granted that the sub-state in $%
\mathcal{H}_{[(y+1)...N]}$ is highly separable if it is expected to
represent everything else in a semi-classical looking Universe.\bigskip

Consider now the next state of the universe, $\psi _{n+1},$ which is one of
the eigenvectors of some Hermitian operator $\hat{\Sigma}_{n+1}.$ There are,
of course, a number of different forms that $\psi _{n+1}$ could take, and it
could potentially be in any one of the many partitions of $\mathcal{H}%
_{[1...y]}.$

For example, this subsequent state could also be in the separation $\mathcal{%
H}_{AB},$ such that it may be of the form
\begin{equation}
\psi _{n+1}=|\phi _{n+1}\rangle _{\lbrack 1...x]}\otimes |\Lambda
_{n+1}\rangle _{\lbrack (x+1)...y]},
\end{equation}
with $|\phi _{n+1}\rangle _{\lbrack A]}\neq |\phi _{n}\rangle _{\lbrack A]},$
or $|\Lambda _{n+1}\rangle _{\lbrack B]}\neq |\Lambda _{n}\rangle _{\lbrack
B]}$ if, by axiom, $\hat{\Sigma}_{n+1}$ is assumed not to be a null test. In
this case, and from the conclusions of the previous sub-section, no
information may be said to have been exchanged between the components in
Hilbert sub-space $\mathcal{H}_{[1...x]}$ and the components in $\mathcal{H}%
_{[(x+1)...y]}$ during the transition, because the state did not change in
separability relative to these: $\psi _{n}\in \mathcal{H}_{AB}$ \textit{and}
$\psi _{n+1}\in \mathcal{H}_{AB}.$

So from the point of view of the separation $\mathcal{H}_{AB},$ the
sub-state in $\mathcal{H}_{[1...x]}$ and the sub-state in $\mathcal{H}%
_{[(x+1)...y]}$ are developing independently of one another, with no sort of
influence or interaction occurring between them. To all intents and
purposes, during this jump from $\psi _{n}$ to $\psi _{n+1}$ the factor in $%
\mathcal{H}_{[1...x]}$ and the factor $\mathcal{H}_{[(x+1)...y]}$ would be
developing like distinct `mini-universes' separate from each other, though
care is needed not to take this interpretation too far because the operator $%
\hat{\Sigma}_{n+1}$ is still acting across the entire Hilbert space, $%
\mathcal{H}_{[1...N]}.$

In this case, it may be convenient to represent the transition $\psi
_{n}\rightarrow \psi _{n+1}$ as
\begin{equation}
\mathbf{\{}|\phi _{n}\rangle _{\lbrack 1...x]}\rightarrow |\phi
_{n+1}\rangle _{\lbrack 1...x]}\mathbf{\}}\otimes \mathbf{\{}|\Lambda
_{n}\rangle _{\lbrack (x+1)...y]}\rightarrow |\Lambda _{n+1}\rangle
_{\lbrack (x+1)...y]}\mathbf{\}}.  \label{Ch4Caus}
\end{equation}

Of course, the preservation of the separability of the state relative to $%
\mathcal{H}_{[AB]}$ does not automatically imply that the factors themselves
have developed in trivial ways. After all, recall that $\mathcal{H}%
_{[1...x]} $ and $\mathcal{H}_{[(x+1)...y]}$ may each be a product of very
many subregisters, and this gives rise to the possibility of many different
types of internal transitions within these individual spaces. For example, $%
|\phi _{n}\rangle _{\lbrack A]}$ may be completely separable relative to $%
\mathcal{H}_{[1...x]},$ whereas $|\phi _{n+1}\rangle _{\lbrack A]}$ may be
completely entangled, assuming that $x>1.$

In fact, it is this type of possibility that provides the most manifest
difference between information change and information exchange: the
information content of the factors of the state in sub-spaces $\mathcal{H}%
_{A}$ and $\mathcal{H}_{B}$ may have changed during the transition from $%
\psi _{n}$ to $\psi _{n+1},$ even though no information was exchanged
between these two sub-states. Of course, due to the internal transitions,
information may still potentially have been exchanged between any of the
factors of $|\phi _{n}\rangle _{\lbrack 1...x]},$ and, similarly, also
between any of the individual components of $|\Lambda _{n}\rangle _{\lbrack
(x+1)...y]}.$

Indeed, these points may be applied to the context of the quantum universe
by observing that both the `subject' and `device' described in the present
discussion may be arbitrarily large, and might therefore incorporate many
different levels of sub-subjects and sub-devices within themselves. For
instance, the `subject' could be a large `black-box' containing an electron
of unknown spin, a loaded gun, and a Cat, etc. Moreover, this is what
generally happens in laboratory physics, where a sample is often prepared
and left to undergo many different `internal interactions' before it is
eventually measured by an apparatus at some later time; an experiment in the
field of chemistry is a good example of this.

Continuing, because the factors in each of these sub-spaces are effectively
developing like independent mini-universes, then if this isolation remains
for many more transitions they might also begin to develop their own
internal causal set type relationships, as apparent from (\ref{Ch4Caus}).
This could, in turn, give rise to concepts such as internal measures of
space and differing notions of endo-time.

\bigskip

Assume instead, however, that the jump from $\psi _{n}$ to $\psi _{n+1}$
represents the pregeometric equivalent of a device measuring a subject.
Indeed, given that scientists do seem to be able to perform experiments, and
that these scientists and their equipment are fundamentally just sub-states
of the quantum Universe, there must be some sort of origin for this emergent
effect.

Now, although it may be difficult at this stage to say exactly how such a
pregeometric experiment occurs, by appealing to the consequences of actual
physical measurements it is possible to make inferences about their
microscopic counterparts. For example, an experiment necessarily involves an
extraction of information, because the purpose of a measurement is
ultimately to obtain information about the subject under investigation.
Thus, the information content of the sub-state representing the device must
necessarily change during the measurement: its information afterwards must
be different from its information before, because it must incorporate the
newly acquired information regarding the measured subject.

Likewise, and for two similar reasons, the information content of the
sub-state representing the subject must also change. Firstly, because there
are no non-invasive measurements in quantum physics, any observation
automatically affects the sub-state being observed. The only exceptions to
this rule are null tests, and these are considered unobservable. Secondly,
and encompassing the first point, the symmetry of the situation implies an
equivalence between observer and observed. From the point of view of a fully
quantum Universe, both the observer and the system under investigation are
just factors of the overall state, and so it is not really valid to say
`who' is actually doing the measuring, nor are there any real grounds to
make such a choice. So, if a device is measuring a subject, symmetry implies
that the subject is equally measuring the device. In fact, in the emergent
limit it is only ever possible to discuss a physical observer performing a
measurement on a subject (instead of vice versa) because observer states are
often taken to be very much larger that the systems under investigation.
Thus, if these `observer' states do not change very much during the process,
it may be valid to make the approximation of a `constant' observer measuring
a changing quantum system. This point is discussed later.\bigskip

The conclusion of the above discussion is therefore that an endo-physical
measurement relies on an exchange of information between the device and the
subject. Moreover, and by the results of the previous sub-sections, for an
exchange of this type to occur it is necessary for the relationship between
one of the components of the state in $\mathcal{H}_{[1...x]}$ to change its
relationship with one of the state's components in $\mathcal{H}%
_{[(x+1)...y]} $ during the jump from $\psi _{n}$ to $\psi _{n+1}.$
Furthermore, this in turn implies that the state $\psi _{n+1}$ must be
entangled relative to the split $\mathcal{H}_{A}\otimes \mathcal{H}_{B},$
and hence a member of $\mathcal{H}^{AB}.$

Of course, exactly how this partition change physically affects the state is
a greater question. Indeed, the resolution of this issue involves the actual
choice of the operators $\hat{\Sigma}_{n+1}$ themselves, and this requires a
knowledge of exactly how the self-referential nature of the Universe's
development might work (a point that is discussed in Chapter 8). Elaborating
on this, presumably the sub-state of the device must be changed in a way
that depends on the sub-state of the subject if the jump from $\psi _{n}$ to
$\psi _{n+1}$ is to represent the type of measurement familiar to
experimental physics. Indeed, if this were not the case the process could
hardly be called a measurement at all, because no useful information would
have been extracted about the subject by the device.

Similarly, and by the symmetry of the situation, the sub-state of the
subject must also be changed during the transition from $\psi _{n}$ to $\psi
_{n+1}$ in a way that depends on the sub-state of the device. This point is
also echoed in empirical science, where the state a subject is projected
into upon measurement may depend very much on the object that was measuring
it; for example, in an experiment involving the measurement of an electron's
spin, the electron is projected into a spin-state that depends on the
orientation of the Stern-Gerlach apparatus. Such `feed-back' mechanisms,
however, are beyond the scope of the present discussion.

What can be concluded, though, is still an important point regarding the
nature of experimentation in the quantum universe:

\begin{itemize}
\item  \textit{An endo-physical measurement necessarily implies an exchange
of information between the subject and the device. Further, this necessarily
implies a partition change involving some of those components of the state
representing the subject and some of those components of the state from
which the (emergent) description of the device arises}.\bigskip
\end{itemize}

Specifically, if $\psi _{n}\in \mathcal{H}_{AB},$ then a subject sub-state
in $\mathcal{H}_{A}$ is `measured' by a device sub-state in $\mathcal{H}_{B}$
during the transition from $\psi _{n}$ to $\psi _{n+1}$ iff $\psi _{n+1}\in
\mathcal{H}^{AB}.$ Clearly, for such a measurement the partition $\mathcal{P}%
_{n}$ of $\mathcal{H}_{[1...y]}$ containing $\psi _{n}$ cannot be equal to
the partition $\mathcal{P}_{n+1}$ of $\mathcal{H}_{[1...y]}$ containing $%
\psi _{n+1}.$\bigskip

It is this point that justifies the earlier simplification of discussing a
`toy-universe' containing just a subject and a device. If the original case
is again considered, i.e. when $y\neq N$ and the Universe's Hilbert space $%
\mathcal{H}_{[1...N]}$ is split as (\ref{Ch4Split}) allowing the state $\Psi
_{n}\in \mathcal{H}_{[1...N]}$ to have a `rest of the Universe' factor $%
|r_{n}\rangle _{\lbrack (y+1)...N]}$ in $\mathcal{H}_{[(y+1)...N]},$ then as
long as $|r_{n}\rangle _{\lbrack (y+1)...N]}$ does not interact or exchange
information with any of the components of the state in $\mathcal{H}_{A}$ or $%
\mathcal{H}_{B},$ then this sub-state is effectively existing in its own
isolated mini-universe. Paraphrasing, if the development of the components
of the state in spaces $\mathcal{H}_{A}$ or $\mathcal{H}_{B}$ is restricted
such that they can only interact with other components in $\mathcal{H}_{A}$
or $\mathcal{H}_{B},$ then these are also effectively existing as a
mini-universe separate from the components of the state in $\mathcal{H}%
_{[(y+1)...N]}.$

In fact, if $\mathcal{H}_{[AB]}$ is itself suitably factorisable into sets
of sub-spaces and sub-sub-spaces, it is additionally possible that within
this mini-universe whole levels of sub-measurements could simultaneously
occur as it develops from one state to the next. This type of process would
thus be equivalent to various sub-devices measuring sub-subjects, and
sub-sub-devices measuring sub-sub-subjects, etc., each of which is contained
in its own factor sub-space of $\mathcal{H}_{[AB]}.$ As before, a strong
parallel is drawn here with the Schr\"{o}dinger's Cat paradox, where within
the Hilbert space of the `black-box' containing the Cat, the gun, and the
electron, numerous levels of endo-measurement could be occurring.

These points may be stated more formally: because $\mathcal{H}_{[AB]}=%
\mathcal{H}_{A}\otimes \mathcal{H}_{B}$ it just a vector space in its own
right it can in some sense be treated as an independent entity. It is then
always possible to tensor product $\mathcal{H}_{[AB]}$ with additional
Hilbert spaces without affecting the physics as long as sub-states contained
in $\mathcal{H}_{[AB]}$ do not become entangled with components in these new
spaces. Moreover, because any factors of $\mathcal{H}_{A}$ and $\mathcal{H}%
_{B}$ are themselves also vector spaces, each of these may too be granted an
independent existence.

This again reinforces the point that it is acceptable to consider just a
`subject and device' toy-universe without loss of generality.\bigskip

The caveat to this discussion involves the operators. In the quantum
Universe, the entire system is tested by a global, Hermitian operator $\hat{%
\Sigma}_{n+1}$ acting self-referentially according to the current stage. It
is this fact that may prevent the separate mini-universes from being real,
physical, isolated and independent universes, because the overall choice of
operator affecting the sub-state in one mini-universe may be influenced by
the sub-state of another mini-universe. In other words, the operator $\hat{%
\Sigma}_{n+1},$ which is obviously responsible for developing sub-states in $%
\mathcal{H}_{[1...y]},$ may be dependent on the sub-state contained in $%
\mathcal{H}_{[(y+1)...N]}.$ This point is addressed again by Chapter 8, but
in the present discussion involving just the \textit{principles} of an ideal
physics experiment such a technicality is not too drastic.

Note that a jump from one state to the next could contain many different
`isolated universes' if $N\gg y.$ Indeed, if the probability amplitude $%
\langle \Psi _{n+1}|\Psi _{n}\rangle $ factorises into $F_{P}$ factors, then
each of these is effectively representing a separate mini-universe during
that transition. So, every factor of $\langle \Psi _{n+1}|\Psi _{n}\rangle $
that contains an initial product of factors of $\Psi _{n}$ entangling with
one another during the jump to $\Psi _{n+1}$ implies an endo-physical
measurement occurring between these factors of the initial product. It is
hence possible that a jump from one state to the next may permit many
different sets of device sub-states independently and simultaneously
appearing to measure their own subject sub-states.\bigskip

Information exchanging partition changes need not actually be too dramatic,
a point that can be illustrated when the above spaces $\mathcal{H}_{[A]}$
and $\mathcal{H}_{[B]}$ are written as $\mathcal{H}_{[1...x]}$ and $\mathcal{%
H}_{[(x+1)...y]},$ with $y=N$ again for simplicity. For example, assume that
$|\phi _{n}\rangle _{\lbrack 1...x]}$ is completely separable relative to $%
\mathcal{H}_{[1...x]},$ i.e. $|\phi _{n}\rangle _{\lbrack 1...x]}=|\phi
_{n}\rangle _{1...x},$ and $|\Lambda _{n}\rangle _{\lbrack (x+1)...y]}$ is
completely separable relative to $\mathcal{H}_{[(x+1)...y]},$ such that $%
\psi _{n}$ is in the partition $\mathcal{P}_{n}\equiv \mathcal{H}_{1...y}.$
Assume further that the next state of the Universe $\psi _{n+1}$ is given by
\begin{equation}
\psi _{n+1}=|\varphi _{n+1}\rangle _{\lbrack 1...(i-1)(i+1)...x]}\otimes
|\Upsilon _{n+1}\rangle _{\lbrack i(x+1)...y]}  \label{Ch4Init}
\end{equation}
where $|\varphi _{n+1}\rangle _{\lbrack 1...(i-1)(i+1)...x]}$ is completely
separable relative to $\mathcal{H}_{[1...(i-1)(i+1)...x]},$ i.e.
\begin{equation}
|\varphi _{n+1}\rangle _{\lbrack 1...(i-1)(i+1)...x]}\in \mathcal{H}%
_{1...(i-1)(i+1)...x}.
\end{equation}

Finally, assume that the component of $\psi _{n+1}$ in $\mathcal{H}_{i}$ is
entangled with just one other component, namely the component of $\psi
_{n+1} $ in $\mathcal{H}_{j},$ where $(x+1)\leq j\leq y,$ and that $%
|\Upsilon _{n+1}\rangle _{\lbrack i(x+1)...y]}$ is completely separable
relative to $\mathcal{H}_{[i(x+1)...y]}$ apart from this one entangled
factor. Evidently, $|\Upsilon _{n+1}\rangle _{\lbrack i(x+1)...y]}$ has $%
(y-x)$ factors. Then, $\psi _{n+1}$ is in the partition $\mathcal{P}_{n+1}$
given by
\begin{equation}
\mathcal{P}_{n+1}\equiv \mathcal{H}_{1...(i-1)(i+1)...(j-1)(j+1)...y}^{ij}
\end{equation}
and may be written
\begin{equation}
\psi _{n+1}=|\varphi _{n+1}\rangle _{1...(i-1)(i+1)...x}\otimes |\Upsilon
_{n+1}\rangle _{(x+1)...(j-1)(j+1)...y}^{ij}  \label{Ch4Onecomp}
\end{equation}

Clearly, the jump from $\psi _{n}$ to $\psi _{n+1}$ represents an
information exchanging process between the factor of $\psi _{n}$ in $%
\mathcal{H}_{[A]}\equiv \mathcal{H}_{[1...x]}$ and the factor of $\psi _{n}$
in $\mathcal{H}_{[B]}\equiv \mathcal{H}_{[(x+1)...y]},$ because the state
has changed its separability relative to $\mathcal{H}_{[AB]}.$ That is, $%
\psi _{n}\in \mathcal{H}_{AB}$ whereas $\psi _{n+1}\in \mathcal{H}^{AB},$
such that $\mathcal{P}_{n}\neq \mathcal{P}_{n+1}.$ This conclusion, however,
is despite the fact that nearly all of the components of the state
representing the subject (in the individual subregisters $\mathcal{H}_{m},$
for $1\leq m\leq x)$ and nearly all of the components of the state
representing the device (in the individual subregisters $\mathcal{H}_{m},$
for $(x+1)\leq m\leq y)$ did not interact with anything or each other.

\bigskip

The above points regarding different sets and levels of subjects and devices
in the quantum Universe call for the definition of an endo-physical
measurement to be refined. Consider a Universe in a Hilbert space $\mathcal{H%
}_{[1...N]},$ where $N$ is large such that states in $\mathcal{H}_{[1...N]}$
may be highly separable. Consider also a particular $k$-partite split $%
\mathcal{H}_{[K_{1}K_{2}...K_{k}]}$ of $\mathcal{H}_{[1...N]}$ as
\begin{equation}
\mathcal{H}_{[1...N]}\equiv \mathcal{H}_{[K_{1}K_{2}...K_{k}]}=\mathcal{H}%
_{[K_{1}]}\otimes \mathcal{H}_{[K_{2}]}\otimes ...\otimes \mathcal{H}%
_{[K_{k}]}
\end{equation}
where the $k$ factor spaces of the split need not be fundamentally
factorised relative to the individual subregisters, and so need not be of
prime dimension; the $(K_{a})^{th}$ factor space $\mathcal{H}_{[K_{a}]}$
could be the tensor product of a number of elementary subregisters, such
that for example $\mathcal{H}_{[K_{1}]}=\mathcal{H}_{6}\otimes \mathcal{H}%
_{14}\otimes \mathcal{H}_{23},$ $\mathcal{H}_{[K_{2}]}=\mathcal{H}%
_{4}\otimes \mathcal{H}_{10},$ etc.

Assume now that the $n^{th}$ state $\Psi _{n}$ of the Universe has a factor
in the product sub-space $\mathcal{H}_{[K_{X}K_{Y}]}$ of $\mathcal{H}%
_{[1...N]},$ where $1\leq X,Y\leq k,$ and moreover that this factor is
separable relative to $\mathcal{H}_{[K_{X}]}\otimes \mathcal{H}_{[K_{Y}]}.$
Immediately, this implies that $\Psi _{n}$ must have a factor in the
sub-space $\mathcal{H}_{[K_{X}]}$ and a factor in the sub-space $\mathcal{H}%
_{[K_{Y}]}.$ Without loss of generality, the factor in $\mathcal{H}%
_{[K_{X}]} $ may be called the `device' whereas the factor in $\mathcal{H}%
_{[K_{Y}]}$ may be called the `subject', though of course a vice versa
description would be equally true, and exactly how well these labelled
sub-states represent actual physical and macroscopic objects is a question
for a theory of emergence.

Now, from an extension of the earlier definition given for a toy-universe
containing just an isolated device measuring an isolated subject, the factor
of $\Psi _{n}$ in sub-space $\mathcal{H}_{[K_{X}]}$ may be said to `measure'
(i.e. exchange information with) the factor of $\Psi _{n}$ in sub-space $%
\mathcal{H}_{[K_{Y}]}$ during the transition from $\Psi _{n}$ to $\Psi
_{n+1} $ if there exists a factor of $\Psi _{n+1}$ that is entangled
relative to \textit{at least} $\mathcal{H}_{[K_{X}]}\otimes \mathcal{H}%
_{[K_{Y}]}.$ In other words, for such a measurement the state $\Psi _{n+1}$
must either have a factor that is entangled relative to $\mathcal{H}%
_{[K_{X}]}\otimes \mathcal{H}_{[K_{Y}]},$ or else a factor that is entangled
relative to the larger tensor product sub-space $\mathcal{H}%
_{[K_{X}]}\otimes \mathcal{H}_{[K_{Y}]}\otimes \mathcal{H}_{[C]},$ where $%
\mathcal{H}_{[C]}$ is an arbitrary factor Hilbert space of dimension
\begin{equation}
2\leq \dim (\mathcal{H}_{[C]})\leq \dim (\mathcal{H}_{[1...N]})\div \left[
\dim (\mathcal{H}_{[K_{X}]})\times \dim (\mathcal{H}_{[K_{Y}]})\right]
\end{equation}
in some given split of $\mathcal{H}_{[1...N]}$ that includes $\mathcal{H}%
_{[K_{X}]}$ and $\mathcal{H}_{[K_{Y}]}$ as sub-spaces. Clearly, $\mathcal{H}%
_{[C]}$ must satisfy
\begin{equation}
\mathcal{H}_{[C]}\cap \mathcal{H}_{[K_{X}]}=\mathcal{H}_{[C]}\cap \mathcal{H}%
_{[K_{Y}]}=\emptyset .
\end{equation}

The above assertion defines a necessary condition for an endo-physical
measurement to occur between any two given factors of a state during a
transition, whether or not these factors may be further separated relative
to a more fundamental splitting of the Hilbert space. Paraphrasing this
definition:

\begin{itemize}
\item  \textit{Given a particular split }$\mathbf{S}$\textit{\ of }$\mathcal{%
H}_{[1...N]},$\textit{\ two factors }$X$\textit{\ and }$Y$\textit{\ of a
state }$\Psi _{n}$\textit{\ that is separable relative to }$\mathbf{S}$%
\textit{\ exchange information during a jump from }$\Psi _{n}$\textit{\ to }$%
\Psi _{n+1}$\textit{\ if these factors become entangled relative to this
original split}.\bigskip
\end{itemize}

This definition will be particularly important when discussing measurements
on the emergent level. Consider a state that is fundamentally separated%
\footnote{%
Recall from Chapter 4 that a state may be said to be written in a
`fundamentally separated' form if it cannot be separated further relative to
the fundamental factorisation of the Hilbert space (that is, the split of $%
\mathcal{H}_{[1...N]}$ with $N$ sub-spaces).} relative to $\mathcal{H}%
_{[1...N]};$ clearly, the above definition will apply to endo-physical
measurements occurring between each of these factors. However, as shown in
Chapter 4, given a fundamentally separated state, its factors can often be
`grouped' into larger factors such that it can be re-written as a separable
product of these new sub-states relative to an alternative, less fundamental
split of the Hilbert space (i.e. a split that has less than $N$ factor
spaces). For example, a state $\phi _{123}\in \mathcal{H}_{[1...3]}$ defined
as $\phi _{123}=\phi _{1}\otimes \phi _{2}\otimes \phi _{3}$ with three
factors may be re-written as the state $\phi _{A3}=$ $\phi _{A}\otimes \phi
_{3}$ with two factors, where $\phi _{A}\equiv \phi _{1}\otimes \phi _{2}\in
\mathcal{H}_{[A]}\equiv \mathcal{H}_{[12]},$ relative to the split $\mathcal{%
H}_{[A3]}.$

These large factors may then play an important role in the theory of
emergence, where it may be convenient to consider a particular group of
(more fundamental) sub-states as a single entity in order to describe a
certain macroscopic `object'. The point is that the above assertion is
always applicable when discussing whether or not an endo-physical
measurement has occurred between two given factor sub-states, regardless of
which particular split they are being defined separate relative to. Thus for
the simple example in $\mathcal{H}_{[1...3]},$ the rule could be applied to
define a potential measurement occurring between $\phi _{1}$ and $\phi _{2},$
or between $\phi _{1}$ and $\phi _{3},$ or between $\phi _{2}$ and $\phi
_{3},$ or, crucially, also between $\phi _{A}$ and $\phi _{3},$ depending
upon which particular situation is of current interest.

In other words, it has been possible to define information exchange on any
given `level' of separation, and the method is equally valid both on the
most fundamental level, i.e. between components in the $N$ individual
subregisters, and on more `macroscopic' scales between factors of $\Psi _{n}$
that may themselves be separable relative to a more fundamental split of the
Hilbert space. Of course, exactly which level or split is the most
appropriate in a given situation to describe an endo-physical measurement in
an actual physical, laboratory experiment depends entirely on how these
macroscopic objects emerge from the underlying pregeometric structure.

Thus, it is evident that an endo-physical measurement occurring on one level
may not necessarily occur on another. Equivalently, an exchange occurring
between factors defined relative to one particular split of the Hilbert
space does not necessarily occur between every factor defined relative to a
different split. This point resonates strongly with the Schr\"{o}dinger's
Cat paradox, and is equally apparent in the second example of above where
the completely separable state (\ref{Ch4Init}) jumped to the `nearly'
completely separable state (\ref{Ch4Onecomp}).

\bigskip

As concluded above, in order to argue that any endo-physical measurements
have occurred during a transition from an arbitrary initial state to a final
one, there must be a change of partition\footnote{%
Noting that similar reverse statements are not necessarily true: a change in
partition does not automatically imply that every arbitrary pair of factors
of the state must have exchanged information during the jump. To judge
whether a particular `device' has measured a particular `subject', changes
in the separability of the state relative to the given split are of issue.}.
However, different degrees of fundamental separability of different initial
and final states lead to different degrees of partition change, in which $%
\mathcal{P}_{n+1}$ bears different degrees of `resemblance' to $\mathcal{P}%
_{n},$ and this inevitably leads to different degrees of information
exchange. For instance, given an initial state $\Theta _{1...N}\in \mathcal{H%
}_{[1...N]}$ that is completely separable and in the partition $\mathcal{H}%
_{1...N},$ it is intrinsically obvious that a next state $\Phi _{3...N}^{12}$
in $\mathcal{H}_{3...N}^{12}$ is `more similar' to $\Theta _{1...N}$ than an
alternative next state $\tilde{\Phi}^{1...N}$ in $\mathcal{H}^{1...N},$ and
so less information might be expected to be exchanged during a transition
from $\Theta _{1...N}$ to $\Phi _{3...N}^{12}$ than from $\Theta _{1...N}$
to $\tilde{\Phi}^{1...N}.$

It is therefore useful to define the concept of `\textit{partition overlap}'
that attempts to account for how `congruent' $\mathcal{P}_{n+1}$ is to $%
\mathcal{P}_{n},$ and is related to the factorisability of the probability
amplitude discussed previously. Specifically, this partition overlap may
then help to quantify just `how much' information is exchanged during a
particular transition from one state to a next, as shown below.

Clearly, partition overlaps will be appropriate for discussions involving
information exchanges occurring on the most fundamental level (i.e. between
components of the state in the individual subregisters themselves), because
partitions are defined relative to the fundamental factorisation of the
Hilbert space $\mathcal{H}_{[1...N]}$ into its $N$ subregisters, and not
just an arbitrary `higher order' split. Of course, this is the most natural
stance to adopt anyway, because it is changes occurring on the most
fundamental, non-emergent level that are assumed ultimately responsible for
changes on every other scale.\bigskip

However, a definition similar to the partition overlap given below could
easily be formulated for any other given split if the `partitions' are
replaced by `\textit{split-partitions}', which are defined in an obvious
manner in terms of the number of different ways an arbitrary state could
potentially be separable relative to a particular split of the Hilbert space.

As an illustrative example of this last idea, the split $\mathcal{H}%
_{[A3]}\equiv \mathcal{H}_{[A]}\otimes \mathcal{H}_{[3]}$ of the Hilbert
space $\mathcal{H}_{[123]},$ where $\mathcal{H}_{[A]}\equiv \mathcal{H}%
_{[12]},$ has two split-partitions $\mathcal{H}_{A3}$ and $\mathcal{H}^{A3},$
even though the overall space $\mathcal{H}_{[123]}$ has five partitions: $%
\mathcal{H}_{123},$ $\mathcal{H}_{3}^{12},$ $\mathcal{H}_{2}^{13},$ $%
\mathcal{H}_{1}^{23}$ and $\mathcal{H}^{123}.$

Similarly, the split $\mathcal{H}_{[AB5]}\equiv \mathcal{H}_{[A]}\otimes
\mathcal{H}_{[B]}\otimes \mathcal{H}_{[5]}$ of the Hilbert space $\mathcal{H}%
_{[1...5]},$ where $\mathcal{H}_{[A]}\equiv \mathcal{H}_{[12]}$ and $%
\mathcal{H}_{[B]}\equiv \mathcal{H}_{[34]},$ clearly has five
split-partitions, $\mathcal{H}_{AB5},$ $\mathcal{H}_{A}^{B5},$ $\mathcal{H}%
_{B}^{A5},$ $\mathcal{H}_{5}^{AB}$ and $\mathcal{H}^{AB5},$ even though the
alternative split $\mathcal{H}_{[A345]}$ has $15$ split-partitions, whilst
equation (\ref{Ch3Iter}) of Section 5.3 showed that $\mathcal{H}_{[1...5]}$
itself has $52$ partitions.

In fact, the concept of split-partitions is evidently more general than that
of partitions: a partition is just a special case of a split-partition where
the split in question is the fundamental factorisation of the Hilbert space $%
\mathcal{H}_{[1...N]}$ into its $N$ subregister sub-spaces. Ultimately,
`split-partition overlaps' could therefore be employed to compare how
`similar' one state is to the next relative to the same split, just as
partition overlaps will be shown below to provide a comparison of how
similar one partition is to another relative to the fundamental
factorisation of the Hilbert space. Again, this concept may be useful from
the perspective of emergence, and for discussions of how much information is
exchanged during a particular transition from the point of view of sets of
`macroscopic' sub-states that are themselves further separable relative to a
more fundamental split of the Hilbert space.\bigskip

Consider a Hilbert space $\mathcal{H}_{[1...N]},$ and a partition $\mathcal{P%
}_{\alpha }$ of this with $a$ blocks $B_{\alpha }^{(1)},B_{\alpha
}^{(2)},...,$ $B_{\alpha }^{(a)}.$ Consider also a second partition $%
\mathcal{P}_{\beta }$ of $\mathcal{H}_{[1...N]}$ which has $b$ blocks $%
B_{\beta }^{(1)},...,B_{\beta }^{(b)}.$ Now, the block $B_{\alpha }^{(S)},$
for $S=1,2,...a,$ implies that every state in the partition $\mathcal{P}%
_{\alpha }$ of $\mathcal{H}_{[1...N]}$ possesses a factor that is entangled
relative to the sub-space $\mathcal{H}^{(S)}$ defined as the product of the $%
N(S)$ subregisters:
\begin{equation}
\mathcal{H}^{(S)}\equiv \mathcal{H}_{S_{1}}\otimes \mathcal{H}%
_{S_{2}}\otimes ...\otimes \mathcal{H}_{S_{N(S)}}=\mathcal{H}%
_{[S_{1}S_{2}...S_{N(S)}]}\subset \mathcal{H}_{[1...N]}
\end{equation}
where each $S_{s}$ is uniquely one of the set $\{1,2,...,N\}$ for all $%
s=1,2,...,N(S)$ and $S=1,2,...,a.$ Clearly, the partition $\mathcal{P}%
_{\alpha }$ necessarily defines some split of $\mathcal{H}_{[1...N]},$ such
that
\begin{equation}
\mathcal{H}_{[1...N]}\equiv \prod_{S=1}^{a}(\otimes \mathcal{H}%
^{(S)})=\prod_{S=1}^{a}\prod_{s=1}^{N(S)}(\otimes \mathcal{H}_{S_{s}})
\end{equation}
with the obvious condition
\begin{equation}
N=N(1)+N(2)+...+N(a).
\end{equation}

Similarly, the block $B_{\beta }^{(T)}$ for $T=1,2,...,b$ implies that every
state in the partition $\mathcal{P}_{\beta }$ of $\mathcal{H}_{[1...N]}$
possesses a factor that is entangled relative to the sub-space $\mathcal{H}%
^{(T)}$ defined as the product of $N(T)$ subregisters, i.e.
\begin{equation}
\mathcal{H}^{(T)}\equiv \mathcal{H}_{T_{1}}\otimes \mathcal{H}%
_{T_{2}}\otimes ...\otimes \mathcal{H}_{T_{N(T)}}
\end{equation}
where each $T_{t}$ is uniquely one of the set $\{1,2,...,N\}$ for all $%
t=1,2,...,N(T)$ and $T=1,2,...,b,$ and this defines another split of $%
\mathcal{H}_{[1...N]}.$

Consider now a function $F(B_{\alpha }^{(S)},B_{\beta }^{(T)})$ that
effectively compares block $B_{\alpha }^{(S)}$ to block $B_{\beta }^{(T)},$
with the result that $F(B_{\alpha }^{(S)},B_{\beta }^{(T)})=1$ if $B_{\alpha
}^{(S)}=B_{\beta }^{(T)},$ but $F(B_{\alpha }^{(S)},B_{\beta }^{(T)})=0$
otherwise. The equality $B_{\alpha }^{(S)}=B_{\beta }^{(T)}$ is taken to
hold only if there is a one-to-one equivalence between the subregisters $%
\mathcal{H}_{S_{s}}$ in $B_{\alpha }^{(S)}$ and the subregisters $\mathcal{H}%
_{T_{t}}$ in $B_{\beta }^{(T)};$ that is, $B_{\alpha }^{(S)}=B_{\beta
}^{(T)} $ if for each $s=1,2,...,N(S)$ there is one, and only one, $t$ such
that
\begin{equation}
\mathcal{H}_{S_{s}}=\mathcal{H}_{T_{t}}
\end{equation}
for $t=1,2,...,N(T).$

Then, the partition overlap $\mathcal{P}(\alpha ,\beta )$ between $\mathcal{P%
}_{\alpha }$ and $\mathcal{P}_{\beta }$ may be defined as
\begin{equation}
\mathcal{P}(\alpha ,\beta )=\sum_{S=1}^{a}\sum_{T=1}^{b}\frac{F(B_{\alpha
}^{S},B_{\beta }^{T})}{\max (a,b)}
\end{equation}
with normalising factor $\max (a,b).$ Clearly, $\mathcal{P}(\alpha ,\beta )=%
\mathcal{P}(\beta ,\alpha ),$ as would be expected from symmetry.

As an example, consider three states $\lambda ,\mu $ and $\omega $ in $%
\mathcal{H}_{[1...4]}$ that are elements of the separations $\lambda \in
\mathcal{H}_{1234},$ $\mu \in \mathcal{H}_{12}^{34}$ and $\omega \in
\mathcal{H}_{4}^{123}.$ Evidently, the partition $\mathcal{P}_{\lambda }$ of
$\mathcal{H}_{[1...4]}$ of which $\lambda $ is a member contains four
blocks: $1,2,3$ and $4.$ Similarly, $\mathcal{P}_{\mu }$ has three blocks, $%
1,2,$ and $34,$ whilst $\mathcal{P}_{\omega }$ has only two blocks, $123$
and $4.$ In this case, then, the partition overlap $\mathcal{P}(\lambda ,\mu
)$ between $\mathcal{P}_{\lambda }$ and $\mathcal{P}_{\mu }$ is clearly $%
(1+1)/4=%
%TCIMACRO{\UNICODE[m]{0xbd}}%
%BeginExpansion
{\frac12}%
%EndExpansion
,$ whilst $\mathcal{P}(\lambda ,\omega )=1/4$ and $\mathcal{P}(\mu ,\omega
)=(0/3)=0.$\bigskip

This idea can now be incorporated into the discussion of the quantum
universe. To this end, it is asserted that more information is exchanged
during a transition from $\Psi _{n}\in \mathcal{P}_{n}\subset \mathcal{H}%
_{[1...N]}$ to $\Psi _{n+1}\in \mathcal{P}_{n+1}\subset \mathcal{H}%
_{[1...N]} $ if the partition overlap $\mathcal{P}(n,n+1)$ is small than if
the partition overlap is large. This follows immediately from the
observation that in order for $\mathcal{P}(n,n+1)$ to be large, most of the
components of the state must not have changed their block during the jump
from $\Psi _{n}$ to $\Psi _{n+1},$ and so have not exchanged information.
Thus, the case where $\mathcal{P}(n,n+1)=0$ represents maximum information
exchange, whereas when $\mathcal{P}(n,n+1)=1$ no information has been
exchanged. Note that for $\mathcal{P}(n,n+1)=1,$ a necessary but
insufficient condition is that $a=b.$

So, it is expected that a jump from a state contained in $\mathcal{H}_{1234}$
to a state contained in $\mathcal{H}_{12}^{34}$ results in less information
being exchanged than an alternative jump to a state contained in $\mathcal{H}%
_{4}^{123}.$

It is now obvious how the partitions used in the definition of the partition
overlap could easily be replaced by split-partitions to provide an analogous
definition for a \textit{split-partition overlap}. For example, given a
particular split $\mathcal{H}_{[ABC]}$ of $\mathcal{H}_{[1...N]}$ and two of
its split-partitions $\mathcal{H}_{ABC}$ and $\mathcal{H}_{A}^{BC},$ the
split-partition overlap between $\mathcal{H}_{ABC}$ and $\mathcal{H}%
_{A}^{BC} $ is clearly $(1/3);$ this is despite the fact that each of $%
\mathcal{H}_{A}, $ $\mathcal{H}_{B}$ and $\mathcal{H}_{C}$ may itself be
factorisable into very many subregisters relative to the fundamental
factorisation $\mathcal{H}_{[1...N]}.$ Of course, and as before, which
particular split is of interest depends very much on the `level' required to
describe a given emergent, physical situation.

The idea of partition overlap is related to the statement that, generally,
the more factorisable a transition amplitude $\langle \Psi _{n+1}|\Psi
_{n}\rangle $ is, the less information is exchanged.\bigskip

The above points lead to the question as to when a state such as $|\varphi
_{n+1}\rangle $ can really still be said to represent just a `later version'
of the same `subject' as $|\phi _{n}\rangle $ if these factors are in
different Hilbert sub-spaces of different dimensions. In other words, how
similar is the physical object that emerges from $|\varphi _{n+1}\rangle $
to the object that emerged from the earlier factor $|\phi _{n}\rangle ?$ An
analogous comment arises concerning the similarity of the `device's' states
at different times, such as $|\Lambda _{n}\rangle _{\lbrack (x+1)...y]}$ and
$|\Upsilon _{n+1}\rangle _{\lbrack i(x+1)...y]};$ can both of these be said
to represent the `before' and `after' configurations of the same piece of
physical apparatus?

Furthermore, it is also an important question to ask about just how much
`pregeometric information' (i.e. component information) needs to be
exchanged to constitute the sort of real, physical measurements occurring in
laboratories. For instance, is the one component exchange of (\ref
{Ch4Onecomp}) enough, or are more exchanges required for the device to
really `learn' about the subject? Indeed, is it alternatively necessary for
the subject and device to become completely entangled with each other in
order for a physical measurement to take place? This last point is
presumably not the case, since in real physical experiments the device at
least generally seems to possess a classical identity after the interaction,
and this alternative would ultimately lead to all sorts of Schr\"{o}dinger's
Cat type paradoxes. The issue still remains, however, as to how much
entanglement is either `allowed' or required.

Of course, it is in practice very difficult to say exactly how `similar' one
sub-state physically is to another. Indeed, even if differences could easily
be quantified, any resulting argument would then have to rely on knowing
exactly how classical objects emerge from the underlying pregeometric
description, and a theory of this has not yet been completed. In other
words, only once it is understood how the factor state $|\phi _{n}\rangle $
gives rise to a physical description of the `subject' can it be compared to
whatever semi-classical object emerges from a similar treatment of $|\varphi
_{n+1}\rangle .$

That said, it is a natural speculation to suggest that for large macroscopic
objects represented by sub-states (or groups of sub-states, depending on the
split being discussed) in factor Hilbert spaces of very large dimension, the
`addition or subtraction of just one or two components' may not be expected
to affect their emergent appearance too much, and this justifies the earlier
argument of why it is usually reasonable to accept the approximation that a
large semi-classical observer often seems unaffected by an observation.
Conversely, for microscopic sub-states in factor Hilbert spaces of very low
dimension, the `loss of one or two components' might be much more severe,
and may lead to an object that looks completely different on emergent
scales. These, however, are just heuristic arguments, and a great deal of
work on the issues of emergent and persistence is required in order to fully
justify them.\bigskip

The ideas of the above discussions may now be summarised. When a factor
remains in the same block during a jump from one state to the next, it has
not exchanged information with any of the others. Such factors are
effectively de-coupled and isolated from everything else, and so appear to
develop independently. It is only when the partition of the state changes in
such a way that the relationship between a component of it contained in one
factor changes with respect to a component of the state contained in another
factor that an exchange of information occurs between them. Then, these two
factors may be said to have interacted during the transition. This type of
partition changing process is thus viewed as the pregeometric origin of an
endo-physical measurement of one particular factor by another.

On the emergent scale, of course, actual physical measurements are highly
complex sets of events. Real experiments involving real devices extracting
real information from real subjects may well take place over very many jumps
of the state of the Universe, and may incorporate devices with perhaps very
many constituent parts each experiencing their own passages of internal
endo-time and giving rise to whole different levels of sub-measurements. In
fact, particle detections in high energy collider physics provide the
perfect example of this point. However, if quantum mechanics does indeed
hold for a consistent picture of physical reality, such emergent experiments
should fundamentally result from chains and sets of endo-measurements
occurring on the pregeometric scale, and as such might ultimately be hoped
to be governed by the principles discussed in this section.

\bigskip

As a final remark, it is worth commenting on the roles of the operators in
the above types of development. In the type of development that was
discussed first, where both $\psi _{n}$ and $\psi _{n+1}$ were in the same
separation $\mathcal{H}_{AB},$ the factorisability of the operator $\hat{%
\Sigma}_{n+1}$ is not important. This is because both factorisable and
entangled operators can have separable outcomes, as discussed in Chapter 5.
The only circumstance where the factorisability of $\hat{\Sigma}_{n+1}$
would matter is if, somehow and for some reason, it is known in advance that
$\psi _{n+1}$ \textit{must} be in the same separation as $\psi _{n};$ in
this case either a factorisable operator would have to chosen, or a
particular entangled operator would have to be selected that has only
separable outcomes.

For the second type of development, however, where $\psi _{n}$ and $\psi
_{n+1}$ were in different separations and $\psi _{n+1}$ was entangled
relative to the split $\mathcal{H}_{[AB]},$ the operator $\hat{\Sigma}_{n+1}$
must also be entangled because factorisable operators can only have
separable outcomes. In other words, for the outcome of $\hat{\Sigma}_{n+1}$
to be in the entanglement $\mathcal{H}^{AB},$ i.e. entangled relative to $%
\mathcal{H}_{A}\otimes \mathcal{H}_{B},$ this operator must also be
entangled relative to $\mathcal{H}_{A}\otimes \mathcal{H}_{B}.$ This
observation then leads to the result that for a measurement to occur between
a `subject' and a `device', i.e. for previously separate factors to become
entangled, the operator must be entangled.

These last points may be placed in context by remembering that operators are
used in conventional quantum mechanics to denote physical tests. Moreover,
these physical tests are themselves often also associated with sets of
physical apparatus. Recall, however, that in a Universe free from external
observers, and as discussed more fully in Chapter 8, its development depends
upon operators self-referentially chosen according to the current stage.
Thus, since every physical `object' is expected to emerge from the
underlying pregeometric description, and because human physicists do appear
to be able to construct sets of apparatus in order to measure things, it
must be the case that groups of factors representing devices are somehow
able to influence the Universe's decision about which particular operator is
chosen to test the state. In other words, the presence of a given set of
factors in $\Psi _{n}$ may result in a certain choice of operator $\hat{%
\Sigma}_{n+1},$ and so the existence of a particular emergent device and
subject may consequently lead to a particular `action' being taken by the
Universe. So, in the quantum Universe, groups of factors representing a
physical apparatus may also hence be labelled by the action of the
particular operator they induce.

Of course as noted before, exactly how and why particular operators are
chosen to develop the Universe's state is an interesting question for the
future. How this decision might be sufficiently self-referential to give the
impression that physical devices are measuring physical subjects, however,
will be addressed later.

\bigskip \newpage

\section{Quantum Field Theory from Quantum Computation}

\renewcommand{\theequation}{7.\arabic{equation}} \setcounter{equation}{0} %
\renewcommand{\thetheorem}{7.\arabic{theorem}} \setcounter{theorem}{0}

\bigskip

A central theme of this thesis is to investigate how the semi-classical
picture of physics familiar to science may begin to arise from a fundamental
quantum state description. Specifically, one matter of particular interest
is the question of when it might be possible to argue that ``this object
with these properties is here''. Now, two thirds of this issue have already
been addressed: Chapter 4 discussed when it is possible to describe
something as a distinct and independent looking object, whereas Chapter 5
investigated the concept of spatial location. It is therefore time to
examine the remainder of these three points, namely, how a state represented
by a vector in a Hilbert space may give rise to objects with particular
physical properties.\bigskip

Clearly, the idea of a `property' is very vague, and the word is often used
in science to describe almost any number of the physical characteristics
exhibited by a semi-classical object, for example its size, or shape, or
weight, or appearance, or odour. However, ever since the philosophy of
Democritus \cite{Wiki-Democritus}, a reductionist viewpoint has generally
been accepted in which each of these qualities is ultimately a feature
resulting from a more fundamental picture of reality, such that every
macroscopic object comprises of enormous numbers of microscopic `indivisible
elements'. Furthermore, it is the different ways that these individual
entities interact and group together that are expected to eventually account
for the types of phenomena observed in the everyday world.

Of course, over time this picture has been refined, and it is now known that
Democritus' ``atoms'' should really be associated with elementary particles
of given mass, charge, colour, spin etc. Going further, these particles are
themselves in turn associated with the various excitation modes of quantum
field theory (QFT), and are hence directly determined by the laws,
symmetries and formalisms of the equations governing this. Thus, in
conventional physics it is the theory of quantised fields that is often
ultimately taken to provide an explanation as for why a particular object
has the properties it does.\bigskip

The objective attempted in this present chapter is to go one step further.
Because the notion of operators and states in a Hilbert space is taken in
this thesis to be \textit{the} fundamental description of physical reality,
if it may be shown how quantum field theory might emerge from such a
picture, it might consequently be argued that semi-classical properties will
also equally arise as a natural continuation, just as in the conventional
case. To this end, it is the `link' between quantum field theory and the
model proposed in this thesis that is now explored.

The mechanism suggested to achieve this result follows naturally from the
work of the preceding Chapter 6. That is, a treatment of the Universe with
the principles of quantum computation will be shown to reproduce the desired
field theoretic concepts. Of course, such an approach may not be too
surprising; after all, given the suggestion that the Universe is running as
a giant quantum automaton, the application of quantum computational
procedures might in some sense naturally be expected to account for every
physical effect.

In fact, the emergence of QFT from the type of quantum computation discussed
in this thesis is not just desirable for the completeness and consistency of
the proposed quantum Universe paradigm. Further to this hope, it might also
be expected that such an analysis could give rise to a slightly modified
(and hence improved) version of quantum field theory that is free from some
of problems inherent in the traditional case. Indeed, this might be hoped
for immediately: given that the Hilbert space of the quantum Universe is
taken to be very large but finite, and consists of a discrete set of
subregisters, the problems associated with infinite dimensional and
continuous theories may be expected not to arise. As an example of this, it
could be hoped that in the finite case the infra-red and ultra-violet
divergences may not occur. This, too, would clearly be an additional success
for the model.\bigskip

Note that the idea of generating particle field theories from this type of
analysis is not completely new or unconventional. Feynman, for example,
envisaged a description of quantum field theory resulting from quantum
computation \cite{Feynman1}, and Bjorken and Drell similarly demonstrated
how QFT may be derived from a set of objects, each of which is positioned at
a unique and well specified site \cite{Bjorken} (and referenced therein to
\cite{Jordan}-\cite{Fock}). This latter model will be seen to have strong
analogies with the method presented in the following. More recently, Wu and
Lidar \cite{Wu} explored the algebraic relationships existing between qubits
and parafermions, and Deutsch \cite{Deutsch1} discussed a qubit field theory
embedded in a fixed background spacetime.

\bigskip

\subsection{Preliminary Structure}

\bigskip

The proposal starts from the premise that the usual quantum field theory
familiar to physicists is really an effective, emergent view of a more
fundamental mechanism at work. The overall approach will be to use to the
rules and principles of quantum computation to construct a model of QFT from
the basic, underlying structure of operators and statevectors in a
factorisable Hilbert space.\bigskip

Consider the Hilbert space $\mathcal{H}_{[1...N]}$ factorisable into $N$
subregisters $\mathcal{H}_{m}$ of prime dimension, for $m=1,2,...,N.$ As
always, the sub-script $m$ is used merely as a convenient label; the overall
Hilbert space is assumed invariant to any left-right re-positioning of the
subregister spaces, and hence $m$ it is not in any way meant to denote
physical location. This last point should perhaps be emphasised by
remembering that positions and distances have only been defined on the
emergent scale, due in fact to considerations of causal set relationships as
discussed in Chapter 5, and may be further reinforced by observing that
\textit{if} the $N$ `sites' were taken to be directly equivalent to physical
locations, it would be difficult to envisage how the three dimensional
Minkowski space of physics could be translated into the one dimensional
lattice that would result. These issues will be important to recall later.

The model suggested below may be greatly simplified by assuming that each
subregister $\mathcal{H}_{m}$ is two dimensional, such that the overall
space $\mathcal{H}_{[1...N]}$ is a product of $N$ qubit sub-spaces. In this
instance, it is possible to define an orthonormal basis $\mathcal{B}_{m}$
for $\mathcal{H}_{m}$ in the usual way as
\begin{equation}
\mathcal{B}_{m}\equiv \{|0\rangle _{m},|1\rangle _{m}\}
\end{equation}
where, as before, $|0\rangle _{m}$ may be referred to as `down' and $%
|1\rangle _{m}$ as `up'. Thus, a basis $\mathcal{B}_{1...N}$ for $\mathcal{H}%
_{[1...N]}$ may be defined as
\begin{equation}
\mathcal{B}_{1...N}\equiv \{|i_{1}\rangle _{1}\otimes |i_{2}\rangle
_{2}\otimes ...\otimes |i_{N}\rangle _{N}:i_{m}=0,1;m=1,2,...N\}
\end{equation}
where the representation $|0\rangle _{m}\equiv \binom{1}{0}_{m}$ and $%
|1\rangle _{m}\equiv \binom{0}{1}_{m}$ could also be adopted.

Recall now the work of Section 6.1, where equation (\ref{CH4Trans}) defined
the `Transformation' operator $\hat{T}_{m}^{ij}$ acting in the space $%
\mathcal{H}_{m}$ as $\hat{T}_{m}^{ij}=|i\rangle _{mm}\langle j|,$ for $%
i,j=0,1.$ As was discussed, any operator acting locally in $\mathcal{H}_{m}$
may be built up of complex sums of these four operators. It is beneficial
now to enlarge this definition to describe `extended transformation
operators', which have the same resulting effects as their local
counterparts, but act globally in the whole space $\mathcal{H}_{[1...N]}.$
As alluded to in Section 6.1, this extension may be accomplished by taking
the tensor product of the local transformation operator $\hat{T}_{m}^{ij}$
with $N-1$ identity operators $\hat{I}_{m^{\prime }}$ in $\mathcal{H}%
_{m^{\prime }}$ where $m,m^{\prime }=1,2,...,N$ but $m\neq m^{\prime }.$ So,
the extended transformation operator $\mathcal{\hat{T}}_{m}^{ij}$ that has
the same effect as the local transformation $\hat{T}_{m}^{ij}$ is defined as
\begin{equation}
\mathcal{\hat{T}}_{m}^{ij}=\hat{I}_{1}\otimes \hat{I}_{2}\otimes ...\otimes
\hat{I}_{m-1}\otimes \hat{T}_{m}^{ij}\otimes \hat{I}_{m+1}\otimes ...\otimes
\hat{I}_{N}.
\end{equation}

Clearly, these extended operators satisfy the product rule
\begin{equation}
\mathcal{\hat{T}}_{m}^{ij}\mathcal{\hat{T}}_{m}^{kl}=\delta _{jk}\mathcal{%
\hat{T}}_{m}^{il}
\end{equation}
and the commutation relation
\begin{equation}
\lbrack \mathcal{\hat{T}}_{r}^{ij},\mathcal{\hat{T}}_{s}^{kl}]=0\text{ \ \ ,
\ \ }r\neq s.  \label{Ch5Comm}
\end{equation}

Note that by using these types of transformation operator, the model of
quantum field theory to be presented encodes ideas such as information and
logic from the outset, as discussed in Section 6.1.

Just as arbitrary local operators can be constructed from local
transformation operators, the extended transformation operators may be used
to generate arbitrary operators that act globally on the whole state.
Crucially, however, this also includes the construction of arbitrary global
operators that appear to act locally on a particular sub-space. For example,
the `extended Pauli operators', written $\underline{\hat{\sigma}}_{m}^{a}$
for $a=1,2,3,$ may be given by
\begin{eqnarray}
\underline{\hat{\sigma}}_{m}^{1} &\equiv &\hat{I}_{1}\otimes \hat{I}%
_{2}\otimes ...\otimes \hat{I}_{m-1}\otimes \hat{\sigma}_{m}^{1}\otimes \hat{%
I}_{m+1}\otimes ...\otimes \hat{I}_{N} \\
&=&\hat{I}_{1}\otimes \hat{I}_{2}\otimes ...\otimes \hat{I}_{m-1}\otimes %
\left[ \hat{T}_{m}^{01}+\hat{T}_{m}^{10}\right] \otimes \hat{I}_{m+1}\otimes
...\otimes \hat{I}_{N}  \notag \\
&=&\mathcal{\hat{T}}_{m}^{01}+\mathcal{\hat{T}}_{m}^{10}  \notag
\end{eqnarray}
and similarly
\begin{eqnarray}
\underline{\hat{\sigma}}_{m}^{2} &\equiv &-i\mathcal{\hat{T}}_{m}^{01}+i%
\mathcal{\hat{T}}_{m}^{10} \\
\underline{\hat{\sigma}}_{m}^{3} &\equiv &\mathcal{\hat{T}}_{m}^{00}-%
\mathcal{\hat{T}}_{m}^{11}.  \notag
\end{eqnarray}
where these \textit{appear} to act only upon components of the state in the
sub-space $\mathcal{H}_{m}$ of $\mathcal{H}_{[1...N]}.$

The three extended operators $\underline{\hat{\sigma}}_{m}^{a}$ obey an
algebra analogous to their local equivalents, $\hat{\sigma}_{m}^{a}.$
Assuming the\ Einstein summation convention over $c$ only, then
\begin{equation}
\underline{\hat{\sigma}}_{r}^{a}\underline{\hat{\sigma}}_{s}^{b}=\delta
_{rs}\left( \delta _{ab}\underline{\hat{\sigma}}_{r}^{0}+i\epsilon _{abc}%
\underline{\hat{\sigma}}_{r}^{c}\right) +\left( 1-\delta _{rs}\right)
\underline{\hat{\sigma}}_{s}^{b}\underline{\hat{\sigma}}_{r}^{a}
\label{Ch5ExtAlg}
\end{equation}
for $a,b,c=1,2,3$ and $r,s=1,2,...,N,$ with the extended identity operator $%
\underline{\hat{\sigma}}_{m}^{0}$ defined as
\begin{equation}
\underline{\hat{\sigma}}_{m}^{0}\equiv \hat{\sigma}_{1}^{0}\otimes \hat{%
\sigma}_{2}^{0}\otimes ...\otimes \hat{\sigma}_{N}^{0}\equiv \hat{I}%
_{1}\otimes \hat{I}_{2}\otimes ...\otimes \hat{I}_{N}\equiv \hat{I}_{1...N}
\end{equation}
and where the first set of bracketed terms in (\ref{Ch5ExtAlg}) account for
local products when $\hat{\sigma}_{r}^{a}$ and $\hat{\sigma}_{s}^{b}$ act in
the same sub-register, and the second set is due to the commutation relation
(\ref{Ch5Comm}) for local operators acting in different sub-registers.
Moreover, it also follows that
\begin{equation}
\underline{\hat{\sigma}}_{r}^{a}\underline{\hat{\sigma}}_{s}^{0}=\underline{%
\hat{\sigma}}_{s}^{0}\underline{\hat{\sigma}}_{r}^{a}=\underline{\hat{\sigma}%
}_{r}^{a}
\end{equation}
noting how these results compare with the `usual' local Pauli algebra of (%
\ref{Ch3pauli1}), as given in the standard literature \cite{Rae}.\bigskip

As an aside, note that the extended Pauli operators may be used to
demonstrate the group symmetry of the extended transformation operators.
Consider a unitary operator $\hat{U}(\mathbf{\theta })$ defined as
\begin{equation}
\hat{U}(\mathbf{\theta })\equiv \exp \left( i\theta _{m}^{1}\underline{\hat{%
\sigma}}_{m}^{1}+i\theta _{m}^{2}\underline{\hat{\sigma}}_{m}^{2}+i\theta
_{m}^{3}\underline{\hat{\sigma}}_{m}^{3}\right)  \label{Ch5Rot}
\end{equation}
that appears to act locally in the $m^{th}$ Hilbert sub-space $\mathcal{H}%
_{m},$ where $\theta _{m}^{1},\theta _{m}^{2},\theta _{m}^{2}\in \mathbb{R}.$
Then, the algebra of the extended transformation operators is invariant to
`rotations' under the $SU(2)$ group, such that the operator $(\mathcal{\hat{T%
}}_{m}^{ij})^{\prime }$ defined as
\begin{equation}
(\mathcal{\hat{T}}_{m}^{ij})^{\prime }\equiv \hat{U}^{\ast }(\mathbf{\theta }%
)\mathcal{\hat{T}}_{m}^{ij}\hat{U}(\mathbf{\theta })
\end{equation}
where $\hat{U}^{\ast }(\mathbf{\theta })$ is the conjugate transpose of $%
\hat{U}(\mathbf{\theta }),$ obeys
\begin{equation}
(\mathcal{\hat{T}}_{m}^{ij})^{\prime }(\mathcal{\hat{T}}_{m}^{kl})^{\prime
}=\delta _{jk}(\mathcal{\hat{T}}_{m}^{il})^{\prime }
\end{equation}
as expected.

\bigskip

It is here that a discussion of the algebraic structure necessary for field
theory can begin. Consider the extended transformation operators $\mathcal{%
\hat{T}}_{m}^{10}$ and $\mathcal{\hat{T}}_{m}^{01}.$ If the `ground state'
in $\mathcal{H}_{[1...N]},$ written as $|\underline{0}\rangle ,$ is chosen
to be
\begin{equation}
|\underline{0}\rangle =|\underline{0}\rangle _{1...N}=|00...0\rangle
_{1...N}=|0\rangle _{1}\otimes |0\rangle _{2}\otimes ...\otimes |0\rangle
_{N}
\end{equation}
then the operator $\mathcal{\hat{T}}_{m}^{10}$ acting on $|\underline{0}%
\rangle $ results in the transition of the state of the $m^{th}$ qubit (i.e.
the component of $|\underline{0}\rangle $ in $\mathcal{H}_{m})$ of $\mathcal{%
H}_{[1...N]}$ from $|0\rangle _{m}$ to $|1\rangle _{m}.$ That is
\begin{equation}
\mathcal{\hat{T}}_{m}^{10}|\underline{0}\rangle =|0\rangle _{1}\otimes
|0\rangle _{2}\otimes ...\otimes |0\rangle _{m-1}\otimes |1\rangle
_{m}\otimes |0\rangle _{m+1}\otimes ...\otimes |0\rangle _{N}.
\end{equation}

This transformation may be reversed by the operator $\mathcal{\hat{T}}%
_{m}^{01}.$ That is, if the $m^{th}$ qubit is in the state $|1\rangle _{m},$
then $\mathcal{\hat{T}}_{m}^{01}$ leads to a transition of this to $%
|0\rangle _{m},$ such that
\begin{equation}
\mathcal{\hat{T}}_{m}^{10}\left( |0\rangle _{1}\otimes |0\rangle _{2}\otimes
...\otimes |0\rangle _{m-1}\otimes |1\rangle _{m}\otimes |0\rangle
_{m+1}\otimes ...\otimes |0\rangle _{N}\right) =|\underline{0}\rangle .
\end{equation}

The operator $\mathcal{\hat{T}}_{m}^{10}$ can be applied in products that
effectively act on different sub-registers; viz,
\begin{eqnarray}
\mathcal{\hat{T}}_{r}^{10}\mathcal{\hat{T}}_{s}^{10}|\underline{0}\rangle
&=&|0\rangle _{1}\otimes |0\rangle _{2}\otimes ...\otimes |0\rangle
_{r-1}\otimes |1\rangle _{r}\otimes |0\rangle _{r+1}\otimes ... \\
&&...\otimes |0\rangle _{s-1}\otimes |1\rangle _{s}\otimes |0\rangle
_{s+1}\otimes ...\otimes |0\rangle _{N}  \notag
\end{eqnarray}
for $r\neq s,$ noting that
\begin{equation}
\mathcal{\hat{T}}_{r}^{10}\mathcal{\hat{T}}_{r}^{10}|\underline{0}\rangle =0
\label{Ch5Prod}
\end{equation}
as expected from an analogy with the local transformation operators, and as
discussed in Section 6.1. Similar results apply in obvious ways for higher
order products of $\mathcal{\hat{T}}_{r}^{10}\mathcal{\hat{T}}_{s}^{10}%
\mathcal{\hat{T}}_{t}^{10}...,$ for products of the `opposite' operators $%
\mathcal{\hat{T}}_{m}^{01},$ or for various `mixtures' of the $\mathcal{\hat{%
T}}_{r}^{10}$'s and $\mathcal{\hat{T}}_{s}^{01}$'s.

The last result (\ref{Ch5Prod}) follows naturally from the rule that if the
operator $\mathcal{\hat{T}}_{m}^{10}$ is applied to a qubit that is in the
state $|1\rangle _{m},$ then the outcome is $0.$ Similarly, if the operator $%
\mathcal{\hat{T}}_{m}^{01}$ is applied to a qubit that is in the state $%
|0\rangle _{m},$ then the outcome of this is $0.$ Clearly, then,
\begin{equation}
\mathcal{\hat{T}}_{m}^{01}|\underline{0}\rangle =0.
\end{equation}

The operators $\mathcal{\hat{T}}_{m}^{10}$ and $\mathcal{\hat{T}}_{m}^{01}$
are in some sense analogous to the creation and annihilation ladder
operators of quantum field theory. Specifically, $\mathcal{\hat{T}}_{m}^{10}$
may be interpreted as an operator that creates a $|1\rangle _{m}$ state from
the `vacuum' $|\underline{0}\rangle ,$ whereas $\mathcal{\hat{T}}_{m}^{01}$
destroys this to return the ground state.

In addition to this comparison, $\mathcal{\hat{T}}_{m}^{10}$ and $\mathcal{%
\hat{T}}_{m}^{01}$ are seen (\ref{Ch5Comm}) to obey some of the commutation
relations familiar to bosonic ladder operators. It might be suggested,
therefore, that they could hence be used to construct a pregeometric theory
that might reproduce the properties of physical bosons in the emergent
limit. However, a problem with this hypothesis is that conventional bosonic
theories permit `multi-occupation number states' \cite{Mandl}. That is,
given a conventional, bosonic creation operator $\hat{a}_{B}^{\dagger }$
acting on a conventional vacuum ground state $\varphi _{0}=|0\rangle ,$ then
\begin{equation}
\hat{a}_{B}^{\dagger }|0\rangle =|1_{B}\rangle
\end{equation}
produces the single boson particle state $|1_{B}\rangle ,$ whilst
\begin{equation}
\hat{a}_{B}^{\dagger }\hat{a}_{B}^{\dagger }|0\rangle =\hat{a}%
_{B}|1_{B}\rangle =|2_{B}\rangle
\end{equation}
yields the two boson particle state, and so on.

Conversely, an immediate difficulty that would be faced in attempts to
reconstruct bosonic field theory from the transformation operators of above
is that although $\mathcal{\hat{T}}_{m}^{10}|\underline{0}\rangle \neq 0,$ a
second application of $\mathcal{\hat{T}}_{m}^{10}$ to this new state gives $%
\mathcal{\hat{T}}_{m}^{10}\mathcal{\hat{T}}_{m}^{10}|\underline{0}\rangle
=0. $ So, even though $\mathcal{\hat{T}}_{m}^{10}|\underline{0}\rangle $ is
not being directly interpreted here as a single particle state \textit{per se%
}, it is still difficult to see how multiple particle states could
ultimately be generated in this manner if the underlying qubit operator
algebra is so contrary to that employed to describe bosons.\bigskip

However, note that the relationship $\mathcal{\hat{T}}_{m}^{10}\mathcal{\hat{%
T}}_{m}^{10}|\underline{0}\rangle =0$ is instead similar to the Exclusion
Principle condition used in standard quantum field theory for particles
obeying fermi-dirac statistics. Thus, it is this connection that is now
explored, and an attempt is made to recover fermionic field theory from the
pregeometric framework.

To this end, consider first the non-local operator $\eta _{m}$ defined as
\begin{eqnarray}
\eta _{m} &\equiv &\underline{\hat{\sigma}}_{1}^{3}\underline{\hat{\sigma}}%
_{2}^{3}...\underline{\hat{\sigma}}_{m-1}^{3}\hat{I}_{m}\text{ \ \ , \ \ }%
2\leq m\leq N \\
&=&\hat{\sigma}_{1}^{3}\otimes \hat{\sigma}_{2}^{3}\otimes ...\otimes \hat{%
\sigma}_{m-1}^{3}\otimes \hat{I}_{m}\otimes \hat{I}_{m+1}\otimes ...\otimes
\hat{I}_{N}  \notag
\end{eqnarray}
with $\eta _{1}\equiv \hat{I}_{1}\otimes \hat{I}_{2}\otimes ...\otimes \hat{I%
}_{N}.$

Now, consider the operator $\hat{\alpha}_{m}$ defined as
\begin{equation}
\hat{\alpha}_{m}=\eta _{m}\mathcal{\hat{T}}_{m}^{01}
\end{equation}
and its adjoint (i.e. Hermitian conjugate) operator $\hat{\alpha}%
_{m}^{\dagger }$%
\begin{equation}
\hat{\alpha}_{m}^{\dagger }=\eta _{m}\mathcal{\hat{T}}_{m}^{10}.
\end{equation}

Clearly,
\begin{equation}
\hat{\alpha}_{m}|\underline{0}\rangle =0
\end{equation}
and
\begin{equation}
\hat{\alpha}_{m}\left( |0\rangle _{1}\otimes |0\rangle _{2}\otimes
...\otimes |0\rangle _{m-1}\otimes |1\rangle _{m}\otimes |0\rangle
_{m+1}\otimes ...\otimes |0\rangle _{N}\right) =|\underline{0}\rangle
\end{equation}
whilst
\begin{equation}
\hat{\alpha}_{m}^{\dagger }|\underline{0}\rangle =|0\rangle _{1}\otimes
|0\rangle _{2}\otimes ...\otimes |0\rangle _{m-1}\otimes |1\rangle
_{m}\otimes |0\rangle _{m+1}\otimes ...\otimes |0\rangle _{N}
\end{equation}
and
\begin{equation}
\hat{\alpha}_{m}^{\dagger }\hat{\alpha}_{m}^{\dagger }|0\rangle =0.
\label{Ch5Two}
\end{equation}

Moreover, due to the presence of the operators $\underline{\hat{\sigma}}%
_{1}^{3},$ $\underline{\hat{\sigma}}_{2}^{3},...,$ $\underline{\hat{\sigma}}%
_{m-1}^{3},$ and hence unlike the transformation operators $\mathcal{\hat{T}}%
_{m}^{01}$ and $\mathcal{\hat{T}}_{m}^{10},$ the operators $\hat{\alpha}_{m}$
and $\hat{\alpha}_{m}^{\dagger }$ obey anti-commutation relations
\begin{eqnarray}
\{\hat{\alpha}_{r},\hat{\alpha}_{s}\} &=&0  \label{Ch5anti-pre} \\
\{\hat{\alpha}_{r}^{\dagger },\hat{\alpha}_{s}^{\dagger }\} &=&0  \notag \\
\{\hat{\alpha}_{r},\hat{\alpha}_{s}^{\dagger }\} &=&\delta _{rs}\hat{I}%
_{1...N}  \notag
\end{eqnarray}
as may be readily demonstrated.

\begin{proof}
A proof of the first of these is presented as follows. Given
\begin{equation}
\{\hat{\alpha}_{r},\hat{\alpha}_{s}\}=\hat{\alpha}_{r}\hat{\alpha}_{s}+\hat{%
\alpha}_{s}\hat{\alpha}_{r}=\eta _{r}\mathcal{\hat{T}}_{r}^{01}\eta _{s}%
\mathcal{\hat{T}}_{s}^{01}+\eta _{s}\mathcal{\hat{T}}_{s}^{01}\eta _{r}%
\mathcal{\hat{T}}_{r}^{01}
\end{equation}
then expanding produces
\begin{eqnarray}
&=&\left( \hat{\sigma}_{1}^{3}\hat{\sigma}_{2}^{3}...\hat{\sigma}_{r-1}^{3}%
\hat{I}_{r}...\hat{I}_{N}\right) \left( \hat{I}_{1}\hat{I}_{2}...\hat{T}%
_{r}^{01}...\hat{I}_{N}\right) \left( \hat{\sigma}_{1}^{3}\hat{\sigma}%
_{2}^{3}...\hat{\sigma}_{s-1}^{3}\hat{I}_{s}...\hat{I}_{N}\right) \left(
\hat{I}_{1}\hat{I}_{2}...\hat{T}_{s}^{01}...\hat{I}_{N}\right)  \notag \\
&&+\left( \hat{\sigma}_{1}^{3}\hat{\sigma}_{2}^{3}...\hat{\sigma}_{s-1}^{3}%
\hat{I}_{s}...\hat{I}_{N}\right) \left( \hat{I}_{1}\hat{I}_{2}...\hat{T}%
_{s}^{01}...\hat{I}_{N}\right) \left( \hat{\sigma}_{1}^{3}\hat{\sigma}%
_{2}^{3}...\hat{\sigma}_{r-1}^{3}\hat{I}_{r}...\hat{I}_{N}\right) \left(
\hat{I}_{1}\hat{I}_{2}...\hat{T}_{r}^{01}...\hat{I}_{N}\right)  \notag \\
&=&\left( \hat{\sigma}_{1}^{3}\hat{\sigma}_{2}^{3}...\hat{\sigma}_{r-1}^{3}%
\hat{T}_{r}^{01}...\hat{I}_{N}\right) \left( \hat{\sigma}_{1}^{3}\hat{\sigma}%
_{2}^{3}...\hat{\sigma}_{s-1}^{3}\hat{T}_{s}^{01}...\hat{I}_{N}\right)
\notag \\
&&+\left( \hat{\sigma}_{1}^{3}\hat{\sigma}_{2}^{3}...\hat{\sigma}_{s-1}^{3}%
\hat{T}_{s}^{01}...\hat{I}_{N}\right) \left( \hat{\sigma}_{1}^{3}\hat{\sigma}%
_{2}^{3}...\hat{\sigma}_{r-1}^{3}\hat{T}_{r}^{01}...\hat{I}_{N}\right)
\end{eqnarray}
where tensor product symbols have been omitted. Assuming, without loss of
generality, that $r<s$ gives
\begin{eqnarray}
&=&\left( \hat{\sigma}_{1}^{3}\hat{\sigma}_{1}^{3}\right) \left( \hat{\sigma}%
_{2}^{3}\hat{\sigma}_{2}^{3}\right) ...\left( \hat{\sigma}_{r-1}^{3}\hat{%
\sigma}_{r-1}^{3}\right) \left( \hat{T}_{r}^{01}\hat{\sigma}_{r}^{3}\right)
\left( \hat{I}_{r+1}\hat{\sigma}_{r+1}^{3}\right) ...\left( \hat{I}_{s}\hat{T%
}_{s}^{01}\right) ...\left( \hat{I}_{N}\hat{I}_{N}\right) \\
&&+\left( \hat{\sigma}_{1}^{3}\hat{\sigma}_{1}^{3}\right) \left( \hat{\sigma}%
_{2}^{3}\hat{\sigma}_{2}^{3}\right) ...\left( \hat{\sigma}_{r-1}^{3}\hat{%
\sigma}_{r-1}^{3}\right) \left( \hat{\sigma}_{r}^{3}\hat{T}_{r}^{01}\right)
\left( \hat{\sigma}_{r+1}^{3}\hat{I}_{r+1}\right) ...\left( \hat{T}_{s}^{01}%
\hat{I}_{s}\right) ...\left( \hat{I}_{N}\hat{I}_{N}\right) .  \notag
\end{eqnarray}

Now, from the algebra (\ref{Ch3pauli1}) of the Pauli operators, $\hat{\sigma}%
_{m}^{3}\hat{\sigma}_{m}^{3}=\hat{\sigma}_{m}^{0}=\hat{I}_{m},$ it is
evident that
\begin{eqnarray}
\{\hat{\alpha}_{r},\hat{\alpha}_{s}\} &=&\hat{I}_{1}\hat{I}_{2}...\left(
\hat{T}_{r}^{01}\hat{\sigma}_{r}^{3}\right) \left( \hat{\sigma}%
_{r+1}^{3}\right) ...\left( \hat{T}_{s}^{01}\right) \hat{I}_{s+1}...\hat{I}%
_{N} \\
&&+\hat{I}_{1}\hat{I}_{2}...\left( \hat{\sigma}_{r}^{3}\hat{T}%
_{r}^{01}\right) \left( \hat{\sigma}_{r+1}^{3}\right) ...\left( \hat{T}%
_{s}^{01}\right) \hat{I}_{s+1}...\hat{I}_{N}  \notag \\
&=&\hat{I}_{1}\hat{I}_{2}...\left( \hat{T}_{r}^{01}\hat{\sigma}_{r}^{3}+\hat{%
\sigma}_{r}^{3}\hat{T}_{r}^{01}\right) \left( \hat{\sigma}_{r+1}^{3}\right)
...\left( \hat{T}_{s}^{01}\right) \hat{I}_{s+1}...\hat{I}_{N}.  \notag
\end{eqnarray}

Recall, however, that (like all local operators) the Pauli operators can be
written as a complex sum of local transformation operators. So, by using (%
\ref{Ch4 PAULI}), i.e. $\hat{\sigma}_{r}^{3}=\hat{T}_{r}^{00}-\hat{T}%
_{r}^{11},$ it is evident that
\begin{eqnarray}
\left( \hat{T}_{r}^{01}\hat{\sigma}_{r}^{3}+\hat{\sigma}_{r}^{3}\hat{T}%
_{r}^{01}\right) &=&\{\hat{T}_{r}^{01},\hat{\sigma}_{r}^{3}\} \\
&=&\left( \hat{T}_{r}^{01}[\hat{T}_{r}^{00}-\hat{T}_{r}^{11}]+[\hat{T}%
_{r}^{00}-\hat{T}_{r}^{11}]\hat{T}_{r}^{01}\right)  \notag \\
&=&-\hat{T}_{r}^{01}+\hat{T}_{r}^{01}=0  \notag
\end{eqnarray}
where the last line follows from the usual $\hat{T}_{r}^{ij}$ algebra (\ref
{Ch4ts}). So, substituting in gives
\begin{equation}
\{\hat{\alpha}_{r},\hat{\alpha}_{s}\}=0
\end{equation}
as expected.

Clearly, the proof holds also for $r\geq s.$

Moreover, the relations $\{\hat{\alpha}_{r}^{\dagger },\hat{\alpha}%
_{s}^{\dagger }\}=0$ and $\{\hat{\alpha}_{r},\hat{\alpha}_{s}^{\dagger
}\}=\delta _{rs}\hat{I}_{r}$ can be readily verified via analogous
methods.\bigskip
\end{proof}

Due to their similarities to conventional theory, the fermionic-looking
operators $\hat{\alpha}_{m}^{\dagger }$ and $\hat{\alpha}_{m}$ will be
called pregeometric (fermionic) creation and annihilation operators
respectively, or equivalently, qubit ladder operators. The extent of this
similarity will be investigated fully in due course.

As an aside, however, note that the result (\ref{Ch5Two}) can also be given
by the relation
\begin{equation}
\left( \hat{\alpha}_{m}^{\dagger }\right) ^{D}|0\rangle \text{ \ \ }%
\QATOPD\{ \} {\neq 0\text{ \ \ , \ \ }D=1}{=0\text{ \ \ , \ \ }D>1}
\end{equation}
in anticipation of higher order generalisations in the future; the algebra (%
\ref{Ch5Two}) obeys parafermionic statistics of order $1$ \cite{Green}\cite
{Greenberg}.\bigskip

Just as the transformation operators were invariant under $SU(2)$ rotations,
so too are the pregeometric creation and annihilation operators.
Specifically, using the rotation operator $\hat{U}(\mathbf{\theta })$ given
in (\ref{Ch5Rot}), then the operators $(\hat{\alpha}_{m}^{\dagger })^{\prime
}$ and $(\hat{\alpha}_{m})^{\prime }$ defined as
\begin{eqnarray}
(\hat{\alpha}_{m}^{\dagger })^{\prime } &\equiv &\hat{U}^{\ast }(\mathbf{%
\theta })\hat{\alpha}_{m}^{\dagger }\hat{U}(\mathbf{\theta }) \\
(\hat{\alpha}_{m})^{\prime } &\equiv &\hat{U}^{\ast }(\mathbf{\theta })\hat{%
\alpha}_{m}\hat{U}(\mathbf{\theta })  \notag
\end{eqnarray}
also obey the fermionic algebra. That is
\begin{eqnarray}
\{(\hat{\alpha}_{r})^{\prime },(\hat{\alpha}_{s})^{\prime }\} &=&\{(\hat{%
\alpha}_{r}^{\dagger })^{\prime },(\hat{\alpha}_{s}^{\dagger })^{\prime }\}=0
\\
\{(\hat{\alpha}_{r})^{\prime },(\hat{\alpha}_{s}^{\dagger })^{\prime }\}
&=&\delta _{rs}\hat{I}_{1...N}.  \notag
\end{eqnarray}

This result may lead to important consequences for the gauge symmetry of the
emergent theory.\bigskip

As with separable states and factorisable Hilbert spaces, it is evident that
the pregeometric creation and annihilation operators are also invariant to
any left-right re-positioning of their factors. This, of course, is because
it is always assumed that the factor sub-operator with subscript $t$ still
acts in the sub-register $\mathcal{H}_{t},$ for $1\leq t\leq N,$ such that
the imposed left-right ordering of the equations is merely typographical.
So, $\hat{\sigma}_{a}^{3}$ always acts in $\mathcal{H}_{a},$ $\hat{I}_{b}$
always acts in $\mathcal{H}_{b},$ $\hat{T}_{c}^{10}$ always acts in $%
\mathcal{H}_{c},$ and so on, such that a sub-operator's location within the
tensor product is immaterial. As an example, the pregeometric creation
operator
\begin{equation}
\hat{\alpha}_{r}^{\dagger }=\hat{\sigma}_{1}^{3}\hat{\sigma}_{2}^{3}...\hat{%
\sigma}_{14}^{3}\hat{T}_{15}^{10}\hat{I}_{16}...\hat{I}_{28}
\end{equation}
may be rewritten as
\begin{equation}
\hat{\alpha}_{r}^{\dagger }=\hat{I}_{23}\hat{I}_{17}\hat{\sigma}_{14}^{3}%
\hat{\sigma}_{1}^{3}\hat{I}_{28}\hat{\sigma}_{6}^{3}\hat{T}_{15}^{10}\hat{I}%
_{18}...
\end{equation}
without affecting the anti-commutation algebra.

\bigskip

\subsection{Dirac Field Theory}

\bigskip

It is now possible to begin to construct quantum field theories from the
basic principles described above. Specifically, attention will be focused on
the emergence of Dirac theory from the underlying pregeometric structure,
because this field is often taken to be one of the most basic (and hence
important) ingredients of elementary particle physics. Indeed, it is even
possible to describe many boson species in terms of groups of fermions
obeying the Dirac theory; the pion of particle phenomenology, and the
existence of Cooper pairs in superconductivity provide good physical
examples of this point.

So, in this section a description of spin-$%
%TCIMACRO{\UNICODE[m]{0xbd}}%
%BeginExpansion
{\frac12}%
%EndExpansion
$ fermions will be given.

For convenience, the two types of particle and two types of anti-particle
associated with the field's excitations will be referred to below as spin-up
and spin-down electrons and positrons. In should be noted, however, that
this is merely for linguistic advantage, and in principle the presented
analysis is not restricted to any particular particle species.\bigskip

In order to justify the enormous simplification involved in considering just
a single, isolated Dirac field, recall the types of situation in physics in
which such circumstances are generally encountered. In conventional particle
accelerator experiments, for example, scientists often go to great lengths
to construct apparatus that effectively `shuts an area off' from the rest of
the Universe, such that the region inside the collider can be treated as an
isolated system in which only a few basic fields are present.

Now, in a fully quantum Universe, this sort of experimental arrangement is
taken to arise in the large-scale limit when the state $\Psi _{n}$ is
sufficiently and suitably separable so that its factors give rise to such an
emergent, semi-classical picture. In this case, various factors and groups
of factors may be used to represent the detector, the particles it contains,
the physicist, the laboratory, and, indeed, everything else. Moreover, in
fact, the operators chosen to develop the Universe are assumed to be
carefully and self-referentially chosen such that this semi-classical
description\ appears to persist over a number of jumps, as has been
discussed previously.

So, it should therefore be reasonable to meaningfully discuss parts of the
Universe that seem to contain nothing but isolated, fundamental fields,
because this is what scientists tend to be able to do in real, physical
experiments. Furthermore, since it has been conjectured that there is a
strong link between the `parts' of the Universe and the factors of its
state, it is equally reasonable to assert the possibility of discussing
factors that `contain' just the particles inside the detector. Indeed, by
rephrasing this argument for the perspective advocated in this thesis, it
might be possible to generate isolated quantum fields out of a consideration
of the pregeometric `properties' of a particular part of the Universe.
Specifically, and in the language of the previous chapter, the information
content of such regions might somehow be expected to include notions of
quantum fields, though exactly how this might be achieved is what is to be
examined below.

Of course, the factors that represent the insides of particle detectors will
also possess many other types of features. For instance, they will have a
well defined location in emergent physical space because of their familial
relationships with other factors (c.f. Chapter 5), and since the
separability of the Universe may change as it jumps from one state to the
next, it may be possible to discuss `observers' appearing to measure the
sub-state representing the colliding particles (c.f. Chapter 6). Ultimately,
then, it should be possible to envisage a typical particle physics
experiment from a pregeometric point of view, where isolated particles
appear to collide and be scattered, before being measured by various
components, detectors and scientists.

For now, however, just the isolated particle fields shall be discussed, with
the implicit assumption being made that any such procedures could also
eventually be applied to more `complicated' situations.\bigskip

So consider just that factor $\varphi ,$ of the state $\Psi \in \mathcal{H}%
_{[1...N]}$ of the Universe, that represents the `inside' of the collider.
Thus in the following, the label $\varphi $ will be used to denote the part
of the Universe's state $\Psi $ responsible for a description of everything
of interest that occurs inside the detector during a collision event. By
re-labelling the subregisters of the overall Hilbert space $\mathcal{H}%
_{[1...N]}$ of $\Psi $ in a convenient way, the factor $\varphi $ may be
said to be contained in a factor Hilbert sub-space $\mathcal{H}%
_{[1...N^{\prime }]},$ where $\mathcal{H}_{[1...N^{\prime }]}\subset
\mathcal{H}_{[1...N]}.$ Of course, $\varphi $ may or may not itself be
highly separable relative to $\mathcal{H}_{[1...N^{\prime }]}.$ Clearly, the
remaining sub-space $\mathcal{H}_{[(N^{\prime }+1)...N]}$ contains factors
and groups of factors that represent the Physicist, $P,$ the Apparatus, $A,$
and the Rest of the Universe, $R.$

Overall, the goal therefore becomes to investigate the circumstances in
which the Hilbert space $\mathcal{H}_{[1...N^{\prime }]}$ may be described
as `containing' a single, isolated Dirac field.

In conventional quantum field theory, and in particular in the S-Matrix
approach to particle scattering \cite{S-mat}, it is assumed that what
actually occurs during the collision may be represented by a type of `Black
Box'; only the initial `In' particle state, $|\psi _{in}\rangle ,$ and final
`Out' particle state, $|\psi _{out}\rangle ,$ are of interest to physicists.
Moreover, in the Heisenberg picture of dynamics traditionally used in
quantum field theory, it is asserted that the initially prepared state $%
|\psi _{in}\rangle $ is effectively `frozen in time' until its later
measurement by an observer, at which point it is collapsed into $|\psi
_{out}\rangle .$ Thus, a typical particle physics experiment proceeds by the
scientist preparing an initial $a$-particle In state, before measuring it at
some time later time with some sort of Hermitian operator (representing an
observable), thereby collapsing it into a final $b$-particle Out state.
Consequently, the time evolution of the system is enforced by transforming
the Observables, in a way that depends, in fact, upon the time experienced
by the physicist between the initial preparation of $|\psi _{in}\rangle $
and its eventual measurement. Amplitudes between initial and particular
final states may therefore be considered.

Now, in order to recreate standard physics from the pregeometric perspective
aimed at in this thesis, the above type of setup must be reproducible in the
quantum Universe paradigm. So, the principles employed in conventional
particle theory are used to guide the present analysis.

To this end, consider an initial state $\Psi _{n}$ defined as
\begin{equation}
\Psi _{n}=|\varphi _{in}\rangle \otimes |P_{i}\rangle \otimes |A_{i}\rangle
\otimes |R_{i}\rangle
\end{equation}
where $|\varphi _{in}\rangle \in \mathcal{H}_{[1...N^{\prime }]}$ represents
the initially prepared sub-state of the particles in the detector prior to
the collision (i.e. before anything has happened), and $|P_{i}\rangle
,|A_{i}\rangle $ and $|R_{i}\rangle $ the initial sub-states of the
Physicist, Apparatus and Rest of Universe respectively.

Moreover, assume that the series of jumps from $\Psi _{n}\longrightarrow
\Psi _{n+1}\longrightarrow ...\longrightarrow \Psi _{n+n^{\prime }},$ where $%
n^{\prime }\gg n,$ represent, on the emergent level, a
observer-apparatus-environment system performing a particle collision
experiment. In this case, $\Psi _{n+n^{\prime }}$ may be taken to be of the
approximate form
\begin{equation}
\Psi _{n+n^{\prime }}=|\varphi _{out}\rangle \otimes |P_{f}\rangle \otimes
|A_{f}\rangle \otimes |R_{f}\rangle
\end{equation}
where $|\varphi _{out}\rangle $ represents the final sub-state of the
particles in the detector after the collision, and $|P_{f}\rangle
,|A_{f}\rangle $ and $|R_{f}\rangle $ the respective final sub-states of the
Physicist, Apparatus and Rest of Universe. Of course, the operator $\hat{%
\Sigma}_{n+n^{\prime }},$ of which $\Psi _{n+n^{\prime }}$ is an eigenstate,
must be carefully defined such that the Physicist factor in $\Psi
_{n+n^{\prime }-1}$ believes herself to be choosing (with apparent
free-will) a particular laboratory test to measure the particle sub-state
with; this general issue is discussed more fully in Chapter 8.\bigskip

By attempting to keep the dynamics congruent to the situation familiar to
conventional physics, a number of inferences may be drawn about what the
above sequence of states might be like. Firstly, because the sub-state
representing the particles is assumed `frozen in time' between its
preparation as part of $\Psi _{n}$ and measurement as part of $\Psi
_{n+n^{\prime }},$ whatever the operators in the sequence $\hat{\Sigma}%
_{n+1}\longrightarrow \hat{\Sigma}_{n+1}\longrightarrow ...\hat{\Sigma}%
_{n+n^{\prime }-1}$ might actually be, each must be taken to result in a
null test on the factor of the universe in $\mathcal{H}_{[1...N^{\prime }]}.$
In such a case, $\varphi _{in}$ would consequently appear unchanged during
this period. Moreover, and in the language of Chapter 6, no information
would therefore be exchanged between the components of the state in $%
\mathcal{H}_{[1...N^{\prime }]}$ and the components of the state in $%
\mathcal{H}_{[(N^{\prime }+1)...N]}$ during this time; as desired, the
sub-state in $\mathcal{H}_{[1...N^{\prime }]}$ representing the inside of
the detector is effectively isolated from the remainder of the Universe. Of
course, the Physicist, Apparatus and Rest of Universe do interact, entangle
and exchange information with one another throughout this duration.

Overall, therefore, it is asserted that the Universe jumps through a series
of states of the form
\begin{eqnarray}
\Psi _{n+1} &=&|\varphi _{in}\rangle \otimes |P^{\prime }\rangle \otimes
|A^{\prime }\rangle \otimes |R^{\prime }\rangle \\
\Psi _{n+2} &=&|\varphi _{in}\rangle \otimes |P^{\prime \prime }\rangle
\otimes |A^{\prime \prime }\rangle \otimes |R^{\prime \prime }\rangle  \notag
\\
&&\vdots  \notag \\
\Psi _{n+n^{\prime }-1} &=&|\varphi _{in}\rangle \otimes |P^{\prime \prime
\prime ...\prime }\rangle \otimes |A^{\prime \prime \prime ...\prime
}\rangle \otimes |R^{\prime \prime \prime ...\prime }\rangle  \notag
\end{eqnarray}
where $P^{\prime \prime },$ for example, is a factor representing a
physicist that has evolved and developed since initial `time' $n.$

As expected, the operators $\hat{\Sigma}_{n+1},\hat{\Sigma}_{n+2},...,\hat{%
\Sigma}_{n+n^{\prime }-1}$ must be very carefully constrained, defined and
selected, in order to ensure that this type of pattern occurs.\bigskip

A second feature that may be concluded about the above system by drawing
parallels to the laboratory case is that the factors of the Universe
representing the physicist will bestow certain properties upon $\varphi .$
For example, because they know it represents what is going on inside a
physical machine, they may assume it occupies a particular volume, or
represents a certain set of conditions. Consider just this volume: if $%
\varphi $ is ultimately taken to represent the state inside a physical
detector, whatever spatial degrees of freedom emerge from a causal set
description of the operators acting in its Hilbert sub-space must match the
observed spatial properties of the object (i.e. the detector) appearing to
physically contain it.

Moreover, recall from Section 5.7.5 that, under certain conditions,
particular sub-operators acting locally in particular subregisters may be
mapped to positions situated on a three dimensional lattice; of course, this
was not too say that the subregisters are actually located at these sites
\textit{per se}, but that in the large scale limit such a description may be
effective. Such an observation then places an important constraint on the
size of the Hilbert sub-space $\mathcal{H}_{[1...N^{\prime }]}$ containing $%
\varphi $: if the inside of a detector is modestly assumed to occupy a
volume of one cubic metre, $\mathcal{H}_{[1...N^{\prime }]}$ must be
sufficiently large such that a causal set description can give rise to one
cubic metre's `worth' of volume.

In fact going further, and following the estimation of Section 3.2 for the
minimum number of degrees of freedom of the quantum Universe (in which
continuous spatial resolutions were assumed valid down to at least distance
scales of the order of the Planck length, $l_{P}),$ if the minimum number $%
n_{m^{3}}$ of discrete `points' in a cubic metre is given by
\begin{equation}
n_{m^{3}}=\QOVERD( ) {1}{l_{P}}^{3}\sim \left( 10^{+35}\right) ^{3}=10^{105}
\end{equation}
and if each of these $n_{m^{3}}$ points is associated with just a single,
two-dimensional qubit degree of freedom, then the dimension of $\mathcal{H}%
_{[1...N^{\prime }]}$ must be at least $2^{10^{105}}.$

In actual fact, the model of field theory presented in the following will be
simplified by restricting the discussion to dynamics occurring in a one
dimensional volume. Thus, from this viewpoint it is assumed that if an
external scientist considers $\varphi ,$ she would conclude that it is of
emergent length $L$ because it is contained in a one-dimensional detector
with a `known' internal volume of $L.$ This simplification will be justified
later by restricting the field theory analysis to one dimension of momentum,
and noting that it is hoped that an extension to $D$ dimensional space could
be achieved merely by using a formalism of greater complexity. For now,
however, note that by using such an assumption, the number of spatial points
discussed is given by
\begin{equation}
n_{m^{1}}=10^{35}
\end{equation}
where this value is still sufficiently enormous such that a causal set
approximation of continuous space is expected to be valid.\bigskip

The physicist factor will also be able to make statements regarding the
duration of the sub-state $|\varphi _{in}\rangle .$ For example, by
recording various changes in her surroundings as the Universe developed from
$\Psi _{n}$ to $\Psi _{n+1}$ to... to $\Psi _{n+n^{\prime }},$ she may be
able to argue that a certain quantity of (emergent) time elapsed between the
preparation of $\varphi _{in}$ in $\Psi _{n}$ and the measurement of $%
\varphi _{out}$ as part of $\Psi _{n+n^{\prime }}.$ In fact, such a
determination is vital: measures of time are necessary for many quantum
field calculations to be performed and probability amplitudes to be
evaluated, because these are generally dependent on the extent of dynamical
unitary evolutions.

It is possible to estimate a value for the length of time perceived by the
physicist between $\Psi _{n}$ and $\Psi _{n+n^{\prime }}$ by again appealing
to what actually occurs in laboratory collider experiments. If the colliding
particles are assumed to be travelling at the speed of light, $c,$ and are
assumed to interact only when in the cubic metre enclosed inside the
detector, then the physicist could conclude that they interact for a
duration of $\left( 1/c\right) \sim 10^{-8}$ seconds. In other words, if the
interaction is taken to endure between jumps $n$ and $n+n^{\prime },$ then
this period will last approximately $10^{-8}$ seconds according to the clock
of the emergent observer who had defined $c$ and the metre (and hence also
the size of the detector).

Moreover, if it is also assumed that physicists are able to resolve a
continuous temporal parameter down to at least durations of the order of the
Planck time $t_{P},$ then an extrapolation indicates that the time taken for
the Universe to jump discretely from a state $\Psi _{n}$ to the next state $%
\Psi _{n+1}$ could not be greater than this value. So, from the external
point of view of the emergent observer, the number $N_{J}$ of jumps during
which the Universe will possess a factor $\varphi _{in}$ is given by
\begin{equation}
N_{J}\equiv (n+n^{\prime }-1)-n\geq \QOVERD( ) {1}{ct_{P}}\sim \QOVERD( )
{1}{(10^{8})(10^{-43})}=10^{35}
\end{equation}
which may again be expected to be large enough to generate the type of
causal set structures required.

\bigskip

Attention may now be turned to the actual field theory. An immediate
question is: given that physical particle states are conventionally taken to
be governed by annihilation and creation operators, and also the assumption
of the fundamental, underlying pregeometric structure advocated in this
thesis, then is there a mechanism by which these conventional operators
could emerge from a consideration of their pregeometric counterparts? In
other words, given a pregeometric creation operator of the form\ $\hat{\alpha%
}_{r}^{\dagger },$ how might it be possible to relate this to the operator $%
\hat{c}^{\dagger }(p,\kappa )$ that creates a physical electron with
momentum $p$ and spin $\kappa ,$ or to the operator $\hat{d}^{\dagger
}(p^{\prime },\kappa ^{\prime })$ that creates a physical positron with
momentum $p^{\prime }$ and spin $\kappa ^{\prime }?$ Complicated functions
of such operators could then ultimately be used to relate the physical In
state $|\varphi _{in}\rangle $ to the final Out state $|\varphi
_{out}\rangle ,$ as is typically done in the Heisenberg picture approach to
particle collision experiments.

To begin to answer this, recall that it is well known that momentum and
position space variables may be related to one another in quantum theory by
the use of Fourier transforms \cite{Townsend}. This relationship follows
from their reciprocal dependence in the uncertainty principle, and is often
exploited in conventional quantum field theory. So, if it is possible to
discuss ladder operators in position space, it is equally possible to
discuss ladder operators in momentum space by using Fourier methods.

Now, since an origin of spatial position has already be suggested for the
quantum Universe, it might be possible to use this as a starting point in
order to obtain ladder operators that are a function of momentum. Thus, it
may be possible to employ Fourier transform methods to generate momentum
space relations from the positional degrees of freedom that emerge from the
causal set structure. It is this type of procedure that is consequently
investigated below.

However, because space is assumed to be discrete and finite in the paradigm
envisaged in this thesis, the conventional Fourier transform must be
replaced with a discretised version. This then places an additional
constraint on the proposal: the standard results of the continuous theory
must still emerge in the large scale limit if the model suggested is to
accurately represent the observed physics. This issue will be discussed in
due course.\bigskip

Given that the model in hand has been restricted to a single spatial
dimension, assume that a causal set analysis\ of the local operators in the
sub-space $\mathcal{H}_{[1...N^{\prime }]}$ `containing' the Dirac field
concludes that the region of space that ultimately emerges from it is of
length $L.$ Moreover, assume that this length may be associated with a
discrete, one-dimensional lattice consisting of $(2M+1)$ points, where $M$
is very large, such that $(2M+1)\geq n_{m^{1}}=10^{35}.$ Finally, assume
that $\mathcal{H}_{[1...N^{\prime }]}$ may be factorised into at least $%
(2M+1)$ subspaces, such that $N^{\prime }\geq (2M+1)\geq n_{m^{1}}.$ Then,
it may be argued that each local operator acting in each of these factor
sub-spaces of $\mathcal{H}_{[1...N^{\prime }]}$ could somehow map to one of
these $(2M+1)$ possible `positions' or `sites', analogously to the mechanism
presented in Section 5.7.5 for relating sub-operators to the vertices of a
three dimensional lattice. Equivalently, each subregister of $\mathcal{H}%
_{[1...N^{\prime }]}$ may be asserted to somehow correspond to a certain one
of the $(2M+1)$ sites.

Now, for the sake of simplicity, assume that $\mathcal{H}_{[1...N^{\prime
}]} $ is a tensor product of just $(2M+1)$ sub-spaces (this assumption will
be enlarged later). In this case, it might be possible to relate each local
sub-operator (or subspace) in a one-to-one way to a unique position along
the one-dimensional lattice consisting of $(2M+1)$ points. Thus overall, and
from the approximate point of view of emergence, it might therefore be
possible to imagine that each of the $(2M+1)$ subregisters (or local
sub-operator acting in it) in\ $\mathcal{H}_{[1...N^{\prime }]}$ may in some
sense be considered to exist at a definite location along this line of
emergent length $L.$

\textbf{(}To illustrate the proposed perspective further, a slightly
different analysis could instead be schematically discussed. Assume that the
Hilbert sub-space $\mathcal{H}_{[1...N^{\prime }]}$ containing the particles
is a tensor product of at least $(2M+1)$ subregisters, where $M$ is very
large, and that a causal set analysis concludes that the region of space
that ultimately emerges from it (by considering the sub-states and
sub-operators acting in it) is of length $L.$ Then, given that the model has
been restricted to a single spatial dimension, it may be argued from the
approximate point of view of emergence that each subregister of $\mathcal{H}%
_{[1...N^{\prime }]}$ is `responsible', somehow, for one of $(2M+1)$
possible `positions' or `sites'. Equivalently, the particular sub-operator
that acts locally in a particular subregister of $\mathcal{H}%
_{[1...N^{\prime }]}$ may be asserted to correspond to a certain one of
these $(2M+1)$ sites. Reversing this argument, then, each local sub-operator
(or subregister) may effectively be mapped to a unique position along a
one-dimensional lattice consisting of $(2M+1)$ points, similarly to the
mechanism of Section 5.7.5., and it is therefore possible to imagine that
each subregister (or local sub-operator) in\ $\mathcal{H}_{[1...N^{\prime
}]} $ may in some sense be considered to exist at a definite location when
this lattice is associated with a continuous one-dimensional volume of
length $L.\mathbf{)}$

Note that the presented approach differs greatly from the work of Deutsch
\cite{Deutsch1}, in which qubit subregisters are instead actually embedded
into a fixed and independently existing background space whose nature or
origin is not further questioned or justified. Note also that the
pregeometric ladder operators discussed in the previous section act locally
in particular subregister spaces; this observation will be crucial in the
following.\bigskip

Overall, a vision is imagined in which each subregister of $\mathcal{H}%
_{[1...N^{\prime }]}$ is effectively placed at a unique site along a locus
of $(2M+1)$ points\ that correspond to a spatial length of $L$ in the
continuum limit. Each sub-operator (for example, the pregeometric ladder
operators) acting locally in one of these subregisters may therefore also be
envisaged to act at a definite one of these $(2M+1)$ positions.

From the above, it is possible to define a length scale $X$ as the average,
effective distance between these subregisters. If the overall length arising
out of $\mathcal{H}_{[1...N^{\prime }]}$ is taken to be $L,$ and because
there are precisely $2M$ intervals between $(2M+1)$ points, then
\begin{equation}
X=\QOVERD( ) {L}{2M}
\end{equation}
noting that $X$ is defined only by extrapolating backwards from the overall
length experienced on the emergent scale and is meaningless on the actual
pregeometric level.

Paraphrasing these ideas: from an emergent perspective the vision is of a
one dimensional lattice of length $L$ comprising of $(2M+1)$ points
separated by a distance of $X,$ but on the pregeometric level the $(2M+1)$
subregisters are still just factors of the overall Hilbert space and are not
embedded into any sort of spatial background. As with the causal set
discussions of Chapter 5, it is the overall network of relationships between
operators and states that leads to the generation of effective concepts such
as distance on the macroscopic scale.

Note that the above considerations immediately imply a splitting of the
Hilbert space $\mathcal{H}_{[1...N^{\prime }]}$ containing $\varphi ,$ which
may be denoted as
\begin{eqnarray}
\mathcal{H}_{[1...N^{\prime }]} &=&\mathcal{H}_{[\varphi _{(-M)}]}\otimes
\mathcal{H}_{[\varphi _{(-M+1)}]}\otimes ...\otimes \mathcal{H}_{[\varphi
_{(M)}]} \\
&=&\prod_{R=-M}^{M}\otimes \mathcal{H}_{[\varphi _{(R)}]}.  \notag
\end{eqnarray}
where $R$ runs from $-M$ to $M$ for reasons that will become apparent.
Moreover, from this split a suitable relabelling of the subregisters may be
performed in the obvious way for visual convenience, such that from a given
emergent perspective the $(-M)^{th}$ sub-space $\mathcal{H}_{[\varphi
_{(-M)}]}$ of $\mathcal{H}_{[1...N^{\prime }]}$ may be thought of as being
located at the `left-most' position, whereas sub-operators acting locally in
the $(+M)^{th}$ subregister $\mathcal{H}_{[\varphi _{(M)}]}$ of $\mathcal{H}%
_{[1...N^{\prime }]}$ may be thought of as being responsible for the
`right-most' position, with the remaining sub-spaces positioned at the
obvious locations in between.

\bigskip

Now consider the components of the state in each of these $(2M+1)$
subregisters. The nature of this set of components may be thought of as
defining a type of field across the one dimensional lattice. Moreover,
concepts of information can be applied in this context, such that the
information contained in the $R^{th}$ factor space can be seen as analogous
to the value of the field at spatial position $R.$

Assuming that this field is not trivially constant, the minimum dimension of
each subregister $\mathcal{H}_{[\varphi _{(R)}]}$ is evidently two.
Defining, then, the basis vectors for each subregister in the usual way as $%
|0\rangle $ and $|1\rangle ,$ it implies that at each of the locations it is
possible to assign a `value' for the field in terms of these bases. This
information will play an important part in describing the physical
properties of the particle system, as shown later.

However, matters are straight away seen as being considerably more involved
than a simple two dimensional scenario permits. After all, a single Dirac
field allows the existence of four distinct particle species (both spin-up
and spin-down electrons and positrons), so each of the $(2M+1)$ factor
sub-spaces of $\mathcal{H}_{[1...N^{\prime }]}$ will be required to contain
information regarding four different types of particle at that site if such
a field is ultimately to be modelled. Thus, in order to consistently
describe all of these possibilities at least eight degrees of freedom will
be required for each of the $(2M+1)$ factors of\ $\mathcal{H}%
_{[1...N^{\prime }]}.$ These may in turn be grouped into four sets, each
corresponding to a sub-space of\ $\mathcal{H}_{[\varphi _{(R)}]}$ relating
to a given particle species.

The simplest possible model for the Dirac field may consequently be achieved
by associating four qubits to each of the $(2M+1)$ factor spaces, one for
each particle type. In this case
\begin{equation}
N^{\prime }=4(2M+1)=4\times 10^{35}
\end{equation}
and it is possible to consider a further split of each of the $(2M+1)$
sub-spaces of $\mathcal{H}_{[1...N^{\prime }]}$ into the products of
subregisters they comprise. That is, the sub-space $\mathcal{H}_{[\varphi
_{(R)}]}$ may be written
\begin{equation}
\mathcal{H}_{[\varphi _{(R)}]}=\mathcal{H}_{R_{(1)}}\otimes \mathcal{H}%
_{R_{(2)}}\otimes \mathcal{H}_{R_{(3)}}\otimes \mathcal{H}_{R_{(4)}}
\end{equation}
such that
\begin{equation}
\mathcal{H}_{[1...N^{\prime }]}=\prod_{R=-M}^{M}\prod_{z=1}^{4}\otimes
\mathcal{H}_{R_{(z)}}.  \label{Ch5Reg}
\end{equation}

Evidently, the dimension of the Hilbert space $\mathcal{H}_{[1...N^{\prime
}]}$ is $2^{4\times 10^{35}}.$

As desired, each of the four qubits in each $\mathcal{H}_{[\varphi _{(R)}]}$
is to be associated with one particle type. So, by arbitrarily relabelling
the subregisters for convenience, the sub-register $\mathcal{H}_{R_{(1)}}$
will be used in the following to contain information about the spin-down
electron, whereas $\mathcal{H}_{R_{(2)}}$ is chosen to be the space
containing information for a spin-up electron, whilst $\mathcal{H}_{R_{(3)}}$
regards the spin-down positron, and $\mathcal{H}_{R_{(4)}}$ is the space to
be used for the construction of the spin-up positron, for all $R=-M,...,M.$

Moreover, by choosing an orthonormal $\mathcal{B}_{R_{(z)}}$ basis for $%
\mathcal{H}_{R_{(z)}}$ to be $\mathcal{B}_{R_{(z)}}\equiv \{|0\rangle
_{R_{(z)}},|1\rangle _{R_{(z)}}\}$ it is possible to specify the component
of $\varphi $ in a particular one of the $4(2M+1)$ sub-spaces of $\mathcal{H}%
_{[1...N^{\prime }]}.$\bigskip

The pregeometric ladder operators discussed in the previous section may be
employed to manipulate the components of the state $\varphi \in \mathcal{H}%
_{[1...N^{\prime }]}$ at various emergent positions.

For example, by defining the completely separable vector $|\underline{0}%
\rangle _{1...N^{\prime }}$ as the `vacuum' or `ground' state in $\mathcal{H}%
_{[1...N^{\prime }]},$ the operation
\begin{equation}
\hat{\alpha}_{R_{(z)}}^{\dagger }|\underline{0}\rangle =|0\rangle
_{1}\otimes |0\rangle _{2}\otimes ...\otimes |0\rangle _{R_{(z-1)}}\otimes
|1\rangle _{R_{(z)}}\otimes |0\rangle _{R_{(z+1)}}\otimes ...\otimes
|0\rangle _{N}  \label{Ch5Cyc}
\end{equation}
evidently changes the component of the state $\varphi =|\underline{0}\rangle
$ in $\mathcal{H}_{R_{(z)}}$ from $|0\rangle _{R_{(z)}}$ to $|1\rangle
_{R_{(z)}},$ whilst the operator $\hat{\alpha}_{R_{(z)}}$ reverses this
`occupation' of subregister\footnote{%
With the notation of (\ref{Ch5Cyc}) defined `cyclically', such that $%
R_{(0)}\equiv (R-1)_{(4)}$ and $R_{(5)}\equiv (R+1)_{(1)}.$} $\mathcal{H}%
_{R_{(z)}}.$ Equivalently, it may be possible to imagine that the operation $%
\hat{\alpha}_{R_{(z)}}^{\dagger }|\underline{0}\rangle $ creates a
particular `pregeometric particle' at location $R_{(z)}$ from the vacuum,
whereas the operator $\hat{\alpha}_{R_{(z)}}$ annihilates this result. So,
in this case a component of the form $|1\rangle _{R_{(z)}}$ is being chosen
to relate to the existence of a particular `pregeometric particle' at a
given position $R_{(z)},$ whereas the value $|0\rangle _{R_{(z)}}$ is
implying its absence; the similarities to the type of binary logic discussed
in quantum computation are obvious.

Overall, it is the information contained in the individual factor spaces
that will be used to control the physics of the situation.

Of course,
\begin{eqnarray}
\hat{\alpha}_{R_{(z)}}^{\dagger }\hat{\alpha}_{R_{(z)}}^{\dagger }|%
\underline{0}\rangle &=&0 \\
\hat{\alpha}_{R_{(z)}}|\underline{0}\rangle &=&0  \notag
\end{eqnarray}
as expected.

The introduction of ladder operators to the current analysis is actually
more fundamental than the above simple manipulation of the components of the
state seems to suggest. After all, recall that the second quantisation
procedure of conventional field theory involves re-writing the field as a
field operator, before this is then typically expressed in terms of
annihilation and creation operators. Thus in the present case, it is
expected that pregeometric field operators acting across the space of
interest could ultimately be described in terms of the pregeometric ladder
operators.

The overall aim is therefore to investigate how such a `pregeometric field'
could relate to the physical particle field used in conventional quantum
theory. In particular, it is considered how `real' Dirac particles in
momentum space might emerge from the simple logic of\ the pregeometric
framework. The issue effectively becomes one of relating the momentum space
ladder operators of familiar physics to the pregeometric annihilation and
creation operators that act at definite positions along the emergent spatial
lattice. Concepts normally expressed in conventional particle physics as
functions of momentum space ladder operators, such as momentum space
Hamiltonian or field operators, may then be recovered from the underlying
pregeometric framework by substituting the usual momentum space annihilation
and creation operators for their pregeometric definitions, thereby
re-writing the Hamiltonian or field in terms of these pregeometric ladder
operators.

\bigskip

In fact, it is in principle possible to conjecture many different ways of
bridging the gap between the pregeometric picture and emergent momentum
space. However, due to the conventional relationship that exists between
momentum and position in terms of Fourier analysis, it is this type of
method that presents itself as a natural candidate to provide such a
mechanism here. It is this that is now proposed.

Consider the standard Fourier Transform \cite{Kufner} of an arbitrary
function $f(\tau )$ with the variable $\omega $%
\begin{equation}
f(\tau )=\int_{-\infty }^{\infty }[F(\omega )\exp (i\tau \omega )]\text{ }%
d\omega
\end{equation}
where $F(\omega )$ is an amplitude of value
\begin{equation}
F(\omega )=\QOVERD( ) {1}{2\pi }\int_{-\infty }^{\infty }[f(\tau )\exp
(-i\tau \omega )]\text{ }d\tau .
\end{equation}

The present goal is to write the ladder operators in momentum space as a
Fourier transform of terms in pregeometric space. Thus, by temporarily
introducing the continuous position variable $x,$ then a `first guess' for
the form of an annihilation operator $\hat{a}(p,\kappa )$ of known spin $%
\kappa $\ may initially be given by
\begin{equation}
\hat{a}(p,\kappa )=\int_{-\infty }^{\infty }[F(x)\exp (ipx)]dx
\end{equation}
with modifications required as follows:

\begin{itemize}
\item  Firstly, the parameter $x$ must be discretised if it is to fit into
the proposed paradigm, because space in the quantum Universe is assumed
ultimately to be non-continuous. Assuming from above that the minimum
spatial resolution of the volume is $X,$ then $x$ is given by
\begin{equation}
x=mX\text{ \ \ , \ \ }m=0,\pm 1,\pm 2,...
\end{equation}
\end{itemize}

and similarly
\begin{equation}
F(x)=F(mX)=F_{m}
\end{equation}

such that $\hat{a}(p,\kappa )$ is given by a sum of discrete terms instead
of as a continuous integral.

\begin{itemize}
\item  Next, the limits need to be constrained. Since the method is being
used to describe a finite model (i.e. a part of a finite dimensional Quantum
Universe) of length $L$ with a finite number of possible positions, the
Fourier sum must also be finite. Specifically, since the integral is
normally performed across all space, the sum in this case must be taken over
all the positions for which the system is defined. This is hence bounded by
the number of `sites' contained in $\mathcal{H}_{[1...N^{\prime }]},$ namely
$(2M+1).$ So, by preserving symmetry around the origin, the minimum and
maximum positions are given by
\begin{equation}
x_{\min }=-L/2=(-M)(X)\text{ \ \ , \ \ }x_{\max }=+L/2=(+M)(X)
\end{equation}
from which it follows that $-M\leq m\leq M$ as expected. Clearly, the index $%
m$ is equivalent to the index $R$ used above to label the $(2M+1)$ factors.
Moreover, for\ $M\gg 1$ it follows that $x_{\max }\gg 1$ and $x_{\min }\ll
-1,$ such that the infinite integral is well approximated by the sum.

\item  Finally, the amplitudes $F_{m}=F_{R}$ are hoped to be functions of
the pregeometric ladder operators, since this is after all the point of the
current work. Taking proportionality here gives
\begin{equation}
F_{R}=A_{p}\hat{\alpha}(R)
\end{equation}
where $A_{p}$ is a type of `form function' to be investigated, and the
pregeometric annihilation operator has been temporarily written as $\hat{%
\alpha}(R)$ to indicate that it is a function of $R.$
\end{itemize}

Overall, $\hat{a}(p,\kappa )$ becomes of the form
\begin{equation}
\hat{a}(p,\kappa )=\sum_{R=-M}^{M}A_{p}\hat{\alpha}(R)\exp \left(
iRpX\right) .  \label{Ch5generica}
\end{equation}

The argument of the exponential in equation (\ref{Ch5generica}) introduces a
periodicity into the analysis; specifically,
\begin{equation}
\left( \exp \left( ipX\right) \right) ^{R}=\left( \exp \left( i[pX\pm 2j\pi
]\right) \right) ^{R}\text{ \ \ , \ \ }j=0,1,2...
\end{equation}

Moreover, defining the resulting maximum positive momentum as $p_{\max },$
and assuming symmetry about the lowest value $p=0$ such that $\left|
p\right| =\left| -p\right| ,$ it implies that $p_{\max }$ is in practice
bounded by the half-period
\begin{equation}
\left| p_{\max }\right| X=\pi
\end{equation}
so $p_{\max }=\pi /X.$ Note that this relationship indicates how the largest
momentum is related to the reciprocal of the smallest spatial resolution, as
analogous to the uncertainty principle. Importantly, observe also that since
$M\gg 1,$ it follows that $X\ll 1,$ so $p_{\max }\gg 1$ as required.

The energy $E_{p}$ of the mode of momentum $p$ may now be defined in the
conventional way. It follows, then, that this too is bounded, such that
\begin{equation}
(E_{p})_{\max }\equiv \left( p_{\max }^{2}c^{2}+\mu ^{2}c^{4}\right) ^{1/2}
\end{equation}
where $\mu $ is the particle's mass and $c$ is the speed of light. Again, it
is remarked that $(E_{p})_{\max }$ is clearly very large.\bigskip

The operator $\hat{\alpha}(R)$ of equation (\ref{Ch5generica}) is assumed to
be one of four possible types because it could act on one of four possible
qubits for each $R.$ Thus, equation (\ref{Ch5generica}) must be refined
further to account for the four different sorts of particle that could be
annihilated by the generic operator $\hat{a}(p,\kappa ).$ If $\hat{c}%
(p,\kappa )$ is defined as the annihilation operator for electrons of
momentum $p$ and spin $\kappa ,$ for $\kappa =1,2$ where $\kappa =1$
represents spin-down and $\kappa =2$ represents spin-up, and if $\hat{d}%
(p,\kappa )$ conversely annihilates positrons of momentum $p$ and spin $%
\kappa ,$ then from the discussion following (\ref{Ch5Reg}) and the relation
(\ref{Ch5generica}) it is evident that
\begin{equation}
\hat{c}(p,\kappa )=\sum_{R=-M}^{M}A_{p}\hat{\alpha}_{4(M+R)+\kappa }\exp
\left( iRpX\right)  \label{Ch5ane}
\end{equation}
and
\begin{equation}
\hat{d}(p,\kappa )=\sum_{R=-M}^{M}A_{p}\hat{\alpha}_{4(M+R)+2+\kappa }\exp
\left( iRpX\right) .
\end{equation}

The corresponding creation operators may be defined in an analogous way as
the Hermitian conjugates of the annihilation operators; viz.
\begin{equation}
\hat{c}^{\dagger }(p,\kappa )=\sum_{R=-M}^{M}A_{p}^{\ast }\hat{\alpha}%
_{4(M+R)+\kappa }^{\dagger }\exp \left( -iRpX\right)
\end{equation}
and
\begin{equation}
\hat{d}^{\dagger }(p,\kappa )=\sum_{R=-M}^{M}A_{p}^{\ast }\hat{\alpha}%
_{4(M+R)+2+\kappa }^{\dagger }\exp \left( -iRpX\right) .  \label{Ch5crp}
\end{equation}

\bigskip

For the conclusions defined in (\ref{Ch5ane})-(\ref{Ch5crp}) to be accepted
as physically valid, they must be able to reproduce the standard results
obtained for conventional fermionic ladder operators. For example, the
anti-commutation relations of momentum space annihilation and creation
operators must be obeyed, such that all the anti-commutators vanish apart
from the results $\{\hat{c}(p,\kappa ),\hat{c}^{\dagger }(p^{\prime },\kappa
^{\prime })\}$ and $\{\hat{d}(p,\kappa ),\hat{d}^{\dagger }(p^{\prime
},\kappa ^{\prime })\}$ which should give
\begin{equation}
\{\hat{c}(p,\kappa ),\hat{c}^{\dagger }(p^{\prime },\kappa ^{\prime })\}=\{%
\hat{d}(p,\kappa ),\hat{d}^{\dagger }(p^{\prime },\kappa ^{\prime
})\}=\delta _{pp^{\prime }}\delta _{\kappa \kappa ^{\prime }}.
\label{Ch5Anti-USUAL}
\end{equation}

The vanishing terms are clearly satisfied: for the results defined in (\ref
{Ch5ane})-(\ref{Ch5crp})
\begin{eqnarray}
\{\hat{c}(p,\kappa ),\hat{c}(p^{\prime },\kappa ^{\prime })\} &=&\{\hat{c}%
^{\dagger }(p,\kappa ),\hat{c}^{\dagger }(p^{\prime },\kappa ^{\prime })\}=0
\\
\{\hat{d}(p,\kappa ),\hat{d}(p^{\prime },\kappa ^{\prime })\} &=&\{\hat{d}%
^{\dagger }(p,\kappa ),\hat{d}^{\dagger }(p^{\prime },\kappa ^{\prime })\}=0
\notag \\
\{\hat{c}(p,\kappa ),\hat{d}(p^{\prime },\kappa ^{\prime })\} &=&\{\hat{c}%
(p,\kappa ),\hat{d}^{\dagger }(p^{\prime },\kappa ^{\prime })\}=0  \notag \\
\{\hat{c}^{\dagger }(p,\kappa ),\hat{d}(p^{\prime },\kappa ^{\prime })\}
&=&\{\hat{c}^{\dagger }(p,\kappa ),\hat{d}^{\dagger }(p^{\prime },\kappa
^{\prime })\}=0  \notag
\end{eqnarray}
which follow immediately from the relations (\ref{Ch5anti-pre}) for the
pregeometric ladder operators.

Considering instead the relations $\{\hat{c}(p,\kappa ),\hat{c}^{\dagger
}(p^{\prime },\kappa ^{\prime })\}$ and $\{\hat{d}(p,\kappa ),\hat{d}%
^{\dagger }(p^{\prime },\kappa ^{\prime })\},$ gives
\begin{eqnarray}
\{\hat{c}(p,\kappa ),\hat{c}^{\dagger }(p^{\prime },\kappa ^{\prime })\}
&=&\sum_{R=-M}^{M}A_{p}\hat{\alpha}_{4(M+R)+\kappa }\exp \left( iRpX\right)
\sum_{S=-M}^{M}A_{p^{\prime }}^{\ast }\hat{\alpha}_{4(M+S)+\kappa ^{\prime
}}^{\dagger }\exp \left( -iSp^{\prime }X\right)  \notag \\
&&+\sum_{S=-M}^{M}A_{p^{\prime }}^{\ast }\hat{\alpha}_{4(M+S)+\kappa
^{\prime }}^{\dagger }\exp \left( -iSp^{\prime }X\right) \sum_{R=-M}^{M}A_{p}%
\hat{\alpha}_{4(M+R)+\kappa }\exp \left( iRpX\right)  \notag \\
&=&\sum_{R=-M}^{M}\sum_{S=-M}^{M}A_{p}A_{p^{\prime }}^{\ast }\exp \left(
i(Rp-Sp^{\prime })X\right) \left[
\begin{array}{c}
\hat{\alpha}_{4(M+R)+\kappa }\hat{\alpha}_{4(M+S)+\kappa ^{\prime
}}^{\dagger } \\
+\hat{\alpha}_{4(M+S)+\kappa ^{\prime }}^{\dagger }\hat{\alpha}%
_{4(M+R)+\kappa }
\end{array}
\right]  \notag \\
&=&\sum_{R=-M}^{M}\sum_{S=-M}^{M}A_{p}A_{p^{\prime }}^{\ast }\{\hat{\alpha}%
_{4(M+R)+\kappa },\hat{\alpha}_{4(M+S)+\kappa ^{\prime }}^{\dagger }\}\exp
\left( i(Rp-Sp^{\prime })X\right)  \notag \\
&&
\end{eqnarray}

So
\begin{eqnarray}
\{\hat{c}(p,\kappa ),\hat{c}^{\dagger }(p^{\prime },\kappa ^{\prime })\}
&=&\sum_{R=-M}^{M}\sum_{S=-M}^{M}A_{p}A_{p^{\prime }}^{\ast }\delta
_{RS}\delta _{\kappa \kappa ^{\prime }}\exp \left( i(Rp-Sp^{\prime })X\right)
\label{CH5aneexp} \\
&=&\sum_{R=-M}^{M}A_{p}A_{p^{\prime }}^{\ast }\delta _{\kappa \kappa
^{\prime }}\exp \left( iR(p-p^{\prime })X\right)  \notag
\end{eqnarray}
with a similar result for $\{\hat{d}(p,\kappa ),\hat{d}^{\dagger }(p^{\prime
},\kappa ^{\prime })\}.$

Now, consider the Fourier expansion of the continuous space Dirac delta
function $\delta (p)$ of period $2\infty ,$ defined in the usual way as \cite
{Kufner}\cite{Cartwright}
\begin{equation}
\delta (p)\equiv \QOVERD( ) {1}{2\pi }\int_{-\infty }^{\infty }(1)\exp
\left( -ipx\right) dx.  \label{Ch5Dir}
\end{equation}

By making the same type of approximation as before, i.e. associating the
function with a large, but finite, period of at least $2\QOVERD( ) {\pi }{X}$
defined over a discretised background space of emergent length $L=2MX,$ the
expression (\ref{Ch5Dir}) may be truncated, and re-written as
\begin{equation}
\delta _{p}\equiv \QOVERD( ) {1}{2\pi }\sum_{R=-M}^{M}C_{R}\exp \left(
-iRpX\right) .  \label{ch5Rearr}
\end{equation}

Multiplying both sides by $e^{iR^{\prime }pX}$ and integrating over all
momentum $\left| p\right| \leq p_{\max }$ gives
\begin{eqnarray}
\int_{-\pi /X}^{\pi /X}\delta _{p}\exp \left( iR^{\prime }pX\right) \text{ }%
dp &\equiv &\QOVERD( ) {1}{2\pi }\sum_{R=-M}^{M}C_{R}\int_{-\pi /X}^{\pi
/X}\exp \left( -i(R-R^{\prime })pX\right) \text{ }dp  \notag \\
e^{0} &=&\QOVERD( ) {1}{2\pi }\sum_{R=-M}^{M}C_{R}\delta _{RR^{\prime
}}\QOVERD( ) {2\pi }{X}  \notag \\
1 &=&C_{R^{\prime }}\QOVERD( ) {1}{X}
\end{eqnarray}
so $C_{R}=X$ such that $\sum_{R=-M}^{M}\exp \left( -iRpX\right) =\QOVERD( )
{2\pi }{X}\delta _{p}$ from (\ref{ch5Rearr}).

Clearly then
\begin{equation}
\sum_{R=-M}^{M}\exp \left( -iR(p-p^{\prime })X\right) =\QOVERD( ) {2\pi
}{X}\delta _{pp^{\prime }}.
\end{equation}
and substituting this into (\ref{CH5aneexp}) gives
\begin{eqnarray}
\{\hat{c}(p,\kappa ),\hat{c}^{\dagger }(p^{\prime },\kappa ^{\prime })\}
&=&\sum_{R=-M}^{M}A_{p}A_{p^{\prime }}^{\ast }\delta _{\kappa \kappa
^{\prime }}\QOVERD( ) {2\pi }{X}\delta _{p^{\prime }p}  \label{Ch5anti-preg}
\\
&=&\QOVERD( ) {2\pi }{X}(A_{p}A_{p^{\prime }}^{\ast })\delta _{\kappa \kappa
^{\prime }}\delta _{pp^{\prime }}  \notag
\end{eqnarray}
which is equal in form to the usual anti-commutation algebra (\ref
{Ch5Anti-USUAL}). So, from a comparison of (\ref{Ch5Anti-USUAL}) and (\ref
{Ch5anti-preg}) it follows that the expressions are equal if $A_{p}\in
\mathbb{R}$ may be defined as
\begin{equation}
A_{p}=\sqrt{\frac{X}{2\pi }.}
\end{equation}

Collecting these solutions defines the momentum space annihilation and
creation operators in terms of pregeometric ladder operators. For electrons
these are
\begin{eqnarray}
\hat{c}(p,\kappa ) &=&\sqrt{\frac{X}{2\pi }}\sum_{R=-M}^{M}\hat{\alpha}%
_{4(M+R)+\kappa }\exp \left( iRpX\right)  \label{Ch5elec} \\
\hat{c}^{\dagger }(p,\kappa ) &=&\sqrt{\frac{X}{2\pi }}\sum_{R=-M}^{M}\hat{%
\alpha}_{4(M+R)+\kappa }^{\dagger }\exp \left( -iRpX\right)  \notag
\end{eqnarray}
whilst for positrons the results are
\begin{eqnarray}
\hat{d}(p,\kappa ) &=&\sqrt{\frac{X}{2\pi }}\sum_{R=-M}^{M}\hat{\alpha}%
_{4(M+R)+2+\kappa }\exp \left( iRpX\right)  \label{Ch5posit} \\
\hat{d}^{\dagger }(p,\kappa ) &=&\sqrt{\frac{X}{2\pi }}\sum_{R=-M}^{M}\hat{%
\alpha}_{4(M+R)+2+\kappa }^{\dagger }\exp \left( -iRpX\right) .  \notag
\end{eqnarray}

These expressions may be substituted into the standard equations of
fermionic field theory, to give, for example, the momentum space field
operators. More importantly, perhaps, they may also be used to construct the
actual observables familiar to the conventional theory, as discussed next.

\bigskip

\subsubsection{The Hamiltonian}

\bigskip

The results of above can be used to formulate a version of the Hamiltonian
in terms of pregeometric ladder operators. Such a formulation is important,
as is the derivation of the momentum and charge operators discussed later,
because it is operators like these that form the basis for actual
observables in physics.

Of course, such operators require careful interpretation from the
perspective of the quantum Universe paradigm proposed in this thesis.
Specifically, the operators of below are assumed to be associated with the
part of the operator $\hat{\Sigma}_{n+n^{\prime }}$ (which is used to
develop the entire Universe) that appears to test the sub-state $|\varphi
_{in}\rangle $ of $\Psi _{n+n^{\prime }-1}.$ Thus, the Hamiltonian, momentum
and charge operators discussed below are expected to ultimately be
represented by different factors of different possible tests $\hat{\Sigma}%
_{n+n^{\prime }}.$

Moreover, given that the quantum Universe is taken to be completely
self-contained and autonomous, the operators it self-referentially chooses
must be very carefully controlled if emergent endo-observers are to gain the
impression that they can detect electrons and positrons in the medley of
ways familiar to physicists. This again is emphatic of the point that a
quantum state cannot really be said to exist independently of the tests used
to observe it, and consequently that different choices of test (e.g. energy
or charge) lead to different `experiences' of physical reality by emergent
endo-observers.\bigskip

Consider the conventional Hamiltonian operator $\hat{H}$ for the free-field
theory of spin-$%
%TCIMACRO{\UNICODE[m]{0xbd}}%
%BeginExpansion
{\frac12}%
%EndExpansion
$ fermions, defined \cite{Bjorken} in three dimensional momentum space as
\begin{equation}
\hat{H}=\sum_{\kappa }\int_{-\infty }^{\infty }E_{p}\left[ \hat{c}^{\dagger
}(\mathbf{p},\kappa )\hat{c}(\mathbf{p},\kappa )+\hat{d}^{\dagger }(\mathbf{p%
},\kappa )\hat{d}(\mathbf{p},\kappa )\right] \text{ }d^{3}p
\end{equation}
noting that this equation has been derived in Appendix B for completeness.
Here, $E_{p}$ is the energy of the particle, $\mathbf{p}$ is its momentum
3-vector, and the sum is over both spin states $\kappa =1,2.$

Now, noting that in the present chapter the momentum $\mathbf{p}$ has been
restricted to a one dimensional variable $p,$ it is possible to rewrite the
conventional Hamiltonian in terms of the pregeometric operators defined in (%
\ref{Ch5elec}) and (\ref{Ch5posit}). So,
\begin{eqnarray}
\hat{H} &=&\sum_{\kappa }\int_{-\pi /X}^{\pi
/X}E_{p}\sum_{R=-M}^{M}\sum_{S=-M}^{M}\QOVERD( ) {X}{2\pi } \\
&&\times \left[
\begin{array}{c}
\hat{\alpha}_{4(M+R)+\kappa }^{\dagger }\exp \left( -iRpX\right) \hat{\alpha}%
_{4(M+S)+\kappa }\exp \left( iSpX\right) \\
+\hat{\alpha}_{4(M+R)+2+\kappa }^{\dagger }\exp \left( -iRpX\right) \hat{%
\alpha}_{4(M+S)+2+\kappa }\exp \left( iSpX\right)
\end{array}
\right] \text{ }dp  \notag
\end{eqnarray}
with the obvious imposition that the integral limits $\pm \infty $ have been
constrained to $\pm \pi /X$ as before. Recalling that the energy is defined
as $E_{p}=(p^{2}c^{2}+\mu ^{2}c^{4})^{1/2},$ it follows that
\begin{eqnarray}
\hat{H} &=&\QOVERD( ) {X}{2\pi }\sum_{\kappa
}\sum_{R=-M}^{M}\sum_{S=-M}^{M}\int_{-\pi /X}^{\pi /X}(p^{2}c^{2}+\mu
^{2}c^{4})^{1/2}e^{i(S-R)pX}\text{ } \\
&&\times \left[
\begin{array}{c}
\hat{\alpha}_{4(M+R)+\kappa }^{\dagger }\hat{\alpha}_{4(M+S)+\kappa } \\
+\hat{\alpha}_{4(M+R)+2+\kappa }^{\dagger }\hat{\alpha}_{4(M+S)+2+\kappa }
\end{array}
\right] \text{ }dp.  \notag
\end{eqnarray}

Erd\'{e}lyi \textit{et al} \cite{Erdelyi} list no known explicit solution
for this integral, suggesting that the Hamiltonian may only be evaluated as
a numerical approximation. Whilst on the surface this may appear
unsatisfactory, it does evidence the fact that the Hamiltonian is a highly
non-trivial function of pregeometric variables, as might perhaps be expected
for an operator defined in the emergent limit.

\bigskip

\subsubsection{The Momentum Operator}

\bigskip

Just as for the Hamiltonian, it is also possible to write the emergent
momentum operator in terms of the pregeometric ladder operators.

Recall the conventional momentum operator $\mathbf{\hat{P},}$ defined \cite
{Bjorken} as
\begin{equation}
\mathbf{\hat{P}}=\sum_{\kappa }\int_{-\infty }^{\infty }\mathbf{p}\left[
\hat{c}^{\dagger }(\mathbf{p},\kappa )\hat{c}(\mathbf{p},\kappa )+\hat{d}%
^{\dagger }(\mathbf{p},\kappa )\hat{d}(\mathbf{p},\kappa )\right] \text{ }%
d^{3}p
\end{equation}
and derived also in Appendix B. By restricting again the analysis to one
finite dimension, and substituting in the relations (\ref{Ch5elec}) and (\ref
{Ch5posit}), the momentum operator becomes
\begin{equation}
\mathbf{\hat{P}}=\QOVERD( ) {X}{2\pi }\sum_{\kappa
}\sum_{R=-M}^{M}\sum_{S=-M}^{M}\left[
\begin{array}{c}
\hat{\alpha}_{4(M+R)+\kappa }^{\dagger }\hat{\alpha}_{4(M+S)+\kappa } \\
+\hat{\alpha}_{4(M+R)+2+\kappa }^{\dagger }\hat{\alpha}_{4(M+S)+2+\kappa }
\end{array}
\right] \int_{-\pi /X}^{\pi /X}pe^{i(S-R)pX}\text{ }dp.
\end{equation}

In order to solve the integral, it will prove useful to separate this last
expression into a `diagonal' part for which $R=S$ and an off-diagonal part
for which $R\neq S.$ Thus
\begin{equation}
\mathbf{\hat{P}}=\mathbf{\hat{P}}_{D}+\mathbf{\hat{P}}_{OD}
\end{equation}
where
\begin{equation}
\mathbf{\hat{P}}_{D}=\QOVERD( ) {X}{2\pi }\sum_{\kappa }\sum_{R=-M}^{M}\left[
\begin{array}{c}
\hat{\alpha}_{4(M+R)+\kappa }^{\dagger }\hat{\alpha}_{4(M+R)+\kappa } \\
+\hat{\alpha}_{4(M+R)+2+\kappa }^{\dagger }\hat{\alpha}_{4(M+R)+2+\kappa }
\end{array}
\right] \int_{-\pi /X}^{\pi /X}p\text{ }dp  \label{Ch5PDIAG}
\end{equation}
and
\begin{eqnarray}
\mathbf{\hat{P}}_{OD} &=&\QOVERD( ) {X}{2\pi }\sum_{\kappa
}\sum_{R=-M}^{M}\sum_{S=-M}^{M}(1-\delta _{RS})  \label{Ch5OffD} \\
&&\times \left[
\begin{array}{c}
\hat{\alpha}_{4(M+R)+\kappa }^{\dagger }\hat{\alpha}_{4(M+S)+\kappa } \\
+\hat{\alpha}_{4(M+R)+2+\kappa }^{\dagger }\hat{\alpha}_{4(M+S)+2+\kappa }
\end{array}
\right] \int_{-\pi /X}^{\pi /X}pe^{i(S-R)pX}\text{ }dp.  \notag
\end{eqnarray}

The integral in (\ref{Ch5PDIAG}) gives
\begin{equation}
\int_{-\pi /X}^{\pi /X}p\text{ }dp=\left( \frac{\pi ^{2}}{2X^{2}}-\frac{\pi
^{2}}{2X^{2}}\right) =0
\end{equation}
such that
\begin{equation}
\mathbf{\hat{P}}_{D}=0.
\end{equation}

Turning now to the off-diagonal case, and defining the integral in (\ref
{Ch5OffD}) to be $\Lambda _{R\neq S},$ it follows that
\begin{eqnarray}
\Lambda _{R\neq S} &=&\int_{-\pi /X}^{\pi /X}pe^{i(S-R)pX}\text{ }dp\text{ \
\ , \ \ }R\neq S \\
&=&\QOVERD( ) {1}{(S-R)^{2}X^{2}}\left\{
\begin{array}{c}
i\pi (R-S)\left\{ e^{i\pi (S-R)}+e^{-i\pi (S-R)}\right\} \\
+\left\{ e^{i\pi (S-R)}-e^{-i\pi (S-R)}\right\}
\end{array}
\right\}  \notag \\
&=&\QOVERD( ) {2i}{(S-R)^{2}X^{2}}\left\{ \sin (\pi (S-R))+\pi (R-S)\cos
(\pi (S-R))\right\} .  \notag
\end{eqnarray}

However, because $R,S\in \mathbb{Z}$ and $R\neq S$%
\begin{equation}
\sin (\pi (S-R))=0\text{ \ \ , \ \ }\forall R,S
\end{equation}
and
\begin{equation}
\cos (\pi (S-R))=(-1)^{(S-R)}\text{ \ \ , \ \ }\forall R,S.
\end{equation}

So finally
\begin{eqnarray}
\mathbf{\hat{P}}_{OD} &=&\QOVERD( ) {-i}{X}\sum_{\kappa
}\sum_{R=-M}^{M}\sum_{S=-M}^{M}(1-\delta _{RS})\QOVERD( )
{(-1)^{(S-R)}}{(S-R)} \\
&&\times \left[
\begin{array}{c}
\hat{\alpha}_{4(M+R)+\kappa }^{\dagger }\hat{\alpha}_{4(M+S)+\kappa } \\
+\hat{\alpha}_{4(M+R)+2+\kappa }^{\dagger }\hat{\alpha}_{4(M+S)+2+\kappa }
\end{array}
\right] .  \notag
\end{eqnarray}

Overall,
\begin{eqnarray}
\mathbf{\hat{P}} &=&\QOVERD( ) {1}{X}\sum_{\kappa
}\sum_{R=-M}^{M}\sum_{S=-M}^{M}\left[
\begin{array}{c}
\hat{\alpha}_{4(M+R)+\kappa }^{\dagger }\hat{\alpha}_{4(M+S)+\kappa } \\
+\hat{\alpha}_{4(M+R)+2+\kappa }^{\dagger }\hat{\alpha}_{4(M+S)+2+\kappa }
\end{array}
\right] \\
&&\times \QOVERD( ) {-i(-1)^{(S-R)}}{(S-R)}(1-\delta _{RS}).  \notag
\end{eqnarray}

Again, note that this emergent construct is also a complicated function of
basic pregeometric logic operators.

\bigskip

\subsubsection{Charge}

\bigskip

To conclude this section, it is shown how the conventional charge operator
may also be written as a function of pregeometric ladder operators.

Recall the operator $\hat{Q},$ defined as
\begin{equation}
\hat{Q}=q\int_{-\infty }^{\infty }d^{3}p\sum_{\kappa =1}^{2}\left[ \hat{c}%
^{\dagger }(\mathbf{p},\kappa )\hat{c}(\mathbf{p},\kappa )-\hat{d}^{\dagger
}(\mathbf{p},\kappa )\hat{d}(\mathbf{p},\kappa )\right]
\end{equation}
and again derived in Appendix B. Here, $q$ is an arbitrary scalar constant,
but of course for `real' electrons and positrons it is known to have the
value $q=-e\sim -1.602\times 10^{-19}$ \cite{PDG}.

By once again considering only a finite, one dimensional volume, and
substituting in the relations (\ref{Ch5elec}) and (\ref{Ch5posit}), it
follows that $\hat{Q}$ becomes
\begin{equation}
\hat{Q}=\QOVERD( ) {X}{2\pi }q\sum_{\kappa }\sum_{R=-M}^{M}\sum_{S=-M}^{M}%
\left[
\begin{array}{c}
\hat{\alpha}_{4(M+R)+\kappa }^{\dagger }\hat{\alpha}_{4(M+S)+\kappa } \\
-\hat{\alpha}_{4(M+R)+2+\kappa }^{\dagger }\hat{\alpha}_{4(M+S)+2+\kappa }
\end{array}
\right] \int_{-\pi /X}^{\pi /X}e^{i(S-R)pX}\text{ }dp.
\end{equation}

Solving the integral gives
\begin{eqnarray}
\hat{Q} &=&\QOVERD( ) {Xq}{2\pi }\sum_{\kappa }\sum_{R=-M}^{M}\sum_{S=-M}^{M}%
\left[
\begin{array}{c}
\hat{\alpha}_{4(M+R)+\kappa }^{\dagger }\hat{\alpha}_{4(M+S)+\kappa } \\
-\hat{\alpha}_{4(M+R)+2+\kappa }^{\dagger }\hat{\alpha}_{4(M+S)+2+\kappa }
\end{array}
\right]  \label{Ch5ChargePre} \\
&&\times \QOVERD( ) {1}{i(S-R)X}(e^{i(S-R)\pi }-e^{-i(S-R)\pi })  \notag \\
&=&\QOVERD( ) {q}{\pi }\sum_{\kappa }\sum_{R=-M}^{M}\sum_{S=-M}^{M}\left[
\begin{array}{c}
\hat{\alpha}_{4(M+R)+\kappa }^{\dagger }\hat{\alpha}_{4(M+S)+\kappa } \\
-\hat{\alpha}_{4(M+R)+2+\kappa }^{\dagger }\hat{\alpha}_{4(M+S)+2+\kappa }
\end{array}
\right] \QOVERD( ) {\sin ((S-R)\pi )}{(S-R)}.  \notag
\end{eqnarray}

Now, when $(S-R)\neq 0,$ the relation
\begin{equation}
\QOVERD( ) {\sin ((S-R)\pi )}{(S-R)}=0
\end{equation}
holds for all $S,R.$ However, when $(S-R)=0$ it becomes
\begin{equation}
\lim_{(S-R)=0}\QOVERD( ) {\sin ((S-R)\pi )}{(S-R)}=\pi
\end{equation}
as may be readily verified by Maclaurin expansion.

Clearly, the expression only has non-zero values when $R=S.$ So,
substituting this result into (\ref{Ch5ChargePre}) gives the final
expression for the charge operator
\begin{equation}
\hat{Q}=q\sum_{\kappa }\sum_{R=-M}^{M}\left[
\begin{array}{c}
\hat{\alpha}_{4(M+R)+\kappa }^{\dagger }\hat{\alpha}_{4(M+R)+\kappa } \\
-\hat{\alpha}_{4(M+R)+2+\kappa }^{\dagger }\hat{\alpha}_{4(M+R)+2+\kappa }
\end{array}
\right] .
\end{equation}

\bigskip

\subsection{Field Theory and CNOT}

\bigskip

Instead of the pregeometric ladder operators, it is both possible and useful
to write the expressions for the above Hamiltonian, momentum and charge in
terms of a set of more conventional quantum operators acting on the qubits.
One such set comprises of extended local unitary operators and the extended
CNOT operator defined\footnote{%
Repeated here from equation (\ref{Ch4CNOT}).} as
\begin{eqnarray}
\hat{C}_{(a,b)} &=&\hat{P}_{a}^{0}\otimes \hat{\sigma}_{b}^{0}+\hat{P}%
_{a}^{1}\otimes \hat{\sigma}_{b}^{1} \\
&=&\hat{P}_{a}^{0}\otimes (\hat{P}_{b}^{0}+\hat{P}_{b}^{1})+\hat{P}%
_{a}^{1}\otimes (\hat{Q}_{a}+\hat{Q}_{a}^{\dagger })  \notag
\end{eqnarray}
where the tensor product of identity operators $\hat{I}_{c},$ $c\neq a,b$
and $c=1,2,...,N^{\prime },$ in the extension has been omitted for brevity.
Of course, this definition could easily be extended even further, such that
the tensor product is also taken with the identity operators acting in every
other sub-space of the Universe's total Hilbert space $\mathcal{H}%
_{[1...N]}\supset \mathcal{H}_{[1...N^{\prime }]},$ but this is an
unnecessary amendment here.

As has been discussed previously, $\hat{C}_{(a,b)}$ acts on every qubit in
the Hilbert space $\mathcal{H}_{[1...N^{\prime }]},$ but only changes the
value of the qubit sub-state in sub-space $\mathcal{H}_{b}$ depending on the
value of the qubit sub-state in sub-space $\mathcal{H}_{a}.$

The motivation for choosing this particular set of operators is two-fold.
Firstly, such a possibility provides an immediate bridge between the work of
this chapter and the discussions of quantum computation in Chapter 6.
Secondly, and perhaps more importantly, this choice follows directly from
the suggestion of Feynman \cite{Feynman} that all of physics could in
principle originate from quantum computation, and then from the work of
Barenco \textit{et al} \cite{Barenco} that every qubit quantum computation
may be achieved by the use of local unitary operations and the CNOT gate.
Thus by amalgamating this second idea into the model proposed currently, the
result is demonstrated that fermionic field theory in momentum space can be
obtained from these standard quantum computational operators acting on
qubits defined on the pregeometric level. Feynman's prediction is hence
confirmed.\bigskip

In order to convert the equations for the Hamiltonian, momentum and charge
operators into this chosen set of operators, the goal would be to express
the sums of products of pregeometric ladder operators present in their
constructions in terms of local unitary operators and CNOT. By way of a
demonstration of how this can be achieved, consider a typical such sum of
products given by
\begin{equation}
\Pi _{RS}=\hat{\alpha}_{4(M+R)+\kappa }^{\dagger }\hat{\alpha}%
_{4(M+S)+\kappa }+\hat{\alpha}_{4(M+R)+2+\kappa }^{\dagger }\hat{\alpha}%
_{4(M+S)+2+\kappa }.
\end{equation}

For convenience in this example, it is possible to restrict attention to
just the first of these products. Furthermore, it is also possible to
re-label the sub-registers featured, and define the product $\Pi _{rs}$ as
\begin{equation}
\Pi _{rs}=\hat{\alpha}_{r}^{\dagger }\hat{\alpha}_{s}
\end{equation}
where $r=4(M+R)+\kappa $ and $s=4(M+S)+\kappa .$ Writing out the expression
in full extended notation, this product becomes
\begin{eqnarray}
\Pi _{rs} &=&\left( \hat{\sigma}_{1}^{3}\hat{\sigma}_{2}^{3}...\hat{\sigma}%
_{r-1}^{3}\hat{T}_{r}^{10}\hat{I}_{r+1}...\hat{I}_{N}\right) \left( \hat{%
\sigma}_{1}^{3}\hat{\sigma}_{2}^{3}...\hat{\sigma}_{s-1}^{3}\hat{T}_{s}^{01}%
\hat{I}_{s+1}...\hat{I}_{N}\right) \\
&=&\hat{I}_{1}\hat{I}_{2}...\hat{I}_{r-1}(\hat{T}_{r}^{10}\hat{\sigma}%
_{r}^{3})\hat{\sigma}_{r+1}^{3}...\hat{\sigma}_{s-1}^{3}\hat{T}_{s}^{01}\hat{%
I}_{s+1}...\hat{I}_{N}  \notag
\end{eqnarray}
using $\hat{\sigma}_{t}^{0}\equiv \hat{I}_{t}$ with the usual $SU(2)$
product algebra (\ref{Ch3pauli1}), and assuming $r<s$ without loss of
generality. So,
\begin{eqnarray}
\Pi _{rs} &=&(\hat{T}_{r}^{10}\hat{\sigma}_{r}^{3})\hat{T}_{s}^{01}\hat{I}%
_{1}\hat{I}_{2}...\hat{I}_{r-1}\hat{\sigma}_{r+1}^{3}...\hat{\sigma}%
_{s-1}^{3}\hat{I}_{s+1}...\hat{I}_{N}  \label{Ch5Pirs} \\
&=&\hat{T}_{r}^{10}\hat{T}_{s}^{01}\hat{I}_{1}\hat{I}_{2}...\hat{I}_{r-1}%
\hat{\sigma}_{r+1}^{3}...\hat{\sigma}_{s-1}^{3}\hat{I}_{s+1}...\hat{I}_{N}
\notag
\end{eqnarray}
using the assumed `rearrangement' property of the tensor product and
recalling the definition (\ref{Ch4 PAULI}) that $\hat{\sigma}_{r}^{3}=\hat{T}%
_{r}^{00}-\hat{T}_{r}^{11}.$

Now consider just the product of local transformation operators, $\hat{T}%
_{r}^{10}\hat{T}_{s}^{01}.$ The question becomes: what combinations of local
unitary operators and CNOT gates will give $\hat{T}_{r}^{10}\hat{T}_{s}^{01}$
as a result? To begin to answer this, consider the product of CNOT operators
$\hat{C}_{(r,s)}\hat{C}_{(s,r)}$ given by
\begin{eqnarray}
\hat{C}_{(r,s)}\hat{C}_{(s,r)} &=&\left( \hat{P}_{r}^{0}\hat{P}_{s}^{0}+\hat{%
P}_{r}^{0}\hat{P}_{s}^{1}+\hat{P}_{r}^{1}\hat{Q}_{s}+\hat{P}_{r}^{1}\hat{Q}%
_{s}^{\dagger }\right) \\
&&\times \left( \hat{P}_{r}^{0}\hat{P}_{s}^{0}+\hat{P}_{r}^{1}\hat{P}%
_{s}^{0}+\hat{Q}_{r}\hat{P}_{s}^{1}+\hat{Q}_{r}^{\dagger }\hat{P}%
_{s}^{1}\right)  \notag
\end{eqnarray}
where tensor product symbols are omitted and the notations $\hat{T}%
^{00}\equiv \hat{P}^{0},$ $\hat{T}^{11}\equiv \hat{P}^{1},$ $\hat{T}%
^{01}\equiv \hat{Q}$ and $\hat{T}^{10}\equiv \hat{Q}^{\dagger }$ used in
Section 6.1 have been adopted instead for clarity. By recalling the algebra
of transformation operators (\ref{Ch4ts}), this becomes
\begin{equation}
\hat{C}_{(r,s)}\hat{C}_{(s,r)}=\hat{P}_{r}^{0}\hat{P}_{s}^{0}+\hat{Q}_{r}%
\hat{P}_{s}^{1}+\hat{Q}_{r}^{\dagger }\hat{Q}_{s}+\hat{P}_{r}^{1}\hat{Q}%
_{s}^{\dagger }.  \label{Ch5Cprod}
\end{equation}

It is possible to multiply the product $\hat{C}_{(r,s)}\hat{C}_{(s,r)}$ with
unitary operators that act locally upon the individual qubits in sub-spaces $%
\mathcal{H}_{r}$ and $\mathcal{H}_{s}.$ Three such operators are $\hat{\sigma%
}_{r}^{3}\otimes \hat{\sigma}_{s}^{0},$ $\hat{\sigma}_{r}^{0}\otimes \hat{%
\sigma}_{s}^{3}$ and $\hat{\sigma}_{r}^{3}\otimes \hat{\sigma}_{s}^{3},$ and
these lead to the results
\begin{eqnarray}
\left( \hat{\sigma}_{r}^{3}\otimes \hat{\sigma}_{s}^{0}\right) \left[ \hat{C}%
_{(r,s)}\hat{C}_{(s,r)}\right] &=&\left( \hat{P}_{r}^{0}\hat{I}_{s}-\hat{P}%
_{r}^{1}\hat{I}_{s}\right) \left[ \hat{P}_{r}^{0}\hat{P}_{s}^{0}+\hat{Q}_{r}%
\hat{P}_{s}^{1}+\hat{Q}_{r}^{\dagger }\hat{Q}_{s}+\hat{P}_{r}^{1}\hat{Q}%
_{s}^{\dagger }\right]  \notag \\
&=&\hat{P}_{r}^{0}\hat{P}_{s}^{0}+\hat{Q}_{r}\hat{P}_{s}^{1}-\hat{Q}%
_{r}^{\dagger }\hat{Q}_{s}-\hat{P}_{r}^{1}\hat{Q}_{s}^{\dagger }
\label{Ch5CSr}
\end{eqnarray}
and
\begin{eqnarray}
\left( \hat{\sigma}_{r}^{0}\otimes \hat{\sigma}_{s}^{3}\right) \left[ \hat{C}%
_{(r,s)}\hat{C}_{(s,r)}\right] &=&\left( \hat{I}_{r}\hat{P}_{s}^{0}-\hat{I}%
_{r}\hat{P}_{s}^{1}\right) \left[ \hat{P}_{r}^{0}\hat{P}_{s}^{0}+\hat{Q}_{r}%
\hat{P}_{s}^{1}+\hat{Q}_{r}^{\dagger }\hat{Q}_{s}+\hat{P}_{r}^{1}\hat{Q}%
_{s}^{\dagger }\right]  \notag \\
&=&\hat{P}_{r}^{0}\hat{P}_{s}^{0}-\hat{Q}_{r}\hat{P}_{s}^{1}+\hat{Q}%
_{r}^{\dagger }\hat{Q}_{s}-\hat{P}_{r}^{1}\hat{Q}_{s}^{\dagger }
\label{Ch5CSs}
\end{eqnarray}
and
\begin{eqnarray}
\left( \hat{\sigma}_{r}^{3}\otimes \hat{\sigma}_{s}^{3}\right) \left[ \hat{C}%
_{(r,s)}\hat{C}_{(s,r)}\right] &=&\left( \hat{P}_{r}^{0}\hat{P}_{s}^{0}-\hat{%
P}_{r}^{0}\hat{P}_{s}^{1}-\hat{P}_{r}^{1}\hat{P}_{s}^{0}+\hat{P}_{r}^{1}\hat{%
P}_{s}^{1}\right)  \notag \\
&&\times \left[ \hat{P}_{r}^{0}\hat{P}_{s}^{0}+\hat{Q}_{r}\hat{P}_{s}^{1}+%
\hat{Q}_{r}^{\dagger }\hat{Q}_{s}+\hat{P}_{r}^{1}\hat{Q}_{s}^{\dagger }%
\right]  \notag \\
&=&\hat{P}_{r}^{0}\hat{P}_{s}^{0}-\hat{Q}_{r}\hat{P}_{s}^{1}-\hat{Q}%
_{r}^{\dagger }\hat{Q}_{s}+\hat{P}_{r}^{1}\hat{Q}_{s}^{\dagger }.
\label{Ch5CSrSs}
\end{eqnarray}

Now, adding (\ref{Ch5Cprod}) to (\ref{Ch5CSs}) and then subtracting (\ref
{Ch5CSr}) and (\ref{Ch5CSrSs}) gives $4\hat{Q}_{r}^{\dagger }\hat{Q}_{s},$
so it is evident that
\begin{eqnarray}
\hat{Q}_{r}^{\dagger }\hat{Q}_{s} &=&\frac{1}{4}\left( 1-\hat{\sigma}%
_{r}^{3}\otimes \hat{\sigma}_{s}^{0}+\hat{\sigma}_{r}^{0}\otimes \hat{\sigma}%
_{s}^{3}-\hat{\sigma}_{r}^{3}\otimes \hat{\sigma}_{s}^{3}\right) \hat{C}%
_{(r,s)}\hat{C}_{(s,r)} \\
&=&\frac{1}{4}\left[ \left( \hat{\sigma}_{r}^{0}-\hat{\sigma}_{r}^{3}\right)
\otimes \left( \hat{\sigma}_{s}^{0}+\hat{\sigma}_{s}^{3}\right) \right] \hat{%
C}_{(r,s)}\hat{C}_{(s,r)}.  \notag
\end{eqnarray}

Substituting this into (\ref{Ch5Pirs}) gives
\begin{equation}
\Pi _{rs}=\frac{1}{4}\left[ \left( \hat{\sigma}_{r}^{0}-\hat{\sigma}%
_{r}^{3}\right) \otimes \left( \hat{\sigma}_{s}^{0}+\hat{\sigma}%
_{s}^{3}\right) \right] \hat{C}_{(r,s)}\hat{C}_{(s,r)}\hat{\sigma}_{1}^{0}%
\hat{\sigma}_{2}^{0}...\hat{\sigma}_{r-1}^{0}\hat{\sigma}_{r+1}^{3}...\hat{%
\sigma}_{s-1}^{3}\hat{\sigma}_{s+1}^{0}...\hat{\sigma}_{N}^{0}
\end{equation}
with the usual interchangeability between $\hat{\sigma}_{m}^{0}$ and $\hat{I}%
_{m}.$\bigskip

Similarly, products of local transformation operators of the form $\hat{T}%
_{r}^{01}\hat{T}_{s}^{10}\equiv \hat{Q}_{r}\hat{Q}_{s}^{\dagger }$ may be
obtained from alternative products of CNOT and Pauli operators. Viz, from
\begin{eqnarray}
\hat{C}_{(s,r)}\hat{C}_{(r,s)} &=&\left( \hat{P}_{r}^{0}\hat{P}_{s}^{0}+\hat{%
P}_{r}^{1}\hat{P}_{s}^{0}+\hat{Q}_{r}\hat{P}_{s}^{1}+\hat{Q}_{r}^{\dagger }%
\hat{P}_{s}^{1}\right) \\
&&\times \left( \hat{P}_{r}^{0}\hat{P}_{s}^{0}+\hat{P}_{r}^{0}\hat{P}%
_{s}^{1}+\hat{P}_{r}^{1}\hat{Q}_{s}+\hat{P}_{r}^{1}\hat{Q}_{s}^{\dagger
}\right)  \notag \\
&=&\hat{P}_{r}^{0}\hat{P}_{s}^{0}+\hat{P}_{r}^{1}\hat{Q}_{s}+\hat{Q}_{r}\hat{%
Q}_{s}^{\dagger }+\hat{Q}_{r}^{\dagger }\hat{P}_{s}^{1}  \notag
\end{eqnarray}
it follows that
\begin{equation}
\left( \hat{\sigma}_{r}^{3}\otimes \hat{\sigma}_{s}^{0}\right) \left[ \hat{C}%
_{(s,r)}\hat{C}_{(r,s)}\right] =\hat{P}_{r}^{0}\hat{P}_{s}^{0}-\hat{P}%
_{r}^{1}\hat{Q}_{s}+\hat{Q}_{r}\hat{Q}_{s}^{\dagger }-\hat{Q}_{r}^{\dagger }%
\hat{P}_{s}^{1}
\end{equation}
with
\begin{equation}
\left( \hat{\sigma}_{r}^{0}\otimes \hat{\sigma}_{s}^{3}\right) \left[ \hat{C}%
_{(s,r)}\hat{C}_{(r,s)}\right] =\hat{P}_{r}^{0}\hat{P}_{s}^{0}+\hat{P}%
_{r}^{1}\hat{Q}_{s}-\hat{Q}_{r}\hat{Q}_{s}^{\dagger }-\hat{Q}_{r}^{\dagger }%
\hat{P}_{s}^{1}
\end{equation}
and
\begin{equation}
\left( \hat{\sigma}_{r}^{3}\otimes \hat{\sigma}_{s}^{3}\right) \left[ \hat{C}%
_{(s,r)}\hat{C}_{(r,s)}\right] =\hat{P}_{r}^{0}\hat{P}_{s}^{0}-\hat{P}%
_{r}^{1}\hat{Q}_{s}-\hat{Q}_{r}\hat{Q}_{s}^{\dagger }+\hat{Q}_{r}^{\dagger }%
\hat{P}_{s}^{1}
\end{equation}
so
\begin{eqnarray}
\hat{Q}_{r}\hat{Q}_{s}^{\dagger } &=&\frac{1}{4}\left[ 1+\hat{\sigma}%
_{r}^{3}\otimes \hat{\sigma}_{s}^{0}-\hat{\sigma}_{r}^{0}\otimes \hat{\sigma}%
_{s}^{3}-\hat{\sigma}_{r}^{3}\otimes \hat{\sigma}_{s}^{3}\right] \hat{C}%
_{(s,r)}\hat{C}_{(r,s)} \\
&=&\frac{1}{4}\left[ \left( \hat{\sigma}_{r}^{0}+\hat{\sigma}_{r}^{3}\right)
\otimes \left( \hat{\sigma}_{s}^{0}-\hat{\sigma}_{s}^{3}\right) \right] \hat{%
C}_{(s,r)}\hat{C}_{(r,s)}.  \notag
\end{eqnarray}

\bigskip

With results such as these, it is easy to see how it is possible to write
the Hamiltonian, momentum and charge operators in terms of local unitary
operators and the two-qubit CNOT gate, as expected and desired. Thus, the
emergence of physics from universal quantum computation acting upon
pregeometric qubit structure is shown.

\bigskip

\subsection{Discussion}

\bigskip

The aim of this chapter has been to demonstrate how quantum field theoretic
descriptions of physical particles might begin to emerge from the underlying
pregeometric structure proposed in this thesis. Whilst some success may
therefore be claimed from the point of view of fermions, due to a derivation
of emergent ladder, Hamiltonian, momentum and charge operators for particles
obeying the Dirac equation, a number of issues still remain and the overall
programme behind this work is far from being complete. These points are
highlighted now.\bigskip

Firstly, the observation is made that elementary particles are objects
possessing more than just spin. The fundamental fermions currently
understood in the Standard Model, namely quarks and leptons, are known to
possess an array of different physical characteristics in addition to
angular momentum, examples being colour, flavour, chirality, etc. A truly
complete theory of matter would therefore have to explain how each of these
degrees of freedom emerge from the sub-register picture. Indeed, this
problem was stated in its large-scale entirety by the original brief of the
chapter: how can the pregeometric description advocated in this thesis
account for the enormous variety of `properties' exhibited by
classical-looking objects?

An obvious direction to proceed therefore involves an extension of the
presented discussion of the Dirac field to multi-field theories
incorporating, say, colour and flavour degrees of freedom.

Exactly how this should best be accomplished remains a question for the
future, but it is however possible to speculate that the principles and
types of methods used in the previous section would not be wholly
inappropriate in an implementation of alternative fields. After all,
historically the theory of colour gauge symmetry grew from a foundation
based on an initial study of the Dirac field, so it might be expected that
any future pregeometric description of hadrons could equally be derived from
a set of underlying logic operations, just as the Dirac field was shown to
be in the present work. Of course, this is a viewpoint prejudiced by
tradition, but a consideration of this type of argument might at least
provide a reasonable starting point for more advanced models, or
alternatively perhaps suggest the need for a novel approach to the
problem.\bigskip

Following on from the above point regarding additional degrees of freedom,
it is noted that one other important property of an elementary particle is
its rest mass. In a fully quantum Universe this too would be expected to
have a pregeometric origin, an issue that is presumably related to how a
Higgs-type mechanism could emerge from the described sub-register picture.
However, this extension is actually more pertinent than the questions of the
origin of colour or flavour, because the mass of the particle was implicitly
assumed in the formalism of the above discussion, specifically as the
parameter $\mu $ introduced to arrive at the Hamiltonian. So, without
explicitly knowing the mechanism for generating mass in the quantum
Universe, a question then arises as to whether it is valid to make such an
insertion.

The introduction of $\mu ,$ however, may be justified on two grounds.
Firstly, physicists often ignore various degrees of freedom when discussing
particular effects. The conventional Dirac equation, for example, provides a
perfectly good description of spin-$%
%TCIMACRO{\UNICODE[m]{0xbd}}%
%BeginExpansion
{\frac12}%
%EndExpansion
$ particles of a given mass, even though the actual mechanism that produces
this mass is neglected\footnote{%
And indeed was totally unknown at the time of Dirac. Perhaps this is further
support for the argument that the development of quantum field theory in
terms of pregeometry could follow an `historical' route.}. It seems
reasonable, then, that a similar approximation is equally valid in a
pregeometric discussion, certainly at least as a first step towards a more
complete picture of field theory.

Secondly, the mass term $\mu $ was only used anyway to formulate the
Hamiltonian. So, even if its introduction does involve an element of
`cheating', the results found for the ladder, momentum and charge operators
still provide a valid description of fermionic objects.

Nevertheless, for a consistent and thorough understanding of physical
reality, an account of the pregeometric origin of mass is a further
necessary direction to take.

In addition, it is interesting to speculate how or whether any such
hypothetical mechanism might influence the types of causal set structure
exhibited by the state of the Universe as it changes its separability
through a sequence of jumps. The answer to this question would itself
provide useful insight into the origin of general relativity and the
apparent curvature of space by mass in the quantum Universe.

\bigskip

Perhaps the most obvious extension to an understanding of how fermionic
particles emerge from the sub-register picture involves asking the question
of how bosons might also.

This, however, immediately presents a difficulty. To demonstrate why,
consider a conventional fermionic ladder operator $\hat{a}_{F}^{\dagger }(p)$
that creates a particle of momentum $p$ from the vacuum $|0\rangle ,$ to
give the single particle state
\begin{equation}
\hat{a}_{F}^{\dagger }(p)|0\rangle =|1_{F}^{(p)}\rangle
\end{equation}
where the actual spin of the particle is temporarily ignored. As expected, $%
\hat{a}_{F}^{\dagger }(p)$ satisfies the usual relationship
\begin{equation}
\hat{a}_{F}^{\dagger }(p)\hat{a}_{F}^{\dagger }(p)|0\rangle =0
\end{equation}
with
\begin{equation}
\hat{a}_{F}^{\dagger }(p^{\prime })\hat{a}_{F}^{\dagger }(p)|0\rangle
=(1-\delta _{pp^{\prime }})|1_{F}^{(p^{\prime })}1_{F}^{(p)}\rangle
\end{equation}
and so on. The point is that the maximum occupancy of a given particle state
is one, as predicted for objects obeying Fermi-Dirac statistics.
Furthermore, these relations imply that the operator $\hat{a}_{F}^{\dagger
}(p)$ may be seen, in some ways, as being analogous to a transition operator
between the qubit states $|0\rangle _{m}$ and $|1\rangle _{m},$ because
these too follow the product algebra of fermions. Of course, it was this
type of association that formed the basis for the work of the previous
section.

Consider instead, however, a conventional bosonic ladder operator $\hat{a}%
_{B}^{\dagger }(p)$ that creates a boson of momentum $p$ from the vacuum $%
|0\rangle ,$ such that
\begin{equation}
\hat{a}_{B}^{\dagger }(p)|0\rangle =|1_{B}^{(p)}\rangle
\end{equation}
where $|1_{B}^{(p)}\rangle $ represents a single boson particle state.

Now, because $\hat{a}_{B}^{\dagger }(p)$ governs the creation of bosons, it
is taken to obey the relationships
\begin{eqnarray}
\hat{a}_{B}^{\dagger }(p)\hat{a}_{B}^{\dagger }(p)|0\rangle &=&\hat{a}%
_{B}^{\dagger }(p)|1_{B}^{(p)}\rangle =|2_{B}^{(p)}\rangle \\
\hat{a}_{B}^{\dagger }(p)\hat{a}_{B}^{\dagger }(p)\hat{a}_{B}^{\dagger
}(p)|0\rangle &=&|3_{B}^{(p)}\rangle  \notag \\
\vdots &=&\vdots  \notag \\
(\hat{a}_{B}^{\dagger }(p))^{n}|0\rangle &=&|n_{B}^{(p)}\rangle  \notag
\end{eqnarray}
that is, the theory allows multi-occupancy of states: the state $%
|n_{B}^{(p)}\rangle $ contains $n$ identical bosons. Moreover, the
statistics are assumed to be valid for all $n$ up to $n=\infty .$

But, it is difficult to envisage how such multi-occupancy states could be
incorporated into the binary `on/off' logic of qubits. Not only that, but it
is also difficult to imagine how infinite occupancies could arise at all
from any pregeometric structure based on finite dimensional Hilbert spaces.
Ultimately, then, these two comments might perhaps lead to the criticism
that only half of particle physics could ever emerge from the subregister
model suggested.

Whilst the origin of bosons in the quantum Universe remains an unresolved
issue, a number of points may be made against the above conclusion, and
which should therefore provide a guide for future research. For example, it
is noted that in the Standard Model of particle physics the four fundamental
bosons may not so much be interpreted as observable entities, but should
perhaps instead be viewed in terms of representations of interactions. In
other words, it is remarked that these bosons are not necessarily directly
observed \textit{per se}, but that it is only their effect on the fermions
comprising physical matter that is actually seen. Thus, the apparent
existence of bosonic particles in conventional physics could perhaps be
viewed simply as an artifact of a misunderstanding of how fermions interact
with one another, based itself on the mistake of assuming that fermionic
particles are representative of the most fundamental level of reality. Real,
physical bosonic particles such as photons, gluons and gravitons may not
actually exist, or at least not in the sense that they are conventionally
assumed to.

So, the suggestion here is that a discussion of bosons may therefore
naturally be postponed until the questions of Section 6.2 surrounding
pregeometric interactions, reactions and measurements are better understood.
In short, the argument is that it may never be possible to understand the
ultimate nature of the fundamental forces and the apparent existence of
bosons simply by devising ever more complicated particle theories and
experiments (as conventional physics attempts to do), but will instead only
be resolved when a fuller cognition of the issues present at the very heart
of quantum mechanics' measurement problem has been achieved. Of course, an
admission is also made here that an element of ``sweeping under the carpet''
may appear to be present in this argument.\bigskip

Alternatively, another potential way around the highlighted problem might be
to relax the `infinity condition' of bosonic occupancy. As has been done
consistently throughout this work, infinities have often been removed in the
quantum Universe paradigm by assuming finite, but very large, degrees of
freedom. This may then be the solution for the present case involving
bosons, and is reinforced by finite energy arguments in favour of
restricting the number of particles of given energy-momentum to non-infinite
values. Paraphrasing this last point, and echoing the words of Feynman \cite
{Feynman}, the question is again asked as to whether it is really physically
possible to have an infinite number of positive energy particles in a
Universe of finite size and mass.

So from this perspective, and as speculated upon by Lu and Widar \cite{Wu},
it might be suggested that bosons should instead really obey a rule of the
form
\begin{equation}
(\hat{a}_{B}^{\dagger }(p))^{q}|0\rangle =|q_{B}^{(p)}\rangle
\end{equation}
but
\begin{equation}
(\hat{a}_{B}^{\dagger }(p))^{q+1}|0\rangle =0
\end{equation}
where $q\gg 1.$ In the language of Green \cite{Green} and Greenberg \cite
{Greenberg}, bosons would then be viewed as parafermions of order $q,$ and
would be governed by ladder operators of the form
\begin{equation}
\hat{a}_{B}^{\dagger }(p)|n\rangle =|(n+1)_{B}^{(p)}\rangle
\label{Ch5ParaCre}
\end{equation}
for $0\leq n\leq (q-1),$ and
\begin{equation}
\hat{a}_{B}(p)|n\rangle =|(n-1)_{B}^{(p)}\rangle  \label{Ch5ParaAnn}
\end{equation}
for $1\leq n\leq q,$ with
\begin{equation}
\hat{a}_{B}(p)|0\rangle =0
\end{equation}

In principle, as long as $q$ is sufficiently large such that the maximum
occupancy $|q_{B}^{(p)}\rangle $ is never actually reached physically, the
mechanism would, to all intents and purposes, appear equivalent to the
conventional theory of bosons that allows any number of particles to be in
the same state.

Of course, this solution is still highly speculative, and a great deal of
work is required to investigate whether the usual laws applying to bosons
may be extended to such high-order parafermions.\bigskip

Emergent operators possessing the relations inherent to (\ref{Ch5ParaCre})
and (\ref{Ch5ParaAnn}) may not actually be able to arise from a qubit
subregister structure, especially if the procedure described in Section 7.2
is used. To illustrate why this might be the case, recall that qubit Hilbert
(sub-)spaces are spanned by just two orthogonal basis states, and can hence
only exhibit two possible transition operators relating them. Furthermore,
these two possible transition operators were taken above to be the ultimate
origin of ladder operators in momentum space that governed physical particle
states that could only exist in either one of two possible occupancies (i.e.
fermions). Thus, it is difficult to see how this result\ could be
generalised to account\ for emergent ladder operators that cause apparent
multi-occupancy of states, if these emergent operators are a direct
consequence of pregeometric transition operators confined to act only in
two-level spaces.

Consequently, in order to incorporate higher order occupancies it might be
necessary to remove the constraint of using two-level, qubit Hilbert
sub-spaces and instead consider subregisters of higher dimension. Of course,
this removal may be justified anyway by recalling that qubit spaces were
only ever employed in the first place because they provided the simplest
starting point for the ensuing discussion, and not because of any physical
constraints.

So, assuming that the $m^{th}$ subregister $\mathcal{H}_{m}^{(d)}$ of the
Hilbert space $\mathcal{H}_{\varphi }$ containing the field $\varphi $ is
not a qubit space but is of dimension $d,$ then $\mathcal{H}_{m}^{(d)}$ may
be spanned by an orthonormal basis set $\mathcal{B}_{m}^{(d)}$ defined as $%
\mathcal{B}_{m}^{(d)}\equiv \{|i\rangle _{m}:i=0,1,...,(d-1)\}.$ In this
case, it is now possible to define transformation operators between the
bases of $\mathcal{H}_{m}^{(d)}$ in an analogous manner to that used for
qubits. That is, an operator of the form
\begin{equation}
\lbrack \hat{T}^{(d)}]_{m}^{(i+1)i}=|i+1\rangle _{mm}\langle i|
\end{equation}
acting on the state $|i\rangle _{m}\in $ $\mathcal{H}_{m}^{(d)}$ changes it
to
\begin{equation}
\lbrack \hat{T}^{(d)}]_{m}^{(i+1)i}|i\rangle _{m}=|i+1\rangle _{m}
\end{equation}
for $i=0,1,...,(d-2),$ whereas an operator of the form
\begin{equation}
\lbrack \hat{T}^{(d)}]_{m}^{i(i+1)}=|i\rangle _{mm}\langle i+1|
\end{equation}
acting on the state $|i+1\rangle _{m}\in $ $\mathcal{H}_{m}^{(d)}$ changes
it to
\begin{equation}
\lbrack \hat{T}^{(d)}]_{m}^{i(i+1)}|i+1\rangle _{m}=|i\rangle _{m}.
\end{equation}

Of course, due to the orthogonality of the bases, any operator $[\hat{T}%
^{(d)}]_{m}^{kj}$ acting on the basis state $|i\rangle _{m}$ gives
\begin{equation}
\lbrack \hat{T}^{(d)}]_{m}^{kj}|i\rangle _{m}=0
\end{equation}
unless $j=i.$ So, it is therefore useful to define `general ladder
operators' such as $\hat{A}_{m}^{(d)}$ and $\hat{A}_{m}^{(d)\dagger }$
acting in the space $\mathcal{H}_{m}^{(d)};$ assuming that the current state
contained in $\mathcal{H}_{m}^{(d)}$ is in one of the $d$ basis states $%
|i\rangle _{m},$ then the operator $\hat{A}_{m}^{(d)}$ given by
\begin{equation}
\hat{A}_{m}^{(d)}\equiv \lbrack \hat{T}^{(d)}]_{m}^{01}+[\hat{T}%
^{(d)}]_{m}^{12}+...+[\hat{T}^{(d)}]_{m}^{(d-2)(d-1)}
\end{equation}
`lowers' it to $|i-1\rangle _{m},$ for $i\geq 1,$ whereas the general
operator $\hat{A}_{m}^{(d)\dagger }$ of the form
\begin{equation}
\hat{A}_{m}^{(d)\dagger }\equiv \lbrack \hat{T}^{(d)}]_{m}^{10}+[\hat{T}%
^{(d)}]_{m}^{21}+...+[\hat{T}^{(d)}]_{m}^{(d-1)(d-2)}
\end{equation}
may be used to `raise' it from this state $|i\rangle _{m}$ to $|i+1\rangle
_{m},$ for $i\leq (d-2).$ Clearly, $\hat{A}_{m}^{(d)\dagger }$ is behaving
analogously to the pregeometric creation operator defined previously for
qubits, whilst the generalisation $\hat{A}_{m}^{(d)}$ is evidently related
to the qubit annihilation operator.

Such higher dimensional subregisters and ladder operators may play an
important part in accounting for the large number of fields, both fermionic
and bosonic, familiar in the emergent world of high-energy physics.

\bigskip

As a final comment to this chapter, it is observed that the treatment given
in Section 7.2 is for a free-field theory. However, almost all `interesting'
physics, and certainly that occurring in a real collider experiment, arises
out of the interactions existing between particles. Moreover, it is also
worth recalling that is the actual results of the interactions, and hence
ultimately the observations of the particles, that are considered the only
physically real phenomena in quantum mechanics. It is therefore an important
question to ask exactly how an interacting field theory could emerge from
the underlying structure of the proposed paradigm.

Of course, this issue involves a huge research program in its own right.
From the pregeometric point of view advocated in this thesis, the question
involves the arduous task of describing the mechanism by which information
may be exchanged and extracted from the emergent field theory picture, and
hence relies upon a full integration of the principles introduced in this
and the preceding chapter. Putting this in perspective, the question could
equally be phrased as analogous to how the measurement problem of standard
quantum mechanics may be reconciled with the awesome formalism of
conventional quantum field theory.

This point may be continued. One eventual goal of future research would be
to apply the types of ideas presented here to describe real collider physics
experiments from the point of view of the underlying pregeometric structure.
So, further to the above question regarding the incorporation of the work of
the previous chapter into the results of the current one, such an
investigation would then require non-trivial extensions in order to explain
how:

\begin{enumerate}
\item  large collections of pregeometric particles organise themselves into
the `elementary' particles familiar to the Standard Model;

\item  how these elementary particles interact and organise themselves to
form atoms and molecules;

\item  how these then accumulate into large, semi-classical looking objects,
some of which are called human beings;

\item  and how these scientists can then manipulate other large collections
of emergent objects such that they resemble apparatus, laboratories,
isolated particles etc., and then perform experiments with them.
\end{enumerate}

Point $``4."$ itself incorporates an enormous number of different questions,
such as how humans gain the illusion of free-will, and hence why particular
experiments are actually performed. This is particularly important in the
context of a collider physics experiment, because certain particles may only
appear to exist when they are being looked for. Indeed, whilst this last
comment may appear meta-physical, it should be recalled that it is a central
tenet of the principles of quantum measurement theory, and follows naturally
from the conclusion of Wheeler \cite{Wheeler1} that no phenomenon can really
be said to exist independently of observation.\bigskip

The comments $``1."$ to $``4."$ also highlight the importance of the role of
observables in the above dynamics: after all, it is these that are
responsible for the resulting states. As before, the conclusion is again
that these operators must be selected very carefully if large, separate
groups of factors can form and persist over enormous numbers of jumps, such
that classical-looking collider experiments may be carried out on the
emergent scale.

Furthermore, and as has been discussed previously, recall that from an
endo-physical point of view the process of a scientist performing an
experiment on a subject is equivalent to one part of the Universe's state
appearing to perform an experiment on another part of it. Moreover, in a
fully quantum Universe free from external influences, this implies a very
careful and intricate interplay between the state representing the
scientists and subjects, and the Hermitian operator used to develop it. How
this type of mechanism might work is therefore also an important question.

It is this interplay that is the focus of the final chapter.

\bigskip \newpage

\section{The Developing Quantum Universe}

\renewcommand{\thefigure}{8.\arabic{figure}} \setcounter{figure}{0} %
\renewcommand{\theequation}{8.\arabic{equation}} \setcounter{equation}{0} %
\renewcommand{\thetheorem}{8.\arabic{theorem}} \setcounter{theorem}{0}

\bigskip

Three of the major conclusions that have played an important role in the
paradigm proposed in this thesis are the following:

\begin{enumerate}
\item  The Universe is a fully quantum system, and as such may be
represented by a pure state $\Psi _{n}$ in a Hilbert space $\mathcal{H}%
^{(D)} $ of enormous dimension $D,$ where this vector is one of the $D$
orthonormal eigenstates of the Hermitian operator $\hat{\Sigma}_{n}$ chosen
by the dynamics;

\item  The Universe is developed by testing its state with the Hermitian
operator $\hat{\Sigma}_{n+1}.$ It consequently jumps to the new state $\Psi
_{n+1};$

\item  The Universe is `closed' in the sense that it is self-contained,
because by definition the Universe contains everything. Thus, there cannot
be any notion of an external observer.
\end{enumerate}

In conventional quantum mechanics, however, these points might appear
somewhat contradictory. In laboratory quantum mechanics, a wavefunction (for
example a single, free electron) may be developed by testing it with a given
Hermitian operator, which, in general, collapses the state into one of its
eigenvectors. Moreover, these\ Hermitian operators represent the observables
of the system, and as such denote the various physical tests that a
scientist may perform on the quantum subject.

Now, in conventional quantum mechanics the choice of the test is usually
made by the physicist, and is hence often assumed to be a result of the
free-will processes of a semi-classical observer. So, and perhaps\ most
importantly, these processes are generally assumed to be occurring
externally to the quantum system under investigation.\bigskip

Herein lies the problem. If the quantum Universe does not possess observers
external to it, yet is governed by state reduction dynamics according to
Hermitian operators $\hat{\Sigma}_{n+1},$ the question remains as to how
these tests are actually chosen.

In other words, if there are no external agents to decide the Universe's
fate, and if the existence of the states $\Psi _{n},$ $\Psi _{n+1},$ $\Psi
_{n+2},...$ may be explained as being the results of the operators $\hat{%
\Sigma}_{n},$ $\hat{\Sigma}_{n+1},$ $\hat{\Sigma}_{n+2},...,$ what mechanism
accounts for, governs, or explains the choice of the these tests?
Specifically, if the actual state $\Psi _{n}$ of the Universe depends
(albeit stochastically) upon the choice of test, how is this operator $\hat{%
\Sigma}_{n}$ actually chosen? Ultimately, then, in the case where the
quantum state encompasses the entire Universe, what takes the place of the
external observer familiar to conventional quantum mechanics? Thus, how may
the traditional exo-observer be replaced by any sort of endo-physical
equivalent?

It is these issues that are addressed in this chapter.

\bigskip

Before beginning to see how these questions might be answered, however, a
number of points should be noted.

Firstly, if the entire Universe is described by quantum principles, yet
cannot support any sort of external observers, the conclusion must be that
it is somehow able to prepare, evolve and test itself. Moreover, the result
of this test could then be associated with the preparation of a new state,
which could itself subsequently be evolved and tested, in an automatic
process. Thus, and as has been discussed a number of times throughout this
work, the Universe could ultimately be envisaged to be a giant,
self-developing \textit{quantum\ automaton}.

Overall, then, the suggestion is that the Universe itself somehow chooses a
sequence of tests $\hat{\Sigma}_{n},$ $\hat{\Sigma}_{n+1},$ $\hat{\Sigma}%
_{n+2},...$ to test its state with.\bigskip

Secondly, and as has also been discussed a number of times, the exact way
the Universe develops appears to be highly constrained. Equivalently,
therefore, this statement implies that whatever mechanism is actually
responsible for choosing the operator used for the state's development, it
must itself be highly constrained. The existence of continuous looking
space, the nature of particle physics experiments, and the apparent
persistence of semi-classical objects all support this conclusion.

However, the actual nature of these observations suggests something else
about whatever mechanism is responsible for choosing operators. Namely,
since the Universe appears to look so very similar from one stage to the
next, part of the mechanism must be responsible for ensuring that this is
the case. For example, the mechanism might be such that the Universe
examines, somehow, its current state $\Psi _{n},$ and then only chooses
operators $\hat{\Sigma}_{n+1}$ that have an eigenstate that is `almost
identical' to $\Psi _{n}$ in some sense. Thus, and by the usual rules of
probability amplitudes, it would then be highly likely that the next state $%
\Psi _{n+1}$ is virtually the same as $\Psi _{n},$ especially when the
enormous dimensionality of the Hilbert space of $\Psi _{n}$ is taken into
account.

Continuing, if this process repeats for all $n,$ the overall result would
therefore be of a Universe that appears to change only very slightly from
one jump to the next, such that over a large number of steps it could appear
to evolve smoothly and continuously, just as appears to be the situation in
physics.

Such a mechanism would thus be an example of a `\textit{self-referential}'
system, because the dynamics are such that the Universe examines itself
before deciding upon a choice of test. Specifically, in the above example,
the selection of $\hat{\Sigma}_{n+1}$ depends somehow upon the state $\Psi
_{n}.$\bigskip

In fact, the suggestion that the Universe might actually develop in this
self-referential manner is reinforced further by the fact that human
physicists, who are themselves just parts of the Universe's state, do appear
to be able to manipulate and test the sub-states around them. In physics,
the next state of the Universe does seem to depend on the current state,
because scientists do appear able to perform physical experiments, and
consequently `determine' future states of reality (within the limits set by
quantum probability, of course). In essence, because the initial
wavefunction of a subject, and the outcome of any laboratory quantum
experiment upon it, are both viewed in the proposed paradigm as just factors
of states of the Universe, physicists must actually be changing the overall
state of the entire Universe whenever they perform an experiment.

Furthermore, in fact, the physicist, the apparatus, and the laboratory
should themselves all really be viewed simply as groups of factors of the
Universe's state. So, because the factor that represents a piece of physical
apparatus does appear to be able to examine the sub-state representing the
sample under investigation, such an experiment should really be interpreted
as one part of the Universe's state being developed according to another
part of it. The conclusion must therefore be that the real, physical
Universe is developing self-referentially.\bigskip

As an aside and an illustration, note that an example of a self-referential
system is a modern computer. Many computers possess software that is able to
run diagnostic checks upon themselves; the software may examine the internal
state of the machine, and `decide' upon a course of action depending on
which particular configuration the computer is currently in. This decision
may be based entirely upon the current state of the computer according to
the `rules' of action programmed into the software.

In fact, in the context of the current work a better example may involve a
quantum computer, but the point remains the same. To continue the parallel
drawn throughout his work, the development of the Universe is therefore
again viewed as an enormous quantum computation, such that overall the
Universe is envisaged to be an enormous, self-referential, quantum automaton.

\bigskip

The structure of this chapter is as follows. Firstly, definitions will be
given for some of the different types of way in which a quantum system free
from external agents may be able to decide upon a choice of operator. For
the reasons given above, particular attention will then be given to the
self-referential mechanisms in which the development of the universe depends
somehow upon its current state. The different sorts of such `state
self-referential' mechanisms will then be discussed, and examples given in
order to demonstrate and explain some of the issues surrounding the dynamics
that result.

Many of the examples will take the form of `toy-universe' models that are
represented by states in Hilbert spaces of low dimension, with the general
assumption being made that the reduction in complexity of these models from
the real Universe\footnote{%
In general in the following, a capitalised `Universe' will indicate The
Real, physical Universe, whereas a lower case `universe' will imply the
low-dimensional, toy-model case. The context of the word should nevertheless
make it clear anyway which U/universe is being discussed.} has not been made
at the expense of discarding any overriding physical principles. This is
analogous to how similar such low dimensional examples have been used
throughout this thesis to illustrate various points when it has been
necessary to not become overwhelmed by the enormous number of degrees of
freedom inherent in the real, physical situation.

\bigskip

\subsection{Types of Development}

\bigskip

The dynamics of the developing Universe are, at least in principle, fairly
simple.

At the $n^{th}$ stage of the Universe's development, its state may be
represented by the unique vector $\Psi _{n}.$ This wavefunction may be
`tested' by some non-degenerate Hermitian operator $\hat{\Sigma}_{n+1},$ or
might first be `evolved' with some sort of unitary operator $\hat{U}_{n},$
i.e. $\Psi _{n}\rightarrow \Psi _{n}^{\prime }=\hat{U}_{n}\Psi _{n},$ which
is equivalent to a `rotation' of the vector within its Hilbert space, before
then being `tested' by $\hat{\Sigma}_{n+1}.$ Moreover, this testing process
is irreversible, and the state $\Psi _{n}$ (or $\Psi _{n}^{\prime })$
consequently collapses to a next state $\Psi _{n+1},$ which is one of the
eigenstates of $\hat{\Sigma}_{n+1}.$ Thus, the development of the universe
is a discontinuous process, with jumps from one state to the next occurring
due to state reduction. The exact forms of the operators $\hat{U}_{n}$ and $%
\hat{\Sigma}_{n+1}$ are dictated by the Rules governing the system.

In the following, various sets of Rules are investigated, where the aim is
to generate a dynamics that develops the Universe in an automatic way. A
goal of this chapter is therefore to examine some of the different
mechanisms by which a decision regarding the choice of next operator $\hat{%
\Sigma}_{n+1}$ may be made, without appealing to any sort of external
observer with free-will.\bigskip

Now, in the paradigm proposed in this thesis, a Stage $\Omega _{n}$ of the
Universe is parameterised by (amongst others) a state $\Psi _{n},$ the
operator $\hat{\Sigma}_{n}$ of which $\Psi _{n}$ is an eigenstate, and the
Rules $R_{n}$ governing the dynamics. Of course, the presence of an operator
$\hat{\Sigma}_{n}$ equivalently implies the presence of a unique basis set $%
\frak{B}_{n}^{(D)}$ of orthonormal vectors\footnote{%
Noting that the reverse of this is not strictly true: a given basis set $%
\frak{B}_{n}^{(D)}$ does not specify a unique operator $\hat{\Sigma}_{n},$\
but instead describes an entire equivalence class.}, which spans the Hilbert
space $\mathcal{H}^{(D)}$ of $\Psi _{n}$ and comprises of the $D$
eigenstates of $\hat{\Sigma}_{n}.$

The general question of this chapter then becomes: how can some, all, or
none of these parameters be used to decide what happens next? In other
words, given a state $\Psi _{n}$ and an operator $\hat{\Sigma}_{n},$ what
Rules $R_{n}$ could govern the mechanism for generating\ $\Psi _{n+1}?$ In
short, what are the different types of mechanism that could relate the
present state $\Psi _{n}$ to the next state $\Psi _{n+1}?$\bigskip

In fact, it turns out that there are a number of different types of Rule
that could be used to determine a state's development. One possible such
list of types of mechanism is now introduced, with its members described in
turn. The list is written in order of increasing (intuitive) complexity:

\begin{enumerate}
\item[Type 0:]  \textit{Free Will}. The state is developed by an operator
chosen by an external observer.

\item[Type I:]  \textit{Evolution}. $\Psi _{n}$ is mapped to $\Psi _{n+1}$
by continuous evolution, say by an unitary function $\mathit{U},$\ such that
$\Psi _{n}\rightarrow \Psi _{n+1}=\Psi _{n}^{\prime }=\mathit{U}\Psi _{n}.$

\item[Type II:]  \textit{Deterministic Choice}. The Rules $R_{n}$ select an
operator $\hat{\Sigma}_{n+1}$ by some sort of deterministic algorithm
without reference to the current state or last operator. Thus, the choice of
operator $\hat{\Sigma}_{n+1}$ does not depend in any way upon $\Psi _{n}$ or
$\hat{\Sigma}_{n}.$

\item[Type IIa:]  \textit{Probabilistic Choice}. The Rules $R_{n}$ select an
operator $\hat{\Sigma}_{n+1}$ according to some sort of probabilistic
algorithm, without reference to the current state or last operator. Thus, $%
\hat{\Sigma}_{n+1}=\hat{A}$ with probability $P_{A},$ whilst $\hat{\Sigma}%
_{n+1}=\hat{B}$ with probability $P_{B},$ etc., where the various
probabilities sum to unity.

\item[Type III:]  \textit{State Self-Referential}. The Rules dictate that
the algorithm used to generate the next operator $\hat{\Sigma}_{n+1}$ (or
equivalently $\frak{B}_{n+1})$ refers, somehow, to the vector $\Psi _{n}.$
Thus, the choice of the next operator depends on the current state.

\item[Type IIIa:]  \textit{Probabilistic State Self-Referential}. As with
Type $III,$ the Rules still dictate that the algorithm used to generate the
next operator $\hat{\Sigma}_{n+1}$ (or equivalently $\frak{B}_{n+1})$ refers
to the current state $\Psi _{n},$ but there is now an element of probability
in the algorithm. For instance, a given state $\Psi _{n}$ could imply a
probabilistic choice of one of a number of possible operators, or
alternatively, the way in which the next operator $\hat{\Sigma}_{n+1}$
depends on the state $\Psi _{n}$ could be probabilistic. As a schematic
example of this last possibility, given two `functions' $f$ and $g$ defined
somehow by the Rules, it could follow that $\hat{\Sigma}_{n+1}=f(\Psi _{n})$
with probability $P_{f},$ whereas $\hat{\Sigma}_{n+1}=g(\Psi _{n})$ with
probability $P_{g},$ with $P_{f}+P_{g}=1.$

\item[Type IV:]  \textit{Basis Self-Referential}. The next operator $\hat{%
\Sigma}_{n+1}$ is generated in a way that makes reference to the current
basis, such that in effect $\hat{\Sigma}_{n+1}$ depends somehow on $\hat{%
\Sigma}_{n}$ (or equivalently $\frak{B}_{n+1}$ depends somehow on $\frak{B}%
_{n}).$

\item[Type IVa:]  \textit{Probabilistic Basis Self-Referential}. The choice
of the next operator depends probabilistically on the current basis. As
analogously to the Type $III$ case, a given basis $\frak{B}_{n}$ could imply
a number of potential bases $\frak{B}_{n+1},$ or there could alternatively
perhaps be a number of different ways in which the basis $\frak{B}_{n+1}$
depends on the basis $\frak{B}_{n},$ with the actual choice in both
circumstances made probabilistically.

\item[Type V:]  \textit{Fully Self-Referential}. The next operator $\hat{%
\Sigma}_{n+1}$ is generated in a way that makes reference to the current
basis $\frak{B}_{n}$ \textit{and} whichever of its eigenvectors was actually
chosen to be the current state $\Psi _{n},$ \ Thus, the choice of the next
operator depends deterministically on both $\Psi _{n}$ and $\frak{B}_{n}.$

\item[Type Va:]  \textit{Probabilistic Fully Self-Referential}. The choice
of the next operator depends probabilistically on the current basis \textit{%
and} whichever of its eigenvectors was actually chosen to be the current
state $\Psi _{n}.$ As before, this probabilistic algorithm could take a
number of forms, one schematic example of which would be if $\hat{\Sigma}%
_{n+1}=F_{n}(\Psi _{n},\hat{\Sigma}_{n})$ with probability $P_{F}$ whereas $%
\hat{\Sigma}_{n+1}=G_{n}(\Psi _{n},\hat{\Sigma}_{n})$ with probability $%
P_{G},$ where $F_{n}$ and $G_{n}$ are `functions' defined in the Rules, and $%
P_{F}+P_{G}=1.$
\end{enumerate}

Of course, further Types are possible if more parameters are specified. For
example, if the current Stage $\Omega _{n}$ also contains information
regarding previous states $\Psi _{n-1},$ $\Psi _{n-2},...$ it would
additionally be possible to consider higher order `Historic' Rules, such as
an `Historic Type $III$ mechanism' in which the choice of operator $\hat{%
\Sigma}_{n+1}$ somehow depends on $\Psi _{n}$ \textit{and} $\Psi _{n-1}$
\textit{and}... etc. Variants on this theme may also be imagined, in obvious
ways, for this and other Types by including or excluding particular
permutations of the parameters.\bigskip

This chapter focuses primarily on Type $III$ and $IIIa$ mechanisms, for the
reasons given now.

Firstly, any Type $0$ mechanism is clearly unsatisfactory for the
development of a Universe free from external observers. It is also wrong
from the epistemological viewpoint, in which every phenomenon, including
free will, is ultimately hoped to be explainable from a consistent set of
underlying physical laws.

The Type $I$ mechanisms are effectively the same as decoherence and Many
Worlds type dynamics. They would hence be accompanied by all the problems
and contortions associated with these. So, since such difficulties are hoped
to be avoided, not least for the reasons outlined in Section 4.3, the Type $%
I $ mechanisms will be neglected in the following discussions.

Type $II$ and $IIa$ mechanisms are indicative of the current state of play
in conventional quantum mechanics, and acknowledge the fact that although
physicists are able to obtain accurate answers to quantum questions (via
probability amplitudes etc.), they cannot say why these questions were posed
in the first place. In the language of causal sets, the causal set generated
over a series of stages by operators chosen according to Type $II$ Rules
would be completely independent of the causal set generated by the states
over the same period. Thus, in Type $II$ and $IIa$ models it is difficult to
see how the very finely tuned constraints discussed throughout this thesis
could be achieved, because the mechanisms do not possess any of the
`feed-back' processes that appear necessary for the emergence of classical
looking physics.

Lastly, although Types $IV,$ $IVa,$ $V$ and $Va$ do, in principle, describe
possible Rules for the development of the physical Universe, they are also
anticipated to give rise to mechanisms that are considerably more
complicated than the Type $III$ cases. In short, the dynamics of systems
based upon Rules in which the next operator $\hat{\Sigma}_{n+1}$ depends
upon the current state $\Psi _{n}$ and/or the current basis $\frak{B}_{n}$\
are expected to be far richer than those in which the next operator depends
only upon the current state, and as such a full investigation of the Types $%
IV,$ $IVa,$ $V$ and $Va$ will be left as a necessary avenue of inquiry for
future research.\bigskip

Type $III$ and $IIIa$ Rules hence provide a suitable starting point, useful
to investigate some of the ways in which the quantum Universe could develop.
The central focus of this chapter is therefore to investigate operator
selection mechanisms that permit the Universe to develop self-referentially
according to the current state.

To this end, the description of a Type $III$ mechanism may now be expanded.
Specifically, if the quantum state of the Universe after $n$ jumps is
represented by $\Psi _{n},$ then a state self-referential mechanism is
defined as one in which if $\Psi _{n}$ has one particular `property' then it
is tested with one particular operator (say, $\hat{\Sigma}_{n+1}=\hat{X}%
_{n+1}),$ but if it has another `property' it is instead tested with a
different operator (i.e. by $\hat{\Sigma}_{n+1}=\hat{Y}_{n+1}).$ And so on,
including of course what happens if it instead has a third, or a fourth, or
a..., particular `property'. Of course, the Rules must be sufficiently well
defined so that whatever the current state $\Psi _{n}$ may be, or whatever
`properties' it might have, it will definitely lead to a certain and unique
selection of the next operator $\hat{\Sigma}_{n+1},$ for all $n.$

Thus, in such state self-referential mechanisms, the way in which the
universe develops depends entirely upon which state it is in. Clearly, this
is analogous to the type of computer described earlier that is able to
`examine' its state in order to decide upon the next course of action.
Again, the physical Universe may be compared in this fashion to a type of
self-examining quantum computer.\bigskip

In fact, it is even possible to imagine extending the above definitions, and
consider a universe in which if $\Psi _{n}$ has one particular `property'
then it is \textit{evolved} in one way (say, by $\hat{U}_{n}=\hat{u}%
_{n}^{A}),$ but if it has another property it is instead evolved in a second
way (i.e. by $\hat{U}_{n}=\hat{u}_{n}^{B}).$ So, in these universes it is
the next unitary operator $\hat{U}_{n}$ that depends on the current state $%
\Psi _{n}.$ Such a dynamics could be defined as obeying Type $III^{\ast }$
Rules\footnote{%
Similar `starred' extensions are readily imagined for other Types of Rule.},
and is an issue discussed further in Section 8.5. The evolved state $\Psi
_{n}^{\prime }=\hat{U}_{n}\Psi _{n}$ could then be tested by an operator $%
\hat{\Sigma}_{n+1},$ itself chosen by a particular Type of Rule.

In the general case, the choice of both the next unitary operator $\hat{U}%
_{n}$ and the next test $\hat{\Sigma}_{n+1}$ may depend on their own
individual Types of Rule.

\bigskip

The Type $III$ and $IIIa$ categories of mechanism can in fact themselves be
further subdivided, because there are very many different sorts of way in
which the actual choice of operator $\hat{\Sigma}_{n+1}$ could depend on $%
\Psi _{n}.$

One important subdivision involves the question as to whether the operators
are defined independently of the states, or whether they are somehow
`created' by them. In the first instance, it is possible to imagine a
dynamics associated with a large, fixed set of possible operators defined at
the outset, with the decision about which member of this set is actually
chosen to be $\hat{\Sigma}_{n+1}$ made in reference to the state $\Psi _{n}.$
In the second scenario, no such pre-existing set of potential operators is
present, and the state $\Psi _{n}$ is assumed to somehow `generate' the
operator $\hat{\Sigma}_{n+1}$ itself according to whatever Rules $R_{n}$
govern the universe's dynamics. In other words, in the former case the state
$\Psi _{n}$ may be said to `pick' an operator $\hat{\Sigma}_{n+1}$ from a
pre-existing list, whereas in the latter circumstance the state $\Psi _{n}$
is directly responsible for `creating' the operator $\hat{\Sigma}_{n+1}.$

Clearly, the difference between these two sorts of Rule is most manifest
when $\Psi _{n}$ is unspecified: in the first sort of dynamics it cannot now
be said which operator $\hat{\Sigma}_{n+1}$ will be chosen out of the list
of known possibilities, whilst in the second case there is now simply no way
of knowing what $\hat{\Sigma}_{n+1}$ might be like at all.

For obvious reasons, these two different sorts of Type $III$ mechanism may
be labelled \textit{List-Sort} and \textit{Generated-Sort} respectively, and
are discussed more fully in the next few sections.

\bigskip

\subsection{List-Sort Dynamics}

\bigskip

Consider a universe represented at `time' $n$ by a state $\Psi _{n}$ in a
Hilbert space $\mathcal{H}$ of enormous dimension $D.$ As has been asserted
previously, it is assumed that the next state $\Psi _{n+1}$ will be one of
the $D$ orthonormal eigenstates of whichever Hermitian operator $\hat{\Sigma}%
_{n+1}$ is chosen by the Rules $R_{n}$ to develop the system. The question
is: what will this operator actually be?\bigskip

In List-Sort dynamics, it is conjectured that along with the overall Hilbert
space defining the system, a universe is also provided with an enormous List
$\frak{L}$ of different Hermitian operators
\begin{equation}
\frak{L}\equiv \{\hat{B}_{1},\hat{B}_{2},...,\hat{B}_{l}\}=\{\hat{B}%
_{i}:i=1,2,...,l\}=\{\hat{B}_{i}\}  \label{Ch6List}
\end{equation}
where $l$ may be called the `List Length'.

Then, the central principle behind the development of this universe
according to Type $III$ List-Sort dynamics is that a particular state $\Psi
_{n}$ leads to the selection of one, and only one, operator from the List $%
\frak{L}.$ This chosen operator is consequently identified with $\hat{\Sigma}%
_{n+1},$ and can hence be used to test the state $\Psi _{n}.$ The next state
$\Psi _{n+1}$ will thus be one of the eigenvectors of this operator\footnote{%
As an aside, note that other Types of List-Sort dynamics are permitted. A
Type $II$ List-Sort mechanism, for example, could choose the next operator $%
\hat{\Sigma}_{n+1}$ from a List according to some sort of deterministic
algorithm that does not refer in any way to the current state $\Psi _{n}.$%
\par
Type $III$ List-Sort dynamics could similarly be expected to have their
analogies in universes developing according to Type $IV,$ $IVa,$ $V$ and $Va$
mechanisms.}.\bigskip

In fact, each operator $\hat{B}_{i}$ in $\frak{L}$ is actually a member of
an equivalence class of Hermitian operators, which could be written $\hat{B}%
_{i}\equiv \{\hat{B}_{i}^{1},$ $\hat{B}_{i}^{2},$ $\hat{B}_{i}^{3},...\},$
with identical eigenvectors but different eigenvalues. Moreover, this
equivalence class may be used to specify a unique basis $\frak{B}_{i}$
containing a set of $D$ orthonormal vectors spanning the Hilbert space $%
\mathcal{H},$ such that $\frak{B}_{i}$ may be given by
\begin{equation}
\frak{B}_{i}\equiv \{\Theta _{i}^{a}:a=1,2,...,D\}
\end{equation}
with $\langle \Theta _{i}^{a}|\Theta _{i}^{b}\rangle =\delta _{ab},$ where
the set $\{\Theta _{i}^{a}\}$ are the eigenvectors of $\hat{B}_{i}.$

Effectively, then, the List $\frak{L}$ also implies a List of different
possible basis sets
\begin{equation}
\frak{L}\equiv \{\frak{B}_{1},\frak{B}_{2},...,\frak{B}_{l}\}=\{\frak{B}%
_{i}:i=1,2,...,l\}=\{\frak{B}_{i}\}.
\end{equation}
such that bases $\frak{B}_{i}$ and operators $\hat{B}_{i}$ may be used
interchangeably in the following, depending on context.

So, the state $\Psi _{n}$ of a universe developing according to List-Sort
dynamics implies the selection of one, and only one, basis set from the List
$\frak{L},$ and this in turn implies the selection of a unique equivalence
class of operators. The next state $\Psi _{n+1}$ will hence be one of the
elements of whichever basis set was chosen from the List.\bigskip

Of course, exactly which basis set is chosen from the List depends on the
actual Rules $R_{n}$ that govern the universe's dynamics, and the current
state $\Psi _{n}.$ However, once a basis set has been picked, it
automatically fixes the set of possible next states. So, if the Rules
conspire such that a particular state $\Psi _{n}=\Phi _{A}$ leads to a
selection of the basis $\frak{B}_{A}$ from $\frak{L},$ then it is assumed
that $\Psi _{n}$ will be tested by the operator $\hat{\Sigma}_{n+1}=\hat{B}%
_{A},$ so that the next state $\Psi _{n+1}$ will be one of the vectors $%
\{\Theta _{A}^{a}\}.$

Furthermore, whichever one of these eigenvectors will actually be realised
is given in the usual, stochastic way of the Born probability rule, such
that the probability $P_{n+1}^{a}$ that the next state of the universe is $%
\Psi _{n+1}=\Theta _{A}^{a}$ is clearly
\begin{equation}
P_{n+1}^{a}=\left| \langle \Psi _{n+1}=\Theta _{A}^{a}|\Psi _{n}\rangle
\right| ^{2}.
\end{equation}

Note that there is no form of `compound' probability involved here: the
selection of the basis $\frak{B}_{A}$ is assumed to be fully deterministic
given the state $\Psi _{n}=\Phi _{A},$ such that under this circumstance, $%
\hat{B}_{A}$ is chosen from the List $\frak{L}$ with probability $1.$ This
rule will be modified later.\bigskip

As discussed, it is hoped in this section to investigate state
self-referential Rules, that is, mechanisms in which the actual selection of
a particular basis from the List $\{\frak{B}_{i}\}$ depends upon the current
state $\Psi _{n}.$

To this end, it is possible to schematically envisage a situation in which
if $\Psi _{n}$ has one particular `property', then the basis set $\frak{B}%
_{A}$ is chosen from the List, whereas if $\Psi _{n}$ has instead a
different such `property', then the basis set $\frak{B}_{B}$ is
alternatively chosen from the List, whilst if $\Psi _{n}$ has neither of
these but has instead a third particular `property', then the basis set $%
\frak{B}_{C}$ is chosen from the List, and so on.

Clearly, the Rules must be defined such that whichever state the universe is
in, one, and only one, operator is picked from the List. Effectively, then,
this implies that the various `properties' made reference to by the Rules
must be mutually exclusive: that is, if the state $\Psi _{n}$ has a property
$A,$ it cannot also have the properties $B,C,...,$ and so there is therefore
no ambiguity in which of $\frak{B}_{A},\frak{B}_{B},\frak{B}_{C},...$ is
picked from the List to give $\hat{\Sigma}_{n+1}.$ Moreover, the set of
possible properties made reference to must also be exhaustive, such that
whatever $\Psi _{n}$ may be, it will definitely have one (and only one) of
the properties $A,B,C....,$ so that the Rules can definitely select one of $%
\frak{B}_{A},\frak{B}_{B},\frak{B}_{C},...$

The question dominating the dynamics them becomes one concerning which
particular `property' the Rules $R_{n}$ deem important. Again,\ of course,
this depends entirely on how these Rules are defined, and many different
types of Rule are possible.\bigskip

The above general definition of a Type $III$ List-Sort mechanism is perhaps
best demonstrated by illustration. As a simple example of these ideas,
consider a universe represented by a state $\Psi _{n}$ and governed by
List-Sort dynamics according to the List $\frak{L}=\{\frak{B}_{1},\frak{B}%
_{2},...,\frak{B}_{l}\}.$ Assume also that the Hilbert space $\mathcal{H}$
of this universe may be written as a product of $l$ sub-registers
\begin{equation}
\mathcal{H}\equiv \mathcal{H}_{[12...l]}=\mathcal{H}_{1}\otimes \mathcal{H}%
_{2}\otimes ...\otimes \mathcal{H}_{l}.
\end{equation}

Then, the state $\Psi _{n}$ is separable into $F_{n}$ factors relative to
this fundamental factorisation, where $F_{n}$ is clearly an integer between $%
1$ and $l.$

Now, the dynamics of this toy-universe could be such that it is governed by
the simple, algorithmic rule

\begin{itemize}
\item  Given a state $\Psi _{n}$ separable into $F_{n}$ factors, the basis
set $\frak{B}_{F_{n}}$ is chosen from the list $\frak{L}.$
\end{itemize}

So, $\Psi _{n}$ is tested by the Hermitian operator $\hat{\Sigma}_{n+1}=\hat{%
B}_{F_{n}},$ such that the next state $\Psi _{n+1}$ is one of the
eigenvectors $\{\Theta _{F_{n}}^{a}:a=1,...,D\},$ with individual
probabilities summing as
\begin{equation}
\sum_{a=1}^{D}|\langle \Theta _{F_{n}}^{a}|\Psi _{n}\rangle |^{2}=1.
\end{equation}

Thus, the development of this toy-universe proceeds in a manner that depends
on a particular `property' of its state. In this instance, the `property' in
question regards the universe's separability. Specifically, if $\Psi _{n}$
has $F_{n}$ factors relative to the fundamental factorisation of $\mathcal{H}%
_{[12...l]},$ the Rules conspire such that the state is tested by the $%
(F_{n})^{th}$ operator $\hat{B}_{F_{n}}$ from the List $\frak{L}=\{\hat{B}%
_{1},\hat{B}_{2},...,\hat{B}_{l}\}.$ Evidently, and as required, the
particular properties of the states referred to by the Rules are mutually
exclusive, because if $\Psi _{n}$ has $F_{n}=A$ factors it cannot also have $%
F_{n}=B$ factors, where $A\neq B;$ they are also exhaustive, because no
matter what $\Psi _{n}$ is, it will definitely possess between $1$ and $l$
factors.

Now, whichever eigenvector of $\hat{B}_{F_{n}}$ becomes the new state $\Psi
_{n+1}$ will possess $F_{n+1}$ factors, where $1\leq F_{n+1}\leq l.$ The
Rules hence dictate that the $(F_{n+1})^{th}$ test $\hat{B}_{F_{n+1}}$ is
selected from the List $\frak{L}$ to become $\hat{\Sigma}_{n+2},$ and the
universe collapses to a state $\Psi _{n+2}$ which is one of the eigenvectors
of this chosen operator.

Clearly, the procedure may be iterated indefinitely.\bigskip

As an aside, note that if each operator $\hat{B}_{F_{n}}$ in the above List
is itself factorisable into $F_{n}$ sub-operators, the described Rules could
give rise to a universe in which the state $\Psi _{n+1}$ may have the same
number of factors as the state $\Psi _{n}$ (or, at least, no fewer: $%
F_{n+1}\geq F_{n};$ recall from Chapter 5 that factorisable operators only
have separable eigenvectors), exactly as required for both apparent
persistence and a quantum causal set description of classical spacetime to
begin to arise.

Furthermore, the particular List-Sort mechanism discussed here is also
attractive from the point of view of generating the types of pregeometric
lightcone structure necessary for the emergence of classical spacetime.
Recall from Section 5.4 that such basic causal relationships are expected to
arise from considering how the individual factors of $\Psi _{n+1}$ are
affected by counterfactual changes in the factors of $\Psi _{n}.$ In the
present case, moreover, a separability preserving operator $\hat{\Sigma}%
_{n+1},$ that is selected solely according to the number of factors of $\Psi
_{n},$ could permit such counterfactual arguments to be applied, if, for
example, $\hat{B}_{i+1}$ is factorisable relative to the same split $%
\mathcal{H}_{[x_{1}x_{2}...x_{i}]}$ of the Hilbert space as $\hat{B}_{i}$
for all $i=1,...,(l-1)$ and $\left( \tprod\nolimits_{y=1}^{i}\otimes
\mathcal{H}_{x_{y}}\right) =\mathcal{H}_{[12...l]}.$

Thus, List-Sort Rules might effectively be used to generate potential
lightcone-like causal relationships between the factors of successive states
and operators, and could therefore possibly be employed to develop a
universe containing classical-looking spacetime.

Inflationary scenarios could also potentially be accounted for by modifying
the Rules in similar, suitable ways.\bigskip

The above illustrative model is just one very simple way of obtaining
List-Sort dynamics; many alternative Rules are, of course, possible. In
fact, although the state of a universe in a $D$ dimensional Hilbert space
governed by a List of List Length $l$ will always be one of only $D\times l$
eigenvectors, there are in general very many ways in which the various
different `properties' of the members of this set could be used to specify
particular choices of\ next operator.

As a second illustration, then, it could be imagined that if these $D\times
l $ \ different vectors are arbitrarily labelled $\{\Phi
_{j}:j=1,2,...(D\times l)\},$ a universe could be considered that follows
the Rules

\begin{itemize}
\item  If $\Psi _{n}\in \{\Phi _{1},\Phi _{2},...,\Phi _{a_{1}}\},$ then
choose the operator $\hat{B}_{1}$ from the List $\frak{L}=\{\hat{B}_{1},\hat{%
B}_{2},...,$ $\hat{B}_{l}\}$ to be $\hat{\Sigma}_{n+1};$

\item  Alternatively, if $\Psi _{n}\in \{\Phi _{a_{1}+1},\Phi
_{a_{1}+2},...,\Phi _{a_{2}}\},$ then instead choose the operator $\hat{B}%
_{2}$ from the List $\frak{L}$ to be $\hat{\Sigma}_{n+1};$

\item  $\vdots $

\item  And so on, up to the case where if $\Psi _{n}\in \{\Phi
_{a_{l-1}+1},\Phi _{a_{l-1}+2},...,\Phi _{a_{l}}\},$ the operator $\hat{B}%
_{l}$ is chosen from the List $\frak{L}$ to be $\hat{\Sigma}_{n+1};$
\end{itemize}

where the vectors have arbitrarily re-labelled in ascending order for
simplicity in this example, and where
\begin{equation}
a_{1}+a_{2}+...+a_{l}=D\times l.
\end{equation}

As before and as required, the choice of next operator $\hat{\Sigma}_{n+1}$
depends on the current state $\Psi _{n}.$ Also as before, the algorithm may
be repeated indefinitely, because $\Psi _{n+1}\in \{\Phi
_{j}:j=1,2,...(D\times l)\}$ for all $n.$\bigskip

Two points are immediately obvious from this second example. First, it is
evident that more that one state can lead to the same choice of operator. In
fact, this observation was also paralleled in the first example, because if $%
F_{n}\neq 1$ or $F_{n}\neq l,$ the state $\Psi _{n}$ could be in a number of
different partitions, each with $F_{n}$ factors, and each resulting in the
same operator\ $\hat{B}_{F_{n}}.$ The general conclusion is therefore that
although a given state $\Psi _{n}$ must result in the definite choice of a
specific operator $\hat{\Sigma}_{n+1},$ the reverse need not be true: an
operator $\hat{\Sigma}_{n+1}$ could have been chosen due to the universe
being in any one of a variety of different states.

Secondly, it is apparent that the same List $\frak{L}$ was appropriate in
both examples, even though the `property' of the state that is of interest
to the Rules was completely different in the two cases. This highlights the
overall point that is the actual Rules relating the state $\Psi _{n}$ to the
operator $\hat{\Sigma}_{n+1}$ that are of most importance in List-Sort
mechanisms.

\bigskip

The List-Sort mechanisms discussed up until now have been manifestly
deterministic, at least as far as the choice of operators is concerned: a
given state $\Psi _{n}$ results in the selection of a unique operator $\hat{%
\Sigma}_{n+1}$ from the List, because the universe is obeying Type $III$
List-Sort Rules. These types of mechanism, however, may be extended to
\textit{Probabilistic List-Sort} dynamics (governed by Type $IIIa$ List-Sort
Rules) in a straightforward manner.

One way of achieving this would be to relax the `uniqueness' condition of
the basis $\frak{B}_{i}$ chosen by the Rules in reference to $\Psi _{n}.$
Specifically, given a state $\Psi _{n},$ then instead of this
deterministically implying the definite selection of a single basis $\frak{B}%
_{i}$ from the list $\frak{L},$ it could instead be used to imply the
selection of a set of possible bases $\frak{B}_{i},$ $\frak{B}_{j},$ $\frak{B%
}_{k},...$ from $\frak{L},$ with probabilities $p_{i},$ $p_{j},$ $p_{k},...$
respectively. Of course, in this case the condition must hold that
\begin{equation}
p_{i}+p_{j}+p_{k}+...=1.
\end{equation}

In other words, a given state (or more specifically, perhaps, a given
`property' of a given state) could give rise to a number of potential next
tests $\hat{B}_{i},\hat{B}_{j},\hat{B}_{k},...,$ but which of these
operators actually becomes $\hat{\Sigma}_{n+1}$ is determined randomly.
Thus, given a state $\Psi _{n},$ the next test $\hat{\Sigma}_{n+1}$ could be
the operator $\hat{B}_{i}$ with probability $p_{i},$ or the operator $\hat{B}%
_{j}$ with probability $p_{j},$ and so on.

Equally, then, the probability that the next state $\Psi _{n+1}$ will be any
one of the set of eigenstates\ $\{\Theta _{i}^{a}\}$ of $\hat{B}_{i}$ is
given by $p_{i},$ whereas the probability it will instead be any one of the
set of eigenstates\ $\{\Theta _{j}^{a}\}$ of $\hat{B}_{j}$ is given by $%
p_{j},$ etc.\bigskip

The probability of choosing a particular operator compounds with the
standard quantum mechanical probability governing the state collapse
mechanism to give the overall probability that the universe will jump to a
particular next state. Thus, given a state $\Psi _{n},$ the probability $%
P_{n+1}^{a}$ that the next state $\Psi _{n+1}$ will be a particular
eigenstate $\Theta _{A}^{a}$ of $\hat{B}_{A}$ is the product of the
probability that the operator $\hat{B}_{A}$ is chosen, with the probability
that the state will then collapse to this eigenvector; viz,
\begin{equation}
P_{n+1}^{a}=p_{A}\left| \langle \Theta _{A}^{a}|\Psi _{n}\rangle \right| ^{2}
\end{equation}
noting that a measure of entropy could be applied in this type of situation,
as analogous to Section 5.6.

Actually, the probability that the next state $\Psi _{n+1}$ will be a
particular vector may in fact be much higher than this individual result,
because it is possible that a given eigenstate may be a member of more than
one of the potential basis sets. Of course, this point is a mere
technicality in the present discussion, and does not affect the general
principles being suggested; it would only be important if just `before' and
`after' states were being considered, with the `route' (i.e. the operator)
taken by the state being ignored.

\bigskip

As a final comment, note that exactly how the universe is provided with the
particular List $\frak{L}$ containing the particular members $\{\frak{B}%
_{i}\}$ is unexplained. Indeed, it is perhaps unexplainable, and in a
universe developing according to List-Sort Rules, the List $\frak{L}$ may
simply have to be accepted along with the laws of physics as just one of the
necessary pre-requisites for dynamics to occur. Paraphrasing, it would be as
if the universe needs to be provided with an enormous `data-bank' of
possible operators at the outset, just as a (quantum) computer must be
provided with all of the logic gates necessary for its functioning in order
for it to work.

In fact, this quantum computational analogy may be continued: the Rules $%
R_{n}$ governing the universe could similarly be associated with the
algorithm or program governing the steps taken during a computation, whilst
the state $\Psi _{n}$ may be related to the current internal configuration
of the machine. Again, the universe is viewed as a giant, self-referential
quantum automaton.

Of course, if the actual Universe develops according to List-Sort dynamics,
it would be a task for physicists to attempt to discover what its List is,
and hence what the total set of possible operators available to the physical
Universe actually are. \textit{Why} it has whatever List it has, however, is
perhaps a question for either the Anthropic Principle or for philosophy.

\bigskip

\subsection{Examples of List-Sort Dynamics}

\bigskip

The aim of this section is to provide a set of examples that illustrate how
toy-universes obeying List-Sort Rules might develop. For reasons of
simplicity, attention will be restricted to low dimensional systems, with
the usual assumption being made that the underlying principles are not
entirely unrelated to those existing in more complicated situations.
Specifically, a two qubit model is discussed.\bigskip

Consider a universe represented by a state in a four dimensional Hilbert
space\ $\mathcal{H}$ fundamentally factorised into two qubit subregisters,
i.e.
\begin{equation}
\mathcal{H}\equiv \mathcal{H}_{[12]}^{(4)}=\mathcal{H}_{1}^{(2)}\otimes
\mathcal{H}_{2}^{(2)}
\end{equation}
where dimensional superscripts shall, again, generally be dropped from now
on.

Define also an orthonormal basis for $\mathcal{H}_{a},$ where $a=1,2,$ as $%
\mathcal{B}_{a}=\{|0\rangle _{a},|1\rangle _{a}\},$ such that any state $%
\Psi $ of the two qubit universe may be written in the form
\begin{equation}
\Psi =\sum_{ij}C_{ij}|i\rangle _{1}\otimes |j\rangle _{2}\text{ \ \ , \ \ }%
i,j=0,1  \label{gen}
\end{equation}
\qquad where the four coefficients $C_{ij}\in \mathbb{C}$ produce a $2$ $%
\times $ $2$ matrix.

As before, and when no confusion is likely to occur, the subscripts denoting
subregisters $1$ and $2$ will be omitted in the following in favour of the
convention that vectors to the left of the tensor product represent the
state of qubit $1$ in $\mathcal{H}_{1},$ whereas vectors to the right of the
tensor product represent the state of qubit $2$ in $\mathcal{H}_{2}.$
Further, the additional simplification may be made that the tensor product
is always implied, so that its symbol is consequently omitted: $|i\rangle
_{1}\otimes |j\rangle _{2}\equiv |i\rangle \otimes |j\rangle \equiv
|i\rangle |j\rangle \equiv |ij\rangle .$\bigskip

For a universe represented by a vector in a factorisable Hilbert space, it
is possible that its state may be entangled. Moreover, and as has been
discussed previously, the issue of entanglement versus separability is of
fundamental importance in any discussion of quantum theory, and has played a
central role in this thesis.

In Chapter 4 the issue of separability versus entanglement was introduced,
with a goal being to classify the different ways an arbitrary vector could
be entangled or separable relative to a given factorisation of its Hilbert
space. In the current situation, however, any state in $\mathcal{H}%
_{12}^{(4)}$ is either fully entangled or fully separable, because it can
only possess either $1$ or $2$ factors. This simplification will be useful
in the following.\bigskip

Consider an arbitrary set $\frak{B}$ of orthonormal vectors that forms a
basis for a given Hilbert space. Specifically, in fact, it is a standard
theorem of vector algebra that for a Hilbert space of dimension $D,$ each
basis set of such vectors contains $D$ elements.

Now, the elements of these basis sets will each possess a certain degree of
separability or entanglement relative to the fundamental factorisation of
the Hilbert space, and this may be classified by referring to the `type' of
the basis. For example, and returning to the current four dimensional case
of a two qubit system, a basis set $\frak{B}_{(p,q)}$ can be said to be of
type $(p,q)$ if it contains $p$ entangled and $q$ separable elements, where $%
p+q=4.$ Furthermore, these $p+q$ vectors may be associated with the $p+q$
orthogonal eigenstates of a set of operators acting upon the two qubit
universe, and so to define the basis $\frak{B}_{(p,q)}$ also defines the
equivalence class of operators $\hat{B}_{(p,q)}.$

Of course, the above analysis is highly simplistic, and may neglect a number
of important points. Indeed, it was shown in Section 4.2 that only type $%
(0,4),$ $(2,2),$ $(3,1),$ and $(4,0)$ bases can be found to span the Hilbert
space $\mathcal{H}_{[12]}^{(4)},$ and that no example of a type $(1,3)$
basis set can exist.

Nevertheless, from these elementary ideas it is now possible to begin to
construct toy-model universes that develop according to state
self-referential List-Sort Rules, as is shown by the following examples.

\bigskip

\subsubsection{Example I}

\bigskip

Consider as above a two qubit universe represented by a state $\Psi _{n}$ in
the factorisable Hilbert space $\mathcal{H}_{[12]}.$ Consider further the
particular $\left( 2,2\right) $ type basis set $\frak{B}_{(2,2)}$ of
orthonormal vectors described by
\begin{equation}
\frak{B}_{(2,2)}=\left\{ |00\rangle \text{ , }|11\rangle \text{ , }\frac{1}{%
\sqrt{2}}\left( |01\rangle +|10\rangle \right) \text{ , }\frac{1}{\sqrt{2}}%
(|01\rangle -|10\rangle )\right\}  \label{B22}
\end{equation}
and\ also the type $\left( 4,0\right) $ basis set $\frak{B}_{(4,0)}$ defined
by
\begin{equation}
\frak{B}_{(4,0)}=\left\{
\begin{array}{c}
\frac{1}{\sqrt{2}}\left( |00\rangle +|11\rangle \right) \text{ , }\frac{1}{%
\sqrt{2}}(|00\rangle -|11\rangle )\text{ ,} \\
\text{ }\frac{1}{\sqrt{2}}\left( |01\rangle +|10\rangle \right) \text{ , }%
\frac{1}{\sqrt{2}}(|01\rangle -|10\rangle )
\end{array}
\right\} .  \label{B40}
\end{equation}

For convenience and brevity, $\frak{B}_{(2,2)}$ may be written as $\frak{B}%
_{(2,2)}=\{a,b,c,d\},$ where the order as defined in (\ref{B22}) is
preserved such that $c=\frac{1}{\sqrt{2}}(|01\rangle +|10\rangle )$ etc.
Similarly, $\frak{B}_{(4,0)}$ may be written as $\frak{B}_{(4,0)}=\{e,f,g,h%
\},$ where for example $f=\frac{1}{\sqrt{2}}(|00\rangle -|11\rangle ).$

The basis set $\frak{B}_{(2,2)}$ corresponds to the set of orthonormal
eigenvectors of an (equivalence class of) operator $\hat{B}_{(2,2)}$ of the
form given in (\ref{B22op}), where $A,$ $B,$ $C,$ $D$ are real,
non-degenerate and non-zero eigenvalues
\begin{eqnarray}
\hat{B}_{(2,2)} &=&A|00\rangle \langle 00|+B|11\rangle \langle 11|+\frac{C}{2%
}(|01\rangle +|10\rangle )(\langle 01|+\langle 10|)  \label{B22op} \\
&&+\frac{D}{2}(|01\rangle -|10\rangle )(\langle 01|-\langle 10|)  \notag
\end{eqnarray}
with a similar construction of $\hat{B}_{(4,0)}$ from $\frak{B}_{(4,0)}.$

Now, in the language of the List-Sort dynamics, it is possible to define a
List $\frak{L}_{X}$ of list length $2$ as
\begin{equation}
\frak{L}_{X}\equiv \{\frak{B}_{(2,2)},\frak{B}_{(4,0)}\}
\end{equation}
or equivalently, $\frak{L}_{X}\equiv \{\hat{B}_{(2,2)},\hat{B}_{(4,0)}\}.$

Moreover, and for the sake of example, it is also possible to define the set
of Rules governing the development of this toy-universe such that they make
reference to the above List. A dynamics based upon a Type $III$ List-Sort
mechanism is thus defined.\bigskip

On possible such mechanism is the following. Suppose the universe develops
according to the Rule that

\begin{itemize}
\item  If the state $\Psi _{n}$ is separable, then it is tested by an
operator whose eigenstates form the basis $\frak{B}_{(4,0)};$

\item  But if instead $\Psi _{n}$ is entangled, it is alternatively tested
by an operator whose eigenstates form the basis $\frak{B}_{(2,2)}.$
\end{itemize}

Then, the development of the universe involves an operator chosen from the
List $\frak{L}_{X}$ in a manner that depends upon a `property' of the
current state. Specifically, if $\Psi _{n}$ is separable then an operator $%
\hat{B}_{(4,0)}$ is chosen from the List to be $\hat{\Sigma}_{n+1}$ and the
state $\Psi _{n+1}$ will be one of the eigenstates $\{e,f,g,h\},$ whereas if
$\Psi _{n}$ is entangled then an operator $\hat{B}_{(2,2)}$ is instead
picked from the List and the state $\Psi _{n+1}$ will be one of the vectors $%
\{a,b,c,d\}.$ Note that these two properties are mutually exclusive and
exhaustive, as required: every state is either separable or entangled.

Of course, \textit{which} of the four eigenstates is actually chosen in each
case depends entirely on the random nature of the quantum collapse
process.\bigskip

As an illustration of how such a model could develop, consider a universe in
an initial state $\Psi _{0}$ given by $\Psi _{0}=|00\rangle =a.$ This state
is separable, so by following the Rules, the next state $\Psi _{1}$ will be
one of the eigenvectors of an operator $\hat{\Sigma}_{1}=\hat{B}_{(4,0)},$
such that
\begin{equation}
\hat{B}_{(4,0)}|\Psi _{1}\rangle =\lambda ^{4,0}|\Psi _{1}\rangle .
\label{eigen}
\end{equation}

$\Psi _{1}$ will be one of the states $e,$ $f,$ $g,$ or $h,$ with
corresponding eigenvalues $\lambda _{e}^{4,0},$ $\lambda _{f}^{4,0},$ $%
\lambda _{g}^{4,0},$ or $\lambda _{h}^{4,0}$ respectively, where the actual
values of the $\lambda ^{4,0}$ need play no further part in the discussion,
save to say that they are real, non-degenerate and non-zero (c.f. the
discussion of Strong operators in Chapter 5).

As in conventional quantum mechanics, the probability $P(\Psi _{1},\Psi
_{0}) $ of jumping from the state $\Psi _{0}$ to a potential state $\Psi
_{1} $ is given by the square of the amplitude, that is
\begin{equation}
P(\Psi _{1},\Psi _{0})=\left| \langle \Psi _{1}|\Psi _{0}\rangle \right|
^{2}.  \label{amp}
\end{equation}

So, from an examination of (\ref{B22}) and (\ref{B40}), the relationship (%
\ref{amp}) clearly leads to the amplitudes: $\langle e|a\rangle =\langle
f|a\rangle =1/\sqrt{2}$ and $\langle g|a\rangle =\langle h|a\rangle =0.$
Thus, if the initial state of the universe is $\Psi _{0}=a=|00\rangle $ then
the subsequent state $\Psi _{1}$ will be either
\begin{equation}
\Psi _{1}=\Psi _{1}^{e}=e=\frac{1}{\sqrt{2}}\left( |00\rangle +|11\rangle
\right)
\end{equation}
or
\begin{equation}
\Psi _{1}=\Psi _{1}^{f}=f=\frac{1}{\sqrt{2}}(|00\rangle -|11\rangle )
\end{equation}
with equal probabilities of $%
%TCIMACRO{\UNICODE[m]{0xbd}}%
%BeginExpansion
{\frac12}%
%EndExpansion
.$\bigskip

Now, both of the states $\Psi _{1}^{e}$ and $\Psi _{1}^{f}$ are entangled,
so no matter what happens, $\Psi _{1}$ will be entangled. Consequently, and
according to the defined algorithm, the Rules next pick out the basis set $%
\frak{B}_{(2,2)}$ from the list $\frak{L}_{X},$ such that the successive
state $\Psi _{2}$ will be one of the eigenvectors of an operator $\hat{\Sigma%
}_{2}=\hat{B}_{(2,2)};$ viz.,
\begin{equation}
\hat{B}_{(2,2)}|\Psi _{2}\rangle =\lambda ^{2,2}|\Psi _{2}\rangle .
\label{eigen2}
\end{equation}

Thus, $\Psi _{2}$ will be one of the states $a,$ $b,$ $c,$ or $d,$ with
corresponding eigenvalues $\lambda _{a}^{2,2},$ $\lambda _{b}^{2,2},$ $%
\lambda _{c}^{2,2}$ or $\lambda _{d}^{2,2},$ which are again ignored.

This time, the relevant amplitudes in the transition from $\Psi _{1}$ to $%
\Psi _{2}$ are $^{i)}$ $\langle a|e\rangle =\langle b|e\rangle =1/\sqrt{2}$
with $\langle c|e\rangle =\langle d|e\rangle =0,$ or $^{ii)}$ $\langle
a|f\rangle =\langle b|f\rangle =1/\sqrt{2}$ with $\langle c|f\rangle
=\langle d|f\rangle =0,$ depending on whether $\Psi _{1}$ is $e$ or $f.$

So, if the state $\Psi _{1}$ of the universe after the transition $\Psi
_{0}\rightarrow \Psi _{1}$ was measured and found to be $\Psi _{1}^{e},$
then the next state will either be $\Psi _{2}=\Psi _{2}^{e,a}=a=|00\rangle $
or alternatively $\Psi _{2}=\Psi _{2}^{e,b}=b=|11\rangle ,$ each with equal
probability of $%
%TCIMACRO{\UNICODE[m]{0xbd}}%
%BeginExpansion
{\frac12}%
%EndExpansion
.$

However, if instead the collapse from $\Psi _{0}$ to $\Psi _{1}$ had left
the state at time $n=1$ as $\Psi _{1}^{f},$ then the subsequent state will
be either $\Psi _{2}=\Psi _{2}^{f,a}=a=|00\rangle $ or $\Psi _{2}=\Psi
_{2}^{f,b}=b=|11\rangle ,$ again each with equal probability of $%
%TCIMACRO{\UNICODE[m]{0xbd}}%
%BeginExpansion
{\frac12}%
%EndExpansion
.$

So even after two jumps, the random nature of the quantum state reduction
and the chosen Rules for the List-Sort mechanism have led to four different
`histories' for the development of the state from $\Psi _{0}\rightarrow \Psi
_{1}\rightarrow \Psi _{2}.$ Namely, the four possible `routes' are either $%
a\rightarrow e\rightarrow a,$ or $a\rightarrow e\rightarrow b$ or $%
a\rightarrow f\rightarrow a$ or $a\rightarrow f\rightarrow b.$

\bigskip

Although it may appear trivial in this case, note that the probability of
going from $\Psi _{1}=e$ to $\Psi _{2}=a$ is $%
%TCIMACRO{\UNICODE[m]{0xbd}}%
%BeginExpansion
{\frac12}%
%EndExpansion
,$ and not $%
%TCIMACRO{\UNICODE[m]{0xbc}}%
%BeginExpansion
{\frac14}%
%EndExpansion
$ as would be the case if the sum
\begin{equation}
\langle a|e\rangle +\langle b|e\rangle +\langle a|f\rangle +\langle
b|f\rangle
\end{equation}
had to be normed to unity. This is because although the transition from one
state to the next makes use of quantum probability amplitudes, once a jump
has happened it is possible to say with certainty which state the system is
in. In other words, if the state $\Psi _{1}$ is measured and found to be $%
\Psi _{1}=\Psi _{1}^{e}=e,$ it is no longer valid to discuss the probability
of jumping from the alternative state $\Psi _{1}=\Psi _{1}^{f}=f$ to any
possible future state $\Psi _{2},$ because $\Psi _{1}^{f}$ does not exist.

In fact, the ability to describe the state with certainty is a fundamental
difference between Schr\"{o}dinger evolution and state reduction, and
arguably leads to the single valued nature of reality. Once state reduction
has occurred, the universe is in a unique state $\Psi _{n}^{x},$ and it is
therefore meaningless to discuss the probability of transition from any
other state $\Psi _{n}^{y}.$ This is a central principle of quantum theory,
which holds that although $\Psi _{n}^{x}$ and $\Psi _{n}^{y}$ may both have
been potential futures of the state $\Psi _{n-1},$ once state reduction has
selected the state $\Psi _{n}^{x},$ no other state $\Psi _{n}^{y}$ can be
said to exist. Over a series of jumps it is therefore necessary to discuss
the \textit{classical} probability that the universe will jump from one
state to another, and then from that new, now `known' state to the next.

This type of reasoning was at the heart of the discussion of Section 6.1.4
concerning the qubit Bell inequalities.\bigskip

The difference between a dynamics based on state reduction and one acting
without it may be highlighted by appealing again to the above example. If $%
\Psi _{0}=a$ and also $\Psi _{2}=a$ then the quantum probability $P(\Psi
_{2},\Psi _{0})$ of jumping directly from $\Psi _{2}$ to $\Psi _{0}$ (if
this were allowed by the Rules) would be $\left| \langle \Psi _{2}|\Psi
_{0}\rangle \right| ^{2}=1,$ which is effectively equivalent to a null test
from the point of view of the universe. If, however, it is specified that
the universe develops from $\Psi _{0}$ to $\Psi _{1}$ to $\Psi _{2},$ where $%
\Psi _{1}=\Psi _{1}^{e}$ and so $\Psi _{2}=a=\Psi _{2}^{e,a},$ then the
probability $P(\Psi _{2}^{e,a},\Psi _{1}^{e},\Psi _{0})$ is instead given by
\begin{equation}
P(\Psi _{2}^{e,a},\Psi _{1}^{e},\Psi _{0})=\left| \langle \Psi
_{2}^{e,a}|\Psi _{1}^{e}\rangle \right| ^{2}\left| \langle \Psi
_{1}^{e}|\Psi _{0}\rangle \right| ^{2}=%
%TCIMACRO{\UNICODE[m]{0xbd}}%
%BeginExpansion
{\frac12}%
%EndExpansion
\times
%TCIMACRO{\UNICODE[m]{0xbd}}%
%BeginExpansion
{\frac12}%
%EndExpansion
=%
%TCIMACRO{\UNICODE[m]{0xbc}}%
%BeginExpansion
{\frac14}%
%EndExpansion
\end{equation}
and this indicates that information has been extracted from the system. As
discussed previously, the state reduction mechanism is essential in order to
associate the subscript on $\Psi _{n}$ with a temporal-like parameter.

Generalising, the probability $P(\Psi _{n+N},\Psi _{n+N-1},...,\Psi _{n})$
of the universe developing from the state $\Psi _{n}$ to a given state $\Psi
_{n+N}$ via a series of \textit{specified} intermediate states $\Psi _{n+1,}$
$\Psi _{n+2},$ $...,$ $\Psi _{n+N-1}$ is given by the classical product of
the squared moduli of the\ $N$ appropriate quantum probability amplitudes,
such that
\begin{equation}
P(\Psi _{n+N},\Psi _{n+N-1},...,\Psi _{n})=\left| \langle \Psi _{n+N}|\Psi
_{n+N-1}\rangle \right| ^{2}\left| \langle \Psi _{n+N-1}|\Psi
_{n+N-2}\rangle \right| ^{2}...\left| \langle \Psi _{n+1}|\Psi _{n}\rangle
\right| ^{2}.  \label{class}
\end{equation}

Note that this generalisation may be related to the concept of entropy
discussed in Section 5.6, where a measure of entropy is associated with
different sets of possible futures of a quantum system.\bigskip

Note also that the inclusion of state reduction shows the inherent asymmetry
and irreversibility of time. Given that the universe is in a state $\Psi
_{n,}$ it is reasonable to ask the question: what is the probability that
the universe will jump to a specific state $\Psi _{n+1,}$ and will then jump
to another specified state $\Psi _{n+2,}$ and so on through a chain of
specified states up to $\Psi _{n+N}?$ However, the reverse question is
different. If it is known that the universe is in a state $\Psi _{n+N},$
then the probability that it jumped from the previous state $\Psi _{n+N-1}$
is $1,$ assuming that no information has been lost during the transition
such that $\Psi _{n+N-1}$ is also known. It is meaningless to ask in this
context about the probability of arriving at the current state from a given
alternative past, because state reduction ensures that only one past
actually occurred. Although it may not always be possible in the present to
retrodict with certainty what the past actually \textit{was}, a physicist
can be sure that a unique past \textit{did} occur. This is different from
discussions of the future, because it is never possible to predict what
\textit{will} happen, but only what \textit{might} happen as a potential
outcome of a potential quantum test. Again, strong parallels are drawn here
with the discussion of the qubit Bell inequalities in Section 6.1.4.

\bigskip

Returning now to the example in hand, the universe described by this model
would continue to be tested by the operators $\hat{B}_{(2,2)}$ or $\hat{B}%
_{(4,0)}$ according to whether its state is respectively entangled or
separable.

In fact, ignoring the `route' by which it got there, it can easily be shown
that after $2n$ steps, for $n\in \mathbb{N},$ the universe is in either the
state\ $\Psi _{2n}=a=|00\rangle $ or $\Psi _{2n}=b=|11\rangle ,$ each with
probability $%
%TCIMACRO{\UNICODE[m]{0xbd}}%
%BeginExpansion
{\frac12}%
%EndExpansion
,$ whereas after $2n-1$ steps the universe is in either the state\ $\Psi
_{2n-1}=e=\frac{1}{\sqrt{2}}\left( |00\rangle +|11\rangle \right) $ or $\Psi
_{2n-1}=f=\frac{1}{\sqrt{2}}\left( |00\rangle -|11\rangle \right) ,$ also
each with probability $%
%TCIMACRO{\UNICODE[m]{0xbd}}%
%BeginExpansion
{\frac12}%
%EndExpansion
.$ The system will `oscillate' between having a state that is separable and
one that is entangled, though which particular separable (i.e. $a$ or $b)$
or entangled (i.e. $e$ or $f)$ state it is actually in depends on the random
nature of quantum state reduction.\bigskip

It is perhaps surprising to note that a different choice of initial
condition, $\Psi _{0},$ would not drastically affect the subsequent
development of the model. To illustrate why, observe that, according to the
Rules governing the dynamics, if the universe were ever to collapse to any
of the vectors $c,$ $d,$ $g$ or $h,$ it would remain in that state from then
on, because these states are eigenvectors of both operators. So, if the
initial state was an arbitrary normalised vector of the form
\begin{equation}
\Psi _{0}=\alpha |00\rangle +\beta |01\rangle +\gamma |10\rangle +\delta
|11\rangle
\end{equation}
where $\alpha ,\beta ,\gamma ,\delta \in \mathbb{C},$ it is possible to
conclude that after a period of $n$ steps the universe would either be
following the above pattern of $...[a/b]\longrightarrow \lbrack
e/f]\longrightarrow \lbrack a/b]...,$ where $``[a/b]"$ denotes $``a$ or $b"$
etc., or else it would be `stuck' in one of the states $c$ $(=g)$ or $d$ $%
(=h).$ Of course, exactly which course of action has the highest propensity
for occurring depends on the actual values of the complex coefficients $%
\alpha ,\beta ,\gamma ,\delta ,$ because these determine whether $\Psi _{0}$
is entangled or separable, and the nature of the potential probability
amplitudes $\langle \Psi _{1}|\Psi _{0}\rangle .$

Actually, the details of the above `extension' are in fact considered fairly
unimportant anyway. After all, it is the principles behind the mechanisms
investigated in this section that are of interest, specifically those
concerning the question of how a universe might develop according to
List-Sort Rules. Thus, in the remaining examples it is to be recalled that a
different choice of initial condition would not add anything significant to
the discussion, and so the possibility of choosing alternative states as $%
\Psi _{0}$ is omitted.

\bigskip

\subsubsection{Example II}

\bigskip

A toy-universe model that is perhaps more interesting than that of Example $%
I $ could add to the List $\frak{L}_{X}$ a type $\left( 0,4\right) $ basis
set $\frak{B}_{(0,4)},$ where this completely separable basis is defined as
\begin{equation}
\frak{B}_{(0,4)}=\left\{
\begin{array}{c}
\frac{1}{2}\left( |00\rangle +|10\rangle +|01\rangle +|11\rangle \right)
\text{ , }\frac{1}{2}(|00\rangle +|10\rangle -|01\rangle -|11\rangle ) \\
\text{ }\frac{1}{\sqrt{2}}\left( |00\rangle -|10\rangle \right) \text{ , }%
\frac{1}{\sqrt{2}}(|01\rangle -|11\rangle )
\end{array}
\right\}  \label{b04}
\end{equation}

Analogously to before, where the bases $\frak{B}_{(2,2)}$ and $\frak{B}%
_{(4,0)}$ were redefined as $\{a,b,c,d\}$ and $\{e,f,g,h\}$ respectively,
the elements of $\frak{B}_{(0,4)}$ written in the above order may be
labelled as $\frak{B}_{(0,4)}=\{j,k,l,m\}$ for simplicity, with for example $%
l=\frac{1}{\sqrt{2}}(|0\rangle -|1\rangle )\otimes |0\rangle .$

Clearly, the addition of $\frak{B}_{(0,4)}$ to $\frak{L}_{X}$ defines a new
List $\frak{L}_{Y},$ such that
\begin{equation}
\frak{L}_{Y}\equiv \{\frak{B}_{(2,2)},\frak{B}_{(4,0)},\frak{B}_{(0,4)}\}
\end{equation}
or equivalently $\frak{L}_{Y}\equiv \{\hat{B}_{(2,2)},\hat{B}_{(4,0)},\hat{B}%
_{(0,4)}\},$ where evidently $\frak{L}_{Y}\supset \frak{L}_{X}.$\bigskip

In the mechanism proposed now, it is supposed that the universe follows the
Rule

\begin{itemize}
\item  If the state $\Psi _{n}$ is separable then the basis $\frak{B}%
_{(2,2)}=\{a,b,c,d\}$ is picked from the List $\frak{L}_{Y},$ and the next
operator $\hat{\Sigma}_{n+1}$ to test the state will be $\hat{B}_{(2,2)};$

\item  Whereas if $\Psi _{n}$ is entangled then the basis $\frak{B}%
_{(0,4)}=\{j,k,l,m\}$ is instead chosen from $\frak{L}_{Y},$ and the next
state $\Psi _{n+1}$ is one of the eigenstates of an alternative operator $%
\hat{\Sigma}_{n+1}=\hat{B}_{(0,4)}.$
\end{itemize}

So, and as desired, a particular `property' of the state $\Psi _{n}$ of the
universe (again, its separability) is being used to select a particular
operator from the List $\frak{L}_{Y}$ to become $\hat{\Sigma}_{n+1};$ the
universe is hence governed by Type $III$ List-Sort dynamics.\bigskip

As as aside, note that there is no general constraint in the List-Sort
method for the Rules to use \textit{every} operator contained in the
specified List. Indeed, the current mechanism could equally be realised with
the shorter List $\frak{L}_{Z}$ defined as $\frak{L}_{Z}\equiv \{\frak{B}%
_{(2,2)},\frak{B}_{(0,4)}\},$ where $\frak{L}_{Z}\subset \frak{L}_{Y}.$

However, if the actual physical Universe runs according to List-Sort Rules,
a question would remain in this case as to why any operator should be
included on its List if it could never be used. In general, then, such
`non-essential' operators should perhaps best be removed for efficiency.

That said, in the present example, the inclusion of the non-used $\frak{B}%
_{(4,0)}$ to the List does not make any real difference, and so the longer
List $\frak{L}_{Y}$ will be retained; this is in preparation for Examples $%
III$ and $IV,$ where the whole of $\frak{L}_{Y}$ will be employed.\bigskip

At any time $n$ the state of the universe in question is constrained to be
one of the eight possible vectors: $a,b,c,d,j,k,l,m.$ Furthermore, which of
these it actually \textit{is} at time $n$ determines the basis set at time $%
n+1,$ according to the above Rules.

In addition, and as in Example $I,$ the likelihood of transition from each
of these states to the next depends on the probability amplitudes.
Specifically, the probability $\left| \langle y|x\rangle \right| ^{2}$ of
transition from the state $\Psi _{n}=|x\rangle $ to a potential state $\Psi
_{n+1}=|y\rangle $ is given as the element in column $x,$ row $y$ of Table
8.1
\begin{equation}
\begin{tabular}{|l|l|l|l|l|l|l|l|l|}
\hline
{*} & \textbf{a} & \textbf{b} & \textbf{c} & \textbf{d} & \textbf{j} &
\textbf{k} & \textbf{l} & \textbf{m} \\ \hline
\textbf{a} & 1 & 0 & - & - & 1/4 & 1/4 & 1/2 & 0 \\ \hline
\textbf{b} & 0 & 1 & - & - & 1/4 & 1/4 & 0 & 1/2 \\ \hline
\textbf{c} & 0 & 0 & - & - & 1/2 & 0 & 1/4 & 1/4 \\ \hline
\textbf{d} & 0 & 0 & - & - & 0 & 1/2 & 1/4 & 1/4 \\ \hline
\textbf{j} & - & - & 1/2 & 0 & - & - & - & - \\ \hline
\textbf{k} & - & - & 0 & 1/2 & - & - & - & - \\ \hline
\textbf{l} & - & - & 1/4 & 1/4 & - & - & - & - \\ \hline
\textbf{m} & - & - & 1/4 & 1/4 & - & - & - & - \\ \hline
\end{tabular}
\tag*{Table 8.1}
\end{equation}
noting that transitions which are forbidden by the Rules, for example a
separable state jumping to one of the members of $\frak{B}_{(0,4)},$ are
indicated by a dash.

Moreover, note that the probabilities are normed in such a way that
\begin{equation}
\sum_{i}\left| \langle i|x\rangle \right| ^{2}=1\text{ \ \ , \ \ }%
i,x=a,b,c,d,j,k,l,m
\end{equation}
because if the universe is in a state $\Psi _{n}=|x\rangle $ it must
certainly be able to jump to \textit{something}.

\bigskip

By way of example, let the initial state $\Psi _{0}$ of the universe be $%
\Psi _{0}=c=$ $\frac{1}{\sqrt{2}}\left( |01\rangle +|10\rangle \right) .$
This state is entangled, so according to the Rules the next state $\Psi _{1}$
will be one of the members of the basis set $\frak{B}_{(0,4)}=\{j,k,l,m\},$
and will hence be an eigenstate of the equation
\begin{equation}
\hat{B}_{(0,4)}|\Psi _{1}\rangle =\lambda ^{0,4}|\Psi _{1}\rangle
\label{eigen04}
\end{equation}
where $\lambda ^{0,4}$ is an eigenvalue that is subsequently ignored.

It is possible to develop the state of this toy-universe model over a number
of steps, just as it was in Example $I.$ So, if $\Psi _{0}=c,$ then $\Psi
_{1}$ must be either $j,l$ or $m$ because the probability that the universe
will collapse to the state $\Psi _{1}=k$ is zero.

Moreover, by applying the same logic, and from the results given in Table
8.1, it follows that if $\Psi _{1}=j$ then $\Psi _{2}$ must be either $a,b$
or $c.$ Alternatively, if instead $\Psi _{1}=l,$ it implies that $\Psi _{2}$
must be either $a,c$ or $d,$ whereas if $\Psi _{1}=m$ it implies that $\Psi
_{2}$ must be either $b,c$ or $d.$

This process may be continued to generate a set of possible transitions from
$\Psi _{2}$ to $\Psi _{3},$ and then from $\Psi _{3}$ to $\Psi _{4},$ etc.,
until a `family tree' of different possible chains of states are created.

Of course, quantitatively some states or patterns are more likely to occur
than\ others due to the list of probabilities given in Table 8.1. For
example, once the universe has jumped into the state $a,$ then according to
the Rules it will remain in this state forever.\bigskip

It is easy to write a computer program that will iterate this two qubit
universe over $N$ steps according to the specified Rules. Furthermore, a
number of questions can then be asked of the system's development. For
example, what is the probability that after $N=3$ jumps the universe will
have proceeded through the history $\Psi _{0}=c,$ $\Psi _{1}=m,$ $\Psi
_{2}=d,$ $\Psi _{3}=k?$ (Answer: $1/32).$ Alternatively, what is the
probability that after $N=57$ jumps the universe is in the state $j?$
(Answer: $\sim 9.3132\times 10^{-10}).$

One interesting question is: what is the probability that after $N$ steps
the universe is in an entangled state, given that $\Psi _{0}=c?$
Paraphrasing, what is the probability that $\Psi _{N}=c$ or $\Psi _{N}=d?$
The result of this is shown in Figure 8.1, where the x-axis is $n$ and the
y-axis is the probability $P(\Psi _{n}=[c/d]).$

%\FRAME{fhFU}{347.125pt}{266.1875pt}{0pt}{\Qcb{Probability, $P(\Psi
%_{n}=[c/d]),$ of an entangled universe after $n$ steps.}}{}{Figure 8.1}{%
%\special{language "Scientific Word";type "GRAPHIC";maintain-aspect-ratio
%TRUE;display "USEDEF";valid_file "T";width 347.125pt;height 266.1875pt;depth
%0pt;original-width 340.25pt;original-height 260.5pt;cropleft "0";croptop
%"1";cropright "1";cropbottom "0";tempfilename
%'I1KOB107.wmf';tempfile-properties "XPR";}}

\begin{figure}[th]
\begin{center}
\includegraphics[height=4in]{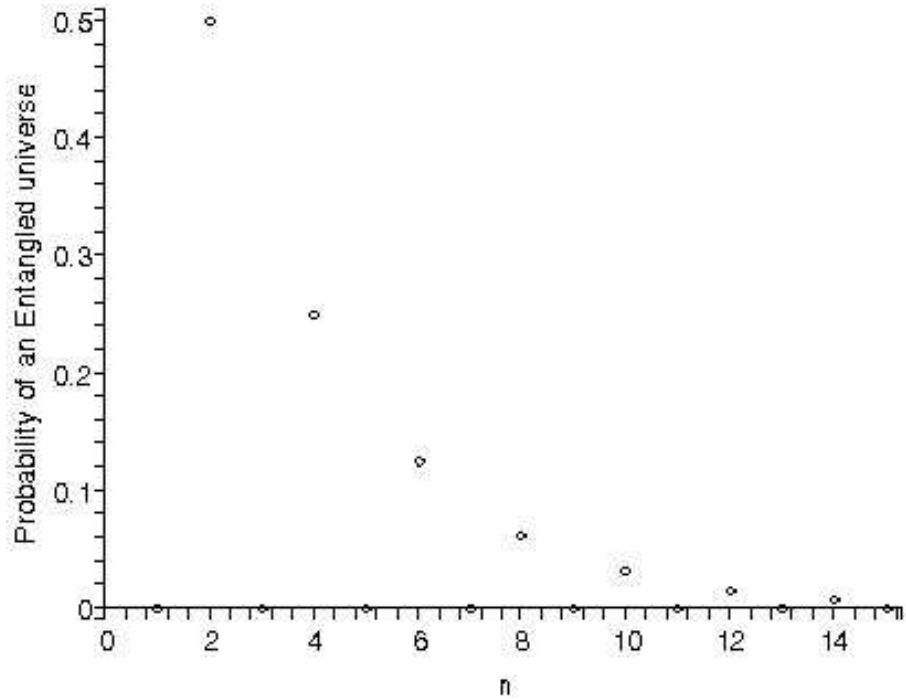}
\caption{Probability, $P(\Psi
_{n}=[c/d]),$ of an entangled universe after $n$ steps.}\label{Figure 8.1}
\end{center}
\end{figure}

As is evident from the graph, the outcome of the presented Rules governing a
universe in an initial state $\Psi _{0}=c$ is similar to the result of the
previous toy-model universe (for the case when $\Psi _{0}$ equalled $a),$ in
the respect that the wavefunction $\Psi _{n}$ will definitely be separable
at periodic intervals: in this example, $\Psi _{n}$ will always be separable
when $n$ is odd, whereas in Example $I$ the state was always separable when $%
n$ was even. The major difference between these two examples, however, is
that in the current case the probability of getting an entangled state at
the remaining times decreases as the number of jumps, $n,$ increases,
whereas in Example $I$ this probability remained at unity.

This model therefore has an important physical interpretation. The situation
here is of a universe that begins in an initial entangled state, but is
driven to a greater likelihood of separability as it develops. Obvious
comparisons can hence consequently be made with the discussions of Chapters
3, 4 and 5 in which it was suggested how the actual Universe may have
developed in an analogous manner, from an initial entangled state at the
quantum Big Bang, to one that now appears to possess an enormous amount of
semi-classical looking separability and persistence.

\bigskip

\subsubsection{Example III}

\bigskip

The dynamics of the models in sub-sections 8.3.1 and 8.3.2 could be
described as `semi-deterministic'. That is, although the quantum reduction
of the state into one of the eigenvectors of the operator is a stochastic
process, if it is known which state $\Psi _{n}$ the universe is in, it is
always possible to say with certainty what the next test $\hat{\Sigma}_{n+1}$
will be. The models in the previous two sub-sections provided examples of
deterministic List-Sort dynamics, that is, Type $III$ Rules.

It is, however, possible to consider a development mechanism based on
probabilistic List-Sort, Type $IIIa$ dynamics, as alluded to in Section 8.2.
Under such circumstances, a given state $\Psi _{n}$ may imply a number of
potential `candidates' to become the next operator $\hat{\Sigma}_{n+1},$ but
which is actually chosen depends on some kind of random factor. Thus, each
potential operator is associated with a particular probability of being
chosen, given the presence of a certain state.

Of course, there are many forms that these various operator probabilities
could take. Firstly, for example, they could simply be fixed `weighting
factors', where each operator on the list is associated with a fixed
probability of being chosen, given a particular property of the state (for
instance, whether it is entangled or separable). Secondly, however, they
could instead involve a Rule in which these probabilities themselves are a
function of the current state, as will be explained later. Thirdly, even,
the choice of operator probability could actually depend somehow on some
sort of higher order quantum process that would be in need of
definition.\bigskip

For the time being, attention will be restricted to the simplest possible
type of situation, and a model will be considered in which the `random
factor' is constant. In the following case of Example $IV$ in Sub-section
8.3.4, however, this factor will instead be a variable that depends upon
which particular state the universe is currently in.

Thus for the example at hand, an elementary dynamics is suggested in which
if the state of the universe has one particular `property' then it is
developed in one way, but if it has another `property' then it will instead
be developed in one of two potential ways, though which of these ways is
actually chosen is a random process. Specifically, the probability that it
will be tested by the first of the two potential operators is defined to be
a constant, $Q,$ whereas the probability that the next state will be one of
the eigenstates of the alternative potential operator is given by $(1-Q),$
where $0\leq Q\leq 1.$\bigskip

Consider as before a two qubit system, and consider again the above List $%
\frak{L}_{Y},$ defined as $\frak{L}_{Y}\equiv \{\frak{B}_{(2,2)},\frak{B}%
_{(4,0)},\frak{B}_{(0,4)}\}.$ Suppose further that the universe follows
probabilistic List-Sort dynamics, and is governed by the Rules

\begin{itemize}
\item  If the state of the universe is entangled, then the basis $\frak{B}%
_{(0,4)}$ is picked from the list $\frak{L}_{Y},$ and $\Psi _{n}$ is tested
by an operator $\hat{\Sigma}_{n+1}=\hat{B}_{(0,4)},$ such that the next
state $\Psi _{n+1}$ is one of the members of $\frak{B}_{(0,4)};$

\item  However, if $\Psi _{n}$ is separable there is a probability $Q$ that
the operator $\hat{B}_{(2,2)}$ will be chosen from the list to be $\hat{%
\Sigma}_{n+1},$ but also a probability $(1-Q)$ that $\hat{B}_{(4,0)}$ will
be selected instead. Thus if $\Psi _{n}$ is separable, there is a
probability $Q$ that the next state $\Psi _{n+1}$ will be one of the
elements of $\frak{B}_{(2,2)},$ but a probability $(1-Q)$ that it will
instead be one of the elements of $\frak{B}_{(4,0)}.$
\end{itemize}

Here $\frak{B}_{(2,2)}=\{a,b,c,d\},$ $\frak{B}_{(4,0)}=\{e,f,g,h\}$ and $%
\frak{B}_{(0,4)}=\{j,k,l,m\}$ are as defined previously, and $0\leq Q\leq 1.$
Clearly, such a universe is governed by a Type $IIIa$ mechanism.

As in the previous examples, the two qubit universe described here will
develop in an automatic way, with its state jumping from one vector to the
next according to the Rules that govern the model's dynamics.

Also as in the previous examples, the individual probability amplitudes will
play a crucial role in determining the propensity for a given state $\Psi
_{n}$ to jump to a particular future state $\Psi _{n+1},$ within, of course,
the boundaries set by the Rules. In fact, it is easy to generalise Table 8.1
for the 144 probabilities given by $|\langle y|x\rangle |^{2}$ for $\Psi
_{n}=|x\rangle ,$ $\Psi _{n+1}=|y\rangle $ and $x,y=a,b,c,...,m,$ noting
again that many of the transitions would be `dashed' because they are
forbidden in the current mechanism.\bigskip

To illustrate the type of dynamics proposed here, assume that as in Example $%
II,$ the two qubit universe may be prepared in the initial state $\Psi
_{0}=c=\frac{1}{\sqrt{2}}\left( |01\rangle +|10\rangle \right) ,$ where it
is again noted that alternative initial conditions would not give rise to
significantly different or interesting outcomes.

Clearly, $\Psi _{0}=c$ is entangled, so the Rules dictate that the next
operator $\hat{\Sigma}_{1}$ will be $\hat{B}_{(0,4)},$ such that the next
state $\Psi _{1}$ will be either $j,k,l$ or $m$ with relative probabilities
of $1/2,$ $0,$ $1/4$ and $1/4$ respectively.

Now, if $\Psi _{1}$ turns out to be $\Psi _{1}=j,$ it is evident that the
universe has collapsed to a separable state (actually, the same would be
true whether it had collapsed to $l$ or $m,$ but that is beside the point).
So, according to the Rules, the next operator $\hat{\Sigma}_{2}$ to test the
state will either be $\hat{\Sigma}_{2}=\hat{B}_{(2,2)}$ with probability $Q,$
or else $\hat{\Sigma}_{2}=\hat{B}_{(4,0)}$ with probability $(1-Q).$
Overall, then, the next state $\Psi _{2}$ will be one of eight
possibilities: it will be one of the vectors $a,$ $b,$ $c,$ $d,$ $e,$ $f,$ $%
g,$ $h$ with relative probabilities given in Table 8.2.
\begin{equation}
\begin{tabular}{|l|l|l|l|l|l|l|l|l|}
\hline
$\mathbf{\Psi }_{2}$ & \ $a$ & $\ b$ & $\ c$ & $d$ & $\ \ \ \ e$ & $f$ & $\
\ \ \ g$ & $h$ \\ \hline
\textbf{Prob.} $P(\mathbf{\Psi }_{2},\mathbf{\Psi }_{1}=j)$ & $Q/4$ & $Q/4$
& $Q/2$ & $0$ & $(1-Q)/2$ & $0$ & $(1-Q)/2$ & $0$ \\ \hline
\end{tabular}
\tag*{Table 8.2}
\end{equation}

Of course, similar tables would be generated for the probabilities of
jumping to a particular state $\Psi _{2}$ from the alternative states $\Psi
_{1}=l$ or $\Psi _{1}=m.$ In these instants, the same set $\{a,...,h\}$ of
eight possible vectors would be present, because $l$ and $m$ are both
separable and would hence both imply that $\hat{\Sigma}_{2}=\hat{B}_{(2,2)}$
with probability $Q$ or $\hat{\Sigma}_{2}=\hat{B}_{(4,0)}$ with probability $%
(1-Q),$ but the various quantum probability amplitudes that result would now
be different.

As in the previous examples, the above model could be developed through an
arbitrary number of steps to give rise to complicated `trees' of possible
histories for the system, each with a particular probability of occurring.
Also as before, various questions can be asked regarding the possible nature
of the system after a given number of jumps.\bigskip

The number $Q$ is seen as a free parameter in the model. Two particular
situations, however, are the special cases when $Q=0$ and $Q=1.$ For $Q=0,$
the development proceeds in a similar way to that experienced by the system
described in Example $I,$ that is, a universe with a state that oscillates
between being certainly entangled and certainly separable.

In the converse case of $Q=1,$ however, the model is instead identical to
the system described in Example $II,$ that is, a universe with a state that
it increasingly more likely to be separable as it develops.

But, a more novel situation occurs for $0<Q<1.$ In these cases the
probability that the universe is separable or entangled after $n$ jumps
tends to some fixed value as $n$ becomes large. Additionally, unlike for $%
Q=1,$ in which for $n=odd$ the state is always separable and it is only the $%
n=even$ states whose probability of being entangled is driven to zero, for $%
0<Q<1$ the probability that the state is entangled (or separable) tends to
the same fixed value for both odd and even values of $n;$ at any `time' $n$
there is always a possibility that the state could be entangled. In such a
universe, the likelihood of the $n^{th}$ state being separable for $n\gg 1$
is approximately the same as the likelihood of the $(n-1)^{th}$ state also
being separable. This point is illustrated in Figure 8.2, which is a plot of
the probability (y-axis) of getting an entangled state after $n$ jumps
(x-axis) for $Q=1/2.$

Such a model may have an important physical interpretation in terms of
discussions regarding the emergence of persistence.

%\FRAME{fhFU}{331.875pt}{257.75pt}{0pt}{\Qcb{Probability of an entangled
%universe after $n$ steps for $Q=1/2.$}}{}{Figure 8.2}{\special{language
%"Scientific Word";type "GRAPHIC";maintain-aspect-ratio TRUE;display
%"USEDEF";valid_file "T";width 331.875pt;height 257.75pt;depth
%0pt;original-width 325.1875pt;original-height 252.1875pt;cropleft
%"0";croptop "1";cropright "1";cropbottom "0";tempfilename
%'I1KOB208.wmf';tempfile-properties "XPR";}}

\begin{figure}[th]
\begin{center}
\includegraphics[width=4in]{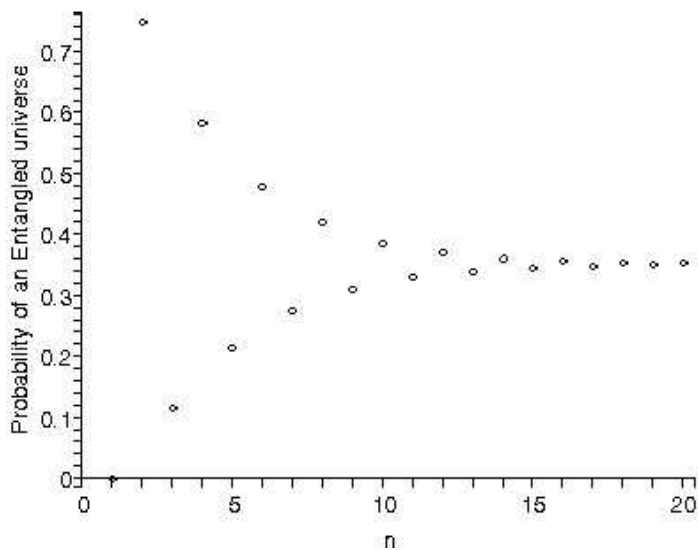}
\caption{Probability of an entangled
universe after $n$ steps for $Q=1/2.$}\label{Figure 8.2}
\end{center}
\end{figure}

Of course, it is not in principle difficult to determine what the
probability of obtaining an entangled state tends to as $n$ becomes large,
for the given initial condition, value of $Q,$ and set of Rules. It may also
be interesting to consider the rate at which the probability tends to this
fixed value. In fact, since from the above graph the convergence to this
value appears `smooth' (in some sense), it is not impossible to suggest that
the probability of obtaining an entangled universe might begin to be
approximated by a continuous function of $n,$ particularly as $n$ increases.
In short, it might be possible to fit a continuous curve to the above data,
and the equation of this curve might play an important role in discussions
regarding the emergence of continuous physics from the underlying and
discrete pregeometric structure.

The same argument may also be true for the results indicated by Figure 8.1
of Example II: a continuous `decay' curve could be fitted to the results
when $n$ is even, and the equation of this curve might be useful in
approximating average properties of the system.\bigskip

Continuing this train of thought, and by considering the various
probabilities that the universe will be in a particular state $\Psi _{n}\in
\{a,...,m\}$ after $n$ steps, it is noted that it is possible to describe
the likely `trajectories' or `histories' of the system between times $0$ to $%
n.$ In other words, by considering the probabilities of obtaining various
`histories' $\Psi _{0}\rightarrow \Psi _{1}\rightarrow ...\rightarrow \Psi
_{n},$ a discrete and probabilistic `equation of motion' could be determined
for the development of the universe. Moreover, it might then be the case
that this too could be approximated using continuous looking laws and
functions of $n.$ In this case, therefore, the discrete process of jumps of
the system would, in some sense, effectively begin to be described by
continuous equations, exactly as required for the emergence of continuous
physics.

It is intriguing to speculate on the potential links between this type of
analysis of the presented models, and the types of dynamics discussed in
models of quantum stochastic calculus (e.g. \cite{Bouten}).

\bigskip

\subsubsection{Example IV}

\bigskip

Example $III$ is a probabilistic List-Sort universe that develops in a way
that depends on a fixed probability: the choice of operator $\hat{B}_{(2,2)}$
or $\hat{B}_{(4,0)}$ is influenced by the value of $Q,$ and $Q$ remains
constant throughout. A natural extension to this type of mechanism is
therefore to allow the probability of using $\hat{B}_{(2,2)}$ or $\hat{B}%
_{(4,0)}$ to depend on the current state.

It is important to clarify the difference between these two types of
mechanism. In the `fixed' case, the Rules select a set of potential
operators from the List based upon a particular `property' of the state, and
there is then a fixed probability as to which of these operators is actually
used. So, in the case of Example $III,$ the probability of picking a given
operator from the List depended only on a `property' of the state: if the
state was separable, then there was a probability $Q$ of choosing the
operator $\hat{B}_{(2,2)},$ but a probability $(1-Q)$ of instead choosing
the operator $\hat{B}_{(4,0)}.$ Moreover, these probabilities were
independent of what the details of the state actually were: all that was
important was whether it was entangled or separable, because this was the
only property used in the selection process.

In the type of mechanism proposed in this sub-section, however, although the
Rules are such that a particular `property' of the state is still used to
select a set of potential operators, which operator from this set actually
gets chosen does now depend on the precise details of the state. In other
words, the probability of picking a particular operator from the set of
potentials is not a fixed number defined at the outset, but is instead a
variable defined as a particular function of the state. Thus, and as will be
shown below, in the type of mechanism proposed here it is not sufficient to
simply say whether the state $\Psi _{n}$ is entangled or separable in order
to determine the propensity of using a particular next operator; it is also
necessary to know exactly what the state is in order to determine the
probability of what the next operator $\hat{\Sigma}_{n+1}$ will be.\bigskip

The general case of this type of idea is therefore the following. Given a
state $\Psi _{n}$ in a $D$ dimensional Hilbert space $\mathcal{H}$ with a
List of bases $\frak{L}$ defined as $\frak{L}\equiv \{\frak{B}%
_{a}:a=1,2,...,l\},$ the Rules could pick out a set of potential operators $%
\{\hat{B}_{i},$ $\hat{B}_{j},$ $\hat{B}_{k},...\}$ to be the next test $\hat{%
\Sigma}_{n+1},$ with respective probabilities $p_{i},$ $p_{j},$ $p_{k},...$
However, unlike in the previous, fixed probability case for which $p_{i},$ $%
p_{j},$ $p_{k},...$ were constants, a mechanism is now considered where
\begin{equation}
p_{i}=f_{i}(\Psi _{n})\text{ \ \ , \ \ }p_{j}=f_{j}(\Psi _{n})\text{ \ \ , \
\ }\ldots
\end{equation}
with the actual functions $\{f_{i},$ $f_{j},...\}$ defined in the Rules
governing the dynamics, and where
\begin{equation}
f_{i}(\Psi _{n})+f_{j}(\Psi _{n})+...=1.
\end{equation}
\qquad \qquad

The idea is perhaps best illustrated by example.\bigskip

Consider again a two qubit universe, and the basis sets $\frak{B}%
_{(2,2)}=\{a,b,c,d\},$ $\frak{B}_{(4,0)}=\{e,f,g,h\}$ and $\frak{B}%
_{(0,4)}=\{j,k,l,m\}$ defining the List $\frak{L}_{Y}$ as before. In this
illustration, the Rules governing the system are chosen to be analogous to
those used in Example $III;$ viz.

\begin{itemize}
\item  If $\Psi _{n}$ is entangled then the basis $\frak{B}_{(0,4)}$ is
picked from the list $\frak{L}_{Y},$ and the next state $\Psi _{n+1}$ is one
of the eigenstates of $\hat{B}_{(0,4)};$

\item  However, if $\Psi _{n}$ is separable then there is a probability $R$
that the basis $\frak{B}_{(2,2)}$ is picked from the list $\frak{L}_{Y},$
such that the next state $\Psi _{n+1}$ is one of the eigenstates of $\hat{B}%
_{(2,2)},$ but a probability $(1-R)$ that the basis $\frak{B}_{(4,0)}$ is
instead picked from the list $\frak{L}_{Y},$ such that the next state $\Psi
_{n+1}$ is one of the eigenstates of $\hat{B}_{(4,0)}.$
\end{itemize}

In this case, however, $R$ is not a constant, but is a function of $\Psi
_{n}.$ Specifically, $R$ could be defined in this example by
\begin{equation}
R=|\langle \Psi _{n}|X\rangle |^{2}=|\langle X|\Psi _{n}\rangle |^{2}
\label{R}
\end{equation}
where $X$ is some fixed `reference' vector that is normed such that $0\leq
R\leq 1.$

Clearly, the value of $R$ depends upon which state the universe is currently
in, thereby making the dynamics strongly self-referential. As a consequence,
it turns out that some separable states $\Psi _{n}$ are more likely to be
tested by the operator $\hat{B}_{(2,2)}$ whilst others are more likely to be
tested by $\hat{B}_{(4,0)},$ depending of course on the magnitude of the
inner product of $\Psi _{n}$ with $X.$\bigskip

For the sake of this illustration, $X$ could arbitrarily be chosen as
\begin{equation}
X=\frac{1}{2}\left( |00\rangle +|10\rangle +|01\rangle +|11\rangle \right)
\end{equation}
so that in fact $X=j.$ Thus, $R$ will be given by one of the square
amplitudes $R=|\langle y|X=j\rangle |^{2}$ for $y=a,b,j,k,l,m,$ noting that $%
\{a,b,j,k,l,m\}$ are the only separable states, and hence the only states of
relevance here. Clearly, the values of $\langle a|j\rangle $ and $\langle
b|j\rangle $ may be readily extracted from Table 8.2 (by putting $Q=1),$
whilst $\langle j|j\rangle =1$ with $\langle k|j\rangle =\langle l|j\rangle
=\langle m|j\rangle =0,$ and this highlights the above point that some
states are considerably more likely to be tested by, say, $\hat{B}_{(2,2)}$
than others.

So as an example, if the universe is known to be in the separable state $%
\Psi _{n}=a,$ the Rules dictate that the next operator $\hat{\Sigma}_{n+1}$
will be $\hat{B}_{(2,2)}$ with probability $|\langle a|j\rangle |^{2},$ but
will be $\hat{B}_{(4,0)}$ with probability $(1-|\langle a|j\rangle |^{2}).$

Furthermore, given the state $\Psi _{n}=a,$ the `compound' probability $%
P(e,a)$ that the next state $\Psi _{n+1}$ will be the element $e$ of the
basis set $\frak{B}_{(4,0)}$ is clearly given by
\begin{eqnarray}
P(e,a) &=&(1-|\langle a|j\rangle |^{2})|\langle e|a\rangle |^{2}
\label{prod} \\
&=&(3/4)\text{ }(1/2)\text{ }=3/8  \notag
\end{eqnarray}
which is just the product of the probability that $a$ will be tested by the
operator $\hat{B}_{(4,0)}$ multiplied by the probability that the outcome of
this test will be $e.$

Of course, if $\Psi _{n+1}$ is indeed the entangled state $e,$ then the next
test will be $\hat{\Sigma}_{n+2}=\hat{B}_{(0,4)},$ and the subsequent state\
$\Psi _{n+2}$ will be one of the elements $\{j,k,l,m\}$ of the basis set $%
\frak{B}_{(0,4)}.$ Evidently, and as in previous models, it is easy to
continue this process indefinitely and generate a `tree' of sets of possible
`histories' from a given initial state. It is also possible to ask questions
of the system, such as the probability of obtaining a certain $\Psi _{N}$ at
time $N,$ or whether the universe at time $M$ is likely to be entangled or
separable.

\bigskip

As with Examples $II$ and $III,$ the model presented here also has an
important physical interpretation. In the system described in this
sub-section, a dynamics is presented in which the presence of certain states
leads to a greater propensity that the universe will be developed by a
particular operator. In other words, some states are more likely to be
tested by certain operators than others.

This, however, is generally what occurs in the real Universe. Given a state
that is separable into a number of particular factors (representing an
apparatus, a subject, a physicist etc.), it is often possible to predict
what the next operator may be like, because scientists are generally able to
set up certain experiments in the laboratory, and represent them by
Hermitian operators. Furthermore, the presence of a particular set of
initial sub-states does generally seem to make some choices of test
considerably more likely than others. As an example, if a Stern-Gerlach
machine and an electron are present as factors of $\Psi _{n},$ it might be
expected that the Universe will select an operator that appears to represent
a spin-measurement; indeed, this course of action certainly seems more
likely than an alternative choice of operator being made, where, perhaps,
the next state $\Psi _{n+1}$ of the universe appears to contain the results
of some sort of position measuring experiment.

Thus, the point is that certain initial conditions, i.e. certain states $%
\Psi _{n},$ do appear to constrain the Universe to develop in certain ways;
a particular state $\Psi _{n}$ does seem to make a particular test more
likely than others.\bigskip

These issues themselves lead onto a general philosophical point.
Conventional quantum mechanics generally deals with statements of the form:
``if a given quantum system is tested in a certain way, what is the
probability that a certain outcome will be measured?''. However, this view
ignores the more fundamental question that should perhaps be asked first:
``what is the probability that a physicist will choose to apply that
particular test to the system anyway?''.

Such a question is presumably an important feature in a fully quantum
universe. As has been discussed previously, if the Universe contains
everything, there can be no external agent acting as ``The Physicist''
deciding which test to apply to its state at any particular time. The choice
of operator acting upon the wavefunction must therefore be a result of
something going on inside the Universe. Further, assuming that human
physicists are themselves quantum systems (or at least are comprised of
quantum systems), they must be subject to quantum laws and are the outcomes
of quantum tests. Thus, any `decision' they appear to make, regarding the
selection of a particular operator to test their surroundings with, is the
result of an earlier quantum process\footnote{%
It is intended here to ignore arguably metaphysical notions that involve
free-will or consciousness.}. This conclusion was very much the stance of
Feynman \cite{Feynman1}.

So, given that the Universe is represented by a quantum state, and that this
state is one of the eigenvectors of a quantum operator, physicists are left
with the question of why this particular operator was selected. Exactly how
this selection mechanism might work, and whether it is based on a
deterministic algorithm or the stochastic result of quantum probability, are
interesting questions seldom addressed in a science normally concerned with
predicting the answers to specific, well defined questions. In a
self-referential, quantum Universe featuring endo-physical observers,
however, they must be unavoidable considerations.

\bigskip

\subsection{Generated-Sort Dynamics}

\bigskip

In Type $III$ List-Sort dynamics, the universe possesses an enormous set $%
\frak{L}$ of `pre-ordained' operators $\{\hat{B}_{1},\hat{B}_{2},...,\hat{B}%
_{l}\},$ and the Rules $R_{n}$ select just one of these to be the next test $%
\hat{\Sigma}_{n+1}$ based on the `properties' of the current state $\Psi
_{n}.$

A converse to this sort of mechanism would therefore be one in which there
is no pre-existing set of operators waiting to be picked to test the state.
Under circumstances such as these, a dynamics could be imagined in which the
next operator $\hat{\Sigma}_{n+1}$ is somehow `created' at time $n$ in a
manner that is based entirely upon the current state. In other words, in
these scenarios the next operator $\hat{\Sigma}_{n+1}$ is not selected from
an already existing List, but is instead generated from $\Psi _{n}$
according to\ the Rules governing the universe in question.

Thus, the operator $\hat{\Sigma}_{n+1}$ could be taken to be some sort of
function $f_{n}$ of the current state, and it would be possible to write
\begin{equation}
\hat{\Sigma}_{n+1}=f_{n}(\Psi _{n})
\end{equation}
or equivalently
\begin{equation}
\frak{B}_{n+1}=f_{n}(\Psi _{n})
\end{equation}
where $\frak{B}_{n+1}\equiv \{\Phi _{n+1}^{1},\Phi _{n+1}^{2},...,\Phi
_{n+1}^{D}\}$ for a $D$ dimensional Hilbert space, and $\langle \Phi
_{n+1}^{i}|\Phi _{n+1}^{j}\rangle $ $=\delta _{ij}$ for $i,j=1,2,...D.$

Such a mechanism may be called a `\textit{Generated-Sort}' dynamics.\bigskip

Generated-Sort dynamics could lead to a `phase space' of possible states
that is much larger than that available in List-Sort dynamics, where the
phase space is defined in terms of the number of \textit{different} states
the universe could exist in over all time $n.$ In particular, in
Generated-Sort dynamics this set of different states could be unbounded,
whereas in List-Sort dynamics the total set of possible states will always
be constrained according to the size of the List.

To justify this last point, note that a finite List $\frak{L}$ of operators
acting over a finite Hilbert space implies a finite number of possible
states. Specifically, given that every possible Hermitian operator in a
Hilbert space of dimension $D$ possesses $D$ orthogonal eigenstates, then if
the number $l$ of possible equivalent classes of operators contained in $%
\frak{L}$ is finite (i.e. $\frak{L}$ \ is of List Length $l),$ there can be
no more than $D\times l$ different states for the universe, such that the
universe's phase space is of `size' $D\times l.$ In other words, as the
universe develops, its state will always be one of these $D\times l$
different possibilities. Of course, quantum stochastics does still ensure
that it is generally impossible to say in advance \textit{which} of this set
the $n^{th}$ state will be.

In fact, this point is evident from the examples of Section 8.3: for any two
qubit model with the List $\frak{L}_{X}$ of List Length $2,$ the universe
could only ever be in one out of no more than eight possible states, $%
\{a,b,...,h\}.$

In Generated-Sort dynamics, however, the set of different possible states
could potentially be limitless, depending of course on the exact details of
the function $f_{n}(\Psi _{n}).$ Specifically, the set of different possible
futures for a state obeying Generated-Sort dynamics could, in principle,
increase exponentially with $n.$

\bigskip

There are two obvious ways that may be introduced in order to achieve a
Generated-Sort dynamics according to the Rule $\frak{B}_{n+1}=f_{n}(\Psi
_{n}).$ These may be called

\begin{enumerate}
\item  Basis Method;

\item  One-to-Many Method;
\end{enumerate}

and are each described in turn.

In the Basis Method, the individual elements $\{\Phi _{n+1}^{1},\Phi
_{n+1}^{2},...,\Phi _{n+1}^{D}\}$ of $\frak{B}_{n+1}$ are different
functions of the state $\Psi _{n}.$ Thus, in this mechanism it is assumed
that (for each $n)$ the function $f_{n}$ really implies a set of $D$
`sub-functions'
\begin{equation}
f_{n}=\{f_{n}^{(1)},f_{n}^{(2)},...,f_{n}^{(D)}\}
\end{equation}
such that the basis set $\frak{B}_{n+1}$ is given according to the Rule
\begin{equation}
\frak{B}_{n+1}=f_{n}(\Psi _{n})=\{f_{n}^{(1)}(\Psi _{n}),f_{n}^{(2)}(\Psi
_{n}),...,f_{n}^{(D)}(\Psi _{n})\}
\end{equation}
where clearly
\begin{equation}
\Phi _{n+1}^{i}=f_{n}^{(i)}(\Psi _{n})\text{ \ \ , \ \ }i=1,2,...,D
\end{equation}
with the constraints that the sub-functions $\{f_{n}^{(i)}\}$ are defined so
that $f_{n}^{(i)}(\Psi _{n})$ is orthogonal to $f_{n}^{(j)}(\Psi _{n})$ for
all $i\neq j,$ $\left| f_{n}^{(i)}(\Psi _{n})\right| =1,$ and $%
f_{n}^{(i)}(\Psi _{n})\neq \Psi _{n}.$ Clearly, each $f_{n}^{(i)}$ is a $%
1\rightarrow 1$ function that maps a given state $\Psi _{n}$ to a different,
unique vector $\Phi _{n+1}^{i}.$

Conversely, in the One-to-Many Method it is assumed that $f_{n}$ is instead
defined as some sort of $1\rightarrow D$ function that maps the state $\Psi
_{n}$ into $D$ different, orthogonal vectors. These $D$ states are then
taken to form the basis set $\frak{B}_{n+1}.$

The exact mechanics and viability of these two potential methods is explored
in the following sub-sections.\bigskip

Note first, however, that a Rule of the form $\frak{B}_{n+1}=f_{n}(\Psi
_{n}) $ is manifestly deterministic: given a state $\Psi _{n},$ it is
assumed that the function $f_{n}$ is used to generate a unique basis set $%
\frak{B}_{n+1}.$ The Generated-Sort mechanisms could, though, be extended to
a probabilistic (Type $IIIa)$ dynamics in the obvious way, by re-writing the
Rule for generating $\frak{B}_{n+1}$ as
\begin{equation}
\frak{B}_{n+1}=\left\{
\begin{array}{c}
f_{n}(\Psi _{n})\text{ with Probability }P_{f} \\
g_{n}(\Psi _{n})\text{ with Probability }P_{g} \\
\vdots
\end{array}
\right\}
\end{equation}
where the probabilities $P_{f},P_{g},...$ of using the various functions $%
f_{n},$ $g_{n},...$ sum as
\begin{equation}
P_{f}+P_{g}+...=1.
\end{equation}

As with their analogies in List-Sort dynamics, the forms of the above
probabilities could themselves be fixed or variable, depending of course on
the Rules governing the system.

However, a modification from a deterministic (Type $III)$ Rule to a
probabilistic (Type $IIIa)$ one does not significantly add to the discussion
presented in this section. The possibility of the above extension will
therefore be taken for granted from now on, and will hence not be explored
further.

\bigskip

\subsubsection{The Basis Method}

\bigskip

As indicated above, in the Basis Method the next basis set $\frak{B}%
_{n+1}=\{\Phi _{n+1}^{1},\Phi _{n+1}^{2},...,\Phi _{n+1}^{D}\}$ of
orthogonal eigenstates is generated from the vector $\Psi _{n}$ according to
the rule $\Phi _{n+1}^{i}=f_{n}^{(i)}(\Psi _{n})$ for $i=1,...,D.$ The
question then becomes: what sort of functions $f_{n}^{(i)}$ are able to give
rise to such a mechanism?

To begin to answer this, note that each $f_{n}^{(i)}$ is necessarily a
function that maps a vector $\Psi _{n}$ in $\mathcal{H}^{(D)}$ uniquely into
another vector $\Phi _{n+1}^{i}$ in $\mathcal{H}^{(D)}.$ It is hence
possible to associate with $f_{n}^{(i)}$ a unitary operator $\hat{U}%
_{n}^{(i)}$ that achieves the same end, that is
\begin{equation}
\hat{U}_{n}^{(i)}\Psi _{n}=\Phi _{n+1}^{i}.
\end{equation}

Thus, the procedure that creates the basis set $\frak{B}_{n+1}$ from the
state $\Psi _{n}$ may be performed by defining a set of unitary operators, $%
\{\hat{U}_{n}^{(i)}:i=1,...D\}.$

Now, the actual forms of these unitary operators are generally seen as free
parameters in the model, defined, perhaps, by whatever Rules govern the
system. But, because the basis $\frak{B}_{n+1}$ must contain a set of
orthonormal vectors, that is
\begin{equation}
\langle \Phi _{n+1}^{i}|\Phi _{n+1}^{j}\rangle =\delta _{ij}\text{ \ \ , \ \
}i,j=1,2,...,D  \label{Ch6Condition}
\end{equation}
it is necessary that whatever definition is chosen, the set of operators $\{%
\hat{U}_{n}^{(i)}\}$ must satisfy the constraint
\begin{equation}
\langle \Psi _{n}\hat{U}_{n}^{(i)}|\hat{U}_{n}^{(j)}\Psi _{n}\rangle =\delta
_{ij}\text{ \ \ , \ \ }i,j=1,2,...,D  \label{Ch6Cons}
\end{equation}
such that the product operation $(\hat{U}_{n}^{(i)\ast }\hat{U}_{n}^{(j)})$
acting on the state $\Psi _{n}$ maps it to an orthogonal vector\footnote{%
Note throughout that the `starred' operator $\hat{U}^{\ast }$ denotes the
complex conjugate transpose of the operator $\hat{U};$ this is equivalently
represented in some textbooks as $\overline{\hat{U}^{T}},$ or $\hat{U}%
^{\dagger },$ or even $\hat{U}^{+}.$\ Clearly, for unitary operators $\hat{U}%
^{\ast }=\overline{\hat{U}^{T}}=\hat{U}^{-1},$ where the inverse operator $%
\hat{U}^{-1}$ gives $\hat{U}^{-1}\hat{U}=\hat{I},$ with $\hat{I}$ the
identity.} for all $i\neq j.$\bigskip

In fact, from the above discussion\ it turns out that only $(D-1)$ of the
operators $\{\hat{U}_{n}^{(i)}\}$ can be chosen arbitrarily at each time $n,$
and not $D$ as might be expected. Specifically, after the definition of $%
(D-1)$ operators $\{\hat{U}^{(i)}:i=1,2,...,(D-1)\},$ the remaining operator
is immediately defined by constraint.

To demonstrate this explicitly, observe that, without loss of generality, if
$\{\hat{U}_{n}^{(1)},\hat{U}_{n}^{(2)},...,$ $\hat{U}_{n}^{(D-1)}\}$ are
freely chosen according to the constraint that $\langle \Psi _{n}\hat{U}%
_{n}^{(i)}|\hat{U}_{n}^{(j)}\Psi _{n}\rangle =\delta _{ij}$ for\ $%
i,j=1,2,...,(D-1),$ the $``D^{th}"$ operator $\hat{U}_{n}^{(D)}$ is
automatically defined by the condition that only one state is orthogonal to
all of the vectors $\{\hat{U}_{n}^{(1)}\Psi _{n},\hat{U}_{n}^{(2)}\Psi
_{n},...,\hat{U}_{n}^{(D-1)}\Psi _{n}\}.$ Assuming, then, that this `last'
vector is given by $\hat{U}_{n}^{(D)}\Psi _{n},$ it consequently follows
that there can be no freedom in the definition of the $D^{th}$ operator $%
\hat{U}_{n}^{(D)}.$

So, in $D$ dimensional Hilbert spaces $\mathcal{H}^{(D)},$ an orthonormal
basis set $\frak{B}_{n+1}=\{\hat{U}_{n}^{(i)}\Psi _{n}:i=1,...,D)$ cannot be
specified by using $D$ unitary operators if each $\hat{U}_{n}^{(i)}$ is
viewed as a free parameter. In reality, only $(D-1)$ of the operators $\{%
\hat{U}^{(i)}\}$ may actually be chosen freely.

Continuing, in fact, there is\ actually no need to define an operator $\hat{U%
}_{n}^{(D)}$ that maps $\Psi _{n}$ to $\Phi _{n+1}^{D}=$ $\hat{U}%
_{n}^{(D)}\Psi _{n}$ at all; the `remaining' vector $\Phi _{n+1}^{D}$ is
immediately determined by the operations $\{\hat{U}_{n}^{(j)}\Psi
_{n}:i=1,...,(D-1)\}$ and by appealing to the mutual orthogonality of the
elements of $\frak{B}_{n+1}.$

Note that for clarity and to avoid confusion, from now on in this
sub-section, Latin indices $i,j,...$ will generally be used to run from $%
1,2,...,D,$ whereas Greek indices $\mu ,\nu ,...$ will be assumed to run
from $1,2,...,(D-1).$\bigskip

With these comments in mind, it is possible to restate and clarify the Basis
Method Rules. Specifically

\begin{itemize}
\item  Given a state $\Psi _{n}$ and a set of $(D-1)$ unitary operators $\{%
\hat{U}_{n}^{(\mu )}:\mu =1,2,...(D-1)\}$ defined arbitrarily but obeying
the rule
\begin{equation}
\langle \Psi _{n}\hat{U}_{n}^{(\mu )}|\hat{U}_{n}^{(\nu )}\Psi _{n}\rangle
=\delta _{\mu \nu }\text{ \ \ , \ \ }\mu ,\nu =1,2,...,(D-1)
\label{Ch6RCond}
\end{equation}
it is possible to construct a unique basis set of vectors, $\frak{B}_{n+1},$
as
\begin{equation}
\frak{B}_{n+1}=\{\hat{U}_{n}^{(1)}\Psi _{n},\hat{U}_{n}^{(2)}\Psi _{n},...,%
\hat{U}_{n}^{(D-1)}\Psi _{n},\Phi _{n+1}^{D}\}
\end{equation}
where the $D^{th}$ vector is defined according to the constraint
\begin{equation}
\langle \Phi _{n+1}^{D}|\hat{U}_{n}^{(\mu )}\Psi _{n}\rangle =0\text{ \ \ ,
\ \ }\mu =1,2,...,(D-1).  \label{Ch6OrthoCond}
\end{equation}

\item  From this basis set $\frak{B}_{n+1},$ an equivalence class of
Hermitian operators $\hat{B}_{n+1}$ are implied, with eigenstates $\{\Phi
_{n+1}^{i}\}$ equal to
\begin{equation}
\{\Phi _{n+1}^{i}\}=\{\{\hat{U}_{n}^{(\mu )}\Psi _{n}:\mu
=1,...,(D-1)\},\Phi _{n+1}^{D}\}.
\end{equation}
The next test, $\hat{\Sigma}_{n+1},$ of the universe is then taken to be one
of these operators $\hat{B}_{n+1},$ and the universe collapses to the state $%
\Psi _{n+1},$ which is an element of the set $\frak{B}_{n+1}.$
\end{itemize}

Of course, since by definition $\Psi _{n+1}\neq \Psi _{n},$ none of the
unitary operators $\hat{U}_{n}^{(i)}$ may be defined as the identity
operator.\bigskip

It will be useful in the following to consider the `\textit{Reduced basis
set'} $\frak{B}_{n+1}^{R}$ at time $n+1.$ Specifically, $\frak{B}%
_{n+1}^{R}\subset \frak{B}_{n+1}$ is defined as the set of vectors
\begin{equation}
\frak{B}_{n+1}^{R}=\{\hat{U}_{n}^{(1)}\Psi _{n},\hat{U}_{n}^{(2)}\Psi
_{n},...,\hat{U}_{n}^{(D-1)}\Psi _{n}\},
\end{equation}
where the unitary operators $\{\hat{U}_{n}^{(\mu )}:\mu =1,...,(D-1)\}$
acting on the state\ $\Psi _{n}$ obey the condition (\ref{Ch6RCond}). The
actual next basis set $\frak{B}_{n+1}$ consequently comprises of this
Reduced basis set $\frak{B}_{n+1}^{R}$ and a vector $\Phi _{n+1}^{D}$
obeying (\ref{Ch6OrthoCond}) that is orthogonal to every element of $\frak{B}%
_{n+1}^{R}.$ Thus
\begin{equation}
\frak{B}_{n+1}=\{\frak{B}_{n+1}^{R},\Phi _{n+1}^{D}\}.
\end{equation}

Clearly, because the $D^{th}$ vector $\Phi _{n+1}^{D}$ is provided by
constraint, the central task for the Basis Method Rules lies in defining a
set of $(D-1)$ unitary operators $\{\hat{U}_{n}^{(\mu )}:\mu =1,...,(D-1)\}$
that can be used to generate the Reduced basis set $\frak{B}_{n+1}^{R}.$%
\bigskip

Before discussing this further, however, note as an aside that in a single
qubit universe governed by Basis Method Rules, there is only one free
parameter: $\hat{U}_{n}^{(1)}.$ For a single qubit universe in a two
dimensional Hilbert space $\mathcal{H}^{(2)},$ only one unitary operator $%
\hat{U}_{n}^{(1)}$ needs be specified in order to generate a unique basis
set of vectors, because if $\frak{B}_{n+1}=\{\Phi _{n+1}^{1},\Phi
_{n+1}^{2}\}$ and $\Phi _{n+1}^{1}$ is defined as $\Phi _{n+1}^{1}=\hat{U}%
_{n}^{(1)}\Psi _{n},$ the remaining vector $\Phi _{n+1}^{2}$ is given
immediately from the orthogonality condition
\begin{equation}
\langle \Phi _{n+1}^{2}|\Phi _{n+1}^{1}\rangle =\langle \Phi _{n+1}^{2}|\hat{%
U}_{n}^{(1)}\Psi _{n}\rangle .
\end{equation}

Equivalently, it is evident that the Reduced basis set for a single qubit
universe contains only one member. This discussion is analogous to that
presented later in Sub-section 8.5.2 regarding unitary rotation in single
qubit spaces.

Of course, $\hat{U}_{n}^{(1)}$ is chosen freely, and could be any unitary
operator in $\mathcal{H}^{(2)},$ obviously excluding the identity.

\bigskip

Similarly to every other mechanism used in this chapter to develop the
universe, an important principle of the Basis Method dynamics is that the
Rules should be repeatable. In the present case, such a principle implies
that the Reduced basis set $\frak{B}_{n+2}^{R}$ for the next step must be
generated from the function $f_{n+1}$ acting on $\Psi _{n+1},$ i.e. $\frak{B}%
_{n+2}^{R}=\{\hat{U}_{n+1}^{(\mu )}\Psi _{n+1}:\mu =1,...,(D-1)\}.$ The
universe would then develop in an automatic, iterative way.

So, an issue that is immediately faced concerns how the $(D-1)$ unitary
operators $\{\hat{U}_{n}^{(\mu )}\}$ used to determine $\frak{B}_{n+1}^{R}$
from $\Psi _{n}$ could relate to the set $\{\hat{U}_{n+1}^{(\mu )}\}$ that
will be used to determine $\frak{B}_{n+2}^{R}$ from $\Psi _{n+1}.$
Specifically, a question of particular interest is whether the same set of
operators $\{\hat{U}_{n}^{(\mu )}\}$ could be used in both cases, such that $%
\{\hat{U}_{n}^{(\mu )}\}\equiv \{\hat{U}_{n+1}^{(\mu )}\}.$ In other words,
this question is effectively asking whether it is possible to have Basis
Method dynamics based upon functions $f_{n}^{(i)}=f^{(i)}$ that are constant
for all $n,$ or whether they have to change with $n$ in order for the
proposed mechanism to work.

In fact, if both possibilities are valid, it would consequently lead to two
classes of Basis Method Rules:

\begin{enumerate}
\item[Class 1:]  The set $\{\hat{U}_{n}^{(\mu )}\}$ is fixed for all $n,$
such that
\begin{equation}
\Phi _{n+1}^{\mu }=\hat{U}_{n}^{(\mu )}\Psi _{n}\text{ \ \ , \ \ }\mu
=1,...,(D-1)
\end{equation}

and
\begin{equation}
\Phi _{n+2}^{\mu }=\hat{U}_{n+1}^{(\mu )}\Psi _{n+1}=\hat{U}_{n}^{(\mu
)}\Psi _{n+1}\text{ \ \ , \ \ }\mu =1,...,(D-1)
\end{equation}

and so on. By dropping the now redundant subscripts, such a fixed set of
operators could be denoted by $U_{F}=\{\hat{U}^{(1)},\hat{U}^{(2)},...,\hat{U%
}^{(D-1)}\},$ and, like the List in List-Sort dynamics, would be defined for
all time at the outset. Moreover, the definition of $U_{F}$ would be taken
as a necessary pre-requisite without further justification, just as, for
example, the existence of the underlying Hilbert space is assumed to be.

\item[Class 2:]  The operators $\{\hat{U}_{n}^{(\mu )}\}$ do change with $n,$
such that
\begin{equation}
\Phi _{n+1}^{\mu }=\hat{U}_{n}^{(\mu )}\Psi _{n}\text{ \ \ , \ \ }\mu
=1,...,(D-1)
\end{equation}

but
\begin{equation}
\Phi _{n+2}^{\mu }=\hat{U}_{n+1}^{(\mu )}\Psi _{n+1}\text{ \ \ , \ \ }\mu
=1,...,(D-1)
\end{equation}

where $\hat{U}_{n}^{(\mu )}$ is not (necessarily) equal to $\hat{U}%
_{n+1}^{(\mu )}.$
\end{enumerate}

Evidently, Class $1$ Rules are a special case of Class $2$ Rules, in which $%
\hat{U}_{n+1}^{(\mu )}=\hat{U}_{n}^{(\mu )}$ for all $n$ and $\mu .$

These two possible cases are now discussed in turn.\bigskip

\paragraph{Class 1 Basis Method}

\begin{equation*}
\end{equation*}

The short answer to the above question is that it does not seem likely that
a universe\ is able to develop according to Basis Method Rules that
incorporate a fixed set $U_{F}$ of unitary operators. In other words, whilst
its non-existence has not yet been proved rigorously, no (Type $III)$ Class $%
1$ mechanism has been found that can be used to self-referentially develop
the state of the universe from $\Psi _{n}\rightarrow \Psi _{n+1}\rightarrow
\Psi _{n+2}\rightarrow ...,$ continuing indefinitely, and no such mechanism
is expected to be found.

This conclusion arises because in order for Class $1$ Basis Method Rules to
be valid, it is required that if $\frak{B}_{n+1}^{R}=\{\hat{U}^{(\mu )}\Psi
_{n}\}$ with $\langle \Psi _{n}\hat{U}^{(\mu )}|\hat{U}^{(\nu )}\Psi
_{n}\rangle =\delta _{\mu \nu },$ then $\frak{B}_{n+2}^{R}=\{\hat{U}^{(\mu
)}\Psi _{n+1}\}$ with $\langle \Psi _{n+1}\hat{U}^{(\mu )}|\hat{U}^{(\nu
)}\Psi _{n+1}\rangle =\delta _{\mu \nu },$ for all\ $\mu ,\nu
=1,2,...,(D-1). $ This validity therefore rests on the assumption that the
set $U_{F}=\{\hat{U}^{(\mu )}\}$ defined `initially' to ensure the mutual
orthogonality of the vectors $\{\hat{U}^{(\mu )}\Psi _{n}\}$ may also be
used generate a set of orthogonal vectors $\{\hat{U}^{(\mu )}\Psi _{n+1}\}$
from $\Psi _{n+1}.$

However, a set of unitary operators $U_{F}=\{\hat{U}^{(\mu )}:\mu
=1,...,(D-1)\}$ obeying the constraint $\langle \Psi _{n}\hat{U}^{(\mu )}|%
\hat{U}^{(\nu )}\Psi _{n}\rangle =\delta _{\mu \nu }$ for all $\mu ,\nu $
will not in general also satisfy the relationship $\langle \Theta \hat{U}%
^{(\mu )}|\hat{U}^{(\nu )}\Theta \rangle =\delta _{\mu \nu },$ where $\Theta
\neq \Psi _{n}$ is an arbitrary vector in $\mathcal{H}^{(D)}.$

Specifically, in fact, given an `initial' state $\Psi _{0},$ then if the
operators $\{\hat{U}^{(\mu )}\}$ are in the first instance defined so that
they obey $\langle \Psi _{0}\hat{U}^{(\mu )}|\hat{U}^{(\nu )}\Psi
_{0}\rangle =\delta _{\mu \nu }$ to give the Reduced basis set $\frak{B}%
_{1}^{R},$ the same set $\{\hat{U}^{(\mu )}\}$\ will not then in general
also satisfy the relationship $\langle \Psi _{1}\hat{U}^{(\mu )}|\hat{U}%
^{(\nu )}\Psi _{1}\rangle =\delta _{\mu \nu }$ required for the following
step of the dynamics, where $\Psi _{1}\in \frak{B}_{1}$ and recalling that $%
\Psi _{1}\neq \Psi _{0}$ by definition. So, the set of vectors $\{\hat{U}%
^{(1)}\Psi _{1},\hat{U}^{(2)}\Psi _{1},...,\hat{U}^{(D-1)}\Psi _{1}\}$ will
not in general be orthogonal, and so cannot be used to determine the next
basis set $\frak{B}_{2}.$

In fact, in order for such a set of vectors $\{\hat{U}^{(1)}\Psi _{1},\hat{U}%
^{(2)}\Psi _{1},...,\hat{U}^{(D-1)}\Psi _{1}\}$ to be orthogonal, the set $%
U_{F}$ must satisfy the following condition.\bigskip

\begin{proof}[Condition]
Assume an initial state $\Psi _{n}$ and a set $U_{F}$ of unitary operators $%
\{\hat{U}^{(\mu )}\}$ defined such that $\langle \Psi _{n}\hat{U}^{(\mu )}|%
\hat{U}^{(\nu )}\Psi _{n}\rangle =\delta _{\mu \nu }$ for\ $\mu ,\nu
=1,2,...,(D-1).$ Each $\hat{U}^{(\mu )}$ generates a unique vector $\Phi
_{n+1}^{\mu },$ given by $\Phi _{n+1}^{\mu }=\hat{U}^{(\mu )}\Psi _{n},$
such that the set $\{\Phi _{n+1}^{\mu }\}$ defines the Reduced basis set $%
\frak{B}_{n+1}^{R}.$

Consider also the $``D^{th}"$ operator $\hat{U}^{(D)},$ defined according to
the constraint that it maps $\Psi _{n}$ to the vector $\Phi _{n+1}^{D}$ that
is orthogonal to every $\Phi _{n+1}^{\mu }.$ Then, the set $\{U_{F},\hat{U}%
^{(D)}\}$ acting on the state $\Psi _{n}$ can be used to generate an
orthonormal basis set $\frak{B}_{n+1}=\{\Phi _{n+1}^{i}:i=1,...,D\}.$\bigskip

Now, consider an additional set of $D$ mutually orthogonal vectors $\{\Psi
^{k}:k=1,...,D\},$ defined arbitrarily apart from the condition that the
`first' of these,\ $\Psi ^{1},$ is identical to the state $\Psi _{n},$ i.e. $%
\Psi ^{1}=\Psi _{n}.$ Then, the set $\{\Psi ^{k}\}$ is effectively
equivalent to some basis in $\mathcal{H}^{(D)},$ which may be labelled $%
\frak{B}_{K}.$

Clearly, the subset $\{\Psi ^{j}:j=2,...,D\}$ of $\frak{B}_{K}$ contains an
arbitrary set of vectors that are orthogonal to the current state $\Psi _{n}$
and to each other.

It is possible to find a transformation that maps each of the vectors $\Psi
^{k},$ $k=1,...,D,$ to the state $\Psi _{n}=\Psi ^{1}.$ One such map
involves an operator $V$ defined as
\begin{equation}
V=|\Psi ^{1}\rangle \langle \Psi ^{2}|+|\Psi ^{2}\rangle \langle \Psi
^{3}|+...+|\Psi ^{D-1}\rangle \langle \Psi ^{D}|+|\Psi ^{D}\rangle \langle
\Psi ^{1}|
\end{equation}

with the rule
\begin{equation}
(V)^{k-1}\Psi ^{k}=\Psi ^{1}
\end{equation}

where $(V)^{k-1}$ implies the operator $V$ raised to the $(k-1)^{th}$ power.

So, as an example
\begin{eqnarray}
(V)^{2}\Psi ^{3} &=&VV|\Psi ^{3}\rangle \\
&=&\left(
\begin{array}{c}
{\large (}|\Psi ^{1}\rangle \langle \Psi ^{2}|+|\Psi ^{2}\rangle \langle
\Psi ^{3}|+...+|\Psi ^{D}\rangle \langle \Psi ^{1}|{\large )} \\
\times {\large (}|\Psi ^{1}\rangle \langle \Psi ^{2}|+|\Psi ^{2}\rangle
\langle \Psi ^{3}|+...+|\Psi ^{D}\rangle \langle \Psi ^{1}|{\large )}
\end{array}
\right) |\Psi ^{3}\rangle  \notag \\
&=&{\Large (}|\Psi ^{1}\rangle \langle \Psi ^{3}|+|\Psi ^{2}\rangle \langle
\Psi ^{4}|+...+|\Psi ^{D}\rangle \langle \Psi ^{2}|{\Large )}|\Psi
^{3}\rangle =|\Psi ^{1}\rangle =\Psi _{n}  \notag
\end{eqnarray}

and note that $(V)^{1-1}\Psi ^{1}=\hat{I}\Psi ^{1}$ as expected, where $\hat{%
I}$ is the identity.

Evidently, $V$ is unitary
\begin{eqnarray}
VV^{\ast } &=&\left(
\begin{array}{c}
{\large (}|\Psi ^{1}\rangle \langle \Psi ^{2}|+|\Psi ^{2}\rangle \langle
\Psi ^{3}|+...+|\Psi ^{D}\rangle \langle \Psi ^{1}|{\large )} \\
\times {\large (}|\Psi ^{2}\rangle \langle \Psi ^{1}|+|\Psi ^{3}\rangle
\langle \Psi ^{2}|+...+|\Psi ^{1}\rangle \langle \Psi ^{D}|{\large )}
\end{array}
\right) \\
&=&{\large (}|\Psi ^{1}\rangle \langle \Psi ^{1}|+|\Psi ^{2}\rangle \langle
\Psi ^{2}|+...+|\Psi ^{D}\rangle \langle \Psi ^{D}|{\large )}=\hat{I}.
\notag
\end{eqnarray}

Moreover, $(V)^{k-1}$ is therefore also unitary for all $k,$ as may be
readily shown.

So as an aside, note that the vectors $\Psi ^{k}$ with the operators $%
(V)^{k-1}$ are defined very much in the spirit of the Basis Method from the
state $\Psi ^{1};$ that is, from the unitary transformation
\begin{equation}
\Psi ^{k}=(V^{\ast })^{k-1}\Psi ^{1}.
\end{equation}

Recall now that it is always possible to rotate any orthogonal basis set of
vectors into a second orthogonal basis set of vectors, by using a suitably
defined unitary operator. So, the basis set $\frak{B}_{K}=\{\Psi ^{k}\}$ may
be rotated into the basis set $\frak{B}_{n+1}=\{\Phi _{n+1}^{i}\}$ by a
unitary operator, which may be labelled $\hat{U}^{(n+1,K)}.$

In fact, this possibility implies that each element $\Psi ^{k}\in \frak{B}%
_{K}$ may be uniquely mapped into an element $\Phi _{n+1}^{i}\in \frak{B}%
_{n+1}$ by $\hat{U}^{(n+1,K)},$ such that
\begin{equation}
\Phi _{n+1}^{i}=\hat{U}^{(n+1,K)}\Psi ^{k}
\end{equation}
with the reverse also, of course, holding: $\Psi ^{k}=(\hat{U}%
^{(n+1,K)})^{\ast }\Phi _{n+1}^{i}.$

Now, rewriting $\Psi ^{k}$ as $\Psi ^{k}=(V^{\ast })^{k-1}\Psi ^{1}$ by
definition, and associating the $i^{th}$ vector $\Phi _{n+1}^{i}$\ of $\frak{%
B}_{n+1}$ with the $k^{th}$ vector $\Psi ^{k}$ of $\frak{B}_{K}$ without
loss of generality, gives
\begin{equation}
\Phi _{n+1}^{i}=\hat{U}^{(n+1,K)}(V^{\ast })^{i-1}\Psi ^{1}.
\end{equation}

Moreover, because the product operation $(\hat{U}^{(n+1,K)}(V^{\ast
})^{i-1}) $ is unitary
\begin{eqnarray}
{\large (}\hat{U}^{(n+1,K)}(V^{\ast })^{i-1}{\large )(}\hat{U}%
^{(n+1,K)}(V^{\ast })^{i-1}{\large )}^{\ast } &=&{\large (}\hat{U}%
^{(n+1,K)}(V^{\ast })^{i-1}{\large )(}(V)^{i-1}(\hat{U}^{(n+1,K)})^{\ast }%
{\large )}  \notag \\
&=&\hat{U}^{(n+1,K)}(V^{\ast }V)^{i-1}(\hat{U}^{(n+1,K)})^{\ast }  \notag \\
&=&\hat{U}^{(n+1,K)}\hat{I}(\hat{U}^{(n+1,K)})^{\ast }=\hat{I}
\end{eqnarray}

it may be associated with a single unitary operator, and by denoting this
single operator as $\hat{U}^{(i)},$ it is evident that
\begin{equation}
\Phi _{n+1}^{i}=\hat{U}^{(n+1,K)}(V^{\ast })^{i-1}\Psi ^{1}=\hat{U}%
^{(i)}\Psi _{n}.
\end{equation}

Clearly, then,
\begin{eqnarray}
\langle \Phi _{n+1}^{i}|\Phi _{n+1}^{j}\rangle &=&\langle \Psi _{n}\hat{U}%
^{(i)}|\hat{U}^{(j)}\Psi _{n}\rangle \\
&=&\langle \Psi ^{1}|(V)^{i-1}(\hat{U}^{(n+1,K)})^{\ast }\hat{U}%
^{(n+1,K)}(V^{\ast })^{j-1}|\Psi ^{1}\rangle  \notag \\
&=&\langle \Psi ^{1}|(V)^{i-1}(V^{\ast })^{j-1}|\Psi ^{1}\rangle =\langle
\Psi ^{i}|\Psi ^{j}\rangle =\delta _{ij}  \notag
\end{eqnarray}

as required, so that the product $(\hat{U}^{(n+1,K)}(V^{\ast })^{i-1})$ can
clearly be used to perform the unitary rotation $\hat{U}^{(i)}$ used in the
`standard' Basis Method, assuming of course that $\hat{U}^{(n+1,K)}\neq \hat{%
I}.$

Furthermore, because the vectors $\{\Psi ^{j}:j=2,...,D\}$ were originally
chosen arbitrarily, such an association of $\hat{U}^{(i)}$ with $\hat{U}%
^{(n+1,K)}(V^{\ast })^{i-1}$ can always be found for any definition of the
fixed set $U_{F}$ of unitary operators $\{\hat{U}^{(\mu )}:\mu
=1,...,(D-1)\},$ with the $D^{th}$ operator $\hat{U}^{(D)}$ defined by
constraint. In short, there is always a transformation $\hat{U}^{(n+1,K)}$
and a basis $\frak{B}_{K}$ that defines an operator $V,$ from which a set of
unitary operators $\hat{U}^{(i)}$ may be defined that act on $\Psi _{n}$ to
generate an orthogonal basis $\frak{B}_{n+1};$ similarly, any set of unitary
operators $\hat{U}^{(i)}$ that relate $\Psi _{n}$\ to $\frak{B}_{n+1}$ may
be constructed in the above manner by defining a suitable transformation $%
\hat{U}^{(n+1,K)}$ and basis $\frak{B}_{K}.$\bigskip

Without loss of generality, assume that on testing $\Psi _{n}$ by an
operator $\hat{\Sigma}_{n+1}$ with eigenvectors $\{\Phi
_{n+1}^{i}:i=1,...,D\},$ the universe collapses into the state $\Psi
_{n+1}=\Phi _{n+1}^{a}=\hat{U}^{(a)}\Psi _{n},$ where $a\in \{1,...,D\}.$
The question faced by the Class $1$ Basis Method Rules is: can the same set $%
U_{F}=\{\hat{U}^{(\mu )}:\mu =1,...,(D-1)\}\subset \{\hat{U}%
^{(i)}:i=1,...,D\}$ of unitary operators be used to construct a new \textit{%
orthonormal} Reduced basis set of vectors from\ $\Psi _{n+1}?$ In other
words, given a set of vectors $\{\Theta ^{\mu }:\mu =1,...,(D-1)\}$ defined
as
\begin{equation}
\Theta ^{\mu }=\hat{U}^{(\mu )}\Psi _{n+1}=\hat{U}^{(\mu )}\hat{U}^{(a)}\Psi
_{n}
\end{equation}

does it in general follow that
\begin{equation}
\langle \Theta ^{\mu }|\Theta ^{\nu }\rangle =\delta _{\mu \nu }
\end{equation}

for all $\mu ,\nu =1,...,(D-1)?$

Now, the condition is obviously satisfied for the case when $\mu =\nu ,$%
\begin{equation}
\langle \Theta ^{\mu }|\Theta ^{\mu }\rangle =\langle \Psi _{n+1}|\hat{U}%
^{(\mu )\ast }\hat{U}^{(\mu )}|\Psi _{n+1}\rangle =\langle \Psi _{n+1}|\hat{I%
}|\Psi _{n+1}\rangle =1.
\end{equation}

For $\mu \neq \nu ,$ however, it follows that
\begin{eqnarray}
\langle \Theta ^{\mu }|\Theta ^{\nu }\rangle &=&\langle \Psi _{n+1}|\hat{U}%
^{(\mu )\ast }\hat{U}^{(\nu )}|\Psi _{n+1}\rangle =\langle \Psi
_{n+1}|(V)^{\mu -1}(\hat{U}^{(n+1,K)})^{\ast }\hat{U}^{(n+1,K)}(V^{\ast
})^{\nu -1}|\Psi _{n+1}\rangle  \notag \\
&=&\langle \Psi _{n+1}|(V)^{\mu -1}(V^{\ast })^{\nu -1}|\Psi _{n+1}\rangle
=\langle \Psi _{n}|\hat{U}^{(a)\ast }(V)^{\mu -1}(V^{\ast })^{\nu -1}\hat{U}%
^{(a)}|\Psi _{n}\rangle  \notag \\
&=&\langle \Psi ^{1}|(V)^{a-1}(\hat{U}^{(n+1,K)})^{\ast }(V)^{\mu
-1}(V^{\ast })^{\nu -1}\hat{U}^{(n+1,K)}(V^{\ast })^{a-1}|\Psi ^{1}\rangle
\notag \\
&=&\langle \Psi ^{a}|(\hat{U}^{(n+1,K)})^{\ast }(V)^{\mu -1}(V^{\ast })^{\nu
-1}\hat{U}^{(n+1,K)}|\Psi ^{a}\rangle .
\end{eqnarray}

Now, the product $(V)^{\mu -1}(V^{\ast })^{\nu -1}$ gives
\begin{eqnarray}
(V)^{\mu -1}(V^{\ast })^{\nu -1} &=&\left(
\begin{array}{c}
\left( |\Psi ^{1}\rangle \langle \Psi ^{\mu }|+|\Psi ^{2}\rangle \langle
\Psi ^{\mu \oplus 1}|+...+|\Psi ^{D}\rangle \langle \Psi ^{\mu \oplus
(D-1)}|\right) \\
\times \left( |\Psi ^{\nu }\rangle \langle \Psi ^{1}|+|\Psi ^{\nu \oplus
1}\rangle \langle \Psi ^{2}|+...+|\Psi ^{\nu \oplus (D-1)}\rangle \langle
\Psi ^{D}|\right)
\end{array}
\right)  \notag \\
&=&\sum_{x=1}^{D}|\Psi ^{x}\rangle \langle \Psi ^{\mu \oplus
(x-1)}|\sum_{y=1}^{D}|\Psi ^{\nu \oplus (y-1)}\rangle \langle \Psi ^{y}|
\notag \\
&=&\sum_{x=1}^{D}\sum_{y=1}^{D}|\Psi ^{x}\rangle \langle \Psi ^{y}|\delta
_{\lbrack \mu \oplus (x-1)],[\nu \oplus (y-1)]}
\end{eqnarray}

where $\oplus $ implies addition modulo $D,$ i.e. $D\oplus z=z$ for $1\leq
z\leq D.$

So overall, the condition, $\langle \Theta ^{\mu }|\Theta ^{\nu }\rangle =0,$
for orthogonality is only satisfied iff
\begin{equation}
\sum_{x=1}^{D}\sum_{y=1}^{D}\delta _{\lbrack \mu \oplus x],[\nu \oplus
y]}\langle \Psi ^{a}|(\hat{U}^{(n+1,K)})^{\ast }|\Psi ^{x}\rangle \langle
\Psi ^{y}|\hat{U}^{(n+1,K)}|\Psi ^{a}\rangle =0
\end{equation}

for $\mu \neq \nu .$\bigskip
\end{proof}

Now, noting that
\begin{equation}
\hat{U}^{(n+1,K)}|\Psi ^{a}\rangle \notin \{\Psi ^{1},\Psi ^{2},...,\Psi
^{D}\}
\end{equation}
for a valid development from\ $\Psi _{n}\rightarrow \Psi _{n+1}$ to occur in
this quantum universe, where $\Psi ^{1}=\Psi _{n}\neq \Psi _{n+1}$ by
definition, such that the inner products $\langle \Psi ^{a}|(\hat{U}%
^{(n+1,K)})^{\ast }|\Psi ^{x}\rangle $ and $\langle \Psi ^{y}|\hat{U}%
^{(n+1,K)}|\Psi ^{a}\rangle $\ are not equal to zero, the above condition is
only fulfilled if the sum of products of amplitudes equals zero. Clearly,
this will not in general be the case.

In fact, following on from this, it is observed that since the dimension of
the Hilbert space of the real Universe is expected to be enormous, for the
above result to be obtained in reality a truly remarkable level of
cancellation must occur between the probability amplitudes in the relevant
sum of terms.

It is expected, then, that in general $\langle \Theta ^{\mu }|\Theta ^{\nu
}\rangle \neq \delta _{\mu \nu }$ for every $\mu ,\nu ,$ so that a set of
vectors given by $\{\hat{U}^{(\mu )}\Psi _{n+1}:\mu =1,...,(D-1)\}$ are
unlikely to be orthogonal, where\ the unitary operators $\{\hat{U}^{(\mu
)}\} $ are defined such that $\langle \Psi _{n}\hat{U}^{(\mu )}|\hat{U}%
^{(\nu )}\Psi _{n}\rangle =\delta _{\mu \nu }$ for\ $\mu ,\nu
=1,2,...,(D-1). $

An arbitrarily defined set $U_{F}$ cannot therefore be expected to specify
an orthonormal Reduced basis set $\frak{B}_{n+2}^{R}.$\bigskip

Furthermore, note that the above condition is only defined for the `first'
potential transition, that is from $\Psi _{n+1}$ to $\Psi _{n+2}.$ However,
for the set $U_{F}$ to provide valid Class $1$ Basis Method dynamics,
similar conditions must also hold for the indefinite series of transitions $%
\Psi _{n+2}\longrightarrow \Psi _{n+3}\rightarrow \Psi _{n+4}\rightarrow ...$
Thus, the set $U_{F}$ must be defined such that it actually satisfies an
`infinite tower' of conditions, with a fortuitous level of cancellation
required at each stage.

It is these observations that prompt the conclusion that such a set is
unlikely to exist.

\bigskip

Summarising, given an arbitrary set $U_{F}$ of unitary operators $\{\hat{U}%
^{(\mu )}:\mu =1,...,(D-1)\}$ defined such that they satisfy (\ref{Ch6RCond}%
) for $\Psi _{n},$ and a vector $\Phi _{n+1}^{D}$ defined `orthogonally'
such that it satisfies (\ref{Ch6OrthoCond}), then whichever member $\Phi
_{n+1}^{i}$ of the set $\frak{B}_{n+1}=\{\{\Phi _{n+1}^{\mu }=\hat{U}^{(\mu
)}\Psi _{n}\},\Phi _{n+1}^{D}\}$ the universe collapses into when it becomes
$\Psi _{n+1},$ the elements of a new set of vectors defined as $\{\hat{U}%
^{(1)}\Phi _{n+1}^{i},\hat{U}^{(2)}\Phi _{n+1}^{i},....,\hat{U}^{(D-1)}\Phi
_{n+1}^{i}\}$ are not expected to be orthogonal.

Thus, the set $\{\hat{U}^{(\mu )}\Psi _{n+1}:\mu =1,...,(D-1)\}$ is unlikely
to form an orthogonal Reduced basis $\frak{B}_{n+2}^{R}$ for $\mathcal{H}%
^{(D)},$ and so cannot be used to specify a unique basis set $\frak{B}%
_{n+2}, $ or, consequently, an equivalence class of operators $\hat{B}%
_{n+2}. $ The next operator $\hat{\Sigma}_{n+2}$ cannot therefore be
generated in this manner, from the state $\Psi _{n+1}$ being rotated by the
members of a fixed set $U_{F}$ of unitary operators.

Concluding, the Type $III$ Class $1$ Basis Method is expected to be invalid.

\bigskip

As a caveat to this conclusion, note that it is in fact always possible for
a single qubit universe in a two dimensional Hilbert space to be governed by
Type $III$ Class $1$ Basis Method Rules. This result follows because any
Reduced basis set $\frak{B}_{n+1}^{R}$ for $\mathcal{H}^{(2)}$ has only one
member; there is therefore no `orthogonality problem' for its elements.

So, given any arbitrary vector $\theta \in \mathcal{H}^{(2)}$ and any
unitary operator $\hat{U}^{(1)},$ it is always possible to specify an
orthonormal basis for $\mathcal{H}^{(2)}$ that contains the vector $\hat{U}%
^{(1)}\theta $ and whichever vector in $\mathcal{H}^{(2)}$ is orthogonal to $%
\hat{U}^{(1)}\theta .$ Effectively, the vector $\hat{U}^{(1)}\theta $
single-handedly implies a unique basis set for $\mathcal{H}^{(2)}.$

Thus, given $\hat{U}^{(1)}$ and a state $\Psi _{n}\in \mathcal{H}^{(2)},$
the basis set $\frak{B}_{n+1}$ is readily generated, and either element of
this can be used with $\hat{U}^{(1)}$ to generate a new basis set $\frak{B}%
_{n+2}.$ In fact, this process may be repeated indefinitely.

Such a possibility is unique to two dimensional Hilbert spaces.\bigskip

\paragraph{Class 2 Basis Method}

\begin{equation*}
\end{equation*}

In the Class 2 Basis Method, the set of unitary operators $\{\hat{U}%
_{n}^{(\mu )}:\mu =1,...,(D-1)\},$ and hence the functions $f_{n}^{(i)},$
change as the universe jumps from one stage to the next. In other words, a
new set of $(D-1)$ unitary operators is chosen at each time step, $n.$
Clearly, this Class of Basis Method mechanism is immediately valid, because
it is always possible to define a set of unitary operators that provides the
next orthogonal Reduced basis set of vectors when acting upon a given state,
for all $n.$

Summarising, then, a universe developing according to a Type $III$ (state
self-referential) Basis Method mechanism is governed by the general Rule
\begin{equation}
\frak{B}_{n+1}=\{\frak{B}_{n+1}^{R},\Phi _{n+1}^{D}\}=\left\{ \{\hat{U}%
_{n}^{(1)}\Psi _{n},\hat{U}_{n}^{(2)}\Psi _{n},...,\hat{U}_{n}^{(D-1)}\Psi
_{n}\},\Phi _{n+1}^{D}\right\}
\end{equation}
with the set of unitary operators $\{\hat{U}_{n}^{(\mu )}\}$ defined such
that they satisfy the conditions

\begin{equation}
\langle \Psi _{n}\hat{U}_{n}^{(\mu )}|\hat{U}_{n}^{(\nu )}\Psi _{n}\rangle
=\delta _{\mu \nu }
\end{equation}
and the vector $\Phi _{n+1}^{D}$ defined as
\begin{equation}
\langle \Phi _{n+1}^{D}|\hat{U}_{n}^{(\mu )}\Psi _{n}\rangle =0
\end{equation}
for all $\mu ,\nu =1,...,(D-1)$ and $n.$\bigskip

Of course, in order to ensure the above conditions, the actual choice of the
operators $\{\hat{U}_{n}^{(\mu )}\}$ must rely to some extent on some sort
of `knowledge' of what the current state $\Psi _{n}$ is. Certainly, it is
difficult to imagine how $\hat{U}_{n}^{(\mu )}$ could be defined obeying $%
\langle \Psi _{n}\hat{U}_{n}^{(\mu )}|\hat{U}_{n}^{(\nu )}\Psi _{n}\rangle
=\delta _{\mu \nu }$ if $\Psi _{n}$ is unknown, especially considering the
conclusion from the Class $1$ case that no generalised set of such operators
is expected to exist that can give orthogonal results when acting on
arbitrary vectors. The suggestion, then, could be that the definition of
these unitary operators might itself depend on self-referential rules, such
that the members of the set $\{\hat{U}_{n}^{(\mu )}\}$ might themselves be
some unknown function, $F_{n},$ of the state $\Psi _{n}.$ Of course, $F_{n}$
could perhaps be a $1\rightarrow (D-1)$ function, such that $\{\hat{U}%
_{n}^{(\mu )}\}=F_{n}(\Psi _{n}),$ or could maybe `contain' $(D-1)$
sub-functions, such that $\hat{U}_{n}^{(\mu )}=F_{n}^{(\mu )}(\Psi _{n}).$
Thus, the function $F_{n}$ would be defined such that $F_{n}(\Psi _{n})$
generates a set of unitary operators $\{\hat{U}_{n}^{(\mu )}\}$ that satisfy
the conditions $\langle \Psi _{n}\hat{U}_{n}^{(\mu )}|\hat{U}_{n}^{(\nu
)}\Psi _{n}\rangle =\delta _{\mu \nu }$ for $\mu ,\nu =1,...,(D-1)$ and all $%
n.$

Overall, therefore, the Rules governing a universe that develops according
to the Class $2$ Basis Method could rely on a choice of Hermitian operator
that is a result of unitary operators acting upon the state, that are
themselves functions of the state.\bigskip

Such a mechanism would overcome the non-orthogonality problem experienced by
the Class $1$ dynamics, but does lead to the question as to exactly how the
unitary operators $\{\hat{U}_{n}^{(\mu )}\}$ are defined at each time step:
what exactly is the form of the function $F_{n}?$

Indeed, at first glance it appears from such a question that very little
progress is actually gained from analysing the presented method. After all,
recall that the original aim of this chapter was to investigate how the
universe chooses which operator it uses to develop itself with. So, if an
answer to this question is that the actual mechanism relies on a particular
choice of unitary operators that is itself unexplainable, or at least relies
upon some higher-order self-referential process $F_{n}(\Psi _{n}),$ it is
still unclear as to how the dynamics of a universe developing according to
the Basis Method might actually proceed.

Nevertheless, the mathematical possibility of such a Class $2$ Basis Method
dynamics, and hence the possibility that the physical Universe itself
develops according to such a mechanism, does imply that a full investigation
into operators that are a result of operators that depend on the state is a
necessary direction for future research.\bigskip

\paragraph{Summary}

\begin{equation*}
\end{equation*}

Summarising, in order for a set of $(D-1)$ unitary operators $\{\hat{U}%
_{n}^{(\mu )}\}$ to generate a Reduced basis $\frak{B}_{n+1}^{R}$ from a
state $\Psi _{n}$ in a universe governed by Type $III$ Basis Method Rules,
it is expected that the set $\{\hat{U}_{n}^{(\mu )}\}$ must be defined at
each $n$ in a way that depends upon this current state $\Psi _{n}.$ In other
words, Class $1$ Basis Method Rules are unlikely to be valid for the
development of a quantum universe (for $D>2).$ Class $2$ Basis Method Rules,
however, are.

Overall, then, in a universe governed by Type $III$ Class $2$ Basis Method
Rules, the state develops as follows.

Given a state $\Psi _{n}$ and a set of $(D-1)$ unitary operators $\{\hat{U}%
_{n}^{(\mu )}\}$ defined such that $\langle \Psi _{n}\hat{U}_{n}^{(\mu )}|%
\hat{U}_{n}^{(\nu )}\Psi _{n}\rangle =\delta _{\mu \nu }$ for $\mu ,\nu
=1,...,(D-1),$ a Reduced basis set of vectors $\frak{B}_{n+1}^{R}$ is
determined, where $\frak{B}_{n+1}^{R}=\{\hat{U}_{n}^{(\mu )}\Psi _{n}\}.$

Moreover, given also a vector $\Phi _{n+1}^{D}$ defined such that $\langle
\Phi _{n+1}^{D}|\hat{U}_{n}^{(\mu )}\Psi _{n}\rangle =0,$ a basis set $\frak{%
B}_{n+1}$ of mutually orthogonal states can then be generated, where $\frak{B%
}_{n+1}=\{\frak{B}_{n+1}^{R},\Phi _{n+1}^{D}\}.$

Equally, therefore, an equivalence class of operators $\{\hat{B}_{n+1}\}$ is
also specified, and these may be associated with the operator $\hat{\Sigma}%
_{n+1}$ used to develop the state $\Psi _{n}.$ The universe consequently
jumps to the state $\Psi _{n+1},$ which is an eigenvector of $\hat{\Sigma}%
_{n+1},$ with probability given in the usual way.\bigskip

The process may then be continued, so that a new set of unitary operators $\{%
\hat{U}_{n+1}^{(\mu )}\},$ defined such that $\langle \Psi _{n+1}\hat{U}%
_{n+1}^{(\mu )}|\hat{U}_{n+1}^{(\nu )}\Psi _{n+1}\rangle =\delta _{\mu \nu }$
for $\mu ,\nu =1,...,(D-1),$ is used to generate the next Reduced basis set
of vectors $\frak{B}_{n+2}^{R}$ from the new state $\Psi _{n+1}$ according
to the Rule: $\frak{B}_{n+2}^{R}=\{\hat{U}_{n+1}^{(\mu )}\Psi _{n+1}\}.$

Consequently, the next basis set $\frak{B}_{n+2}=\{\frak{B}_{n+2}^{R},\Phi
_{n+2}^{D}\}$ may immediately be determined from the conditions $\langle
\Phi _{n+2}^{D}|\hat{U}_{n+1}^{(\mu )}\Psi _{n+1}\rangle =0,$ and this is
turn implies an equivalence class of operators $\{\hat{B}_{n+2}\}.$ The
universe then jumps to the state $\Psi _{n+2},$ which is one of the members
of $\frak{B}_{n+2}.$

And so on; the procedure may be iterated indefinitely.

\bigskip

\subsubsection{The One-to-Many Method}

\bigskip

The development of a state $\Psi _{n}\in \mathcal{H}^{(D)}$ according to the
Basis Method relies on $D$ functions $f_{n}^{(i)},$ for $i=1,...,D,$ each of
which maps the state $\Psi _{n}$ to a unique vector $f_{n}^{(i)}(\Psi _{n}).$
Assuming that these new vectors are orthogonal, a condition ensured by the
actual definitions of $\{f_{n}^{(i)}\},$ they are then taken to comprise the
next preferred basis set $\frak{B}_{n+1}.$ The universe subsequently jumps
to one of these possible states.

In the One-to-Many Method, however, it is instead postulated that there
exists a single function $f_{n}$ that maps the state\ $\Psi _{n}$ to $D$
different, orthogonal vectors. In other words, $f_{n}$ is defined to be a $%
1\rightarrow D$ function which, when applied to $\Psi _{n},$ has $D$
outcomes: $\Theta ^{1},\Theta ^{2},...,\Theta ^{D}.$

Now, because the function $f_{n}$ is defined such that these $D$ outcomes
are all mutually orthogonal, then the application of $f_{n}$ to $\Psi _{n}$
effectively defines a basis set. Labelling this basis set $\frak{B}_{n+1},$
it is possible to write
\begin{equation}
\frak{B}_{n+1}\equiv \{\Theta _{n+1}^{1},\Theta _{n+1}^{2},...,\Theta
_{n+1}^{D}\}=f_{n}(\Psi _{n})
\end{equation}
where the `temporal' subscript has now been added for completeness. Thus,
the next basis set $\frak{B}_{n+1}$ is generated from the current state by
applying the $1\rightarrow D$ function $f_{n}$ to $\Psi _{n};$ such a Rule
may be called a One-to-Many Method.

Evidently, the determination of the basis set $\frak{B}_{n+1}$ implies the
determination of an equivalence class of operators $\hat{B}_{n+1},$ and
these may be associated in the usual way with the next test $\hat{\Sigma}%
_{n+1}$ used to develop the state.\bigskip

In order to provide a consistent mechanism for the automatic development of
a universe from $\Psi _{n}\rightarrow \Psi _{n+1}\rightarrow \Psi
_{n+2}\rightarrow ...,$ it is expected that the next basis set $\frak{B}%
_{n+2}$ may be determined by applying a One-to-Many function $f_{n+1}$ to
the vector $\Psi _{n+1},$ where $\Psi _{n+1}$ is defined as whichever member
of the set $\{\Theta _{n+1}^{1},\Theta _{n+1}^{2},...,\Theta _{n+1}^{D}\}$
the state of the universe collapsed into. As with the earlier Basis Method,
an immediate question then arises regarding how the function $f_{n+1}$ might
be related to $f_{n}.$ Also as before, two different classes of Rule
consequently become apparent:

\begin{enumerate}
\item[Class 1:]  The function $f_{n}$ is constant for all $n,$ so that $%
f_{n}=f.$ Thus, the same $1\rightarrow D$ function is used to generate $%
\frak{B}_{n+2}$ from $\Psi _{n+1}$ as was used to generate $\frak{B}_{n+1}$
from $\Psi _{n},$ such that
\begin{eqnarray}
\frak{B}_{n+1} &=&f(\Psi _{n}) \\
\frak{B}_{n+2} &=&f(\Psi _{n+1})  \notag \\
&&\vdots  \notag
\end{eqnarray}

\item[Class 2:]  The function $f_{n}$ changes with $n.$ Thus, a different $%
1\rightarrow D$ function may be used to generate $\frak{B}_{n+2}$ from $\Psi
_{n+1}$ than was used to generate $\frak{B}_{n+1}$ from $\Psi _{n},$ such
that
\begin{eqnarray}
\frak{B}_{n+1} &=&f_{n}(\Psi _{n}) \\
\frak{B}_{n+2} &=&f_{n+1}(\Psi _{n+1})  \notag \\
&&\vdots  \notag
\end{eqnarray}
\end{enumerate}

Of course, and as with the Basis Method dynamics, the first Class is
evidently a special example of the second Class in the case where $%
f_{n+1}=f_{n}$ for all $n.$\bigskip

It is expected that there are many different functions $f_{n}$ that could be
used to generate a basis set of $D$ orthogonal vectors from a given state $%
\Psi _{n};$ there might be many ways in which $f_{n}$ might generally be
constructed, and many forms it could then take. Indeed, this point may be
reinforced by recalling that there are an infinite number of basis sets of
orthogonal vectors spanning a Hilbert space $\mathcal{H}^{(D)},$ and hence,
at least in principle, an infinite number of functions $f_{n}$ relating them
to a particular state $\Psi _{n}.$

However, it must be recalled that not every conceivable function acting in a
vector space will provide $D$ orthogonal outcomes when applied to a given
state. In fact, the set of valid One-to-Many functions is a tiny subset of
the set of all possible functions. Moreover, there are no obvious guidelines
to suggest what a `typical' such function should look like, and it is
difficult to predict exactly how suitable One-to-Many relationships should
in general be constructed. Clearly, then, it is a task for future research
to attempt to discover what the various types and forms of valid One-to-Many
functions might actually be.

For now, though, it is remarked that the validity of any proposed
One-to-Many method depends entirely on the definition of the function in
question, and hence relies on the underlying choice of Rules governing the
system. This point is particularly important in\ regard to the question of
when it is possible to construct a valid dynamics from a constant function $%
f_{n}=f.$

\bigskip

It is possible to provide a simple example of the Generated-Sort,
One-to-Many Method using a single qubit universe represented by a state in a
two dimensional Hilbert space. To demonstrate this, however, observe first
that the application of the function $f_{n}$ to $\Psi _{n}$ to give the
basis set $\frak{B}_{n+1}=\{\Theta _{n+1}^{1},\Theta _{n+1}^{2},...,\Theta
_{n+1}^{D}\}$ is, as expected, effectively the same as generating an
equivalence class of operators $\hat{B}_{n+1}$ with eigenstates $\{\Theta
_{n+1}^{1},\Theta _{n+1}^{2},...,\Theta _{n+1}^{D}\}.$ Moreover, the reverse
of this is also clearly true: if an operator $\hat{B}_{n+1}$ could be
constructed that is a function $f_{n}^{\prime }$ of $\Psi _{n},$ then this
process also automatically defines the basis set $\frak{B}_{n+1}.$

Thus, there are two equivalent ways of specifying One-to-Many Method Rules:
either a $1\rightarrow D$ function $f_{n}$ should be defined that maps a
state $\Psi _{n}$ directly to $D$ orthogonal vectors; or else a mechanism
for obtaining an operator $\hat{B}_{n+1}=f_{n}^{\prime }(\Psi _{n})$ may be
provided, where $\hat{B}_{n+1}$ is a Strong operator with $D$ orthogonal
eigenstates.

It is this latter possibility that will prove useful in the following
examples.\bigskip

\paragraph{Example A}

\begin{equation*}
\end{equation*}

Consider a state $\Psi _{n}$ in the qubit Hilbert space $\mathcal{H}^{(2)}$
spanned by the `usual' qubit basis
\begin{equation}
\mathcal{B}=\{|0\rangle ,|1\rangle \}=\left\{ \left( \QATOP{1}{0}\right)
,\left( \QATOP{0}{1}\right) \right\} .
\end{equation}

Moreover, assume that the universe is governed by a One-to-Many Method
mechanism, and develops according to the Class $1$ Rule that the next
operator $\hat{\Sigma}_{n+1}$ is defined simply as the projection of the $%
n^{th}$ state. In other words, the dynamics of the universe proceeds by the
general algorithm

\begin{itemize}
\item  $\Psi _{n}$ is tested by the Hermitian operator $\hat{\Sigma}_{n+1},$
to give the next state $\Psi _{n+1}$ which is one of the eigenvectors of $%
\hat{\Sigma}_{n+1},$ where $\hat{\Sigma}_{n+1}$ is given by
\begin{equation}
\hat{\Sigma}_{n+1}=|\Psi _{n}\rangle \langle \Psi _{n}|.  \label{Ch6rule}
\end{equation}
\end{itemize}

Consider also an arbitrary state encountered in the universe's development.
In fact, assume that at time $n$ the universe may be described by the most
general vector possible, that is, $\Psi _{n}=\frac{1}{\sqrt{\kappa }}(\alpha
|0\rangle +\beta |1\rangle ),$ where $\alpha ,\beta \in \mathbb{C}$ and $%
\sqrt{\kappa }=\sqrt{\alpha \bar{\alpha}+\beta \bar{\beta}}.$ Clearly, in
the representation employed in this example, $\hat{\Sigma}_{n+1}$ is then
given by the matrix
\begin{equation}
\hat{\Sigma}_{n+1}=\frac{1}{\kappa }\left(
\begin{array}{ll}
\alpha \bar{\alpha} & \alpha \bar{\beta} \\
\beta \bar{\alpha} & \beta \bar{\beta}
\end{array}
\right)
\end{equation}
which has orthonormal eigenstates $\Theta _{n+1}^{1}$ and $\Theta _{n+1}^{2}$
equal to
\begin{eqnarray}
\Theta _{n+1}^{1} &=&\frac{1}{\sqrt{\kappa }}(\alpha |0\rangle +\beta
|1\rangle ) \\
\Theta _{n+1}^{2} &=&\frac{1}{\sqrt{\kappa }}(\bar{\beta}|0\rangle -\bar{%
\alpha}|1\rangle )  \notag
\end{eqnarray}

So, the eigenstates\footnote{%
Strictly, $\hat{\Sigma}_{n+1}$ as given here is not a strong operator
because it possesses a zero eigenvalue, so should technically not be allowed
to develop the state according to the paradigm proposed in this thesis.
However the reasons given in Chapter 5, for ruling out such operators in
general, do not actually apply in the special case of two dimensional
universes, and so the current example may still be validly discussed.} of $%
\hat{\Sigma}_{n+1}$ define an orthogonal basis set $\frak{B}_{n+1}$ given by
$\frak{B}_{n+1}=\{\Theta _{n+1}^{1},\Theta _{n+1}^{2}\}.$ Moreover, because $%
\hat{\Sigma}_{n+1}$ is a function of $\Psi _{n},$ i.e. $\hat{\Sigma}_{n+1}=%
\hat{\Sigma}_{n+1}(\Psi _{n}),$ a single qubit universe developing according
to the Rule $\hat{\Sigma}_{n+1}=|\Psi _{n}\rangle \langle \Psi _{n}|$
provides a potential example of a Generated-Sort, One-to-Many Method
mechanism.

Furthermore, because applying the same Rule (\ref{Ch6rule}) to whichever of
the eigenstates $\Theta _{n+1}^{1}$ and $\Theta _{n+1}^{2}$ becomes $\Psi
_{n+1}$ gives rise to an operator $\hat{\Sigma}_{n+2}=|\Psi _{n+1}\rangle
\langle \Psi _{n+1}|$ which also has two orthogonal eigenstates, and because
this process may be continued indefinitely, where the next operator is
always the same function of the current state for all $n,$ a single qubit
universe developing according to the Rule $\hat{\Sigma}_{n+1}=|\Psi
_{n}\rangle \langle \Psi _{n}|$ actually provides an example of a Class $1$
One-to-Many Method dynamics. Such a universe will develop with the next
operator always dependent on the current state.\bigskip

Of course, it must immediately be noted at this point that a Rule of the
form $\hat{\Sigma}_{n+1}=|\Psi _{n}\rangle \langle \Psi _{n}|$ only really
gives trivial dynamics. Clearly, $\Psi _{n}$ is an eigenstate of $\hat{\Sigma%
}_{n+1},$ so the operator is ultimately equivalent to a null test, and $\Psi
_{n}=\Psi _{n+1}$ for all $n.$ Nevertheless, this example does, at least in
principle, provide an illustration of the proposed mechanism.\bigskip

\paragraph{Example B}

\begin{equation*}
\end{equation*}

A dynamics for a single qubit system that is perhaps more interesting than
that of the above could instead be governed by the Class $1$ One-to-Many
Method Rule

\begin{itemize}
\item  $\Psi _{n}$ is tested by the Hermitian operator $\hat{\Sigma}_{n+1},$
to give the next state $\Psi _{n+1}$ which is one of the eigenvectors of $%
\hat{\Sigma}_{n+1},$ where $\hat{\Sigma}_{n+1}$ is given by
\begin{equation}
\hat{\Sigma}_{n+1}=\hat{U}^{\ast }|\Psi _{n}\rangle \langle \Psi _{n}|\hat{U}
\label{Ch6SigB}
\end{equation}
\end{itemize}

where $\hat{U}$ is a unitary operator of the form $\hat{U}=\exp
(-i\varepsilon \hat{\sigma}_{1}),$ for $\varepsilon \in \mathbb{R}^{+}$ a
real parameter and $\hat{\sigma}_{1}$ the usual Pauli operator. Clearly,
such a Rule leads to a universe governed by an operator $\hat{\Sigma}_{n+1}$
with eigenvectors different from its current state.

As a visual interpretation of how this mechanism proceeds, it may be
possible to imagine projecting the state $\Psi _{n}$ onto a fixed vector $%
\hat{U}^{\ast }\Psi _{n}$ to get the next state $\Psi _{n+1},$ which is one
of the eigenvectors of $\hat{U}^{\ast }|\Psi _{n}\rangle \langle \Psi _{n}|%
\hat{U}.$ Continuing, a new `fixed vector' may then be generated by slightly
rotating $\Psi _{n+1}$ to $\hat{U}^{\ast }\Psi _{n+1},$ and the state $\Psi
_{n+1}$ may be projected onto this new fixed vector $\hat{U}^{\ast }\Psi
_{n+1}$ to get the subsequent state $\Psi _{n+2},$ which is one of the
eigenvectors of $\hat{U}^{\ast }|\Psi _{n+1}\rangle \langle \Psi _{n+1}|\hat{%
U}.$ And so on.\bigskip

As a simple illustration, consider without loss of generality a universe
initially in the state $\Psi _{0}=|0\rangle .$ According to the above Rule,
the subsequent state $\Psi _{1}$ will be an eigenvector of the operator $%
\hat{\Sigma}_{1}$ given by
\begin{eqnarray}
\hat{\Sigma}_{1} &=&\hat{U}^{\ast }|\Psi _{0}\rangle \langle \Psi _{0}|\hat{U%
}=e^{i\varepsilon \hat{\sigma}_{1}}|0\rangle \langle 0|e^{-i\varepsilon \hat{%
\sigma}_{1}} \\
&=&\left(
\begin{array}{cc}
\cos ^{2}\varepsilon & -i\cos \varepsilon \sin \varepsilon \\
i\cos \varepsilon \sin \varepsilon & \sin ^{2}\varepsilon
\end{array}
\right)  \notag
\end{eqnarray}
where the last line follows from the usual representation $|a\rangle =\left(
\QATOP{1-a}{a}\right) ,$ for $a=0,1,$ and from the identity $%
e^{-i\varepsilon \hat{\sigma}_{1}}=\binom{\cos \varepsilon \text{ \ }-i\sin
\varepsilon }{-i\sin \varepsilon \text{ \ }\cos \varepsilon }=\hat{\sigma}%
_{0}\cos \varepsilon -i\hat{\sigma}_{1}\sin \varepsilon ,$ which itself
follows from the standard algebra (\ref{Ch3pauli1}) of the Pauli operators.

The operator $\hat{\Sigma}_{1}$ defined above has eigenvectors $\Theta
_{1}^{a}$ and $\Theta _{1}^{b}$ given by
\begin{eqnarray}
\Theta _{1}^{a} &=&i\sin \varepsilon |0\rangle +\cos \varepsilon |1\rangle \\
\Theta _{1}^{b} &=&-i\cos \varepsilon |0\rangle +\sin \varepsilon |1\rangle .
\notag
\end{eqnarray}

So, the next state $\Psi _{1}$ of this universe will be either $\Psi
_{1}=\Psi _{1}^{a}=\Theta _{1}^{a}$ with probability $|\langle \Theta
_{1}^{a}|\Psi _{0}\rangle |^{2}=\sin ^{2}\varepsilon ,$ or else\ $\Psi
_{1}=\Psi _{1}^{b}=\Theta _{1}^{b}$ with probability $|\langle \Theta
_{1}^{b}|\Psi _{0}\rangle |^{2}=\cos ^{2}\varepsilon ,$ noting that $%
|\langle \Theta _{1}^{a}|\Psi _{0}\rangle |^{2}+|\langle \Theta
_{1}^{b}|\Psi _{0}\rangle |^{2}=1$ as expected.

Now, because $\Psi _{1}$ will be one of two possibilities, $\Psi _{1}^{a}$
or $\Psi _{1}^{b},$ then according to the rule (\ref{Ch6SigB}) it is evident
that the subsequent test $\hat{\Sigma}_{2}$ will take one of two possible
forms. Labelling these $\hat{\Sigma}_{2}^{a}$ and $\hat{\Sigma}_{2}^{b},$ it
is clear that they are given by
\begin{eqnarray}
\hat{\Sigma}_{2}^{a} &=&\hat{U}^{\ast }|\Psi _{1}^{a}\rangle \langle \Psi
_{1}^{a}|\hat{U} \\
\hat{\Sigma}_{2}^{b} &=&\hat{U}^{\ast }|\Psi _{1}^{b}\rangle \langle \Psi
_{1}^{b}|\hat{U}.  \notag
\end{eqnarray}

Of course, which one of these is actually used to test the universe depends
entirely upon which state, $\Psi _{1}^{a}$ or $\Psi _{1}^{b},$ the system
collapsed into when it became $\Psi _{1}.$

The operators $\hat{\Sigma}_{2}^{a}$ and $\hat{\Sigma}_{2}^{b}$ will
themselves each possess two orthogonal eigenvectors. For $\hat{\Sigma}%
_{2}^{a}$ these may be labelled $\Psi _{2}^{ac}$ and $\Psi _{2}^{ad},$ and
are given by
\begin{eqnarray}
\Psi _{2}^{ac} &=&i\sin 2\varepsilon |0\rangle +\cos 2\varepsilon |1\rangle
\label{Ch6Psi2a} \\
\Psi _{2}^{ad} &=&-i\cos 2\varepsilon |0\rangle +\sin 2\varepsilon |1\rangle
\notag
\end{eqnarray}
whereas for $\hat{\Sigma}_{2}^{b}$ they may be labelled $\Psi _{2}^{be}$ and
$\Psi _{2}^{bf},$ with
\begin{eqnarray}
\Psi _{2}^{be} &=&-i\cos 2\varepsilon |0\rangle +\sin 2\varepsilon |1\rangle
\label{Ch6Psi2b} \\
\Psi _{2}^{bf} &=&i\sin 2\varepsilon |0\rangle +\cos 2\varepsilon |1\rangle .
\notag
\end{eqnarray}

So, given an initial state $\Psi _{0},$ then according to the rule (\ref
{Ch6SigB}) the wavefunction $\Psi _{2}$ after two steps will be one of these
four possible states $\{\Psi _{2}^{ac},\Psi _{2}^{ad},\Psi _{2}^{be},\Psi
_{2}^{bf}\}$ with appropriate probabilities given by, for example,
\begin{eqnarray}
P(\Psi _{2}^{ac}) &=&P(\Psi _{2}^{ac}|\Psi _{1}^{a}|\Psi _{0})=P(\Psi
_{2}^{ac}|\Psi _{1}^{a})\times P(\Psi _{1}^{a}|\Psi _{0}) \\
&=&|\langle \Psi _{2}^{ac}|\Psi _{1}^{a}\rangle |^{2}\times |\langle \Psi
_{1}^{a}|\Psi _{0}\rangle |^{2}  \notag \\
&=&|(\sin \varepsilon \sin 2\varepsilon +\cos \varepsilon \cos 2\varepsilon
)|^{2}\sin ^{2}\varepsilon =\cos ^{2}\varepsilon \sin ^{2}\varepsilon  \notag
\end{eqnarray}
in obvious notation.

Further, the next test $\hat{\Sigma}_{3}$ will be one of four possibilities
(one of which is $\hat{U}^{\ast }|\Psi _{2}^{ac}\rangle \langle \Psi
_{2}^{ac}|\hat{U},$ etc.), each of which possesses two eigenstates. Clearly,
the process continues such that the $n^{th}$ state $\Psi _{n}$ will be one
of the eigenstates of one of $2^{n-1}$ possible operators $\hat{\Sigma}_{n}.$%
\bigskip

The development of a universe described by a Rule such as $\hat{\Sigma}%
_{n+1}=\hat{U}^{\ast }|\Psi _{n}\rangle \langle \Psi _{n}|\hat{U}$ may be
modelled by a simple computer simulation looping through a program a finite
number of times. One\ successful method of achieving this has been to supply
the program with an input containing an initial vector $\Psi _{0},$ an
unitary matrix $\hat{U},$ and a number of iterations $N$ to perform. The $%
i+1^{th}$ iteration, for $i=0,1,..,(N-1),$ has two parts: the first step of
the program is to compute the matrix $\hat{U}^{\ast }|\Psi _{i}\rangle
\langle \Psi _{i}|\hat{U},$ determine its two eigenvectors, and fill an
array with the results of this evaluation. In the second step, a `random
number generator' is introduced that, when called, produces a number $%
r_{i+1} $ that has either the value $0$ or $1$ with equal likelihood. If a $%
``0"$ is found then the first element of the eigenvector array is recovered
and set to $\Psi _{i+1};$ the second element is discarded from further
discussion. If a $``1"$ if found then the converse occurs and the second
element is chosen instead to be $\Psi _{i+1}.$

The program loops through the process $N$ times to yield a unique `history'
of states $\{\Psi _{0},\Psi _{1},\Psi _{2},...,\Psi _{N}\},$ where there are
$2^{N}$ possible such histories, corresponding to the $2^{N}$ chains $%
r_{1},r_{2},...,r_{N}$ of numbers produced by the random number generator,
for $r_{i+1}=0,1$ for $i=0,1,..,(N-1).$ Of course, obtaining $2^{N}$ sets of
results after time $N$ is to be expected: any string of $N$ characters $\{$ $%
r_{i+1}:i=0,...,(N-1)\},$ where each character may take one of two values,
may be thought of as representing a binary number $<2^{N}.$ These binary
numbers may thus be used to effectively label the quantum history of the
system.

\bigskip

As with the model described in Example $A,$ it is to be noted that the types
of development resulting from the above Rule are extremely limited. In fact,
this conclusion follows from two observations.

Firstly, and as is evident from (\ref{Ch6Psi2a}) and (\ref{Ch6Psi2b}), the
eigenvectors of $\hat{\Sigma}_{2}^{a}$ are degenerate with those of $\hat{%
\Sigma}_{2}^{b},$ that is, $\Psi _{2}^{ac}\equiv \Psi _{2}^{bf}$ and $\Psi
_{2}^{ad}\equiv \Psi _{2}^{be}.$ It can be shown, moreover, that this is a
trend that is continued throughout the universe's development, such that the
state $\Psi _{n}$ after $n$ steps will not be one out of $2^{n}$ \textit{%
different} possibilities, but will instead be one of only $2$ different
vectors. Specifically, in fact, it may be shown that if $\Psi _{n}^{X}$ and $%
\Psi _{n}^{Y}$ are the two possible outcomes of a test $\hat{\Sigma}_{n},$
then the two subsequent potential operators $\hat{\Sigma}_{n+1}^{X}=\hat{U}%
^{\ast }|\Psi _{n}^{X}\rangle \langle \Psi _{n}^{X}|\hat{U}$ and $\hat{\Sigma%
}_{n+1}^{Y}=\hat{U}^{\ast }|\Psi _{n}^{Y}\rangle \langle \Psi _{n}^{Y}|\hat{U%
}$ share the same set of eigenstates.

This conclusion follows from the observations that $\Psi _{n}^{X}$ and $\Psi
_{n}^{Y}$ are necessarily orthogonal, that the eigenstates of $\hat{\Sigma}%
_{n+1}^{X}$ must also be orthogonal, and that the eigenstates of $\hat{\Sigma%
}_{n+1}^{Y}$ must be orthogonal too. Now, labelling the eigenstates of $\hat{%
\Sigma}_{n+1}^{X}$ as $\{\psi ,\varphi \},$ it is evident that the vector $%
\hat{U}^{\ast }|\Psi _{n}^{X}\rangle $ is an eigenstate of $\hat{\Sigma}%
_{n+1}^{X}$ with eigenvalue $1,$ because $\hat{\Sigma}_{n+1}^{X}(\hat{U}%
^{\ast }|\Psi _{n}^{X}\rangle )=\hat{U}^{\ast }|\Psi _{n}^{X}\rangle ;$
thus, $\psi $ may be chosen as $\psi =\hat{U}^{\ast }|\Psi _{n}^{X}\rangle .$

Now, the other (unknown) eigenvector, $\varphi ,$ of $\hat{\Sigma}_{n+1}^{X}$
must be orthogonal to $\hat{U}^{\ast }|\Psi _{n}^{X}\rangle ,$ such that
\begin{equation}
\langle \varphi |\hat{U}^{\ast }\Psi _{n}^{X}\rangle =0.
\end{equation}

The question becomes: what is this vector $\varphi ?$ Clearly, one possible
candidate for $\varphi $ is the vector $\hat{U}^{\ast }\Psi _{n}^{Y},$
because $\hat{U}\hat{U}^{\ast }=\hat{I}$ and $\langle \Psi _{n}^{Y}|\Psi
_{n}^{X}\rangle =0,$ where $\hat{I}$ is the identity operator. Furthermore,
because the Hilbert space of the system is two dimensional, this candidate
is the only choice. So, the eigenvectors of $\hat{\Sigma}_{n+1}^{X}$ must be
$\hat{U}^{\ast }\Psi _{n}^{X}$ and $\hat{U}^{\ast }\Psi _{n}^{Y}.$

A similar analysis of $\hat{\Sigma}_{n+1}^{Y}$ can readily be used to
demonstrate that the eigenstates of this operator are also $\hat{U}^{\ast
}\Psi _{n}^{X}$ and $\hat{U}^{\ast }\Psi _{n}^{Y};$ the conclusion is
shown.\bigskip

Secondly, the above types of mechanism, where the tests $\hat{\Sigma}_{n+1}$
depend on projection operators $|\Psi _{n}\rangle \langle \Psi _{n}|,$ can
only provide a suitable dynamics for single qubit universes. This is because
projection operators are not strong, as discussed in Section 5.5, on account
of them possessing (degenerate) eigenvalues of zero. So, an operator of the
form $\hat{U}^{\ast }|\Psi \rangle \langle \Psi |\hat{U},$ where $\Psi $ is
a vector in a $D>2$ dimensional Hilbert space $\mathcal{H}^{(D)},$ does not
specify a unique basis set of orthogonal eigenvectors, and hence cannot be
used to provide valid dynamics in the scheme proposed here.

The problem is in fact symptomatic of the result that in situations with
dimensions greater than $2,$ it is difficult to find a mechanism that uses a
single vector $\Psi $ to uniquely specify $D-1$ other vectors orthogonal to $%
\Psi .$ This follows because there are very many sets of $D-1$ vectors in $%
\mathcal{H}^{(D)}$ that are mutually orthogonal, whilst also being
orthogonal to $\Psi ,$ and so in general it is hard to find a Rule that
effectively picks just one of these out.

For instance, consider the case when $D=3,$ and define an orthonormal basis $%
\mathcal{B}^{(3)}$ for $\mathcal{H}^{(3)}$ as $\mathcal{B}^{(3)}\equiv
\{|0\rangle ,|1\rangle ,|2\rangle \}.$ Now, without loss of generality,
given a state $\Psi =|0\rangle $ it is possible to find many different pairs
of vectors $[\phi ,\theta ]$ that are mutually orthogonal to one another,
and also orthogonal to $|0\rangle .$\ One example is $[\phi ,\theta
]=[|1\rangle ,|2\rangle ],$ but of course
\begin{eqnarray}
\lbrack \phi ,\theta ] &=&\left[ \frac{1}{\sqrt{2}}(|1\rangle +|2\rangle ),%
\frac{1}{\sqrt{2}}(|1\rangle -|2\rangle )\right] , \\
\lbrack \phi ,\theta ] &=&\left[ \frac{1}{5}(4|1\rangle -3i|2\rangle ),\frac{%
1}{5}(-3i|1\rangle +4|2\rangle )\right] ,  \notag
\end{eqnarray}
and so on, also satisfy this criterion. Thus, specifying just the vector $%
\Psi =$ $|0\rangle $ does not imply an automatic choice for a unique
orthonormal basis set $\{\Psi ,\phi ,\theta \}$ of states spanning $\mathcal{%
H}^{(3)},$ because $\phi $ and $\theta $ could take many different forms. In
fact, more information needs to be provided in order to select a particular
pair of orthonormal vectors from the infinite set of possibilities. In this
case,\ only when, say, $\Psi =$ $|0\rangle $ and $\phi $ are both given is
it then possible to specify what $\theta $ must be.

In the context of this chapter, a unique basis set $\frak{B}_{n+1}$ of
states is hoped to be generated from $\Psi _{n}$ in order for the universe
to develop. So, in any One-to-Many mechanism based upon a Rule of the form $%
\frak{B}_{n+1}=f_{n}(\Psi _{n}),$ it would clearly be unsatisfactory if the
result of $f_{n}$ acting on $\Psi _{n}$ gave a number of orthonormal basis
sets. Instead, $f_{n}$ must be sufficiently well defined such that this
process `pins down' just one unique set.

A dynamics based on projection operators, however, cannot in general achieve
this.

Fortunately, though, an exception to this conclusion occurs in two
dimensional Hilbert spaces. In this circumstance, specifying just a single
vector $\Psi $ does imply a unique orthonormal basis set of states, because
there is \textit{only} one\ other vector that is orthogonal to $\Psi $ in $%
\mathcal{H}^{(2)}.$ So, when $D=2$ it is possible to label an orthonormal
basis set by using just one of its two states $\Psi ,$ and this principle
may ultimately be exploited to formulate rules determining the system's
development.

So, in the dynamics described in Examples $A$ and $B,$ a given $\Psi _{n}\in
\mathcal{H}^{(2)}$ is able to generate a unique, preferred basis set $\frak{B%
}_{n+1}$ for the next jump. Thus, a dynamics based upon projection operators
may be justified in $\mathcal{H}^{(2)},$ and a unique basis set of states
can still generated, even though an eigenvalue of zero is present.

Of course, despite potential objections questioning how physically
`interesting' a single qubit universe models might be, it does not detract
from the overall point that the proposed examples show how a state in a
Hilbert space can be developed according to Type $III,$ Generated-Sort
One-to-Many method Rules.\bigskip

\paragraph{Example C}

\begin{equation*}
\end{equation*}

Examples $A$ and $B$ involve universes developing according to a
deterministic (Type $III)$ Class $1$ One-to-Many Method mechanism. However,
just as in previous sections of this chapter, it is also possible to
consider probabilistic (Type $IIIa)$ dynamics by extending the Rules in the
obvious way.

For instance, instead of generating the next basis set $\frak{B}_{n+1}$ from
the current state $\Psi _{n}$ according to the deterministic rule $\frak{B}%
_{n+1}=f_{n}(\Psi _{n}),$ it is alternatively possible to consider Rules in
which $\frak{B}_{n+1}=f_{n}(\Psi _{n})$ with probability $P_{f_{n}},$ but $%
\frak{B}_{n+1}=g_{n}(\Psi _{n})$ with probability $P_{g_{n}},$ whilst\ $%
\frak{B}_{n+1}=h_{n}(\Psi _{n})$ with probability $P_{h_{n}},$ and so on,
where $f_{n},$ $g_{n}$ and $h_{n}$ are different functions. As before,
conservation of probability requires that $P_{f}+P_{g}+...=1.$

So, Example $B$ may readily be augmented to a probabilistic Class $1$
One-to-Many Method mechanism for a state $\Psi _{n}$ in a two dimensional
Hilbert space $\mathcal{H}^{(2)},$ by considering a Rule such as

\begin{itemize}
\item  $\Psi _{n}$ is tested by the Hermitian operator $\hat{\Sigma}_{n+1},$
to give the next state $\Psi _{n+1}$ which is one of the eigenvectors of $%
\hat{\Sigma}_{n+1},$ where $\hat{\Sigma}_{n+1}$ is given by
\begin{equation}
\hat{\Sigma}_{n+1}=\left\{
\begin{array}{c}
\hat{U}^{(1)\ast }|\Psi _{n}\rangle \langle \Psi _{n}|\hat{U}^{(1)}\text{
with Probability }P^{(1)} \\
\hat{U}^{(2)\ast }|\Psi _{n}\rangle \langle \Psi _{n}|\hat{U}^{(2)}\text{
with Probability }P^{(2)}
\end{array}
\right\}
\end{equation}
\end{itemize}

where $P^{(1)}+P^{(2)}=1,$ and
\begin{equation}
\hat{U}^{(1)}=e^{-i\varepsilon \hat{\sigma}_{1}}\text{ \ \ , \ \ }\hat{U}%
^{(2)}=e^{-i\mu \hat{\sigma}_{2}}
\end{equation}
with $\hat{\sigma}_{1}$ and $\hat{\sigma}_{2}$ Pauli operators and $%
\varepsilon ,\mu \in \mathbb{R}^{+}.$

As was the case in previous sections, extending a Type $III$ Rule to a Type $%
IIIa$ one does not add anything significantly new to the discussion.\bigskip

\paragraph{Example D}

\begin{equation*}
\end{equation*}

Up until now, attention has been focused on the Class $1$ One-to-Many
Method. It is, however, also possible to consider Class $2$ models.

As an example of such a dynamics, consider as before a state $\Psi _{n}$ in
a two dimensional, single qubit Hilbert space $\mathcal{H}^{(2)},$ and
assume that the universe is governed by the Rule

\begin{itemize}
\item  $\Psi _{n}$ is tested by the Hermitian operator $\hat{\Sigma}_{n+1},$
to give the next state $\Psi _{n+1}$ which is one of the eigenvectors of $%
\hat{\Sigma}_{n+1},$ where $\hat{\Sigma}_{n+1}$ is given by
\begin{equation}
\hat{\Sigma}_{n+1}=(\hat{U}^{\ast })^{n+1}|\Psi _{n}\rangle \langle \Psi
_{n}|(\hat{U})^{n+1}
\end{equation}
\end{itemize}

where $\hat{U}$ is an arbitrary unitary operator which, for the sake of
illustration, could be defined again as $\hat{U}=\exp (-i\varepsilon \hat{%
\sigma}_{1}).$

So, from a given state $\Psi _{0}$ at initial `time' $n=0,$ it follows that
\begin{eqnarray}
\hat{\Sigma}_{1} &=&\hat{U}^{\ast }|\Psi _{0}\rangle \langle \Psi _{0}|\hat{U%
} \\
\hat{\Sigma}_{2} &=&\hat{U}^{\ast }\hat{U}^{\ast }|\Psi _{1}\rangle \langle
\Psi _{1}|\hat{U}\hat{U}  \notag \\
\hat{\Sigma}_{3} &=&\hat{U}^{\ast }\hat{U}^{\ast }\hat{U}^{\ast }|\Psi
_{2}\rangle \langle \Psi _{2}|\hat{U}\hat{U}\hat{U}  \notag
\end{eqnarray}
and so on.

A universe developing according to this type of Rule would proceed
analogously to the models discussed in Example $A$ and $B;$ the universe
always collapses to one of the eigenstates of $\hat{\Sigma}_{n},$ and the
next operator $\hat{\Sigma}_{n+1}$ is then given as a function of this new
state $\Psi _{n}.$ The major difference, however, is that in the present
case the function that generates the next operator $\hat{\Sigma}_{n+1}$ from
the current state $\Psi _{n}$ is not constant, but is instead a dynamic
relationship that depends on the parameter $n.$

Thus, the above Rule provides an example of a universe that is developed
according to operators that depend on both the current state and the current
`time': a Class $2$ mechanism.\bigskip

The difference between Class $1$ and Class $2$ dynamics might perhaps be
likened to the differences encountered in laboratory quantum mechanics
between systems that are evolved by a constant Hamiltonian, $\hat{H},$ and
those that are alternatively evolved by a time dependent Hamiltonian $\hat{H}%
(t).$ After all, a jump dependent operator on the pregeometric level is
directly analogous to a time dependent operator in conventional physics,
because the parameter $n$ is ultimately assumed to be the pregeometric
origin of emergent time $t.$

However, this similarity should not of course be taken too literally:
Hamiltonians are viewed in the proposed paradigm as emergent constructs, and
as such are not defined on the pregeometric level discussed here.

\bigskip

\subsubsection{A Type IV Extension}

\bigskip

As an extension to this section, note that both List-Sort and Generated-Sort
mechanisms are expected to have their analogies in universes developing
according to Type $IV,$ $IVa,$ $V$ and $Va$ Rules. Moreover, in fact, such
analogies may provide richer possibilities for dynamics than their Type $III$
counterparts.

For example, note that although no Type $III$ Class $1$ Basis Method Rule
has been found that can provide a valid dynamics for a universe developing
self-referentially according to the current state (i.e. where the Reduced
basis set $\frak{B}_{n+1}^{R}=\{\hat{U}^{(1)}\Psi _{n},\hat{U}^{(2)}\Psi
_{n},...,$ $\hat{U}^{(D-1)}\Psi _{n}\}$ is generated by a \textit{fixed} set
of $(D-1)$ unitary operators $\{\hat{U}^{(\mu )}\}$ acting on $\Psi _{n}),$
such a Class $1$ Basis Method mechanism may easily be implemented in
universes developing according to Type $IV$ rules.\bigskip

To demonstrate this last point, recall from Section 8.1 that a Type $IV$
Rule is defined as one in which the next basis set $\frak{B}_{n+1}$ depends
on the current basis set $\frak{B}_{n}$ (unlike, of course, a Type $III$
mechanism, where the next basis set $\frak{B}_{n+1}$ depends on the current
state $\Psi _{n},$ which is just one element of $\frak{B}_{n}).$

So, by denoting the orthonormal elements of these two preferred bases as
\begin{eqnarray}
\frak{B}_{n} &\equiv &\{\Phi _{n}^{1},\Phi _{n}^{2},...,\Phi _{n}^{D}\} \\
\frak{B}_{n+1} &\equiv &\{\Phi _{n+1}^{1},\Phi _{n+1}^{2},...,\Phi
_{n+1}^{D}\}  \notag
\end{eqnarray}
it is clear that the analogy of the Class $1$ Basis Method mechanism for a
Type $IV$ universe requires a fixed set of unitary operators $\{\hat{U}%
^{(i)}\}$ to be found, where the operator $\hat{U}^{(i)}$ maps the element $%
\Phi _{n}^{i}\in \frak{B}_{n}$ to the element $\Phi _{n+1}^{i}\in \frak{B}%
_{n+1}$ in the manner
\begin{equation}
\Phi _{n+1}^{i}=\hat{U}^{(i)}\Phi _{n}^{i}
\end{equation}
where the elements of the bases\ have been indexed in the simplest way,
without loss of generality.

For completeness, note how this compares with the Type $III$ Rule
\begin{equation}
\Phi _{n+1}^{i}=\hat{U}^{(i)}\Psi _{n}.
\end{equation}

Of course, just as in the Type $III$ situation, the constraint
\begin{equation}
\langle \Phi _{n+1}^{i}|\Phi _{n+1}^{j}\rangle =\langle \Phi _{n}^{i}\hat{U}%
^{(i)}|\hat{U}^{(j)}\Phi _{n}^{j}\rangle =\delta _{ij}  \label{Ch6Cons1b}
\end{equation}
must be enforced on the definition of the unitary operators $\{\hat{U}%
^{(i)}\}$ to ensure that the vectors $\{\Phi _{n+1}^{i}\}$ are orthogonal.

Moreover, of course, in actuality only $(D-1)$ unitary operations $\{\hat{U}%
^{(\mu )}\}$ can be freely defined: as with the Type $III$ case, a
specification of the Reduced set $\{\Phi _{n+1}^{1},\Phi _{n+1}^{2},...,\Phi
_{n+1}^{D-1}\}$ automatically defines the $``D^{th}"$ vector $\Phi
_{n+1}^{D} $ because of the required orthogonality.\bigskip

Now, in order for Type $IV$ Class $1$ Basis Method Rules to be accepted as
providing a suitable mechanism for the universe's development from $\Psi
_{n}\rightarrow \Psi _{n+1}\rightarrow \Psi _{n+2}\rightarrow ...,$ the
subsequent Reduced basis set $\frak{B}_{n+2}^{R}$ defined as $\frak{B}%
_{n+2}^{R}\equiv \{\Phi _{n+2}^{1},\Phi _{n+2}^{2},...,\Phi _{n+2}^{D-1}\}$
must contain $(D-1)$ orthogonal elements $\Phi _{n+2}^{\mu }$ that are
generated from those of $\frak{B}_{n+1}^{R}$ according to the map\footnote{%
In principle, the $(D-1)$ `free parameter' operators $\{\hat{U}^{(\mu )}\}$
could be defined such that they act on any of the $D$ vectors in $\frak{B}%
_{n}.$ The permutations that result, however, do not seriously affect the
situation discussed. In short, it does not matter which member of $\frak{B}%
_{n}$ is `left out' of the Reduced basis $\frak{B}_{n+1}^{R}.$}: $\Phi
_{n+2}^{\mu }=\hat{U}^{(\mu )}\Phi _{n+1}^{\mu },$ where $\mu =1,...,(D-1).$
Of course, these conditions must hold for all $n.$

However, unlike for the Type $III$ case, in which no \textit{fixed} set $\{%
\hat{U}^{(\mu )}\}$ has been found that can develop the system from $\Psi
_{n}\rightarrow \Psi _{n+1}\rightarrow \Psi _{n+2}\rightarrow ...$ according
to the Basis Method algorithm, in Type $IV$ universes it is trivially easy
to find a constant set of unitary operators that provide a valid mechanism
for dynamics. In fact one such set occurs for the Rule
\begin{equation}
\hat{U}^{(\mu )}=\hat{U}_{f}\text{ \ \ , \ \ }\forall \mu
\end{equation}
where $\hat{U}_{f}$ is an arbitrary, fixed unitary operator, such that the
set $\{\hat{U}^{(\mu )}\}$ becomes a set of $(D-1)$ equal members, $\{\hat{U}%
_{f},\hat{U}_{f},...,\hat{U}_{f}\}.$

Under this circumstance, the constraint $\langle \Phi _{n}^{\mu }\hat{U}%
^{(\mu )}|\hat{U}^{(\nu )}\Phi _{n}^{\nu }\rangle =\delta _{\mu \nu }$ is
clearly satisfied by definition, because
\begin{equation}
\hat{U}_{f}^{\ast }\hat{U}_{f}=\hat{I}
\end{equation}
where $\hat{I}$ is the identity operator in $\mathcal{H}^{(D)},$ and the
Reduced basis set $\frak{B}_{n+1}^{R}$ is given by
\begin{equation}
\frak{B}_{n+1}^{R}=\{\hat{U}_{f}\Phi _{n}^{1},\hat{U}_{f}\Phi _{n}^{2},...,%
\hat{U}_{f}\Phi _{n}^{D-1}\}.
\end{equation}

Moreover, for the `remaining vector' $\Phi _{n+1}^{D},$ it turns out in this
case that
\begin{equation}
\Phi _{n+1}^{D}=\hat{U}_{f}\Phi _{n}^{D}
\end{equation}
because in this instance
\begin{equation}
\langle \Phi _{n+1}^{D}|\hat{U}^{(\mu )}\Phi _{n}^{\mu }\rangle =\langle
\Phi _{n}^{D}\hat{U}_{f}|\hat{U}_{f}\Phi _{n}^{\mu }\rangle =\langle \Phi
_{n}^{D}|\Phi _{n}^{\mu }\rangle =0
\end{equation}
for all $\mu =1,...,(D-1),$ as required.

So, assuming that $\hat{U}_{f}\neq \hat{I}$ it is evident that
\begin{equation}
\frak{B}_{n+1}=\hat{U}_{f}\frak{B}_{n}\neq \frak{B}_{n}
\end{equation}
such that $\frak{B}_{n+1}$ does provide a suitable basis set of states for a
jump from $\Psi _{n}$ to $\Psi _{n+1}$ to occur.

Furthermore, the same set of operators $\{\hat{U}^{(i)}=\hat{U}%
_{f}:i=1,...,D\}$ can then be applied to the elements of $\frak{B}_{n+1}$ to
give a new orthogonal basis, which may be labelled as $\frak{B}_{n+2},$ and
the process may be continued. Thus, the proposed mechanism may be described
as valid, and the above conclusion is justified: Type $IV$ Class $1$ Basis
Method Rules are indeed allowed.\bigskip

Overall, then, the development of a universe according to this type of Type $%
IV$ Basis Method Rule proceeds by the state $\Psi _{n}$ collapsing to the
state $\Psi _{n+1},$ which is one of the elements $\Phi _{n+1}^{i}$ of the
basis set
\begin{equation}
\frak{B}_{n+1}=\hat{U}_{f}\frak{B}_{n}=\{\Phi _{n+1}^{i}=\hat{U}_{f}\Phi
_{n}^{i}:i=1,...,D\}
\end{equation}
with probability
\begin{equation}
P(\Psi _{n+1}=\Phi _{n+1}^{i})=|\langle \hat{U}_{f}\Phi _{n}^{i}|\Psi
_{n}\rangle |^{2}
\end{equation}
where $\Phi _{n+1}^{i}$ may equally be viewed as one of the eigenstates of $%
\hat{\Sigma}_{n+1},$ which is a member of the equivalence class of operators
$\hat{B}_{n+1}$ implied by $\frak{B}_{n+1}.$ The universe then collapses to
the state $\Psi _{n+2},$ which is one of the elements $\Phi _{n+2}^{i}$ of
the basis set
\begin{equation}
\frak{B}_{n+2}=\hat{U}_{f}\frak{B}_{n+1}=\hat{U}_{f}\hat{U}_{f}\frak{B}%
_{n}=\{\Phi _{n+2}^{i}=\hat{U}_{f}\hat{U}_{f}\Phi _{n}^{i}:i=1,...,D\}
\end{equation}
with probability $P(\Psi _{n+2}=\Phi _{n+2}^{i})=|\langle \hat{U}_{f}\hat{U}%
_{f}\Phi _{n}^{i}|\Psi _{n+1}\rangle |^{2}.$ And so on.\bigskip

Of course, the particular deterministic (Type $IV)$ Rules presented above
can be generalised to probabilistic (Type $IVa)$ cases in the obvious way.
For instance, the next basis $\frak{B}_{n+1}$ could be given by
\begin{equation}
\frak{B}_{n+1}=\hat{U}_{f}\frak{B}_{n}
\end{equation}
with probability $P_{f},$ or instead by
\begin{equation}
\frak{B}_{n+1}=\hat{U}_{g}\frak{B}_{n}
\end{equation}
with probability $P_{g},$ and so on, where $\hat{U}_{f}$ and $\hat{U}_{g}$
are different unitary operators, and $P_{f}+P_{g}+...=1.$\bigskip

In effect, in the above mechanisms the unitary operator $\hat{U}_{f}$ (or $%
\hat{U}_{g}$ etc.) may be thought of as `rotating' the entire basis set $%
\frak{B}_{n}$ into the set $\frak{B}_{n+1},$ and then subsequently rotating
this new basis set $\frak{B}_{n+1}$ into the set $\frak{B}_{n+2},$ and so
on. In addition, because $\hat{U}_{f}$ is effectively behaving globally on
the whole basis set, that is, because $\hat{U}_{f}$ \ is rotating each
member of the basis set in the same way, its application automatically
preserves the orthogonality between the individual elements, as required.

So, the proposed Rule clearly provides a simple, but valid, mechanism for
dynamics, where the next operator $\hat{\Sigma}_{n+1}=\hat{B}_{n+1}$ chosen
by the universe to test the state $\Psi _{n}$ is strongly related to the
previous operator $\hat{\Sigma}_{n}=\hat{B}_{n}$ of which $\Psi _{n}$ is an
eigenstate. Moreover, the procedure is valid for all $n.$ The above
mechanism therefore provides an example of a Basis Self-Referential, Class $%
1 $ Basis Method dynamics, a conclusion made particularly significant by the
lack of any analogous State Self-Referential, Class $1$ Basis Method
dynamics.\bigskip

Of course, analogies of the other sorts of mechanism discussed in this
chapter are naturally expected to exist within the frameworks of\ Type $IV$
and $V$ Rules.

\bigskip

Note that the type of Rule proposed above could have an important physical
consequence. Consider as before a universe developing according to the Rule $%
\frak{B}_{n+1}=\hat{U}_{f}\frak{B}_{n},$ but this time impose the additional
condition that
\begin{equation}
\hat{U}_{f}=\hat{I}+\epsilon \hat{U}^{\prime }
\end{equation}
where $\hat{I}$ is the identity operator, $\epsilon $ a small parameter, and
$\hat{U}^{\prime }$ an operator chosen according to the constraint that $%
\hat{U}_{f}$ obeys the conditions required for the dynamics (i.e. $\hat{U}%
_{f}$ is unitary).

Now because $\epsilon $ is small, it follows that $\hat{U}_{f}$ approximates
to $\hat{I},$ i.e. $\hat{U}_{f}\approx \hat{I}.$ So, in this case
\begin{equation}
\frak{B}_{n+1}=\hat{U}_{f}\frak{B}_{n}\approx \hat{I}\frak{B}_{n}=\frak{B}%
_{n}
\end{equation}
such that the preferred basis at time $n+1$ is `roughly' the same as the
preferred basis at time $n.$ Moreover, this then implies that the next basis
set $\frak{B}_{n+1}$ will contain a member that is very `similar', in some
sense, to the current state $\Psi _{n}.$ So, and due to the Born probability
rule, the universe is highly likely to jump to this `very similar' state,
where the probability that the universe collapses to this vector is expected
to approach unity in models with Hilbert spaces of high dimensionality.
Overall, then, the outcome from such a Rule is that $\Psi _{n+1}\approx \Psi
_{n}$ for all $n.$

In other words, in a universe developing according to this type of Rule, the
state changes only very `slightly' from one jump to the next. Importantly,
then, such a mechanism might be useful to describe a possible origin of
apparent persistence in the quantum universe. Moreover, it might also
provide a dynamics in which the universe's development appears almost
deterministic, just as seems to be the case in classical physics: given a
state $\Psi _{n},$ it would be possible to predict what the next state $\Psi
_{n+1}$ will be like with near certainty, because $\Psi _{n+1}\approx \Psi
_{n}.$

Of course, much work is required to fully justify these assertions, and to
define exactly what the notion of `similarity' might imply.

\bigskip

\subsubsection{Summary}

\bigskip

As a final remark to this section, it should be mentioned that it is also
possible to envisage universes governed by Rules that are themselves subject
to change. Indeed, the Type of Rule used to select the next operator $\hat{%
\Sigma}_{n+1}$ could actually depend on $n,$ such that for example at `time'
$m$ a particular Type $III$ List-Sort mechanism could be used to select $%
\hat{\Sigma}_{m+1},$ whereas at time $m^{\prime }$ a different Type $III$
List-Sort Rule might be employed instead to give $\hat{\Sigma}_{m^{\prime
}+1},$ whilst at time $m^{\prime \prime }$ the universe could adopt a Type $%
III$ Generated-Sort dynamics, but at time $m^{\prime \prime \prime }$an
entirely different Type $IV$ Rule could be used. And so on.

Under such circumstances, it might be expected that there is some sort of
`Meta-Rule' governing the dynamics of how the Rules change with $n,$ a
concept that is analogous to Buccheri's idea of the ``Rules of the rules''
\cite{Buccheri} which determines how the laws of physics may develop with
time.\bigskip

It is possible now to sum up the various Types of fixed Rule dynamics
discussed so far in this chapter, and compare these with universes governed
by Rules that change and develop over time.

Starting with classical physics, for example, if a scientist knows the
current `state' of the Universe (i.e. the position and momentum of every
particle it contains) and the laws of physics, she is able to determine with
certainty what its entire future will be. This is the deterministic physics
of Newton's clockwork universe.

In Type $0$ quantum universes, on the other hand, if the scientist knows the
current state of the universe and decides upon a particular operator to test
it with, she is able to determine with certainty what the next basis set of
eigenstates will be, and hence estimate the next state within the bounds
imposed by quantum probability.

Furthermore, the same comment is broadly true for the Type $III$ List-sort
dynamics discussed in Section 8.2: if the List $\frak{L}\equiv \{\frak{B}%
_{i}:i=1,2,...,l\}$ of potential next operators is specified, and the Rules
governing which of these is chosen to test the state are understood, then
given a state $\Psi _{n}$ it is always possible to determine what the next
set of eigenstates will be. Moreover, under such circumstances the state of
the universe will always be an element of one of the basis sets from the
list $\frak{L},$ and this state will always be tested by one of the
operators $\{\hat{B}_{i}\}.$

So, it is consequently possible not only to predict the probability of
obtaining a particular next state, but also to ask questions of the form:
``if the next state turns out to be $X,$ what is the probability that the
subsequent state will be $Y?";$ or, ``what is the probability that the test
after $n$ steps will be a particular operator $\hat{B}_{a}?".$\bigskip

In Class $1$ Type $III$ Generated-Sort dynamics, if the current state of the
universe is known, it is only ever possible to determine the next operator
that acts. Unlike the List-sort dynamics, there is now no pre-existing List
of basis sets specifying every operator that will ever be used in the
universe's development, because the universe is `making-up' its tests as it
goes along. Additionally, unlike for List-Sort dynamics, in which the number
of different states the universe could ever potentially exist in (its phase
space) is fixed by the length of the List and the dimension of the Hilbert
space, in Generated-Sort dynamics this number of different states could be
unbounded. However, whilst it may not be known in advance what every future
operator will be, because the Rules are known and the function $f$ relating $%
\frak{B}_{n+1}$ to $\Psi _{n}\in \mathcal{H}^{(D)}$ is fixed, it is possible
to say that if the universe were ever in a particular state $Z,$ the
subsequent state would be one of the $D$ eigenvectors of an operator $\hat{%
\Sigma}=f(Z)$ determined by $Z.$

Conversely, in Class $2$ Type $III$ Generated-Sort dynamics it is not
possible to determine what the next operator will be if only the current
state is known. In this case, if the universe were ever in a state $Z,$ it
could not immediately be inferred what the next set of eigenvectors must be,
because the relationship $f_{n}$ between states and operators is always
changing. Specifically, if a universe is governed by a Rule of the form $%
\hat{\Sigma}_{n+1}=f_{n}(\Psi _{n}),$ it is necessary to know both the state
$\Psi _{n}$ and the `time' $n$ in order to determine the next operator $\hat{%
\Sigma}_{n+1}.$ In this instance, information additional to the knowledge of
the current state is required (for example how many jumps have taken place
since a particular `reference state' $\Psi _{0},$ or at least what the
previous operator was), because even if it is known how the operator changes
with `time', it is still necessary to specify what the time is in order to
say how much it has changed. However, given $\Psi _{n}$ and $n$ it is then
possible to determine $\hat{\Sigma}_{n+1},$ because the Rule $\hat{\Sigma}%
_{n+1}=f_{n}(\Psi _{n})$ relating $\hat{\Sigma}_{n+1}$ to $\Psi _{n}$ is
defined for all time at the outset.\bigskip

Finally, in a universe governed by Rules which also change, it is not
sufficient just to know the current Rule, state and time in order to
determine $\hat{\Sigma}_{n+1},$ but it is also necessary to know the Rules
of the Rules. Such `Meta-Rules' could then be used to select a particular
Type of Rule, which could then choose a particular operator $\hat{\Sigma}%
_{n+1}$ based somehow, perhaps, on the current state $\Psi _{n}$ and/or the
last basis $\frak{B}_{n}$ and/or the current time $n.$ Of course, these
choices could also even depend upon some sort of additional variable
previously indiscussed.

In reality, it would be very difficult for endo-physical observers to ever
ascertain what the Rules of the Rules governing their universe actually are.
After all, a physicist could (at best) only ever really be sure of what the
current Rule is, and the Rules are assumed to be constantly changing. So,
although such universes will not be discussed further in this work, note
that this point has analogies with some of the recent speculations in
fundamental physics regarding whether the speed of light or the electron
charge have remained constant throughout the history of the Universe (e.g.
\cite{Albrecht} and \cite{Peres2}, respectively). In both of these cases, it
is difficult to reconstruct what the laws of physics were like in the
distant past when only the current state of the Universe is available for
study.\bigskip

It is also far beyond the scope of this thesis to take the logically greater
step and consider a completely `free' universe, that is, one in which
neither the Rules, nor the Rules governing the Rules, are fixed and
specified in advance. Indeed, it is difficult to imagine how such a model
could even be envisaged that required no order or direction, at least at the
outset. After all, even a universe incorporating Meta-Rules relies on a
definition of what this Meta-Rule is, and additionally on what the boundary
or initial condition $\Psi _{0}$ was. In fact, in any such `free universe',
order, Rules, and even Rules of Rules would have to be defined or `created'
somehow on their own account as the universe develops, and it is almost
impossible to comprehend how this process could occur. A mechanism governed
by Rules of the Rules might therefore represent the `final level' that can
be used to describe a fully quantum universe.

Ultimately, then, the actual definition of the Rules governing a universe
(or at least the choice of the Rules of the Rules) could have no origin that
is explainable in terms of any sort of higher order mechanism. Consequently,
their presence and form in any given model may have to be accepted merely as
a fundamental pre-requisite, just as the existence of the underlying Hilbert
space or the List $\frak{L}$ is taken to be.

Of course, this is similar to the philosophical problem faced in the real
Universe: physicists might one day be able to determine \textit{what} the
Theory of Everything is, but to say \textit{why} it is like this without
appealing to blind chance, the Anthropic Principle, or a Higher Being may be
beyond the scope of empirical physics. Scientists may never be able to say
why the constants of nature have the values that they do, but just that they
are predicted by a theory that happens to describe the reality they exist in.

\bigskip

\subsection{Examine-Decision Mechanisms}

\bigskip

Each of List-Sort and Generated-Sort dynamics attracts an obvious comment.

The List-Sort Rules rely on the decision of operator $\hat{\Sigma}_{n+1}$
being made based upon a particular `property' of the state $\Psi _{n},$ for
example its separability. No explanation is given, however, as to how the
Rules actually get to `know' what this property is, such that they can then
make the selection. Paraphrasing, there is no `self-examining' part of the
List-Sort algorithm that allows the universe to explicitly investigate its
state for a particular property (e.g. count how many factors\ $\Psi _{n}$
has), such that the next operator may then be chosen appropriately from the
List.

The same remark is not necessary in Generated-Sort mechanisms, because in
these the next operator is a direct function of the current state. However,
the types of Rules discussed in Section 8.4 suffer instead from the fact
that the dynamics which results is rather inflexible: once a state $\Psi
_{n} $ is specified, there is no choice about how the next operator will be
defined. In other words, there is no `decision making' part of the
Generated-Sort algorithm applying logic of the form: \textit{if} the
universe finds itself in state $\Psi _{n}=x$ \textit{then} use an operator $%
\hat{\Sigma}_{n+1}=f(x);$ but \textit{if} \textit{instead} the universe is
in state $y$ \textit{then instead} use an operator $\hat{\Sigma}_{n+1}=g(y)$
generated in an alternative way, where $f\neq g;$ and so on. For example,
there is no freedom in the mechanism to allow the state to be tested by,
say, $\hat{\Sigma}_{n+1}=h(\Psi _{n})$ if $\Psi _{n}$ has $F_{H}$ factors,
but instead by $\hat{\Sigma}_{n+1}=k(\Psi _{n})$ if it has $F_{K}$ factors;
according to Generated-Sort Rules, the actual `properties' of $\Psi _{n}$
are not used to decide how the state develops.\bigskip

Whilst these two comments are not serious problems, such that both List-Sort
and Generated-Sort Rules can still be taken to provide valid dynamics for
the quantum universe, it is natural to speculate on whether mechanisms could
exist that appear to develop the state without these limitations. After all,
on the small scale this type of development is what real physicists tend to
experience in the laboratory: in general, scientists do believe themselves
to be able to examine and investigate quantum sub-systems, and then choose
how to develop them, from a huge number of different ways, based upon what
they have learnt.

In this section, therefore, it is hoped to explore the possibility of
universes that are somehow able to `examine' their state themselves for a
particular property, and then develop it in a way that depends on what this
property is.\ The desire, then, is to investigate sets of Rules that could
provide a fully automatic and self-referential mechanism that leads to a
dynamics equivalent to a process of examination, decision, development,
examination, decision, development,..., continuing indefinitely. Universes
governed by such mechanisms could be described as obeying \textit{%
Examine-Decision} (ED) Rules.

\bigskip

\subsubsection{Preliminary Considerations}

\bigskip

In order to be valid, the Rules governing ED dynamics must define a single
quantum computation that, in one time step, `examines' the universe's state $%
\Psi _{n},$ `decides' upon a course of action according to the result of the
initial examination, and then consequently develops it to the next state $%
\Psi _{n+1}.$ So, the question becomes: how might it be possible to
construct mechanisms that examine the state, and then develop it in a way
that depends on the outcome of this investigation?

The overall goal of this section is to investigate how such computations
might be achieved.\bigskip

The first point to note, however, is that any \textit{measurement} of any
property of the state\ $\Psi _{n}$ of the universe necessarily involves a
process of information extraction. This fact would not cause any problems in
a classical universe, because in classical physics it is possible to observe
an object and expect it to remain unchanged. In a fully quantum universe, on
the other hand, such non-invasive techniques are forbidden: it is not
possible to measure a quantum state for one property, and then test the same
state for something else. The first measurement destroys the original state
and creates a new one that is an eigenstate of whichever operator was used,
and it is this new state that then has to be tested in subsequent
measurements.\bigskip

Now, \textit{if} a hypothetical ED mechanism was proposed that followed the
general algorithm

\begin{enumerate}
\item  test the state $\Psi _{n}$ for a particular `property' $p^{(i)}$ from
the set $\{p^{(i)},i=1,2,...\},$ by applying an operator $\hat{P};$

\item  then collapse the state $\Psi _{n}$ into one of the eigenvectors of a
particular Hermitian operator $\hat{\Sigma}_{n+1},$ where, if the result of $%
``1."$ is $p^{(1)}$ the operator $\hat{\Sigma}_{n+1}=\hat{O}^{(1)}$ is used;
but, if instead the result of $``1."$ is $p^{(2)}$ then the operator $\hat{%
\Sigma}_{n+1}=\hat{O}^{(2)}$ is used instead; and so on, accounting for
every possible outcome $p^{(i)}$ of $\hat{P};$
\end{enumerate}

the above discussion would consequently cause a problem. Specifically, if
the examination by $\hat{P}$ of the property $p^{(i)}$ of the state $\Psi
_{n}$ is taken to be a measurement process, then it would lead to a
wavefunction collapse, and so after this examination the universe will be in
a new state $\Psi _{n+1},$ which is one of the eigenvectors of $\hat{P}.$
Clearly, it is now too late to test the `old' state $\Psi _{n}$ by whichever
operator $\hat{\Sigma}_{n+1}\in \{\hat{O}^{(1)},\hat{O}^{(2)},...\}$ is
implied by Rule $``2."$ from the information gained as a result of this
measurement.

In fact, the universe would next have to be developed by some operator $\hat{%
\Sigma}_{n+2},$ and would accordingly jump to a new state $\Psi _{n+2}.$%
\bigskip

This\ conclusion highlights the fact that any method used to measure the
state $\Psi $ for a particular property must be seen as equivalent to the
operators $\hat{\Sigma}$ used in the universe's development. Indeed, this is
not surprising: recall that the Hermitian operators $\hat{\Sigma}$ have been
regarded throughout this thesis as being synonymous with physical tests
anyway. Thus, in this case the test $\hat{P}$ was effectively used as the
test $\hat{\Sigma}_{n+1},$ so in this universe $\hat{\Sigma}_{n+1}=\hat{P}.$

So, any examination procedure that involves a measurement does not fulfil
the intention of finding an ED computation that proceeds in a single time
step. Consequently, in any suggested ED mechanism the examination part of
the algorithm cannot rely on any sort of physical measurement or information
extraction process: two tests per time step are not allowed in the proposed
paradigm.

It is therefore necessary to choose the ED Rules very carefully, such that
whatever `examination' procedure is employed avoids an actual physical
measurement of the state.

\bigskip

\subsubsection{Selective Global Evolution}

\bigskip

In an attempt to find a way around the above difficulty, recall that
operators used in standard quantum mechanics are generally one of two types,
namely, either Hermitian or unitary. Hermitian operators represent physical
measurements, and are used to test the state, thereby resulting in its
collapse into one of the operator's eigenvectors. Unitary operators,
conversely, are used to evolve the state, or, in the sense discussed in this
thesis, `rotate' it into a new vector in its Hilbert space. Unlike Hermitian
operators, unitary operators do not extract any physical information from
the state, and are hence traditionally used in quantum theory to describe
the (Schr\"{o}dinger) development of the system in the absence of
observation.

Now, since it has been shown that the examination part of an ED mechanism
cannot be based upon a physical extraction of information from the state, if
Rules are to be proposed that do provide suitable Examine-Decision dynamics
for the development of a quantum universe, they must rely on non-invasive
techniques.

Particularly, if the examination part cannot rely on Hermitian tests, the
suggestion might be that it should instead be based upon unitary operators.

So, as an alternative to a dynamics based upon a universe that is \textit{%
tested} by an operator chosen according to some property of its state (as
hypothesised in Section 8.5.1), it might instead be possible to conject a
dynamics in which the universe is \textit{evolved} in a way that depends on
some property of its state.\bigskip

The suggestion, then, is that the examination and decision parts of an ED
mechanism could be governed by a unitary operator. Consequently, the
existence of a unitary operator $\hat{U}_{n}$ is hypothesised that appears
to `examine' the state $\Psi _{n}$ for a particular property, `decides' how
it is to be developed, and then accordingly develops it into $\Psi
_{n}^{\prime }=\hat{U}_{n}\Psi _{n}$ in a way that depends upon this
`property'.

So, Examine-Decision Rules are considered that involve both unitary
operators $\hat{U}_{n}$ and Hermitian tests $\hat{\Sigma}_{n+1}.$
Specifically, when a certain, carefully defined unitary operator $\hat{U}%
_{n} $ is applied to the state $\Psi _{n},$ if $\Psi _{n}$ has one
particular property the universe will be evolved in one way, whilst if
instead $\Psi _{n}$ has a second property it is rotated in a different way,
and so on. The evolved state $\Psi _{n}^{\prime }$ may then be tested in the
usual manner by some Hermitian operator $\hat{\Sigma}_{n+1}$ that is chosen,
perhaps, by one of the various Types of Rule discussed in Section 8.1, and
the universe subsequently collapses to $\Psi _{n+1}.$

Clearly, because no physical information gets extracted from the state $\Psi
_{n}$ during the application of $\hat{U}_{n}$ (i.e. regarding what its
properties actually are; the choice of $\hat{\Sigma}_{n+1}$ does not depend
on this initial examination), no quantum collapse occurs, and the overall
development of the state using the operators $\hat{U}_{n}$ and $\hat{\Sigma}%
_{n+1}$ proceeds in one time step, as required. Of course, this condition is
ensured because the `examination' process is part of $\hat{U}_{n},$ and is
hence not a Hermitian test.

Such a mechanism may be described as \textit{Selective Global Evolution}
(SGE), because properties of the state are being used to select the way in
which $\Psi _{n}$ is globally evolved.\bigskip

The development of a universe according to a SGE mechanism would therefore
proceed in two distinct parts: an evolution part and a test part. Thus, the
general Rules governing such a universe may be of the form

\begin{itemize}
\item  Apply $\hat{U}_{n}$ to $\Psi _{n}$ to give the `rotated' state $\Psi
_{n}^{\prime }=\hat{U}_{n}\Psi _{n};$

\item  Test $\Psi _{n}^{\prime }$ with $\hat{\Sigma}_{n+1}$ to give the new
state $\Psi _{n+1},$ which is one of the eigenvectors of $\hat{\Sigma}%
_{n+1}; $
\end{itemize}

where $\hat{U}_{n}$ is defined such that it can Selectively Globally Evolve
the state $\Psi _{n}.$ The process may then be repeated, such that $\Psi
_{n+1}$ is selectively rotated by $\hat{U}_{n+1}.$

The actual development of a SGE governed universe will become clearer in the
following.

\bigskip

The overall goal of SGE dynamics is to use an operator $\hat{U}_{n}$ that
evolves the universe in a manner that depends upon some property of its
state. So, the obvious first question is: what sort of unitary operators
could give rise to such a selective development? How could $\hat{U}_{n}$ be
constructed such that it evolves different vectors in different ways,
depending on what this vector is?\bigskip

Perhaps the most obvious answer to this question would be to suggest that $%
\hat{U}_{n}$ is an operator that somehow `contains' (in a sense to be
defined) many other unitary operators $\{\hat{u}_{n}^{(i)}:i=1,2,...\},$
each of which can be somehow `turned on or off' depending on the particular
properties of the state $\Psi _{n}$ on which $\hat{U}_{n}$ acts. In other
words, a mechanism might schematically be suggested in which under some
circumstances the operator $\hat{U}_{n}$ `looks' like the unitary operator $%
\hat{u}_{n}^{(1)},$ whereas under different circumstances it instead
effectively behaves like the unitary operator $\hat{u}_{n}^{(2)},$ etc.,
where $\hat{u}_{n}^{(1)},\hat{u}_{n}^{(2)},...$ are different unitary
operators. Clearly, the ``actual circumstances'' would be dictated by the
Rules governing the universe's development, and it would be a `property' of
the state $\Psi _{n}$ that actually causes $\hat{U}_{n}$ to resemble one
particular operator, $\hat{u}_{n}^{(i)},$ over another, $\hat{u}_{n}^{(j)}.$

Continuing this schematic viewpoint, it is possible to imagine a mechanism
in which $\hat{U}_{n}$ is defined such that if $\Psi _{n}$ has one
particular form, then the application of $\hat{U}_{n}$ to $\Psi _{n}$ gives
the same result as an application of\ $\hat{u}_{n}^{(1)}$ to $\Psi _{n},$
whereas if $\Psi _{n}$ has a different property, then $\hat{U}_{n}\Psi _{n}$
is instead effectively equivalent to $\hat{u}_{n}^{(2)}\Psi _{n},$ and so
on. For example, $\hat{U}_{n}$ could be such that if $\Psi _{n}$ is the
state $\Psi _{n}=\Theta ,$ then $\hat{U}_{n}\Psi _{n}=\hat{U}_{n}\Theta =%
\hat{u}_{n}^{(i)}\Theta ,$ whereas if $\Psi _{n}\ $is instead the state $%
\Psi _{n}=\Phi ,$ then $\hat{U}_{n}\Psi _{n}=\hat{U}_{n}\Phi =\hat{u}%
_{n}^{(j)}\Phi ,$ where $\hat{u}_{n}^{(i)}\neq \hat{u}_{n}^{(j)}.$

Overall, then, different states are taken to cause the same operator $\hat{U}%
_{n}$ to behave differently.\bigskip

Of course, even in such a schematic model the issue of how the various
operators $\{\hat{u}_{n}^{(i)}\}$ are actually ``turned on or off'' by the
properties of the state remains to be addressed.

So, one potential suggestion might be to propose that $\hat{U}_{n}$ could in
fact `contain' (again, in a sense to be defined) many pairs of operators $\{%
\hat{u}_{n}^{(i)}\hat{S}_{n}^{(i)}\},$ where the $\{\hat{S}_{n}^{(i)}\}$ are
defined as `asking' operators. Thus, the idea, again schematic, is that when
$\hat{U}_{n}$ acts on $\Psi _{n},$ the operator $\hat{S}_{n}^{(i)}$ in each
pair `asks' a question of the state, and the remaining operator $\hat{u}%
_{n}^{(i)}$ either rotates $\Psi _{n}$ if the answer to this question is
``Yes'', but is not applied to $\Psi _{n}$ if the answer is instead ``No''.
It is this potential mechanism that is explored now.

Each `asking' operator $\hat{S}_{n}^{(i)}$ must obey the condition that it
provides a definite answer, either ``Yes'' or ``No'', for a given state $%
\Psi _{n},$ such that there is no ambiguity in whether the corresponding
unitary operator $\hat{u}_{n}^{(i)}$ is applied or not. Furthermore, note
that if the set $\{\hat{S}_{n}^{(i)}\}$ is constrained such that the
individual `questions' are mutually exclusive of the others, that is
\begin{equation}
\hat{S}_{n}^{(i)}\Psi _{n}\rightarrow \left\{
\begin{array}{c}
\text{``Yes'' for }i=a \\
\text{``No'' for all }i\neq a
\end{array}
\right\}
\end{equation}
it follows in this case that $\{\hat{u}_{n}^{(i)}\hat{S}_{n}^{(i)}\}$ acting
on $\Psi _{n}$ is equivalent to just $\hat{u}_{n}^{(a)}$ acting on $\Psi
_{n},$ as desired; no other operator $\hat{u}_{n}^{(j)},$ $j\neq a,$ is
applied.

Continuing, it is also observed that the Yes/No answers to the `questions' $%
\{\hat{S}_{n}^{(i)}\}$ could be associated with binary logic of the form
``Yes''$\Rightarrow 1$ and ``No''$\Rightarrow 0.$

So, one choice for the operators $\{\hat{S}_{n}^{(i)}\}$ could be to define
them according to the rule

\begin{equation}
\hat{S}_{n}^{(i)}\Psi _{n}=\left\{
\begin{array}{c}
1\times \Psi _{n}\text{ if }\hat{S}_{n}^{(i)}\Psi _{n}\text{ }\rightarrow
\text{ ``Yes''} \\
0\times \Psi _{n}\text{ if }\hat{S}_{n}^{(i)}\Psi _{n}\text{ }\rightarrow
\text{ ``No''}
\end{array}
\right\}
\end{equation}

In this case, $\hat{U}_{n}$ could be written as a linear sum of the pairs $\{%
\hat{u}_{n}^{(i)}\hat{S}_{n}^{(i)}\},$ that is, in the form
\begin{equation}
\hat{U}_{n}=\hat{u}_{n}^{(1)}\hat{S}_{n}^{(1)}+\hat{u}_{n}^{(2)}\hat{S}%
_{n}^{(2)}+\hat{u}_{n}^{(3)}\hat{S}_{n}^{(3)}+...  \label{Ch6Choices}
\end{equation}
and this imposes an additional, obvious constraint on the operators $\{\hat{u%
}_{n}^{(i)}\hat{S}_{n}^{(i)}\}$: the pairs $\hat{u}_{n}^{(i)}\hat{S}%
_{n}^{(i)}$ must be chosen such that the unitarity of the overall operator $%
\hat{U}_{n}$ is preserved.

Under the above circumstances, for each pair $\hat{u}_{n}^{(i)}\hat{S}%
_{n}^{(i)}$ acting on $\Psi _{n},$ if the operator $\hat{S}_{n}^{(i)}$
applied to $\Psi _{n}$ gives the result $1,$ the product $\hat{u}_{n}^{(i)}%
\hat{S}_{n}^{(i)}\Psi _{n}$ equals $\hat{u}_{n}^{(i)}\Psi _{n},$ and so $%
\hat{u}_{n}^{(i)}$ is used to evolve the state $\Psi _{n}.$ However, if
instead $\hat{S}_{n}^{(i)}\Psi _{n}$ gives the result $0,$ the combination $%
\hat{u}_{n}^{(i)}\hat{S}_{n}^{(i)}\Psi _{n}$ also becomes $0,$ and the
unitary operator $\hat{u}_{n}^{(i)}$ is effectively `removed' from the
equation. In essence, the state $\Psi _{n}$ only `sees' the unitary operator
$\hat{u}_{n}^{(i)}$ if $\hat{S}_{n}^{(i)}\Psi _{n}=1\times \Psi _{n};$ the
operator $\hat{S}_{n}^{(i)}$ is effectively being used to turn $\hat{u}%
_{n}^{(i)}$ `on or off' depending entirely on the properties of the state $%
\Psi _{n}.$ As throughout this chapter, the parallels between the above type
of logic and that exhibited in (quantum) computational gates are evident.

So, if $i=a$ is the only value for which $\hat{S}_{n}^{(i)}\Psi
_{n}\rightarrow ``Yes",$ it consequently follows that
\begin{equation}
\hat{U}_{n}\Psi _{n}=\hat{u}_{n}^{(a)}\Psi _{n}
\end{equation}
as desired.

Note that $\{\hat{S}_{n}^{(i)}\}$ could be associated with a suitable set of
projection operators, as suggested below, because these can conventionally
be interpreted as Yes/No operators in quantum\ mechanics.

\bigskip

As an illustration of how a possible such SGE mechanism might be
constructed, consider a universe represented by a state $\Psi _{n}$ in a $D$
dimensional Hilbert space $\mathcal{H}^{(D)}$ spanned by the orthonormal
basis $\mathcal{B}^{(D)}=\{|i\rangle :i=0,...,(D-1)\}.$ Further, assume that
the dynamics of the universe are governed by the SGE two part Rules: $\Psi
_{n}$ is evolved to $\Psi _{n}^{\prime }$ by applying the unitary operator $%
\hat{U}_{n},$ i.e. $\Psi _{n}\rightarrow \Psi _{n}^{\prime }=\hat{U}_{n}\Psi
_{n},$ where $\hat{U}_{n}$ acts selectively; and then $\Psi _{n}^{\prime }$
is collapsed to $\Psi _{n+1}$ by a test with the Hermitian operator $\hat{%
\Sigma}_{n+1},$ where $\Psi _{n+1}$ is an eigenstate of $\hat{\Sigma}_{n+1}.$
Finally, consider for all $n$ defining $\hat{\Sigma}_{n+1}$ as an operator
with a basis set of eigenstates $\frak{B}_{n+1}=$ $\{|i\rangle
:i=0,...,(D-1)\},$ such that the collapsed state is always a member of the
set $\{|i\rangle \}.$

Now, assume that $\hat{U}_{n}$ is of the form given in (\ref{Ch6Choices}),
but that each $\hat{S}_{n}^{(i)}$ is defined as the projection operator $%
\hat{S}_{n}^{(i)}=|i\rangle \langle i|,$ for $i=0,...,(D-1).$ That is

\begin{itemize}
\item  $\hat{U}_{n}=\hat{u}_{n}^{(0)}|0\rangle \langle 0|+\hat{u}%
_{n}^{(1)}|1\rangle \langle 1|+...+\hat{u}_{n}^{(D-1)}|D-1\rangle \langle
D-1|$
\end{itemize}

where the $\{\hat{u}_{n}^{(i)}\}$ are particular unitary operators to be
defined in due course.

Clearly, each pair of operators $\hat{u}_{n}^{(i)}|i\rangle \langle i|$ acts
sequentially on $\Psi _{n};$ first the operator $|i\rangle \langle i|$ is
applied to the state $\Psi _{n},$ then the resulting vector $(|i\rangle
\langle i|\Psi _{n})$ is rotated by the unitary operator $\hat{u}_{n}^{(i)}.$
However, since $\Psi _{n}\in \frak{B}_{n}=$ $\{|i\rangle \}$ for all $n,$
the expression
\begin{equation}
|i\rangle \langle i|\Psi _{n}=1\times \Psi _{n}
\end{equation}
is true for only one value of $i;$ for all other values, the application of $%
|i\rangle \langle i|$ to $\Psi _{n}$ gives $|i\rangle \langle i|\Psi
_{n}=0\times \Psi _{n}.$ Labelling this one value $i=a,$ it implies that $%
\Psi _{n}=|a\rangle ,$ and so it consequently follows that
\begin{equation}
\hat{U}_{n}\Psi _{n}=\left[ \sum_{i=0}^{D-1}\hat{u}_{n}^{(i)}|i\rangle
\langle i|\right] |a\rangle =0+0+...+\hat{u}_{n}^{(a)}|a\rangle +0+...+0.
\end{equation}

Thus, the state is evolved globally by an operator $\hat{u}_{n}^{(a)},$
selected from the set $\{\hat{u}_{n}^{(i)}\}$ `contained' in $\hat{U}_{n},$
because the universe is initially in the state $\Psi _{n}=|a\rangle .$

Overall, the projection operator $|i\rangle \langle i|$ is effectively
`asking' whether or not $\Psi _{n}$ is in the state $\Psi _{n}=|i\rangle .$
The interpretation of the above type of Selection mechanism is that the
projection operators cause $\hat{U}_{n}$ to act like a set of ``\textit{If}%
'' statements: if the state $\Psi _{n}$ is $|0\rangle ,$ then it is rotated
by $\hat{u}_{n}^{(0)},$ but if the state $\Psi _{n}$ is instead $|1\rangle ,$
then it is instead rotated by $\hat{u}_{n}^{(1)},$ and so on. In other
words, given a universe prepared as $\Psi _{n}=|i\rangle ,$ the $|i\rangle
\langle i|$ part of the operator $\hat{U}_{n}$ effectively selects the
operator $\hat{u}_{n}^{(i)}$ to evolve the state.\bigskip

The overall SGE mechanism is then concluded by the second part of the Rules.
So, the evolved state $\Psi _{n}^{\prime }=\hat{U}_{n}\Psi _{n}=\hat{u}%
_{n}^{(a)}|a\rangle $ is collapsed back into one of the vectors $\{|i\rangle
\}$ by the operator $\hat{\Sigma}_{n+1},$ with the usual quantum
probabilities.

The two part process may then be repeated, noting that because $\Psi
_{n+1}\in \frak{B}_{n}=$ $\{|i\rangle \},$ the projection operators $%
|i\rangle \langle i|$ in the next SGE operator $\hat{U}_{n+1}=\left[
\sum_{i=0}^{D-1}\hat{u}_{n+1}^{(i)}|i\rangle \langle i|\right] $ are still
able to provide a mutually exclusive and exhaustive set of `questions' for
the new state $\Psi _{n+1}.$\ Clearly, this point will be true for all $n.$%
\bigskip

Whilst the methodology behind the mechanism suggested above is sound, it
turns out that it cannot actually accomplish non-trivial SGE development.\
In particular, in order for the proposed mechanism to provide valid
dynamics, the set of operators $\{\hat{u}_{n}^{(i)}\}$ must be so
restrictively defined that any non-trivial selection is effectively removed.
Specifically, the necessary constraint of choosing the set $\{\hat{u}%
_{n}^{(i)}\}$ such that the overall unitarity of the operator $\hat{U}_{n}$
is preserved prevents the suggested SGE mechanism from developing the
universe in a non-trivial way.

Phrasing this more mathematically, it can be shown that the overall operator
$\hat{U}_{n}$ is only unitary if
\begin{equation}
\langle i|\hat{u}_{n}^{(i)\ast }\hat{u}_{n}^{(j)}|j\rangle =\delta _{ij}
\end{equation}
for all $i,j=0,1,...,(D-1),$ and this is only achieved if $\hat{u}_{n}^{(i)}=%
\hat{u}_{n}^{(j)}.$

This result is derived now. Note how the present ideas may be related to the
discussions given in Section 8.4.1 regarding Class $1$ and $2,$ Type $III$
Basis Method dynamics.\bigskip

\begin{proof}
Consider an operator $\hat{U}_{n}$ defined, as above, as
\begin{equation}
\hat{U}_{n}=\hat{u}_{n}^{(0)}|0\rangle \langle 0|+\hat{u}_{n}^{(1)}|1\rangle
\langle 1|+...+\hat{u}_{n}^{(D-1)}|D-1\rangle \langle D-1|
\end{equation}

where the $\hat{u}_{n}^{(i)}$ are unitary, for $i=0,1,...,(D-1).$

In order for $\hat{U}_{n}$ to be unitary it must be the case that $\hat{U}%
_{n}^{\ast }\hat{U}_{n}=\hat{I},$ where $\hat{I}$ is the identity operator,
and $\hat{U}_{n}^{\ast }$ is the transpose conjugate of $\hat{U}_{n}$ given
by
\begin{equation}
\hat{U}_{n}^{\ast }=|0\rangle \langle 0|\hat{u}_{n}^{(0)\ast }+|1\rangle
\langle 1|\hat{u}_{n}^{(1)\ast }+...+|D-1\rangle \langle D-1|\hat{u}%
_{n}^{(D-1)\ast }
\end{equation}

with $\hat{u}_{n}^{(i)\ast }\hat{u}_{n}^{(i)}=\hat{I}.$ So,
\begin{eqnarray}
\hat{I} &=&\left( |0\rangle \langle 0|\hat{u}_{n}^{(0)\ast }+|1\rangle
\langle 1|\hat{u}_{n}^{(1)\ast }+...+|D-1\rangle \langle D-1|\hat{u}%
_{n}^{(D-1)\ast }\right)  \label{Ch6IdentUU} \\
&&\times \left( \hat{u}_{n}^{(0)}|0\rangle \langle 0|+\hat{u}%
_{n}^{(1)}|1\rangle \langle 1|+...+\hat{u}_{n}^{(D-1)}|D-1\rangle \langle
D-1|\right)  \notag \\
&=&\tsum_{i=0}^{D-1}\tsum_{j=0}^{D-1}|i\rangle \langle i|\hat{u}%
_{n}^{(i)\ast }\hat{u}_{n}^{(j)}|j\rangle \langle j|.  \notag
\end{eqnarray}

The sum of terms $\tsum_{i=0}^{D-1}\tsum_{j=0}^{D-1}|i\rangle \langle i|\hat{%
u}_{n}^{(i)\ast }\hat{u}_{n}^{(j)}|j\rangle \langle j|$ may be used to
generate a $D\times D$ matrix. Moreover, the specific term $|i\rangle
\langle i|\hat{u}_{n}^{(i)\ast }\hat{u}_{n}^{(j)}|j\rangle \langle j|$ gives
the value $\langle i|\hat{u}_{n}^{(i)\ast }\hat{u}_{n}^{(j)}|j\rangle \in
\mathbb{C}$ of the component in the $[(i+1)^{th}$ row, $(j+1)^{th}$ column]
of this matrix.

Now, recall that the identity operator $\hat{I}$ may also be represented by
a $D\times D$ matrix, which contains components of zero everywhere apart
from the leading diagonal, where the values are one. So, for the above
equality (\ref{Ch6IdentUU}) to hold, it must be the case that
\begin{equation}
\langle i|\hat{u}_{n}^{(i)\ast }\hat{u}_{n}^{(j)}|j\rangle =\delta _{ij}
\end{equation}

for all $i$ and $j.$

Clearly, the equation $\langle i|\hat{u}_{n}^{(i)\ast }\hat{u}%
_{n}^{(j)}|j\rangle =\delta _{ij}$ is equivalent to the matrix elements of
the identity $\hat{I}$ operator: $\langle i|\hat{I}|j\rangle =\delta _{ij}.$
It is therefore the case that $\hat{u}_{n}^{(i)\ast }\hat{u}_{n}^{(j)}=\hat{I%
}.$ Moreover, since $\hat{u}_{n}^{(i)}$ and $\hat{u}_{n}^{(j)}$ are unitary
by definition, the relations $\hat{u}_{n}^{(i)\ast }\hat{u}_{n}^{(i)}=\hat{I}
$ and $\hat{u}_{n}^{(j)\ast }\hat{u}_{n}^{(j)}=\hat{I}$ must also be true.
Thus, because each unitary operator has one, and only one, inverse, it must
follow that $\hat{u}_{n}^{(j)\ast }=\hat{u}_{n}^{(i)\ast },$ such that $\hat{%
u}_{n}^{(i)}=\hat{u}_{n}^{(j)}$ for all $i$ and $j.$

Summarising, $\hat{U}_{n}$ is only unitary if $\hat{u}_{n}^{(i)}=\hat{u}%
_{n}^{(j)}$ for all $i,j=0,1,...,(D-1).$\bigskip
\end{proof}

Thus, $\hat{U}_{n}$ is only unitary if $\hat{u}_{n}^{(i)}=\hat{u}_{n}^{(j)}.$
In this case,
\begin{equation}
\langle i|\hat{u}_{n}^{(i)\ast }\hat{u}_{n}^{(j)}|j\rangle =\langle i|\hat{u}%
_{n}^{(i)\ast }\hat{u}_{n}^{(i)}|j\rangle =\langle i|\hat{I}|j\rangle
=\langle i|j\rangle =\delta _{ij}
\end{equation}
and it then follows that $\hat{U}_{n}=\hat{u}_{n}^{(i)},$ because
\begin{eqnarray}
\hat{U}_{n} &=&\hat{u}_{n}^{(0)}|0\rangle \langle 0|+\hat{u}%
_{n}^{(1)}|1\rangle \langle 1|+...+\hat{u}_{n}^{(D-1)}|D-1\rangle \langle
D-1| \\
&=&\hat{u}_{n}^{(i)}{\Large (}|0\rangle \langle 0|+|1\rangle \langle
1|+...+|D-1\rangle \langle D-1|{\Large )}=\hat{u}_{n}^{(i)}\hat{I}=\hat{u}%
_{n}^{(i)}.  \notag
\end{eqnarray}

Evidently, such a choice of $\{\hat{u}_{n}^{(i)}\}$ or $\hat{U}_{n}$ does
not lead to any Selective Global Evolution of the sort aimed at in this
section, because the universe would be evolved by the same operator\ $\hat{u}%
_{n}^{(i)}$ regardless of which state it is in.

So, the conclusion is that it is not possible to use the mechanism proposed
above to define a SGE dynamics for a universe based upon the selection of
one of a number of \textit{different} unitary operators $\{\hat{u}%
_{n}^{(i)}:i=0,...,(D-1)\};$ the procedure can only work if $\hat{u}%
_{n}^{(i)}=\hat{u}_{n}^{(j)}$ for all $i,j,$ because only under this
circumstance is the overall operator $\hat{U}_{n}$ unitary. Clearly, then,
such a $\hat{U}_{n}$ is not selectively evolving the universe in a way that
non-trivially depends on what $\Psi _{n}$ is: whichever of the set $%
\{|i\rangle :i=0,...,(D-1)\}$ the state is in, it will always be rotated in
the same way.

It appears that Selective Global Evolution mechanisms of the sort described
above cannot be used to self-referentially develop the state of the universe.

\bigskip

To demonstrate this point explicitly, consider a single qubit universe
represented by a state $\Psi _{n}$ in a two dimensional Hilbert space
spanned by the basis $\mathcal{B}^{(2)}=\{|0\rangle ,|1\rangle \},$ where $%
\mathcal{B}^{(2)}$ may equally be given in the usual representation as $%
\left\{ \tbinom{1}{0},\tbinom{0}{1}\right\} .$

If this universe was to be governed by the above type of Selective Global
Evolution mechanism, two unitary operators, $\hat{u}_{n}^{(0)}$ and $\hat{u}%
_{n}^{(1)},$ would be required for its development. Moreover, these
operators would in turn define a third operator $\hat{U}_{n}$ according to
the Rule $\hat{U}_{n}=\hat{u}_{n}^{(0)}|0\rangle \langle 0|+\hat{u}%
_{n}^{(1)}|1\rangle \langle 1|,$ where $\hat{U}_{n}$ is also required to be
unitary, which would be used to evolve the state $\Psi _{n}.$

The three unitary operators $\hat{u}_{n}^{(0)},$ $\hat{u}_{n}^{(1)}$ and $%
\hat{U}_{n}$ may be given by the general matrices
\begin{equation}
\hat{U}_{n}=\left(
\begin{array}{cc}
A & B \\
C & D
\end{array}
\right) \hspace{0.25cm},\hspace{0.25cm}\hat{u}_{n}^{(0)}=\left(
\begin{array}{cc}
a & b \\
c & d
\end{array}
\right) \hspace{0.25cm},\hspace{0.25cm}\hat{u}_{n}^{(1)}=\left(
\begin{array}{cc}
e & f \\
g & h
\end{array}
\right)
\end{equation}
with the values of $A,...,D,a,...,h\in \mathbb{C}$ to be investigated.

Now, in order for $\hat{U}_{n},$ $\hat{u}_{n}^{(0)}$ and $\hat{u}_{n}^{(1)}$
to fulfil the unitarity conditions $\hat{U}_{n}^{\ast }\hat{U}_{n}=\hat{u}%
_{n}^{(0)\ast }\hat{u}_{n}^{(0)}=\hat{u}_{n}^{(1)\ast }\hat{u}_{n}^{(1)}=%
\hat{I},$ it must be the case that $B=-\bar{C},$ $D=\bar{A},$ $b=-\bar{c},$ $%
d=\bar{a},$ $f=-\ \bar{g}$ and $h=\bar{e}$ (ignoring row exchange
permutations), where the bar denotes complex conjugation, with $%
AD-BC=ad-bc=eh-gh=1.$

So by substituting these, the relation $\hat{U}_{n}=\hat{u}%
_{n}^{(0)}|0\rangle \langle 0|+\hat{u}_{n}^{(1)}|1\rangle \langle 1|$
becomes
\begin{eqnarray}
\left(
\begin{array}{cc}
A & -\bar{C} \\
C & \bar{A}
\end{array}
\right) &=&\left(
\begin{array}{cc}
a & -\bar{c} \\
c & \bar{a}
\end{array}
\right) \left(
\begin{array}{cc}
1 & 0 \\
0 & 0
\end{array}
\right) +\left(
\begin{array}{cc}
\bar{h} & f \\
-\bar{f} & h
\end{array}
\right) \left(
\begin{array}{cc}
0 & 0 \\
0 & 1
\end{array}
\right) \\
&=&\left(
\begin{array}{cc}
a & 0 \\
c & 0
\end{array}
\right) +\left(
\begin{array}{cc}
0 & f \\
0 & h
\end{array}
\right) =\left(
\begin{array}{cc}
a & f \\
c & h
\end{array}
\right)  \notag
\end{eqnarray}
such that clearly $a=A,$ $c=C,$ $f=-\bar{C}$ and $h=\bar{A}.$ Thus, by
inspection of their components' complex conjugates, it immediately follows
that $\hat{U}_{n}=\hat{u}_{n}^{(0)}=\hat{u}_{n}^{(1)}.$

Evidently, there can be no choice of operator, $\hat{u}_{n}^{(0)}$ or $\hat{u%
}_{n}^{(1)},$ in this single qubit universe; it does not matter what the
state $\Psi _{n}$ actually is, it will always be rotated in the same way.

\bigskip

\subsubsection{Selective Local Evolution and Endophysics}

\bigskip

In universes developing according to the Rules described in Section 8.5.2,
every part of the state would be evolved at the same time and in the same
way.

This, however, would lead to a physical limitation. An outcome of Chapter 6
was that the\ result of a global application of a unitary operator to a
state is effectively unobservable, at least from the point of view of an
endo-observer who is only able to witness the universe developing if it
changes relative to herself. But, real endo-physical observers do appear to
be able to witness relative changes occurring in the real physical Universe.

As an illustration of this issue, consider again the notion of an idealised
physics experiment. Traditionally, in such an experiment a physicist
prepares an apparatus and some sort of sample to be investigated. She then
decides what she wants to measure, tests the sample, and records the result.
She may then go on to do any number of further investigations on the sample
based upon what she has learnt.

In the context of a fully quantum reality, however, it should be recalled
that the physicist, the sample, and the apparatus are just sub-systems
within the state $\Psi _{n}$ of the universe. Each may be represented by a
factor of the universe's state, such that $\Psi $ may be written in the form
\begin{equation}
|\Psi \rangle =|\text{\textit{Physicist}}\rangle \otimes |\text{\textit{%
Sample}}\rangle \otimes |\text{\textit{Apparatus}}\rangle \otimes |\text{%
\textit{Rest of Universe}}\rangle .  \label{univ}
\end{equation}

Of course, these sub-states of $\Psi $ are likely to change as the
experiment proceeds; they must also necessarily entangle with one another as
new information is exchanged. Thus, the separability of the state changes as
the universe develops, and this gives rise to apparent measurements of the
sample's sub-state, movement of the apparatus' pointer, changes in the
scientist's brain as she learns the result, etc.

So, as the universe develops from state $\Psi _{n}$ to $\Psi _{n+1}$ to $%
\Psi _{n+2}$ etc., the factors representing the sub-systems also develop.
However, if this sequence is to be consistent with the reality experienced
by scientists, the development must be such that, overall and from the
endo-physical perspective of the physicist, the factor representing the
physicist appears to use the factor representing the apparatus to
independently prepare and test the factor representing the sample.

Thus in the context of a quantum universe, the above experiment may be
viewed as one part of the universe's wavefunction apparently developing
another part of the universe's wavefunction. Further, in such a universe the
decision about which operator is used to develop one part of the universe's
state may be made by considering changes in a different part of the
universe's state. In other words, what the scientist decides to do next to
the sample may be based upon the result contained in the position of the
apparatus' pointer.\bigskip

This then leaves a question. If physicists are part of the Universe, yet
appear to be able to develop their surroundings with an apparent freedom
that depends (literally) on their current `state of mind', how does this
procedure actually work? How are physicists able to get the impression that
the evolution of the Universe around them actually depends on their
sub-state and what they are doing? In other words, what mechanisms could be
used to self-referentially develop the universe in a manner that appears to
depend on one or some of its sub-states, but not on others? Thus, how might
it be possible for the universe to develop in a way such that some parts of
it are evolved, whilst other parts appear to remain unaffected, so that
changes and relative differences may be observed and catalogued?

So, to begin to answer to these questions, and as an extension to the
previous Selective Global Evolution Rules, it may be possible to consider
Examine-Decision mechanisms in which one factor of the state is examined,
and where the outcome of this is then used to determine how a different
factor is evolved. Such a mechanism may be called a \textit{Selective Local
Evolution} (SLE).\bigskip

Care must be taken when interpreting this type of evolution. Emphatically,
the suggestion is not to introduce a dynamics in which only a portion of the
state $\Psi _{n}$ is evolved, measured or collapsed at any one time. After
all, recall that quantum mechanics is a holistic theory, and so the entire
state of a universe based upon quantum principles must be evolved and tested
at the same time, and not just sub-states or factors of it. Indeed, even in
the simple toy-universes described in this chapter as tensor products of
qubits, it has not been possible to evolve or test just one of these qubits
on its own, but has instead relied on the projection of the entire
wavefunction $\Psi _{n}$ into one of members of the basis sets of
eigenvectors of $\hat{\Sigma}_{n+1}$ that span the overall Hilbert space of
the universe's state.

So in the hope of proposing Rules for a quantum universe\ that appears to
evolve according to examinations of parts of its state, the `trick' is
therefore to devise a mechanism in which although every part of the universe
is evolved or tested at the same time, it appears as if, after each
evolution or state reduction, only certain factors have changed whereas
others have been unaffected. Moreover, the way in which a changed factor is
rotated appears to depend on the sub-state of an unaffected factor.

Putting this in context, the aim is to investigate mechanisms in which the
`Physicist' factor remains unchanged during the evolution of the `Sample'
factor under investigation, and where the sub-state of the Physicist somehow
determines how the Sample sub-state is actually rotated.\bigskip

Universes incorporating SLE Rules necessarily require a separable state, and
hence a factorisable Hilbert space. Without loss of generality, and for
simplicity, consider a bi-partite factorisation of the Hilbert space $%
\mathcal{H}$ of the universe, such that $\mathcal{H}\equiv \mathcal{H}%
_{[AB]}=\mathcal{H}_{A}\otimes \mathcal{H}_{B}.$ Consider also an arbitrary
separable state $\Psi _{n}\in \mathcal{H}_{AB}$ of the form
\begin{equation}
\Psi _{n}=|a\rangle _{A}\otimes |b\rangle _{B}.
\end{equation}

In order to generate the SLE dynamics desired, a unitary operator $\hat{U}%
_{n}$ is sought that, when acting upon the state $\Psi _{n},$ examines the
`properties' of one of the factors (i.e. the ``Physicist''), and then
evolves the other factor (i.e. the ``Sample'') in a way that depends upon
the result of this examination. Moreover, the sub-state of the physicist
must be left unchanged by the application of $\hat{U}_{n}.$ The resulting
vector $\Psi _{n}^{\prime }=\hat{U}_{n}\Psi _{n}$ (consisting of the
unchanged Physicist and the evolved Sample) may then be tested and collapsed
by some Hermitian operator $\hat{\Sigma}_{n+1},$ to give the next state $%
\Psi _{n+1}.$ For obvious reasons, the particular SLE mechanism described
here may hence be labelled a Physicist-Sample (PS) mechanism.

Overall, then, this type of Physicist-Sample, Selective Local Evolution,
Examine-Decision mechanism is based upon general Rules of the form

\begin{enumerate}
\item  Apply a particular unitary operator $\hat{U}_{n}$ to $\Psi _{n}$ to
give the `rotated' state $\Psi _{n}^{\prime }=\hat{U}_{n}\Psi _{n};$

\item  Test $\Psi _{n}^{\prime }$ with $\hat{\Sigma}_{n+1}$ to give the new
state $\Psi _{n+1},$ which is one of the eigenvectors of $\hat{\Sigma}%
_{n+1}; $
\end{enumerate}

Of course, the actual mechanism employed to choose the Hermitian operator $%
\hat{\Sigma}_{n+1}$ is left unspecified in this discussion, and may,
perhaps, involve any of the Types of Rule described in Section 8.1. In fact,
to ensure that the state $\Psi _{n+1}$ is also separable relative to $%
\mathcal{H}_{A}\otimes \mathcal{H}_{B},$ as would be necessary for the
procedure to repeat, the choice of operator $\hat{\Sigma}_{n+1}$ could
actually be constrained such that it is itself factorisable relative to this
split, as discussed in Chapter 5.

In order to provide the type of Physicist-Sample dynamics hoped for, the
unitary operator $\hat{U}_{n}$ must be defined such that
\begin{equation}
\Psi _{n}^{\prime }=\hat{U}_{n}\Psi _{n}=\hat{U}_{n}(|a\rangle _{A}\otimes
|b\rangle _{B})=|a\rangle _{A}\otimes |c^{(a)}\rangle _{B}
\end{equation}
where $|c^{(a)}\rangle \in \mathcal{H}_{B}$ is a vector in $\mathcal{H}_{B}$
whose form depends somehow on $|a\rangle .$ Thus\ $|c^{(a)}\rangle
=|b^{\prime }\rangle =\hat{u}_{B}^{(a)}|b\rangle ,$ where $\hat{u}_{B}^{(a)}$
is a unitary operator acting locally in the sub-space $\mathcal{H}_{B}$ that
is chosen according to some property of $|a\rangle .$

A schematic description of the SLE dynamics discussed here is analogous to
the SGE case introduced in Sub-section 8.5.2. In particular, a mechanism is
similarly imagined in which the operator $\hat{U}_{n}$ is taken to `contain'
within it a further set of unitary operators, each of which may be
appropriately `turned on or off'. This time, however, which one of these
unitary operators is actually `activated' depends on the properties of one
particular factor of $\Psi _{n},$ and not on overall properties of the whole
state. Thus, the factor $|a\rangle $ in $\mathcal{H}_{A}$ is used to select
how the factor $|b\rangle $ in $\mathcal{H}_{B}$ is evolved, such that the
selection of the particular operator $\hat{u}_{B}^{(a)}$ depends somehow on $%
|a\rangle .$

Also similarly to the SGE mechanism suggested previously, the proposal here
is to construct $\hat{U}_{n}$ from a linear sum of pairs of operators. This
time, however, one member of the pair is used to `ask' about the factor of $%
\Psi _{n}$ in $\mathcal{H}_{A},$ whilst the other operator is used to evolve
the factor of $\Psi _{n}$ in $\mathcal{H}_{B}$ in a way that depends on the
result of this `question'. Consequently, in the present circumstance the
operator $\hat{U}_{n}$ is given by a linear sum of tensor products of such
pairs of operators, and may be given in the form
\begin{equation}
\hat{U}_{n}=\hat{A}_{A}^{(1)}\otimes \hat{u}_{B}^{(1)}+\hat{A}%
_{A}^{(2)}\otimes \hat{u}_{B}^{(2)}+...  \label{Ch6Tens}
\end{equation}
where $\hat{A}_{A}^{(i)}$ is an operator that `asks' about the factor in $%
\mathcal{H}_{A},$ and $\hat{u}_{B}^{(i)}$ evolves the factor in $\mathcal{H}%
_{B}$ according to the answer. As before, the `questions' $\{\hat{A}%
_{A}^{(i)}\}$ are assumed to be mutually exclusive and exhaustive, and are
taken to provide a definite `Yes' or `No' (i.e. $1$ or $0)$ for each $\Psi
_{n}.$ So, if the answer to $\hat{A}_{A}^{(i)}$ `asking' about the factor in
$\mathcal{H}_{A}$ is `Yes' (or $1),$ then the factor in $\mathcal{H}_{B}$ is
rotated by $\hat{u}_{B}^{(i)};$ otherwise, if the answer is `No' (or $0),$ $%
\hat{u}_{B}^{(i)}$ is not applied.

The actual details of the Selective Local Evolution mechanism are perhaps
best illustrated by example, as given in the following sub-sections. As will
be seen, the asking operators may again be associated with projection
operators.

\bigskip

First, however, note that in any SLE governed universe, the apparent
dynamics depends very much upon point of view. In fact, this comment itself
reflects the opposing viewpoints of exo- and endo-physics.

From the exo-physical point of view of an observer standing outside the
quantum universe and examining the system as a whole, the dynamics describes
a single state $\Psi _{n}$ evolving as $\Psi _{n}\rightarrow \Psi
_{n}^{\prime }=\hat{U}_{n}\Psi _{n}$ according to a global operator $\hat{U}%
_{n},$ before undergoing collapse to one of the non-degenerate eigenvectors
of a particular operator $\hat{\Sigma}_{n+1}.$ The new state $\Psi _{n+1}$
is then evolved and collapsed, and the process continued.

This `external' point of view is generally the most convenient way to
discuss the development of a universe, and is the one that has been used
almost exclusively throughout this thesis. Of course, such a perspective is
also inherently unphysical, because by definition nothing can stand outside
of the universe. However whilst this may be the case, it is still valid to
discuss this hypothetical point of view if it is specified that such an
external `observer' does not interact with the universe in any way; it is
merely a privileged vantage point illustrative when discussing the
development of the state as a whole. Thus, such observers do not actually
observe anything, in the true quantum sense of the word.\bigskip

Now, a feature of the Physicist-Sample mechanism suggested above is that
some sub-states that were present as factors of $\Psi _{n}$ may still exist
as factors of $\Psi _{n+1}.$ Say, for example, that the operator $\hat{\Sigma%
}_{n+1}$ is such that the next state $\Psi _{n+1}$ is of the form
\begin{equation}
\Psi _{n+1}=|a\rangle _{A}\otimes |d\rangle _{B}
\end{equation}
where $|d\rangle _{B}\in \mathcal{H}_{B}.$ Clearly, then, $|a\rangle _{A}$
is a factor of each of $\Psi _{n},$ $\Psi _{n}^{\prime }$ and $\Psi _{n+1}.$

Now, from the endo-physical point of view of this unchanged factor $%
|a\rangle _{A},$ it would look like nothing had been done to it during the
transition from $\Psi _{n}$ to $\Psi _{n+1}$ whilst other, apparently
isolated, parts of the universe have changed. From such a factor's
perspective, it would appear as if the rest of the universe had evolved
`around' it, whilst it had been unaffected by either the evolution from $%
\Psi _{n}$ to $\Psi _{n}^{\prime }$ or the state reduction from $\Psi
_{n}^{\prime }$ to $\Psi _{n+1}.$

Alternatively, from the point of view of the rest of the universe, it
appears that the unchanged factor has been `frozen in time'. The frozen
factor is a part of the universe that seems to have been created at some
time in the past, but has since appeared to have been left alone in the
subsequent development.\bigskip

So, if the way in which the universe evolves depends somehow on properties
of one of these unchanged factors, it could appear, again from the
perspective of such a factor, that it was these properties that caused the
change in the rest of the universe. From this point of view, it is as if the
way in which the universe is developed depends on one of the factors of its
state.\bigskip

Summarising, from the point of view of an endo-physical observer, it is
possible to devise a mechanism in which the dynamical evolution of the
universe appears to depend on parts of its state, as will be shown in the
following. From such a perspective, it would seem that the unchanged factor
(i.e. the Physicist) `chooses' how the Sample evolves, according to which
sub-state this `endo-observer' is in.

From the external point of view of the entire universe, however, the state $%
\Psi _{n}$ will be seen to evolve to $\Psi _{n}^{\prime }$ in a
deterministic, global fashion, as expected from the unitary relationship $%
\Psi _{n}\rightarrow \Psi _{n}^{\prime }=\hat{U}_{n}\Psi _{n}.$

SLE rules are therefore a variant of SGE dynamics in which local
examinations and relative evolutions seem to become apparent from an
endo-physical perspective.

The exact details behind such SLE mechanisms will be introduced, elaborated
upon, and demonstrated in the following examples.

\bigskip

\subsubsection{A Two Qubit `Physicist-Sample' Universe}

\bigskip

To illustrate the type of Selective Local Evolution, Physicist-Sample
dynamics suggested in the previous subsection, consider a two qubit universe
represented by a state $\Psi _{n}$ in the factorisable Hilbert space $%
\mathcal{H}_{[12]}=\mathcal{H}_{1}\otimes \mathcal{H}_{2}.$ Further, assume
that $\Psi _{n}$ is separable relative to $\mathcal{H}_{[12]},$ such that $%
\Psi _{n}\in \mathcal{H}_{12}\subset \mathcal{H}_{[12]},$ and label the
factor in $\mathcal{H}_{1}$ as `qubit $1$', or $q_{1},$ and the factor in $%
\mathcal{H}_{2}$ as `qubit $2$', or $q_{2}.$ Additionally, consider the
usual orthonormal basis sets $\mathcal{B}_{1},$ $\mathcal{B}_{2}$ and $%
\mathcal{B}_{12}$ for the Hilbert spaces $\mathcal{H}_{1},$ $\mathcal{H}_{2}$
and $\mathcal{H}_{[12]}$ respectively, defined as
\begin{eqnarray}
\mathcal{B}_{1} &=&\{|0\rangle _{1},|1\rangle _{1}\}\text{ \ \ , \ \ }%
\mathcal{B}_{2}=\{|0\rangle _{2},|1\rangle _{2}\} \\
\mathcal{B}_{12} &=&\{|00\rangle _{12},|01\rangle _{12},|10\rangle
_{12},|11\rangle _{12}\}.  \notag
\end{eqnarray}

As before, the matrix representation $\{|0\rangle _{a},|1\rangle
_{a}\}\equiv \left\{ \tbinom{1}{0}_{a},\tbinom{0}{1}_{a}\right\} $ may be
adopted for $a=1,2.$

The intention of this sub-section is to introduce a model in which the
development of qubit $1$ is controlled somehow by the state of qubit $2.$ In
particular, and to illustrate the general principle, the aim will be to
analyse a system in which if $q_{2}=|0\rangle $ then $q_{1}$ is evolved
using an unitary operator $\hat{u}_{0},$ whereas if $q_{2}=|1\rangle $ then $%
q_{1}$ is evolved using a different unitary operator $\hat{u}_{1}.$

As should be evident from before, such a mechanism is analogous to
introducing a `physicist' into the universe. The physical interpretation is
that qubit $2$ acts like the decision making scientist: if the `physicist'
is in one particular state then a certain experiment is performed on qubit $%
1,$ but if `she' is in another state then something completely different is
done to qubit $1.$\bigskip

An important comment, however, must first be made at this point. Note that
in the following, a sub-script on an \textit{operator} is used to
distinguish it, whereas its super-script denotes which qubit(s) Hilbert
space(s) it is acting upon. Thus, for example, the operator written $\hat{u}%
_{0}^{1}$ indicates the operator $\hat{u}_{0}$ acting in $\mathcal{H}_{1},$
whereas $\hat{u}_{0}^{2}$ implies the same operator $\hat{u}_{0}$ acting
instead in $\mathcal{H}_{2}.$ Similarly, the operator $\hat{U}_{n}\equiv
\hat{U}_{n}^{[12]}$ acts across the entire Hilbert space $\mathcal{H}%
_{[12]}. $ This notation is converse to both the usual convention adopted
generally throughout this thesis to label operators, and to the usual
reservation of sub-scripts for labelling Hilbert spaces, and results from a
desire to keep the sub-script $n$ on $\hat{U}_{n}$ as a `temporal' parameter.

Sub-scripts on \textit{states} are still used to denote Hilbert space
affiliation, such that $|0\rangle _{1}\in \mathcal{H}_{1}$ etc., apart from
on $\Psi _{n}$ where it indicates the state of the universe at time $n$ in
the usual way.

Note also that labels may be omitted for clarity when no confusion is likely
to occur, such that for example $\hat{U}_{n}\equiv \hat{U}_{n}^{[12]},$ and $%
|0\rangle _{1}\otimes |1\rangle _{2}\equiv |0\rangle \otimes |1\rangle
\equiv |01\rangle .$ Further, note that the `lower case' unitary operators $%
\hat{u}_{0}$ and $\hat{u}_{1}$ will be defined\ constantly for all time: $%
\hat{u}_{0}$ is the `zeroth' operator, and not an operator $\hat{u}$ at
`time' $n=0.$ In general, of course, the purpose of a particular sub- or
super-script in any individual case in the following should be fairly
obvious from context.\bigskip

As expected from (\ref{Ch6Tens}), the evolution of a two qubit PS universe
is taken to be governed by a unitary operator $\hat{U}_{n}$ of the form%
\footnote{%
Noting the now changed sub- and super-script convention.} $\hat{U}_{n}=\hat{u%
}_{0}^{1}\otimes \hat{A}_{0}^{2}+\hat{u}_{1}^{1}\otimes \hat{A}_{1}^{2}+...,$
where $\hat{A}_{j}^{i}$ is an operator that `asks' the $j^{th} $ `question'
of the $i^{th}$ qubit. In fact for simplicity in the current two qubit
universe, attention may be restricted to operators $\hat{U}_{n}$ of the form
\begin{equation}
\hat{U}_{n}=\hat{u}_{0}^{1}\otimes \hat{A}_{0}^{2}+\hat{u}_{1}^{1}\otimes
\hat{A}_{1}^{2}  \label{Ch6Tens1}
\end{equation}
where $\hat{A}_{0}^{2}$ and $\hat{A}_{1}^{2}$ provide mutually exclusive and
exhaustive `questions': if $\hat{A}_{0}^{2}$ acting on $q_{2}$ is `Yes',
then $\hat{A}_{1}^{2}$ acting on $q_{2}$ must be `No', and vice versa.

In order to define a valid Physicist-Sample mechanism, unitary operators $%
\hat{U}_{n},$ $\hat{u}_{0}$ and $\hat{u}_{1}$ are sought such that, from the
exo-physical point of view, the entire state of the universe is evolved
globally by a single unitary operator $\hat{U}_{n},$ i.e. $\Psi
_{n}\rightarrow \Psi _{n}^{\prime }=\hat{U}_{n}\Psi _{n},$ but from the
endo-physical viewpoint of qubit $2$ this operation $\hat{U}_{n}$ appears to
be equivalent to a selection of either $\hat{u}_{0}$ or $\hat{u}_{1}$ to act
locally upon $q_{1},$ by a decision made in reference to the state of $%
q_{2}. $ Clearly, a definition for the operators $\hat{A}_{j}^{2},$ that
`ask' whether qubit $2$ is in the state $|0\rangle $ or $|1\rangle ,$ will
therefore also be required.

This observation provokes a second important comment; namely, observe that
there is immediately an inherent difference between the operators $\hat{U}%
_{n},$ $\hat{u}_{0}$ and $\hat{u}_{1}.$ The operator $\hat{u}_{0}^{1}$ may
be represented by a $2\times 2$ matrix, because it is to act locally on
qubit $1$ in the two dimensional Hilbert space $\mathcal{H}_{1}.$ Similarly,
$\hat{u}_{1}^{1}$ may also be represented by a $2\times 2$ matrix because it
also acts just on $q_{1}$ in $\mathcal{H}_{1}.$ However, $\hat{U}_{n}$ \
acts globally on the state of the entire universe, i.e. on the state $\Psi
_{n}$ of both qubits in the four dimensional Hilbert space $\mathcal{H}%
_{[12]},$ and so must be represented by a $4\times 4$ matrix. Clearly, if $%
\hat{U}_{n}$ is given in the form (\ref{Ch6Tens1}), its dimension must equal
the product of the dimensions of the operator acting in $\mathcal{H}_{1}$
(i.e. $\hat{u}_{0}^{1}$ or $\hat{u}_{1}^{1})$ and the operator $\hat{A}%
_{j}^{2}$ acting in $\mathcal{H}_{2}$ that `asks' which state qubit $2$ is
in. Thus, the `asking' operator must also be represented by a $2\times 2$
matrix, as expected from the observation that it is to `ask' about the state
of a single qubit.

As before, suitably defined projection operators are obvious candidates for
the $\hat{A}_{j}^{i}.$

\bigskip

The following example illustrates how a two qubit Physicist-Sample mechanism
might be constructed. The development of the presented model will proceed
through two different steps, each incorporating an evolution part and a
state reduction. Thus, one `cycle' of the dynamics takes place in two time
steps, as will become evident. Moreover, under this circumstance it is
necessary to define an initial stage as a type of `reference', so that the
Rules `know' which of the first or second steps should be applied to the
current state.

Defining the initial state as $\Psi _{N}$ at initial `time'\ $n=N,$ the
development of the proposed two qubit PS universe is governed by the Rules

\begin{enumerate}
\item  Evolve the initial state $\Psi _{N}$ with the particular unitary
operator $\hat{U}_{N}=\hat{U}_{X},$ such that $\Psi _{N}\rightarrow \Psi
_{N}^{\prime }=\hat{U}_{X}\Psi _{N};$

\item  This evolved state $\Psi _{N}^{\prime }$ is then collapsed into one
of the eigenstates of a particular Hermitian operator $\hat{\Sigma}_{N+1}=%
\hat{B},$ and this vector may now be associated with the next state $\Psi
_{N+1},$ with the usual probability amplitudes $\langle \Psi _{N+1}|\Psi
_{N}^{\prime }\rangle ;$

\item  The new state $\Psi _{N+1}$ is then evolved with a different unitary
operator, $\hat{U}_{N+1}=\hat{U}_{Y},$ into the state $\Psi _{N+1}^{\prime
}, $ such that $\Psi _{N+1}\rightarrow \Psi _{N+1}^{\prime }=\hat{U}_{Y}\Psi
_{N+1};$

\item  Finally, this new evolved state is collapsed back into one of the
eigenstates of $\hat{\Sigma}_{N+2}=\hat{B}$ to give the next state $\Psi
_{N+2},$ with probabilities $|\langle \Psi _{N+2}|\Psi _{N+1}^{\prime
}\rangle |^{2}.$
\end{enumerate}

The `first step' hence contains procedures $``1."$ and $``2.",$ whilst the
second step is parts $``3."$ and $``4.".$ The two step mechanism then
repeats, such that $\Psi _{N+2}$ is next rotated by $\hat{U}_{N+2}=\hat{U}%
_{X}=\hat{U}_{N},$ and so on. Of course, the operators $\hat{U}_{X},$ $\hat{U%
}_{Y}$ and $\hat{B}$ must be carefully defined in order for the universe to
follow SLE dynamics.

In general, then, the Rules of the model are such that
\begin{eqnarray}
\hat{\Sigma}_{N+m} &=&\hat{\Sigma}_{N+m+1}=\hat{B}\text{ \ \ , \ \ }%
m=0,1,2,... \\
\hat{U}_{N+m} &=&\left\{
\begin{array}{c}
\hat{U}_{N}=\hat{U}_{X}\text{ if }m=0,2,4,... \\
\hat{U}_{N+1}=\hat{U}_{Y}\text{ if }m=1,3,5,...
\end{array}
\right\}  \notag
\end{eqnarray}
noting the unavoidable clash of notation: $\hat{U}_{N}$ is taken to indicate
the relevant unitary operator at time $N,$ whereas $\hat{U}_{X}$ denotes a
fixed unitary operator. Similarly, $\hat{U}_{Y}$ is also a fixed operator,
and is not meant to imply $\hat{U}$ at time $Y.$

In order to provide a suitable mechanism for a universe developing according
to Physicist-Sample, SLE Rules, the unitary operators $\hat{U}_{X}$ and $%
\hat{U}_{Y}$ are defined to be
\begin{equation}
\hat{U}_{X}=\hat{u}_{0}^{1}\otimes \hat{P}_{0}^{2}+\hat{u}_{1}^{1}\otimes
\hat{P}_{1}^{2}
\end{equation}
where $\hat{P}_{s}^{r}\equiv |s\rangle _{rr}\langle s|$ for $s=0,1$ is the $%
s^{th}$ projection operator acting in $\mathcal{H}_{r},$ and
\begin{equation}
\hat{U}_{Y}=\hat{I}^{1}\otimes \hat{u}_{2}^{2}
\end{equation}
where $\hat{I}^{t}$ is the identity operator in $\mathcal{H}_{t}.$ The exact
reasons for these choices will become apparent, noting immediately, however,
that $\hat{U}_{X}$ is in the form expected from (\ref{Ch6Tens1}).

Similarly, the Hermitian operator $\hat{B}$ is chosen to be one that
possesses a completely separable basis set $\frak{B}_{(0,4)}$ of orthonormal
eigenvectors,
\begin{equation}
\frak{B}_{(0,4)}=\{|00\rangle _{12},|01\rangle _{12},|10\rangle
_{12},|11\rangle _{12}\}  \label{Ch6Basis}
\end{equation}
such that in fact $\hat{B}\equiv \hat{B}_{(0,4)}$ and $\frak{B}_{(0,4)}=%
\mathcal{B}_{12},$ again for reasons given below.\bigskip

The significance of this model, and in particular the appearance of
Selective Local Evolution dynamics, can be demonstrated from a comparison of
the exo- and endo-physical interpretations of the universe's development.

From the point of view of an observer external to the system, the first step
involves the global rotation of the entire initial state $\Psi _{N}$ by $%
\hat{U}_{X},$ followed by its subsequent measurement with $\hat{B},$ whilst
the second step involves the new state $\Psi _{N+1}$ being globally rotated
in a different way by $\hat{U}_{Y},$ before the whole universe is again
measured with $\hat{B}.$ So, from the exo-physical point of view, the
development of the universe proceeds in a semi-deterministic globalised
fashion according to the operators $\hat{U}_{X},\hat{B},\hat{U}_{Y}$ and $%
\hat{B}$ being applied in turn, and the only randomness occurs as a result
of the stochastic nature of the wavefunction collapse process, when a
particular state $\Psi _{N+m}$ is obtained from the set of eigenvectors $%
\frak{B}_{(0,4)}$ of $\hat{\Sigma}_{N+m}=\hat{B}.$\bigskip

However, to see the apparent Selective Local Evolution present in the model,
it is necessary to examine the endo-physical point of view of one of the
qubits. From such a perspective, it appears that the unitary operators $\hat{%
U}_{X}$ and $\hat{U}_{Y}$ are evolving the universe in a manner that depends
on one of its sub-states; this conclusion is demonstrated now.

Firstly, and for simplicity, assume that the universe has been prepared such
that its initial state $\Psi _{N}$ is separable (i.e. the qubits are not
entangled with one another), that the factor of $\Psi _{N}$ representing $%
q_{1}$ in $\mathcal{H}_{1}$ is either in the state $|0\rangle _{1}$ or the
state $|1\rangle _{1},$ and that the factor of $\Psi _{N}$ representing $%
q_{2}$ in $\mathcal{H}_{2}$ is either in the state $|0\rangle _{2}$ or the
state $|1\rangle _{2}$ (i.e. neither qubit is in an arbitrary superposition
of its basis vectors). These assumptions will be justified later, noting
that the former has already been taken to be essential if a classically
\textit{distinct} `Physicist' and `Sample' are to be discussed.

Consider now the application of the operator $\hat{U}_{X}$ to this state $%
\Psi _{N}.$ A projection operator $\hat{P}_{s}^{2}\equiv |s\rangle
_{22}\langle s|,$ where $s=0,1,$ may be interpreted as an operator that
`asks' which state qubit $2$ is in: if $q_{2}$ is in the state $%
|q_{2}\rangle _{2}=|s\rangle _{2},$ then $\hat{P}_{s}^{2}|q_{2}\rangle _{2}$
gives the `answer'
\begin{equation}
\hat{P}_{s}^{2}|q_{2}\rangle _{2}=|s\rangle _{22}\langle s||s\rangle
_{2}=1\times |s\rangle _{2}\text{ \ \ , \ \ }s=0,1
\end{equation}
whereas if $q_{2}$ is in a state orthogonal to $|s\rangle _{2}$ then $\hat{P}%
_{s}^{2}|q_{2}\rangle _{2}$ gives the `answer' $0\times |s\rangle _{2}.$

Thus, the combined operator $\hat{u}_{t}^{1}\otimes \hat{P}_{s}^{2}$ acting
on the general qubit product state $|q_{1}\rangle _{1}\otimes |q_{2}\rangle
_{2},$ where $|q_{2}\rangle _{2}$ is $|0\rangle _{2}$ or $|1\rangle _{2}$ by
design and $t=0,1,$ gives the result
\begin{equation}
\lbrack \hat{u}_{t}^{1}\otimes \hat{P}_{s}^{2}]|q_{1}\rangle _{1}\otimes
|q_{2}\rangle _{2}=\left\{
\begin{array}{c}
\lbrack \hat{u}_{t}^{1}|q_{1}\rangle _{1}]\otimes \lbrack 1\times
|q_{2}\rangle _{2}]\text{ if }|q_{2}\rangle _{2}=|s\rangle _{2} \\
\lbrack \hat{u}_{t}^{1}|q_{1}\rangle _{1}]\otimes \lbrack 0\times
|q_{2}\rangle _{2}]\text{ if }|q_{2}\rangle _{2}\neq |s\rangle _{2}
\end{array}
\right\} \text{ \ \ ,\ \ \ }s,t=0,1.
\end{equation}

Overall, then, the evolution $\Psi _{N}\rightarrow \Psi _{N}^{\prime }$ $=%
\hat{U}_{X}\Psi _{N}$ is consequently equivalent to an operation that leaves
$q_{2}$ in its initial state (either $|0\rangle $ or $|1\rangle )$ whilst
rotating the state of $q_{1}.$ Further, if qubit $2$ is initially in the
state $q_{2}=|0\rangle $ then qubit $1$ is evolved by $\hat{u}_{0}^{1},$
whereas if qubit $2$ is in the state $q_{2}=$ $|1\rangle $ then qubit $1$ is
evolved by $\hat{u}_{1}^{1}.$

So, from the endo-physical point of view of the `Physicist' qubit, $q_{2},$
it appears that the rest of the universe, $q_{1},$ has developed in a way
that depends on `her' state, whilst `she' has remained unchanged. Qubit $2$
could conclude that it was she who determined how the universe evolved, by
`choosing' to apply an operator $\hat{u}_{0}^{1}$ or $\hat{u}_{2}^{1}$ to $%
q_{1};$ she would not automatically assume that it was actually the whole
universe that was globally evolved by an operator $\hat{U}_{X}.$\bigskip

The next part of the first step involves the collapse of the wavefunction $%
\Psi _{N}^{\prime }.$

In many ways, it is possible to consider qubit $2$ as being a classical
`object'; qubit $2$ is still in a `classical' looking state, $|0\rangle $ or
$|1\rangle ,$ because the unitary operators $\hat{u}_{0}^{1}$ and $\hat{u}%
_{1}^{1}$ act solely on $q_{1},$ such that only this first qubit may be in a
superposition. It might therefore be tempting to disregard $q_{2}$ entirely
when discussing the collapse of the statefunction $\Psi _{N}^{\prime }.$\
This follows from the general logic that if the initial state of qubit $2$
is known, no new information may be extracted from the system by the
application of the projection operators $\hat{P}_{0}^{2}$ or $\hat{P}%
_{1}^{2}.$

From the point of view of the physical interpretation of the current model,
this disregarding seems a natural conclusion, and represents the normal,
`everyday' exo-physical approach to quantum theory in which it appears
possible to perform some sort of quantum experiment on one particular
sub-system of the universe whilst leaving other sub-systems alone.

So, following on from this (erroneous) perspective, a suitable next operator
$\hat{\Sigma}_{N+1}=\hat{B}_{suit?}$ used to test the universe's state $\Psi
_{N}^{\prime }$ might therefore be expected to be of the form
\begin{equation}
\hat{B}_{suit?}=\hat{\sigma}_{3}^{1}\otimes \hat{I}^{2}
\end{equation}
which could be interpreted as the product of a Pauli operator $\hat{\sigma}%
_{3}^{1}$ that collapses qubit $1$ into either $|0\rangle $ or $|1\rangle ,$
with the identity operator $\hat{I}^{2}$ doing nothing to qubit $2.$

Such an analysis, however, is incorrect. The operator $\hat{B}_{suit?}$ can
be shown to have degenerate eigenvalues, and as such does not possess a
unique basis set of orthogonal eigenvectors; consequently it is not a valid
operator to use when discussing tests in the proposed fully quantum paradigm.

The general problem associated with operators such as $\hat{B}_{suit?}$
results from the fact that the universe cannot be tested simply as a product
of classical objects: in quantum theory it is \textit{not} possible to
isolate a sub-state from everything else. Phrasing this more mathematically,
it is not always possible to test the universe by independently testing its
parts: an operator $\hat{\Lambda}^{1}$ acting on qubit $1$ in $\mathcal{H}%
_{1}$ may have two, unique orthogonal eigenvectors, as might an operator $%
\hat{\Delta}^{2}$ acting on qubit $2$ in $\mathcal{H}_{2},$ but this does
not imply that the combined operator $\hat{\Lambda}^{1}\otimes \hat{\Delta}%
^{2}$ must necessarily have four, unique orthogonal eigenvectors. Such an
argument was presented in the discussion of `Strong' and `Weak' operators
given in Chapter 5, and is also related to the conclusion of Chapter 4 that
separable states are just a tiny subset of the set of all vectors in a
Hilbert space.

In short, the state of the universe may be described as greater than the sum
of its parts, and so care must therefore always be taken to choose an
operator $\hat{\Sigma}_{N+1}$ that acts on the entire quantum state, yet
also possesses a basis set of four orthogonal eigenvectors.\bigskip

The operator $\hat{B}_{(0,4)},$ however, defined by
\begin{equation}
\hat{B}_{(0,4)}=w|00\rangle \langle 00|+x|01\rangle \langle 01|+y|10\rangle
\langle 10|+z|11\rangle \langle 11|
\end{equation}
in accordance with the basis set $\frak{B}_{(0,4)}$ in (\ref{Ch6Basis}),
does satisfy this condition, where $w,x,y,z\in \mathbb{R}^{+}$ are
non-degenerate eigenvalues of no further consequence to the discussion.
Thus, $\hat{B}_{(0,4)}$ is a valid Hermitian operator, and so\ may be used
to test the overall state $\Psi _{N}^{\prime }$ of the universe.

Moreover, an operator of the form $\hat{\Sigma}_{N+1}=\hat{B}_{(0,4)},$
which defines the basis set $\frak{B}_{N+1}=\frak{B}_{(0,4)}$ for the next
state $\Psi _{N+1},$ also has the important consequence that there is no
eigenstate of $\hat{\Sigma}_{N+1}$ for which $q_{1}$ and $q_{2}$ are
entangled with one another, nor is there an eigenstate containing a linear
superposition of the form $(a|0\rangle _{1}+b|1\rangle _{1})\otimes
(c|0\rangle _{2}+d|1\rangle _{2}),$ for $a,b,c,d\neq 0$ and $a,b,c,d\in
\mathbb{C}.$ Thus after testing $\Psi _{N}^{\prime }$ with $\hat{\Sigma}%
_{N+1}=\hat{B}_{(0,4)},$ both qubits have the `classical' form $%
q_{1}=|0\rangle $ or $q_{1}=|1\rangle $ and $q_{2}=|0\rangle $ or $%
q_{2}=|1\rangle .$

Furthermore, since the Rules are such that $\hat{\Sigma}_{N+m}=\hat{B}%
_{(0,4)}$ for all $m,$ it implies that $\Psi _{N+m}\in \frak{B}_{(0,4)}$ for
all $m,$ and this justifies the assumption made earlier that the `initial'
wavefunction is always a separable state with neither qubit superposed.

So, after the test of $\Psi _{N}^{\prime }$ by $\hat{\Sigma}_{N+1}=\hat{B}%
_{(0,4)},$ the subsequent state $\Psi _{N+1}$ will be one of the eigenstates
of $\hat{B}_{(0,4)},$ with appropriate probability amplitudes given in the
usual way.\bigskip

The above discussion highlights the fact that from the holistic point of
view of a quantum universe, it is not possible to naively segregate the
state into factors under investigation and everything else. In fact, it is
the entire state of the universe that must instead be measured.

The endo-physical observation that qubit $2$ \textit{appears} to be
unaffected by the application of $\hat{U}_{X}$ or $\hat{\Sigma}_{N+1}=\hat{B}%
_{(0,4)},$ however, is because the jump from the state
\begin{equation}
\Psi _{N}^{\prime }=(a|0\rangle _{1}+b|1\rangle _{1})\otimes (c|0\rangle
_{2}+d|1\rangle _{2})
\end{equation}
where $a,b\in \mathbb{C}$ and \textit{either} $c=0$ and $d=1$ \textit{or} $%
c=1$ and $d=0$ (noting that $\Psi _{N}^{\prime }$ is still separable, as $%
\hat{u}_{0}^{1},$ $\hat{u}_{1}^{1},$ $\hat{P}_{0}^{2}$ and $\hat{P}_{1}^{2}$
act locally), to a subsequent state
\begin{equation}
\Psi _{N+1}=(A|0\rangle _{1}+B|1\rangle _{1})\otimes (C|0\rangle
_{2}+D|1\rangle _{2})
\end{equation}
which is an eigenstate of $\hat{B}_{(0,4)}$ (one of which, for example, is $%
|00\rangle ,$ where $A=C=1,$ $B=D=0),$ has a non-zero probability, $|\langle
\Psi _{N+1}|\Psi _{N}^{\prime }\rangle |^{2},$ of occurring if and only if $%
C=c$ and $D=d.$

So, from the point of view of qubit $2,$ the test $\hat{B}_{(0,4)}$
therefore \textit{appears} equivalent to the product of a projection of the
evolved sub-state of qubit$1$ onto the basis set $\{|0\rangle _{1},$ $%
|1\rangle _{1}\},$ with a `null' operator acting upon itself. In reality,
both qubits are actually tested, but no new information is acquired about $%
q_{2}.$\bigskip

It is these apparent null tests, i.e. the observation that every factor of a
state is involved in a quantum test but that some outcomes of this
measurement have a zero probability of being realised, that may be a root
cause of apparent permanence in the Universe. Specifically, this mechanism
gives rise to the ``frozen factors'' described previously, and explains why
it is possible to have sub-states that appear unchanged as the Universe
evolves from $\Psi _{n}$ to $\Psi _{n+1}.$

Note that, of course, this `null process' may in principle continue over
several jumps of the real Universe: a sub-state that was present as a factor
of both $\Psi _{n}$ and $\Psi _{n+1}$ may remain as a factor of $\Psi
_{n+2}, $ $\Psi _{n+3},...$ until some later state $\Psi _{n+m},$ which may
have completely different factors. Again, this could contribute to the
phenomena of persistence and longevity.\bigskip

It should be reiterated how important it is that the null tests are included
in the operators. In the present case, qubit $2$ must still be involved in
the measurement of $\Psi _{N}^{\prime },$ because, apart from the degenerate
eigenvalue problem, if this were not the case the question would remain as
to why only parts of the state are evolved or tested when others are left
alone, and this is contrary to the notion of a single set of rules of
physics applying to the whole universe at the same time. Further, if a test
did include eigenstates in which qubit $2$ could be neglected, that is if,
say, $\Psi _{N+1}\equiv \chi =(A|0\rangle _{1}+B|1\rangle _{1}),$ the
question would remain as to what the amplitude $\langle \chi |\Psi
_{N}^{\prime }\rangle $ may mean mathematically, given that the dimensions
of the Hilbert spaces of $\chi $ and $\Psi _{N}^{\prime }$ differ.

Of course, it is possible to restrict attention to individual factors of the
initial and final states, and correctly evaluate amplitudes such as $\langle
\chi |\varphi \rangle ,$ where $\varphi =(a|0\rangle _{1}+b|1\rangle _{1})$
is a factor of $\Psi _{N}^{\prime }=\varphi \otimes (c|0\rangle
_{2}+d|1\rangle _{2}),$ for $a,b\in \mathbb{C}$ and \textit{either} $c=0$
and $d=1$ \textit{or} $c=1$ and $d=0.$ However whilst this is mathematically
sound, the interpretation is really only valid from the exo-physical point
of view of one sub-system (in this case $q_{2})$ describing changes in
another isolated sub-system (in this case $q_{1}).$ It is therefore a bit
misleading when attempting to consider a universe that is a complete quantum
system, in which the endo-physical `observer' (here $q_{2})$ is itself part
of the state it is trying to measure.

In a fully quantum universe, everything has to be evolved and tested at the
same time, though some factors of the universe's state may be unchanged by
the evolution, and may appear unchanged by the test. Despite appearances, it
is not possible to just evolve or measure part of a fully quantum universe,
though it is possible to discuss effects that appear local by ignoring null
tests and identity evolutions and considering the physical interpretation of
the model.

\bigskip

If the development of the two-qubit universe was governed by just repeating
the first step of the Rules (i.e. just parts $``1."$ and $``2."),$ its
dynamics would be rather restricted. Specifically, if the development from $%
...\Psi _{N}\rightarrow \Psi _{N+1}\rightarrow \Psi _{N+2}\rightarrow ...$
was due solely to an application of the operators $\hat{U}_{X},$ $\hat{\Sigma%
}_{N+1},$ $\hat{U}_{X},$ $\hat{\Sigma}_{N+2},$ $\hat{U}_{X},...,$ its
Evolution would actually contain very little Selection. This is because if
qubit $2$ is initially in the state $|0\rangle ,$ it implies that qubit $1$
will always be evolved by $\hat{u}_{0}^{1}$ from then on, whereas if at
initial `time' $N$ qubit $2$ is instead in the state $|1\rangle ,$ then
qubit $1$ would instead be evolved by $\hat{u}_{1}^{1}$ for all $N+m,$ where
$m=0,1,2,...$

The above conclusion follows because there is currently no mechanism for
changing the state of qubit $2,$ and therefore obtaining a more interesting
dynamics based upon selection.

This problem, however, may be remedied by introducing the second step. The
second step is defined such that it begins by evolving $q_{2}$ whilst
appearing to leave $q_{1}$ unaffected. A unitary operator $\hat{U}_{N+1}$ is
hence used that rotates the factor of $\Psi _{N+1}$ in $\mathcal{H}_{2}$
whilst doing nothing to the factor in $\mathcal{H}_{1}.$

Specifically, this ``doing nothing'' operation may be achieved by the
identity operator $\hat{I}^{1}$ acting locally on $q_{1},$ such that a
suitable unitary operator $\hat{U}_{N+1}=\hat{U}_{Y}$ is given by
\begin{equation}
\hat{U}_{N+1}=\hat{U}_{Y}=\hat{I}^{1}\otimes \hat{u}_{2}^{2}
\end{equation}
as suggested earlier. Clearly, both qubits are involved in the evolution,
but only the state of $q_{2}$ is actually changed.

The final procedure of the second step is then to collapse the wavefunction
in order to obtain the next state $\Psi _{N+2}.$ The operator $\hat{B}%
_{(0,4)}$ may again be used, such that the state after reduction is `reset'
back to one of the members of the basis $\frak{B}_{(0,4)}.$

Evidently, from the endo-physical point of view of the individual qubits, it
appears that during the second step qubit $1$ is not taking any part in the
evolution or collapse process. From the exo-physical point of view of the
entire universe, of course, both qubits are involved.

The two step process may then be repeated, starting with the application of $%
\hat{U}_{X}$ to $\Psi _{N+2}.\bigskip $

The proposed mechanism may now be summarised from the endo-physical
perspective. Given an `initial' state $\Psi _{N}\in \frak{B}_{(0,4)},$ the
application of the operator $\hat{U}_{X}$ evolves this vector in a manner
that appears to depend on whether qubit $2$ is $|0\rangle $ or $|1\rangle .$
The rotated state $\Psi _{N}^{\prime }$ is then tested by an operator $\hat{%
\Sigma}_{N+1}=\hat{B}_{(0,4)},$ which collapses it back into one of the
members $\mathcal{B}_{(0,4)},$ with probabilities given in the usual way.

Then, during the second step, the state $\Psi _{N+1}$ of the universe is
evolved by $\hat{U}_{Y},$ which is equivalent to just rotating the sub-state
of qubit $2$ with $\hat{u}_{2}^{2}.$ Finally, the state $\Psi _{N+1}^{\prime
}$ is tested with $\hat{B}_{(0,4)},$ and the universe collapses back into
one of the members of $\frak{B}_{(0,4)},$ noting that whichever member this
may be, $\Psi _{N+2}$ is separable and $q_{2}$ is definitely in either $%
|0\rangle $ or $|1\rangle ,$ as required for the reapplication of $\hat{U}%
_{X}$ when the first step is repeated.\bigskip

\bigskip
\begin{eqnarray*}
&& \\
&&
\end{eqnarray*}

\paragraph{PS Example}

\begin{equation*}
\end{equation*}

It is now shown implicitly how a\ two qubit universe may develop when
governed by the above Selective Local Evolution, PS Rules incorporating the
operators\ $\hat{U}_{X},$ $\hat{U}_{Y}$ and $\hat{B}_{(0,4)}.$

Consider as an example a two qubit system initially in a state $\Psi _{0}$
given by
\begin{equation}
\Psi _{0}=|00\rangle _{12}=|0\rangle _{1}\otimes |0\rangle _{2}=\binom{1}{0}%
_{1}\otimes \binom{1}{0}_{2}=\binom{\QATOP{1}{0}}{\QATOP{0}{0}}.
\label{Psi0}
\end{equation}

Consider also unitary operators $\hat{u}_{0},$ $\hat{u}_{1}$ and $\hat{u}%
_{2} $ defined as
\begin{equation}
\hat{u}_{0}=e^{-i\varepsilon \hat{\sigma}_{1}}\text{ \ \ , \ \ }\hat{u}%
_{1}=e^{-i\mu \hat{\sigma}_{2}}\text{ \ \ , \ \ }\hat{u}_{2}=e^{-i\nu \hat{%
\sigma}_{1}}
\end{equation}
where $\varepsilon ,\mu ,\nu \in \mathbb{R}^{+}$ are real parameters, and $%
\hat{\sigma}_{i}$ is the $i^{th}$ Pauli operator. Thus the operator $\hat{U}%
_{X}$ becomes
\begin{eqnarray}
\hat{U}_{X} &=&\hat{u}_{0}^{1}\otimes \hat{P}_{0}^{2}+\hat{u}_{1}^{1}\otimes
\hat{P}_{1}^{2} \\
&=&\binom{\cos \varepsilon \text{ \ }-i\sin \varepsilon }{-i\sin \varepsilon
\text{ \ \ }\cos \varepsilon }\otimes \binom{1\text{ \ }0}{0\text{ \ }0}+%
\binom{\cos \mu \text{ \ }-\sin \mu }{\sin \mu \text{ \ \ \ \ }\cos \mu }%
\otimes \binom{0\text{ \ }0}{0\text{ \ }1}  \notag
\end{eqnarray}
whilst $\hat{U}_{Y}$ is given by
\begin{equation}
\hat{U}_{Y}=\hat{I}^{1}\otimes \hat{u}_{2}^{2}=\binom{1\text{ \ }0}{0\text{
\ }1}\otimes \binom{\cos \nu \text{\ }-i\sin \nu }{-i\sin \nu \text{ \ \ }%
\cos \nu }.
\end{equation}

The first step of the mechanism evolves the state of the universe to $\Psi
_{0}^{\prime }$ by an application of $\hat{U}_{0}=\ \hat{U}_{X}$ to $\Psi
_{0}.$ So,
\begin{eqnarray}
\Psi _{0}^{\prime } &=&\hat{U}_{X}\Psi _{0}=\hat{U}_{X}(|0\rangle
_{1}\otimes |0\rangle _{2}) \\
&=&[\hat{u}_{0}^{1}|0\rangle _{1}]\otimes \lbrack |0\rangle _{22}\langle
0|0\rangle _{2}]+[\hat{u}_{1}^{1}|0\rangle _{1}]\otimes \lbrack |1\rangle
_{22}\langle 1|0\rangle _{2}]  \notag \\
&=&[\hat{u}_{0}^{1}|0\rangle _{1}]\otimes \lbrack |0\rangle _{2}\times
1]+0=(\cos \varepsilon |0\rangle _{1}-i\sin \varepsilon |1\rangle
_{1})\otimes |0\rangle _{2}.  \notag
\end{eqnarray}

The universe is then tested with the Hermitian operator $\hat{\Sigma}_{1}=%
\hat{B}_{(0,4)},$ such that its next state $\Psi _{1}$ will be a member of
the basis set $\frak{B}_{(0,4)}\equiv \{|00\rangle ,|01\rangle ,|10\rangle
,|11\rangle \},$ with relative probabilities given by

\begin{equation}
\begin{tabular}{|l|l|l|l|l|}
\hline
$\Psi _{1}$ & $|00\rangle $ & $|10\rangle $ & $|01\rangle $ & $|11\rangle $
\\ \hline
Prob.$=|\langle \Psi _{1}|\Psi _{0}^{\prime }\rangle |^{2}$ & $\cos
^{2}\varepsilon $ & $\sin ^{2}\varepsilon $ & $0$ & $0$ \\ \hline
\end{tabular}
\tag*{Table 8.4}
\end{equation}

As an aside, note that if $\varepsilon \sim 0,$ it is highly probable that $%
\Psi _{1}$ is the same state as $\Psi _{0};$ this type of argument could
play an important role in discussions regarding the origins of apparent
persistence.\bigskip

In accordance with the second step of the proposed Rules, the new state $%
\Psi _{1}$ is now evolved by the unitary operator $\hat{U}_{1}=\ \hat{U}%
_{Y}. $

From Table 8.4 it is clear that\ $\Psi _{1}$ will be one of two possible
states, which may be labelled $\Psi _{1}^{a}=|00\rangle $ and $\Psi
_{1}^{b}=|10\rangle .$ The next evolved state $\Psi _{1}^{\prime }=\hat{U}%
_{Y}\Psi _{1}$ will consequently also be one of two possible states, viz. $%
\Psi _{1}^{a\prime }=$ $\hat{U}_{Y}\Psi _{1}^{a}$ or $\Psi _{1}^{b\prime }=$
$\hat{U}_{Y}\Psi _{1}^{b}.$

Specifically, it can be shown that $\Psi _{1}^{a\prime }$ is given by
\begin{equation}
\Psi _{1}^{a\prime }=\hat{U}_{Y}\Psi _{1}^{a}=|0\rangle _{1}\otimes (\cos
\nu |0\rangle _{2}-i\sin \nu |1\rangle _{2})  \label{Psi1ad}
\end{equation}
whereas $\Psi _{1}^{b\prime }$ is given by
\begin{equation}
\Psi _{1}^{b\prime }=\hat{U}_{Y}\Psi _{1}^{b}=|1\rangle _{1}\otimes (\cos
\nu |0\rangle _{2}-i\sin \nu |1\rangle _{2}).  \label{Psi1bd}
\end{equation}

The second part of the second step involves the collapse of the wavefunction
$\Psi _{1}^{\prime }$ back into one of the eigenstates of the operator $\hat{%
\Sigma}_{2}=\hat{B}_{(0,4)}.$

For the case where the state $\Psi _{1}^{\prime }$ turned out to be $\Psi
_{1}^{\prime }=\Psi _{1}^{a\prime },$ the probabilities that the subsequent
state $\Psi _{2}$ will be a particular member of $\frak{B}_{(0,4)}$ are
given by
\begin{equation}
\begin{tabular}{|l|l|l|l|l|}
\hline
$\Psi _{2}$ & $|00\rangle $ & $|10\rangle $ & $|01\rangle $ & $|11\rangle $
\\ \hline
Prob.$=|\langle \Psi _{2}|\Psi _{1}^{a\prime }\rangle |^{2}$ & $\cos ^{2}\nu
$ & $0$ & $\sin ^{2}\nu $ & $0$ \\ \hline
\end{tabular}
\tag*{Table 8.5}
\end{equation}

So, if the state $\Psi _{1}$ at `time' $n=1$ is $\Psi _{1}=\Psi _{1}^{a},$
the next state $\Psi _{2}$ is clearly going to be one of two possibilities,
which may be labelled $\Psi _{2}^{ac}=|00\rangle $ or $\Psi
_{2}^{ad}=|01\rangle .$

Conversely, if the `first' state $\Psi _{1}$ is instead found to be $\Psi
_{1}=\Psi _{1}^{b}=|10\rangle ,$ and not $\Psi _{1}^{a},$ then the
corresponding probabilities of obtaining a particular eigenstate of $\hat{B}%
_{(0,4)}$ for $\Psi _{2}$ would alternatively be given by
\begin{equation}
\begin{tabular}{|l|l|l|l|l|}
\hline
$\Psi _{2}$ & $|00\rangle $ & $|10\rangle $ & $|01\rangle $ & $|11\rangle $
\\ \hline
Prob.$=|\langle \Psi _{2}|\Psi _{1,b}^{\prime }\rangle |^{2}$ & $0$ & $\cos
^{2}\nu $ & $0$ & $\sin ^{2}\nu $ \\ \hline
\end{tabular}
\tag*{Table 8.6}
\end{equation}

As before, $\Psi _{2}$ would again clearly be one of two possibilities in
this case, which may now be labelled $\Psi _{2}^{be}=|10\rangle $ or $\Psi
_{2}^{bf}=|11\rangle .$

The Rules next dictate that the first step is repeated again, such that $%
\hat{U}_{X}$ is used to evolve whichever of $\Psi _{2}^{ac}$ or $\Psi
_{2}^{ad}$ or $\Psi _{2}^{be}$ or $\Psi _{2}^{bf}$ is actually realised.
Now, if $\Psi _{2}$ turns out to be either $\Psi _{2}^{ac}=|00\rangle $ or $%
\Psi _{2}^{be}=|10\rangle ,$ then $\hat{U}_{X}$ will effectively be
equivalent to $\hat{u}_{0}^{1}\otimes \hat{P}_{0}^{2},$ just as it was for $%
\Psi _{0}.$ However, if it is instead the case that $\Psi _{2}$ is either $%
\Psi _{2}^{ad}=|01\rangle $ or $\Psi _{2}^{bf}=|11\rangle ,$ then $\hat{U}%
_{X}$ instead effectively becomes equivalent to $\hat{u}_{1}^{1}\otimes \hat{%
P}_{1}^{2}.$ In this latter circumstance, $\cos ^{2}\mu $ or $\sin ^{2}\mu $
terms are now consequently introduced into the probability amplitudes, in
the obvious way.

And so on; the two step algorithm may be continued indefinitely.\bigskip

As with the examples of previous sections of this chapter, it is possible to
write elementary computer programs that iterate the above procedure through
a number of cycles. Also as previously it is possible to `interrogate' the
results in a number of ways, as desired.

For example, even after just one iteration it is possible to examine the
probability of proceeding from an initial state $\Psi _{0}$ to a particular
state $\Psi _{2}.$ Specifically, defining
\begin{equation}
P_{ac}=P{\Large (}\Psi _{2}=\Psi _{2}^{ac}=|00\rangle |\Psi _{0}=|00\rangle
{\Large )}
\end{equation}
as the probability that the state $\Psi _{2}$ at `time' $n=2$ will be $\Psi
_{2}^{ac}$ given that the initial state is $\Psi _{0}=|00\rangle $ (which is
equivalent to the product of the probability of jumping from state $\Psi
_{0} $ to $\Psi _{1}^{a}$ and the probability of then jumping from state $%
\Psi _{1}^{a}$ to $\Psi _{2}^{ac}),$ the result (\ref{Ch6Pac}) is readily
obtained.
\begin{eqnarray}
P_{ac} &=&P(\Psi _{2}^{ac}\text{ }|\text{ }\Psi _{1}^{a})\cdot P(\Psi
_{1}^{a}\text{ }|\text{ }\Psi _{0})  \label{Ch6Pac} \\
&=&\cos ^{2}\nu \cos ^{2}\varepsilon .  \notag
\end{eqnarray}

Of course, other `histories' of $\Psi _{0}\rightarrow \Psi _{1}\rightarrow
\Psi _{2}$ may alternatively be chosen. In obvious notation, it may
similarly be found that $P_{ad}=\sin ^{2}\nu \cos ^{2}\varepsilon ,$ $%
P_{be}=\cos ^{2}\nu \sin ^{2}\varepsilon $ and $P_{bf}=\sin ^{2}\nu \sin
^{2}\varepsilon .$ Clearly, $P_{ac}+P_{ad}+P_{be}+P_{bf}=1$ as expected.

Continuing, the probability that the universe will develop from the initial
state $|00\rangle $ through the sequence $|00\rangle \rightarrow |10\rangle
\rightarrow |11\rangle \rightarrow |11\rangle $ (i.e. $|00\rangle
\rightarrow \Psi _{1}^{b}\rightarrow \Psi _{2}^{bf}\rightarrow |11\rangle )$
is given by $\cos ^{2}\mu \sin ^{2}\nu \sin ^{2}\varepsilon ,$ as may be
readily verified. And so on.\bigskip

The physical interpretation of the above model should be emphasised from the
endo-physical point of view of qubit $2.$ Initially $q_{2}$ is in the state $%
|0\rangle ,$ so the projection operator part of $\hat{U}_{X}$ `picks out' $%
\hat{u}_{0}^{1},$ and qubit $1$ is evolved accordingly. In other words the
projection operator ensures that the global transformation $\Psi
_{0}\rightarrow \Psi _{0}^{\prime }$ $=\hat{U}_{X}\Psi _{0}$ is effectively
equivalent to the local transformations $|q_{1}\rangle _{1}\rightarrow
|q_{1}^{\prime }\rangle _{1}=\hat{u}_{0}^{1}|q_{1}\rangle _{1}$ and $%
|q_{2}\rangle _{2}\rightarrow |q_{2}^{\prime }\rangle _{2}=|q_{2}\rangle
_{2}.$

After the subsequent collapse of the state into either $|00\rangle $ or $%
|10\rangle ,$ qubit $2$ is evolved into a superposition by the global
operator $\hat{U}_{Y}.$ This time, the global transformation $\Psi
_{1}\rightarrow \Psi _{1}^{\prime }$ $=\hat{U}_{Y}\Psi _{1}$ is clearly
equivalent to the local transformations $|q_{1}\rangle _{1}\rightarrow
|q_{1}^{\prime }\rangle _{1}=|q_{1}\rangle _{1}$ and $|q_{2}\rangle
_{2}\rightarrow |q_{2}^{\prime }\rangle _{2}=\hat{u}_{2}^{2}|q_{2}\rangle
_{2}.$ So, although the system was initially in the state $\Psi
_{0}=|00\rangle ,$ after a second state reduction the wavefunction $\Psi
_{2} $ of the universe could be any member of the set $\frak{B}%
_{(0,4)}\equiv \{|00\rangle ,|01\rangle ,|10\rangle ,|11\rangle \},$ with
appropriate probabilities.

The dynamics become particularly interesting during the next application of $%
\hat{U}_{X}.$ If $\Psi _{2}$ is either $|00\rangle $ or $|10\rangle ,$ then
the projection operator will again pick out the $\hat{u}_{0}^{1}$ part of $%
\hat{U}_{X},$ and $q_{1}$ will be evolved with this. However, if $\Psi _{2}$
is instead either $|01\rangle $ or $|11\rangle ,$ then the projection
operator will alternatively pick out the $\hat{u}_{1}^{1}$ part of $\hat{U}%
_{X},$ and $q_{1}$ will be evolved in a completely different way.

Thus from the endo-physical point of view, the way in which qubit $1$
develops depends on the sub-state of qubit $2.$ For an observer inside the
universe it appears as if the way in which the universe develops depends
upon a `property' of part of it.

\bigskip

As a final comment to this discussion it should be remarked that, despite
the name, the above two qubit Physicist-Sample mechanism is in no way
imagined to be completely descriptive of a real, physical quantum
experiment. After all, from a practical point of view, real experiments in
real laboratories generally occur over very many jumps, and between
apparatus, equipment, scientists and samples that may each be represented by
enormous groups of factors. Furthermore, real experiments generally involve
extended spatial objects, and so could perhaps only be truly discussed in
the large scale limit of very many subregisters, when a quantum causal set
description of emergent space may be incorporated\footnote{%
See Chapter 5.}. Additionally, real physical objects are generally made from
enormous collections of fundamental physical particles, and this perhaps
implies that a quantum field theoretic description should also ultimately be
employed in any discussion of real measurements in physics\footnote{%
See Chapter 7.}.

More importantly, though, an outcome of Chapter 6 was that a real physical
measurement between a physicist and a sample necessarily requires a degree
of entangling to occur between their sub-states if any physical information
is to be exchanged. Specifically, a conclusion was that endophysical
interactions cannot just be the result of local unitary transformations.
Thus, in the two qubit universes investigated above, the Physicist qubit
would not actually witness the selective evolution of the Sample qubit,
because no information is physically exchanged between them during the
system's development.

Having noted these points, however, they are subsequently ignored in the
present chapter, because it is the principles behind the Selective Evolution
mechanisms that is of interest. In particular, the success of this section
is that even in the absence of physical information extraction or exchange,
it is still possible for the universe to develop in a way in which parts of
it appear to evolve relatively to others, and where the development of one
factor appears to determine the development of another. The fact that
neither part is actually `aware' of how the other qubit is developing is not
currently important; what is important is that the overall universe is able
to develop in this apparently self-referential manner. In short, as long as
the overall universe `knows' what it is doing, it does not matter that its
constituent parts do not.

Indeed, an additional entangling step could be added to the presented SLE
Rules without great conceptual difficulty, and this could effectively enable
the Physicist qubit to investigate, in some sense, how the Sample qubit has
been evolved. `She' could then potentially ascertain how her state
influenced the development of the Sample qubit.

Of course, this type of proposed extension is still highly schematic at this
stage; presumably for a Physicist to really make conscious measurements and
deductions would actually require her to possess enormous numbers of degrees
of freedom and be highly and complexly organised. The conclusion, then, is
that the issue of exactly how the suggested SLE mechanism could be extended
and incorporated into the discussions of the previous chapters, so that real
Physicists perform real measurements on real Samples, remains an important
question for the future.

\bigskip

\subsubsection{A Two Qubit `Double Experiment' Universe}

\bigskip

The Physicist-Sample, Selective Local Evolution Rules of sub-section 8.5.4
govern a universe in which, from an endo-physical point of view, the
development of qubit $1$ (the `Sample') appears to be determined by the
state of qubit $2$ (the `Physicist'), whereas the state of qubit $2$ is
evolved independently: during the second step, $q_{2}$ is rotated by $\hat{u}%
_{2}^{2}$ regardless of the state of $q_{1}.$

A natural extension to this mechanism therefore involves a two-factor system
in which \textit{each} factor appears to be evolved in a way that depends
upon the sub-state of the other. Thus, the development of one individual
sub-state of such a universe seems, from the endo-physical perspective, to
be determined by the factor that comprises the remainder of the universe;
during the second step, $q_{2}$ is now rotated by an unitary operator that
is selected according to the state of $q_{1}.$ This is a truly
self-referential system: at each stage, the universe appears to develop by
examining one part of itself and evolving the other part accordingly.

The suggested Rules may thus be described as giving rise to `\textit{Double
Experiment}' (DE) dynamics; they are still a type of Selective Local
Evolution.\bigskip

The physical interpretation of such a universe is of a `Sample' whose state
is evolved according to the state of\ a `Physicist', and where the state of
the `Physicist' is then influenced by the outcome of this experiment.
Furthermore, on repetition of the procedure, the analogy is of a Physicist
who subsequently `decides' to develop the Sample in a way that is based upon
how `she' has been affected. These ideas are consistent with the notion that
when an experiment is performed in reality on a subject, its result is often
registered as a changed `pointer state' of the apparatus and, ultimately, as
a change in the observing scientist's brain. Moreover, the result of an
initial test often dictates how a scientist may decide to perform further
experiments.

In addition, such a dynamics is also fully compatible with one of the
central tenets of quantum theory: in any quantum measurement, there should
be an element of symmetry between the `observer' and the `observed', because
there is no real criterion for deciding exactly which is which
anyway.\bigskip

The Rules that govern such a Double Experiment universe could be similar to
that of the `single experiment', Physicist-Sample mechanism discussed in the
previous sub-section, but modified in the obvious way. As before, a
separable state is required such that a classical distinction may be made
between the Physicist and the Sample.

Defining again a `reference' time $N$ in order to keep track of which step
is currently applicable, for an `initial' separable state $\Psi
_{N}=|a\rangle _{A}\otimes |b\rangle _{B}$ in a Hilbert space $\mathcal{H}$
factorised in the bi-partite form $\mathcal{H}\equiv \mathcal{H}_{[AB]}=%
\mathcal{H}_{A}\otimes \mathcal{H}_{B},$ the Rules could dictate a two step
sequence of the form

\begin{enumerate}
\item  Evolve $\Psi _{N}$ to $\Psi _{N}^{\prime }=\hat{U}_{N}\Psi _{N}$ by
applying the unitary operator $\hat{U}_{N},$ where $\hat{U}_{N}$ is defined
such that it selectively evolves the factor of $\Psi _{N}$ in $\mathcal{H}%
_{A}$\ according to the factor of $\Psi _{N}$ in $\mathcal{H}_{B}.$ Thus, $%
\hat{U}_{N}$ obeys the relationship
\begin{equation}
\hat{U}_{N}\Psi _{N}=\hat{U}_{N}(|a\rangle _{A}\otimes |b\rangle
_{B})=|a^{\prime }\rangle _{A}\otimes |b\rangle _{B}
\end{equation}
with $|a^{\prime }\rangle _{A}\in \mathcal{H}_{A}$ defined as $|a^{\prime
}\rangle _{A}=\hat{u}_{b}^{A}|a\rangle _{A},$ and where the actual choice of
the unitary operator $\hat{u}_{b}^{A}$ depends somehow on the factor $%
|b\rangle _{B}.$ Note that the sub-script and super-script convention
adopted here is the same as in Sub-section 8.5.4;

\item  Collapse $\Psi _{N}^{\prime }$ to $\Psi _{N+1}$ with an operator $%
\hat{\Sigma}_{N+1}$ that has a separable eigenstate of the form $|c\rangle
_{A}\otimes |b\rangle _{B}.$ Thus, $\Psi _{N+1}\in \mathcal{H}_{AB},$ as
would be ensured if $\hat{\Sigma}_{N+1}$ is chosen such that it is
factorisable relative to $\mathcal{H}_{A}\otimes \mathcal{H}_{B};$

\item  Evolve $\Psi _{N+1}$ to $\Psi _{N+1}^{\prime }=\hat{U}_{N+1}\Psi
_{N+1},$ where $\hat{U}_{N+1}$ is defined such that it selectively evolves
the factor of $\Psi _{N+1}$ in $\mathcal{H}_{B}$\ according to the factor of
$\Psi _{N+1}$ in $\mathcal{H}_{A}.$ Thus, $\hat{U}_{N+1}$ obeys the
relationship
\begin{equation}
\hat{U}_{N+1}\Psi _{N+1}=\hat{U}_{N+1}(|c\rangle _{A}\otimes |b\rangle
_{B})=|c\rangle _{A}\otimes |b^{\prime }\rangle _{B}
\end{equation}
with $|b^{\prime }\rangle _{B}\in \mathcal{H}_{B}$ given by $|b^{\prime
}\rangle _{B}=\hat{u}_{c}^{B}|b\rangle _{B},$ where the actual choice of the
unitary operator $\hat{u}_{c}^{B}$ depends somehow on the sub-state $%
|c\rangle _{A};$

\item  Collapse $\Psi _{N+1}^{\prime }$ to $\Psi _{N+2}$ with an operator $%
\hat{\Sigma}_{N+2}$ that has a separable eigenstate of the form $|c\rangle
_{A}\otimes |d\rangle _{B}.$ Thus, $\Psi _{N+2}\in \mathcal{H}_{AB}.$
\end{enumerate}

Procedures $``1."$ and $``2."$ are taken to comprise the first step, whilst $%
``3."$ and $``4."$ define the second step, as analogous to the PS Rules.

Clearly, because $\Psi _{N+2}\in \mathcal{H}_{AB},$ the first step may now
be repeated, and the overall process continued indefinitely. Also clearly,
the universe is developing according to Selective Local Evolution, where in
any given step of the mechanism the sub-state of one factor determines the
evolution of the other factor, before these roles are reversed in the
subsequent step.\bigskip

To illustrate how a typical DE mechanism might proceed, consider as before a
toy-universe represented by a state in a four dimensional, two qubit Hilbert
space $\mathcal{H}^{(4)}=\mathcal{H}_{1}\otimes \mathcal{H}_{2}$ spanned by
the orthonormal basis $\mathcal{B}_{12}=\{|00\rangle _{12},|01\rangle
_{12},|10\rangle _{12},|11\rangle _{12}\},$ where $\mathcal{B}%
_{1}=\{|0\rangle _{1},|1\rangle _{1}\}$ and $\mathcal{B}_{2}=\{|0\rangle
_{2},|1\rangle _{2}\}$ are bases for $\mathcal{H}_{1}$ and $\mathcal{H}_{2}$
respectively. Consider also the above Rules, but specified by operators of
the form

\begin{itemize}
\item  $\hat{\Sigma}_{N+m}=\hat{\Sigma}_{N+m+1}=\hat{B},$ for all $%
m=0,1,2,...$

\item  $\hat{U}_{N+m}$ given by
\begin{equation}
\hat{U}_{N+m}=\left\{
\begin{array}{c}
\hat{U}_{N}=\hat{U}_{S}\text{ if }m=0,2,4,... \\
\hat{U}_{N+1}=\hat{U}_{T}\text{ if }m=1,3,5,...
\end{array}
\right\}
\end{equation}
\end{itemize}

%TCIMACRO{
%\TeXButton{noindent}{\noindent%
%}}%
%BeginExpansion
\noindent%
%
%EndExpansion
where $\hat{B}$ has four separable eigenstates defining the basis set
\begin{equation}
\frak{B}=\frak{B}_{(0,4)}=\{|00\rangle _{12},|01\rangle _{12},|10\rangle
_{12},|11\rangle _{12}\}=\mathcal{B}_{12}
\end{equation}
and
\begin{eqnarray}
\hat{U}_{S} &=&\hat{u}_{0}^{1}\otimes \hat{P}_{0}^{2}+\hat{u}_{1}^{1}\otimes
\hat{P}_{1}^{2}  \label{Ch6US} \\
\hat{U}_{T} &=&\hat{P}_{0}^{1}\otimes \hat{u}_{2}^{2}+\hat{P}_{1}^{1}\otimes
\hat{u}_{3}^{2}.  \notag
\end{eqnarray}
and a suitable `initial' time, $N,$ has been chosen for reference, noting
that the subscripts $S$ and $T$ on the fixed operators $\hat{U}_{S}$ and $%
\hat{U}_{T}$ are obviously labels, and not temporal parameters. Moreover, $%
\hat{P}_{j}^{i}$ is the projection operator $\hat{P}_{j}^{i}=|j\rangle
_{ii}\langle j|$ for $i=1,2$ and $j=0,1,$ whilst $\hat{u}_{0}^{1}$ and $\hat{%
u}_{1}^{1}$ are different unitary operators acting in $\mathcal{H}_{1},$ but
$\hat{u}_{2}^{2}$ and $\hat{u}_{3}^{2}$ are different unitary operators
acting in $\mathcal{H}_{2}.$\bigskip

As with the Physicist-Sample mechanism of the previous sub-section, the
interpretation of the dynamics of a universe developing according to the
above types of Rule depends very much upon whether an exo-physical or an
endo-physical perspective is being discussed.

From a viewpoint external to the system, an `observer' would witness the
state of the universe changing as follows\footnote{%
Note that as before, the `observer' merely possesses a priviliged vantage
point useful for the discussion, and does not interact with the universe in
any way.}. The initial wavefunction $\Psi _{N}$ is globally evolved into the
state $\Psi _{N}^{\prime }$ by an application of the unitary operator $\hat{U%
}_{N}=\hat{U}_{S},$ i.e. $\Psi _{N}\rightarrow \Psi _{N}^{\prime }=\hat{U}%
_{S}\Psi _{N}.$ This evolved state $\Psi _{N}^{\prime }$ is then tested by
the Hermitian operator $\hat{\Sigma}_{N+1}=\hat{B},$ and consequently
collapses into one of the members of the basis set $\frak{B}_{(0,4)}$ with
probability given by the usual Born rule, thereby becoming the new state,\ $%
\Psi _{N+1}.$

The second step in the universe's development begins by the global evolution
of the state $\Psi _{N+1}$ by the operator $\hat{U}_{N+1}=\hat{U}_{T},$ that
is, $\Psi _{N+1}\rightarrow \Psi _{N+1}^{\prime }=\hat{U}_{T}\Psi _{N+1}.$
Finally, the operator $\hat{B}$ is again used to test the universe, and the
state again collapses into one of the members of $\frak{B}_{(0,4)}$ with a
new set of appropriate probabilities. The resulting eigenvector now becomes
the subsequent state $\Psi _{N+2},$\ and the process begins again with a
repetition of the first step and an application of the operator $\hat{U}_{S}$
to $\Psi _{N+2},$ such that $\Psi _{N+2}\rightarrow \Psi _{N+2}^{\prime }=%
\hat{U}_{S}\Psi _{N+2}.$ The two-step procedure may be iterated indefinitely.

Thus, from the external point of view the universe develops in a
semi-deterministic, globalised fashion, with the choice of unitary operator $%
\hat{U}_{S}$ or $\hat{U}_{T}$ used to globally evolve the state depending
only on whether the procedure is in its first or second step. Of course,
randomness does occur in the model, but only due to the stochastic nature of
the collapse mechanism.

Ultimately,\ then, an exo-physical observer would conclude that the universe
is not developing according to operators chosen as a result of any of the
`properties' of the current state.\bigskip

As in the case of the Physicist-Sample universe, the interesting physics in
the current system's development arises when considering the endo-physical
perspective of the individual qubits. From the point of view of one of these
factors, the unitary operator $\hat{U}_{S}$ described by (\ref{Ch6US}) is an
object that appears to `ask' whether qubit $2$ is in the state $|0\rangle $
or $|1\rangle ,$ whilst locally evolving qubit $1$ with either $\hat{u}%
_{0}^{1}$ or $\hat{u}_{1}^{1}$ according to the `answer' to this question.
Specifically, if $q_{2}$ is in the state $|0\rangle ,$ then $q_{1}$ is
evolved by $\hat{u}_{0}^{1},$ but if $q_{2}$ is instead in the state $%
|1\rangle ,$ then $q_{1}$ is alternatively evolved by $\hat{u}_{1}^{1}.$

Similarly, from this endo-physical point of view, the unitary operator $\hat{%
U}_{T}$ appears to `ask' about the state of qubit $1$ before locally
evolving qubit $2$ appropriately with either $\hat{u}_{2}^{2}$ or $\hat{u}%
_{3}^{2};$ if $q_{1}$ is $|0\rangle ,$ then $q_{2}$ is evolved by $\hat{u}%
_{2}^{2},$ whereas if $q_{1}$ is $|1\rangle ,$ then $q_{2}$ is evolved by $%
\hat{u}_{3}^{2}.$\bigskip

Also congruent to the earlier PS example, the repeated use of the operator $%
\hat{\Sigma}_{N+m}=\hat{B}$ constrains, for all $m,$ every collapsed state $%
\Psi _{N+m}$ to be one of the four separable and non-superposed eigenvectors
defined by the basis set $\frak{B}_{(0,4)}.$

Moreover, as before an important fact is that only those eigenstates with
non-zero probability amplitudes with $\Psi _{N+m-1}^{\prime }$ can actually
be realised physically. Consequently, because during the application of
either $\hat{U}_{S}$ or $\hat{U}_{T}$ only one of the qubits actually
changes, only two of the possible eigenvectors of $\hat{\Sigma}_{N+m}=\hat{B}
$ will give rise to non-vanishing inner products with $\Psi _{N+m-1}^{\prime
}.$ In practice, therefore, the randomly selected $\Psi _{N+m}$ can only
ever be one of these two eigenstates. So, from an internal point of view it
appears that only one of the qubits was actually involved in the development
of the state from $\Psi _{N+m-1}$ to $\Psi _{N+m-1}^{\prime }$ to $\Psi
_{N+m},$ because nothing appears to have been done to the other qubit during
this transition.

Thus, from the endo-physical perspective, the development of the universe
proceeds in a manner that appears to depend on parts of its state.
Paraphrasing, during one of the steps an observer associated with a
particular qubit would believe that `she' was `deciding' how the other qubit
is being evolved, whilst she would then conclude during the remaining step
that she was herself being evolved by an operator chosen according to the
sub-state of the other qubit.\bigskip

The development of this universe may now be summarised. From the point of
view of an external observer looking at the entire state, the overall two
qubit system is globally developed according to a deterministic Rule.
Broadly speaking, this Rule implies the successive application of the
operators $\hat{U}_{S},$ $\hat{B},$ $\hat{U}_{T},$ $\hat{B},$ $\hat{U}%
_{S},...$ to the changing state $\Psi .$

However, the specific construction of the operators $\hat{U}_{S}$ and $\hat{U%
}_{T},$ and the fact that the state prior to evolution is always one of the
members of $\frak{B}_{(0,4)},$ ensures that the individual qubits only ever
`see' half of each of these operators at any one time, that is, either $\hat{%
u}_{0}^{1}\otimes \hat{P}_{0}^{2}$ \textit{or} $\hat{u}_{1}^{1}\otimes \hat{P%
}_{1}^{2}$ for $\hat{U}_{S},$ and either $\hat{P}_{0}^{1}\otimes \hat{u}%
_{2}^{2}$ \textit{or} $\hat{P}_{1}^{1}\otimes \hat{u}_{3}^{2}$ for $\hat{U}%
_{T}.$ For example, if at the beginning of the second step the state has the
form $\Psi =|01\rangle ,$ then the application of $\hat{U}_{T}$ is
effectively equivalent to an application of just the operator $\hat{P}%
_{0}^{1}\otimes \hat{u}_{2}^{2},$ and it would appear that the unitary
operator $\hat{u}_{2}^{2}$\ has been `selected' to evolve qubit $2$
according to the state of qubit $1.$

Thus, from the point of view of an individual qubit, it is the state of the
other factor that appears to determine its evolution. From this perspective,
an internal observer associated with an individual qubit would believe
herself to exist in a fully self-referential universe.\bigskip

\paragraph{DE Example}

\begin{equation*}
\end{equation*}

As with the earlier Physicist-Sample model, it is beneficial to illustrate
the Double-Experiment mechanism by example.

Consider\ a separable state $\Psi _{n},$ in the factorisable two qubit
Hilbert space $\mathcal{H}=\mathcal{H}_{1}\otimes \mathcal{H}_{2},$ whose
development is governed by the operators $\hat{U}_{S},$ $\hat{U}_{T}$ and $%
\hat{B}$ according to the above DE Rules, and where\ $\hat{u}_{0},$ $\hat{u}%
_{1},$ $\hat{u}_{2},$ and $\hat{u}_{3}$ are given by
\begin{eqnarray}
\hat{u}_{0} &=&e^{-i\varepsilon \hat{\sigma}_{1}}\hspace{0.25in},\hspace{%
0.25in}\hat{u}_{1}=e^{-i\mu \hat{\sigma}_{2}} \\
\hat{u}_{2} &=&e^{-i\nu \hat{\sigma}_{1}}\hspace{0.25in},\hspace{0.25in}\hat{%
u}_{3}=e^{-i\tau \hat{\sigma}_{2}}  \notag
\end{eqnarray}
for $\varepsilon ,\mu ,\nu ,\tau \in \mathbb{R}^{+}$ with $\hat{\sigma}_{j}$
the $j^{th}$ Pauli operator.

The unitary operators $\hat{U}_{S}=\hat{u}_{0}^{1}\otimes \hat{P}_{0}^{2}+%
\hat{u}_{1}^{1}\otimes \hat{P}_{1}^{2}$ and $\hat{U}_{T}=\hat{P}%
_{0}^{1}\otimes \hat{u}_{2}^{2}+\hat{P}_{1}^{1}\otimes \hat{u}_{3}^{2}$ are
hence given by
\begin{equation}
\hat{U}_{S}=\left(
\begin{array}{cccc}
\cos {\normalsize \varepsilon } & {\normalsize 0} & {\normalsize -i}\sin
{\normalsize \varepsilon } & {\normalsize 0} \\
{\normalsize 0} & \cos {\normalsize \mu } & {\normalsize 0} & {\normalsize -}%
\sin {\normalsize \mu } \\
{\normalsize -i}\sin {\normalsize \varepsilon } & {\normalsize 0} & \cos
{\normalsize \varepsilon } & {\normalsize 0} \\
{\normalsize 0} & \sin {\normalsize \mu } & {\normalsize 0} & \cos
{\normalsize \mu }
\end{array}
\right) ,\text{ }\hat{U}_{T}=\left(
\begin{array}{cccc}
\cos \nu & -i\sin \nu & 0 & 0 \\
-i\sin \nu & \cos \nu & 0 & 0 \\
0 & 0 & \cos \tau & -\sin \tau \\
0 & 0 & \sin \tau & \cos \tau
\end{array}
\right)
\end{equation}
with the matrices constructed from the usual representations of the bases.
As expected, $\hat{U}_{S}^{\ast }{}\hat{U}_{S}=$ $\hat{U}_{T}^{\ast }{}\hat{U%
}_{T}=\hat{I},$ as may be readily shown.

The development of this universe proceeds as follows. Without loss of
generality, let the initial state $\Psi _{0}$ of the system be $\Psi _{0}=$ $%
|00\rangle .$ Then, the evolved state $\Psi _{0}^{\prime }$ is given by
\begin{eqnarray}
\Psi _{0}^{\prime } &=&\hat{U}_{S}\Psi _{0}=\hat{U}_{S}|00\rangle \\
&=&(\hat{u}_{0}^{1}|0\rangle _{1})\otimes (1\times |0\rangle _{2})+0=(\cos
{\normalsize \varepsilon }|0\rangle _{1}-i\sin {\normalsize \varepsilon }%
|1\rangle _{1})\otimes |0\rangle _{2}.  \notag
\end{eqnarray}

The subsequent state will be one of the eigenvectors of $\hat{\Sigma}_{1}=%
\hat{B},$ with appropriate probabilities given by
\begin{eqnarray}
P(\Psi _{1} &=&|00\rangle ,\Psi _{0})=|\langle 00|{\large (}(\hat{u}%
_{0}^{1}|0\rangle _{1})\otimes |0\rangle _{2}{\large )}|^{2}=|_{1}\langle 0|%
\hat{u}_{0}^{1}|0\rangle _{1}|^{2}\times 1=\cos ^{2}{\normalsize \varepsilon
}  \notag \\
P(\Psi _{1} &=&|10\rangle ,\Psi _{0})=|\langle 10|{\large (}(\hat{u}%
_{0}^{1}|0\rangle _{1})\otimes |0\rangle _{2}{\large )}|^{2}=|_{1}\langle 1|%
\hat{u}_{0}^{1}|0\rangle _{1}|^{2}\times 1=\sin ^{2}{\normalsize \varepsilon
}  \notag \\
P(\Psi _{1} &=&|01\rangle ,\Psi _{0})=P(\Psi _{1}=|11\rangle ,\Psi _{0})=0.
\end{eqnarray}

Whichever eigenvector becomes the new state $\Psi _{1}$ is then evolved by
the unitary operator $\hat{U}_{T}.$ This time, however, it is qubit $1$ that
is used to `select' how $\hat{U}_{T}$ `works'. Specifically, if $\Psi
_{1}=|00\rangle $ then
\begin{eqnarray}
\Psi _{1}^{\prime } &=&\hat{U}_{T}\Psi _{1}=\hat{U}_{T}|00\rangle \\
&=&(1\times |0\rangle _{1})\otimes (\hat{u}_{2}^{2}|0\rangle
_{2})+0=|0\rangle _{1}\otimes (\cos \nu |0\rangle _{2}-i\sin \nu |1\rangle
_{2})  \notag
\end{eqnarray}
whereas if alternatively $\Psi _{1}=|10\rangle $ then
\begin{eqnarray}
\Psi _{1}^{\prime } &=&\hat{U}_{T}\Psi _{1}=\hat{U}_{T}|10\rangle \\
&=&0+(1\times |1\rangle _{1})\otimes (\hat{u}_{3}^{2}|0\rangle
_{2})=|1\rangle _{1}\otimes (\cos \tau |0\rangle _{2}+\sin \tau |1\rangle
_{2}).  \notag
\end{eqnarray}

The evolved state $\Psi _{1}^{\prime }$ is then tested again by the
Hermitian operator $\hat{\Sigma}_{2}=\hat{B}$ to give the next state\ $\Psi
_{2}\in \frak{B}_{(0,4)}$ with a new set of appropriate probabilities, and
the process repeated. From the endo-physical perspective, qubit $1$ could
then be evolved by either $\hat{u}_{0}^{1}$ or $\hat{u}_{1}^{1}$ during this
second application of $\hat{U}_{S},$ depending of course on which particular
element of $\frak{B}_{(0,4)}$ the universe collapses into when it becomes $%
\Psi _{2}.$ And so on.

\bigskip

The Double Experiment Rules featured in this sub-section may be extended to
`higher order' mechanisms in obvious, though non-trivial, ways. Indeed, for
a universe in a Hilbert space factorisable into a large number $F$ of
sub-registers, with a state $\Psi _{n}$ that is constrained to be separable
into $F$ factors for all $n,$ it is possible to imagine Selective Local
Evolution dynamics of many types. Continuing, obvious such mechanisms include

\begin{itemize}
\item  \textit{Many-Physicist} Rules, where the evolution of a Sample
sub-state appears to be determined by the sub-states of a number of
different Physicist factors;

\item  \textit{Many-Sample} Rules, where a single Physicist sub-state
appears to determine how a number of Sample factors evolve;

\item  \textit{Many-Physicist/Sample} Rules, where sets of Physicist
sub-states appear to determine how groups of Sample factors evolve;

\item  \textit{Chain-Experiment} Rules, which are effectively $F$ step
models: in the first step factor $1$ determines how factor $2$ evolves,
whilst in the second step factor $2$ determines how factor $3$ evolves, and
in the third step factor $3$ determines how factor $4$ evolves, and so on.
In each step, only one of the state's factors actually changes, with the
remaining $F-1$ sub-states apparently unaffected;

\item  \textit{Simultaneous-Experiment} Rules, in which different
`experiments' occur simultaneously within separate groups of factors of the
universe's state,\ and where the evolution of a member of one group is
independent of any member in another group. The simplest
Simultaneous-Experiment mechanism would require a four qubit universe, where
perhaps in each two step cycle, qubit $1$ and qubit $2$ are used to
determine the evolution of each other, whilst qubit $3$ and qubit $4$ are
also used to evolve each other. Each set of two qubits (say qubits $1$ and $%
2),$ however, neither influences, nor is influenced by, the other set of
qubits (that is, qubits $3$ and $4).$ Thus, the two groups appear to evolve
self-referentially within themselves, but independently of the other;
\end{itemize}

and many others, including potential hybrid `cross'-variants of those Rules
mentioned above.

Clearly, the level of complexity of the different types of Selective Local
Evolution mechanisms increases rapidly as the factorisability of the Hilbert
space increases. Of course, and as with previous discussions throughout this
thesis, such levels are expected to be echoed in the real, physical
Universe, where many different strata of systems, sub-systems,
sub-sub-systems etc., exist and are known to evolve and interact within
themselves and with each other.

\bigskip

\subsection{Concluding Remarks}

\bigskip

The intention of this chapter has been to investigate a number of closed,
quantum systems whose developments are not reliant on the scrutiny, whim or
decisions of any sort of external agent. The focus has been to investigate
the various types of Rules that could govern such a universe, and could lead
it to develop in a wholly isolated and self-consistent manner.

A number of issues and outcomes related to this, however, still remain to be
discussed.

\bigskip

\subsubsection{Self-referential Quantum Computation}

\bigskip

At the beginning of this chapter, an analogy was drawn between
self-referential quantum universes and the self-diagnostic software of a
hypothetical quantum computer. It is now possible to reinforce this
comparison by observing that the mechanisms described in the above sections
are themselves nothing more than quantum computations. From this standpoint,
the developing quantum universe is therefore viewed as equivalent to a
gigantic, self-referential quantum computation, with the universe's
wavefunction $\Psi _{n}$ at time $n$ analogous to the internal state of the
quantum computer after $n$ steps of its algorithm, and with the various
operators acting upon $\Psi _{n}$ interpreted in the form of elementary
quantum logic `gates'.

Continuing on from this, given that it can be shown that all qubit
evolutions may be re-written in terms of local unitary matrices and the CNOT
operator \cite{Barenco}, it should consequently be possible to break down
some of the mechanisms discussed in this chapter into combinations of these
basic quantum computational gates, and this would be taken to additionally
illustrate the equivalence.\bigskip

Ultimately, the above comments complete a central discussion of this thesis.
Recall that a developing quantum universe is able to give rise to spatial
and quantum field theoretic degrees of freedom, as explored in Chapters\ 5,
6 and 7. So, it has effectively been shown in this body of work that space
and particle physics may be generated by considering the quantum universe as
an enormous self-referential quantum computer.

Furthermore, this particular conclusion is granted additional physical
weight by noting that in the Standard Model of modern physics, the real
Universe is often viewed as containing a vast collection of interacting
quantum fields, and by recalling the idea of Feynman \cite{Feynman} that all
quantum field theories should be reinterpreted in terms of quantum
computation.

Summarising, then, the suggestion is that real physics \textit{is} quantum
computation, and that a developing quantum universe behaves like a quantum
computer. The quantum universe paradigm proposed in this thesis could
therefore be hoped to simulate any physical phenomenon observed in the real
Universe.

\bigskip

When it is remembered that the dimension of the Hilbert space containing the
state $\Psi _{n}$ representing the real physical Universe is at least of the
order of $2^{10^{184}},$ the enormity of the quantum computations required
to model the actual Universe highlights just how limited the qubit examples
described in the chapter actually are.

However, even the modelling of these relatively small dimensional quantum
systems can lead to computational problems, and this would be expected to
become far worse as the number of qubits is increased. Indeed, as a simple
illustration of how even low numbers of qubits can cause severe
computational demands, recall that because a system of just fifty qubits is
represented by a vector of dimension $2^{50}$ (i.e. $\sim 10^{15}),$ even a
modern processor of 1 GHz could require something like $10^{6}$ seconds to
compute just one simple step of its evolution.

Going further, if the physical Universe really does behave like a vast
quantum computation, it could be argued that any device on which a scientist
tried to model it would itself have to be a quantum computer with a
dimension even greater than that of the Universe it is modelling. This would
lead to the amusing consequence that such a computer would presumably have
to be larger than the Universe containing it.\bigskip

The point is that the low-dimensional qubit models discussed in this chapter
provide an illustration of how quantum computational methods can be applied
to a universe as a whole. More importantly, the thrust is that it is
possible to develop these toy-universes in ways that do not require external
guidance, as required for a self-contained and all encompassing view of
physics. The hope, then, is that if the principles applied are valid in the
low dimensional region, it might be possible to extend them to cope with
situations where the complexity of the quantum state is increased. Such
extensions will form a basis for future work.

In fact, it is noted that vectors of increased dimensionality would provide
incredibly rich scopes for potential dynamics, involving, for example, whole
sub-systems of factors evolving and appearing to test one another
independently of other sub-systems that are apparently evolving and testing
yet more sub-systems that are apparently evolving and... just like the
interactions and hierarchies present in the real, classical Universe that
humans seem to perceive.

\bigskip

\subsubsection{The Real Universe}

\bigskip

Nevertheless, despite the fact that many different types of Rules and
mechanisms have been successfully proposed and illustrated, a number of
issues still remain when faced with the question of the dynamics of the real
Universe.

Firstly, and perhaps most obviously, is: what Rules govern the development
of the real quantum Universe? Is real physics best described, for example,
by Type $IIIa$ Probabilistic List-Sort Rules, or is a Type $IV$
Generated-Sort Class $1$ Basis Method mechanism more appropriate? In fact,
would some Rules be favoured over others when constrained by attempts to
recreate the physical phenomena known to empirical science? Is one
particular mechanism, for instance, more able or more likely to provide the
sorts of `persistence' of groups of factors that is generally observed in
physics, or provide the sort of causal set structure required to generate
apparently continuous Minkowski spacetime?

Indeed, could the Rules governing the development of the Universe ever
change, perhaps according to some Rules of the Rules \cite{Buccheri}, as
discussed previously? Such a possibility could provide, potentially, abrupt
changes in values of the constants of Nature over time, an idea favoured by
some theorists when attempting to answer some of the problems of cosmology;
the proposed variable speed of light \cite{Albrecht}\ in the early Universe
is an obvious example here.

Of course, the answers to these questions rely on a better understanding of
the principles that govern emergence in physics, and will only really become
more apparent in the future when states in many sub-register Hilbert spaces
have been properly investigated and modelled. In short, exactly how the real
Universe develops, and hence gives rise to the spatial and particle degrees
of freedom observed in physics, remains an enormous question for the
future.\bigskip

A second issue regarding the choice of operators in the real Universe is:
how do they translate to the types of test familiar to empirical physics? \
How exactly does it arise that physicists appear able to develop their
surroundings in an almost unimaginable number of ways, and with complete
apparent freedom?

This problem is made more pertinent by noting that not every question that a
physicist can ask of a quantum state in the laboratory may be asked of the
state of the Universe. In a laboratory, for example, physicists are often
able to ascertain whether a given state is entangled or not; given a large
number $M$ of identically prepared states $\psi =\{\psi ^{x}:x=1,2,...,M\}$
of the form
\begin{equation}
\psi ^{x}=a|00\rangle _{x_{1}x_{2}}+b|01\rangle _{x_{1}x_{2}}+c|10\rangle
_{x_{1}x_{2}}+d|11\rangle _{x_{1}x_{2}}
\end{equation}
in factorisable Hilbert sub-spaces $\mathcal{H}_{[x_{1}x_{2}]}=\mathcal{H}%
_{x_{1}}\otimes \mathcal{H}_{x_{2}}$ spanned by bases $\mathcal{B}%
_{x_{1}x_{2}}=\{|ij\rangle _{x_{1}x_{2}}:i,j=0,1\}$ (with $\mathcal{B}%
_{x_{1}}=\{|i\rangle _{x_{1}}:i=0,1\}$\ and $\mathcal{B}_{x_{2}}=\{|j\rangle
_{x_{2}}:j=0,1\}),$ the coefficients $a,b,c,d\in \mathbb{C}$ may be
statistically determined, and the separability of $\psi $ in turn
discovered. In particular, each laboratory state $\psi ^{x}$ may be tested
by an operator $\hat{B}_{(0,4)}^{x_{1}x_{2}}$ with eigenstates $\{|ij\rangle
_{x_{1}x_{2}}:i,j=0,1\},$ and the frequencies of particular results used to
determine the values of the coefficients $a,b,c$ and $d.$

There is, however, no known general `entanglement operator' $\hat{E}$ of the
form $\hat{E}\Psi =\lambda \Psi $ in physics, where if an arbitrary state $%
\Psi $ is entangled then $\lambda =\lambda _{1}$ (which might be interpreted
as `yes'), but if $\Psi $ is instead separable then $\lambda =\lambda _{2}$
(or `no'). So, if $\Psi _{n}=\psi ^{x}$ were taken to represent the state of
a two qubit universe, there is no operator $\hat{\Sigma}_{n+1}=\hat{E}$ that
could be directly applied to test its separability.

Indeed, it is difficult to imagine how any such operator $\hat{E}$ could
actually be constructed anyway: aside from the fact that a yes/no response
is `binary', whereas any test of $\psi ^{x}$ should give four possible
results; and the point that there are an infinite number of entangled states
in $\mathcal{H}_{[x_{1}x_{2}]},$ whereas $\hat{E}$ must have a finite number
of eigenvectors; it is difficult to see how the measurements of $a,b,c$ and $%
d$ could ever be achieved in a single jump when a statistical approach is
instead generally required. Apart from anything else, it is not possible to
prepare an ensemble of $M$ identical Universes.\bigskip

So, a situation appears to arise in which endophysical tests of one factor
of the universe by another seem, over a number of jumps, to be able to gain
more information than exophysical tests of its entire state. The obvious
parallels with G\"{o}del type incompleteness \cite{Godel} may be drawn here.

The question is, then: how can the self-referential nature of the Universe
organise itself in such a way so that at one stage a physicist group of
factors is prepared along with $M$ separate entangled factors $\{\psi
^{x}\}, $ and then over a series of jumps the state develops such that the
physicist believes she is applying the local operators $\hat{B}%
_{(0,4)}^{x_{1}x_{2}},$ and is hence determining the values of $a,b,c$ and $%
d?$ In a many subregister universe developing over a number of jumps, how
might it be possible for one persistent set of factors to determine whether
another factor is entangled or not?

\bigskip

\subsubsection{Memory and Information}

\bigskip

In Section 8.5 it was shown that it is possible for a universe to
self-referentially develop if it is governed by certain carefully defined,
two-step, Selective Local Evolution Rules. However, in accordance with the
notion of Process time and the Kochen-Specker theorem \cite{Kochen}\cite
{Peres}, it is noted that once the state $\Psi _{n}$ has been realised, then
the previous state $\Psi _{n-1}$ can no longer be said to exist in any
sense. Given this, a natural question faced by such a two-step mechanism is:
how does the universe `know' which step it should be in? If the universe can
only refer to its current state $\Psi _{n},$ how does it `keep track' of
which part of the two step Rules are applicable at that time? Specifically,
and from the example given at the end of Sub-section 8.5.4, after the state
has collapsed to $\Psi _{n}$ via a test with an operator $\hat{\Sigma}_{n}=%
\hat{B}_{(0,4)},$ how does the universe `know' that the state $\Psi _{n-1}$
was evolved by the operator $\hat{U}_{Y},$ say, and not $\hat{U}_{X},$ such
that the current state $\Psi _{n}$ must now be rotated by the operator $\hat{%
U}_{X}?$\bigskip

One suggestion could therefore be that some sort of `memory' is required in
a completely self-referential universe. This memory could perhaps be used to
record what the previous state was, what it did, and how it was developed
(or even, by extension, what the previous few states were). Alternatively,
the memory could maybe equivalently take the form of a type of `clock' that
counts the number of steps taken since a particular `reference' time, as
alluded to in Sub-section 8.5.4. Either way, the Rules would `consult' this
memory store in order to keep track of which step it is in, and to determine
what the universe should do next.

The question now becomes: what exactly is this memory and where could it be
stored?\bigskip

Up until now, the current state $\Psi _{n}$ has been taken to completely
specify everything about the universe. It might seem natural, therefore, to
somehow try to incorporate the proposed memory store into this vector. In
fact, there are two obvious ways to attempt this, each with associated
problems.

Firstly, if $\Psi _{n}$ is assumed to completely represent every conceivable
current property of the universe \textit{and} everything about the previous
state (i.e. if a mechanism is suggested in which $\Psi _{n-2}$ somehow gets
`absorbed' into $\Psi _{n-1},$ which itself then somehow gets `absorbed'
into $\Psi _{n},$ and so on), then the dimension of the Hilbert space of $%
\Psi _{n}$ must be at least twice that of $\Psi _{n-1}.$ This leads to a
situation in which as the universe proceeds from $\Psi _{n-2}$ to $\Psi
_{n-1}$ to $\Psi _{n}$ etc., the associated vector space is growing at an
exponential rate, and it is unclear what this may mean
mathematically.\bigskip

The second way may be to redefine the actual $n^{th}$ state of the universe
as the larger vector $\Phi _{n},$ where $\Phi _{n}=\Psi _{n}\otimes M_{n},$
which contains the usual `physical' current state $\Psi _{n}$ responsible
for the physically observable universe, and a `memory' factor $M_{n}.$
Particularly, $M_{n}$ might contain some sort of `information' regarding
what the previous states $\Psi _{n-1},\Psi _{n-2},...,$ previous tests $\hat{%
\Sigma}_{n},\hat{\Sigma}_{n-1},...,$ and previous operators $\hat{U}_{n-1},%
\hat{U}_{n-2},...,$ were, or even just be a `clock' that somehow registers
the current `time' $n.$ This memory factor, $M_{n},$ could then be
`examined' somehow to determine how the `physical state' factor $\Psi _{n}$
is developed. Then, and overall, during the development of the universe from
$\Phi _{n}$ to $\Phi _{n+1},$ where $\Phi _{n+1}=\Psi _{n+1}\otimes M_{n+1},$
the `memory' part $M_{n}$ in $\Phi _{n}$ might somehow be erased and
replaced by $M_{n+1}$ in $\Phi _{n+1}$ (now containing information about $%
\Psi _{n}$ and/or $\hat{\Sigma}_{n+1}$ and/or $\hat{U}_{n}$ and/or\ $n+1),$\
whilst the `physical state' part $\Psi _{n}$ of $\Phi _{n}$ would be
replaced by $\Psi _{n+1}$ in $\Phi _{n+1}.$\bigskip

However, segregation of the overall vector $\Phi _{n}$ into a `physical
state' and a `memory' could also lead to difficulties. For example, if the
memory factor $M_{n+1}$ is to be a direct copy of $\Psi _{n},$ i.e. $%
M_{n+1}=\Psi _{n},$ then it is difficult to see how the transition $%
M_{n}\rightarrow M_{n+1}$ could occur. In particular, the `erase and
replacement' procedure may not be governed by unitary evolution, because it
is manifestly irreversible. Whilst this might appear good from the point of
view of a universe developing according to the second law of thermodynamics,
it is forbidden by an argument similar to the No-Cloning theorem which
prevents general unitary evolutions $\hat{U}$ of the form $\hat{U}(\psi
\otimes \phi )\rightarrow \psi \otimes \psi .$

Alternatively, if the `erase and replacement' procedure is to result from a
state reduction, then an operator $\hat{\Sigma}_{n+1}$ with an eigenstate $%
\Psi _{n+1}\otimes \Psi _{n}$ needs to be used to test $\Phi _{n}=\Psi
_{n}\otimes \Psi _{n-1},$ and it is unclear how this should in general be
constructed; the Memory factor of $\Phi _{n+1}$ might be expected to result
from the factor $\Psi _{n}$ of $\Phi _{n},$ whilst the `physical state'
factor of $\Phi _{n+1}$ might equally be expected to result somehow from
details of the memory factor of $\Phi _{n}.$ In short, it is difficult to
see how the memory could be both referred to and changed at the same time.
It would also be required that the probabilistic nature of the collapse from
$\Phi _{n}$ to $\Phi _{n+1}$ is taken into account by the Rules, and
additionally that $\hat{\Sigma}_{n+1}$ is defined such that every physically
realisable outcome of it is separable into a `physical state' factor and a
`memory' factor.

Moreover, observe that the form of $\Phi _{n}$ is similar to the
partitioning of the Double-Experiment universe described in Sub-section
8.5.5. However, given that it was this separation that led to the need for a
two-step dynamics in the first place, it is difficult to envisage how such a
form may then be able to solve the problem of specifying which step of the
mechanism the Rules should follow.\bigskip

Of course, the memory $M_{n}$ does not have to contain the entire state $%
\Psi _{n-1}.$ Indeed, the above problems may not occur if the memory factor
instead takes the form of a type of `clock'. However, one difficulty that
would now arise is that if each possible `time' of this clock is assumed to
be represented by a different basis state in the memory's sub-space, the
dimension of this Hilbert space might be expected to be very large.
Specifically, in an eternally enduring universe, the dimension of the
clock's Hilbert sub-space may be required to be infinite, and this is
clearly undesirable.\bigskip

Evidently, the above suggestions are just embryonic ideas at this stage.
However, if the difficulties encountered are indeed insurmountable problems,
the conclusion might be that in order to account for two (or more) step
Rules, it may be necessary to parameterise the universe with two vectors: a
state vector and a memory vector. Exactly how a two vector mechanism might
be defined, what form the memory vector could take, what its implications
might be for the dynamics, and how it might influence the development of the
state $\Psi _{n},$ are left as questions for the future. However, as
remarked in Chapter 3, it could appear that the state $\Psi _{n}$ and the
rules $R_{n}$ in the $n^{th}$ stage $\Omega _{n}$ of the Universe's
development might only be parts of the story; in a complete and consistent
quantum universe some sort of information store $I_{n}$ might also be vital.

Furthermore, in fact, an information content $I_{n}$ may not just be
necessary in universes governed by Selective Evolution dynamics. In
List-Sort dynamics, for instance, an Information $I_{n}$ may be needed to
`contain' the List of possible operators $\frak{L},$ whilst in Type $IV$
Class $1$ Basis Method mechanisms an Information $I_{n}$ may be required to
specify the set of constant unitary transformations $\{\hat{U}^{(i)}\}.$

Consequently, and continuing the computational analogy central to this
thesis, if the Universe may be described as an enormous quantum computer
whose wavefunction $\Psi _{n}$ describes the state during the $n^{th}$ step
of an algorithm specified by the Rules $R_{n},$ the Information $I_{n}$ is
like a `cosmic hard drive' that keeps track of the time $n$ and stores the
set of possible instructions, $\hat{\Sigma}_{n},$ $\frak{L}$ and $\hat{U}%
_{n} $ etc.

\bigskip

\subsubsection{Reduction without Observers}

\bigskip

One point that has not been addressed so far is the actual cause of the
state reduction process ubiquitously present in all of the mechanisms
discussed in this chapter, and, indeed, throughout this thesis.

Every mechanism has assumed that the $n^{th}$ state $\Psi _{n}$ of the
universe is tested by some Hermitian operator $\hat{\Sigma}_{n+1},$ and
consequently collapses into the next state $\Psi _{n+1}$ which is one of the
eigenvectors of $\hat{\Sigma}_{n+1};$ the dynamics are therefore analogous
to those empirically known to govern quantum systems in the laboratory. With
this analogy in mind, however, the issue remains as to why the application
of an operator $\hat{\Sigma}_{n+1}$ actually causes the state of the
universe to jump from $\Psi _{n}$ to $\Psi _{n+1}.$ In other words, how
exactly does the proposed paradigm view the infamous `Measurement Problem'
of laboratory physics?\bigskip

The Measurement Problem of laboratory quantum mechanics traditionally raises
two main questions: firstly, how is a test actually chosen, and secondly,
why does this lead to a collapse of the wavefunction?

In the conventional Copenhagen interpretation of quantum theory, the first
question is often swept under the carpet by assuming the role of the
observer. In this exo-physical ``solution'', the state reduction process is
initiated by an external physicist deciding to measure the quantum state
with a particular operator, and the collapse then manifests itself as the
quantum state reducing to a classically observed object. In short, in
laboratory quantum theory an external agent or environment is assumed to `do
something' to the quantum system, and the quantum state then reacts by
jumping to the observed (semi-) classical form.

Of course, this `explanation' has a number of problems associated with it.
For example, which observer or observers actually get to choose the test? At
what level of observation does the collapse ultimately occur (i.e. the
Schr\"{o}dinger's Cat paradox \cite{Schrodinger})? How does the physicist
decide to test the state in a particular way in the first place, given that
she is presumably a complex of quantum particles herself?\bigskip

The second question also remains unanswered in conventional quantum theory.
However, whilst there has never been a satisfactory explanation for the
existence of this discontinuous and irreversible process in a Universe that
otherwise seems to run on continuous and reversible laws, in many elementary
texts on quantum mechanics a number of different interpretations are given
that attempt to account for the apparent collapse of the wavefunction in
laboratory physics. For example, some explanations involve particular
deterministic evolutions, suggestions being due to the Decoherence paradigm
discussed in Chapter 3, or because of the Many-Worlds interpretation \cite
{Everett} of Multiverse splitting. Other attempts assume that the state
collapses spontaneously, perhaps because of the non-trivial dynamics of the
Hamiltonian in GRW theory \cite{GRW}, or due to quantum gravitational
effects induced by superposed spacetimes \cite{Penrose1}.

Each of these interpretations, however, is associated with its own set of
problems, inconsistencies and difficulties, and it is by no means clear as
to whether any of them is able to provide a coherent, complete and
verifiable explanation for the phenomenon of state reduction.\bigskip

Now, it is not the intention of this thesis to bias any one of the above
exo-physical `explanations' over the others, nor is there a desire\ to
provide fresh insight into possible solutions for the exo-physical
Measurement Problem. Instead, in the presented fully quantum universe
paradigm it is implicitly assumed that state reduction is a necessary part
of quantum theory, and must hence also be a necessary part of any universe
running according to quantum principles. Generally speaking, the way in
which a state appears to collapse in the laboratory is less important here
than the fact that it does indeed collapse.

Summarising, in fact, in the paradigm proposed in this thesis involving the
measurement of a state $\Psi _{n}$ by a test $\hat{\Sigma}_{n+1},$ the
existence of the operator $\hat{\Sigma}_{n+1}$ is considered to be just as
fundamental as the existence of the vector $\Psi _{n}$ or the Hilbert space $%
\mathcal{H}.$ In other words, the existence of the operator $\hat{\Sigma}%
_{n+1}$ is taken to be an integral feature of the dynamics of the universe,
and it is assumed that it is an automatic application of this test that
results in an automatic collapse of the state. Moreover, the choice of the
operator $\hat{\Sigma}_{n+1}$ is governed by a quantum algorithm, as
suggested by the various Rules discussed in this chapter.

There is consequently no real ``Measurement Problem'' in the quantum
universe para- digm, at least not in the sense generally understood. The
traditional exo-physical difficulties of laboratory physics, which should
now be associated with one group of factors appearing to measure another
group of factors, consequently find a natural solution emerging from the
discussion of information exchange (Chapter 6), and from the operator
selection mechanisms proposed in the present chapter.

\bigskip

In conventional laboratory physics, with the scientist and quantum
experiment standing isolated in a much larger universe, a question naturally
arises concerning the associated timescale between the preparation of the
quantum state by the physicist and its subsequent collapse. This question is
especially important in, for example, the GRW and superposed spacetime
interpretations mentioned above, because in these the quantum wavefunction
is assumed to collapse spontaneously after a period of time that depends on
the model in question.

However, in the case presented here, where the state represents the entire
Universe and not just a tiny sub-system within it, such a concern could not
be an issue. Under this circumstance, and because there is no absolute,
external time in which the universe is developing, the question as to `how
long' it takes for the universe to develop from the state $\Psi _{n}$ to the
state $\Psi _{n}^{\prime }\equiv \hat{U}\Psi _{n}$ to the state $\Psi _{n+1}$
is meaningless.

Conversely, in fact, in the universe described in this thesis, time is
nothing but a concept that emerges from the observation that the state $\Psi
_{n}$ is not the same as the state $\Psi _{n-1},$ which was itself not the
same as $\Psi _{n-2}.$ Thus, in this sense time is viewed as synonymous with
\textit{change.} It is consequently no more pertinent to ask about the
timescale involved between the preparation of $\Psi _{n}$ and the collapse
of $\Psi _{n}^{\prime }$ into $\Psi _{n+1},$ than it is to ask about the
spatial distance between different factors of $\Psi _{n}.$ The actual
process is seen as nothing but the mathematical transformation of a vector
in a Hilbert space; time, space and physics are concepts that somehow emerge
internally from the dynamics of this state.

\bigskip

\subsubsection{Time without Time}

\bigskip

The actual origin of continuous-looking physical time in the quantum
Universe is naturally a complicated process due to the dimensionality of the
Hilbert space involved, but the logic behind it is fairly straightforward.

Consider first an exo-physical perspective, and consider two arbitrary
vectors $\Psi _{a}$ and $\Psi _{b}$ in a Hilbert space $\mathcal{H}.$ If the
vector $\Psi _{b}$ results from an operation on $\Psi _{a},$ such that $\Psi
_{b}$ does not exist without the prior existence of $\Psi _{a},$ it may be
concluded that $\Psi _{a}$ is a `cause' of $\Psi _{b},$ in some sense.

Moreover, if $\Psi _{a}$ and $\Psi _{b}$ cannot both exist simultaneously,
and if $\Psi _{b}$ is known to exist `now', it follows that $\Psi _{a}$
existed at one point, but no longer does. It may hence be said that $\Psi
_{a}$ existed `before' $\Psi _{b}.$

Furthermore, if $\Psi _{b}$ is taken to be an eigenstate of a physical test $%
\hat{\Sigma}_{b}$ on $\Psi _{a},$ such that no intermediate state exists%
\footnote{%
Recall that in the context of quantum mechanics, nothing can be said to
physically exist unless it is observed or measured.} between the existence
of $\Psi _{a}$ and the existence of $\Psi _{b},$ it may be argued that $\Psi
_{a}$ and $\Psi _{b}$ are separated by one `step'. It may therefore be
justifiable to relabel the sub-scripts as $\Psi _{a}=\Psi _{n-1}$ and $\Psi
_{b}=\Psi _{n}.$

In addition, if $\Psi _{a}$ and $\Psi _{b}$ are not orthogonal (such that
the inner product $|\langle \Psi _{b}|\Psi _{a}\rangle |=|\langle \Psi
_{a}|\Psi _{b}\rangle |>0),$ and if $\Psi _{a}\neq \Psi _{b}$ (such that $%
\Psi _{b}$ and $\Psi _{a}$ may be distinguished and $|\langle \Psi _{b}|\Psi
_{a}\rangle |<1),$ the vectors $\Psi _{a}=\Psi _{n-1}$ and $\Psi _{b}=\Psi
_{n}$ may be used, from this perspective, to represent successive states of
a quantum universe.\bigskip

Continuing the logic, if the state $\Psi _{n-1}$ resulted from a test $\hat{%
\Sigma}_{n-1}$ on a different, but not orthogonal, state $\Psi _{n-2}$ (i.e.
$0<|\langle \Psi _{n-1}|\Psi _{n-2}\rangle |<1),$ and if this chain may be
repeatedly extended to the observation that the state $\Psi _{n-N+1}$
resulted from a test $\hat{\Sigma}_{n-N+1}$ on a different, but not
orthogonal, state $\Psi _{n-N}$ (i.e. $0<|\langle \Psi _{n-N+1}|\Psi
_{n-N}\rangle |<1),$ it could be interpolated that the state $\Psi _{n-N}$
appeared to develop into the state $\Psi _{n}$ through a sequence of
intermediate states $\Psi _{n-N+1},$ $\Psi _{n-N+2},...,$ $\Psi _{n-1},$ and
hence through a series of discrete jumps. It might consequently be argued
that the universe developed from $\Psi _{n-N}$ to $\Psi _{n}$ in $N$
distinct changes, or steps.

So, it is now possible to define `exo-time' as a measure of the number of
steps taken to get from one state to another in the chain $\Psi _{n-N},\Psi
_{n-N+1},\Psi _{n-N+2},...$ Thus, the exo-time taken for the universe to
develop from $\Psi _{n-N}$ to $\Psi _{n}$ is $N.$ Paraphrasing, according to
this definition, time is at root a counting process.\bigskip

Once a definition of exo-time has been established, it is possible to
consider notions of endo-time. In particular, endo-time is defined in terms
of the number of changes experienced by a particular factor (representing,
for example, a human endo-observer) as the state of the universe develops
through a series of steps. This definition will be clarified in the
following.

Note first of all, however, that since such an endo-observer can at most
only ever be sure of parts of the current state $\Psi _{n},$ can possibly
recollect factors of a state $\Psi _{n-1}$ that appeared different from the
current state, and can predict other possible states $\Psi _{n+1}$ that are
eigenvectors of subsequent potential operators, they could immediately infer
that $\Psi _{n}$ is later than $\Psi _{n-1},$ but $\Psi _{n+1}$ does not yet
exist. Moreover from this logic, such complex, macroscopic and allegedly
intelligent parts of the universe called human beings, who are aware of
parts of the state $\Psi _{n},$ can remember parts of sets of states $\Psi
_{n-1},$ $\Psi _{n-2},...$ but cannot recall anything about sets such as $%
\Psi _{n+1},$ $\Psi _{n+2},...,$ are moved to construct concepts such as
past, present and future in order to describe things that they think have
happened, are currently happening, and may well happen.\bigskip

Humans could then go on to quantify measures of endo-time by a process that
involves counting the changes of the universe around them. They might, for
instance, define the `second' in terms of how many times a particular part
of the universe changed as the universe developed from the remembered state $%
\Psi _{n-X}$ to the current state $\Psi _{n},$ where $X$ implies a huge
number of jumps. If, for example, they notice that over $Y$ jumps particular
factors of the universe representing photons emitted by Caesium-133 atoms
are able to change $9,192,631,770$ times \cite{SI} under certain
circumstances, then they might define $Y$ to constitute one second.

Going further, physicists could even attempt to define time by using laws of
physics that were themselves discovered by observing changes in the
universe. If, for example, constraints on laboratory quantum mechanics and
general relativity seem to indicate that any time scale less than about $%
\tau _{p}=(\frac{Gh}{c})^{1/2}\sim 1.35\times 10^{-43}$\ seconds is
undefined, they might conclude that in one second the overall state of the
universe is able to change $Y=1/\tau _{p}$ times. However, this is not the
clearest way to view the process: it is not that in one second the universe
may change $Y$ times, but that by changing $Y$ times, parts of the universe
may go on to provide a definition of one second. In essence, one second
\textit{is} the fact that the universe changes $Y$ times.

Of course, this then leads to an apparently paradoxical situation in which
although $Y$ jumps might give rise to the definition of one second, it does
not imply that the change from $\Psi _{i}$ to\footnote{%
Or even from $\Psi _{i}$ to $\Psi _{i}^{\prime }=\hat{U}_{i}\Psi _{i}$ to $%
\Psi _{i+1},$ if the Rules governing the dynamics dictate that the state is
rotated during its development.} $\Psi _{i+1}$ has a duration of $1/Y$ $%
^{th} $ of a second. After all, recall that this development is just a
mathematical procedure, and is therefore without duration. Paraphrasing,
because, for example, the evolution of $\Psi _{n}$ to $\Psi _{n}^{\prime }=%
\hat{U}_{n}\Psi _{n}$ by $\hat{U}_{n}$ is nothing but a mathematical
relation, to question its timescale is effectively equivalent to asking how
long it takes for one plus one to equal two.\bigskip

The resolution of this paradox is to note that it only arises from an
exo-physical perspective, that is, when an observer believes she can stand
isolated from the universe and witness it evolving in her own external time.
From the endo-physical point of view of a scientist who is part of the
universe she is trying to observe, the paradox does not arise because only
relative differences can be granted any real, physical significance. In
fact, even if an external time did exist in which the universe evolves, an
observer who is part of the universe would be unable to say whether the
development of the state from $\Psi _{n-1}$ to $\Psi _{n-1}^{\prime }$ to $%
\Psi _{n}$ took the tiniest fraction of an ``exo-second'' or many billions
of ``exo-years'', because all that she can ever be aware of is that $\Psi
_{n}$ is different from $\Psi _{n-1}$ (c.f. the discussion of passive and
active transformations given in Chapter 6).

Moreover, from a strict quantum mechanical point of view, the actual
evolution of the state from $\Psi _{n-1}$ to $\Psi _{n-1}^{\prime }$ can
have no duration, external or internal, because according to the
interpretation of Wheeler (\cite{Wheeler1}), and as discussed in Chapter 3,
no attributes of the state can even be said to exist until it is measured.
Only the measurements, that is the changes from $\Psi _{n-1}$ to $\Psi _{n},$
are physically relevant, so it is only relative to these changes that
physical phenomena such as time may be discussed.\bigskip

Like the apparent existence of Euclidean space, and as proposed in Chapter
5, the emergence of a linear, temporal dimension is something that appears
under specific circumstances, according to the unique point of view of a
particular endo-observer, as the universe jumps from $\Psi _{n-X}$ to $\Psi
_{n-X+1}$ to... to $\Psi _{n},$ where $X\gg 1.$ Schematically, if $X$ is
very large and if $\Psi _{m-1}$ is sufficiently `similar'\footnote{%
In some sense. Certainly, for example, $\Psi _{m-1}$ could not be in a
vastly different partition to $\Psi _{m}$ if apparent continuity is to
result.} to $\Psi _{m}$ for all $(n-X+1)\leq m\leq n,$ a particular causal
set description might begin to generate flat Minkowski spacetime.

The point is that under these special circumstances (which certainly appear
to be the case from the perspective of physicists in the real Universe),
endo-physical observers might falsely conclude that they live in a universe
that is evolving in an external, continuous time that exists independently
of the state.

Moreover, they might therefore make the mistake in this case of asking how
long it takes for the state of such a universe to develop from $\Psi _{m}$
to $\Psi _{m}^{\prime }$ to $\Psi _{m+1}.$ They might also be surprised
when, under specially controlled laboratory conditions, they witness
discontinuous processes such as quantum state collapse occurring, because
their `everyday', large scale, emergent, classical time and laws of physics
appear continuous. And when they extrapolate this continuous time dimension
to the smallest scales, they might find problems with their classical
theories of general relativity and spacetime.

In short, the problem lies in assuming that the state of the universe is
developing in an external, continuous, background (space)time; the mistake
is to apply the Block universe approach to a system that is running
according to quantum principles, and hence according to Process time.
Instead, physical time might really only be defined in terms of counting the
number of changes of this quantum state, and should hence be considered an
emergent feature that is `created' as the universe develops.

\bigskip

It is at this stage that a simple and schematic discussion of the relative
durations of different systems within the universe is permissible. Suppose
that the Hilbert space of the universe may be split in the form $\mathcal{H}%
_{[OABR]}=\mathcal{H}_{O}\otimes \mathcal{H}_{A}\otimes \mathcal{H}%
_{B}\otimes \mathcal{H}_{R},$ where each factor sub-space needs not be of
prime dimension. Suppose also that $\mathcal{H}_{O}$ may be associated with
the Hilbert sub-space of an `Observer' (i.e. a part of the universe
representing a physicist); $\mathcal{H}_{A}$ may be associated with the
Hilbert sub-space of a particular `sample' (an electron, say, in order to
draw parallels with the EPR system discussed in Chapter 3); $\mathcal{H}_{B}$
may be associated with the Hilbert sub-space of a different `sample' (say, a
positron, for the same reasoning); and $\mathcal{H}_{R}$ may be associated
with the Hilbert sub-space comprising the rest of the Universe.

Suppose further that four successive states, $\Psi _{n-3},$ $\Psi _{n-2},$ $%
\Psi _{n-1},$ and $\Psi _{n},$ in the universe's history are separable in
the forms
\begin{eqnarray}
\Psi _{n-3} &\in &\mathcal{H}_{R}^{OAB}\text{ \ \ , \ \ }\Psi _{n-2}\in
\mathcal{H}_{OABR} \\
\Psi _{n-1} &\in &\mathcal{H}_{BR}^{OA}\text{ \ \ , \ \ }\Psi _{n}\in
\mathcal{H}_{R}^{OAB}.  \notag
\end{eqnarray}

Then, the interpretation of the development of this universe is as follows.

First, the jump from $\Psi _{n-3}$ to $\Psi _{n-2}$ may be schematically
imagined to imply the creation at `time' $n-2$ of an `Observer' sub-state in
$\mathcal{H}_{O},$ an isolated electron sub-state in $\mathcal{H}_{A},$ and
an isolated positron sub-state in $\mathcal{H}_{B}.$

Continuing, from the endo-physical point of view of the sub-state
representing the Observer, the sequence $\Psi _{n-2}\rightarrow \Psi
_{n-1}\rightarrow \Psi _{n}$ schematically appears to represent a
progression of a quantum universe from an initial state $\Psi _{n-2}$
prepared as four separate sub-systems, to the state $\Psi _{n-1}$ in which
the Observer has `measured' the electron, and then to the state $\Psi _{n}$\
in which both the electron and positron sub-states have been `measured' by
the Observer\footnote{%
The actual meaning of the word `measurement' is left deliberately vague
here, with reference made to Chapter 6.}. Of course, and as in Section 8.5,
assuming that the factor of $\Psi _{n-2}$ in $\mathcal{H}_{B}$ is identical
to the factor of $\Psi _{n-1}$ in $\mathcal{H}_{B},$ the operator $\hat{%
\Sigma}_{n-1}$ of which $\Psi _{n-1}$ is an eigenstate must be carefully
chosen to ensure that, from an endo-physical perspective, the positron
sub-state appears to be unaffected by the transition from $\Psi _{n-2}$ to $%
\Psi _{n-1}.$

From the endo-Observer's point of view, a physicist would argue that the
time-scale involved between the preparation of the positron and its
measurement was twice that involved between the preparation of the electron
and its measurement. The reasoning is that although both the electron and
positron are apparently prepared at the same time as factors of $\Psi
_{n-2}, $ the factor in $\mathcal{H}_{A}$ is `measured' by the Observer
during the transition from $\Psi _{n-2}$ to $\Psi _{n-1},$ whereas the
`measurement' of the positron by the Observer does not occur until the
transition from $\Psi _{n-1}$ to $\Psi _{n}.$ In other words, the Observer
witnesses two changes of the universe's state between the preparation of the
positron and its measurement, whilst only one step appears to occur between
the preparation of the electron factor and its subsequent re-entangling. The
Observer could therefore conclude that the positron factor exists for twice
as long as the electron, from her endo-physical perspective.

On the other hand, from the perspective of the positron, only one time step
appears to occur between its creation as a factor of $\Psi _{n-2}$ and its
measurement by (or, indeed, of) the Observer, because it is unchanged during
the transition from $\Psi _{n-2}$ to $\Psi _{n-1}.$ Paraphrasing, because
the factor in $\mathcal{H}_{B}$ does not witness any changes occurring
between exo-times $n-2$ and $n-1,$ time appears not to pass for it; the
factor representing the positron effectively behaves as if it is `frozen' in
time during this period.

So overall, the apparent null test on the factor in $\mathcal{H}_{B}$ during
the jump from $\Psi _{n-2}$ to $\Psi _{n-1}$ leads to the type of `route
dependent' endo-time discussed in Chapter 5, and results in concepts
analogous to the notion of proper time in relativity. In short, the Observer
would believe that two time steps occurred between the preparation of the
positron and its measurement, whereas the positron would contend that only
one step occurred between these two events.\bigskip

Of course, the Observer cannot say \textit{absolutely} how long it took for
the universe to develop from the state $\Psi _{n-2}$ to the state $\Psi
_{n}, $ because such an absolute measure is meaningless when time is only
defined relative to the changes themselves. Time is defined in the proposed
paradigm as nothing but a counting process, and so should not be confused
with the mathematical developments of the individual states themselves, such
as would occur by falsely associating durations to mathematical procedures.
Only relative endo-times have any physical significance in a fully quantum
universe.

Moreover, once a concept of relative time-scales has been established, it is
possible to ignore the definition of time as a counting procedure simply by
appealing to the `duration' of a specific process compared to that of an
accepted standard. From this, scientists are consequently able to say that a
certain factor $|F\rangle $ of the universe's state exists for $Z$ seconds
if it persists for, say, $Z\times (9,192,631,770)$ cycles of the radiation
emitted by a particular atom, where, ultimately, this value is itself only
defined relative to the universe developing through a series of $Y$ states.
The `standard' definition of time used in conventional physics is therefore
recovered, without the need to count actual numbers of jumps.\bigskip

It is hence possible to reinterpret the question faced by proponents of
spontaneous collapse models of quantum mechanics (e.g. the GRW process) for
conventional quantum sub-systems of the universe. Any time-scale involved
between the preparation of a quantum sub-state and its apparent collapse is
only relevant either relative to a number of changes of the universe, or
equivalently relative to the duration of another sub-system that itself only
endures relative to a number of changes of the universe.

Clearly, however, similar such questions are not relevant for the Universe
itself. There is no time in which the state of the Universe is evolving;
rather the Universe is generating time as it develops through a series of
states. This is truly a self-contained perspective.

\bigskip \newpage

\section{Summary, Conclusions and Future Directions}

\renewcommand{\theequation}{9.\arabic{equation}} \setcounter{equation}{0} %
\renewcommand{\thetheorem}{9.\arabic{theorem}} \setcounter{theorem}{0}

\bigskip

The purpose of this thesis has been to propose a perspective on the overall
structure of the Universe that is fully compatible and consistent with the
empirically verified principles of quantum mechanics. In effect, the
proposal resulted in extending the standard principles of quantum mechanics
to the case where the state in question represents the Universe itself, and
not just some microscopic sub-system within it.

Despite such an inevitable conclusion, however, it is noticed that the
Universe observed by physicists does not generally appear to resemble a
quantum wavefunction. So from this viewpoint, and by considering the various
properties of a state developing in a Hilbert space factorisable into an
enormous number of subregisters, attempts have been made to suggest how the
aspects of physics familiar to laboratory science could begin to emerge from
this fundamental, mathematical picture. Some success may therefore be
claimed for the endeavour of investigating the potential bridges between the
quantum computational, pregeometric vision of reality unavoidably proposed,
and the semi-classical world experienced by humans.

These attempts are now summarised in this final chapter, with the
conclusions that may be drawn from such work given, and some of the
remaining questions and future directions for research highlighted.\bigskip

In Chapter 3 it was shown that quantum mechanics is a `valid' theory, in the
sense that empirical results confirm the predictions of quantum physics, but
do not support the conclusions of theories based upon classical Hidden
Variables. Furthermore, from a basic set of observations regarding
experimentally known features of physics, it was then argued that the entire
Universe should in fact be treated according to quantum principles.
Specifically, it was suggested that the Universe may be represented by a
pure state $\Psi _{n}$ in a Hilbert space $\mathcal{H}^{(D)}$ of enormous
dimension $D>2^{10^{184}},$ and that this state is subject to `rotations' by
unitary operators $\hat{U}_{n}$ and `testing' by Hermitian operators $\hat{%
\Sigma}_{n+1};$ moreover, it is this discontinuous process of information
extraction by $\hat{\Sigma}_{n+1}$ that justifies the use of the discrete,
`temporal-like' label, $n.$

From this line of thinking, the concept of a Stage was conjectured. Thus, it
was suggested that the operators $\hat{U}_{n}$ and $\hat{\Sigma}_{n+1}$ used
to develop the state $\Psi _{n}$ are chosen by a quantum algorithm according
to a set of Rules $R_{n},$ possibly making reference to some sort of
Information store, $I_{n}.$ In the paradigm proposed in this thesis, the
development of the Universe is therefore envisaged to be analogous to a
gigantic quantum computation, with its state proceeding eternally through a
sequence of collapse, evolution, testing, collapse, evolution, testing,...
in an automatic and self-referential way.\bigskip

Based on the observation that the Universe that humans perceive generally
appears to be classical, and consequently not indicative of the types of
phenomena typically exhibited by conventional quantum states, Chapter 4
attempted to discuss the necessary requirements for arguing that two
physical objects may be described as classically distinct and
distinguishable. To this end, it was shown that if a Hilbert space $\mathcal{%
H}$ may be factorised into two factor sub-spaces $\mathcal{H}_{A}$ and $%
\mathcal{H}_{B},$ such that $\mathcal{H}=\mathcal{H}_{[AB]}=\mathcal{H}%
_{A}\otimes \mathcal{H}_{B},$ then the components of a state $\Phi \in
\mathcal{H}_{[AB]}$ in $\mathcal{H}_{A}$ are distinguishable from the
components of $\Phi $ in $\mathcal{H}_{B}$ if $\Phi $ is separable relative
to the split $\mathcal{H}_{A}\otimes \mathcal{H}_{B}.$ That is, if $\Phi $
may be written in the form $\Phi =\phi \otimes \varphi ,$ where $\phi \in
\mathcal{H}_{A}$ and $\varphi \in \mathcal{H}_{B},$ then the factor $\phi $
is classically distinct from the factor $\varphi .$

Continuing, a test to determine the separability of an arbitrary state was
then given, and the conclusion thereby drawn that separability should be a
surprisingly uncommon feature in a fully quantum Universe. The fact that
this does not seem to be the case in Nature, however, because the Universe
does seem to possess enormous numbers of classically distinct objects,
therefore strongly suggested that very tight constraints must be placed upon
the operators used to produce the states, such that the occurrence of this
result is ensured.

It was then shown that states that are separable relative to one particular
split of the overall Hilbert space may be entangled relative to an
alternative split. This result in turn suggested that a preferred
factorisation of the Hilbert space may be appropriate for the case of the
Universe.

Chapter 4 also raised the issue of basis sets of vectors, and showed that
even in a four dimensional Hilbert space not every combination of entangled
and separable elements exists; specifically, no type $(1,3)$ basis set $%
\frak{B}_{(1,3)}$ is permitted. This then immediately raised the question of
preferred bases for the Universe, and it is consequently a task for future
research to discover which types of basis are allowed in Hilbert spaces of
higher dimensions, and which are forbidden by vector space mathematics. Is
it possible, for example, to find a $(1,[D-1])$ type basis set for a $D>4$
dimensional Hilbert space split into two factors? What limitations exist in
tri-, quad-,... or $N$-partite splits of a Hilbert space?

The last part of Chapter 4 addressed the question of real classicity. The
suggestion was that classical objects on the macroscopic scale may be
somehow associated with groups of factors of the state $\Psi _{n}$ of the
Universe, not least because if two objects may be described as classically
distinct and distinguishable from one another, they cannot by definition be
entangled. Exactly how these groups of factors translate to the classical
looking objects of the laboratory is however left as an important question
for the future, but it is interesting to speculate on whether paradigms
related to emergent theories of decoherence may play a central role in this
discussion.

So overall, the conclusion was that the\ semi-classical degrees of freedom
of the observed Universe somehow ultimately originate from the factors of
the state $\Psi _{n},$ because the factors of a separable state may be
considered to be distinct and distinguishable, as required for classicity.

\bigskip

One necessary requirement for a system of objects to be described as
classical is that it is possible to argue that ``\textit{this} object with
\textit{these} qualities is \textit{here, }whereas\textit{\ that} object
with \textit{those} properties is\textit{\ there}.'' The issues of `here'
and `there' were therefore addressed in Chapter 5. In response, it was
consequently shown that during the development of a universe represented by
a state contained in a highly factorisable Hilbert space, causal set
relationships may begin to arise between the factors of successive states,
and these may in turn give rise to spatial degrees of freedom in the
emergent limit. Specifically, embryonic lightcone structures were introduced
by considering how counterfactual changes in the factors of $\Psi _{n-1}$
could affect the factors of $\Psi _{n},$ and it was conjectured that, over a
large number of jumps, these could ultimately be used to generate manifolds,
metrics and geodesics. Spatial relationships are therefore introduced
between the physical objects that these factors represent on the emergent
scale.

Furthermore, because the state $\Psi _{n}$ is itself the result of a test $%
\hat{\Sigma}_{n},$ the factorisability of the operators was also
investigated; an important conclusion was that factorisable operators can
only have separable eigenstates. The changing sequence of operators $\hat{%
\Sigma}_{n},\hat{\Sigma}_{n+1},\hat{\Sigma}_{n+2},...$ was therefore also
demonstrated to exhibit causal set type patterns, and this fact was asserted
to be responsible for driving the conditions necessary for continuous space
and time to arise. Moreover, and unlike the states, the operators were also
conjectured to be constrained to obey Einstein locality, and it was
suggested that these assertions may play important parts in discussions
explaining why states in quantum physics may exhibit apparently superluminal
correlations, whilst observables are restricted to follow classical
causality.

A number of physical examples were finally given to illustrate these general
points.

Overall, the outcome of Chapter 5 was to show that the changing
factorisability of the operators, as the Universe proceeds through a series
of Stages, can give rise to a changing separability of the state, and this
in turn might consequently begin to exhibit causal set-like relationships
between its factors. In addition, the presence of local null tests as
factors of the operators generates the appearance of `route dependent'
endo-times for the various factors of the changing state, and this may be
interpreted as analogous to the existence of proper time in relativity. So,
the `fictitious' exo-time parameter $n$ gives way to physical, local
endo-times in terms of counting the changes experienced by particular groups
of factors, and this could in turn provide a possible pregeometric origin
for discussions involving the possession of unique inertial frames of
reference by individual endo-observers.\bigskip

Despite the successes mentioned above, a number of questions still remain
unanswered concerning the origin of space from the proposed fully quantum
paradigm. Perhaps the greatest of these is: how exactly does conventional
general relativity emerge from the underlying statevector description of the
Universe?

Now, this question is not just about how the separations of a state in a $%
D>2^{10^{184}}$ dimensional Hilbert space could give rise to the continuous
looking spacetime experienced in physics, though this task is, of course,
itself an enormous issue to be addressed. Neither is it directly concerned
with the mechanics of exactly how a spacetime existing as an apparently
linear, $3+1$ dimensional Block Universe arena could emerge from the
pregeometric, causal set relationships between factors, though this too is
an important point. More importantly, the question of particular concern is:
what features of the separable state picture could be used to contain
information regarding an emergent object's mass, and how could this be used
to affect and distort the lightcone structure so that it appears to result
in gravitationally curved spacetime? In other words, how can a sort of `mass
parameter' be introduced into the statevector description, such that the
self-referential interplay between the factorisable operators and separable
states results in an apparently curved spacetime in the emergent limit?

In short, the emergence of mass curved, four dimensional spacetime from the
quantum universe is an enormous question for the future.

\bigskip

Chapter 6 explored the links between quantum computation, information, and
the quantum universe paradigm.

Section 6.1 was used to set up the necessary framework for the work of the
following chapters; in particular the issues of logic gates and the CNOT
operator were discussed, and the concept of `Transformation' operators that
act between the basis vectors of the individual subregisters introduced. It
was then shown how elementary computations may be performed, with the
accompanying Bell correlations used to provide an example of how care must
be taken when interpreting the results of quantum questions.

The remaining section of Chapter 6 discussed the definition and role of
information in closed quantum systems. In particular, the notions of active
and passive transformations were discussed, and this was followed by
definitions for information change and exchange. Summarising, the conclusion
was that information changing processes necessarily rely on an active
transformation, and as such cannot be achieved simply by a convenient
transformation (e.g. relabelling) of the basis. Moreover, it was argued that
if $\Psi _{n-1}$ and $\Psi _{n}$ are in different partitions, the change
from $\Psi _{n-1}$ to $\Psi _{n}$ is synonymous with an information changing
process. Such physically significant processes cannot be achieved by passive
transformations, and cannot be removed by unitary rotations of the bases of
the individual subregisters.

Related to information change was information exchange, where a component of
the state in a particular subregister $\mathcal{H}_{i}$ may be said to have
exchanged information with a component of the state in a different
subregister $\mathcal{H}_{j}$ during a jump from $\Psi _{n-1}$ to $\Psi _{n}$
if the `relationship' between them changes. Specifically, if the component
of $\Psi _{n-1}$ in $\mathcal{H}_{i}$ is in a different block of the
partition containing $\Psi _{n-1}$ from the component of $\Psi _{n-1}$ in $%
\mathcal{H}_{j},$ but the component of $\Psi _{n}$ in $\mathcal{H}_{i}$ is
in the same block of the partition containing $\Psi _{n}$ as the component
of $\Psi _{n}$ in $\mathcal{H}_{j},$ then these two components may be said
to have exchanged information during the transition from $\Psi _{n-1}$ to $%
\Psi _{n}.$

Ultimately, the conclusion was that if the number of factors of the
probability amplitude $\langle \Psi _{n}|\Psi _{n-1}\rangle $ is less than
the number of factors of either the initial or final states, then the
transition from $\Psi _{n-1}$ to $\Psi _{n}$ is an information exchanging
process.

Following on from these definitions, the question of endo-physical
measurements was addressed. Specifically, if $\Psi _{n-1}\in \mathcal{H}%
_{ABR}$ but $\Psi _{n}\in \mathcal{H}_{R}^{AB},$ where $\mathcal{H}_{[ABR]}=%
\mathcal{H}_{A}\otimes \mathcal{H}_{B}\otimes \mathcal{H}_{R},$ with $%
\mathcal{H}_{R}$ interpreted as a `rest of the universe' factor space, and $%
\mathcal{H}_{A},\mathcal{H}_{B},\mathcal{H}_{R}$ need not be of prime
dimension, then the factor of $\Psi _{n-1}$ in $\mathcal{H}_{A}$ may be said
to have `measured' the factor of $\Psi _{n-1}$ in $\mathcal{H}_{B}$ (and
vice versa) as the universe jumped to $\Psi _{n}.$

The concepts of `split partition' and `partition overlap' were in turn
usefully introduced.\bigskip

As with the work of Chapter 5, the results of Chapter 6 also raise a number
of questions when attempting to apply them to the real world of laboratory
physics; again, it is the issue of how they relate in the emergent limit
that is of issue. For example, for a semi-classical object consisting of
large numbers of factors, exactly `how much' information needs to be
exchanged to constitute the sort of experiment familiar to physics?
Paraphrasing, if a laboratory apparatus is represented by, say, $10^{6}$
components of the state of the universe, whilst a laboratory subject by,
say, $10^{3},$ how many of these must become entangled during a jump from $%
\Psi _{n-1}$ to $\Psi _{n}$ in order to say that the apparatus has measured
the subject? $1$ of each? $46$ of one, but $23$ of the other? All $10^{3}$
and $10^{6}?$

Additionally, how do actual laboratory measurements, that may in reality
take place over very many jumps of the universe, translate to the
pregeometric, single-jump endo-measurements discussed in this thesis?

\bigskip

In Chapter 7 the remaining part of the statement concerning when ``\textit{%
this} object with \textit{these} qualities is \textit{here''} was addressed,
by considering how particular groups of factors might begin to exhibit the
types of physical property possessed by particle fields in Nature. From the
starting point of the Transformation operators introduced in the previous
chapter, pregeometric ladder operators were defined that caused, under
certain circumstances, qubit states to be `raised' or `lowered' within their
individual Hilbert sub-spaces. Moreover, these pregeometric ladder operators
were shown to obey the same statistics and anti-commutation relations as
fermionic annihilation and creation operators, and also exhibited the
characteristics of $U(2)$ symmetry.

From this definition, it was then shown how the Dirac field may be accounted
for from the suggested pregeometric structure, by considering a Hilbert
space factorisable into $4(2M+1)$ subregisters, where $M$ is large.
Specifically, momentum space creation and annihilation operators were
defined in terms of discrete Fourier transforms of the pregeometric ladder
operators over this large number of subregisters, and these were shown to
obey the anti-commutation relationships necessary for physical spin-half
particles. Moreover, Hamiltonian, momentum and charge operators were then
derived for these particles in terms of the pregeometric ladder operators,
by substituting the usual momentum space operators for their subregister
defined counterparts.

To complete the discussion, it was finally shown how these operators may be
re-written as sums of products of two-qubit CNOT gates and unitary operators
acting locally in the individual subregisters. Feynman's vision of
reinterpreting quantum field theory as a form of quantum computation was
therefore demonstrated from the perspective of the proposed paradigm.

In conclusion, then, it was shown that by considering particular
combinations of pregeometric transformation operators defined in a highly
factorisable Hilbert space, the types of operator familiar to experimental
particle physics may be constructed. From such considerations, the physical
properties exhibited by particular physical objects may therefore begin to
emerge from the quantum universe picture envisaged in this thesis.\bigskip

Of course, real objects in the real Universe tend to exhibit enormous
varieties of physical properties, and these are still generally unaccounted
for in the proposed model. It is therefore a task for future work to attempt
to discover how alternative fields may be encoded into the suggested
paradigm. How, for example, could colour degrees of freedom emerge from the
suggested quantum vision? Or flavour? What about more `exotic' fields and
particle species? Indeed, would the existence of, for example, (the so far
unexplained) Dark Energy matter emerge as an inevitable consequence of the
types of pregeometric treatment employed to obtain quantum fields from the
state of the quantum universe? What about string theory: does this fit into
the proposed paradigm, and if so, how?

On a related note, how could the Higgs field be accounted for, or even the
inflation fields conjectured in the early universe, and how would such
mass-involving processes influence the quantum causal sets produced, and
consequently the emergence of spacetime from the statevector description?

Indeed, how could the suggested approach actually be applied in real space
anyway? After all, Chapter 7 discussed the emergence of the Dirac field from
a set of qubits that map to an emergent, one dimensional `lattice'; how
exactly should this approach best be extended to cope with the three
dimensional volumes present in the real Universe, as accounted for by
conventional quantum field theory?\bigskip

A further extension to the work presented in Chapter 7 asks the question: is
the emergence of quantum field theoretic descriptions limited to qubit
subregisters, or could higher dimensional factor spaces be considered?
Certainly, non-qubit subregisters would intuitively seem to be required for $%
SU(3)$ colour gauge symmetry, but is this actually the case; is it
necessary? Going further, could such a proposal account for the appearance
of bosonic particles, with subregisters of enormous, `near infinite'
dimension playing an important part? If so, would the emergence of bosonic
particles from such a para-fermionic treatment of physics make important
comments regarding the theories of supersymmetry currently hypothesised?

Finally, would the suggested approaches to quantum field theory eventually
be able to explain why some of the parameters of the Standard Model have the
values that they do? In fact, could the proposed links between quantum field
theory and the origin of spacetime in the united paradigm be used to explain
some of the other puzzles of fundamental physics, such as why, for example,
the constants of nature have the values they do, or why the curvature of the
Universe is so close to unity?

All of these are necessary questions for the future development of the
quantum universe vision. Many, it is hoped, may be answered from the type of
analysis discussed in Chapter 7.

\bigskip

Chapter 8 attempted to classify and explore some of the different ways that
a fully quantum universe free from external observers might be able to
develop. First, the various Types of way in which $\Psi _{n}$ could develop
into the next state $\Psi _{n+1}$ were classified. Then it was argued that a
self-referential mechanism is required in order to provide some of the
empirical properties of the physical Universe, and attention was therefore
turned onto situations in which the next operator $\hat{\Sigma}_{n+1}$
depends somehow on the present state $\Psi _{n}.$ Two particular Sorts of
mechanism consequently became apparent: those in which the state $\Psi _{n}$
is used to select a particular operator from a pre-existing List, and those
in which the next operator is instead Generated from the current state at
time $n;$ these two possibilities were explored in turn. Moreover, it was
also concluded that not every method of determining $\hat{\Sigma}_{n+1}$
from $\Psi _{n}$ is necessarily permitted, even though it might be possible
to determine $\hat{\Sigma}_{n+1}$ from $\hat{\Sigma}_{n}$ in an analogous
way.

In all cases it was shown that particular types of Rule could give rise to
various physical features for the developing universe; certain List-Sort
Rules, for example, may be particularly suitable to produce the sorts of
embryonic lightcone structure required for a causal set description of space
to begin to emerge.

Chapter 8 finished by discussing the possibility of allowing the state to be
evolved self-referentially, and unitary operators $\hat{U}_{n}$ were
proposed that could rotate $\Psi _{n}$ in ways that appeared to depend on
what it is. However, the crucial point concluded in this work was that an
interpretation of how $\hat{U}_{n}$ acting on $\Psi _{n}$ behaves depends
very much on whether an exo- or an endo-physical perspective is adopted.
Specifically, although from an external point of view the outcome of $\hat{U}%
_{n}\Psi _{n}$ may not seem to be determined self-referentially, from an
internal viewpoint it could appear that $\hat{U}_{n}$ is examining $\Psi
_{n},$ and then developing it in a manner that depends on the result of this
investigation.

Moreover, it was also shown that by defining the operators such that $\hat{U}%
_{n}$ appears to just examine part of $\Psi _{n},$ endo-physical experiment
type effects could consequently arise, again from an endo-physical
perspective. As before, a conclusion drawn was that local null tests play an
important part in the emergence of real physics from the quantum
universe.\bigskip

As discussed in the final part of Chapter 8, a huge number of questions
still remain regarding the development of the state in the proposed
paradigm. What Type of Rule, for example, is the physical Universe actually
governed by, and what exactly is this Rule? Which sorts of mechanism are
most suitable for the generation of particular physical phenomena, highly
separable states persisting over large numbers of jumps being an obvious
example? Could the Rules governing the development of the state change over
`time', $n,$ and how would this affect the resulting physics? Indeed, if
this is the case, how does the universe keep track of what the current time
is, noting that this is also a question faced in the two-step Selective
Local Evolution mechanisms? Is the suggestion therefore to be accepted that
some sort of information store is required for the actual development of the
Universe, and if so, what form could this `memory' take?

In short, although the conclusion may be made that it is possible to
successfully specify consistent quantum algorithms that automatically
develop the state of a universe through an endless series of stages, without
the need or intervention of any sort of external guidance, a great deal of
future research will be required before it is known what the Rules governing
the development of the real Universe actually are, and how these might
physically operate.

\bigskip

By far the greatest question for future research concerns the issue of
emergence: how exactly does the semi-classical world familiar to physicists
arise from the pregeometric, statevector description proposed in this thesis?

Throughout this work, a number of points that begin to answer this issue
have been addressed. The general conclusion is that each successive state $%
\Psi _{n}$ must be highly separable, such that the vision of reality
perceived by physicists, involving countless numbers of quantum
`micro-systems', may be readily generated. Moreover, it additionally follows
that the operators must also be highly factorisable, such that they can
appear to control the development of these quantum micro-systems in
apparently local and microscopic ways.

However, an enormous number of details still remain to be investigated in
this picture, and it is by no means clear as to exactly how the proposed
vision can give rise to every emergent property known to classical physics.
How, for example, is the continuous time Schr\"{o}dinger equation able to
emerge as an accurate tool useful in describing the apparent evolution of
these quantum microsystems in the absence of measurement by emergent
semi-classical observers? Similarly, how do the emergent operators of
laboratory physics arise from the properties of the `universe operators' $%
\hat{\Sigma}_{n+1},\hat{\Sigma}_{n+2},...?$

Of course, the relationships between the `laboratory tests' and the
`universe tests', $\hat{\Sigma},$ and operators $\hat{U}$ are expected to be
highly complex and non-trivial. For instance, field theoretic operators such
as the Hamiltonian may initially be constructed from incredibly complicated
relations between pregeometric transformations, as was discussed in Chapter
7. But, it both interesting and necessary to speculate on how such a vision
could be incorporated into the picture of the developing universe discussed
in Chapter 8. After all, the interpretations of laboratory experiments are
ultimately expected to rely somehow on subjects and apparatus both being
represented by large groups of factors of a state $\Psi _{n},$ and the
universe then self-referentially choosing operators $\hat{U}_{n}$ and $\hat{%
\Sigma}_{n+1}$ according to these sets of factors;\ the resulting state $%
\Psi _{n+1}$ is then taken to represent the outcome of this experiment. It
is, however, unclear at this stage as to exactly how this mechanism might
work in practice, and so a consistent, self-referential version of field
theory is even further away.\bigskip

In fact, the emergent operators familiar to emergent physicists may be
expected to bear no resemblance at all to the `universe operators' $\hat{%
\Sigma},$ and the operators that represent real laboratory measurements may,
perhaps, really only emerge from considering average properties of the
operators $\hat{\Sigma}_{n+1},\hat{\Sigma}_{n+2},...,\hat{\Sigma}_{n+m}$ as
the universe develops over an enormous number $m$ of jumps. This type of
general point was\ again exemplified in Chapter 7, where it was demonstrated
how conventional field theory annihilation and creation operators may emerge
from Fourier transforms of enormous numbers of pregeometric ladder
operators. It may also go some way towards explaining the ``how much
information needs to be exchanged'' question of Chapter 6: groups of factors
representing subjects and apparatus could become slowly entangled, a few
components at a time, over a large number of jumps, and it could only be
over this set of transitions $\Psi _{n}\rightarrow \Psi _{n+1}\rightarrow
...\rightarrow \Psi _{n+m}$ that a large scale measurement may be said to
have occurred between them. Overall properties of the corresponding set of
operators $\hat{\Sigma}_{n+1},\hat{\Sigma}_{n+2},...,\hat{\Sigma}_{n+m}$ may
then be used, somehow, to describe this single laboratory test.

The real operator structure of the developing universe may therefore be
vastly different from the types of\ laboratory Hermitian operators that
physicists are familiar with. Moreover, the ultimate, self-referential
interplay between the states $\Psi _{n}$ and the operators $\hat{\Sigma}%
_{n+1}$ could provide a vision of reality completely different to that
perceived by scientists on the classical, emergent scale. Indeed, to quote
Jaroszkiewicz \cite{Jaroszkiewicz1}: ``\textit{...almost everything that we
humans believe in is a sort of illusion, a convenient fabrication of the
brain, designed to rationalize the massive amounts of stimuli that we
constantly receive from our immediate environment. This includes space. This
illusion gives us a fighting chance of survival. From this point of view,
nothing is really what it seems. If you have seen the film ``The Matrix'',
you may have some idea of what I mean}.''.

Exactly how the patterns and constructs recognisable to the human brain
emerge from the developing quantum state description therefore remains an
enormous question. Indeed, exploring how the human brain creates this
illusion from the constantly changing state is a task potentially beyond the
scope of neuroscientists and psychologists, despite that fact that some
scientists are already beginning to explore the possible quantum origins of
consciousness (e.g. \cite{Hameroff}).

The point is that the reality humans perceive, consisting of macroscopic
semi-classical objects representing subjects, laboratories and apparatus,
and with laboratory tests represented by single Hermitian operators, could
really bear no apparent similarity to the underlying subregister structure
of states and jumps from which they emerge. The sequence of evolutions,
tests, and partition changes occurring on the pregeometric level as the
Universe's state develops could bear very little resemblance to the large
scale reality perceived by human observers.

Humans may therefore never truly be able to understand the apparently
bizarre properties of the underlying quantum structure, because the notions
of pregeometric states and operators are far beyond their sphere of rational
experience. Indeed, what exactly \textit{is} a quantum state?

\bigskip

The conclusion of this thesis is not that the conjectured quantum universe
paradigm provides a `Theory of Everything'. It is hoped, however, that it
could provide a valid and correct framework for such a theory to begin to be
discussed. Thus, the overall desire is that from this work, it will
eventually be possible to describe an all-encompassing and consistent view
of physics, in which the properties of a quantum state undergoing an endless
series of evolutions and tests in a factorisable Hilbert space of enormous
dimension is able to generate, in the emergent limit, every phenomenon
associated with the observed physical Universe.

A real `Theory of Everything' will therefore take the form of a complete set
of self-referential Rules, which may be used to select particular operators,
which in turn cause the development of this state to give rise to these
particular properties. From the underlying pregeometric structure, such a
Theory will therefore consequently govern the emergent scale presence of
classicity, continuous space, an expanding Universe, particle field
theories, interactions, and even human physicists performing tests in
laboratories. And, of course, the phenomenon of time.

\bigskip \newpage

\appendix

\section{Classical and Quantum Computation}

\renewcommand{\theequation}{A-\arabic{equation}} \setcounter{equation}{0} %
\renewcommand{\thetheorem}{A-\arabic{theorem}} \setcounter{theorem}{0}

\bigskip

Culturally, technologically and epistemologically, the Theory of Computation
was one of the revolutionary successes of Twentieth century science. Now in
the Twenty-first century, a great deal of research time is being spent on
extending the idea of a computer acting classically to one obeying the rules
of quantum mechanics. There are a number of reasons for this, both practical
and theoretical.

Practically, modern computer chip technology is reaching its limits. One
reason is that the overall size of the device is bounded by thermodynamic
concerns over the dissipation of heat, and so there is an increasing need to
manufacture silicon chips of highly compact structure. This, however, can
only go so far, and current technology is rapidly approaching scales where
quantum effects become significant.

On the other hand, as suggested by Manoharan \cite{Manoharan}, it might be
expected that the theory of quantum computation is the natural extension to
the classical case. An analogy is drawn here to mechanics: Newtonian
mechanics is the limiting case of relativity $(c\rightarrow \infty )$ and of
quantum theory $(%
%TCIMACRO{\UNICODE[m]{0x127}}%
%BeginExpansion
h\hskip-.2em\llap{\protect\rule[1.1ex]{.325em}{.1ex}}\hskip.2em%
%EndExpansion
\rightarrow 0),$ which are in turn both special cases of relativistic
quantum field theory. In a similar vein, Classical Computation (CC) might be
a limit to Quantum Computation (QC), itself just a subset of Quantum Field
Computation (QFC). It is even conjectured that quantum field theory itself
is only an approximation of higher order theories, for example supersymmetry
or quantum gravity, and so this area of computation might also eventually
need expanding as subsequent models of reality become better understood.

Furthermore, it has been conjectured throughout this thesis that the entire
Universe may be running as a giant quantum computer. If this viewpoint is
correct, a better understanding of the principles of quantum computation is
essential if any type of `Theory of Everything' is to be achieved.\bigskip

The purpose of this appendix is to introduce and elaborate upon the
conventional ideas of classical and quantum computation. It should be noted,
however, that the type of quantum computation discussed here is not strictly
identical to the mechanism adopted by the quantum Universe, as proposed in
the body of this work. In fact, and as has been the case before, the
difference arises from a conflict between endo- and exo-physics. From the
endo-physical viewpoint of a system free from guiding observers, the
Universe prepares, evolves and tests itself according to the Rules governing
its dynamics. Conversely, in the conventional quantum computations examined
below, the concern is for how semi-classical human observers are able to
manipulate an isolated quantum state in order to perform a specific
calculation, and how they can then test this state to obtain a specific
answer.

Thus, the discussion below provides a summary of the principles of
laboratory computation, and therefore forms a useful completion and
comparison to the ideas of Chapter 6.

\bigskip

\subsection{Classical Computation}

\bigskip

In the 1930s Alan Turing wrote his seminal paper on computation \cite{Turing}
and proposed the Turing Machine (TM), the archetypal Classical Computer and
forerunner to all modern electronic computers. Contained in this work is the
definition of computability: ``\textit{A number is computable if its decimal
can be written down (by a machine)}''. Obvious examples are integer
quantities and rational fractions such as $1/2,$ $3/8$ etc.

In fact, it is possible to extend the definition to numbers that can be
written down as decimals to a given degree of accuracy. This extends
computability to quantities that are the result of Cauchy convergent
sequences if a level of approximation is specified. For example, the number $%
2.7183,$ which is the estimate to five significant figures of the value of $%
e^{1},$ provides an approximation to the sequence
\begin{equation}
e^{1}=1+\frac{1}{2!}+\frac{1}{3!}+...+\frac{1}{r!}+...
\end{equation}
for very large values of $r.$ Hence, $2.7183$ is a good approximation to the
infinite sum generating the exponential, so $e^{1}$ can be considered
computable.\bigskip

Mathematics is generally concerned with the processing of numbers via
specific operations. Calculations, for example, often follow the general
logic: \textit{``what is the output number }$O$\textit{\ resulting from the
operation }$A$\textit{\ acting on the input number }$I$ $?".$ From this, and
the definition of computability, it is possible to conject an Automated
Computing Machine (or TM) which, given an input and a set of rules for
computation, is able to solve (or `compute') a specific problem and return
an output.

Turing provided his hypothetical machine with a certain set of
characteristics and components necessary for it to work. The actual physical
design of the machine is taken to be immaterial, and any particular
hypothesised TM is not automatically assumed to be the only (or even the
best) way of encoding and processing information. It is the hierarchy of how
the characteristics and components interrelate that is important, and how
the algorithm proceeds. This principle is reinforced by Church's Thesis,
which argues that all reasonable models of computation are equivalent \cite
{Church}.

Turing's necessary conditions are as follows:

\begin{enumerate}
\item  A TM\ has a finite number of internal states (called
m-configurations). These are analogous to a set of rules to be followed
during the computation, and consequently define it.

\item  The machine is supplied with a `tape', that is, a medium of infinite
capacity on which the Input is recorded, the Output is displayed, and the
result of any intermediate `rough workings' can be temporarily recorded.

\item  The tape itself is divided into a series of sites or `squares'. Each
square can bear one, and only one, `symbol' from a set of possibilities.

\item  The tape is moved along and is `scanned' by the TM one square at a
time. The TM is `aware' of only one square at any one time.

\item  Scanning the symbol in the square may cause the internal state of the
TM to change.

\item  Define the Configuration of the TM as $(S_{r},i),$ where $S_{r}$ is
the symbol in the $r^{th}$ square, and $i$ is the current internal state of
the machine. The Configuration determines the behaviour of the machine.

\item  The machine may erase, amend or do nothing to the symbol, and may
move the tape one square to the left or right, according to the rules
specified by its current configuration.\bigskip
\end{enumerate}

A simple TM can perform all possible computations using just one of two
possible symbols in each `square'; this includes modern computers which run
on binary logic based on microelectronic components that are either switched
`on' or `off'. Thus, each square has a value corresponding to a binary
digit, and so may be called an individual\textit{\ bit}. Labelling these two
possible values $0$ and $1,$ it is easy to show that any input number can be
represented by a string of these bits according to the rules of binary
mathematics, as demonstrated later.

Additionally, special sequences of $0$'s and $1$'s can also be implemented
to incorporate necessary functions or instructions, for example a code to
inform the TM that the input string has ended, and computation can
begin.\bigskip

The m-configurations contain all of the possible instructions required
during a computation. Of course, \textit{which }instructions are used
depends on the actual calculation to be performed. Four of the eight
simplest instructions that might be used are:
\begin{equation}
a\rightarrow b:L
\end{equation}
where $a,b=0$ or $1.$ In words, this command implies: ``If the current
symbol is $a$ then amend it to $b$ and move the tape one square to the
left''. The complement four instructions, $`a\rightarrow b:R$', would move
the tape to the right.

The next simplest set of instructions are of the form
\begin{equation}
(c)\hspace{0.5in}a\rightarrow b:L
\end{equation}
with $a,b,c=0,1$ and implying: ``If the current symbol is $a$ and the last
encountered symbol was $c$ then amend $a$ to $b$ and move the tape one
square to the left''. The complexity of the instructions can of course be
extended in an obvious way by considering the last $2,3,...,n$ encountered
symbols.\bigskip

A computation is conventionally taken to begin with the Input string (input
number + any instructions) on the far left of the tape. The schematics of
this basic computation can now be written as a generic algorithm:

\begin{enumerate}
\item  The TM has a finite number of m-configurations which determine the
nature of the computation. It is initially in a particular state.

\item  The Input is read (from the left) until a certain sequence of symbols
is encountered. This signals the end of the Input, and the actual
calculating can begin.

\item  The TM scans a particular square with the algorithm: If the current
symbol is $a$ and the last $N$ symbols were $ABC...,$ then:

\begin{itemize}
\item  Amend the symbol to $b,$

\item  Move the tape one step to the left or right,

\item  Change the internal state of the TM from m-configuration $X$ to
m-configuration $Y.$
\end{itemize}
\end{enumerate}

Exactly what action the symbols $a,$ $A,$ $B,$ $C,...$ produce is determined
by the particular m-configuration at that time.

\begin{enumerate}
\item[4.]  Step 3 is looped until symbol $z$ is encountered when the last $%
N^{\prime }$ symbols were $A^{\prime }B^{\prime }C^{\prime }...$ and the TM\
is in m-configuration $Z.$ At this point the computation is halted.

\item[5.]  The `answer' to the problem is encoded as the remaining symbols
on the tape.
\end{enumerate}

It is possible that step 4 may never be encountered, for example if the
internal state $Z$ never arises; in this case, the computation effectively
loops forever without producing an answer. In fact, it is a central problem
in computer science to determine whether a given computation will ever yield
an output or will run on indefinitely. There are many examples of this
`Halting problem', G\"{o}del's Incompleteness Theorem \cite{Godel} being a
famous case.\bigskip

Note that it is additionally possible to encode into the Input (as a series
of $0$'s and $1$'s) the rules telling the TM which m-configurations to use.
This gives a binary representation of all the m-configurations used in a
particular TM calculation, and by transforming this binary number into the
decimal number $n$ it is possible to label the TM as the $n^{th}$-Turing
Machine.

This idea can be extended to the concept of a Universal Turing Machine
(UTM), defined as a TM which has all possible m-configurations inbuilt. Any
particular TM, i.e. any specific computation, can be simulated on this
Universal machine simply by supplying the number $n,$ because this
consequently `informs' the UTM which m-configurations are relevant. A modern
PC is effectively a Universal Turing Machine.

\bigskip

Generally speaking, computations involve mathematics, and mathematics
involves numbers. It is therefore necessary to be able to explicitly encode
numerical concepts into the Turing machines if they are to be useful. This
can be achieved by recalling that any non-negative integer $z<2^{r+1}$ may
be represented in binary notation by the $(r+1)$ bit string
\begin{equation}
z=z_{r}z_{r-1}...z_{2}z_{1}z_{0}
\end{equation}
which is shorthand for
\begin{equation}
z=(z_{r})2^{r}+(z_{r-1})2^{r-1}+...+(z_{2})2^{2}+(z_{1})2^{1}+(z_{0})2^{0}
\end{equation}
where $z_{i}=0,1$ \ for $i=0,1,...,r.$ This obviously extends in a natural
way to incorporate, for example, negative integers where $(z_{t})<0$ for all
$t,$ but this is not important here.

As an illustration, by using this binary notation the results: $2=10,$ $%
3=11, $ $4=100,$ $9=1001,$ $23=10111$ etc. are readily obtained. The
important point for the present discussion on computation is that by
employing this method every number $z$ may be uniquely defined by a string
of $0$'s and $1$'s, exactly as required for the classical Turing machine.
Any number $z<2^{r+1}$ can consequently be written as a sequence of\ $(r+1)$
bits, each of which has a definite value.

In fact, it is also possible to cleverly incorporate instructions for
mathematical operations as particular combinations of bit values \cite
{Penrose}. A full discussion of how this may be achieved, however, is beyond
the scope of this short introduction.\bigskip

It is possible now to provide a simple example of how an idealised Turing
machine might actually perform a calculation. For instance, consider the sum
$1+3=4;$ in binary notation, this equation is equivalent to the relation
\begin{equation}
...00001+...00011=...00100
\end{equation}

By noting that the right-most position in a string may be labelled the
`first' bit (that is, perhaps even more confusingly, $r=0),$ a computer is
able to generate the output string according to the following algorithm:

\begin{enumerate}
\item  The computer is initially provided with the input. In this case, the
input takes the form of information regarding the number `$1$' (i.e. the
string $...00001),$ some sort of code telling the machine that an addition
is required, and a number `$3$' (indicated by another string $...00011).$

\item  The value of the first bit of the output is given by the modulo two
addition of the value of the first bit, $1,$ of the first string $(...00001)$
of the input to the value of the first bit, $1,$ of the second string $%
(...00011)$ of the input. So, in this case the value of the first bit of the
output is $0,$ because $1\oplus 1=0.$

\item  This first addition leaves a `carry' of $1.$ The value of the second
bit of the output is then given by the modulo two addition of the value of
the second bit, $0,$ of the first string of the input, to the value of the
second bit, $1,$ of the second string of the input, followed by the modulo
two addition of this result to the remainder (i.e. the `carry'), $1,$ of the
previous addition used to generate the value of the first bit (i.e. ``Step
2''.). So, in this case the value of the second bit of the output is $0,$
because $(0\oplus 1)\oplus 1=0.$

\item  The procedure is repeated for all of the bits in the string. In
general, the value of the $j^{th}$ bit of the output is given by the modulo
two addition of the values of the $j^{th}$ bits of the two strings $%
(...00001 $ and $...00011)$ of the input, added to the remainder of the
result of the modulo two addition of the values of the $(j-1)^{th}$ bits of
these two strings and the `carry' from the determination of the output's $%
(j-2)^{th}$ bit.
\end{enumerate}

Thus, given an input incorporating the number `$1$' $(\equiv ...00001),$ a
code telling the machine to do addition, and another number `$3$' $(\equiv
...00011),$ the computer outputs the number $4$ (represented by the string $%
...00100).$\bigskip

The actual `mechanics' of the above type of computation may be performed
using particular logic gates to manipulate the values of the bits of a
string in order to produce an output \cite{Feynman2}. Consider the AND, the
OR and the Exclusive-OR (XOR) operations that obey the truth table
\begin{equation}
\begin{tabular}{|l|l|l|l|l|}
\hline
$\mathbf{x}$ & $\mathbf{0}$ & $\mathbf{0}$ & $\mathbf{1}$ & $\mathbf{1}$ \\
\hline
$\mathbf{y}$ & $\mathbf{0}$ & $\mathbf{1}$ & $\mathbf{0}$ & $\mathbf{1}$ \\
\hline
$\mathbf{x}$\textbf{\ }$\mathbf{AND}$\textbf{\ }$\mathbf{y}$ & $0$ & $0$ & $%
0 $ & $1$ \\ \hline
$\mathbf{x}$\textbf{\ }$\mathbf{OR}$\textbf{\ }$\mathbf{y}$ & $0$ & $1$ & $1$
& $1$ \\ \hline
$\mathbf{x}$\textbf{\ }$\mathbf{XOR}$\textbf{\ }$\mathbf{y}$ & $0$ & $1$ & $%
1 $ & $0$ \\ \hline
\end{tabular}
\tag*{Table A-1}
\end{equation}

Given an input that incorporates the two $(r+1)$-bit strings of value $%
x_{r}x_{r-1}...x_{2}x_{1}x_{0}$ and $y_{r}y_{r-1}...y_{2}y_{1}y_{0},$ the
result of the sum $%
(x_{r}x_{r-1}...x_{2}x_{1}x_{0}+y_{r}y_{r-1}...y_{2}y_{1}y_{0})$ is expected
to be a string of the form $z_{r+1}z_{r}z_{r-1}...z_{2}z_{1}z_{0},$ where $%
x,y,z=0,1,$ noting that the additional bit $z_{r+1}$ may be required in the
output to cope with a potential `carry' from the addition of $x_{r}$ and $%
y_{r}.$

Now, it is evident that the calculation of the value $z_{0}$ of the $1^{st}$
bit of the output is given by
\begin{equation}
z_{0}=(x_{0}\text{ }XOR\text{ }y_{0}).
\end{equation}

Then, the calculation of the value $z_{1}$ of the $2^{nd}$ bit of the output
may be given by
\begin{equation}
z_{1}=(x_{1}\text{ }XOR\text{ }y_{1})\text{ }XOR\text{ }(c_{0}),
\end{equation}
where $c_{0}$ is the `carry' from the sum of the values of the first bits of
the input, such that $c_{0}$ is clearly $c_{0}=x_{0}$ $AND$ $y_{0}.$

In general, the value $z_{j}$ of the $j^{th}$ bit of the output is
\begin{equation}
z_{j}=(x_{j}\text{ }XOR\text{ }y_{j})\text{ }XOR\text{ }(c_{j-1})
\end{equation}
where the carry $c_{j-1}$ from the earlier calculation of $z_{j-1}$ is given
by the recursive formula
\begin{equation}
c_{j-1}=(x_{j-1}\text{ }AND\text{ }y_{j-1})\text{ }OR\text{ }\left[ (x_{j-1}%
\text{ }OR\text{ }y_{j-1})\text{ }AND\text{ }c_{j-2}\right] .
\end{equation}

Of course, computers are not simply limited to addition, and alternative
calculations can be achieved by using different combinations of logic
gates.\bigskip

Another function useful in computation is the NOT-AND gate (NAND), which
produces the truth table
\begin{equation}
\begin{tabular}{|l|l|l|l|l|}
\hline
$\mathbf{x}$ & $\mathbf{0}$ & $\mathbf{0}$ & $\mathbf{1}$ & $\mathbf{1}$ \\
\hline
$\mathbf{y}$ & $\mathbf{0}$ & $\mathbf{1}$ & $\mathbf{0}$ & $\mathbf{1}$ \\
\hline
$\mathbf{x}$\textbf{\ }$\mathbf{NAND}$\textbf{\ }$\mathbf{y}$ & $1$ & $1$ & $%
1$ & $0$ \\ \hline
\end{tabular}
\tag*{Table A-2}
\end{equation}

This operation is particularly important because it is a standard result
\cite{Feynman2} that all classical computations can be performed just by
using combinations of NAND gates. Thus, the NAND gate is said to be
`complete', and can consequently form the basis for a truly universal Turing
machine.

Moreover it is easy to physically build a NAND gate, for example by
connecting a set of transistors in a certain way. Thus, by incorporating the
principle that low and high voltages may be used to represent the off/on
states associated with the bit values $0$ and $1,$ it is possible to
construct an electronic device whose output is related to its input
potentials according to the logic of the NAND truth table. In fact, a modern
personal computer is effectively just a ``black-box'' containing many such
devices. Thus by representing the input string of bits of value $0$ or $1$
as a set of low and high voltages, and by instructing the computer as to
which groups of NAND gates are to be used in which order and on which bits
for a particular desired calculation, the values of an input sequence of
electrical pulses may be used to generate a digital output sequence of $0$'s
and $1$'s. From this starting point any classical computation is
theoretically possible.

\bigskip

As noted previously, a Turing Machine can act on any computable number.
Computable numbers, however, are only a subset of the field of real numbers,
because Real analysis contains non-computable quantities, that is, those
which do not possess a sequence that is Cauchy convergent and hence those
for which no level of approximation can be used to specify them to an
acceptable degree. As an example, non-computability is exhibited in
non-recursive sets, for instance the Mandlebrot fractal\footnote{%
Strictly, the Mandlebrot set is based on complex numbers. This is
unimportant for the present discussion since moduli may be taken, and it is
the fact that it is irrational and non-Cauchy convergent that is of issue.}
(described in \cite{Penrose} and others).

For both mathematical and computational completeness, and maybe even for
physics as well, it is necessary to extend the encompass of computations
operating over the set of computable numbers to that of `Real Computations'
that also operate over the set of non-computable numbers. This extension has
recently been analysed \cite{Blum} with an algorithm found in polynomial
time, but is beyond the scope of this short introduction. It is, however,
encouraging to note the completeness of mathematics in classical computation.

\bigskip

\subsection{Quantum Computation}

\bigskip

A number of authors have given accounts of how quantum computation may work,
though much of the original idea is accredited to Feynman \cite{Feynman1}.
Gram\ss\ \textit{et al }have written a good introductory text \cite{Gramss},
from which much of this section of Appendix A is based.

A quantum computer has the same general structure as a classical computer:
there is an Output which is the result of some computation on an initial
Input. The Input of a quantum computer, however, is not a classical series
of bits but a wavefunction represented at time $t=0$ by $\psi (0).$ This
wavefunction is dynamically evolved during the computation into an Output
wavefunction $\psi (T),$ which represents the state of the system at some
later time $t=T.$ The actual evolution is governed by an operator $\hat{U},$
and this determines the type of computation to be performed.\bigskip

\smallskip The information contained in the state $\psi (t)$ may be encoded
in a way analogous to a classical computer that incorporates bits of value $%
0 $ and $1.$ Each quantum bit, or `qubit', $q$ is a component of $\psi (t)$
and is contained in a two dimensional Hilbert space spanned by an
orthonormal basis set of vectors conventionally represented by $|0\rangle $
and $|1\rangle .$ However, unlike the bits of a classical computer which can
only take the values $0$ or $1,$ the state of a qubit can exist as a linear
superposition of the form $\alpha |0\rangle +\beta |1\rangle ,$ for $\alpha
,\beta \in \mathbb{C}.$

In practical terms, qubits could be physically associated with the two
orthogonal eigenstates of a `binary' quantum system. Traditionally, the
qubit is identified with the eigenstates of a spin-$1/2$ particle, where
perhaps $|down\rangle $ represents $|0\rangle $ and $|up\rangle $ represents
$|1\rangle ,$ but of course the ground and first excited states of any other
two-level system could equally well be used, as could, for example, left and
right handed photonic polarisation states. The two states can also be
identified with the column matrices
\begin{equation}
|0\rangle =\binom{1}{0}\text{ \ \ and \ \ }|1\rangle =\binom{0}{1}.
\end{equation}

The most general state $\psi (t)$ of the quantum computer may be represented
in the usual way by a vector in\ a Hilbert space $\mathcal{H}.$ However,
since a desire will be to retain the binary logic common to both classical
and quantum bits, attention is restricted to Hilbert spaces of dimension $%
2^{N},$ where $N$ is the number of qubits chosen to comprise the system.
Thus, $\mathcal{H}$ may be written $\mathcal{H}\equiv \mathcal{H}_{[1...N]},$
where $\mathcal{H}_{m}$ is the two-dimensional subregister containing the $%
m^{th}$ qubit, for $m=1,2,...,N.$ The state $\psi (t)$ is now taken to be an
arbitrary vector in $\mathcal{H}_{[1...N]}$ with all the separability and
entanglement properties familiar to quantum theory and discussed elsewhere
in this work.

This general vector approach may be usefully simplified and constrained in
order to draw further parallels with the classical computers described
previously. For example, just as the state of a classical Turing Machine is
given by a string of classical bits, the wavefunction of the quantum
computer could be the tensor product of $N$ qubit factor sub-states; such a
product may also be called a `string'. Continuing the analogy, the classical
symbol $0$ or $1$ in the $m^{th}$ square of the Input section of the TM's
tape may be seen as related to the particular spin eigenstate of the $m^{th}$
qubit.

Thus a classical $N$ bit string $a_{N}a_{N-1}...a_{1},$ where $a_{m}=0$ or $%
1 $ for $m=1,2,...,N,$ is \textit{analogous} (in some sense) to an $N$ qubit
product represented by a wavefunction $\psi $ defined as $|\psi \rangle
=|a_{1}\rangle \otimes |a_{2}\rangle \otimes ...\otimes |a_{N}\rangle ,$
with the important difference being that in the quantum case each of these
sub-states $|a_{m}\rangle $ may exist as a linear superposition of their
bases.

So, $|a_{m}\rangle =\alpha _{m}|0\rangle _{m}+\beta _{m}|1\rangle _{m}$ for $%
m=1,2,...,N$ and $\alpha _{m},\beta _{m}\in \mathbb{C},$ which clearly gives

\begin{equation}
|\psi \rangle =\left[ \alpha _{1}\binom{1}{0}_{1}+\beta _{1}\binom{0}{1}_{1}%
\right] \otimes \left[ \alpha _{2}\binom{1}{0}_{2}+\beta _{2}\binom{0}{1}_{2}%
\right] \otimes ...\otimes \left[ \alpha _{N}\binom{1}{0}_{N}+\beta _{N}%
\binom{0}{1}_{N}\right]
\end{equation}
noting how the left-right ordering of the state has been reversed between
the classical $(a_{N}\rightarrow a_{1})$ and quantum $(a_{1}\rightarrow
a_{N})$ cases. As throughout this thesis, tensor product symbols may be
emitted for brevity, with the position being used instead as the marker of
distinction. For example, the state $|110...1\rangle $ will be taken to
imply $|1\rangle _{1}\otimes |1\rangle _{2}\otimes |0\rangle _{3}\otimes
...\otimes |1\rangle _{N}$ etc.

If a string of qubit sub-states may be written as a product that involves no
quantum superposition, it may be seen to directly represent a classical
input string. Of course, one way of achieving this would be if either $%
\alpha _{m}$ or $\beta _{m}$ is zero for each $m.$ In such cases, products
of qubits may also be associated with unique numbers according to the rules
of binary mathematics, just as with the strings of classical bits. Thus, in
the instance where either $\alpha _{m}$ or $\beta _{m}$ is zero for each $m,$
the state $|a_{1}a_{2}...a_{N}\rangle $ would be classically equivalent to
the string $a_{N}a_{N-1}...a_{1},$ and may hence be labelled by the number $%
a_{N}2^{N-1}+a_{N-1}2^{N-2}+...+a_{1}2^{0}.$ For instance, the product
quantum state $|101\rangle $ is equivalent to the classical string $101,$
and hence represents the binary number $5.$

This idea may be extended in a way that will become important later. If the
individual products of qubit sub-states may be superposed, the overall
wavefunction may then be interpreted as representing a superposition of
numbers. As an illustrative example, the superposed state
\begin{equation}
|101+111\rangle =|101\rangle +|111\rangle
\end{equation}
which may be thought of as a superposition of the quantum strings $%
|101\rangle $ and $|111\rangle $ and is equivalent to the single qubit
superposition
\begin{equation}
|1\rangle _{1}\otimes \lbrack |0\rangle _{2}+|1\rangle _{2}]\otimes
|1\rangle _{3}
\end{equation}
is analogous to a quantum superposition of the classical strings $101$ and $%
111,$ and consequently represents a superposition of the numbers $5$ and $7.$
Note, however, that the actual superposed state $|1(0+1)1\rangle $ has no
classical equivalent itself, because classical physics does not support
superpositions. Reversing this statement: there is no \textit{single} string
of classical bits $a_{N}a_{N-1}...a_{1}$ that has the quantum equivalent $%
|101+111\rangle .$\bigskip

The evolution of the state is governed by an operator $\hat{U}.$ For a
useful quantum computation, this operator must be: $^{a)}$ Reversible; $%
^{b)} $ Universal, so that all computations can be performed (c.f. NAND in
classical logic). The first condition is important because it implies the
existence of the inverse operator $\hat{U}^{-1},$ ensuring that the operator
is unitary as required for the Schr\"{o}dinger evolution of a state.
Furthermore, the one-to-one mapping that then arises from the reversibility
of $\hat{U}$ implies that a given Output state is the result of a unique
Input state.

The operator $\hat{U}$ could be seen as a type of logic gate, for example
the Fredkin-Toffoli gate \cite{Fredkin} $\hat{U}_{FT}.$ An Input
wavefunction $|\psi (0)\rangle $ would be evolved by such a gate into an
output wavefunction, such that, for example, the state $|\psi (1)\rangle $
after one `application' is given by:
\begin{equation}
\hat{U}_{FT}|\psi (0)\rangle =\text{ }|\psi (1)\rangle .
\end{equation}

In reality, the desired form of the unitary operator $\hat{U}$ is achieved
by carefully modifying the Hamiltonian used to determine the dynamical
evolution of the system. Exactly how this is accomplished is, therefore, an
important technical question. However whilst this might be the case, the
issue should really just be seen as a physical practicality that does not
alter the following theoretical discussion.

\bigskip

A Universal Quantum Turing Machine (i.e. Quantum Computer (QC)) is the
quantum version of the reversible classical Turing Machine. There are,
however, important differences between how the two `devices' work. In
general, for example, classical UTM's operate by performing a series of
computations (`Serial Computation'), i.e. by performing one step after
another, where the TM only `reads' and acts upon one particular bit at any
one time. Conversely, the power of quantum computation lies within `Quantum
Parallelism' (QP), as shown below. The Input wavefunction can exist as a
linear superposition of its qubit sub-states, so the quantum computer can in
principle act on more than one `string' of qubits at a time, where each
string has a different classical equivalent. For example, if the input state
was of the form $\psi (0)=|00\rangle +|11\rangle ,$ the computation could
act simultaneously on the strings $|00\rangle $ and $|11\rangle $ (with
classical equivalents $00$ and $11).$ The general idea is that the
computations of the strings (e.g. $|00\rangle $ and $|11\rangle )$ are
performed in parallel (i.e. as $|00\rangle +|11\rangle ),$ and then brought
together at the end to give a result in less time than would be the case if
the computations had been performed on each string serially (e.g. $%
|00\rangle $ followed by $|11\rangle ).$\bigskip

This power of QC can be illustrated by example. Consider some rule or
function $f(i)$ that, given an input integer $i,$ computes an output integer
$i^{\prime }$ (i.e. $f(i)=i^{\prime }:$ $i,i^{\prime }\in \mathbb{Z}^{\ast
}).$ Because any computable function can be constructed from reversible
logic gates, the function $f(i)$ is described by a particular unitary
operator $\hat{U}_{f}.$

Assume that the state of the system may be labelled by $\Psi .$ This state
must necessarily represent everything that is involved in the computation;
if the quantum computer is viewed as a `black box', the state $\Psi $ must
incorporate the part of the computer's memory that stores the input state,
the part used to perform the calculation, the part used to store the
outcome, etc.

So, in order to abbreviate the notation, consider the sub-state $\psi $ of $%
\Psi $ defined as $|\psi \rangle =$ $|i,j\rangle ,$ where $i$ is the state
of the input and $j$ the current state of the output. Moreover, if $i$ and $%
j $ are both assumed to be integers, they may naturally be written in binary
notation as a string of $0$'s and $1$'s, and so may readily be encoded as a
product of non-superposed qubits of given spin.

Consider now an initial state $\psi _{i}(0)$ defined as $|i,\underline{0}%
\rangle ,$ where $\underline{0}$ is a `ground state' string of $0$'s, i.e. $%
\underline{0}=$ $|000...0\rangle ,$ representing the obvious observation
that there is no output $j$ yet. If the function $f(i)$ is associated with
the computation $\hat{U}_{f},$ the evolution of the initial state $\psi
_{i}(0)$ to the final state $\psi _{i}(f)$ is described by
\begin{equation}
\psi _{i}(f)=\hat{U}_{f}\psi _{i}(0)=\hat{U}_{f}|i,\underline{0}\text{ }%
\rangle =|i,f(i)\rangle .
\end{equation}

Note that the number $i$ features in both the Input $\psi _{i}(0)$ and
Output $\psi _{i}(f)$ wavefunctions. This feature is a result of the
constraint that the evolution is unitary; if the input information had been
overwritten or `forgotten', reversibility would be violated.\bigskip

As required for quantum parallelism, it is desirable to write the input
state as a linear superposition of many alternative classical strings of
qubits. Supposing there are $n$ such possible strings, i.e. $i=1,2,...,n$
corresponding ultimately to the `binary' product states $\psi
_{1}(0)=|10...0,\underline{0}\rangle ,$ $\psi _{2}(0)=|01...0,\underline{0}%
\rangle ,...,$ $\psi _{n}(0)=|11...1,\underline{0}\rangle ,$ the superposed
Input state $|\psi $ $\rangle _{I}$ may be given by the sum
\begin{equation}
|\psi \rangle _{I}=\frac{1}{\sqrt{n}}\sum\limits_{i=1}^{n}\psi _{i}(0)=\text{
}\frac{1}{\sqrt{n}}\sum\limits_{i=1}^{n}|i,\underline{0}\rangle .
\end{equation}

The final state $|\psi \rangle _{F}$ is generated by evolving the input
state $|\psi \rangle _{I}$ with $\hat{U}_{f},$ i.e.
\begin{equation}
|\psi \rangle _{F}=\hat{U}_{f}|\psi \rangle _{I}=\hat{U}_{f}\text{ }\left(
\frac{1}{\sqrt{n}}\sum\limits_{i=1}^{n}|i,\underline{0}\rangle \right)
=\left( \frac{1}{\sqrt{n}}\sum\limits_{i=1}^{n}|i,f(i)\rangle \right) .
\end{equation}

Clearly, this final state $|\psi \rangle _{F}$ contains $n$ `solutions'
corresponding to the $n$ many $f(i)$ for $i=1,2,...,n.$ However, the
generation of $|\psi \rangle _{F}$ from $|\psi \rangle _{I}$ has been
achieved during one time step (evolution) of the calculation on only one
quantum computer, i.e. by one application of the gate $\hat{U}_{f}$ to the
Input state $|\psi \rangle _{I}.$\smallskip\ Conversely, if performed
serially on each of the $n$ states $\psi _{i}(0),$ it would take $n$ time
steps to produce $n$ results for $f(i).$\bigskip

There is, however, unfortunately an inherent problem here: it is not
possible to access more than one of these solutions. As soon as the
superposed Output $|\psi \rangle _{F}$ is observed its state vector
collapses to one of the eigenfunctions of whichever Hermitian operator was
used to measure it. From this perspective all that can be known about $|\psi
\rangle _{F}$ is that it collapses to, say, the eigenstate $|e\rangle $ with
relative probability $|\langle e|\psi \rangle _{F}|^{2}.$ Moreover, assuming
that the Hermitian operator is chosen such that its $n$ eigenvectors are the
`answer' states $|i,f(i)\rangle ,$ then the probability $|\langle e|\psi
\rangle _{F}|^{2}=|\langle e,f(e)|\psi \rangle _{F}|^{2}$ of obtaining the $%
e^{th}$ one of these is given by $1/n,$ with each outcome taken to be
equally likely.

Furthermore, once the state $|\psi \rangle _{F}$ has collapsed, any
additional measurements of the system with the same Hermitian operator
produce the same result. Thus, there is no way to retrieve any information
about any of the other $n-1$ parts of the superposition $|i,f(i)\rangle $ of
$|\psi \rangle _{F}$ for $i\neq e,$ and so the fact that all of this other
information is lost renders the QC described above as no more efficient than
a classical computer.

\bigskip

The problem may be rephrased by emphasising that the quantum state $|\psi
\rangle _{F}$ has been `asked a direct question', thereby forcing it into a
single eigenstate. To avoid this, a more stochastic approach needs to be
employed, where sets of questions are simultaneously posed and the results
are given in terms of the probabilities of ensembles of answers.

As it turns out, this procedure is incredibly difficult, and only a few such
possible solutions to this type of problem have been found where quantum
parallel computation can better classical serial computation. Examples are
Shor's algorithm for the fast (polynomial time) factorisation of a large
number into two primes \cite{Shor}, and the work by Deutsch and Jozsa \cite
{Deutsch2} described below.\bigskip

Deutsch and Jozsa's model (henceforth referred to as DJ) begins by
considering a function $f$ that maps a positive integer $z$ randomly to
either $0$ or $1,$ that is
\begin{equation}
f(z)=0\text{ or }1\text{ \ \ , \ \ }\forall \text{ }z\in \mathbb{Z}^{+}.
\end{equation}

Consider now a string of $n$ numbers $\{n\}=\left\{
\tsum\nolimits_{i=1}^{2N}i\right\} =\{1,2,...,2N\},$ where $n=2N$ is clearly
even. In DJ's model the computation $f$ acts on each of these numbers to
yield a bit string $x$ defined as
\begin{equation}
x=f(1)f(2)...f(2N)
\end{equation}
which is evidently a sequence $2N$ characters long of $0$'s and $1$'s that
will randomly take one of the $2^{2N}$ forms:
\begin{equation}
x=\{(000...0),\text{ }(100...0),\text{ }(010...0),\text{ }...\text{ },\text{
}(000...1),\text{ }(110...0),\text{ }(101...0),\text{ }.........\text{ },%
\text{ }(111...1)\}.
\end{equation}

Given an initial sequence $n,$ the thrust of DJ's task is then to find at
least one true statement about the resulting string $x$ from the following
two assertions:

\begin{enumerate}
\item  The string $x$ is neither just a string of $0$'s nor just a string of
$1$'s (i.e. $x$ is neither $000...0$ or $111...1).$ This is equivalent to
the statement that $f$ is not a constant function.

\item  The number of $0$'s in $x$ is not equal to the number of $1$'s in $x.$
In other words, the function $f$ acting on the $2N$ numbers $1,2,...,2N$
will not give exactly $N$ many $0$'s and $N$ many $1$'s.
\end{enumerate}

Clearly, for a string $x$ picked at random both statements are likely to be
true.\bigskip

A schematic algorithm for a classical computation of this sort could be to
compute $f(1),$ then compute $f(2),$ then compare the values of $f(1)$ and $%
f(2),$ then compute $f(3)$ before comparing its value to $f(1)$ and $f(2),$
then compute $f(4),$ and so on. Assuming that each computation takes one
time step to complete, and that the comparison procedure is effectively
instant, a worst case scenario for the efficiency of such a serial method of
testing the validity of statements $``1."$ and $``2."$ consequently takes $%
N+1$ steps: if the first $N$ bits all turn out to be $0$'s, and if the $%
N+1^{th}$ bit is another $0,$ it implies that Assertion $``2."$ is true,
whereas if the $N+1^{th}$ bit is alternatively a $1$ it follows that
Assertion $``1."$ must be true (and similarly, of course, if the first $N$
bits are all $1$'s). In other words, for a serial classical computer the
computation $f$ may need to be called $N+1$ times before an answer can be
obtained to statements $``1."$ and $``2."$ for an initial sequence of $2N$
numbers.\bigskip

The question is: ``Can a quantum computer improve on this efficiency?''. Is
it possible to find a quantum method that appears to compute every number
simultaneously?

The quantum computation in DJ's proposed method makes use of three distinct
stages: preparation of the Input state; computation via dynamical evolution;
and measurement of the Output state.

In a classical computation, the string $n$ comprises of a `chain' of
(binary) numbers $1,2,3,...,2N,$ and the function $f(i)$ acts on each of
them in turn, i.e. serially. In quantum computation, however, the quantum
strings equivalently representing these $2N$ numbers may be linearly
superposed into a single state. Thus, the Input state $|\psi \rangle _{I}$
for the present calculation is taken to be a linear superposition of the $2N$
numbers $\left\{ \tsum\nolimits_{i=1}^{2N}i\right\} $ and may be written
\begin{equation}
|\psi \rangle _{I}=\frac{1}{\sqrt{2N}}\sum\limits_{i=1}^{2N}\psi _{i}(0)=%
\frac{1}{\sqrt{2N}}\sum\limits_{i=1}^{2N}|i,\underline{0}\rangle
\end{equation}
with the $|i=input,j=output\rangle $ defined as before and the $\underline{0}
$ implying a string of $0$'s. Of course, each $i$ is taken to represent a
binary number between $1$ and $2N,$ and is hence a string of non-superposed
qubits of definite value; this therefore requires at least $R$ qubits, where
$R$ is the smallest integer for which $2^{R}>2N.$

As an aside, note that in order to actually prepare the initial state $|\psi
\rangle _{I}$ it is necessary to consider the `pre-Input' state $|\psi
\rangle _{p}.$ This is taken to be $|\psi \rangle _{p}=|\underline{0},%
\underline{0}\rangle ,$ such that every qubit of that part of the quantum
computer allocated to store the input state $i$ is assumed to be in the
ground state $|0\rangle .$ Now, because any number $i$ can be represented in
binary notation by a unique string of $0$'s and $1$'s, every quantum `number
state' $|i\rangle $ is represented by a product of qubits, each of which is
definitely in the state $|0\rangle $ or $|1\rangle .$ To represent a
particular number it is therefore necessary to transform some of the state $%
|0\rangle $ qubits contained in the input product $\underline{0}$ into state
$|1\rangle $ qubits. Moreover, because the eventual Input state $|\psi
\rangle _{I}$ is defined to be a superposition of all of these different
combinations of product qubit states, this must also be taken into account.

One way of achieving this is therefore to use a suitable unitary operator $%
\hat{A}$ acting on $|\psi \rangle _{p}$ that evolves it into the
superposition state $|\psi \rangle _{I}.$ Thus,
\begin{equation}
|\psi \rangle _{I}=\hat{A}|\psi \rangle _{p}=\hat{A}|\underline{0},%
\underline{0}\rangle
\end{equation}
where $\hat{A}$ may act locally on the individual qubit spaces.\bigskip

The actual quantum computation makes use of two operators: an evolution
operator $\hat{U}_{f}$ that evolves a state in the manner
\begin{equation}
\hat{U}_{f}\psi _{i}(0)=\hat{U}_{f}|i,\underline{0}\text{ }\rangle
=|i,f(i)\rangle
\end{equation}
where $f(i)=0$ or $1,$ and a `parity' operator $\hat{S}$ defined as
\begin{equation}
\hat{S}|i,j\rangle =(-1)^{j}|i,j\rangle .
\end{equation}

The computation is achieved by evolving the state, performing a parity
operation, and then applying the inverse operator $\hat{U}_{f}^{-1}$ to
obtain an Output wavefunction $|\psi \rangle _{F}.$ Specifically,
\begin{eqnarray}
|\psi \rangle _{F} &=&\hat{U}_{f}^{-1}\hat{S}\hat{U}_{f}|\psi \rangle _{I}=%
\hat{U}_{f}^{-1}\hat{S}\hat{U}_{f}\left( \frac{1}{\sqrt{2N}}%
\sum\limits_{i=1}^{2N}|i,\underline{0}\rangle \right) \\
&=&\hat{U}_{f}^{-1}\hat{S}\left( \frac{1}{\sqrt{2N}}\sum%
\limits_{i=1}^{2N}|i,f(i)\rangle \right) =\hat{U}_{f}^{-1}\left( \frac{1}{%
\sqrt{2N}}\sum\limits_{i=1}^{2N}(-1)^{f(i)}|i,f(i)\rangle \right)  \notag \\
&=&\left( \frac{1}{\sqrt{2N}}\sum\limits_{i=1}^{2N}(-1)^{f(i)}|i,\underline{0%
}\rangle \right) .  \notag
\end{eqnarray}

Consequently, the $2N$ results of the computation $f(i)$ are stored as phase
information in the $(-1)^{f(i)}$ part of the Output state $|\psi \rangle
_{F}.$\bigskip

Measurement of the Output wavefunction $|\psi \rangle _{F}$ is achieved by a
Hermitian operator that possesses $\phi $ as an eigenvector, where $\phi
=|\psi \rangle _{I}$ is the initial state. The probability of recording the
eigenvalue associated with this is given in the usual way by $P,$ where $%
P=|\langle \phi |\psi \rangle _{F}|^{2}.$ So
\begin{equation}
P=\left| \frac{1}{2N}\sum\limits_{i=1}^{2N}\sum\limits_{j=1}^{2N}(-1)^{f(i)}%
\langle j,\underline{0}|i,\underline{0}\rangle \right| ^{2},
\end{equation}
and assuming orthogonality of the states, $\langle j,\underline{0}|i,%
\underline{0}\rangle =\delta _{ij},$ gives
\begin{equation}
P=\left| \frac{1}{2N}\sum\limits_{i=1}^{2N}(-1)^{f(i)}\right| ^{2}.
\end{equation}

\smallskip

The statements $``1."$ and $``2."$ can be answered by examining $P.$ Three
distinct cases are present:

\begin{itemize}
\item[i)]  If $P=0$ then the sum must have vanished. This implies that $f(i)$
has produced as many $0$'s as it has $1$'s and consequently Assertion $``1."$
must be true (and $``2."$ must be false).

\item[ii)]  If $P=1$ then all of the $f(i)$'s must be either $0$ or $1$ and
hence Assertion $``2."$ is true (and $``1."$ is false).

\item[iii)]  If $0<P<1$ then $f(i)$ has produced an unequal number of $0$'s
and $1$'s, but has produced at least one of each. Both Assertions $``1."$
and $``2."$ must be true.\bigskip
\end{itemize}

The important point is that the time taken for $\hat{U}_{f}$ to act is
assumed to be the same as the time needed to perform just one computation $%
f(i)$ on a single number $i$ in the classical (serial) case. If $\hat{U}%
_{f}^{-1}$ is assumed to take one time step also, and if $\hat{S}$ is taken
to act comparatively `instantly', the entire quantum computation has
proceeded in just two time steps. Thus the quantum computation is performing
$2N$ computations in parallel in only two time steps. Moreover, the testing
of the validity of statements $``1."$ and $``2."$ has also been achieved in
just two time steps, which compares with a serial, classical computer taking
(at worst) $N+1$ time steps to arrive at the same conclusion.\bigskip

So, quantum computers clearly have an enormous advantage over their
classical counterparts in certain specifically defined computations.
Unfortunately, of course, they also have the even greater disadvantage that
they cannot (currently?) actually be built: the effects of their external
surroundings destroy the superposition of the evolving state before any
significant computation can take place.

Whether or not this technological difficulty will ever be overcome is a
question for the future. However, even if the `decohering' presence of an
environment fundamentally prohibits the construction of a working quantum
computer \textit{inside} the Universe, it does not prevent the principles of
quantum computation being applied to the Universe as a whole, as proposed in
this work. After all, the physical Universe has no external environment to
interfere with it.

\smallskip \bigskip

As relativity is an extension of Newtonian mechanics, and quantum field
theory an extension of quantum mechanics, we might expect the extension of
quantum computation into relativistic quantum field computation. After all,
quantum computation proceeds as the time evolution of an initial state
(generally comprising of products of qubits located at definite sites),
where time is treated as a continuous variable. Lorentz invariance, however,
expects space and time to be interchangeable, so a covariant case of quantum
computation might ultimately be sought: quantum field computation.

Quantum field computation would be a new branch of computation drawing from
both Real and Quantum analogues. At this stage, very little is clear about
exactly how a QFC could work, and authors mainly describe it as a necessary
new direction instead of as a well understood procedure with defined
mathematical structure \cite{Manoharan}\cite{Freedman}\cite{Patel}. The
actual preparation of input states, and the eventual defining of the system
via information encoded into Lagrangian formulations, are both interesting
considerations for the future.

The primary difference between quantum computation and quantum field
computation is that whilst QC permits superpositions of qubits, QFC allows
superpositions of entire fields. So, whereas in quantum computations an
Input exists as linear superpositions of `classical' $n$ qubit strings in a $%
Z^{n}$ dimensional Hilbert space (where $Z$ is the number of states per qubit%
\footnote{%
Strictly speaking, ``qubit'' is an acronym for QUantum Binary digIT, so $Z$
can only ever equal $2,$ but it is in principle possible to build quantum
computers out of suitably named qutrits, ququads, ..., quzits (?),
represented by $3,4,...,Z$ level systems.}), each field in a quantum field
computation would possess an infinite number of degrees of freedom, so the
computation would take place in an infinite dimensional Hilbert space.
Additionally, whereas the results of a quantum computation may be exhibited
as single eigenstates, the output of a quantum field computation might be
given in terms of expectation values of field operators.\bigskip

Quantum field computation is an extension from real computation in that it
includes computation over the continuum. This extension naturally increases
the computational power of the system at the cost of an increasingly complex
mathematical formalism.

Exactly what this might imply for a Universe running as a quantum computer
is an intriguing question for the future.

\bigskip \newpage

\section{The Dirac Field}

\renewcommand{\theequation}{B-\arabic{equation}} \setcounter{equation}{0} %
\renewcommand{\thetheorem}{B-\arabic{theorem}} \setcounter{theorem}{0}

\bigskip

In this appendix, the standard Hamiltonian, momentum and charge operator
representations are derived for spin-$%
%TCIMACRO{\UNICODE[m]{0xbd}}%
%BeginExpansion
{\frac12}%
%EndExpansion
$ fermions. The presented approach follows closely the treatment given in
the text of Mandl and Shaw \cite{Mandl}.

\bigskip

\subsection{Lagrangian Dynamics}

\bigskip

The Dirac equation of motion for free particles of rest mass $m$ is
conventionally given by
\begin{equation}
i\hbar \gamma ^{\mu }\partial _{\mu }\psi (x)-mc\psi (x)=0
\end{equation}
where\ $c$ is the speed of light, $\partial _{\mu }\equiv \frac{\partial }{%
\partial x^{\mu }}$ for $\mu =0,1,2,3$ and $x^{\mu }\equiv (ct,\mathbf{x}%
)\equiv (ct,x^{j})$ for $j=1,2,3,$ and $\gamma ^{\mu }$ are $4\times 4$
matrices satisfying the anti-commutation relations
\begin{equation}
\{\gamma ^{\mu },\gamma ^{\nu }\}=2g^{\mu \nu }
\end{equation}
and the Hermiticity conditions $\gamma ^{0\dagger }=\gamma ^{0}$ and $\gamma
^{j\dagger }=-\gamma ^{j},$ so that
\begin{equation}
\gamma ^{\mu \dagger }=\gamma ^{0}\gamma ^{\mu }\gamma ^{0}
\end{equation}
with $g^{\mu \nu }$ the metric tensor of signature $(+,-,-,-)$ such that $%
x_{\mu }=(ct,-\mathbf{x});$ the gamma matrices form a Clifford algebra \cite
{Zee}.

The `adjoint' field $\bar{\psi}(x)$ is defined as\ $\bar{\psi}(x)\equiv \psi
^{\dagger }(x)\gamma ^{0},$ and this satisfies the `adjoint' Dirac equation $%
i\hbar \partial _{\mu }\bar{\psi}(x)\gamma ^{\mu }+mc\bar{\psi}(x)=0.$ Note
that there is no standardised notation in the texts on quantum field theory,
and is often incongruent with `conventional' mathematics: the field $\psi
^{\dagger }(x)$ is taken here to represent the Hermitian (or transpose)
conjugate of the field $\psi (x),$ whereas in linear algebra such an
operator $\psi ^{\dagger }(x)$ would often be called the adjoint of $\psi
(x) $ and may instead be denoted by $\psi ^{\ast }(x).$

Both the Dirac equation and the adjoint Dirac equation can be derived from
the Euler-Lagrange condition, given the Dirac Lagrangian density $\mathit{L}$
defined\footnote{%
Note that for convenience in this appendix an asymmetric Lagrangian has been
used. As can be readily verified, however, a symmetrised version would lead
to the same results.} as
\begin{eqnarray}
\mathit{L} &=&c\bar{\psi}(x)\left[ i\hbar \gamma ^{\mu }\partial _{\mu }-mc%
\right] \psi (x) \\
&=&c\bar{\psi}(x)\left[ i\hbar \gamma ^{0}\frac{\partial }{\partial (ct)}%
+i\hbar \gamma ^{j}\frac{\partial }{\partial x^{j}}-mc\right] \psi (x)
\notag \\
&=&\bar{\psi}(x)\left[ i\hbar \gamma ^{0}\dot{\psi}(x)+i\hbar c\gamma
^{j}\partial _{j}\psi (x)-mc^{2}\psi (x)\right]  \notag
\end{eqnarray}
where the dot denotes differentiation with respect to time, $t.$ Consider
now the conjugate momenta to $\psi (x)$ and $\bar{\psi}(x),$ written as $\pi
(x)$ and $\bar{\pi}(x)$ respectively. These are given by
\begin{eqnarray}
\pi (x) &\equiv &\frac{\partial \mathit{L}}{\partial \dot{\psi}(x)}=\bar{\psi%
}(x)i\hbar \gamma ^{0}=i\hbar \psi ^{\dagger }(x) \\
\bar{\pi}(x) &\equiv &\frac{\partial \mathit{L}}{\partial \overset{\cdot }{%
\bar{\psi}}(x)}=0.  \notag
\end{eqnarray}

Together with the fields, the conjugate momenta satisfy the canonical
anti-commutation algebra
\begin{eqnarray}
\{\psi (x),\pi (x^{\prime })\} &=&i\hbar \delta (x-x^{\prime })
\label{App2ConCOM} \\
\{\psi (x),\psi (x^{\prime })\} &=&\{\pi (x),\pi (x^{\prime })\}=0.  \notag
\end{eqnarray}

So, the Hamiltonian density $\mathit{\tilde{H}}(x)$ defined as
\begin{equation}
\mathit{\tilde{H}}(x)\equiv \pi (x)\dot{\psi}(x)+\bar{\pi}(x)\overset{\cdot
}{\bar{\psi}}(x)-\mathit{L}
\end{equation}
becomes
\begin{eqnarray}
\mathit{\tilde{H}}(x) &=&i\hbar \psi ^{\dagger }(x)\dot{\psi}(x)+0-\left[
i\hbar \psi ^{\dagger }(x)\dot{\psi}(x)+i\hbar c\bar{\psi}(x)\gamma
^{j}\partial _{j}\psi (x)-mc^{2}\bar{\psi}(x)\psi (x)\right]  \notag \\
&=&mc^{2}\bar{\psi}(x)\psi (x)-i\hbar c\bar{\psi}(x)\gamma ^{j}\partial
_{j}\psi (x)
\end{eqnarray}
producing the Hamiltonian $H$%
\begin{equation}
H=\int \bar{\psi}(x)[mc^{2}-i\hbar c\gamma ^{j}\partial _{j}]\psi (x)\text{ }%
d^{3}x.  \label{App2Ham}
\end{equation}

Similarly, the 3-momentum $\mathbf{P}$ defined as
\begin{equation}
\mathbf{P}\equiv -\int [\pi (x)\partial _{j}\psi (x)+\bar{\pi}(x)\partial
_{j}\bar{\psi}(x)]\text{ }d^{3}x
\end{equation}
becomes
\begin{equation}
\mathbf{P}=-\int [i\hbar \psi ^{\dagger }(x)\partial _{j}\psi (x)+0]\text{ }%
d^{3}x.  \label{App2Mom}
\end{equation}
such that the relativistic energy-momentum vector $P^{\mu }=(H/c,\mathbf{P})$
may be evaluated.

Lastly, from the quantity $Q$ defined as
\begin{equation}
Q\equiv -\frac{iq}{\hbar }\int [\pi (x)\psi (x)-\bar{\pi}(x)\bar{\psi}(x)]%
\text{ }d^{3}x
\end{equation}
for particles possessing `charge' of magnitude $q,$ it follows that
\begin{eqnarray}
Q &\equiv &-\frac{iq}{\hbar }\int [i\hbar \psi ^{\dagger }(x)\psi (x)-0]%
\text{ }d^{3}x  \label{App2Charge} \\
&=&q\int \psi ^{\dagger }(x)\psi (x)\text{ }d^{3}x.  \notag
\end{eqnarray}

The result (\ref{App2Charge}) for $Q$ is associated with the conserved
electric charge, and leads to an invariance of the Lagrangian density $%
\mathit{L}$ under a global phase transformation of the fields $\psi
(x)\rightarrow \psi (x)^{\prime }$ and $\psi ^{\dagger }(x)\rightarrow \psi
^{\dagger }(x)^{\prime }$ by $e^{-i\alpha q},$ where
\begin{eqnarray}
\psi (x) &\rightarrow &\psi (x)^{\prime }\equiv e^{-i\alpha q}\psi (x)\sim
(1-i\alpha q+...)\psi (x)  \label{App2Invar} \\
\psi ^{\dagger }(x) &\rightarrow &\psi ^{\dagger }(x)^{\prime }\equiv
e^{+i\alpha q}\psi ^{\dagger }(x)\sim (1+i\alpha q+...)\psi ^{\dagger }(x)
\notag
\end{eqnarray}
for small $\alpha \in \mathbb{R},$ such that
\begin{equation}
\mathit{L}=\mathit{L}(\psi ,\bar{\psi})=\mathit{L}(\psi ^{\prime },\bar{\psi}%
^{\prime })\equiv \mathit{L}([e^{-i\alpha q}\psi ],[e^{+i\alpha q}\bar{\psi}%
])
\end{equation}
as may be readily shown from the condition of invariance, $\delta \mathit{L}%
=0,$ for a Lagrangian density affected as $\mathit{L}\rightarrow \mathit{L}%
^{\prime }\equiv \mathit{L}+\delta \mathit{L}$ by a change of the fields (%
\ref{App2Invar}), where
\begin{equation}
\delta \mathit{L}\equiv \frac{\partial }{\partial x^{\mu }}\left( \frac{%
\partial \mathit{L}}{\partial \psi _{,\mu }}\delta \psi (x)\right)
\end{equation}
and by using the Dirac equation.

Specifically, the unitary transformation $U$ associated with this
unobservable phase change in the fields is given by
\begin{equation}
U=\exp (i\alpha Q)
\end{equation}
such that from Schr\"{o}dinger's equation
\begin{eqnarray}
\psi (x)^{\prime } &=&e^{i\alpha Q}\psi (x)e^{-i\alpha Q} \\
&=&\psi (x)+i\alpha \lbrack Q,\psi (x)]  \notag
\end{eqnarray}
which is seen to result in\ (\ref{App2Invar}) by using the equations
\begin{eqnarray}
\lbrack Q,\psi (x)] &=&-q\psi (x) \\
\lbrack Q,\psi ^{\dagger }(x)] &=&q\psi ^{\dagger }(x)  \notag
\end{eqnarray}
which themselves follow from the anti-commutation relations of on the field
\cite{Ticciati}.

The conserved nature of the charge $Q$ is thus shown from Noether's theorem
regarding the invariance of the Lagrangian density under a given continuous
transformation. Furthermore, the invariance of the dynamics under this
unitary transformation may be incorporated into Heisenberg's equation of
motion.

\bigskip

\subsection{Quantisation}

\bigskip

The system may now be quantised. Firstly, consider a\ large, cubic region of
space of volume $V$ containing the fields. For every periodic mode of
momentum\textbf{\ }$\mathbf{p}$ and positive energy $E_{p}$ given by
\begin{equation}
E_{p}=\sqrt{(\mathbf{p}\cdot \mathbf{p})c^{2}+m^{2}c^{4}}\text{ \ \ , \ \ }%
E_{p}\geq 0
\end{equation}
that is inside this `box' and is constrained to tend to zero at the
boundary, the Dirac equation has four independent, plane wave solutions
represented by
\begin{equation}
\psi _{r}^{+}(x)=K_{\mathbf{p}}u_{r}(\mathbf{p})\frac{e^{-ip\cdot x/\hbar }}{%
\sqrt{V}}\text{ \ \ and \ \ }\psi _{r}^{-}(x)=K_{\mathbf{p}}v_{r}(\mathbf{p})%
\frac{e^{+ip\cdot x/\hbar }}{\sqrt{V}}\text{ \ \ , \ \ }r=1,2
\label{App2Sol}
\end{equation}
where $r=1,2$ and $K_{\mathbf{p}}$ is a constant, with $p=p_{\mu }=(E_{p}/c,-%
\mathbf{p).}$ The $u_{r}(\mathbf{p})$ and $v_{r}(\mathbf{p})$ are constant,
four-component spinors, which, together with their adjoints $\bar{u}_{r}(%
\mathbf{p})\equiv u_{r}^{\dagger }(\mathbf{p})\gamma ^{0}$ and $\bar{v}_{r}(%
\mathbf{p})\equiv v_{r}^{\dagger }(\mathbf{p})\gamma ^{0},$ satisfy
\begin{eqnarray}
(\gamma ^{\mu }p_{\mu }-mc)u_{r}(\mathbf{p}) &=&0\text{ \ \ , \ \ }(\gamma
^{\mu }p_{\mu }+mc)v_{r}(\mathbf{p})=0 \\
\bar{u}_{r}(\mathbf{p})(\gamma ^{\mu }p_{\mu }-mc) &=&0\text{ \ \ , \ \ }%
\bar{v}_{r}(\mathbf{p})(\gamma ^{\mu }p_{\mu }+mc)=0  \notag
\end{eqnarray}
as may be verified by substituting the solutions (\ref{App2Sol}) into the
original Dirac equation.

The index $r=1,2$ labels two distinct solutions for each momentum $\mathbf{p}%
;$ choosing these to be orthogonal, the solutions are ultimately taken to
represent the two spin components required for a spin-$%
%TCIMACRO{\UNICODE[m]{0xbd}}%
%BeginExpansion
{\frac12}%
%EndExpansion
$ theory. Thus, the states containing $u_{r}(\mathbf{p})$ are interpreted as
corresponding to positive energy particles of momentum $\mathbf{p}$ (e.g.
spin-up and spin-down electrons), whereas the states containing $v_{r}(%
\mathbf{p})$ are interpreted as corresponding to negative energy particles
(e.g. spin-up and spin-down positrons). Note, however, that the negative
energy solutions are also traditionally taken to represent positive energy
anti-particles travelling backwards in time, but from the point of view
advocated in this thesis it is debateable as to whether this interpretation
really makes consistent sense.

The normalisation of the spinors is defined \cite{Schweber} as
\begin{equation}
u_{r}^{\dagger }(\mathbf{p})u_{r}(\mathbf{p})=v_{r}^{\dagger }(\mathbf{p}%
)v_{r}(\mathbf{p})=\frac{E_{p}}{mc^{2}}
\end{equation}
such that
\begin{eqnarray}
u_{r}^{\dagger }(\mathbf{p})u_{s}(\mathbf{p}) &=&v_{r}^{\dagger }(\mathbf{p}%
)v_{s}(\mathbf{p})=\frac{E_{p}}{mc^{2}}\delta _{rs}  \label{App2Spin} \\
u_{r}^{\dagger }(\mathbf{p})v_{s}(-\mathbf{p}) &=&0  \notag
\end{eqnarray}
and
\begin{eqnarray}
\bar{u}_{r}(\mathbf{p})u_{s}(\mathbf{p}) &=&-\bar{v}_{r}(\mathbf{p})v_{s}(%
\mathbf{p})=\delta _{rs} \\
\bar{u}_{r}(\mathbf{p})v_{s}(\mathbf{p}) &=&\bar{v}_{r}(\mathbf{p})u_{s}(%
\mathbf{p})=0.  \notag
\end{eqnarray}

Consider now the expansion of the field $\psi (x)$ into a complete set of
plane wave states, that is $\psi (x)=\psi ^{+}(x)+\psi ^{-}(x),$ where $\psi
^{+}(x)$ represents the solutions propagating forwards (i.e. the
`electrons') and $\psi ^{-}(x)$ the solutions propagating backwards (i.e.
the `positrons'). This expansion is given by
\begin{equation}
\psi (x)=\sum_{r=1}^{2}\sum_{\mathbf{p}=0}^{\infty }\left( \frac{mc^{2}}{%
E_{p}}\right) ^{1/2}\left[ c_{r}(\mathbf{p})u_{r}(\mathbf{p})\frac{%
e^{-ip\cdot x/\hbar }}{\sqrt{V}}+d_{r}^{\dagger }(\mathbf{p})v_{r}(\mathbf{p}%
)\frac{e^{+ip\cdot x/\hbar }}{\sqrt{V}}\right]
\end{equation}
where the sum is over all possible spin states, $r,$ and all possible
momenta, $\mathbf{p,}$ noting that this last point leads to some of the
divergence problems associated with quantum field theory. The variable%
\footnote{%
Noting the interchangeability of the notation between this appendix and
Chapter 7, in which sub-scripts were replaced by bracketed parameters for
clarity.} $c_{r}(\mathbf{p})=c(\mathbf{p},r)$ provides the amplitude of the $%
\mathbf{p}^{th}$ contribution of $u_{r}(\mathbf{p})$ of spin $r$ to $\psi
^{+}(x),$ whilst $d_{r}^{\dagger }(\mathbf{p})=d^{\dagger }(\mathbf{p},s)$
similarly provides the amplitude of the $\mathbf{p}^{th}$ contribution of $%
v_{r}(\mathbf{p})$ of spin $r$ to $\psi ^{-}(x),$ and both are scaled such
that the multiplying constant $\left( mc^{2}/E_{p}\right) ^{%
%TCIMACRO{\UNICODE[m]{0xbd}}%
%BeginExpansion
{\frac12}%
%EndExpansion
}$ is chosen for later convenience.

In a similar manner, the adjoint field $\bar{\psi}(x)$ may be expanded as
\begin{equation}
\bar{\psi}(x)=\sum_{r,\mathbf{p}}\left( \frac{mc^{2}}{VE_{p}}\right) ^{1/2}%
\left[ d_{r}(\mathbf{p})\bar{v}_{r}(\mathbf{p})e^{-ip\cdot x/\hbar
}+c_{r}^{\dagger }(\mathbf{p})\bar{u}_{r}(\mathbf{p})e^{+ip\cdot x/\hbar }%
\right] .
\end{equation}

Taking now the continuous limit \cite{Mandl}, the discrete sum over all
momenta in $V\rightarrow \infty $ may be replaced by an integral, such that
\begin{equation}
\psi (x)=\int_{-\infty }^{\infty }\frac{d^{3}p}{(2\pi \hbar )^{3/2}}%
\sum_{r=1}^{2}\left( \frac{mc^{2}}{E_{p}}\right) ^{1/2}\left[ c_{r}(\mathbf{p%
})u_{r}(\mathbf{p})e^{-ip\cdot x/\hbar }+d_{r}^{\dagger }(\mathbf{p})v_{r}(%
\mathbf{p})e^{+ip\cdot x/\hbar }\right]
\end{equation}
and similarly for $\bar{\psi}(x).$\bigskip

In the standard procedure of quantum field theory, quantisation is achieved
by directly quantising the individual harmonic oscillator modes of the
field. Thus the field amplitudes $c_{r}(\mathbf{p})$ and $d_{r}(\mathbf{p})$
are associated with operators, such that for example the operation $\hat{c}%
_{r}^{\dagger }(\mathbf{p})|0\rangle =|\mathbf{p}_{r}^{u}\rangle $ is taken
to imply the creation of a positive energy particle $|\mathbf{p}%
_{r}^{u}\rangle $ of `type' $u_{r}(\mathbf{p})$ with spin $r$ and momentum $%
\mathbf{p}$ from the vacuum $|0\rangle ,$ whilst $\hat{c}_{r}(\mathbf{p})$
is taken to result in the destruction of this state.\

Moreover, different particles can be created by applying different
operators, such that for example $\hat{c}_{r}^{\dagger }(\mathbf{p})\hat{c}%
_{s}^{\dagger }(\mathbf{p}^{\prime })|0\rangle =|\mathbf{p}_{r}^{u},\mathbf{p%
}_{s}^{\prime u}\rangle $ produces the two particle state $|\mathbf{p}%
_{r}^{u},\mathbf{p}_{s}^{\prime u}\rangle .$ Note that the vacuum is defined
as the lowest possible `occupation' of particles, such that $\hat{c}_{r}(%
\mathbf{p})|0\rangle =0.$

Care is needed, however, when applying such an interpretation, because the
spin-$%
%TCIMACRO{\UNICODE[m]{0xbd}}%
%BeginExpansion
{\frac12}%
%EndExpansion
$ particles of Dirac theory are physically observed to obey the statistics
of Fermi. Paraphrasing, this condition ensures that no two identical
particles can exist in the same state, so that $\hat{c}_{r}^{\dagger }(%
\mathbf{p})\hat{c}_{r}^{\dagger }(\mathbf{p})|0\rangle =0.$ This constraint
leads to the result that $[\hat{c}_{r}^{\dagger }(\mathbf{p})]^{2}=[\hat{d}%
_{r}^{\dagger }(\mathbf{p})]^{2}=0,$ which is ensured by assuming the
anticommutation relations for the operators:
\begin{eqnarray}
\{\hat{c}_{r}^{\dagger }(\mathbf{p}),\hat{c}_{s}^{\dagger }(\mathbf{p}%
^{\prime })\} &=&\{\hat{c}_{r}(\mathbf{p}),\hat{c}_{s}(\mathbf{p}^{\prime
})\}=0  \label{App2Anticom} \\
\{\hat{d}_{r}^{\dagger }(\mathbf{p}),\hat{d}_{s}^{\dagger }(\mathbf{p}%
^{\prime })\} &=&\{\hat{d}_{r}(\mathbf{p}),\hat{d}_{s}(\mathbf{p}^{\prime
})\}=0  \notag \\
\{\hat{c}_{r}(\mathbf{p}),\hat{d}_{s}(\mathbf{p}^{\prime })\} &=&\{\hat{c}%
_{r}(\mathbf{p}),\hat{d}_{s}^{\dagger }(\mathbf{p}^{\prime })\}=0  \notag \\
\{\hat{c}_{r}^{\dagger }(\mathbf{p}),\hat{d}_{s}(\mathbf{p}^{\prime })\}
&=&\{\hat{c}_{r}^{\dagger }(\mathbf{p}),\hat{d}_{s}^{\dagger }(\mathbf{p}%
^{\prime })\}=0  \notag
\end{eqnarray}
and
\begin{equation}
\{\hat{c}_{r}(\mathbf{p}),\hat{c}_{s}^{\dagger }(\mathbf{p}^{\prime })\}=\{%
\hat{d}_{r}(\mathbf{p}),\hat{d}_{s}^{\dagger }(\mathbf{p}^{\prime
})\}=\delta _{rs}\delta _{\mathbf{pp}^{\prime }}.  \label{App2Com}
\end{equation}

It is beneficial to define two further operators $\hat{N}_{r}(\mathbf{p})$
and $\overline{\hat{N}}_{r}(\mathbf{p})$ as
\begin{equation}
\hat{N}_{r}(\mathbf{p})=\hat{c}_{r}^{\dagger }(\mathbf{p})\hat{c}_{r}(%
\mathbf{p})\text{ \ \ },\text{ \ \ }\overline{\hat{N}}_{r}(\mathbf{p})=\hat{d%
}_{r}^{\dagger }(\mathbf{p})\hat{d}_{r}(\mathbf{p}).  \label{App2Num}
\end{equation}

$\hat{N}_{r}(\mathbf{p})$ and $\overline{\hat{N}}_{r}(\mathbf{p})$ are then
seen as analogous to the number operators of the conventional harmonic
oscillator, and have product rules evidently given by
\begin{eqnarray}
\lbrack \hat{N}_{r}(\mathbf{p})]^{2} &=&\hat{c}_{r}^{\dagger }(\mathbf{p})%
\hat{c}_{r}(\mathbf{p})\hat{c}_{r}^{\dagger }(\mathbf{p})\hat{c}_{r}(\mathbf{%
p}) \\
&=&\hat{c}_{r}^{\dagger }(\mathbf{p})[1-\hat{c}_{r}^{\dagger }(\mathbf{p})%
\hat{c}_{r}(\mathbf{p})]\hat{c}_{r}(\mathbf{p})=\hat{N}_{r}(\mathbf{p})-0
\notag
\end{eqnarray}
and so on.

As noted above, the vacuum state $|0\rangle $ is defined as
\begin{equation}
\hat{c}_{r}(\mathbf{p})|0\rangle =\hat{d}_{r}(\mathbf{p})|0\rangle =0
\end{equation}
which is equivalent to
\begin{equation}
\psi ^{+}(x)|0\rangle =\bar{\psi}^{+}(x)|0\rangle =0.
\end{equation}

\bigskip

\subsection{The Hamiltonian Operator}

\bigskip

It is now possible to rewrite the Hamiltonian (\ref{App2Ham}) in terms of
the annihilation and creation operators.

Substituting gives
\begin{eqnarray}
\hat{H} &=&\int d^{3}x\left\{ \int_{-\infty }^{\infty }\frac{d^{3}p}{(2\pi
\hbar )^{3/2}}\sum_{r}\left[ \hat{d}_{r}(\mathbf{p})\bar{v}_{r}(\mathbf{p}%
)e^{-ip\cdot x/\hbar }+\hat{c}_{r}^{\dagger }(\mathbf{p})\bar{u}_{r}(\mathbf{%
p})e^{+ip\cdot x/\hbar }\right] \right\}  \notag \\
&&\times \left( \frac{mc^{2}}{E_{p}}\right) ^{1/2}(mc^{2}-i\hbar c\gamma
^{j}\partial _{j})  \label{App2Ham1} \\
&&\left\{ \int_{-\infty }^{\infty }\frac{d^{3}p^{\prime }}{(2\pi \hbar
)^{3/2}}\sum_{s}\left( \frac{mc^{2}}{E_{p^{\prime }}}\right) ^{1/2}\left[
\hat{c}_{s}(\mathbf{p}^{\prime })u_{s}(\mathbf{p}^{\prime })e^{-ip^{\prime
}\cdot x/\hbar }+\hat{d}_{s}^{\dagger }(\mathbf{p}^{\prime })v_{s}(\mathbf{p}%
^{\prime })e^{+ip^{\prime }\cdot x/\hbar }\right] \right\}  \notag \\
&=&\int_{-\infty }^{\infty }\int_{-\infty }^{\infty }\frac{d^{3}p\text{ }%
d^{3}p^{\prime }}{(2\pi \hbar )^{3}}\sum_{r}\sum_{s}\int d^{3}x\left[ \hat{d}%
_{r}(\mathbf{p})\bar{v}_{r}(\mathbf{p})e^{-ip\cdot x/\hbar }+\hat{c}%
_{r}^{\dagger }(\mathbf{p})\bar{u}_{r}(\mathbf{p})e^{+ip\cdot x/\hbar }%
\right]  \notag \\
&&\times \left( \frac{mc^{2}}{\sqrt{E_{p}E_{p^{\prime }}}}\right)
(mc^{2}-i\hbar c\gamma ^{j}\partial _{j})\left[ \hat{c}_{s}(\mathbf{p}%
^{\prime })u_{s}(\mathbf{p}^{\prime })e^{-ip^{\prime }\cdot x/\hbar }+\hat{d}%
_{s}^{\dagger }(\mathbf{p}^{\prime })v_{s}(\mathbf{p}^{\prime
})e^{+ip^{\prime }\cdot x/\hbar }\right] .  \notag
\end{eqnarray}

By rearranging the Dirac equation as
\begin{equation}
i\hbar \gamma ^{i}\partial _{j}\psi (x)=mc\psi (x)-i\hbar \gamma
^{0}\partial _{0}\psi (x)
\end{equation}
the Laplacian derivatives $c(i\hbar \gamma ^{j}\partial _{j})$ may be
removed, and (\ref{App2Ham1}) may be rewritten as
\begin{eqnarray}
\hat{H} &=&\int_{-\infty }^{\infty }\int_{-\infty }^{\infty }\frac{d^{3}p%
\text{ }d^{3}p^{\prime }}{(2\pi \hbar )^{3}}\sum_{r}\sum_{s}\int
d^{3}x\left( \frac{mc^{2}}{\sqrt{E_{p}E_{p^{\prime }}}}\right) \left[ \hat{d}%
_{r}(\mathbf{p})\bar{v}_{r}(\mathbf{p})e^{-ip\cdot x/\hbar }+\hat{c}%
_{r}^{\dagger }(\mathbf{p})\bar{u}_{r}(\mathbf{p})e^{+ip\cdot x/\hbar }%
\right]  \notag \\
&&\times \left( mc^{2}-mc^{2}+i\hbar c\gamma ^{0}\frac{\partial }{\partial
x^{0}}\right) \left[ \hat{c}_{s}(\mathbf{p}^{\prime })u_{s}(\mathbf{p}%
^{\prime })e^{-ip^{\prime }\cdot x/\hbar }+\hat{d}_{s}^{\dagger }(\mathbf{p}%
^{\prime })v_{s}(\mathbf{p}^{\prime })e^{+ip^{\prime }\cdot x/\hbar }\right]
\notag \\
&=&\int_{-\infty }^{\infty }\int_{-\infty }^{\infty }\frac{d^{3}p\text{ }%
d^{3}p^{\prime }}{(2\pi \hbar )^{3}}\sum_{r}\sum_{s}\int d^{3}x\left( \frac{%
mc^{2}}{\sqrt{E_{p}E_{p^{\prime }}}}\right)  \notag \\
&&\times \left[ \hat{d}_{r}(\mathbf{p})\bar{v}_{r}(\mathbf{p})e^{-ip\cdot
x/\hbar }+\hat{c}_{r}^{\dagger }(\mathbf{p})\bar{u}_{r}(\mathbf{p}%
)e^{+ip\cdot x/\hbar }\right]  \notag \\
&&\times (i\hbar c\gamma ^{0})\left[ (-iE_{p^{\prime }}/\hbar c)\hat{c}_{s}(%
\mathbf{p}^{\prime })u_{s}(\mathbf{p}^{\prime })e^{-ip^{\prime }\cdot
x/\hbar }+(iE_{p^{\prime }}/\hbar c)\hat{d}_{s}^{\dagger }(\mathbf{p}%
^{\prime })v_{s}(\mathbf{p}^{\prime })e^{+ip^{\prime }\cdot x/\hbar }\right]
\notag \\
&=&\int_{-\infty }^{\infty }\int_{-\infty }^{\infty }\frac{d^{3}p\text{ }%
d^{3}p^{\prime }}{(2\pi \hbar )^{3}}\sum_{r}\sum_{s}\int d^{3}x\left( mc^{2}%
\sqrt{\frac{E_{p^{\prime }}}{E_{p}}}\right)  \notag \\
&&\times \left[ \hat{d}_{r}(\mathbf{p})\bar{v}_{r}(\mathbf{p})e^{-ip\cdot
x/\hbar }+\hat{c}_{r}^{\dagger }(\mathbf{p})\bar{u}_{r}(\mathbf{p}%
)e^{+ip\cdot x/\hbar }\right]  \notag \\
&&\times \left[ \gamma ^{0}\hat{c}_{s}(\mathbf{p}^{\prime })u_{s}(\mathbf{p}%
^{\prime })e^{-ip^{\prime }\cdot x/\hbar }-\gamma ^{0}\hat{d}_{s}^{\dagger }(%
\mathbf{p}^{\prime })v_{s}(\mathbf{p}^{\prime })e^{+ip^{\prime }\cdot
x/\hbar }\right] .  \label{App2HAM2}
\end{eqnarray}

Clearly, multiplying out the square brackets gives four terms, which in turn
become

\begin{enumerate}
\item
\begin{eqnarray}
\hat{d}_{r}(\mathbf{p})\bar{v}_{r}(\mathbf{p})e^{-ip\cdot x/\hbar }\gamma
^{0}\hat{c}_{s}(\mathbf{p}^{\prime })u_{s}(\mathbf{p}^{\prime
})e^{-ip^{\prime }\cdot x/\hbar } &=&e^{-i(p+p^{\prime })\cdot x/\hbar }\hat{%
d}_{r}(\mathbf{p})\hat{c}_{s}(\mathbf{p}^{\prime })\bar{v}_{r}(\mathbf{p}%
)\gamma ^{0}u_{s}(\mathbf{p}^{\prime })  \notag \\
&=&e^{-i(p+p^{\prime })\cdot x/\hbar }\hat{d}_{r}(\mathbf{p})\hat{c}_{s}(%
\mathbf{p}^{\prime })\times 0=0  \notag \\
&&
\end{eqnarray}

\item
\begin{eqnarray}
\hat{d}_{r}(\mathbf{p})\bar{v}_{r}(\mathbf{p})e^{-ip\cdot x/\hbar }(-\gamma
^{0}\hat{d}_{s}^{\dagger }(\mathbf{p}^{\prime })v_{s}(\mathbf{p}^{\prime
})e^{+ip^{\prime }\cdot x/\hbar }) &=&-e^{-i(p-p^{\prime })\cdot x/\hbar }%
\hat{d}_{r}(\mathbf{p})\hat{d}_{s}^{\dagger }(\mathbf{p}^{\prime })\bar{v}%
_{r}(\mathbf{p})\gamma ^{0}v_{s}(\mathbf{p}^{\prime })  \notag \\
&=&-e^{-i(p-p^{\prime })\cdot x/\hbar }\hat{d}_{r}(\mathbf{p})\hat{d}%
_{s}^{\dagger }(\mathbf{p}^{\prime })v_{r}^{\dagger }(\mathbf{p})v_{s}(%
\mathbf{p}^{\prime })  \notag \\
&&  \label{App2Two}
\end{eqnarray}

\item
\begin{eqnarray}
\hat{c}_{r}^{\dagger }(\mathbf{p})\bar{u}_{r}(\mathbf{p})e^{+ip\cdot x/\hbar
}\gamma ^{0}\hat{c}_{s}(\mathbf{p}^{\prime })u_{s}(\mathbf{p}^{\prime
})e^{-ip^{\prime }\cdot x/\hbar } &=&e_{r}^{+i(p-p^{\prime })\cdot x/\hbar }%
\hat{c}_{r}^{\dagger }(\mathbf{p})\hat{c}_{s}(\mathbf{p}^{\prime })\bar{u}%
_{r}(\mathbf{p})\gamma ^{0}u_{s}(\mathbf{p}^{\prime })  \notag \\
&=&e^{+i(p-p^{\prime })\cdot x/\hbar }\hat{c}_{r}^{\dagger }(\mathbf{p})\hat{%
c}_{s}(\mathbf{p}^{\prime })u_{r}^{\dagger }(\mathbf{p})u_{s}(\mathbf{p}%
^{\prime })  \notag \\
&&  \label{App2Three}
\end{eqnarray}

\item
\begin{eqnarray}
\hat{c}_{r}^{\dagger }(\mathbf{p})\bar{u}_{r}(\mathbf{p})e^{+ip\cdot x/\hbar
}(-\gamma ^{0}\hat{d}_{s}^{\dagger }(\mathbf{p}^{\prime })v_{s}(\mathbf{p}%
^{\prime })e^{+ip^{\prime }\cdot x/\hbar }) &=&-e^{+i(p+p^{\prime })\cdot
x/\hbar }\hat{c}_{r}^{\dagger }(\mathbf{p})\hat{d}_{s}^{\dagger }(\mathbf{p}%
^{\prime })\bar{u}_{r}(\mathbf{p})\gamma ^{0}v_{s}(\mathbf{p}^{\prime })
\notag \\
&=&-e^{+i(p+p^{\prime })\cdot x/\hbar }\hat{c}_{r}^{\dagger }(\mathbf{p})%
\hat{d}_{s}^{\dagger }(\mathbf{p}^{\prime })\times 0=0  \notag \\
&&
\end{eqnarray}
\end{enumerate}

%TCIMACRO{
%\TeXButton{noindent}{\noindent%
%} }%
%BeginExpansion
\noindent%
%
%EndExpansion
where the simplification $p^{\prime }=p$ has been made in $``1."$ and $``4."$
without affecting the outcome, in anticipation of the assumed orthogonality
of the terms in the Fourier expansion as given below and the appearance of
the Dirac delta $\delta _{\mathbf{pp}^{\prime }}$ in (\ref{App2Com}), and
using \cite{Gasiorowicz}
\begin{eqnarray}
u_{r}^{\dagger }(\mathbf{p})v_{s}(\mathbf{p}) &=&0 \\
v_{r}^{\dagger }(\mathbf{p})u_{s}(\mathbf{p}) &=&0.  \notag
\end{eqnarray}

Substituting, (\ref{App2HAM2}) evidently becomes
\begin{eqnarray}
\hat{H} &=&\int_{-\infty }^{\infty }\int_{-\infty }^{\infty }\frac{d^{3}p%
\text{ }d^{3}p^{\prime }}{(2\pi \hbar )^{3}}\sum_{r}\sum_{s}\int
d^{3}x\left( mc^{2}\sqrt{\frac{E_{p^{\prime }}}{E_{p}}}\right) \\
&&\times \left[
\begin{array}{c}
0-e^{-i(p-p^{\prime })\cdot x/\hbar }\hat{d}_{r}(\mathbf{p})\hat{d}%
_{s}^{\dagger }(\mathbf{p}^{\prime })v_{r}^{\dagger }(\mathbf{p})v_{s}(%
\mathbf{p}^{\prime }) \\
+e^{+i(p-p^{\prime })\cdot x/\hbar }\hat{c}_{r}^{\dagger }(\mathbf{p})\hat{c}%
_{s}(\mathbf{p}^{\prime })u_{r}^{\dagger }(\mathbf{p})u_{s}(\mathbf{p}%
^{\prime })+0
\end{array}
\right]  \notag
\end{eqnarray}
and taking the Fourier transform of the exponential over $d^{3}x$ to give a
Dirac delta function, it follows that
\begin{eqnarray}
\hat{H} &=&\int_{-\infty }^{\infty }\int_{-\infty }^{\infty }d^{3}p\text{ }%
d^{3}p^{\prime }\sum_{r}\sum_{s}\delta _{pp^{\prime }}\left( mc^{2}\sqrt{%
\frac{E_{p^{\prime }}}{E_{p}}}\right) \\
&&\times \left[ \hat{c}_{r}^{\dagger }(\mathbf{p})\hat{c}_{s}(\mathbf{p}%
^{\prime })u_{r}^{\dagger }(\mathbf{p})u_{s}(\mathbf{p}^{\prime })-\hat{d}%
_{r}(\mathbf{p})\hat{d}_{s}^{\dagger }(\mathbf{p}^{\prime })v_{r}^{\dagger }(%
\mathbf{p})v_{s}(\mathbf{p}^{\prime })\right] .  \notag
\end{eqnarray}

Furthermore, from the normalisation of the spinors defined in (\ref{App2Spin}%
) the above relationship becomes
\begin{eqnarray}
\hat{H} &=&\int_{-\infty }^{\infty }d^{3}p\text{ }\sum_{r}\sum_{s}\left(
mc^{2}\sqrt{\frac{E_{p}}{E_{p}}}\right) \left[ \hat{c}_{r}^{\dagger }(%
\mathbf{p})\hat{c}_{s}(\mathbf{p})\left( \frac{E_{p}}{mc^{2}}\right) \delta
_{rs}-\hat{d}_{r}(\mathbf{p})\hat{d}_{s}^{\dagger }(\mathbf{p})\left( \frac{%
E_{p}}{mc^{2}}\right) \delta _{rs}\right]  \notag \\
&=&\int_{-\infty }^{\infty }d^{3}p\text{ }\sum_{r}E_{p}\left[ \hat{c}%
_{r}^{\dagger }(\mathbf{p})\hat{c}_{s}(\mathbf{p})-\hat{d}_{r}(\mathbf{p})%
\hat{d}_{s}^{\dagger }(\mathbf{p})\right] .
\end{eqnarray}

This last equation can be rearranged, by making use of the anti-commutation
relations $\{\hat{d}_{r}(\mathbf{p}),\hat{d}_{r}^{\dagger }(\mathbf{p}%
^{\prime })\}=\delta _{rs}\delta _{\mathbf{pp}^{\prime }}$ of the ladder
operators, to give
\begin{equation}
\hat{H}=\int_{-\infty }^{\infty }d^{3}p\sum_{r}E_{p}\left[ \hat{c}%
_{r}^{\dagger }(\mathbf{p})\hat{c}_{r}(\mathbf{p})+\hat{d}_{r}^{\dagger }(%
\mathbf{p})\hat{d}_{r}(\mathbf{p})-1\right] .  \label{App2Ham3}
\end{equation}

Equation (\ref{App2Ham3}) evidently contains two terms involving operators
and a constant term of the form $\left( \int_{-\infty }^{\infty
}d^{3}p\sum_{r}E_{p}\right) .$ Moreover, because of the integral this
constant provides an infinite contribution to the Hamiltonian. To `overcome'
this problem the constant is, perhaps rather dubiously, ignored in
conventional quantum field theory by arguing that only relative differences
in energy are observable, such that the vacuum is consequently instilled
with a non-zero expectation value. Whilst it is not intended to fully
discuss the issue here, it is noted that this property leads to all sorts of
interpretational difficulties when considering general relativity.

The above procedure is implied by adopting the `Normal Order' convention,
denoted by $\mathcal{N},$ in which it is assumed that all anti-commutators
vanish, such that in a product every creation operator is placed to the left
of the absorption operators. Thus, the normal ordered Hamiltonian may be
rewritten as
\begin{eqnarray}
\hat{H} &=&\int_{-\infty }^{\infty }d^{3}p\sum_{r}E_{p}\mathcal{N}[\hat{c}%
_{r}^{\dagger }(\mathbf{p})\hat{c}_{r}(\mathbf{p})-\hat{d}_{r}(\mathbf{p})%
\hat{d}_{r}^{\dagger }(\mathbf{p})] \\
&=&\int_{-\infty }^{\infty }d^{3}p\sum_{r}E_{p}\left[ \hat{c}_{r}^{\dagger }(%
\mathbf{p})\hat{c}_{r}(\mathbf{p})+\hat{d}_{r}^{\dagger }(\mathbf{p})\hat{d}%
_{r}(\mathbf{p})\right] .  \notag
\end{eqnarray}

Finally, by making use of the number operators defined in (\ref{App2Num}),
the Hamiltonian becomes
\begin{equation}
\hat{H}=\int_{-\infty }^{\infty }d^{3}p\sum_{r}E_{p}[\hat{N}_{r}(\mathbf{p})+%
\overline{\hat{N}}_{r}(\mathbf{p})].
\end{equation}

\bigskip

\subsection{The Momentum Operator}

\bigskip

The momentum operator $\mathbf{P}$ of (\ref{App2Mom}) can also be
investigated. Substituting into this expression the plane wave expansions
for $\psi (x)$ and $\bar{\psi}(x)$ gives
\begin{eqnarray}
\mathbf{\hat{P}} &=&-\int d^{3}x[i\hbar \psi ^{\dagger }(x)\partial _{j}\psi
(x)]=-\int d^{3}x[i\hbar \bar{\psi}(x)\gamma ^{0}\partial _{j}\psi (x)] \\
&=&-\int d^{3}x\int_{-\infty }^{\infty }\frac{d^{3}p}{(2\pi \hbar )^{3/2}}%
\sum_{r}i\hbar \left[ \left( \frac{mc^{2}}{E_{p}}\right) ^{1/2}\left[ \hat{d}%
_{r}(\mathbf{p})\bar{v}_{r}(\mathbf{p})\gamma ^{0}e^{-ip\cdot x/\hbar }+\hat{%
c}_{r}^{\dagger }(\mathbf{p})\bar{u}_{r}(\mathbf{p})\gamma ^{0}e^{+ip\cdot
x/\hbar }\right] \right]  \notag \\
&&\times \partial _{j}\left[ \int_{-\infty }^{\infty }\frac{d^{3}p^{\prime }%
}{(2\pi \hbar )^{3/2}}\sum_{s}\left( \frac{mc^{2}}{E_{p^{\prime }}}\right)
^{1/2}\left[ \hat{c}_{s}(\mathbf{p}^{\prime })u_{s}(\mathbf{p}^{\prime
})e^{-ip^{\prime }\cdot x/\hbar }+\hat{d}_{s}^{\dagger }(\mathbf{p}^{\prime
})v_{s}(\mathbf{p}^{\prime })e^{+ip^{\prime }\cdot x/\hbar }\right] \right] .
\notag
\end{eqnarray}

So,
\begin{eqnarray}
\mathbf{\hat{P}} &=&-\int_{-\infty }^{\infty }\int_{-\infty }^{\infty }\frac{%
d^{3}p\text{ }d^{3}p^{\prime }}{(2\pi \hbar )^{3}}\sum_{r}\sum_{s}\int
d^{3}x(i\hbar )\left( \frac{mc^{2}}{\sqrt{E_{p}E_{p^{\prime }}}}\right) \\
&&\times \left[ \hat{d}_{r}(\mathbf{p})v_{r}^{\dagger }(\mathbf{p})\gamma
^{0}\gamma ^{0}e^{-ip\cdot x/\hbar }+\hat{c}_{r}^{\dagger }(\mathbf{p}%
)u_{r}^{\dagger }(\mathbf{p})\gamma ^{0}\gamma ^{0}e^{+ip\cdot x/\hbar }%
\right]  \notag \\
&&\times \left[ (i\mathbf{p}^{\prime }/\hbar )\hat{c}_{s}(\mathbf{p}^{\prime
})u_{s}(\mathbf{p}^{\prime })e^{-ip^{\prime }\cdot x/\hbar }+(-i\mathbf{p}%
^{\prime }/\hbar )\hat{d}_{s}^{\dagger }(\mathbf{p}^{\prime })v_{s}(\mathbf{p%
}^{\prime })e^{+ip^{\prime }\cdot x/\hbar }\right]  \notag \\
&=&\int_{-\infty }^{\infty }\int_{-\infty }^{\infty }\frac{d^{3}p\text{ }%
d^{3}p^{\prime }}{(2\pi \hbar )^{3}}\sum_{r}\sum_{s}\int d^{3}x\left( \frac{%
mc^{2}}{\sqrt{E_{p}E_{p^{\prime }}}}\right) \mathbf{p}^{\prime }  \notag \\
&&\times \left[ \hat{d}_{r}(\mathbf{p})v_{r}^{\dagger }(\mathbf{p}%
)e^{-ip\cdot x/\hbar }+\hat{c}_{r}^{\dagger }(\mathbf{p})u_{r}^{\dagger }(%
\mathbf{p})e^{+ip\cdot x/\hbar }\right]  \notag \\
&&\times \left[ \hat{c}_{s}(\mathbf{p}^{\prime })u_{s}(\mathbf{p}^{\prime
})e^{-ip^{\prime }\cdot x/\hbar }-\hat{d}_{s}^{\dagger }(\mathbf{p}^{\prime
})v_{s}(\mathbf{p}^{\prime })e^{+ip^{\prime }\cdot x/\hbar }\right] .  \notag
\end{eqnarray}

As before, the product may be expanded into four terms:

\begin{enumerate}
\item
\begin{eqnarray}
\hat{d}_{r}(\mathbf{p})v_{r}^{\dagger }(\mathbf{p})e^{-ip\cdot x/\hbar }\hat{%
c}_{s}(\mathbf{p}^{\prime })u_{s}(\mathbf{p}^{\prime })e^{-ip^{\prime }\cdot
x/\hbar } &=&e^{-i(p+p^{\prime })\cdot x/\hbar }\hat{d}_{r}(\mathbf{p})\hat{c%
}_{s}(\mathbf{p}^{\prime })v_{r}^{\dagger }(\mathbf{p})u_{s}(\mathbf{p}%
^{\prime })  \notag \\
&=&e^{-i(p+p^{\prime })\cdot x/\hbar }\hat{d}_{r}(\mathbf{p})\hat{c}_{s}(%
\mathbf{p}^{\prime })\times 0=0  \notag \\
&&
\end{eqnarray}

\item
\begin{equation}
\hat{d}_{r}(\mathbf{p})v_{r}^{\dagger }(\mathbf{p})e^{-ip\cdot x/\hbar
}(-d_{s}^{\dagger }(\mathbf{p}^{\prime })v_{s}(\mathbf{p}^{\prime
})e^{+ip^{\prime }\cdot x/\hbar })=-e^{-i(p-p^{\prime })\cdot x/\hbar }\hat{d%
}_{r}(\mathbf{p})\hat{d}_{s}^{\dagger }(\mathbf{p}^{\prime })v_{r}^{\dagger
}(\mathbf{p})v_{s}(\mathbf{p}^{\prime })
\end{equation}
\qquad

\item
\begin{equation}
\hat{c}_{r}^{\dagger }(\mathbf{p})u_{r}^{\dagger }(\mathbf{p})e^{+ip\cdot
x/\hbar }\hat{c}_{s}(\mathbf{p}^{\prime })u_{s}(\mathbf{p}^{\prime
})e^{-ip^{\prime }\cdot x/\hbar }=e^{+i(p-p^{\prime })\cdot x/\hbar }\hat{c}%
_{r}^{\dagger }(\mathbf{p})\hat{c}_{s}(\mathbf{p}^{\prime })u_{r}^{\dagger }(%
\mathbf{p})u_{s}(\mathbf{p}^{\prime })
\end{equation}

\item
\begin{eqnarray}
\hat{c}_{r}^{\dagger }(\mathbf{p})u_{r}^{\dagger }(\mathbf{p})e^{+ip\cdot
x/\hbar }(-\hat{d}_{s}^{\dagger }(\mathbf{p}^{\prime })v_{s}(\mathbf{p}%
^{\prime })e^{+ip^{\prime }\cdot x/\hbar }) &=&-e^{+i(p+p^{\prime })\cdot
x/\hbar }\hat{c}_{r}^{\dagger }(\mathbf{p})\hat{d}_{s}^{\dagger }(\mathbf{p}%
^{\prime })u_{r}^{\dagger }(\mathbf{p})v_{s}(\mathbf{p}^{\prime })  \notag \\
&=&-e^{+i(p+p^{\prime })\cdot x/\hbar }\hat{c}_{r}^{\dagger }(\mathbf{p})%
\hat{d}_{s}^{\dagger }(\mathbf{p}^{\prime })\times 0=0  \notag \\
&&
\end{eqnarray}
\end{enumerate}

So, $\mathbf{P}$ becomes
\begin{eqnarray}
\mathbf{\hat{P}} &=&\int_{-\infty }^{\infty }\int_{-\infty }^{\infty }\frac{%
d^{3}p\text{ }d^{3}p^{\prime }}{(2\pi \hbar )^{3}}\sum_{r}\sum_{s}\int
d^{3}x\left( \frac{mc^{2}}{\sqrt{E_{p}E_{p^{\prime }}}}\right) \mathbf{p}%
^{\prime } \\
&&\times \left[
\begin{array}{c}
0-e^{-i(p-p^{\prime })\cdot x/\hbar }\hat{d}_{r}(\mathbf{p})\hat{d}%
_{s}^{\dagger }(\mathbf{p}^{\prime })v_{r}^{\dagger }(\mathbf{p})v_{s}(%
\mathbf{p}^{\prime }) \\
+e^{+i(p-p^{\prime })\cdot x/\hbar }\hat{c}_{r}^{\dagger }(\mathbf{p})\hat{c}%
_{s}(\mathbf{p}^{\prime })u_{r}^{\dagger }(\mathbf{p})u_{s}(\mathbf{p}%
^{\prime })+0
\end{array}
\right]  \notag
\end{eqnarray}
which may be Fourier transformed to give
\begin{eqnarray}
\mathbf{\hat{P}} &=&\int_{-\infty }^{\infty }\int_{-\infty }^{\infty }d^{3}p%
\text{ }d^{3}p^{\prime }\sum_{r}\sum_{s}\left( \frac{mc^{2}}{\sqrt{%
E_{p}E_{p^{\prime }}}}\right) \mathbf{p}^{\prime }\delta _{pp^{\prime }} \\
&&\times \left[ \hat{c}_{r}^{\dagger }(\mathbf{p})\hat{c}_{s}(\mathbf{p}%
^{\prime })u_{r}^{\dagger }(\mathbf{p})u_{s}(\mathbf{p}^{\prime })-\hat{d}%
_{r}(\mathbf{p})\hat{d}_{s}^{\dagger }(\mathbf{p}^{\prime })v_{r}^{\dagger }(%
\mathbf{p})v_{s}(\mathbf{p}^{\prime })\right] .  \notag
\end{eqnarray}

Using again the normalisation of the spinors
\begin{equation}
\mathbf{\hat{P}}=\int_{-\infty }^{\infty }d^{3}p\text{ }\sum_{r}\sum_{s}%
\left( \frac{mc^{2}}{E_{p}}\right) \mathbf{p}\left[ \hat{c}_{r}^{\dagger }(%
\mathbf{p})\hat{c}_{s}(\mathbf{p})\left( \frac{E_{p}}{mc^{2}}\right) -\hat{d}%
_{r}(\mathbf{p})\hat{d}_{s}^{\dagger }(\mathbf{p})\left( \frac{E_{p}}{mc^{2}}%
\right) \right] \delta _{rs}  \notag
\end{equation}
and with the anti-commutation relations of the ladder operators, this may be
written in the normal order convention as
\begin{eqnarray}
\mathbf{\hat{P}} &=&\int_{-\infty }^{\infty }d^{3}p\text{ }\sum_{r}\mathbf{p}%
\mathcal{N}[\hat{c}_{r}^{\dagger }(\mathbf{p})\hat{c}_{r}(\mathbf{p})-\hat{d}%
_{r}(\mathbf{p})\hat{d}_{r}^{\dagger }(\mathbf{p})] \\
&=&\int_{-\infty }^{\infty }d^{3}p\sum_{r}\mathbf{p}[\hat{c}_{r}^{\dagger }(%
\mathbf{p})\hat{c}_{r}(\mathbf{p})+\hat{d}_{r}^{\dagger }(\mathbf{p})\hat{d}%
_{r}(\mathbf{p})].  \notag
\end{eqnarray}

Lastly, expressing this as a sum of number operators gives
\begin{equation}
\mathbf{\hat{P}}=\int_{-\infty }^{\infty }d^{3}p\sum_{r}\mathbf{p}\left[
\hat{N}_{r}(\mathbf{p})+\overline{\hat{N}}_{r}(\mathbf{p})\right] .
\end{equation}

\bigskip

\subsection{The Charge Operator}

\bigskip

Finally, the charge operator $\hat{Q}$ may also be evaluated. From equation (%
\ref{App2Charge}), $\hat{Q}$ becomes
\begin{eqnarray}
\hat{Q} &=&q\int \psi ^{\dagger }(x)\psi (x)\text{ }d^{3}x \\
&=&q\int_{-\infty }^{\infty }\int_{-\infty }^{\infty }\frac{d^{3}p\text{ }%
d^{3}p^{\prime }}{(2\pi \hbar )^{3}}\sum_{r=1}^{2}\sum_{s=1}^{2}\int
d^{3}x\left( \frac{mc^{2}}{\sqrt{E_{p}E_{p^{\prime }}}}\right)  \notag \\
&&\times \left[ d_{r}(\mathbf{p})\bar{v}_{r}(\mathbf{p})e^{-ip\cdot x/\hbar
}+c_{r}^{\dagger }(\mathbf{p})\bar{u}_{r}(\mathbf{p})e^{+ip\cdot x/\hbar }%
\right]  \notag \\
&&\times \gamma ^{0}\left[ c_{s}(\mathbf{p}^{\prime })u_{s}(\mathbf{p}%
^{\prime })e^{-ip^{\prime }\cdot x/\hbar }+d_{s}^{\dagger }(\mathbf{p}%
^{\prime })v_{s}(\mathbf{p}^{\prime })e^{+ip^{\prime }\cdot x/\hbar }\right]
\notag
\end{eqnarray}
which again gives four terms upon multiplication:

\begin{enumerate}
\item
\begin{eqnarray}
\hat{d}_{r}(\mathbf{p})\bar{v}_{r}(\mathbf{p})e^{-ip\cdot x/\hbar }\gamma
^{0}\hat{c}_{s}(\mathbf{p}^{\prime })u_{s}(\mathbf{p}^{\prime
})e^{-ip^{\prime }\cdot x/\hbar } &=&e^{-i(p+p^{\prime })\cdot x/\hbar }\hat{%
d}_{r}(\mathbf{p})\hat{c}_{s}(\mathbf{p}^{\prime })v_{r}^{\dagger }(\mathbf{p%
})u_{s}(\mathbf{p}^{\prime })  \notag \\
&=&e^{-i(p+p^{\prime })\cdot x/\hbar }\hat{d}_{r}(\mathbf{p})\hat{c}_{s}(%
\mathbf{p}^{\prime })\times 0=0  \notag \\
&&
\end{eqnarray}

\item
\begin{equation}
\hat{d}_{r}(\mathbf{p})\bar{v}_{r}(\mathbf{p})e^{-ip\cdot x/\hbar }\gamma
^{0}d_{s}^{\dagger }(\mathbf{p}^{\prime })v_{s}(\mathbf{p}^{\prime
})e^{+ip^{\prime }\cdot x/\hbar }=e^{-i(p-p^{\prime })\cdot x/\hbar }\hat{d}%
_{r}(\mathbf{p})\hat{d}_{s}^{\dagger }(\mathbf{p}^{\prime })v_{r}^{\dagger }(%
\mathbf{p})v_{s}(\mathbf{p}^{\prime })
\end{equation}

\item
\begin{equation}
\hat{c}_{r}^{\dagger }(\mathbf{p})\bar{u}_{r}(\mathbf{p})e^{+ip\cdot x/\hbar
}\gamma ^{0}\hat{c}_{s}(\mathbf{p}^{\prime })u_{s}(\mathbf{p}^{\prime
})e^{-ip^{\prime }\cdot x/\hbar }=e^{+i(p-p^{\prime })\cdot x/\hbar }\hat{c}%
_{r}^{\dagger }(\mathbf{p})\hat{c}_{s}(\mathbf{p}^{\prime })u_{r}^{\dagger }(%
\mathbf{p})u_{s}(\mathbf{p}^{\prime })
\end{equation}

\item
\begin{eqnarray}
\hat{c}_{r}^{\dagger }(\mathbf{p})\bar{u}_{r}(\mathbf{p})e^{+ip\cdot x/\hbar
}\gamma ^{0}\hat{d}_{s}^{\dagger }(\mathbf{p}^{\prime })v_{s}(\mathbf{p}%
^{\prime })e^{+ip^{\prime }\cdot x/\hbar } &=&e^{+i(p+p^{\prime })\cdot
x/\hbar }\hat{c}_{r}^{\dagger }(\mathbf{p})\hat{d}_{s}^{\dagger }(\mathbf{p}%
^{\prime })u_{r}^{\dagger }(\mathbf{p})v_{s}(\mathbf{p}^{\prime })  \notag \\
&=&e^{+i(p+p^{\prime })\cdot x/\hbar }\hat{c}_{r}^{\dagger }(\mathbf{p})\hat{%
d}_{s}^{\dagger }(\mathbf{p}^{\prime })\times 0=0  \notag \\
&&
\end{eqnarray}
\end{enumerate}

So, $\hat{Q}$ becomes
\begin{eqnarray}
\hat{Q} &=&q\int_{-\infty }^{\infty }\int_{-\infty }^{\infty }\frac{d^{3}p%
\text{ }d^{3}p^{\prime }}{(2\pi \hbar )^{3}}\sum_{r=1}^{2}\sum_{s=1}^{2}\int
d^{3}x\left( \frac{mc^{2}}{\sqrt{E_{p}E_{p^{\prime }}}}\right) \\
&&\times \left[
\begin{array}{c}
0+e^{-i(p-p^{\prime })\cdot x/\hbar }\hat{d}_{r}(\mathbf{p})\hat{d}%
_{s}^{\dagger }(\mathbf{p}^{\prime })v_{r}^{\dagger }(\mathbf{p})v_{s}(%
\mathbf{p}^{\prime }) \\
+e^{+i(p-p^{\prime })\cdot x/\hbar }\hat{c}_{r}^{\dagger }(\mathbf{p})\hat{c}%
_{s}(\mathbf{p}^{\prime })u_{r}^{\dagger }(\mathbf{p})u_{s}(\mathbf{p}%
^{\prime })+0
\end{array}
\right]  \notag
\end{eqnarray}
which is Fourier transformed as
\begin{eqnarray}
\hat{Q} &=&q\int_{-\infty }^{\infty }\int_{-\infty }^{\infty }d^{3}p\text{ }%
d^{3}p^{\prime }\sum_{r=1}^{2}\sum_{s=1}^{2}\left( \frac{mc^{2}}{\sqrt{%
E_{p}E_{p^{\prime }}}}\right) \delta _{pp^{\prime }} \\
&&\times \left[ \hat{d}_{r}(\mathbf{p})\hat{d}_{s}^{\dagger }(\mathbf{p}%
^{\prime })v_{r}^{\dagger }(\mathbf{p})v_{s}(\mathbf{p}^{\prime })+\hat{c}%
_{r}^{\dagger }(\mathbf{p})\hat{c}_{s}(\mathbf{p}^{\prime })u_{r}^{\dagger }(%
\mathbf{p})u_{s}(\mathbf{p}^{\prime })\right]  \notag
\end{eqnarray}
and, upon substituting the spinor normalisation, gives
\begin{equation}
\hat{Q}=q\int_{-\infty }^{\infty }d^{3}p\sum_{r=1}^{2}\sum_{s=1}^{2}\left(
\frac{mc^{2}}{E_{p}}\right) \left[ \hat{d}_{r}(\mathbf{p})\hat{d}%
_{s}^{\dagger }(\mathbf{p})\left( \frac{E_{p}}{mc^{2}}\right) +\hat{c}%
_{r}^{\dagger }(\mathbf{p})\hat{c}_{s}(\mathbf{p}^{\prime })\left( \frac{%
E_{p}}{mc^{2}}\right) \right] \delta _{rs}.
\end{equation}

Thus, applying the anti-commutation relations of the ladder operators, $\hat{%
Q}$ may be written in normal order as
\begin{eqnarray}
\hat{Q} &=&q\int_{-\infty }^{\infty }d^{3}p\sum_{r=1}^{2}\mathcal{N}\left[
\hat{d}_{r}(\mathbf{p})\hat{d}_{r}^{\dagger }(\mathbf{p})+\hat{c}%
_{r}^{\dagger }(\mathbf{p})\hat{c}_{r}(\mathbf{p}^{\prime })\right] \\
&=&q\int_{-\infty }^{\infty }d^{3}p\sum_{r=1}^{2}\left[ -\hat{d}%
_{r}^{\dagger }(\mathbf{p})\hat{d}_{r}(\mathbf{p})+\hat{c}_{r}^{\dagger }(%
\mathbf{p})\hat{c}_{r}(\mathbf{p}^{\prime })\right] .  \notag
\end{eqnarray}

So, charge is defined as
\begin{equation}
\hat{Q}=q\int_{-\infty }^{\infty }d^{3}p\sum_{r=1}^{2}\left[ \hat{N}_{r}(%
\mathbf{p})-\overline{\hat{N}}_{r}(\mathbf{p})\right] .
\end{equation}

Clearly, the commutation relation
\begin{equation}
\lbrack \hat{Q},\hat{H}]=0
\end{equation}
vanishes due to the relations (\ref{App2Anticom}), such that $\frac{\partial
\hat{Q}}{\partial t}=0$ as expected from Heisenberg's equation.

\bigskip

\newpage

\bigskip
%TCIMACRO{
%\TeXButton{TeX field}{\clearpage
%\addcontentsline{toc}{section}{References}%
%}}%
%BeginExpansion
\clearpage
\addcontentsline{toc}{section}{References}%
%
%EndExpansion

\end{document}